%% file: thesis.tex
\documentclass[11pt]{ucthesis}
\include{thesispreamble}

\begin{document}

% Declarations for Front Matter
% Declarations for Front Matter

\title{First Evidence For Atmospheric Neutrino-Induced Cascades with the IceCube Detector}
\author{Michelangelo Vincent D'Agostino}
\degreeyear{2009}
\degreesemester{Fall}
\degree{Doctor of Philosophy}
\chair{Professor P. Buford Price}
\othermembers{Professor Kam-Biu Luk\\
Professor Chung-Pei Ma}
\numberofmembers{3}
\prevdegrees{A.B. (Harvard College) 2002\\
M.A. (University of California, Berkeley) 2005}
\field{Physics}
\campus{Berkeley}

\maketitle
%\approvalpage
\copyrightpage

\begin{abstract}
IceCube is an all-flavor, cubic kilometer neutrino telescope currently under construction in the deep glacial ice at the South Pole.  Its embedded optical sensors detect Cherenkov light from charged particles produced in neutrino interactions in the ice.  For several years IceCube has been detecting muon tracks from charged-current muon neutrino interactions. However, IceCube has yet to observe the electromagnetic or hadronic particle showers or ``cascades'' initiated by charged-current or neutral-current neutrino interactions.  The first detection of such an event signature is expected to come from the known flux of atmospheric electron and muon neutrinos. 

A search for atmospheric neutrino-induced cascades was performed using 275.46 days of data from IceCube's 22-string configuration.  Reconstruction and background rejection techniques were developed to reach, for the first time, a signal-to-background ratio $\sim$1.  Above a reconstructed energy of 5~TeV, 12 candidate events were observed in the full dataset.  The signal expectation from the canonical Bartol atmospheric neutrino flux model is $5.63\pm2.25$ events, while the expectation from the atmospheric neutrino flux as measured by IceCube's predecessor array AMANDA is $7.48\pm1.50$ events.  Quoted errors include the uncertainty on the flux only.  

While a conclusive detection can not yet be claimed because of a lack of background Monte Carlo statistics, the evidence that we are at the level of background suppression needed to see atmospheric neutrino-induced cascades is strong.  In addition, one extremely interesting candidate event of energy 133~TeV survives all cuts and shows an intriguing double pulse structure in its waveforms that may signal the ``double bang'' of a tau neutrino interaction.

\abstractsignature

\end{abstract}

\begin{frontmatter}

\begin{dedication}
\null\vfil
{\large
\begin{center}
To Willoughby, plain and simple.  I never would have gotten through it without you.\end{center}}
\vfil\null
\end{dedication}

\tableofcontents
\listoffigures
\listoftables
\begin{acknowledgements}
Over the past five years, I've benefitted from the hard work, wise advice, and constant encouragement of many, many people.  First, I'd like to thank the Berkeley post-docs---Ignacio Taboada, Kurt Woschnagg, Kirill Filimonov, and Ryan Bay.  From tips on how to coax ROOT to do exactly what you need it to do, to ideas on analysis and cut variable selection, to the sometimes harsh words that forced me to stand on my own, I owe them a great debt.  Thank you.

To my office-mate Justin Vandenbroucke---I'll never enjoy bickering with anyone as much as I've enjoyed bickering with you.  Thanks for being there to hold my hand on the flight to Antarctica.

Thanks to all of my IceCube collaborators for building such a wonderful detector.  It's been a pleasure working with such interesting people from around the world.  While we've butted heads on occasion and argued our points forcefully, I think we're all the better for it.  I'm certainly a better scientist because of my interactions with you all.  I wish you thousands and thousands of extraterrestrial neutrinos.

To my parents Dolly and Vince D'Agostino and to my friends---thank you for supporting me and for keeping me sane on this long and winding road.  It hasn't always been easy, and it hasn't always been easy to put up with me.

And last but not least, thanks to my advisor, Buford Price.  Thank you for giving me the freedom to explore my interests, both inside and outside of the lab.  Thank you for giving me the space to make my own choices (the good ones and the bad ones) and for teaching me how to be an independent researcher with a broad range of interests.

\end{acknowledgements}

\end{frontmatter}

\include{chapters/chapter1}

\include{chapters/chapter2}

\include{chapters/chapter3}

\include{chapters/chapter4}

\include{chapters/chapter5}

\include{chapters/chapter6}

\include{chapters/chapter7}
\include{chapters/conclusion}

\bibliographystyle{unsrt_withcaps1}
\bibliography{allrefs_mvd}

\appendix
\include{appendices/appendix1}

\include{appendices/appendix2}

\include{appendices/appendix3}
\include{appendices/appendix4}

%Ancillary material should be put in appendices, which appear after the

\end{document}

%% file: thesispreamble.tex
\usepackage{graphicx}% Include figure files
\usepackage{dcolumn}% Align table columns on decimal point
\usepackage{bm, amsmath, amssymb}% bold math
\usepackage{cancel}%for strikethrough symbols for equations
\usepackage{mathrsfs}
\usepackage[MAKE_WEB_PDF]{optional} %this package allows optional stuff to be turned on and off
\opt{MAKE_WEB_PDF}{
\usepackage[final,                  % override "draft" which means "do nothing"
       		     colorlinks,         % rather than outlining them in boxes
		     linktocpage,      % link just the pages, not the full titles
            	     linkcolor=blue, % override truly awful colour choices
                       citecolor=blue,  %   (ditto)
                       urlcolor=blue,    %   (ditto)
		    breaklinks=true, % needed to allow figures list to break onto multiple lines
                       ]{hyperref}
\usepackage[all]{hypcap} %needed to get figure links right
}
\usepackage{float}
\usepackage{color}
\usepackage{lscape} %to get single landscape pages
\usepackage{url}
\usepackage{breakurl} %to break urls in bibliography

\newcommand{\nue}{\nu_e}
\newcommand{\nuebar}{\bar{\nu}_e}
\newcommand{\numu}{\nu_\mu}
\newcommand{\numubar}{\bar{\nu}_\mu}
\newcommand{\nutau}{\nu_\tau}

\newcommand{\conetwo}{c_{12}}
\newcommand{\ctwothree}{c_{23}}
\newcommand{\conethree}{c_{13}}
\newcommand{\sonetwo}{s_{12}}
\newcommand{\stwothree}{s_{23}}
\newcommand{\sonethree}{s_{13}}
\newcommand{\conetwosq}{c^2_{12}}
\newcommand{\ctwothreesq}{c^2_{23}}

\newcommand{\sonetwosq}{s^2_{12}}
\newcommand{\stwothreesq}{s^2_{23}}

\newcommand{\boxedeqn}[1]{%
  \[\fbox{%
      \addtolength{\linewidth}{-28\fboxsep}%
      \addtolength{\linewidth}{-28\fboxrule}%
      \begin{minipage}{\linewidth}%
      \begin{equation}#1\end{equation}%
      \end{minipage}%
    }\]%
}

\newcommand{\comment}[1]{}

%% file: chapters/chapter1.tex
\chapter{Neutrinos in Physics and Astrophysics}\label{chapter:chapter1}

\section{Introduction: Why Neutrinos?}

For the last half-century, scientists have been traveling to remote, often harsh locations to capture neutrinos.  They've spent countless days in deep underground mines \cite{SuperK, KamLAND, SNO, MINOS} and caverns situated next to nuclear reactor cores \cite{ReinesCowan1, ReinesCowan2, CHOOZ}.  They've ventured to frozen Siberian lakes \cite{Baikal} and to the Amundsen-Scott South Pole Station \cite{AMANDA, IceCubeAlbrecht}.  They've even endured the inhospitable climate of the French Riviera \cite{ANTARES}.  So what's all the fuss about?

Over that same time period, the neutrino has moved from a minor player, invented to save energy conservation but never expected to be detected, to a central research topic in particle physics and a hopeful, emerging research topic in astrophysics.  The discovery of neutrino oscillations has opened the door onto a new and as-yet unknown mechanism beyond the Standard Model that must be responsible for giving neutrinos their tiny masses.  Next-generation oscillation experiments hope to probe CP violation in the neutrino sector, which many think could explain the matter-antimatter asymmetry of the universe.  In the realm of astrophysics, neutrinos have confirmed our understanding of solar physics and given us a small glimpse into supernova explosions.  The scientists who are building large detectors to detect supernova neutrinos and high energy astrophysical neutrinos are hoping to shed light on some of the processes at work inside stellar explosions, supermassive black holes, gamma ray bursts, and other violent astrophysical objects.  

This dissertation is concerned with a small piece of this exciting tapestry of neutrino physics and astrophysics.  The IceCube detector is a next-generation, all-flavor neutrino telescope currently under construction in the deep glacial ice at the South Pole.  IceCube's primary goal is to observe astrophysical neutrinos and, from them, to discover the sources of the highest energy cosmic rays.  In order to achieve these goals, IceCube must prove that it can observe the various interaction signatures of different neutrino flavors using the only calibration source of high energy neutrinos that we have---atmospheric neutrinos.  In particular, this dissertation focuses on a search for neutrino-induced particle showers or ``cascades'' from atmospheric neutrinos.

\section{Atmospheric Neutrinos}
\label{SAtmospheric}

\begin{figure}
\centering
\includegraphics[width=0.8\textwidth]{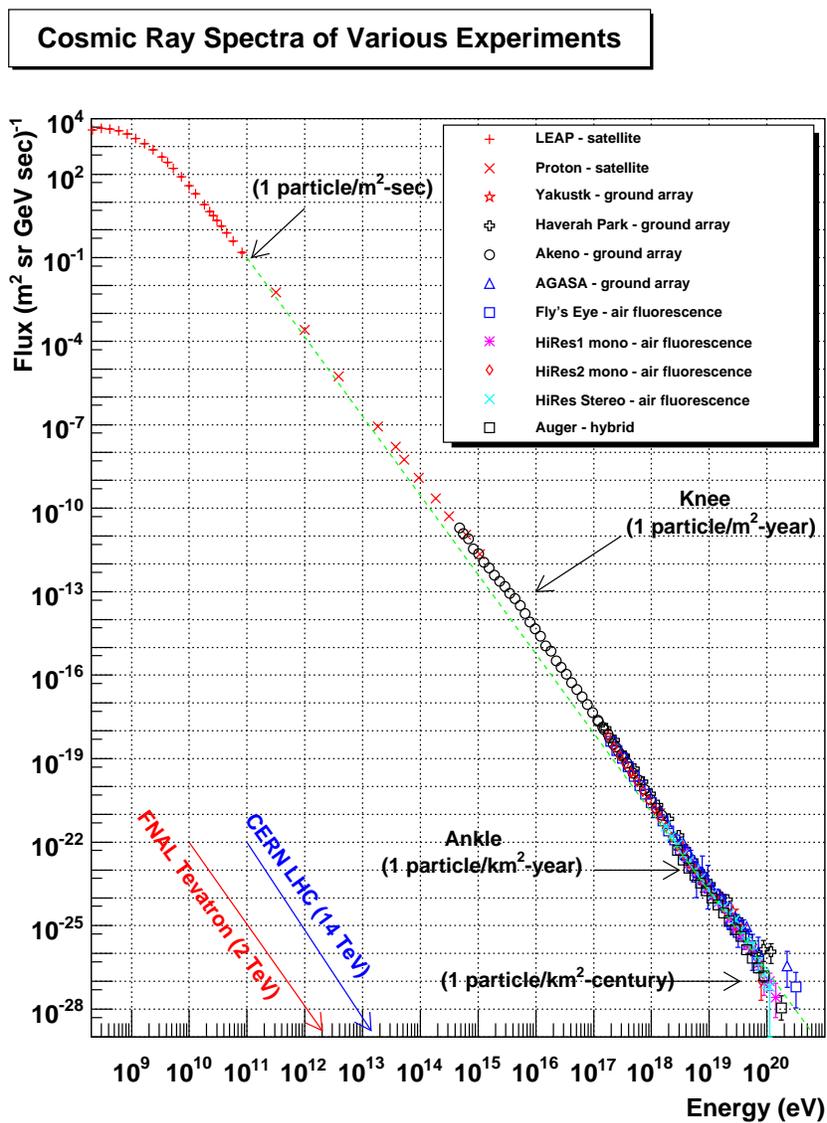}
\vspace{0.95cm}
\caption{Measured cosmic ray energy spectrum.  From \cite{HanlonCRSpectrum}.}
\label{CRSpectrum}
\end{figure}

The earth is constantly bombarded by a stream of high energy particles from space---the so-called cosmic rays.  While the composition of cosmic-rays is energy dependent, roughly $75\%$ are protons, $15\%$ are helium, and $10\%$ are heavier nuclei like carbon, oxygen, and nitrogen up through iron \cite{GaisserBook}.  Cosmic rays reach earth with a tremendous amount of energy.  The spectrum is a power law, $\frac{dN}{dE} \sim E^{-2.7}$ up to energies around $10^3$ TeV (the ``knee''), where the spectrum steepens to $\frac{dN}{dE} \sim E^{-3}$.  At an energy around $10^6$ TeV (the ``ankle") the spectrum hardens again to $\frac{dN}{dE} \sim E^{-2.7}$ before suppression begins around $4 \times 10^7$ TeV \cite{AugerGZK}.  This suppression is due to interactions of the highest energy cosmic rays with the photons of the cosmic microwave background, the so-called GZK effect \cite{GZKG, GZKZK}.  The ankle is thought to signal a transition from galactic to extragalactic sources.  Figure~\ref{CRSpectrum} shows the measured cosmic ray energy spectrum.

Interactions of these primary cosmic ray nuclei with the atmosphere produce unstable mesons which can further interact or decay to secondary muons, muon neutrinos, and electron neutrinos.  Since this dissertation is concerned with a search for neutrino-induced cascades from atmospheric neutrinos, we'll spend some time discussing a few salient features of the atmospheric neutrino spectrum.   

The atmospheric neutrino spectrum is made up of contributions from the decay of muons, pions, and kaons, and its makeup is determined by the competition between interaction and decay.  To quantify this, we follow the atmospheric model of \cite{GaisserBook}.  The pressure is proportional to vertical slant depth $X_v$, where $X_v$ is measured from the top of the atmosphere down in units of $\mbox{g/cm}^2$.   The density is $\rho = -dX_v/dh$, where $h$ is the height in m measured from the ground up.  We can write

$$
\frac{p}{\rho} = \frac{X_v}{-dX_v/dh} \propto T
$$

\noindent For an isothermal atmosphere,

$$
X_v = X_0 e^{-h/h_0} 
$$

\noindent where $X_0 \approx 1030 \mbox{ g/cm}^2$ is the total atmospheric vertical slant depth.  Both the temperature and the scale height actually decrease with altitude, but at sea level $h_0 \approx 8.4$ km.  This model for the atmosphere implies that

$$
\rho = -dX_v/dh = X_v/h_0 \approx \frac{X \cos \theta}{h_0}
$$

\noindent where $X$ is the slant depth along the particle's trajectory measured from the top of the atmosphere down and $\theta$ is the particle's zenith angle.  We want to find the particle decay length $d$ in $\mbox{g/cm}^2$.  This is given by 

\begin{align*}
d = \gamma c \tau \rho = \frac{E}{mc^2} c \tau \frac{X \cos \theta}{h_0} = \frac{EX\cos \theta}{mc^2h_0/c\tau} \equiv \frac{EX\cos \theta}{\epsilon}
\end{align*}

\noindent where the critical energy $\epsilon \equiv \frac{mc^2h_0}{c\tau} = \frac{Eh_0}{\gamma c \tau}$.   The critical energy is the energy at which the particle decay length is equal to the scale height of the atmosphere.  The critical energies for particles important for atmospheric neutrinos are given in table~\ref{CriticalEnergies}.   

At low energies such that $E \ll \epsilon/\cos \theta$, the decay length becomes very short.  Particles will decay before they have a chance to interact.  The neutrinos from these decays will have an energy spectrum that follows the mesons and primary cosmic rays.  At energies $E \gg \epsilon/\cos \theta$, the decay length becomes very long.  Particles are likely to interact before they can decay.  The neutrinos will have an energy spectrum that is one power steeper than that of the mesons and primary cosmic rays because of the decay length dependence on $E$.  The $1/\cos \theta$ dependence also gives atmospheric neutrinos a characteristic zenith angle dependence which peaks towards the horizon, where decay lengths are the longest.

\begin{table}
\caption{Critical energies of various particles.}
\vspace{0.25cm}
\centering
\begin{tabular}{ | c | c | }
\hline
Particle & $\epsilon$ (GeV) \\
\hline
$\mu^\pm$ & 1.0 \\
\hline
$\pi^\pm$ & 115 \\
\hline
$K^\pm$ & 850 \\
\hline
$K^0_L$ & 205 \\
\hline
$D^\pm$ & 4.3 $\times\mbox{ } 10^7$ \\
\hline
$D^0$ & 9.2 $\times\mbox{ } 10^7$ \\
\hline
\end{tabular}
\label{CriticalEnergies}
\end{table}

The most important unstable mesons produced in cosmic ray interactions are the charged pions and the charged and neutral kaons.  The decay of pions and kaons produces the bulk of atmospheric neutrinos.  The pion decay (essentially $100\%$ branching ratio) \cite{PDG} is

\begin{align*}
\pi^\pm \rightarrow &\mu^\pm + \numu (\numubar) \\
 &\downarrow  \\
 & e^\pm + \nue (\nuebar) + \numubar (\numu)  \\
\end{align*}

\noindent  At low neutrino energies, this decay dominates the spectrum.  We therefore expect two muon neutrinos for every electron neutrino at low energies.  As energy increases, however, the muon becomes increasingly boosted and eventually lives long enough to hit the surface of the earth before decaying.  At this point, we lose muon decay as a source of electron neutrinos.  The transition happens first for vertically downgoing muons and last for horizontal muons, which have a longer path over which to decay.

The corresponding charged kaon decay ($63.4\%$ branching ratio \cite{PDG}) is

$$
K^\pm \rightarrow \mu^\pm + \numu (\numubar)
$$

\noindent Since the critical energy for kaons is higher than that for pions (850 GeV vs. 115 GeV), the atmospheric neutrino spectrum becomes dominated by kaon decay at around 100 GeV.  These are the energies most relevant for IceCube.

The charged kaon decays don't produce an appreciable flux of electron neutrinos.  The main source of high energy electron neutrinos is the small contribution from the neutral kaon $K_{\ell3}$ decay mode

$$
K^0_L \rightarrow \pi^\pm + e^\mp + \nuebar (\nue)
$$

\noindent This has a branching ratio of $40.5\%$ \cite{PDG}.  Also, $K^0_L$ has a lower critical energy than the charged kaons (205 GeV vs. 850 GeV).  At high energies, we therefore expect the electron neutrino flux to be about a factor of 20 below the muon neutrino flux \cite{BeacomShowerPower}.

The charged pion and charged kaon decays are two-body processes, so in the rest frame of the decaying parent the energy and momentum of the outgoing particles are completely determined by four-momentum conservation:

\begin{align*}
&p_1 = -p_2 = \frac{\sqrt{M^4-2M^2(m^2_1+m^2_2)+(m^2_1-m^2_2)^2}}{2M} \\
&E_1 = \frac{M^2+m^2_1-m^2_2}{2M} \\
&E_2 = M-\frac{M^2+m^2_1-m^2_2}{2M} \\
\end{align*}

\noindent where $M$ is the mass of the decaying meson and $m_1$,$m_2$ are the outgoing lepton masses.  Several features of the kinematics are important.  If one of the outgoing leptons has a mass close to the decaying meson, as the muon does in pion decay, then it takes most of the energy in the decay.  This is the limit where $m_1\rightarrow M$ above.  In the pion decay chain then, the electron and three resulting neutrinos each carry about $1/4$ of the pion energy.  If, however, both outgoing leptons are light compared to the decaying meson, as is the case for kaon decay, the two outgoing leptons share equally in the energy.  This is the limit where $M \gg m_1,m_2$ above.  The neutrinos which result from charged kaon decay, therefore, are higher in energy than those that result from pion decay.

The energy spectrum in the lab frame is

$$
\frac{dN}{dE_\nu} = \frac{B}{2pP_{\mbox{\fontsize{8}{14}\selectfont lab}}} 
$$

\noindent  where $B$ is the branching ratio and $P_{\mbox{\fontsize{8}{14}\selectfont lab}}$ is the parent momentum in the lab frame.  For the charged kaon decay, for example, this is

$$
\frac{dN}{dE_\nu} = \frac{0.634}{(1-m^2_\mu/M^2_K)P_K}
$$

\begin{figure}
\begin{minipage}[b]{0.48\linewidth} % A minipage that covers half the page
\centering
\includegraphics[width=7.6cm]{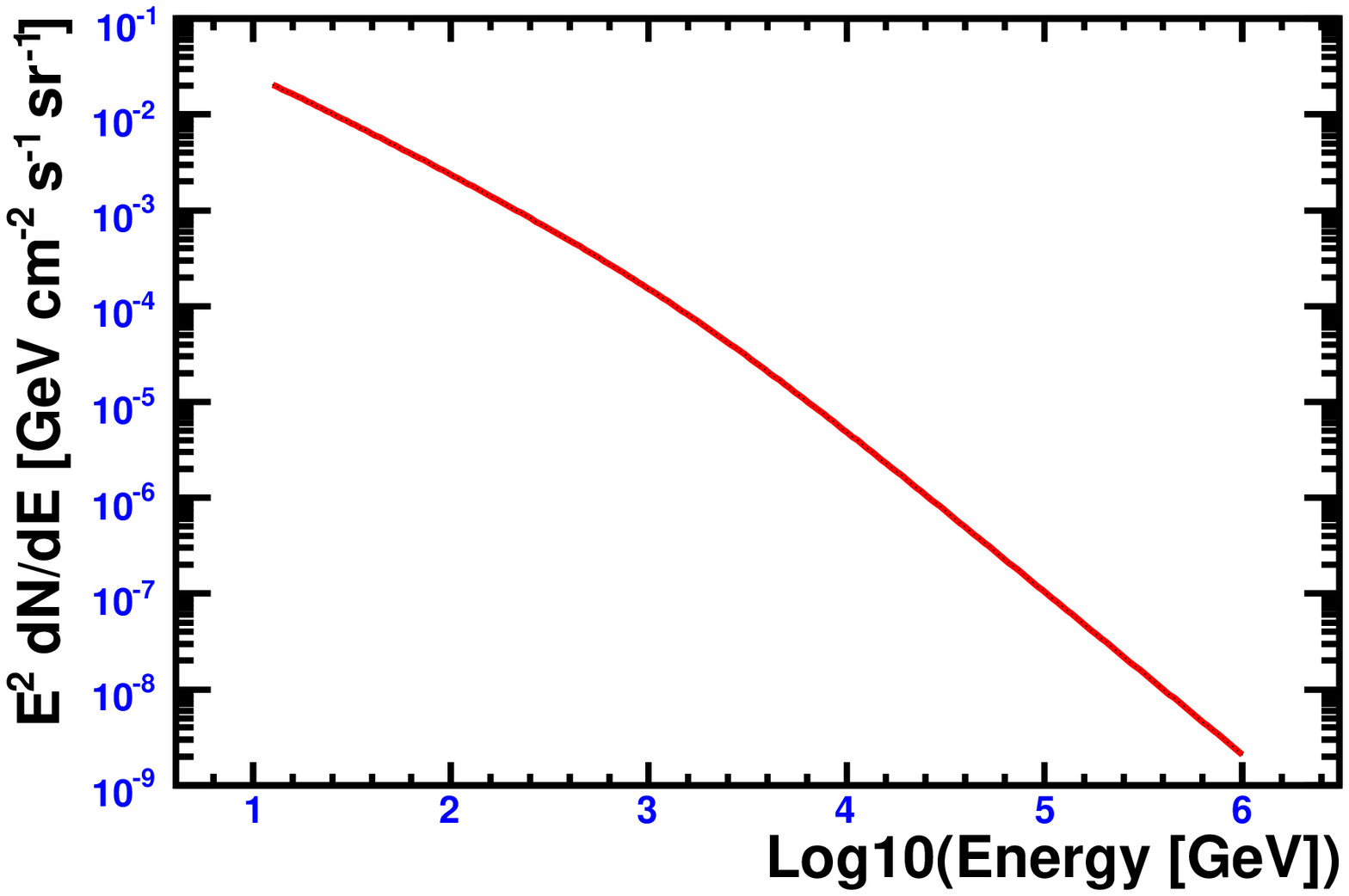}
\end{minipage}
\hspace{0.5cm} %To get a little bit of space between the figures
\begin{minipage}[b]{0.48\linewidth}
\centering
\includegraphics[width=7.6cm]{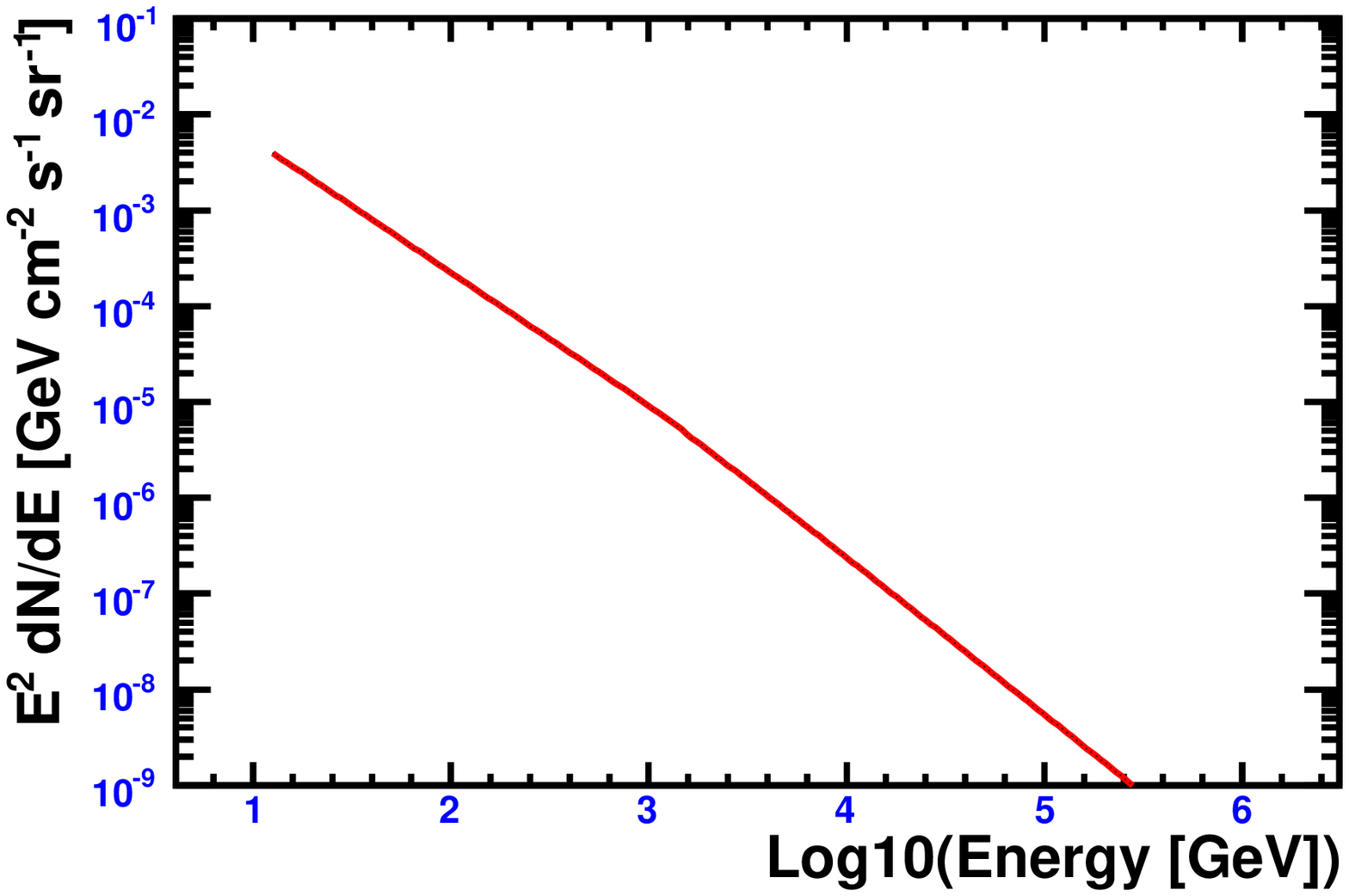}
\end{minipage}
\vspace{0.25cm}
\caption{Atmospheric muon neutrino (left) and electron neutrino (right) fluxes from the Bartol model.  The fluxes have been multiplied by $E^2$ and have units of $\mbox{GeV cm}^{-2}\mbox{ s}^{-1}\mbox{ sr}^{-1}$.}
\label{BartolNuFluxes}
\end{figure}

\begin{figure}
\begin{minipage}[b]{0.48\linewidth} % A minipage that covers half the page
\centering
\includegraphics[width=7.6cm]{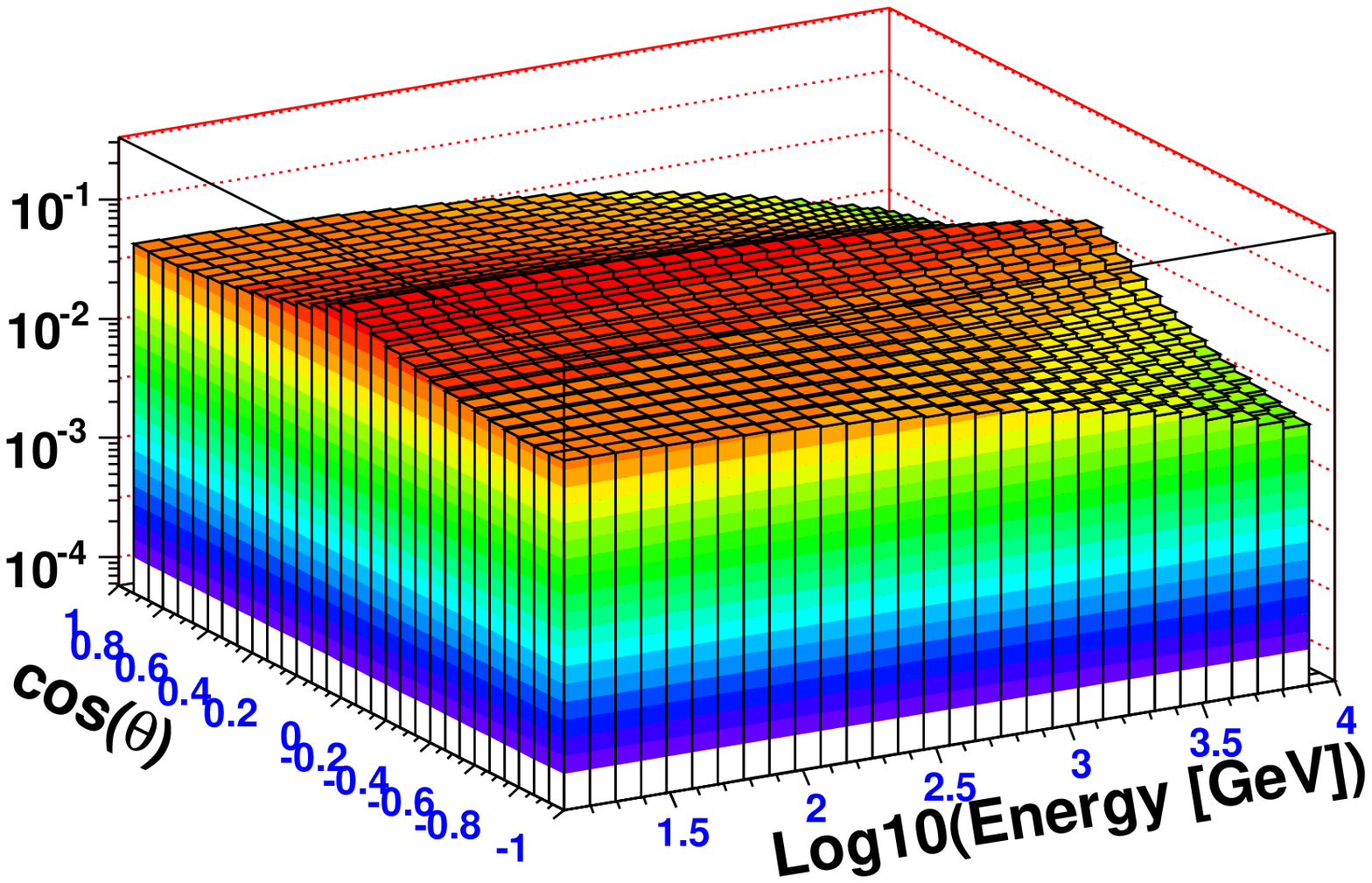}
\end{minipage}
\hspace{0.5cm} %To get a little bit of space between the figures
\begin{minipage}[b]{0.48\linewidth}
\centering
\includegraphics[width=7.6cm]{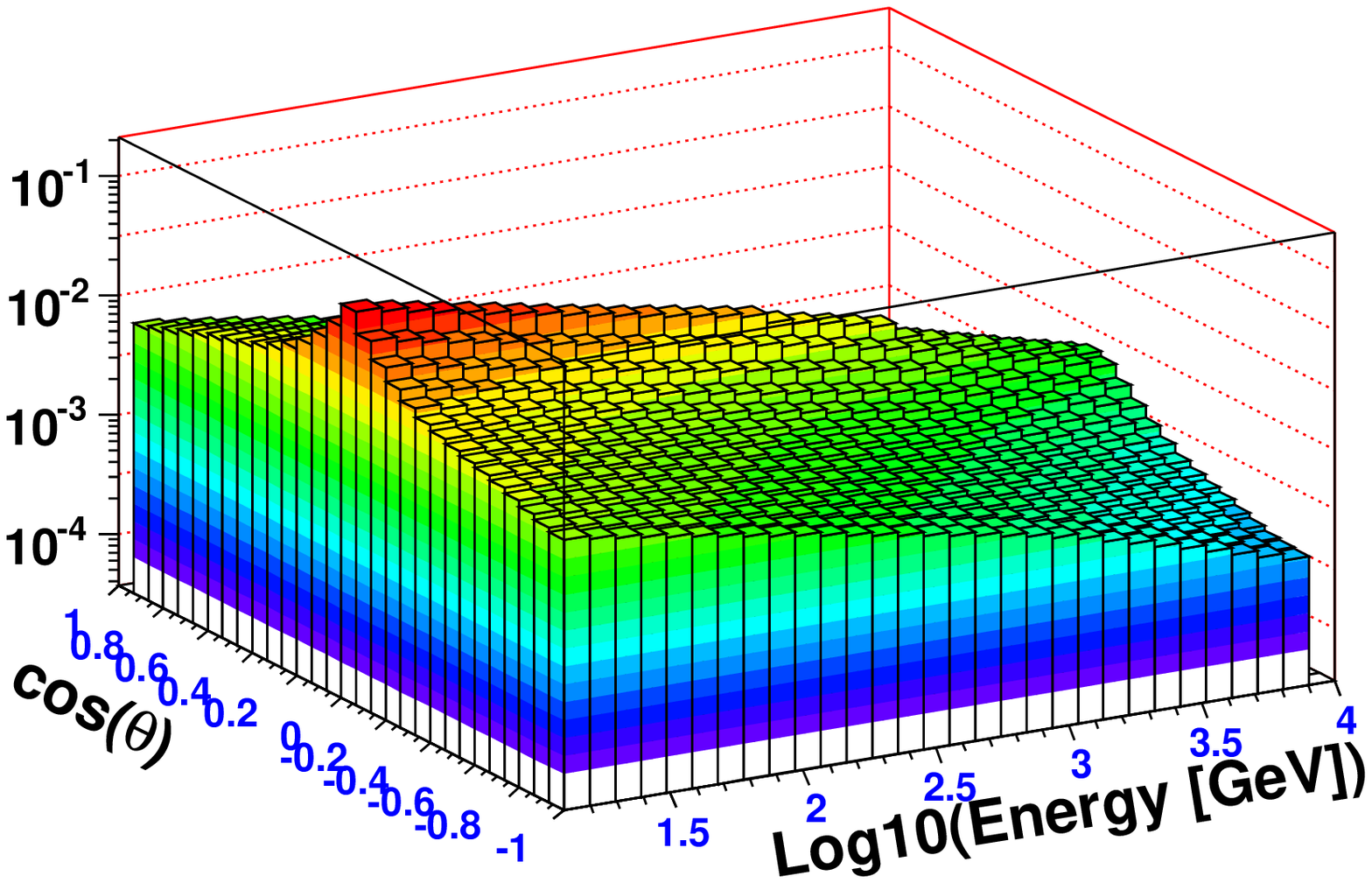}
\end{minipage}
\vspace{0.25cm}
\caption{Atmospheric muon neutrino (left) and electron neutrino (right) fluxes as a function of zenith and energy from the Bartol model.  The fluxes have been multiplied by $E^3$ and have units of $\mbox{GeV}^2\mbox{ cm}^{-2}\mbox{ s}^{-1}\mbox{ sr}^{-1}$.  The fluxes are peaked at the horizon.}
\label{BartolNuFluxes2D}
\end{figure}

To calculate the full flux of atmospheric neutrinos, one needs to solve coupled cascade equations for the nucleon and meson distributions that take into account interaction and decay lengths.  The neutrino energy spectrum from each meson decay, like that for the kaon above, is then folded in with the meson distributions.  This can be done with various analytical approximations (see chapters 3, 4, 6, and 7 of \cite{GaisserBook} for details) or it can be done with Monte Carlo techniques.  The Monte Carlo techniques are most accurate and take into account the geomagnetic effect, which imposes an energy cutoff on primaries and bends the paths of charged secondaries, as well as the three-dimensional nature of the interactions.  See \cite{BarrBartol1, BarrBartol2} for details on the canonical Bartol atmospheric neutrino flux which is used throughout this dissertation.  The atmospheric muon and electron neutrino fluxes from the Bartol calculation are plotted in figure~\ref{BartolNuFluxes}.  Figure~\ref{BartolNuFluxes2D} illustrates their zenith dependence.

\section{Prompt Atmospheric Neutrinos}
\label{SPrompt}

At very high energies, charmed mesons like $D^\pm$ and $D^0$ and charmed baryons like $\Lambda^+_c$ will be produced in the primary cosmic ray interactions.  These charmed particles have very short lifetimes and therefore very high critical energies ($\sim 10^7\mbox{ GeV}$ for the $D\mbox{'s}$).  Because the critical energies are so high, the energy spectrum of these so-called prompt neutrinos follows that of the primary cosmic ray spectrum and is isotropic with respect to zenith angle because the particles always decay before interacting.  The decays also result in roughly equal numbers of muon and electron neutrinos.

The only source of tau neutrinos from the atmosphere is a small component from the charmed, strange meson $D^\pm_s$.  Calculations indicate that this prompt tau neutrino flux is exceedingly small, an order of magnitude lower than the prompt muon neutrino and electron neutrino flux \cite{SarcevicPrompt}.

\begin{figure}
\begin{minipage}[b]{0.48\linewidth} % A minipage that covers half the page
\centering
\includegraphics[width=7.6cm]{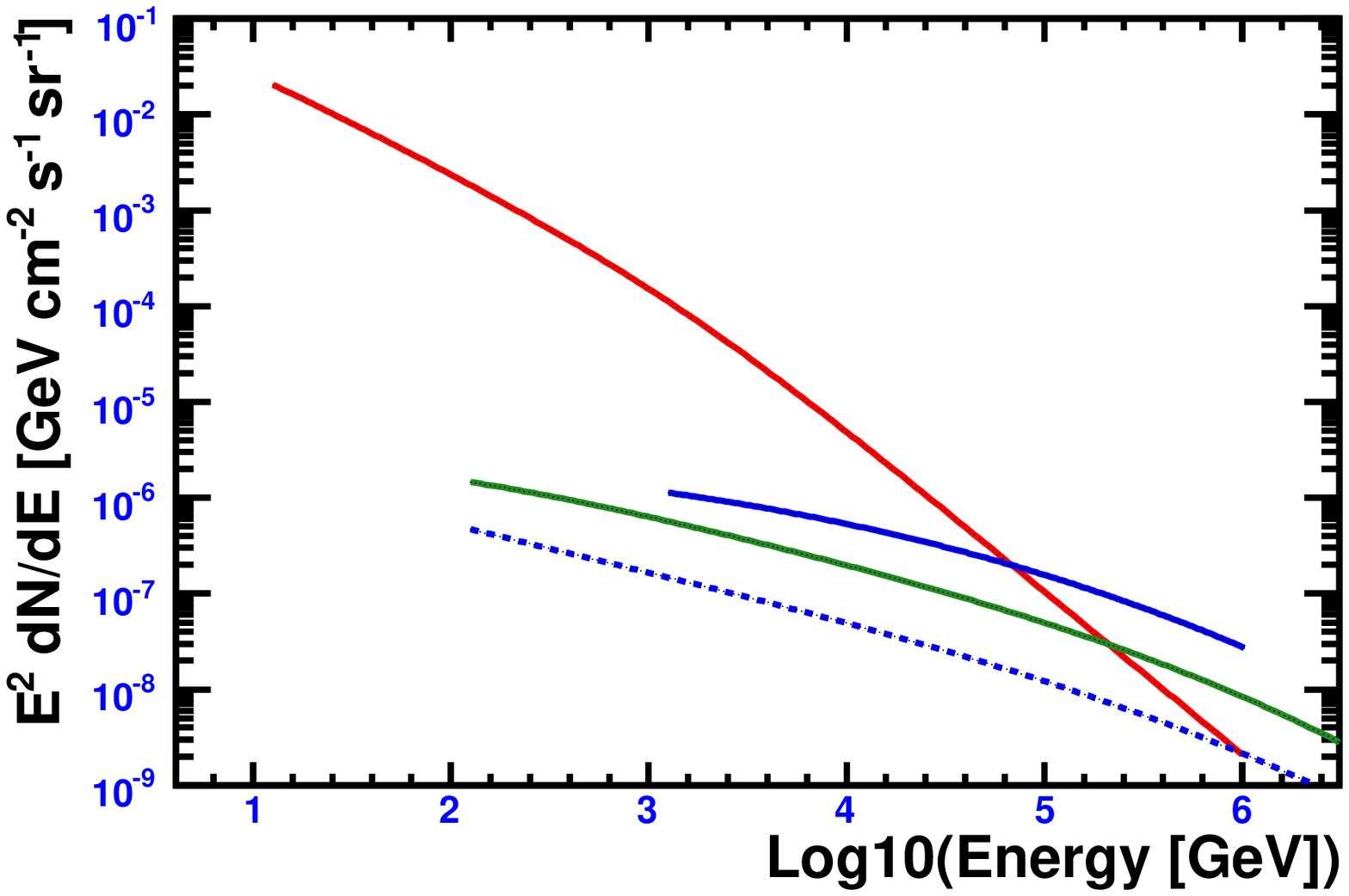}
\end{minipage}
\hspace{0.5cm} %To get a little bit of space between the figures
\begin{minipage}[b]{0.48\linewidth}
\centering
\includegraphics[width=7.6cm]{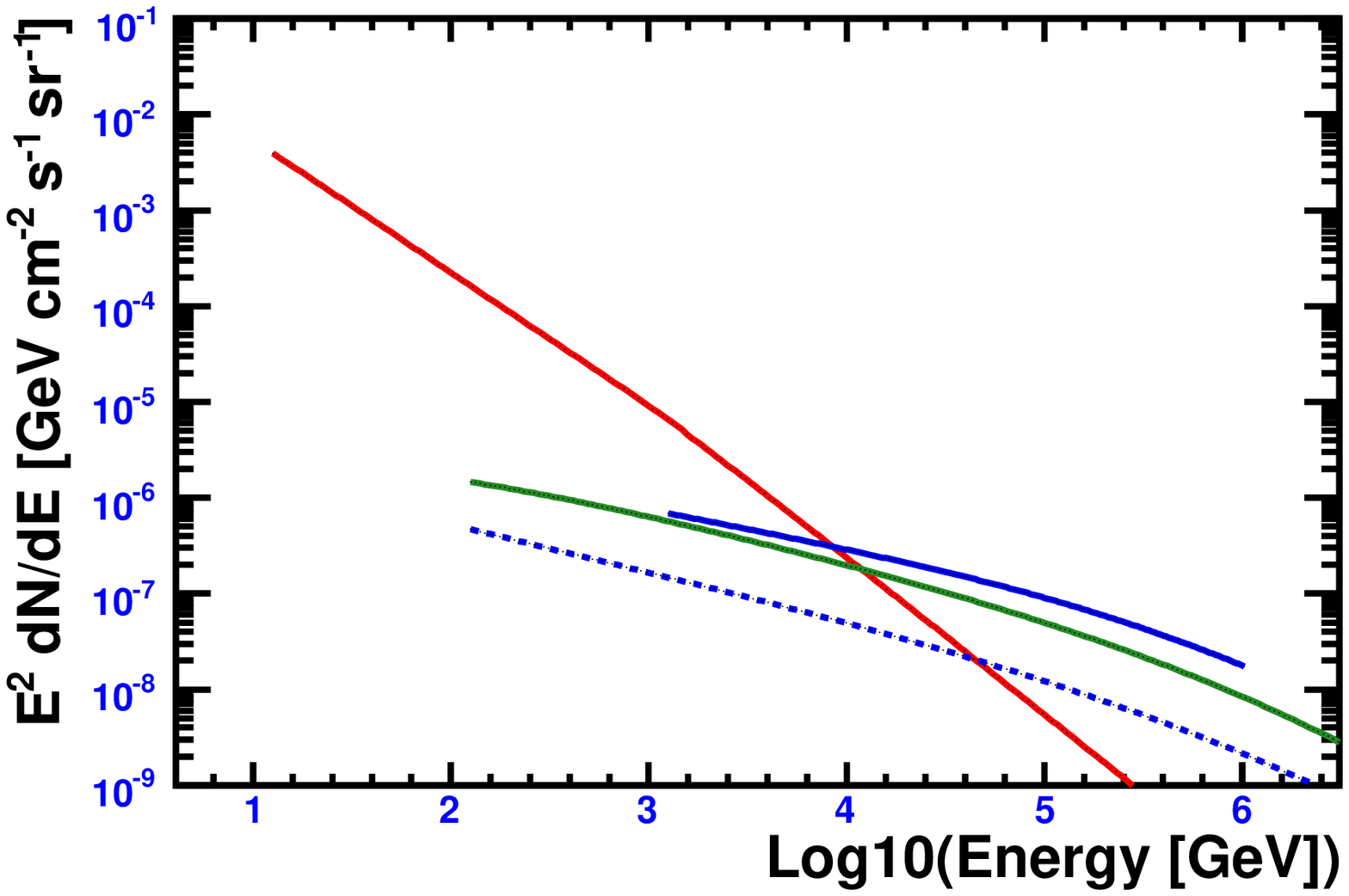}
\end{minipage}
\vspace{0.25cm}
\caption{Conventional Bartol atmospheric neutrino flux (red) and prompt flux models (blue and green) for muon neutrinos (left) and electron neutrinos (right).  The fluxes have been multiplied by $E^2$ and have units of $\mbox{GeV cm}^{-2}\mbox{ s}^{-1}\mbox{ sr}^{-1}$.  Solid blue is the model from \cite{NaumovPrompt}, green is the model from \cite{SarcevicPrompt}, and dashed blue is the model from \cite{MartinPrompt}.}
\label{PromptNuFluxes}
\end{figure}

\begin{figure}
\begin{minipage}[b]{0.48\linewidth} % A minipage that covers half the page
\centering
\includegraphics[width=7.6cm]{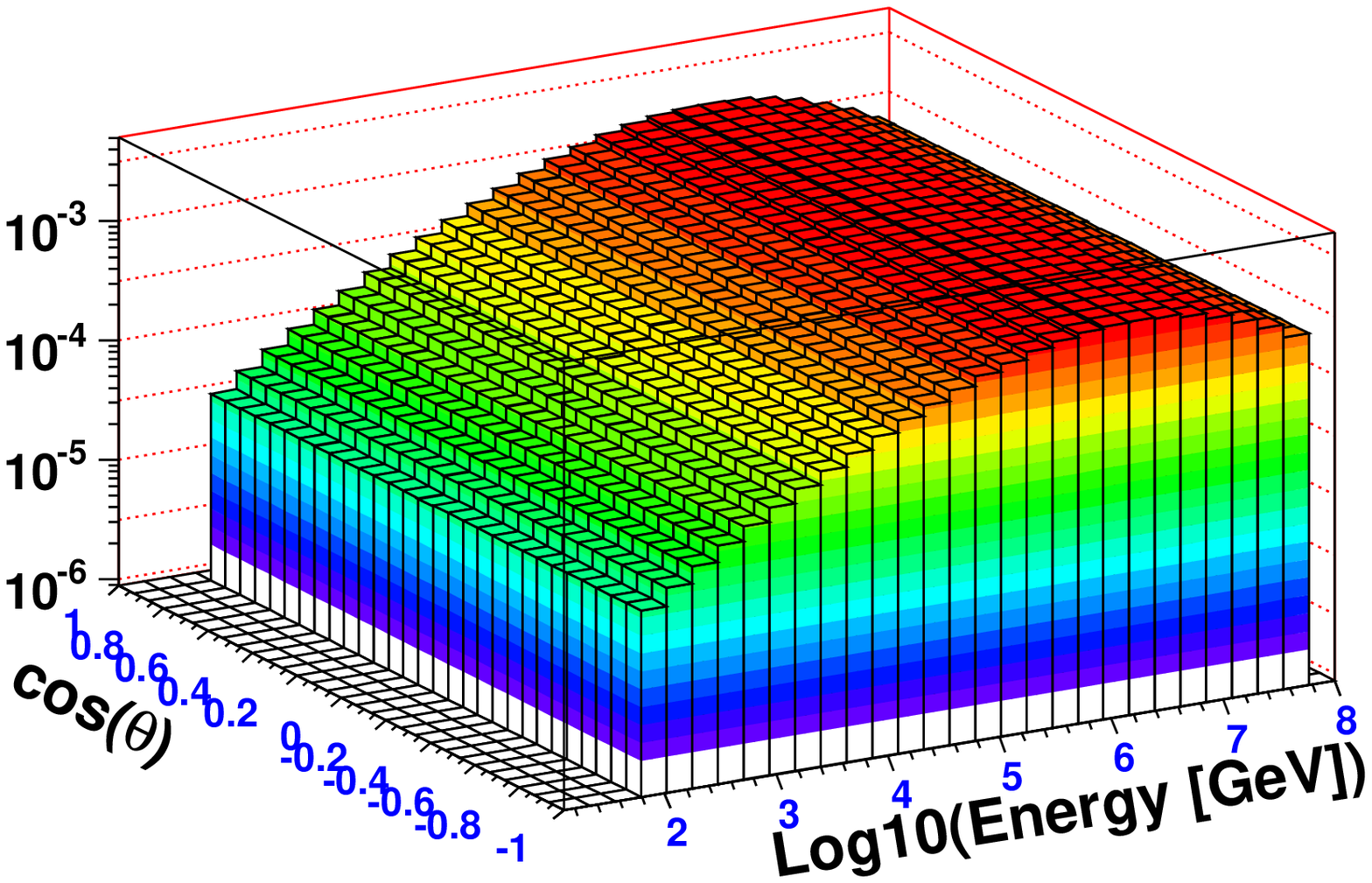}
\end{minipage}
\hspace{0.5cm} %To get a little bit of space between the figures
\begin{minipage}[b]{0.48\linewidth}
\centering
\includegraphics[width=7.6cm]{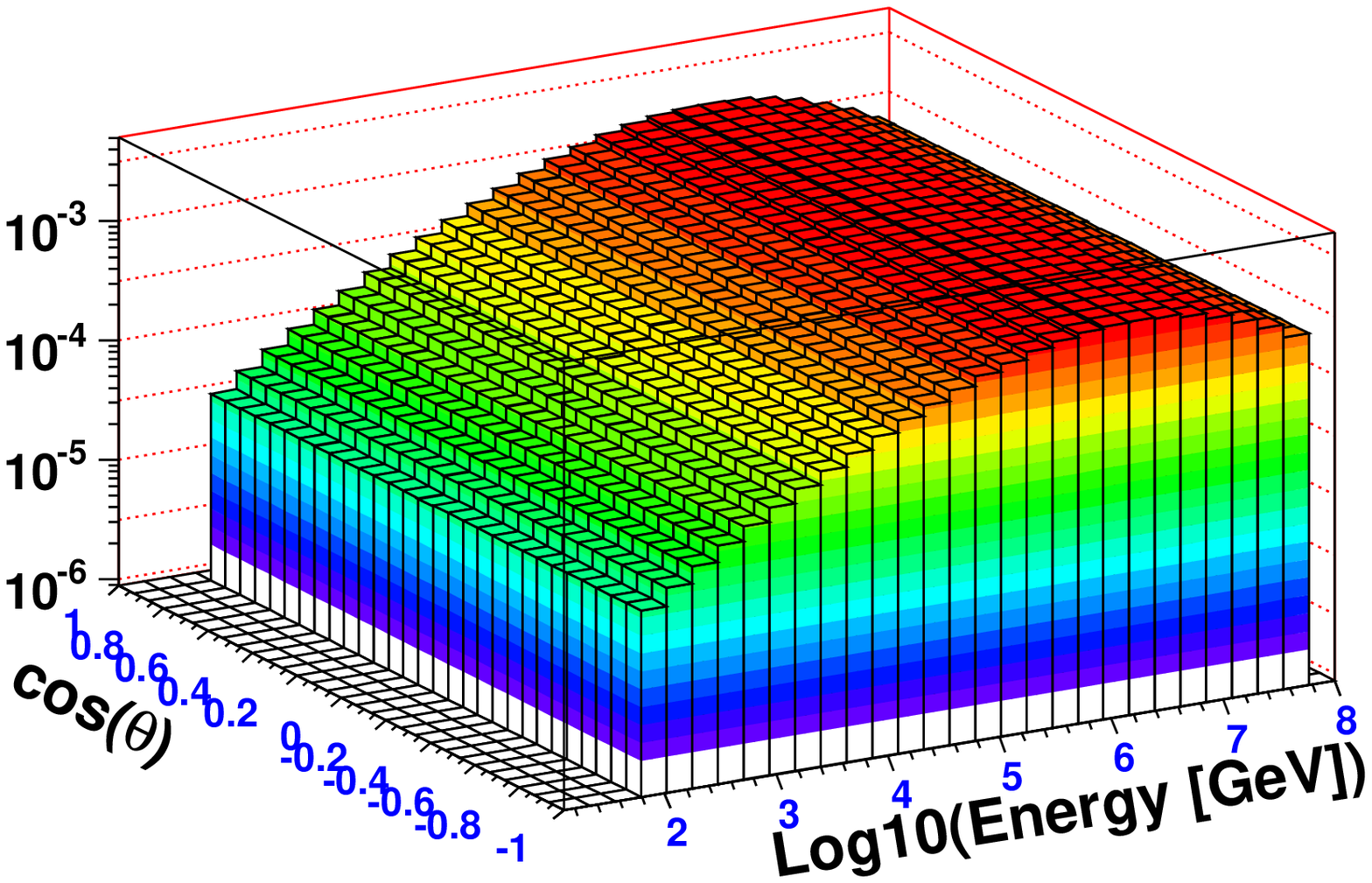}
\end{minipage}
\vspace{0.25cm}
\caption{Prompt flux for muon neutrinos (left) and electron neutrinos (right) as a function of zenith and energy for the model from \cite{SarcevicPrompt}.  The fluxes have been multiplied by $E^3$ and have units of $\mbox{GeV}^2\mbox{ cm}^{-2}\mbox{ s}^{-1}\mbox{ sr}^{-1}$.  The fluxes are isotropic.}
\label{PromptNuFluxes2D}
\end{figure}

The production of charmed mesons at cosmic ray energies involves hadronic physics which is not well-measured at accelerators.  Flux models therefore vary over a large range.  Figure~\ref{PromptNuFluxes} shows several prompt flux models superimposed on the conventional Bartol atmospheric neutrino fluxes.  Figure~\ref{PromptNuFluxes2D} shows the zenith dependence of the prompt flux.  Measuring the prompt atmospheric neutrino flux is a major goal of neutrino telescopes like IceCube.  From figure~\ref{PromptNuFluxes}, it is clear that this should be easier to do with electron neutrinos, where the prompt component rises above the conventional component at almost an order of magnitude lower energy.

\clearpage

\newpage

\clearpage
 
\newpage

\clearpage

\newpage

\section{Neutrino Astrophysics}

As we have seen, cosmic rays are accelerated to enormously high energies in astrophysical sources.  While evidence may be starting to confirm the long-held suspicion that galactic supernova remnants accelerate cosmic rays up to the knee (see section~\ref{SSGalacticSources}), the source of the highest energy cosmic rays is still unknown.  However, the very existence of such high energy accelerators opens the door to the possibility of observing high energy astrophysical neutrinos.  For this reason, neutrino astrophysicists have been designing and constructing large-volume neutrino telescopes in water and ice since the early 1990's.

Astrophysical neutrino production proceeds along the same lines as the neutrino production in the atmosphere described above.  The accelerated protons inside astrophysical accelerators should interact with ambient matter and photon fields to produce high energy astrophysical neutrinos via pion and kaon decay.  This scenario is often described as a ``cosmic beam dump'' in analogy to the production of neutrino beams at particle accelerators on earth.  

The expected sources of these astrophysical neutrinos share several features.  In general, they exhibit non-thermal emission, where the energy comes from the accretion of infalling matter or the gravitational collapse of a massive object.  Sources which show very high energy ($>\mbox{TeV}$) gamma ray emission are also likely candidates.  While gamma rays can be produced via purely leptonic mechanisms (accelerated electrons which synchrotron radiate and inverse Compton scatter), they can also come from hadronic interactions of accelerated protons via neutral pion decay: $\pi^0 \rightarrow \gamma \gamma$.  Such a hadronic mechanism would also produce the charged pions that decay to neutrinos.

\begin{figure}
\centering
\includegraphics[width=0.8\textwidth]{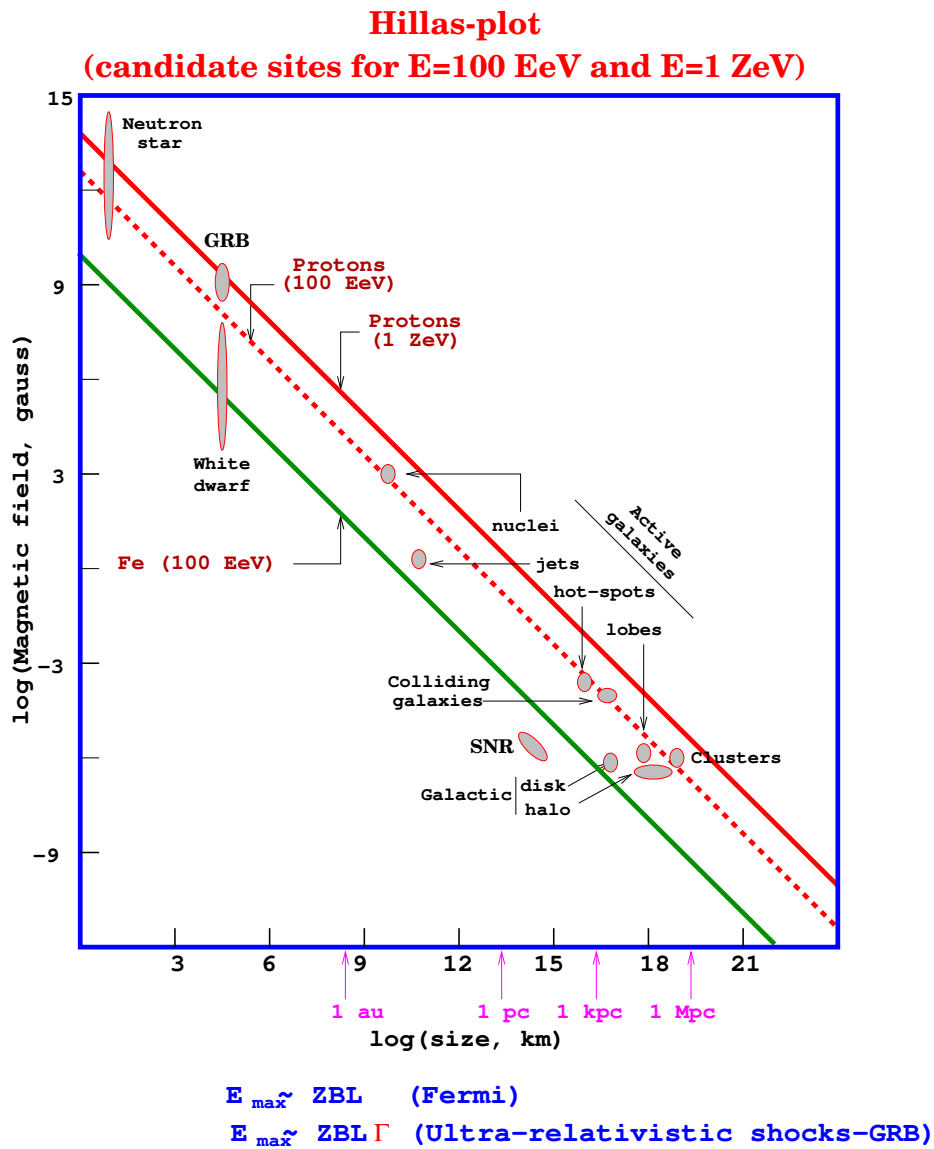}
%\vspace{0.5cm}
\caption{Hillas plot of magnetic field strength versus source size.  The green line indicates what is necessary to accelerate an iron cosmic ray to $10^{20}$~eV, and the red lines indicate what is needed to accelerate a proton to $10^{20}$--$10^{21}$~eV. From~\cite{HillasPlotReference}.}
\label{HillasPlot}
\end{figure}

A general argument can give us a sense of what these astrophysical accelerators might be.  They can be classified by their size $R$ and their magnetic field strength $B$.  In order to be an efficient accelerator, an astrophysical source should be large enough to contain the orbits of the accelerated particles.  This means that the source size should be larger than the gyro radius $R_{\mbox{\fontsize{8}{14}\selectfont gyro}} = \frac{E}{B}$ for a particle of energy $E$.  The maximum energy achievable is then $E = \gamma B R$ where the gamma factor is included because the source may be boosted with respect to earth.  Figure~\ref{HillasPlot} shows the so-called Hillas plot of magnetic field strength as a function of source size needed to accelerate particles to a given energy.  Also displayed are values for various source classes.

If they can be detected from such sources, neutrinos would have several desirable features as astronomical messengers.  They can probe very energetic, distant sources which are not detectable via other channels.  High energy gamma rays from distant sources interact with background photons to produce electron-positron pairs and are hence attenuated.  This occurs above a threshold given by $4E_\gamma E_{\gamma \mbox{\fontsize{8}{14}\selectfont bg}} = (2m_e)^2$.  For photons of energy $10$--$100$~TeV interactions occur on the infrared background, and for PeV photons they occur on the cosmic microwave background.  Because neutrinos interact only weakly, they are not attenuated or blocked by intervening photon fields or matter.  In addition, neutrinos are not deflected by magnetic fields between the source and the earth and should therefore point directly back to their sources.  High energy protons, on the other hand, are scrambled by magnetic fields and are also absorbed on the cosmic microwave background (the GZK effect).  Because neutrinos have the ability to probe sources not readily observable in photons and protons, they may also reveal interesting surprises about the universe.  

This section will give a very brief introduction to the field of neutrino astrophysics.  For good reviews of the subject, see \cite{HalzenHooperReview, LearnedMannheimReview, StanevReview, WaxmanReview}.  We'll begin with a tour of neutrino telescopes, continue with a discussion of Fermi acceleration and astrophysical neutrino production, and then review several likely sources of high energy astrophysical neutrinos.

\subsection{Prototype and Next-Generation Neutrino Telescopes}

\begin{figure}
\begin{minipage}[b]{0.32\linewidth} % A minipage that covers half the page
\centering
\includegraphics[width=4.3cm]{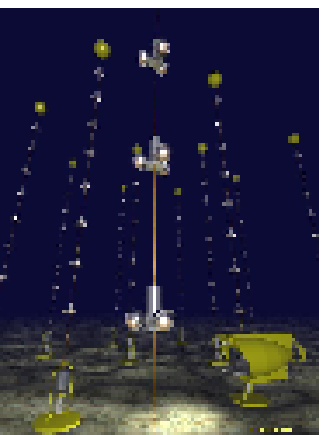}
\end{minipage}
%\hspace{0.25cm} %To get a little bit of space between the figures
\begin{minipage}[b]{0.32\linewidth}
\centering
\includegraphics[width=4.3cm]{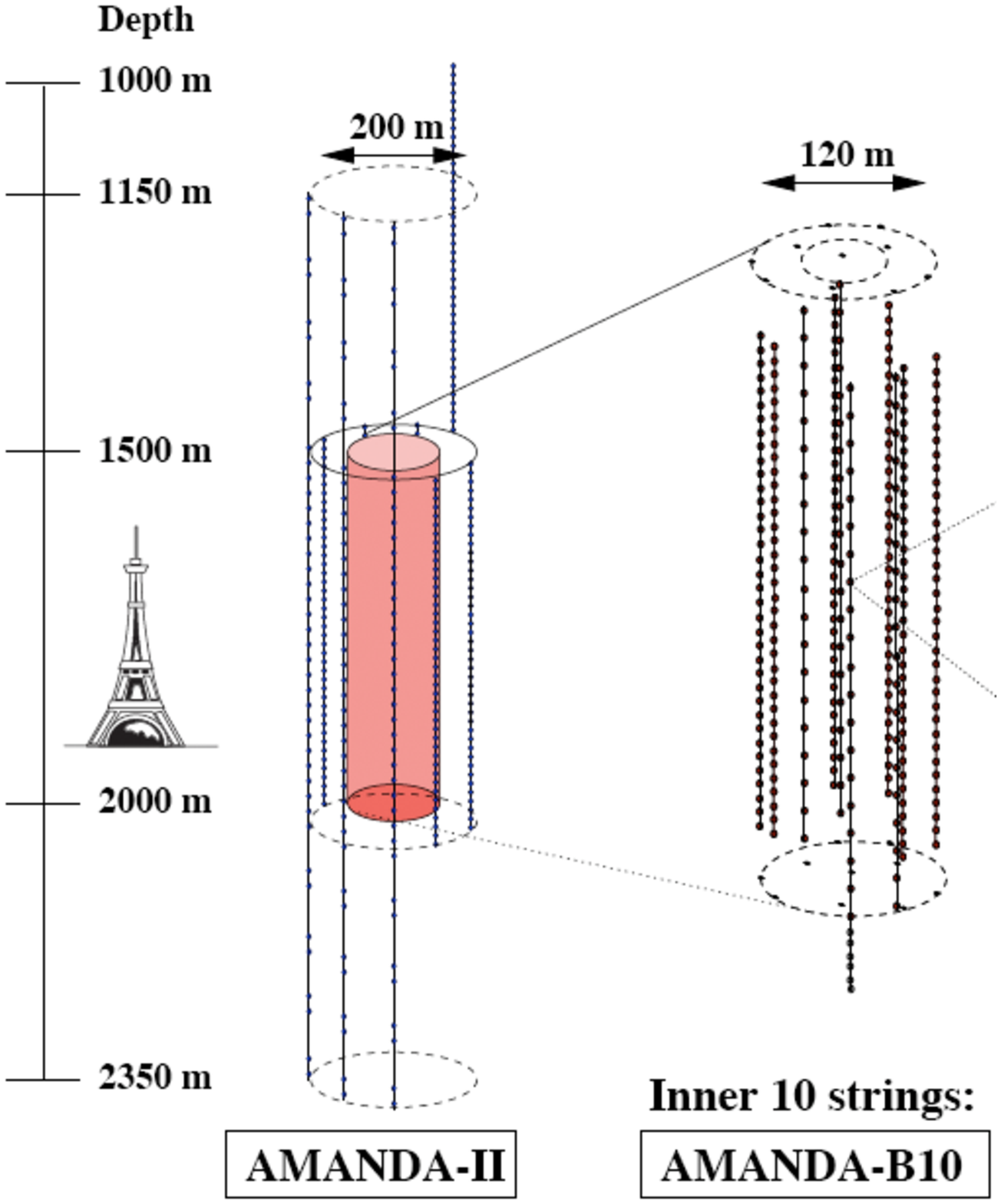}
\end{minipage}
%\hspace{0.25cm} %To get a little bit of space between the figures
\begin{minipage}[b]{0.32\linewidth}
\centering
\includegraphics[width=4.3cm]{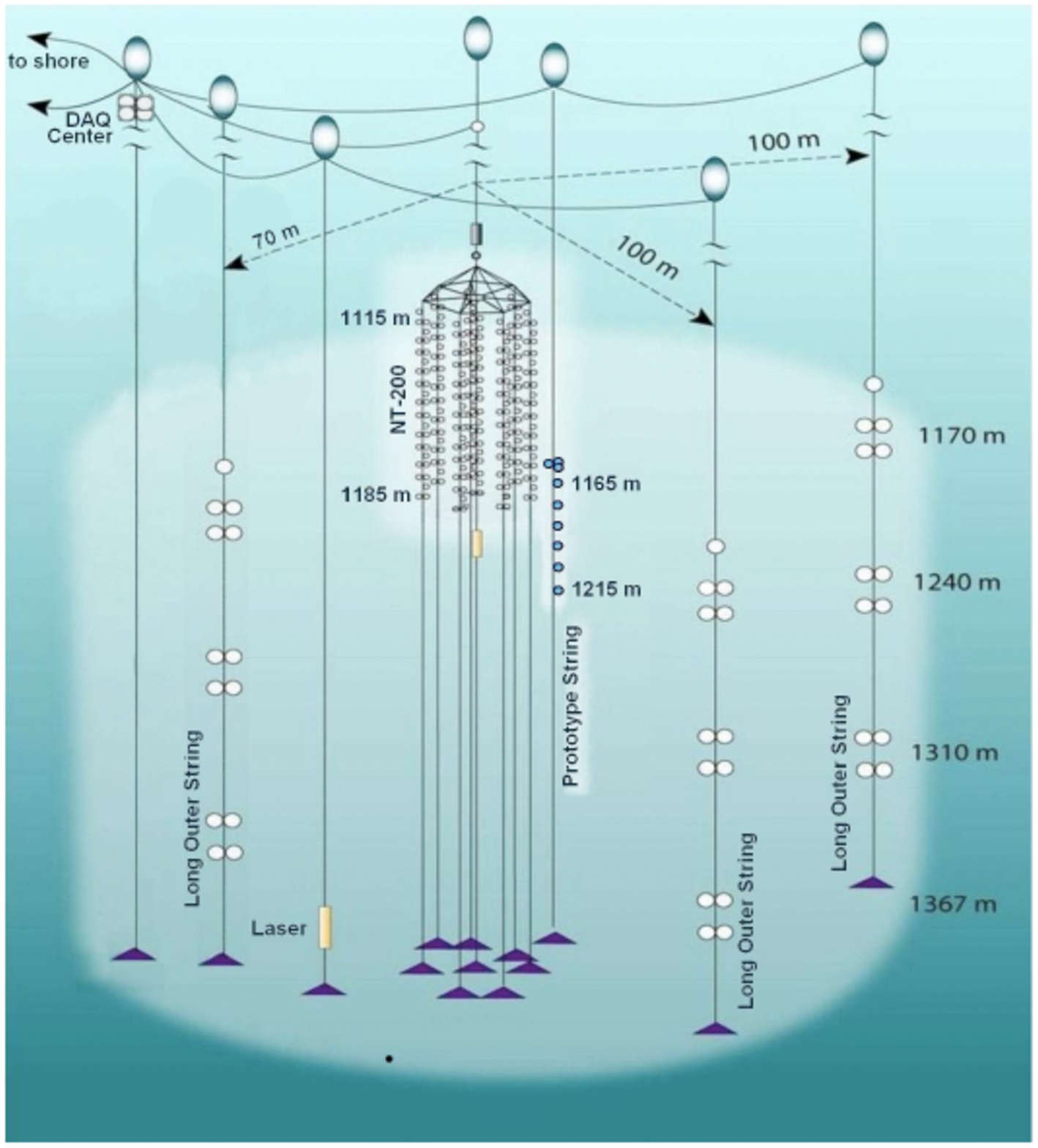}
\end{minipage}
\vspace{0.25cm}
\caption{ANTARES (left, copyright F. Montanet, CNRS/IN2P3 and UJF for Antares), AMANDA (center, from \cite{AMANDAReconstruction}), and Baikal (right, from \cite{BaikalWischnewski}).}
\label{Detectors}
\end{figure}

To observe the low fluxes expected from astrophysical neutrino sources, neutrino telescopes need extremely large detector volumes, eventually reaching the cubic kilometer scale and beyond.  The bulk of the existing and planned neutrino telescopes detect the Cherenkov light emitted by charged secondary particles produced in neutrino interactions.  They therefore make use of large, naturally occurring volumes of transparent Cherenkov media---freshwater, sea water, and glacial ice.  Figure~\ref{Detectors} illustrates several of the completed detector arrays.  

The Baikal experiment \cite{Baikal, BaikalWischnewski} is located 1.1 km below the surface of Lake Baikal in Siberia.  It started operating in 1998 and was the first installed underwater neutrino telescope.  Its NT200 configuration consisted of 8 ``strings'', each of which had 24 optical sensors arranged in pairs to reduce backgrounds from radioactivity and bioluminescence.  NT200+ added 3 additional strings around this dense inner core.  In April 2008, Baikal deployed and successfully tested a prototype string employing new optical sensors and digital technology for full waveform readout \cite{BaikalPrototype}.  The collaboration hopes that this technology will form the basis for a future cubic kilometer scale detector.

Other large-volume neutrino telescopes have been completed or are under construction in the Mediterranean.  The ANTARES array \cite{ANTARES, ANTARESRecentResults} lies 40~km off the coast of Toulon, France at a depth of 2500~m and was completed in May 2008.  It consists of 12 ``lines'', each with 75 optical sensors.  They are arranged into ``storeys'', each of which has three sensors pointed downwards at an angle of $45^\circ$.  Elsewhere in the Mediterranean, the NEMO and NESTOR collaborations are in the prototype and construction stages of building neutrino telescopes off the coasts of Sicily and Greece, respectively \cite{MediterraneanTelescopes}.  

The ANTARES, NEMO, and NESTOR collaborations have joined together to pool their experience and resources to construct a cubic kilometer scale detector  known as KM3NeT \cite{KM3NeTKappes}.  KM3NeT is still in the design, site selection, and planning stages.

At the Amundsen-Scott South Pole Station in Antarctica, the AMANDA array took data from 1995 until 2009.  AMANDA-B10 was composed of 10 strings with 302 optical sensors embedded in the deep glacial ice.  It was eventually upgraded to AMANDA-II, with a total of 19 strings and 677 optical sensors mostly deployed between 1500~m and 2000~m below the surface of the ice.

AMANDA served as the prototype array for IceCube, the first cubic kilometer neutrino telescope.  At its completion, IceCube will have 86 strings, each with 60 optical modules capable of full waveform digitization in ice.  IceCube is the primary subject of this dissertation and will be discussed in detail in chapter 3.

Because of the overwhelming background of downgoing muons from cosmic ray air showers, neutrino telescopes search for upgoing muons induced by neutrinos that have passed through the earth.  For this reason, Baikal and the Mediterranean neutrino telescopes view the southern sky and have the galactic center in their fields of view.  AMANDA and IceCube, on the other hand, view the northern sky.  From the point of view of coverage, it is therefore desirable to have cubic kilometer scale neutrino telescopes in both hemispheres.

\section{Neutrino Production in Astrophysical Sources}

Astrophysical neutrinos are believed to come from the decay of mesons produced in the interactions of accelerated protons with ambient matter and radiation fields in and around astrophysical accelerators.  The accelerated protons can be linked with the observed highest energy cosmic rays, or they can occur in hidden sources where the protons and photons don't make it out of the source but the neutrinos do.  The mechanism of Fermi acceleration is thought to be responsible for accelerating these protons and for giving them their characteristic power law spectrum. 

\subsection{Fermi Acceleration}

The essential idea of Fermi acceleration is that a moving plasma can transfer bulk kinetic energy to individual charged particles.  The charged particles diffuse through turbulent magnetic fields in the moving plasma, scattering elastically in encounters with the magnetic field.

To get the basic contours of Fermi acceleration, we follow the simple argument from chapter 11 of \cite{GaisserBook}.  At each encounter in the magnetized plasma, the charged particle gains an amount of energy proportional to its energy:

$$
\Delta E = \xi E
$$

\noindent After $n$ encounters, the energy is

$$
E_n = E_0(1+\xi)^n
$$

\noindent In order to reach an energy $E$, then, the particle must undergo a number of encounters given by

$$
n = \frac{\ln\left(\frac{E}{E_0}\right)}{\ln(1+\xi)}
$$

\noindent Next, we assume that the particle has some probability $P_{\mbox{\fontsize{8}{14}\selectfont esc}}$ of escaping the accelerator at each encounter.  The probability of a particle remaining after $n$ encounters is then $(1-P_{\mbox{\fontsize{8}{14}\selectfont esc}})^n$.  In order to reach an energy $E$ or greater, the particle has to survive $n$ or more encounters where $n$ is given by the equation above.  So we can write

$$
N(\ge E) \propto \sum_{m=n}^{\infty} (1-P_{\mbox{\fontsize{8}{14}\selectfont esc}})^m = \frac{(1-P_{\mbox{\fontsize{8}{14}\selectfont esc}} )^n}{P_{\mbox{\fontsize{8}{14}\selectfont esc}}}
$$

\noindent Taking the natural log of both sides we have

\begin{align*}
\ln N(\ge E) \propto & \ln\left(\frac{1}{P_{\mbox{\fontsize{8}{14}\selectfont esc}}}\right) + n \ln(1-P_{\mbox{\fontsize{8}{14}\selectfont esc}}) \\
&= \ln\left(\frac{1}{P_{\mbox{\fontsize{8}{14}\selectfont esc}}}\right) + \frac{\ln\left(\frac{E}{E_0}\right)}{\ln(1+\xi)} \ln(1-P_{\mbox{\fontsize{8}{14}\selectfont esc}}) \\
&= \ln\left(\frac{1}{P_{\mbox{\fontsize{8}{14}\selectfont esc}}}\right) + \ln\left(\frac{E}{E_0}\right) \frac{\ln(1-P_{\mbox{\fontsize{8}{14}\selectfont esc}})}{\ln(1+\xi)} \\
\end{align*}

\noindent Exponentiating again we have

$$
N(\ge E) \propto \frac{1}{P_{\mbox{\fontsize{8}{14}\selectfont esc}}} \left(\frac{E}{E_0}\right)^{-\gamma}
$$

\noindent where

$$
\gamma =  \frac{\ln\left(\frac{1}{1-P_{\mbox{\fontsize{8}{14}\selectfont esc}}}\right)}{\ln(1+\xi)} \approx \frac{P_{\mbox{\fontsize{8}{14}\selectfont esc}}}{\xi} = \frac{1}{\xi} \times \frac{T_{\mbox{\fontsize{8}{14}\selectfont cycle}}}{T_{\mbox{\fontsize{8}{14}\selectfont esc}}}
$$

\noindent and $T_{\mbox{\fontsize{8}{14}\selectfont cycle}}$ and $T_{\mbox{\fontsize{8}{14}\selectfont esc}}$ are the characteristic acceleration-cycle and escape times.  Fermi acceleration naturally leads to the power law spectrum that we were trying to reproduce for the observed cosmic rays.

To determine the actual power law index, the fractional energy gain at each encounter must be calculated by considering the scattering of particles from different magnetic field configurations.  Fermi acceleration can be divided into two types.  In first order Fermi acceleration, particles are accelerated by a large, planar shock front which is moving at a velocity $v$.  The fractional energy gain is proportional to $\beta = v/c$.  In second order Fermi acceleration, particles are accelerated by a moving gas cloud, and the fractional energy gain is proportional to $\beta^2$.  First order Fermi acceleration at shock fronts in supernova blast waves is thought to be responsible for accelerating the bulk of the cosmic rays up to the knee, and it is also a possible mechanism for accelerating very high energy protons and electrons in gamma ray bursts and active galactic nuclei, which will be discussed below.

\subsection{Proton-Proton and Proton-Photon Interactions}

To produce neutrinos, these Fermi accelerated protons interact with ambient matter or radiation fields in and around the source.  These interactions can be divided into two types: proton-proton (or simply p-p interactions) and proton-photon (p-$\gamma$) interactions.  The products of these reactions are

$$
p+p, p+\gamma \rightarrow \pi^0, \pi^\pm, K^\pm + X
$$

\noindent If the density of the astrophysical source is sufficiently low, these mesons decay in the familiar chains described above before they have a chance to interact.

Consider p-p interactions first.  The ``volume emissivity'' of pions in units of $\mbox{GeV}^{-1}\mbox{ s}^{-1}\mbox{ cm}^{-3}$ can be expressed as \cite{LearnedMannheimReview}:

$$
Q^{pp}_\pi = \int{n_t ~ c ~ N_\pi ~ \frac{d\sigma_{pp}(E_\pi, E)}{dE} ~ \frac{dN}{dE} ~ dE}
$$

\noindent where $n_t$ is the density of protons in the target, $N_\pi$ is the produced pion multiplicity, $d\sigma_{pp}(E_\pi, E)/dE$ is the differential inclusive cross section for producing $N_\pi$ pions of energy $E_\pi$ from a proton of energy $E$, and $dN/dE$ is the proton spectrum in units of $\mbox{GeV}^{-1}\mbox{ cm}^{-3}$.  We can simply estimate the cross section as $d\sigma_{pp} ~ (E_\pi, E)/dE \approx \sigma_{pp} ~ \delta[E_\pi-\kappa_p E/N_\pi]$ where $\sigma_{pp} \approx 30\mbox{ mb}$ is the proton-proton inelastic cross section at TeV lab energies, $\kappa_p \approx 0.4$ is the inelasticity, and $N_\pi \approx 15$.  This says that the $N_\pi$ pions share the transferred energy equally.  Then the integral above becomes

$$
Q^{pp}_\pi(E_\pi) \approx n_t ~ \sigma_{pp} ~ c ~ \frac{N^2_\pi}{\kappa_p} ~ \frac{dN}{dE}\left( \frac{N_\pi E_\pi}{\kappa_p} \right)
$$

\noindent The neutrino volume emissivity follows from this and from kinematics.  Roughly 1/3 of the proton energy goes into each of the charged and neutral pions.  In a thin source, the pions decay before interacting and three neutrinos are produced, each with roughly 1/4 of the parent pion energy (in dense sources pion and kaon energy loss and interaction must be taken into account).  So we can write

$$
Q^{pp}_\nu \approx 3 \times \frac{1}{4} \times \frac{2}{3} Q^{pp}_\pi(4E_\nu)
$$  

\noindent The neutrino flux then follows by integrating over the spatial extent of the source:

$$
\frac{d\Phi_\nu}{dE_\nu} = \int{Q^{pp}_\nu dr}
$$

\noindent where $\Phi_\nu$ has units of $\mbox{GeV}^{-1}\mbox{ s}^{-1}\mbox{ cm}^{-2}$

For p-$\gamma$ interactions, a similar calculation can be carried out for a target density of photons with a blackbody spectrum or a power law spectrum (which might come, for example, from synchrotron radiation).  We need to impose the threshold condition for pion production in the intgral and to use $\sigma_{p\gamma} \approx 120 ~ \mu \mbox{b}$ well above threshold, $\kappa_p \approx 0.3$, and $N_\pi \approx 3$.  

In p-$\gamma$ interactions, resonant production of the $\Delta$ is particularly important.  This occurs above an energy threshold given by  $E_p E_\gamma = \frac{m^2_\Delta-m^2_p}{4}$.  Flux calculations generally use the fact that $<x_{p \rightarrow \pi}> \approx \frac{1}{5}$ is the average fractional energy transfer from the proton to the pion produced through the delta resonance.

\subsection{Galactic Sources}
\label{SSGalacticSources}

Several galactic candidates for neutrino emission have been considered in the literature.  Pulsars are rapidly spinning neutron stars with strong magnetic fields.  Electrostatic acceleration of particles from the polar surface regions of neutron stars has been discussed as a source of high energy cosmic rays.  If such acceleration occurs, the protons can interact with x-rays, also produced by the pulsar, to produce pions and neutrinos \cite{LinkPulsar}.

Microquasars are several solar mass black holes accreting matter from a binary companion.  They show hard x-ray emission as well as relativistic jets.  Models of neutrino emission include p-$\gamma$ interactions in the jets \cite{WaxmanMicroquasar} and p-p interactions with matter in the accretion disk or in the companion star \cite{BednarekMicroquasar}.  

Perhaps the most exciting galactic candidates are the shell type supernova remnants, which are thought to be the main acceleration sites of galactic cosmic rays.  Recent gamma ray observations by the air Cherenkov telescope H.E.S.S. of the shell type supernova remnants RX J1713.7-3946 and RX J0852.0-4622 show strong TeV gamma ray emission.  Modeling of the multiwavelength spectrum from the radio through x-rays and gamma rays suggests that the gamma ray emission is dominated by the decay of $\pi^0$ produced in the interactions of shock-accelerated protons \cite{BerezhkoSNR1, BerezhkoSNR2, BerezhkoSNR3}.  Observations of neutrinos would be a smoking gun for the acceleration of high energy cosmic rays in these shell type supernova remnants \cite{HalzenSNR}.

\subsection{Extragalactic Sources}
\subsubsection{Active Galactic Nuclei}

Active galactic nuclei (AGN) are million to billion solar mass black holes situated at the centers of galaxies.  They are the brightest objects in the universe, with some radiating as much power as all the stars in the Milky Way from a region smaller than the solar system \cite{LearnedMannheimReview}.  The energy source for these objects is the gravitational energy released from accreting matter.  Many AGN exhibit relativistic jets, TeV gamma ray emission, and short timescale flares.  When the AGN jet points towards the earth, it's known as a blazar.

Shocks in these jets are possible sites for the acceleration of protons.  The protons can interact with matter in the jet itself or with the thermal ultraviolet photons from the heated infalling matter to produce neutrinos.  

\subsubsection{Gamma Ray Bursts}

Gamma ray bursts are extremely bright, cataclysmic explosions distributed over cosmological distances that last on the order of $t_b \sim 10^{-3}$--$10^{3}$ seconds.  The non-thermal gamma ray emission is characterized by broken power laws and can extend into the GeV energy range.  Long gamma ray bursts ($t_b \gtrsim 2\mbox{ s}$) are associated with the collapse of very massive stars to black holes.  Short gamma ray bursts ($t_b \lesssim 2\mbox{ s}$) are thought to arise from the mergers of neutron star binaries or neutron star-black hole binaries.  In both cases, the newly formed black holes can accrete matter and form highly beamed jets with bulk Lorentz factors of $\sim 100$.  See \cite{MeszarosGRBReview} for a good review of the observational and theoretical issues surrounding GRB's.

The jet is an optically thick $e^\pm$,$\gamma$ ``fireball''.  The gamma ray photons can result from synchrotron radiation and inverse Compton scattering from electrons which are Fermi accelerated at internal shocks in the jet.  In addition, the jet may contain accelerated baryons (i.e. protons).  Accelerated protons can interact with these MeV synchrotron photons to photoproduce mesons and neutrinos through the delta resonance described above.  This leads to neutrino energies in the PeV range.  Also, if the jet had to burrow through the collapsing precursor star on its way out, proton-proton and proton-photon interactions can lead to precursor neutrinos 10--100~s before the observed gamma rays.

\subsubsection{Choked Supernovae}

Gamma ray bursts have been associated with type Ib/c  supernovae.  While the jets of these gamma ray bursts have Lorentz factors $\sim 100$, it has been suggested by \cite{RazzaqueMeszarosWaxman} that mildly relativistic jets with Lorentz factors on the order of a few may be a more general feature of supernova collapse.  These mildly relativistic jets may not make it all the way through the stellar envelope, as in a GRB.  However, such ``choked bursts'' will still produce neutrinos without an observable gamma ray signature \cite{AndoBeacom}.

\section{Oscillations and the Astrophysical Flavor Ratio}
\label{SOscillations}

We now have abundant evidence for the phenomenon of neutrino oscillations from atmospheric, solar, and accelerator neutrinos.  That is to say, a neutrino which is produced in a weak interaction in a state of definite flavor has a nonzero probability of being detected as a neutrino of a different flavor in a detector which is situated some distance from the source.  This is only possible if neutrinos have mass.  Oscillations will alter the flavor ratios of astrophysical neutrinos detected on earth.

We can think of two sets of eigenstates for neutrinos---the flavor basis and the mass basis.  These sets of eigenstates are rotated with respect to one other by a unitary mixing matrix $U^*$ \cite{PDG}:

$$
|\nu_\alpha> = \sum_i U^*_{\alpha i} |\nu_i>
$$

\noindent where the index $\alpha$ refers to the flavor eigenstates and the index $i$ refers to the mass eigenstates.  The mixing matrix $U$ is parameterized by three angles and can be written as

\begin{align*}
U &= 
         \begin{pmatrix}
	U_{e1} & U_{e2} & U_{e3} \\
	U_{\mu 1} & U_{\mu 2} & U_{\mu 3} \\
	U_{\tau 1} & U_{\tau 2} & U_{\tau 3} \\ \end{pmatrix} \\
    &=
	\begin{pmatrix}
	1 & 0 & 0 \\
	0 & \ctwothree & \stwothree \\
	0 & -\stwothree & \ctwothree \\ \end{pmatrix}
	\begin{pmatrix}
	\conethree & 0 & \sonethree e^{-\imath\delta} \\
	0 & 1 & 0 \\
	-\sonethree e^{\imath\delta} & 0 & \conethree \\ \end{pmatrix}
	\begin{pmatrix}
	\conetwo & \sonetwo & 0 \\
	-\sonetwo & \conetwo & 0 \\
	0 & 0 & 1 \\ \end{pmatrix}
	\begin{pmatrix}
	e^{\imath\alpha_1/2} & 0 & 0 \\
	0 & e^{\imath\alpha_2/2} & 0 \\
	0 & 0 & 1 \\ \end{pmatrix} \\
\end{align*}

\noindent where $s_{ij} \equiv \sin \theta_{ij}$ and $c_{ij} \equiv \cos \theta_{ij}$, the $\theta_{ij}$ are the mixing angles, $\delta$ is a possible CP-violating phase, and $\alpha_1$, $\alpha_2$ are possible Majorana phases.

Each mass eigenstate picks up a phase $\mbox{exp}(-\imath m_i \tau_i)$ during propagation, where $m_i$ is the mass of the mass eigenstate and $\tau_i$ is proper time.  The probability for $\nu_\alpha$ to oscillate into $\nu_\beta$ can be found by multiplying each term in the sum for $|\nu_\alpha>$ with its phase factor and then projecting the state $<\nu_\beta|$ onto it.  For details see \cite{PDG}.  The formula for the oscillation probability is

\begin{align*}
P(\nu_\alpha \rightarrow \nu_\beta) = &\delta_{\alpha \beta} \\
&- 4\sum_{i>j} \Re(U^*_{\alpha i} U_{\beta i} U_{\alpha j} U^*_{\beta j}) \sin^2[1.27 \Delta m^2_{ij}(L/E)] \\
&+ 2\sum_{i>j} \Im(U^*_{\alpha i} U_{\beta i} U_{\alpha j} U^*_{\beta j}) \sin[2.54 \Delta m^2_{ij}(L/E)] \\
\end{align*}

\noindent where $\Delta m^2_{ij} \equiv m^2_i - m^2_j$ is in eV, $L$ is in km, and $E$ is in GeV.

For astrophysical neutrinos, oscillations mean that the flavor ratio at a detector on earth will be different from the flavor ratio generated at the astrophysical source.  From the meson decay chains described above, we expect a generic flavor ratio at the source of 

$$
(\numu:\nue:\nutau)_{\mbox{source}} \approx 2:1:0
$$.  

\noindent The effect of oscillations over the extremely long baselines of astrophysical sources is to transform this ratio to

$$
(\numu:\nue:\nutau)_{\mbox{earth}} \approx 1:1:1
$$

\noindent That is, we expect an equal ratio of fluxes at our detector on earth.  For details of the calculation, see Appendix I.  

This has an important implication for astrophysical neutrino detection.  Even though we don't expect astrophysical sources to produce tau neutrinos in appreciable numbers, they should be present in an astrophysical neutrino beam because of oscillations.  Since the atmospheric tau neutrino flux is small (no conventional contribution but an exceedingly small contribution from prompt charmed mesons), the detection of a tau neutrino would be an almost smoking gun for an astrophysical source.  

In addition, by detecting the different interaction signatures from different neutrino flavors from an astrophysical neutrino source, neutrino telescopes can hope to probe oscillations over extremely long baselines for effects of exotic physics.  This is described in more detail in the next chapter.

%% file: chapters/chapter2.tex
\chapter{Principles of Neutrino Detection}\label{chapter:chapter2}

Neutrinos are often described as ``ghost particles'' because they are so difficult to detect.  In fact, several of the big names of twentieth-century physics thought neutrino detection would never happen.  ``It is therefore absolutely impossible to observe processes of this kind with the neutrinos created in nuclear transformations,'' Hans Bethe and Rudolph Peierls wrote of the possibility of observing neutrinos via inverse beta decay \cite{NeutrinoHistory}.  Nevertheless, over the course of the last half-century, physicists and astrophysicists have become increasingly adept at building the large-scale detectors needed to observe inverse beta decay and other types of neutrino interactions with reasonably high statistics.  

This chapter will discuss the underlying physics that goes into such detection.  First, we'll give a brief description of Cherenkov light, the basic physical effect that underlies many different neutrino experiments, including IceCube.  Next, we'll discuss the fundamental interactions of neutrinos and how they manifest themselves at different energy scales.  To understand the rates of each interaction we'll discuss the neutrino cross sections.  Finally, we'll go on to talk about charged particle energy loss and conclude with one of the particular signatures of very high energy neutrinos---electromagnetic and hadronic particle showers or ``cascades''.  Cascade detection is the primary subject of this dissertation.

\section{Cherenkov Light}

Cherenkov light is produced when a charged particle travels through a dielectric material faster than light can pass through the same material.  It is the optical analogue of a sonic boom and is responsible for the faint blue glow that emanates from a nuclear reactor core.  The wavefront of Cherenkov light forms a cone whose axis is the particle's trajectory.  The wave vectors make an angle $\theta_c$ with the particle's track:

$$
\cos \theta_c = \frac{1}{\beta n}
$$

\noindent for $\beta > \frac{1}{n}$, where $\beta=v/c$ and $n$ is the index of refraction of the material.  This last inequality expresses the condition that the particle's velocity exceed the speed of light in the material and leads to a momentum threshold below which a particle cannot emit Cherenkov radiation: 

$$
p_{\mbox{\fontsize{8}{14}\selectfont th}} = mc \sqrt{\frac{1}{n^2-1}}
$$ 

\noindent In water or ice with $n=1.33$, the momentum threshold is 0.58 MeV/c for an electron, 120 MeV/c for a muon, 159 MeV/c for a pion, 563 MeV/c for a kaon, and 1070 MeV/c for a proton.  For $\beta \approx 1$, the Cherenkov angle in water and ice is $\theta_c \approx 41^{\circ}$.

The amount of Cherenkov light emitted along a particle's trajectory is given by the Frank-Tamm formula:

$$
\frac{d^2 N}{dx~d\lambda} = \frac{2\pi \alpha z^2}{\lambda^2}\left(1-\frac{1}{\beta^2 n(\lambda)^2}\right)
$$

\noindent where $\alpha$ is the fine structure constant and $z$ is the particle's charge in units of the electron charge.  In ice, an electron or muon traveling with $\beta \approx 1$ emits around 330 photons per centimeter in the 300 to 600 nm range of interest for IceCube's optical sensors.

\section{Neutrino Interactions and Cross Sections}
Depending on the source, neutrino energies vary over some nine orders of magnitude, from MeV-scale reactor, solar, and supernova neutrinos to GeV-scale atmospheric and accelerator neutrinos into the TeV- and PeV-scale astrophysical neutrinos that IceCube hopes to detect.  For different neutrino energies and targets, various classes of reaction are possible between the neutrino and the detector medium.

At their most basic level, neutrino-quark reactions are of two fundamental types:

\begin{align*}
&\nu_l + q \rightarrow l + q' \mbox{  Charged-Current (CC)} \\
&\nu_l + q \rightarrow \nu_l + q \mbox{  Neutral-Current (NC)} \\
\end{align*}

\noindent where $l$ stands for the lepton flavor and $q$, $q'$ are quark species.  The charged-current interaction is mediated via exchange of the charged W boson, and the neutral-current interaction is mediated via exchange of the neutral Z boson.  These fundamental reactions manifest themselves in different ways depending on the neutrino energy.  For a pedagogical introduction, see \cite{ZellerNUSSTalk}.

\subsection{The Cross Section}

In order to predict how many events of each class we expect in a detector, we need to know the cross sections for neutrino reactions.  The number of events will be given by

$$
N_\nu (E) \sim \Phi_\nu (E) \times \sigma_\nu (E) \times N_{\mbox{\fontsize{8}{14}\selectfont target}} \times T
$$

\noindent where $\Phi_\nu (E)$ is the flux of neutrinos of energy $E$, $\sigma_\nu (E)$ is the cross section, $N_{\mbox{\fontsize{8}{14}\selectfont target}}$ is the number of targets (free protons, deuterons, bound nuclei like carbon and oxygen, electrons, etc.), and $T$ is the observation time.  

The next few sections will give a brief introduction to the neutrino reactions that are relevant at different energy scales as well as their cross sections.  Particular attention will be paid to deep inelastic scattering, the regime most important for IceCube.

\subsection{Low Energy}

At the lowest energies relevant for reactor, solar, and supernova neutrinos ($<$ 100 MeV), the important processes are elastic and quasi-elastic scattering.  Since there are no free neutrons  in materials, the basic free nucleon reaction is inverse beta decay:

$$
\nuebar+p \rightarrow e^+ + n
$$

\noindent Inverse beta decay has a threshold around 1.8 MeV and a distinct signature from the delayed-coincidence of the annihilation of the positron and the neutron capture.  It was the basis of the first neutrino detection experiments of Reines and Cowan \cite{ReinesCowan1, ReinesCowan2} and the reactor neutrino experiment KamLAND \cite{KamLAND}.  It is also the basis of upcoming reactor neutrino oscillation experiments like Double Chooz \cite{DoubleCHOOZ} and Daya Bay \cite{DayaBay} that are trying to measure $\theta_{13}$, the last unknown oscillation parameter.

While inverse beta decay dominates the cross section at these energies ($\sigma \sim 10^{-41} \mbox{ cm}^2$ for 10 MeV neutrinos on water), other reactions are possible.  Neutrinos can elastically scatter off electrons in the detector medium.  The cross sections are smaller ($\sigma \sim 10^{-43}-10^{-44} \mbox{ cm}^2$ for 10 MeV neutrinos on water), but the recoiling electron preserves directional information from the incoming neutrino.  This reaction was used to make the famous SuperK images of the sun in neutrinos \cite{SuperKSunImage}.

Neutrinos can also interact with bound nucleons, though the energy thresholds are higher and the cross sections lie between the inverse beta decay cross section and the electron scattering cross section.  For the deuterons in heavy water in experiments like SNO \cite{SNO} the following reactions are important:  

\begin{align*}
&\nue+d \rightarrow e^- + p +p \\
&\nuebar+d \rightarrow e^+ + n +n \\
\end{align*}

\noindent For large water detectors reactions on oxygen need to be taken into account:

\begin{align*}
&\nue+ ^{16}\mbox{O} \rightarrow e^- + ^{16}\mbox{F} \\
&\nuebar+ ^{16}\mbox{O} \rightarrow e^+ + ^{16}\mbox{N} \\
\end{align*}

\noindent And in liquid scintillator there are also contributions from carbon:

\begin{align*}
&\nue+ ^{12}\mbox{C} \rightarrow e^- + ^{12}\mbox{N} \\
&\nuebar+ ^{12}\mbox{C} \rightarrow e^+ + ^{12}\mbox{B} \\
\end{align*}

\noindent Radiochemical neutrino reactions on chlorine provided the first detection of solar $\nue$ by Ray Davis and colleagues \cite{RayDavis}.

\subsection{Intermediate Energy}

At energies on the order of a GeV, quasi-elastic scattering is still important, and there is now enough energy available to produce an outgoing muon in $\numu$ interactions as well:

\begin{align*}
&\numu + n \rightarrow \mu^- + p \\
&\nue + n \rightarrow e^- + p \\
\end{align*}

\noindent plus the corresponding reactions for $\nuebar$ and $\numubar$.

If the neutrino has enough energy, it can also excite resonances (the most important of which is the $\Delta(1232)$ resonance) which lead to the production of charged and neutral pions:

\begin{align*}
&\nu_l + N \rightarrow l + N^* \\
&N^* \rightarrow \pi + N' \\
\end{align*}

\noindent Production of pions can also proceed via coherent interactions with entire nuclei.  Pion production is particularly important for accelerator experiments like NO$\nu$A \cite{NOvA} and T2K \cite{T2K} which are searching for the appearance of $\nue$ in a beam of $\numu$.  Final state $\pi^0$ produced in $\numu$ interactions decay to photon pairs which can mimic the electrons produced in $\nue$ interactions.

As energy increases in this intermediate regime, the total cross sections are linear with neutrino energy and are roughly \cite{PDG}:

\begin{align*}
&\sigma(\nu N) = 0.677\times10^{-38} \mbox{ cm}^2 \times E_\nu/\mbox{(GeV)} \\
&\sigma(\bar{\nu} N) = 0.334\times10^{-38} \mbox{ cm}^2 \times E_\nu/\mbox{(GeV)} \\
\end{align*}

\subsection{High and Ultra-High Energy}

At 100's of GeV and above, we enter the deep-inelastic scattering (DIS) regime, where the neutrino has enough energy to scatter from quarks themselves, breaking up the nucleon in a hadronic shower.  To write down the cross section for the charged-current DIS process $\nu_l+N \rightarrow l+X$, we introduce the conventional kinematic variables.  The four-vectors are given by the following:\\

\noindent {\bf incident neutrino}: $p=(E_\nu, p_\nu)$~{\bf nucleon}: $p_N=(M_N, 0)$ \\ 
\noindent {\bf outgoing lepton}: $p'=(E_l, p_l)$~{\bf outgoing hadronic final state}: $p_X=(E_X, p_X)$ \\

\noindent First, we define the total center-of-mass energy: 

$$
s\equiv (p+p_N)^2=2M_N E_\nu+M^2_N \approx 2M_N E_\nu
$$

\noindent Next, we define $q$ as the four-momentum transfer to the W boson: $q\equiv p -p'\equiv (\nu,q)$.  We define $Q^2$ as the negative magnitude of the four-momentum transfer: 

$$
Q^2 \equiv -q^2 = -(p-p')^2 = 2(E_\nu E_l - p_\nu \cdot p_l) - m^2_\nu - m^2_l \approx 4E_\nu E_l \sin^2\frac{\theta_l}{2}
$$

\noindent where the last approximation holds when we can ignore the lepton masses compared to the energies and $\theta_l$ is the angle of the outgoing lepton to the incoming neutrino.  Next, we define the energy transfer to the W boson in the nucleon rest frame: $\nu=\frac{q \cdot p_N}{M_N} = E_\nu-E_l$.  Finally, we can define the Bjorken scaling variables x and y:

\begin{align*}
&x \equiv \frac{-q^2}{2q \cdot p_N} = \frac{Q^2}{2M_N\nu} \\
&y \equiv \frac{q \cdot p_N}{p \cdot p_N} = \frac{\nu}{E_\nu} = \frac{Q^2}{2M_NE_\nu x} = \frac{Q^2}{(s-M^2_N)x} \\
\end{align*}

\noindent The Bjorken x variable varies from zero to one.  It's zero when $E_l \rightarrow 0$, that is all of the neutrino four-momentum is transferred to the W boson.  It's one when $E_l \rightarrow E_\nu$, that is all of the neutrino four-momentum goes into the outgoing lepton.  x is also the fraction of nucleon momentum carried by the struck quark.  This can be seen via the following argument.  Let $\xi$ be the fraction of nucleon momentum carried by the struck quark.  At the vertex where the W boson interacts with the struck quark and produces an outgoing hadron we have

$$
0= p^2_h = (\xi p_N + q)^2 = q^2 + 2\xi q \cdot p_N = -Q^2 + 2\xi q \cdot p_N \Rightarrow \xi = \frac{Q^2}{2q \cdot p_N} = x
$$ 

\noindent The Bjorken y variable is also called the inelasticity and is the relative energy transfer from the incoming neutrino through the W boson and into the nucleon in its rest frame.

We're now in a position to write down the charged-current DIS cross section on an isoscalar target \cite{GandhiQuigg}:

$$
\frac{d^2 \sigma}{dx~dy} = \frac{2{G_F}^2 ME_\nu}{\pi} \frac{1}{(1+Q^2/{M^2}_{W})^2} \left[xq(x,Q^2)+x\bar{q}(x,Q^2)(1-y)^2 \right]
$$

\noindent where the Fermi constant $G_F=1.16632\times10^{-5} \mbox{ GeV}^{-2}$, $M$ is the nucleon mass, and the quark distribution functions $q(x,Q^2)$ and $\bar{q}(x,Q^2)$ are linear combinations of the valence and sea quark distributions for the different quark types.  To obtain the neutral-current DIS cross section, we replace $M_W$ with $M_Z$ and take another linear combination of the valence and sea quark distributions for the quark types.  

\begin{figure}
\centering
\includegraphics[width=0.8\textwidth]{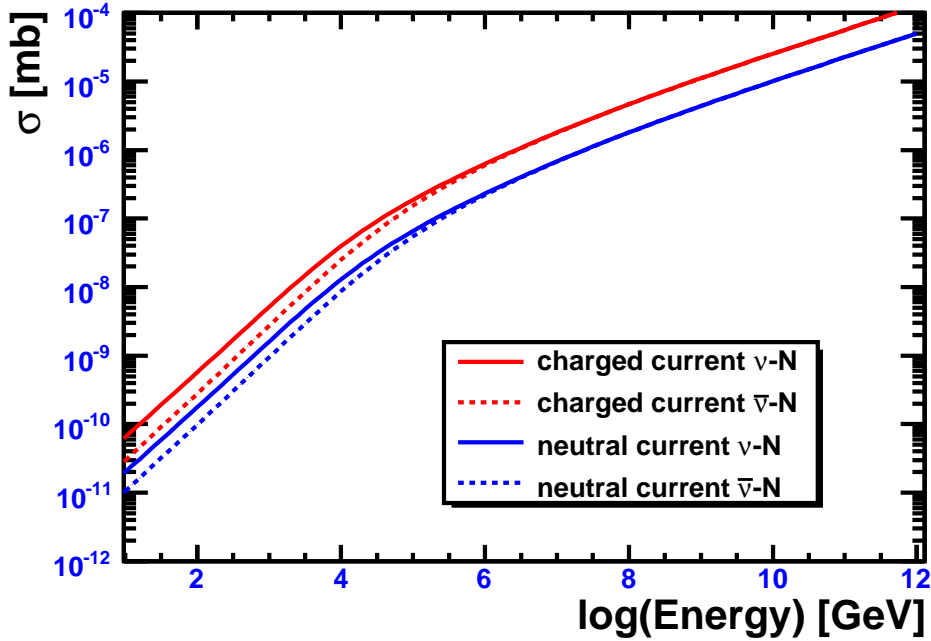}
\vspace{0.95cm}
\caption{Neutrino-nucleon cross sections using the CTEQ5 distribution functions.}
\label{GhandiQuiggCS}
\end{figure}

The cross sections are shown in figure~\ref{GhandiQuiggCS}.  They rise linearly with energy up to several TeV and then decrease in slope because of the $Q^2/{M^2}_W$ term from the propagator.  Extrapolations must be made to carry the quark distributions into unmeasured regions of the small x, large $Q^2$ phase space.  IceCube uses the CTEQ5 (Coordinated Theoretical-Experimental Project on QCD) distribution functions \cite{CTEQ1, CTEQ2} for calculating cross sections.  

One more process must be taken into account in the high energy neutrino cross section.  At $E_{\nuebar} \approx {M^2}_W/2m_e = 6.3 \mbox{ PeV}$, $\nuebar$ can resonantly scatter off electrons to produce the $W^-$.  This is the so-called Glashow resonance \cite{GlashowResonance}.  The cross section for this resonance is several hundred times larger than the charged-current cross section at this energy.  The Glashow resonance is important for high energy astrophysical neutrino searches but doesn't come into play at the energies of atmospheric neutrinos. 

\section{Charged Particle Energy Loss in Matter}

The interactions described above are the fundamental processes that allow for neutrino detection.  In an actual experiment, one makes observations of the products of each reaction---in particular, the muons, electrons, and recoiling nuclei.

Charged particles traversing matter lose energy via ionization and radiation processes.  For heavy particles other than electrons, the ionization loss is given by the Bethe-Bloch equation \cite{PerkinsParticleAstrophysics}:

$$
\left(\frac{dE}{dx}\right)_{\mbox{\fontsize{8}{14}\selectfont ion}} = \left(\frac{4\pi N_A z^2 e^4}{m_e v^2}\right) \left(\frac{Z}{A}\right) \left[ \ln\left( \frac{2m_e v^2\gamma^2}{I}\right) -\beta^2 \right]
$$

\noindent where $m_e$ is the electron mass, $v$ and $z$ are the velocity and charge of the incident particle, $N_A$ is Avogadro's number, and $I\approx10Z\mbox{ eV}$ is the material's mean ionization potential.  This expression goes as $\sim v^{-2}$ at low velocities, reaches a minimum at around $3Mc^2$ where $M$ is the charged particle mass, and then rises logarithmically with energy, eventually flattening out at a value around $2 \mbox{ MeV g}^{-1}\mbox{ cm}^{2}$. 

At higher energies, radiative losses like bremsstrahlung, pair production, and photonuclear interactions start to kick in.  The energy at which the ionization losses equal these radiative losses is the critical energy, which for a muon in a solid or liquid is given by \cite{PDG}:

$$
E_{\mu c} = \frac{5700}{(Z+1.47)^{0.838}}\mbox{ GeV}
$$

\noindent For water and ice this is around 1 TeV.  The main background for all IceCube analyses consists of high energy muons and muon bundles produced by cosmic rays interacting in the atmosphere.  These air shower muons are generally above the critical energy and so lose most of their energy radiatively.  Most of a muon's Cherenkov light will therefore come from the secondary particles produced in these radiative losses.

For high energy electrons, energy loss is dominated by bremsstrahlung.  This energy loss can be expressed as 

$$
\left(\frac{dE}{dx}\right)_{\mbox{\fontsize{8}{14}\selectfont rad}} = -\frac{E}{X_0} 
$$

\noindent where $X_0$ is the radiation length, the distance over which an electron loses on average all but a fraction $1/e$ of its energy.  An approximate formula for the radiation length of an electron in units of $\mbox{g}\mbox{ cm}^{-2}$ is given by \cite{PDG}:

$$
X_0 = \frac{716.4A}{Z(Z+1)\ln(287/\sqrt{Z})}
$$

\noindent which is around $37 \mbox{ g}\mbox{ cm}^{-2}$ or 41 cm for an electron in ice (density $0.9 \mbox{ g}\mbox{ cm}^{-3}$ at $0^\circ$ C).  

The energy at which the ionization losses equal the radiation losses is known as the critical energy $E_c$ and, for an electron in a solid or liquid, is given by \cite{PDG}:

$$
E_c = \frac{610}{(Z+1.24)}\mbox{ MeV}
$$

\noindent This is around 72 MeV for an electron in ice.  

The radiation length also controls the distance scale over which a photon converts to an electron-positron pair.  On average, a photon pair-produces over a distance $\lambda_{\mbox{\fontsize{8}{14}\selectfont pair}} = (9/7)X_0$.

\section{Electromagnetic Showers or ``Cascades''}
\label{SEMCascades}

When a high energy electron passes through a material, it will emit a high energy photon through bremsstrahlung.  This photon can convert to an electron-positron pair which, in turn, can emit photons of their own.  This process goes on and on, forming an electromagnetic particle shower or ``cascade'' of high energy photons, electrons, and positrons.  The high energy electrons that are produced in $\nue$ DIS interactions in a detector like IceCube manifest themselves as electromagnetic cascades.

To understand the main scaling features of an electromagnetic cascade, we can consider an extremely simple branching model \cite{PerkinsParticleAstrophysics}.  We start with an electron of energy $E_0$ which, after one radiation length $X_0$, radiates one bremsstrahlung photon of energy $E_0/2$ (in reality, this fraction varies, but in our simple model we'll always take it to be 1/2).  In the next radiation length, this photon converts to an electron-positron pair, each of energy $E_0/4$.  The original electron also radiates another photon of energy $E_0/4$ in this radiation length.  This process continues to subdivide the energy until the particle energies fall below the critical energy $E_c$, at which point all of the particles simply dissipate their energy through ionization.

After $t$ radiation lengths, each particle has energy $E(t)=E_0/2^t$.  We can define the ``shower maximum'' as the point at which the shower contains the most particles.  In our simple model, this will occur at the point when the particles reach the critical energy at a depth $t_{\mbox{\fontsize{8}{14}\selectfont max}}$ given by

$$
E_c = E_0/2^{t_{\mbox{\fontsize{8}{14}\selectfont max}}} \Rightarrow t_{\mbox{\fontsize{8}{14}\selectfont max}} = \ln(E_0/E_c)/\ln2
$$

\noindent So we see that the depth of shower maximum scales only logarithmically with the energy of the incident particle.  The number of particles at the shower maximum is given by

$$
N_{\mbox{\fontsize{8}{14}\selectfont max}}=2^{t_{\mbox{\fontsize{8}{14}\selectfont max}}}=e^{t_{\mbox{\fontsize{8}{14}\selectfont max}}~\ln2}=E_0/E_c
$$

\noindent The number of particles at shower maximum, however, scales linearly with energy.  The total track length of charged particles is given by

$$
L = \left(\frac{2}{3}\right) \int_0^{t_{\mbox{\fontsize{8}{14}\selectfont max}}} 2^t dt \sim \left(\frac{2}{3~\ln2}\frac{E_0}{E_c}\right)
$$

\noindent where the factor of $2/3$ roughly accounts for the charged particle fraction in our simple model.  The total track length is again proportional to the incident particle energy.  The total track length of a cascade is an important quantity because it will tell us the total amount of Cherenkov light emitted by the cascade. 

Putting in numerical values, we obtain a total track length of $13.4~X_0/\mbox{GeV}$ or $5.5\mbox{ m}/\mbox{GeV}$.  Realistic cascade simulations have determined the actual total track length to be $5.9\mbox{ m}/\mbox{GeV}$ \cite{MarekDissertation}.   However, since the amount of Cherenkov light emitted by a particle depends on its velocity, we define an effective total track length to take into account the velocity distribution of particles within the cascade.  From simulations, this effective total track length is $5.2\mbox{ m}/\mbox{GeV}$ \cite{MarekDissertation}.

The longitudinal profile of an electromagnetic cascade can be fit by the following functional form \cite{PDG}:

$$
\frac{dE}{dt} = E_0 b \frac{(tb)^{a-1} e^{-tb}}{\Gamma (a)}
$$

\noindent where t is the depth in units of radiation lengths, $E_0$ is the electron energy, and $a$ and $b$ are constants which, for ice, were determined by fits to a GEANT simulation \cite{MarekDissertation, WiebuschDissertation}:

\begin{align*}
&a = 2.03 + 0.604 \cdot \mbox{ln}\frac{E_0}{\mbox{GeV}} \\
&b = 0.633 \\
\end{align*}

\noindent The shower maximum is given by $t_{\mbox{\fontsize{8}{14}\selectfont max}}=(a-1)/b$.  Almost all of the energy deposition occurs within 20 radiation lengths.  

In realistic simulations of an electromagnetic cascade, Coulomb scattering of the electrons and positrons as they pass through the medium must be taken into account.  This gives a lateral spread to a shower, which is governed by the Moli\`{e}re radius $R_M$ \cite{PDG}: 

$$
R_M = \left(\frac{21.2\mbox{ MeV}}{E_c}\right) X_0 = 0.0265(Z+1.2)X_0
$$

\noindent  In ice, we have $R_M \approx 8.3 \mbox{ g}\mbox{ cm}^{-2}$ or 9.2 cm.  $90\%$ of the shower energy is deposited within a cylinder of radius $R_M$, and $99\%$ is deposited within a cylinder of radius 3.5 $R_M$.

To summarize then, an electromagnetic cascade in IceCube deposits the bulk of its energy in a thin cylinder $\sim30$ cm in radius and $\sim5$ m in length.

\section{Hadronic Showers or ``Cascades''}
\label{SHadronicCascades}

Energetic hadrons moving through a material result in a shower of baryons and mesons.  These are known as hadronic showers or cascades.  The recoiling nuclei in high energy neutrino DIS will produce a jet of fragmented particles, each of which will shower.  

In general, a hadronic cascade produces less Cherenkov light than an electromagnetic cascade of the same energy.  This happens for several reasons.  First, some of the hadronic cascade energy will go into neutrons, which produce no Cherenkov light.  Second, some of the energy is dissipated in the large nuclear binding energies involved.  Finally, pions, kaons, and protons have higher Cherenkov thresholds than electrons and positrons do \cite{MarekDissertation}.

However, the relative light yield between hadronic and electromagnetic cascades is a function of energy.  The hadronic cascade will produce $\pi^0$'s, which decay to photon pairs.  These photons produce an electromagnetic cascade and represent a ``one-way street'' \cite{Gabriel} by which energy leaves the hadronic sector for the electromagnetic sector.  As the hadron energy increases, more $\pi^0$'s are produced and the cascade becomes more electromagnetic.

Defining the relative total track lengths for a hadronic and electromagnetic cascade as \cite{MarekInternalReport}

$$
F = \frac{L_{\mbox{\fontsize{8}{14}\selectfont hadronic}}}{L_{\mbox{\fontsize{8}{14}\selectfont em}}}
$$

\noindent we can write

$$
F = F_{\mbox{\fontsize{8}{14}\selectfont em}} + (1-F_{\mbox{\fontsize{8}{14}\selectfont em}})\cdot f_0
$$

\noindent where $F_{\mbox{\fontsize{8}{14}\selectfont em}}$ is the electromagnetic part of the cascade fed by $\pi^0$'s, $1-F_{\mbox{\fontsize{8}{14}\selectfont em}}$ is the purely hadronic part of the cascade, and $f_0$ represents the relative total track length for a pure hadronic cascade.  Following \cite{Gabriel}, we model the energy dependence of the electromagnetic fraction $F_{\mbox{\fontsize{8}{14}\selectfont em}}$ as a power law

$$
F_{\mbox{\fontsize{8}{14}\selectfont em}} = 1-(E/E_0)^{-m}
$$  
 
\noindent where E is the hadron energy and $E_0$ and $m$ are fit parameters determined individually from Monte Carlo simulations of injected $p$'s, $n$'s, $\pi$'s, and $K$'s \cite{MarekInternalReport}.  

\begin{figure}
\begin{minipage}[b]{0.48\linewidth} % A minipage that covers half the page
\centering
\includegraphics[width=7.6cm]{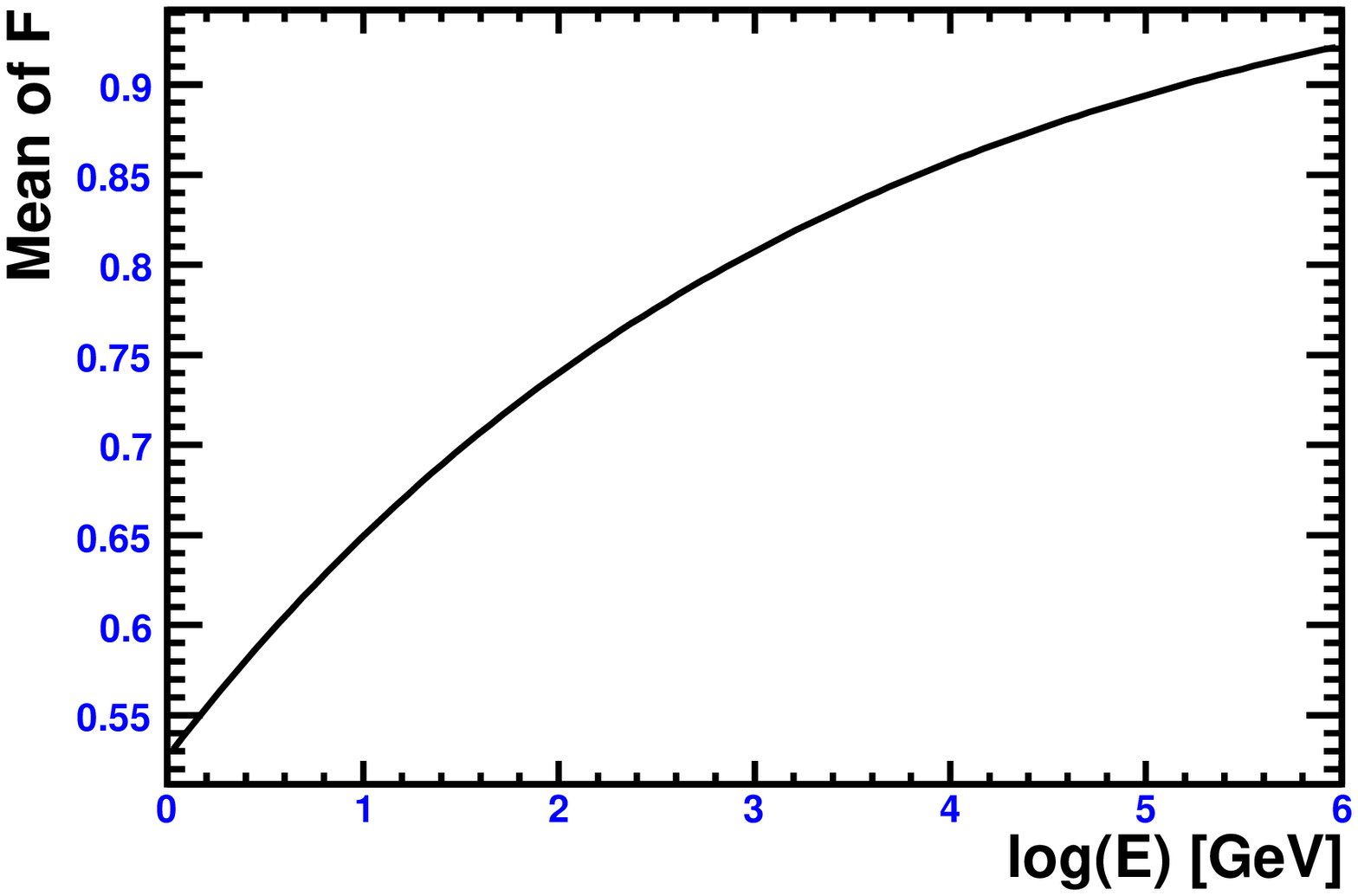}
\end{minipage}
\hspace{0.5cm} %To get a little bit of space between the figures
\begin{minipage}[b]{0.48\linewidth}
\centering
\includegraphics[width=7.6cm]{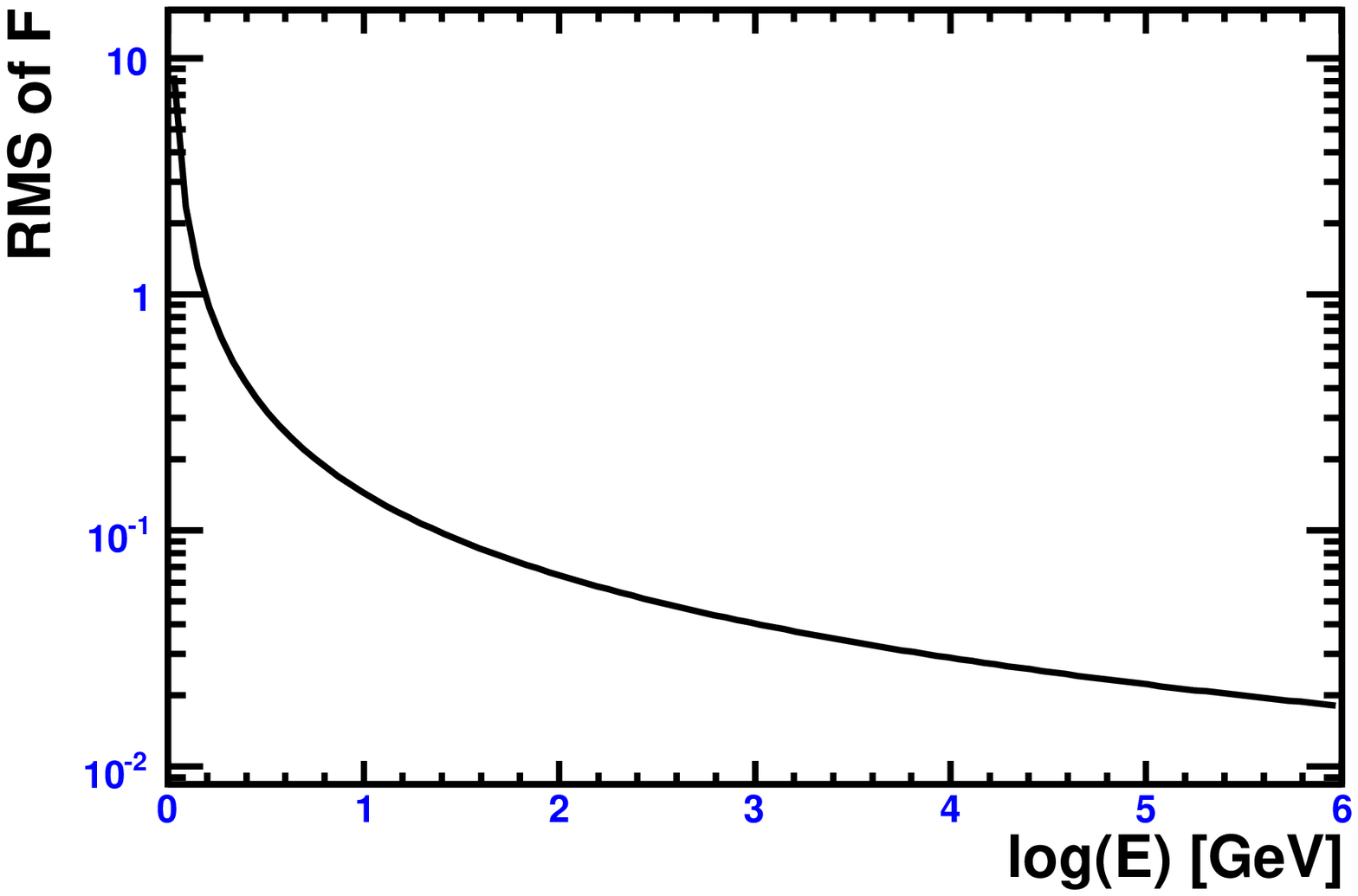}
\end{minipage}
\vspace{0.25cm}
\caption{Mean and RMS of the total track length ratio F for hadronic cascades.}
\label{HadronicFraction}
\end{figure}

In a realistic deep inelastic scattering event, the recoiling nucleus will fragment into a jet of particles, each of which will individually shower.  To determine $F_X$, the total track length fraction for this recoiling nucleus at the neutrino-interaction vertex, simulations were performed by the AMANDA collaboration to determine the composition of the fragmented particles.  $F_X$ was obtained by superimposing the total track length ratios from the individual components of the jet ($p$'s, $n$'s, $\pi$'s, $K$'s, etc.) \cite{MarekInternalReport}.  The mean and RMS values of $F_X$ as a function of the hadronic cascade energy are illustrated in figure~\ref{HadronicFraction}.

\section{Event Signatures in IceCube}

We are now in a position to put everything together to discuss the main neutrino event signatures in the IceCube detector, which detects Cherenkov light from the charged secondary particles from neutrino-nucleon interactions.  The detector itself will be described in detail in the next chapter.  

\begin{figure}
\begin{minipage}[b]{0.32\linewidth} % A minipage that covers half the page
\centering
\includegraphics[width=4.3cm]{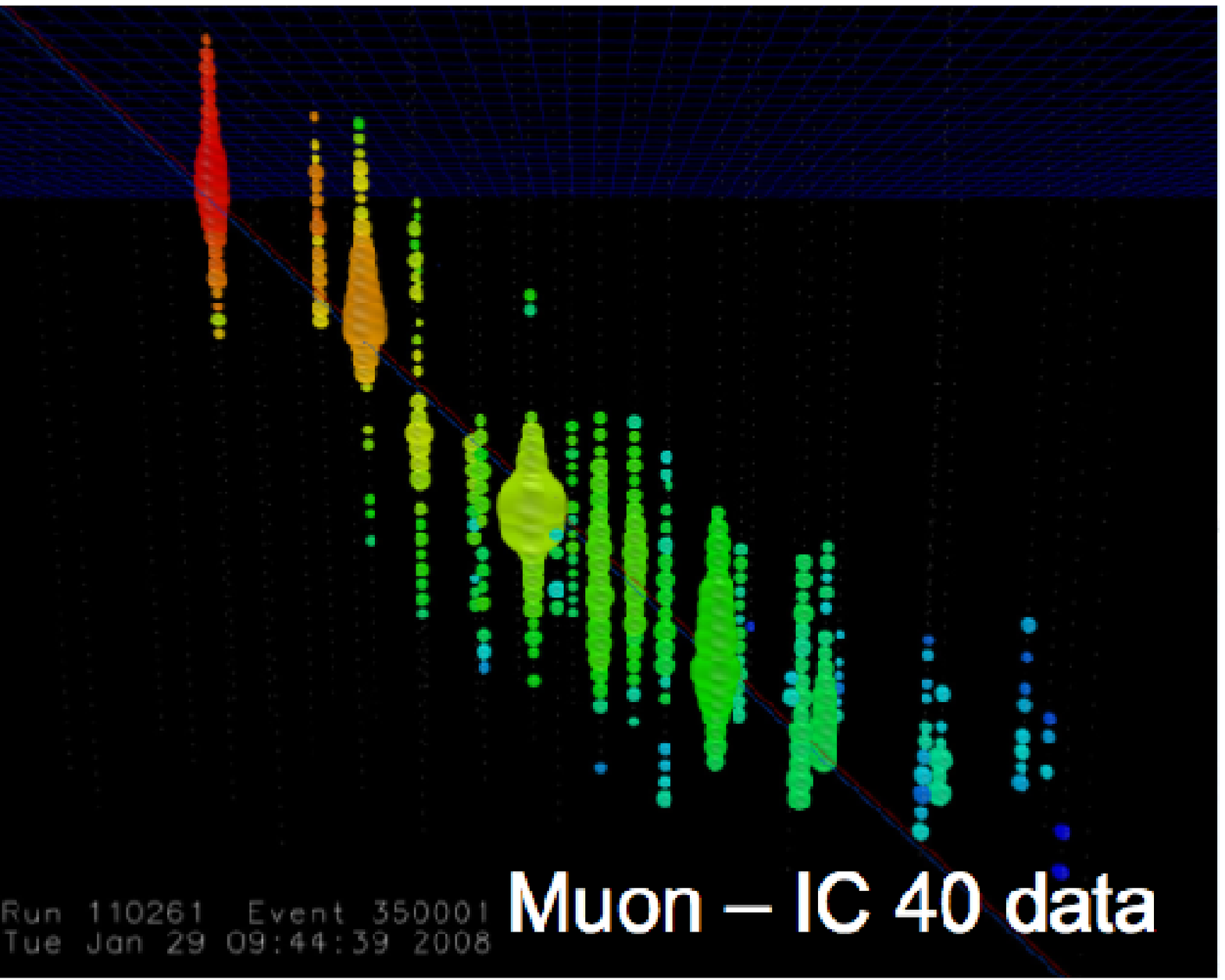}
\end{minipage}
%\hspace{0.25cm} %To get a little bit of space between the figures
\begin{minipage}[b]{0.32\linewidth}
\centering
\includegraphics[width=4.3cm]{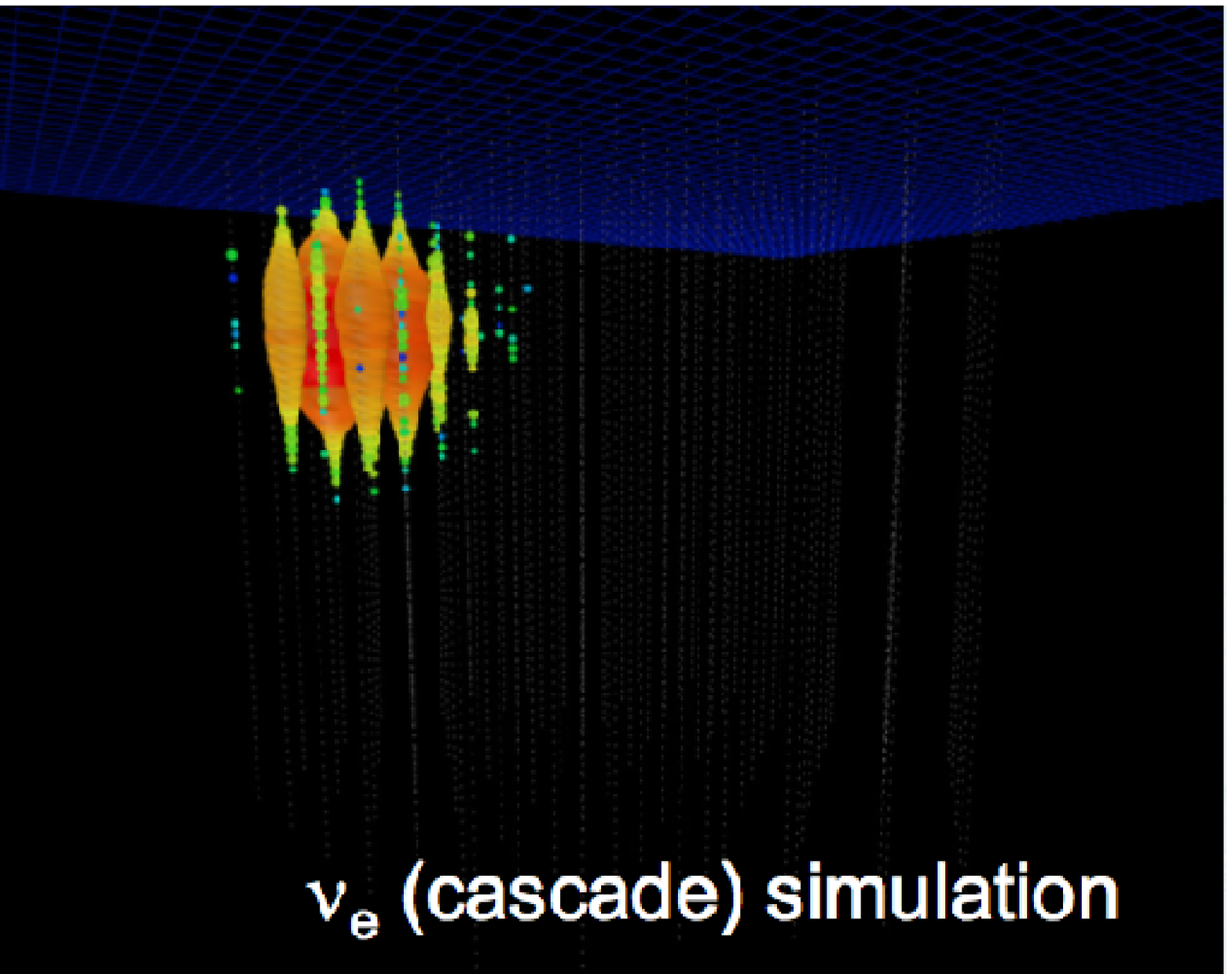}
\end{minipage}
%\hspace{0.25cm} %To get a little bit of space between the figures
\begin{minipage}[b]{0.32\linewidth}
\centering
\includegraphics[width=4.3cm]{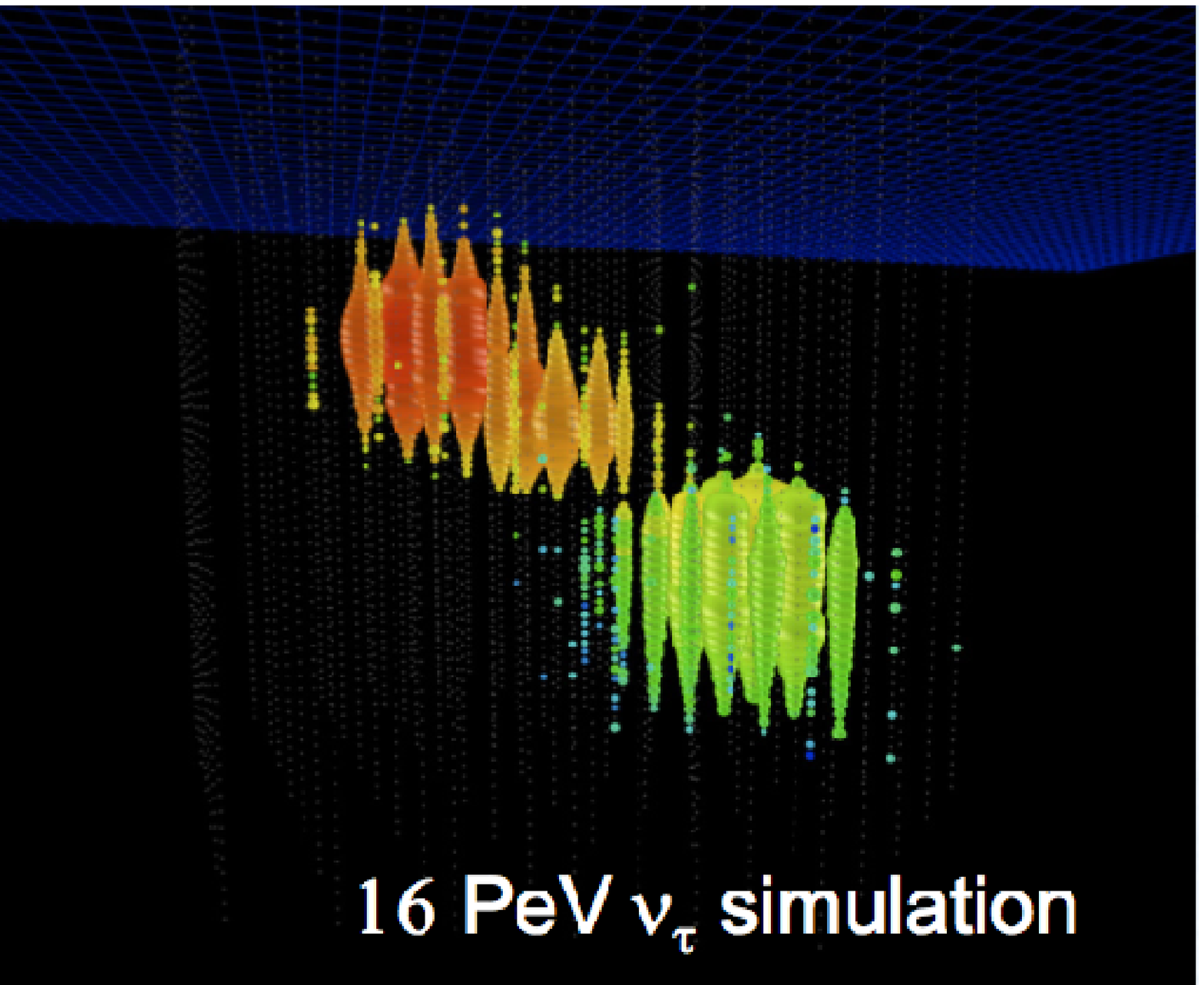}
\end{minipage}
\vspace{0.25cm}
\caption{Different neutrino flavors have different signatures in IceCube.  A long, straight muon track is the hallmark of a muon neutrino (left).  A point like deposition of light from a cascade (center) is produced by electron neutrinos and neutral-current interactions of any neutrino flavor.  Tau neutrinos have a distinctive double bang signature (right). }
\label{Signatures}
\end{figure}

Figure~\ref{Signatures} shows IceCube event-viewer snapshots of the three main neutrino event signatures.  The IceCube event-viewer depicts the amount of detected Cherenkov light in a sensor.  The size of a balloon indicates the amount of detected light, and the colors indicate timing---from the earliest hits in red to the latest hits in blue.

The left panel is a muon track in IceCube.  This particular muon is from a cosmic ray air shower background event and is moving downward through the detector, but a neutrino-induced muon would have the same appearance moving upward through the detector.  Cherenkov light is emitted from the track itself as well as from the radiative energy losses along the track.  

IceCube observes muon neutrinos by looking for muons passing upwards through the detector.  An upward going muon track must have come from a neutrino, since only a neutrino could have traversed the interior of the earth to interact in the rock or ice below the detector.  Cosmic ray air shower muons on the other side of the planet are all absorbed.   

The middle panel shows a simulation of an electron neutrino cascade event.  The Cherenkov light comes from the relatively local energy deposition of the electromagnetic cascade from the outgoing electron plus the hadronic cascade at the interaction vertex.  These two separate cascades cannot be resolved individually.  In a sparsely-instrumented detector like IceCube, cascades look effectively like point sources of Cherenkov light.  

Several other classes of neutrino events also have the cascade topology.  First, the hadronic cascades from neutral-current interactions of all neutrino flavors appear as cascades.  Second, a muon neutrino which has a charged-current interaction inside of the detector will emit light from the hadronic cascade at its interaction vertex.  If the inelasticity is large, the hadronic cascade can dominate the light output from the relatively dim outgoing muon.  These are called ``starting'' muon neutrino events.  Finally, low energy tau neutrino charged-current interactions appear as cascades.

The right panel shows a simulated tau neutrino event.  Tau neutrino charged-current interactions produce a hadronic cascade at the interaction vertex plus an outgoing tau lepton.  The tau lifetime is $2.9\times 10^{-13}\mbox{ s}$, but if it has enough energy it can travel an appreciable distance before decaying (a 1 PeV tau will travel around 50 m before decaying).  Most of the time the tau decay results in another cascade.  This topology, where the two cascades are separated by the long tau track, is known as a ``double bang''.  As the tau neutrino energy decreases, the two cascades move closer together and eventually become indistinguishable from a single cascade.  

As we discussed in section~\ref{SOscillations}, tau neutrinos have one distinct advantage over muon and electron neutrinos: they are not produced in the atmosphere (except for an exceedingly small contribution from prompt charmed meson decay \cite{SarcevicPrompt}).  A detection of tau neutrinos would be a smoking gun for an extraterrestrial neutrino source.  

\section{Why Search for Cascades?}

There are several reasons why we would like to be able to identify the cascade event signature in a high energy neutrino telescope like IceCube.  First and foremost, in the search for an extremely small signal, one wants to be as efficient as possible.  Neutrino oscillations should cause any astrophysical neutrino flux to reach earth in a flavor ratio of $\approx 1:1:1$ (see section~\ref{SOscillations}), and each neutrino flavor can interact with the cascade signature.  All electron neutrinos appear as cascades, muon and tau neutrino neutral-current events manifest themselves as hadronic cascades, and low energy tau neutrino charged-current events look like single cascades.   So an appreciable fraction of any astrophysical signal will make itself known in the cascade channel.  Searching for cascades in conjunction with muon tracks should improve our sensitivity to any small astrophysical signal.

If astrophysical neutrinos are detected, a natural and important next task is to measure their flavor ratio.  In order to tease out the flavor ratio, one needs to be able to measure muon tracks, cascades, and tau signatures.  While this is challenging because of the challenges of cascade and tau detection, the payoff is large.  Several factors can alter the flavor ratio from the expected $1:1:1$ ratio at the detector.  New physics like neutrino decay, CPT violation, sterile neutrinos, and pseudo-Dirac neutrinos are all expected to alter the astrophysical neutrino flavor ratio \cite{BeacomPseudoDirac, KeranenSterileNeutrinos, PakvasaFlavorRatio, QuiggFlavorRatio, BeacomHooperFlavorRatio, BeacomNeutrinoDecay}.  In addition, the flavor ratio is sensitive to the physics and the environment of the astrophysical neutrino source.  The charged pions, kaons, and muons that decay to neutrinos can lose energy in the source due to interactions with radiation and magnetic fields.  This alters the flavor ratio \cite{KashtiWaxman}.  The production and subsequent decay of short-lived charmed mesons like $D^\pm$ and $D^0$ in astrophysical sources will also alter the flavor ratio \cite{GandhiCharm}.  So by measuring cascades and the flavor ratio of astrophysical sources, we can probe new physics and the physics of astrophysical sources.

Next, cascades are useful for astrophysical searches over the full sky.  In this way, cascade searches are quite complementary to muon searches.  Muon neutrino searches have very good angular resolution.  The muon track follows the incoming neutrino direction with a deviation given by $0.7^\circ/(E_\nu/\mbox{TeV})^{0.7}$, and the track directional reconstruction has a resolution $\sim 1^\circ$.  In practice though, the enormous background of cosmic ray air shower muons makes it extremely difficult to search for astrophysical muon neutrinos from the southern sky.  Cascades, on the other hand, have poor angular resolution (see chapter 4).  However, if the cascade event topology can be separated from the cosmic ray muon background, then it can be done over the full sky.  Already, a search for neutrino-induced cascades from gamma ray bursts over the full sky has been conducted with AMANDA \cite{IgnacioGRB}.

Finally, cascades are particularly powerful in searches that look for a diffuse flux.  This diffuse flux could come from the undetected prompt component of the atmospheric neutrino flux (see section~\ref{SPrompt}), or it could come from an unresolved collection of astrophysical sources.  The cascade channel is useful for diffuse searches because, at the energies of interest for IceCube, the atmospheric electron neutrino flux is about a factor of 20 below the atmospheric muon neutrino flux (see section~\ref{SAtmospheric}).  In a search for a harder, diffuse source (for instance an $E^{-2}$ extraterrestrial flux) on top of the atmospheric $E^{-3.7}$ neutrino flux, the new component should manifest itself as a break in the energy spectrum.  Because the atmospheric background is lower for cascades, this break should show up at a lower energy in the cascade channel than the muon channel.  With steeply falling fluxes, there should be more events at this lower energy and it should therefore be easier to resolve the spectral break.  The better intrinsic energy resolution of cascades also makes the search for such a spectral break easier.  Cascades deposit all of their energy in the detector and have an energy resolution in $\log(E)\sim 0.2$.  Muons, on the other hand, deposit energy outside of the detector and have inherently worse energy resolution.

%% file: chapters/chapter3.tex
\chapter{The IceCube Detector}\label{chapter:chapter3}

This chapter discusses the design and implementation of the IceCube detector, which uses the glacial icesheet at the South Pole as both a neutrino target and an optically clear medium in which to observe Cherenkov radiation.  First, we discuss the optical properties of glacial ice at the South Pole.  An understanding of these optical properties is essential for event rate predictions from atmospheric and astrophysical neutrino fluxes and for the reconstruction of event characteristics like direction and energy.  Next, we discuss the IceCube array itself and the Digital Optical Module (DOM), IceCube's basic sensor element.  Finally, we conclude with a description of drilling and sensor installation (deployment) and an update on the current status of construction.

\section{Optical Properties of South Pole Glacial Ice}

\subsection{The Icesheet}

Unlike most particle physics experiments, IceCube does not make use of a well-controlled, man-made detector medium.  In order to achieve the scale needed to observe extremely low astrophysical neutrino fluxes, IceCube has had to make use of a natural material---the glacial icesheet at the South Pole.  This medium, having been subjected to the whims of earth's climatological past, is a complicated and rich one in which to embed a detector.

At the South Pole, the icesheet is 2,820~m thick and was created by snow accumulation over a period of some 165,000 years \cite{AgeVsDepth}.  Icesheets like this have been referred to as ``two-mile time machines'' \cite{AlleyBook}.  As one goes down in depth from the surface, one encounters layers of ice that were deposited as snow during long-distant periods of earth's climatological past.  At the South Pole, a depth of 800-1,000~m corresponds to the onset of the Holocene, earth's current geological epoch which began $\sim$11,000 years ago.  At $\sim$1,300~m depth lies the Last Glacial Maximum, the time period when the glaciers and icesheets last reached their maximum extent on earth $\sim$20,000 years ago.  The lowest deployed optical sensors of the array sit in ice which is $\sim$90,000 years old \cite{AgeVsDepth}. 
 
At the very top of the icesheet lies the firn, a $\sim$100 m layer of loosely packed snow and air.  The firn extends until ``pore closeoff'', where the air no longer exchanges with the atmosphere.  As overburden pressure increases with depth, the firn gives way to compactified polycrystalline ice.  This ice contains impurities which were trapped in the snow when it fell, and the amount and type of impurity are related to the climatological conditions at the time of snowfall.  For the purposes of IceCube, the most important impurity consists of micron-sized dust grains which were carried onto the icesheet as aeolian (wind-borne) aerosols.  Many of these dust particles originated in earth's deserts and were transported around the upper atmosphere.  The amount of this dust is related to how cold and windy a given climatological period is \cite{Petit1, Petit2} and varies strongly as a function of time, and therefore depth, in the icesheet.

\subsection{Basic Definitions}

Two optical properties of the icesheet are of particular importance for IceCube.  The first is the absorption length $\lambda_a$, which is the distance over which a photon's probability of survival drops to $1/e$.  The second is the geometric scattering length $\lambda_s$, which is the average distance a photon travels between scatters.  At each scatter, the photon has an angle change described by a scattering function $p(\theta)$ which depends on the physics of the scattering interaction.  The average of this scattering function is defined to be $\langle \cos \theta \rangle$.  If $\langle \cos \theta \rangle$ is close to one, the scattering is highly peaked towards the forward direction.  

To incorporate the fact that many forward scatters are essentially equivalent to no scattering at all, we define an effective scattering length $\lambda_e$, which is the distance over which a photon has its direction randomized.  To calculate $\lambda_e$, we make use of the following argument from reference \cite{IcePaper}.  A photon travels a distance $\lambda_s$ and is then scattered through an average angle $\langle \cos \theta \rangle$.  It then travels another $\lambda_s$ and is scattered again through an average angle $\langle \cos \theta \rangle$.  Up to this second scatter, the photon distance projected along the original direction of travel is $\lambda_s+\lambda_s \langle \cos \theta \rangle$.  Repeating this procedure over $n$ scatters, the effective length of light travel projected along the original direction of travel is  

$$
\lambda_e = \lambda_s \sum^n_{i=0} \langle \cos \theta \rangle^i
$$

\noindent In the limit of large $n$, this becomes

$$
\lambda_e = \frac{\lambda_s}{1-\langle \cos \theta \rangle}
$$

\noindent If the scattering is forward-backward symmetric, $\langle \cos \theta \rangle = 0$ and $\lambda_e=\lambda_s$.  That is, in one geometric scattering length the direction is effectively randomized.  For forward-scattering, $\langle \cos \theta \rangle > 0$ and so $\lambda_e > \lambda_s$.  The result of the forward-scattering is that one has to travel an effective distance longer than the geometrical scattering length in order to fully randomize the direction.  Calculations of Mie scattering (since the particle size is comparable to the wavelength of light) off micron-sized dust yield a value of $\langle \cos \theta \rangle \approx 0.94$ \cite{IcePaper}. 

It is often convenient to work in terms of the absorptivity $a$ and the scattering coefficient $b_e$ rather than the absorption and scattering lengths themselves.  These are defined by

\begin{align*}
&a=\frac{1}{\lambda_a} \\
&b_e = \frac{1}{\lambda_e} \\
\end{align*}

\noindent Using pulsed and continuous light sources, the AMANDA collaboration measured the absorption and scattering as a function of depth at the South Pole.  Measurements of photon arrival time distributions were fit to full Monte Carlo simulations of photon propagation.  The definitive discussion of the resulting optical properties map can be found in \cite{IcePaper}.

\subsection{Scattering}

A full discussion of scattering mechanisms in glacial ice can be found in \cite{OpticalPropertiesScattering}.  The two most important sources of scattering in South Pole glacial ice are submillimeter-sized air bubbles and micron-sized dust grains.  

At depths shallower than $\sim$1,400 m, submillimeter-sized air bubbles are trapped in the ice and are highly scattering.  The mean free path monotonically increases with depth as the hydrostatic pressure compresses the bubbles.  Eventually, as depth and pressure continue to increase, the air bubbles become unstable against a phase transition to a solid air hydrate phase.  In this phase, which physically resembles ice, the gas molecules are trapped within the crystalline ice.  By $\sim$1,400 m depth at the South Pole, the fraction of bubbles transformed into clathrates has grown to 100$\%$ \cite{BufordClathrateScience}. The refractive index of the clathrates almost perfectly matches that of the ice, making them essentially invisible.

\begin{figure}
\begin{center}
\includegraphics[width=0.5\textwidth]{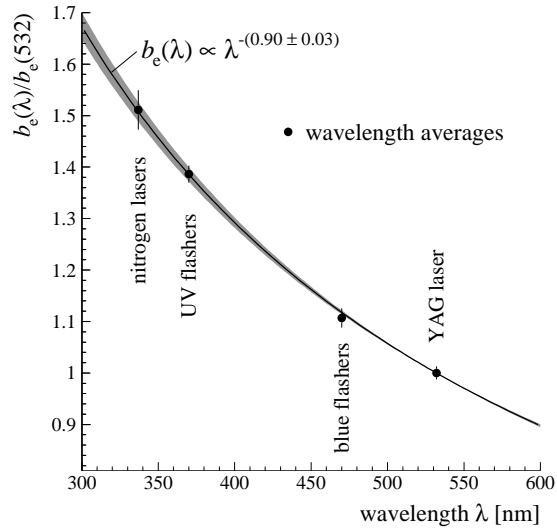}
\caption{Wavelength dependence of the scattering coefficient $b_e$.  From \cite{IcePaper}.}
\label{ScatteringVsWavelength}
\end{center}
\end{figure}

At depths where the air bubbles have disappeared, scattering becomes dominated by micron-sized dust impurities trapped in the ice.  The main components of this dust are mineral grains, sea-salt crystals, liquid acid drops, and soot.  Calculations suggest that scattering off dust should have a power law dependence on wavelength:

$$
b_e \propto \lambda^{-\alpha}
$$      

\noindent where alpha is close to 1.  Figure~\ref{ScatteringVsWavelength} shows the measured wavelength dependence of the scattering coefficient.

\begin{figure}
\begin{center}
\includegraphics[width=0.8\textwidth]{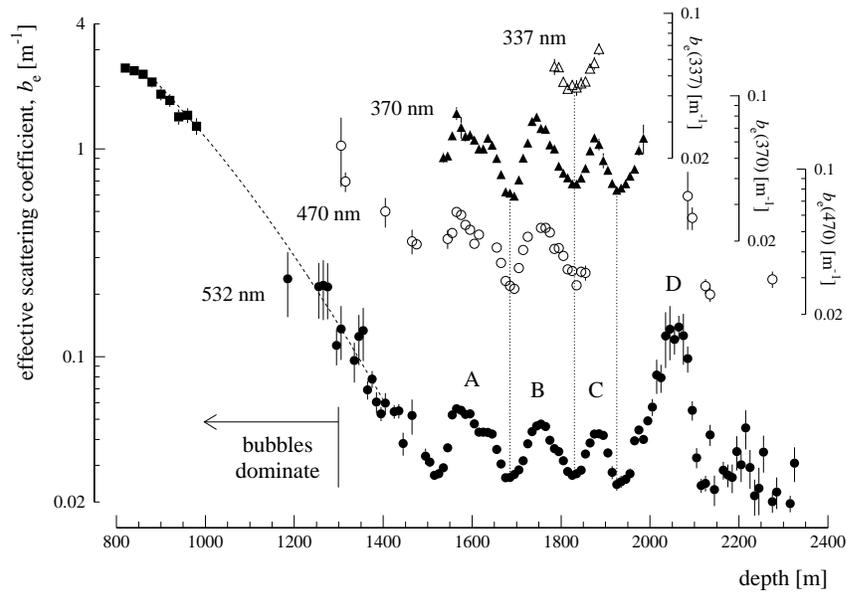}
\caption{Scattering coefficient $b_e$ as a function of depth below the surface of the icesheet.  From \cite{IcePaper}.}
\label{ScatteringVsDepth}
\end{center}
\end{figure}

Figure~\ref{ScatteringVsDepth} shows the measured scattering coefficient as a function of depth below the surface of the icesheet at the South Pole for several wavelengths.  At the depths of interest for IceCube, the effective scattering length varies from around 10 to 50 m.  The dusty peaks labeled A, B, C, and D are related to known climatological features.

\subsection{Absorption}

A full discussion of absorption mechanisms in glacial ice can be found in \cite{OpticalPropertiesAbsorption}.  For near-visible wavelengths, three absorption mechanisms are important for glacial ice at the South Pole.  The absorptivity can be parameterized as \cite{IcePaper}

$$
a(\lambda) = A_{\mbox{\fontsize{8}{14}\selectfont U}} e^{-B_{\mbox{\fontsize{8}{14}\selectfont U}}\lambda} + C_{\mbox{\fontsize{8}{14}\selectfont dust}}\lambda^{-\kappa} + A_{\mbox{\fontsize{8}{14}\selectfont IR}} e^{-\lambda_0/\lambda}
$$ 

\noindent The first and last terms are from intrinsic absorption processes in pure ice itself, and the middle term is due to absorption on dust grains.  The first term is the so-called Urbach tail, which governs the short wavelength behavior of the absorption at wavelengths just above the electronic band-gap energy of the ice crystal.  The last term gives an exponential rise in the red and infrared and comes from molecular stretching and bending modes of the pure ice crystal.  The power law exponent of the second term due to dust was measured to be $\kappa = 1.08 \pm 0.01$.

\begin{figure}
\begin{center}
\includegraphics[width=0.8\textwidth]{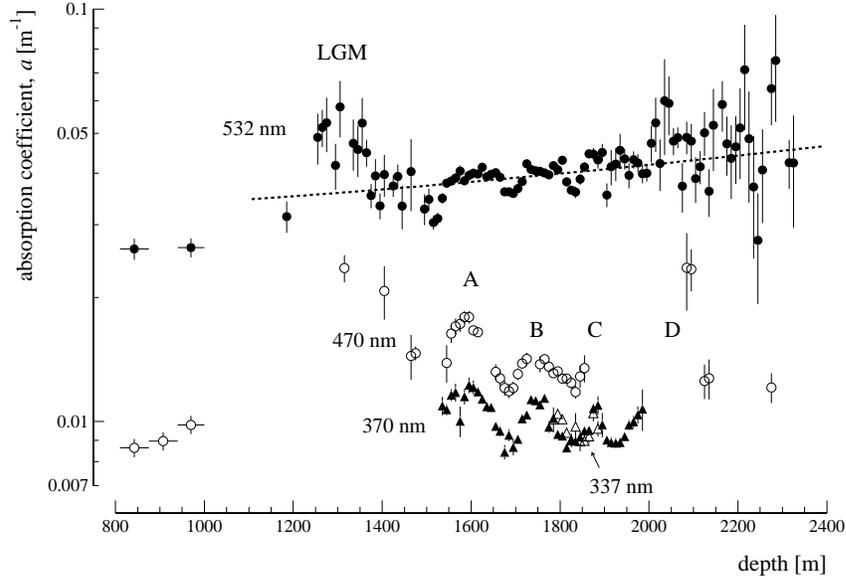}
\caption{Absorption coefficient $a$ as a function of depth below the surface of the icesheet.  From \cite{IcePaper}.}
\label{AbsorptionVsDepth}
\end{center}
\end{figure}

Figure~\ref{AbsorptionVsDepth} shows the measured absorptivity as a function of depth below the surface of the icesheet at the South Pole for several wavelengths.  At the depths of interest for IceCube, the absorption length is around 75 to 100 m.  At a wavelength of 532 nm, the intrinsic absorption of ice dominates the absorption from dust.  The linear rise with depth at this wavelength reflects an increase of ice temperature with depth and the corresponding increase of the intrinsic absorption with temperature.  At the other wavelengths the same climate features as in the scattering coefficient can be observed.

\begin{figure}
\begin{center}
\includegraphics[width=0.5\textwidth]{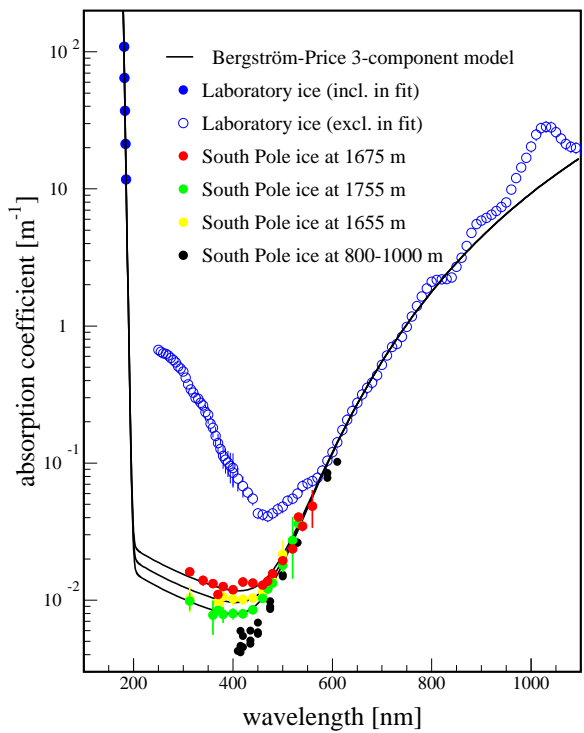}
\caption{Wavelength dependence of the absorptivity $a$ for glacial and laboratory ice.  From \cite{IcePaper}.}
\label{AbsorptionVsWavelength}
\end{center}
\end{figure}

Figure~\ref{AbsorptionVsWavelength} shows the measured wavelength dependence of the absorptivity at several depths.  Note that the glacial ice at the South Pole is an order of magnitude cleaner and more transparent between 200 nm and 400 nm than ice made in a laboratory. 

\begin{figure}
\begin{center}
\includegraphics[width=0.8\textwidth]{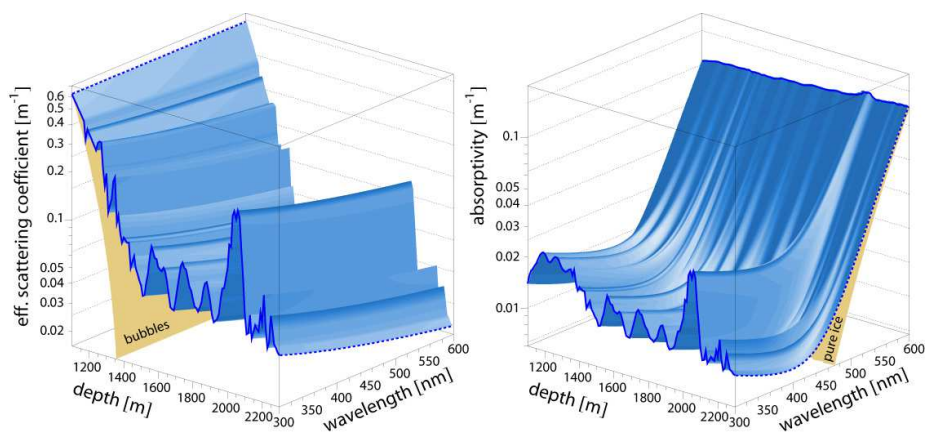}
\caption{Scattering and absorption coefficients as a function of depth and wavelength.  From \cite{IcePaper}.}
\label{ScatteringAndAbsorption3D}
\end{center}
\end{figure}

Figure~\ref{ScatteringAndAbsorption3D} shows the full three-dimensional dependence of scattering and absorption as a function of depth and wavelength.

\subsection{Diffusive Regime}
\label{SDiffusiveRegime}

In the early days of AMANDA, when sensors were deployed in very bubbly ice with scattering lengths on the order of 10 cm, it was realized that photon transport could be described by a diffusion equation with absorption.  For an isotropic point source of light, the photon arrival time distribution at a distance $d$ away is given by \cite{OpticalPropertiesScattering}

$$
u(d,t) = \frac{1}{(4\pi Dt)^{3/2}} \exp\left(\frac{-d^2}{4Dt}\right) \exp\left(\frac{-c_{\mbox{\fontsize{8}{14}\selectfont ice}} t}{\lambda_a}\right)
$$  

\noindent where the diffusion constant $D$ is given by $D=\frac{c_{\mbox{\fontsize{8}{14}\selectfont ice}} \lambda_e}{3}$.  If we integrate this equation over time to get the total number of photons expected at a distance $d$ from the source, we have

$$
N(d) \propto \frac{1}{d} e^{-d/\lambda_p}
$$

\noindent where the propagation length $\lambda_p \equiv \sqrt{\frac{\lambda_e \lambda_a}{3}}$ is typically around 25 m.  This approximation is generally valid in the regime where $d\gtrsim 5\lambda_e$ and has been used in a cascade energy reconstruction which will be discussed in the next chapter.

\section{Hardware}
In essence, the IceCube detector is simply an array of photosensors embedded in the deep glacial ice at the South Pole.  These photosensors detect the Cherenkov light emitted by charged secondary particles produced in high energy neutrino interactions with water molecules.  In practice, the hardware needed to accomplish this task, both in the ice and on the surface, is quite complicated and has required significant research and development.  The digital technology and robust, low-power, low-maintenance design of the IceCube experiment is the culmination of the experience gained by the AMANDA collaboration, which spent more than a decade building and operating a prototype neutrino telescope at the South Pole. 

IceCube's basic sensor element is the Digital Optical Module (DOM), a photomultiplier tube and its accompanying electronics housed in a glass pressure sphere.  DOM's are connected to their neighbors and to the IceCube Lab at the surface via a bundle of twisted-pair copper wires.  A ``string'' consists of 60 DOM's and their cable bundle and is deployed in a vertically-drilled hole in the glacial ice.  DOM's are separated from each other by 17~m on each string, and the string itself is lowered so that all of the DOM's lie at depths between 1,450~m and 2,450~m where the optical properties of the ice are most favorable.

IceCube's baseline plan calls for 80 strings with a nominal inter-string spacing of 125~m.  The 17~m DOM spacing and 125~m string spacing leads to a neutrino energy threshold $\sim$100~GeV.  Recently, approval was granted to install 6 additional densely-instrumented strings at the very center of the array to form an inner detector called DeepCore \cite{DeepCore}.  Each DeepCore string will host 50 DOM's separated from each other by 7~m at depths below 2,100~m where the ice is the clearest.

In addition to this ``in-ice'' detector, IceCube also has a surface array for studying cosmic ray air showers.  IceTop is made up of pairs of frozen water tanks located around 25~m from the top of each in-ice string.  The two tanks are separated from each other by 10~m, and each tank has two DOM's for sensing Cherenkov light from the passage of relativistic charged particles from air showers.

\begin{figure}
\begin{center}
\includegraphics[width=1.0\linewidth]{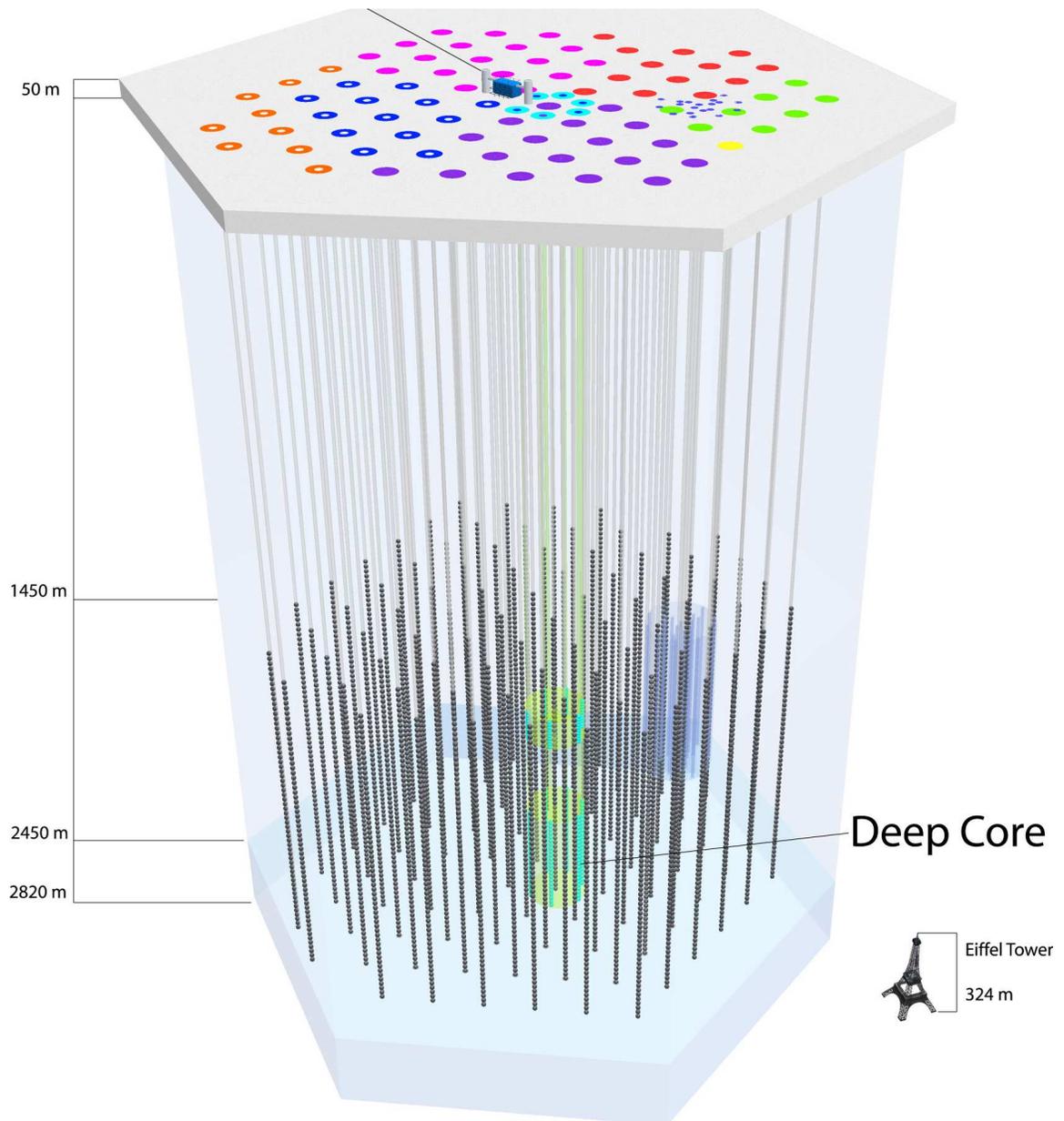}
\caption{Schematic of the IceCube detector.}
\label{ArrayDiagram}
\end{center}
\end{figure}

Figure~\ref{ArrayDiagram} shows a schematic of the IceCube detector.

\newpage

\clearpage

\newpage

\subsection{The Digital Optical Module (DOM)}

\begin{figure}
\begin{minipage}[b]{0.48\linewidth} % A minipage that covers half the page
\centering
\includegraphics[width=7.6cm]{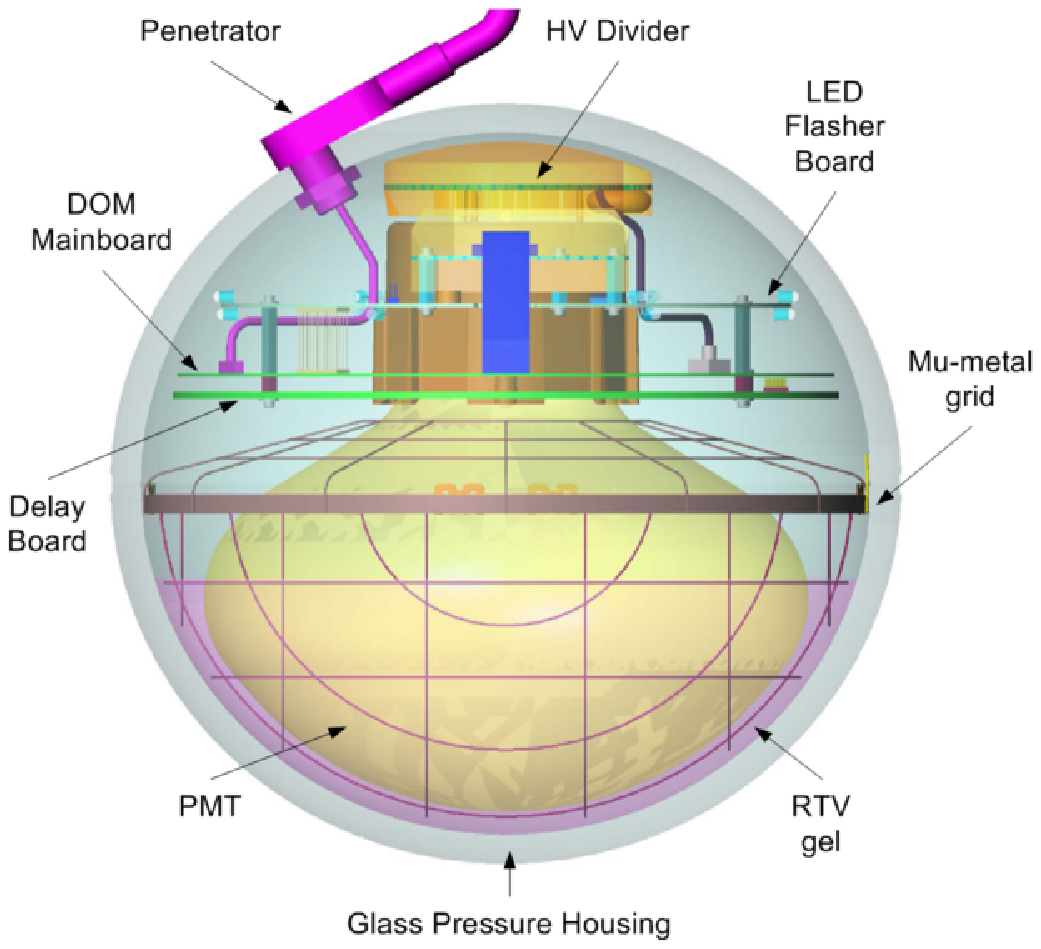}
\end{minipage}
\hspace{0.5cm} %To get a little bit of space between the figures
\begin{minipage}[b]{0.48\linewidth}
\centering
\includegraphics[width=7.6cm]{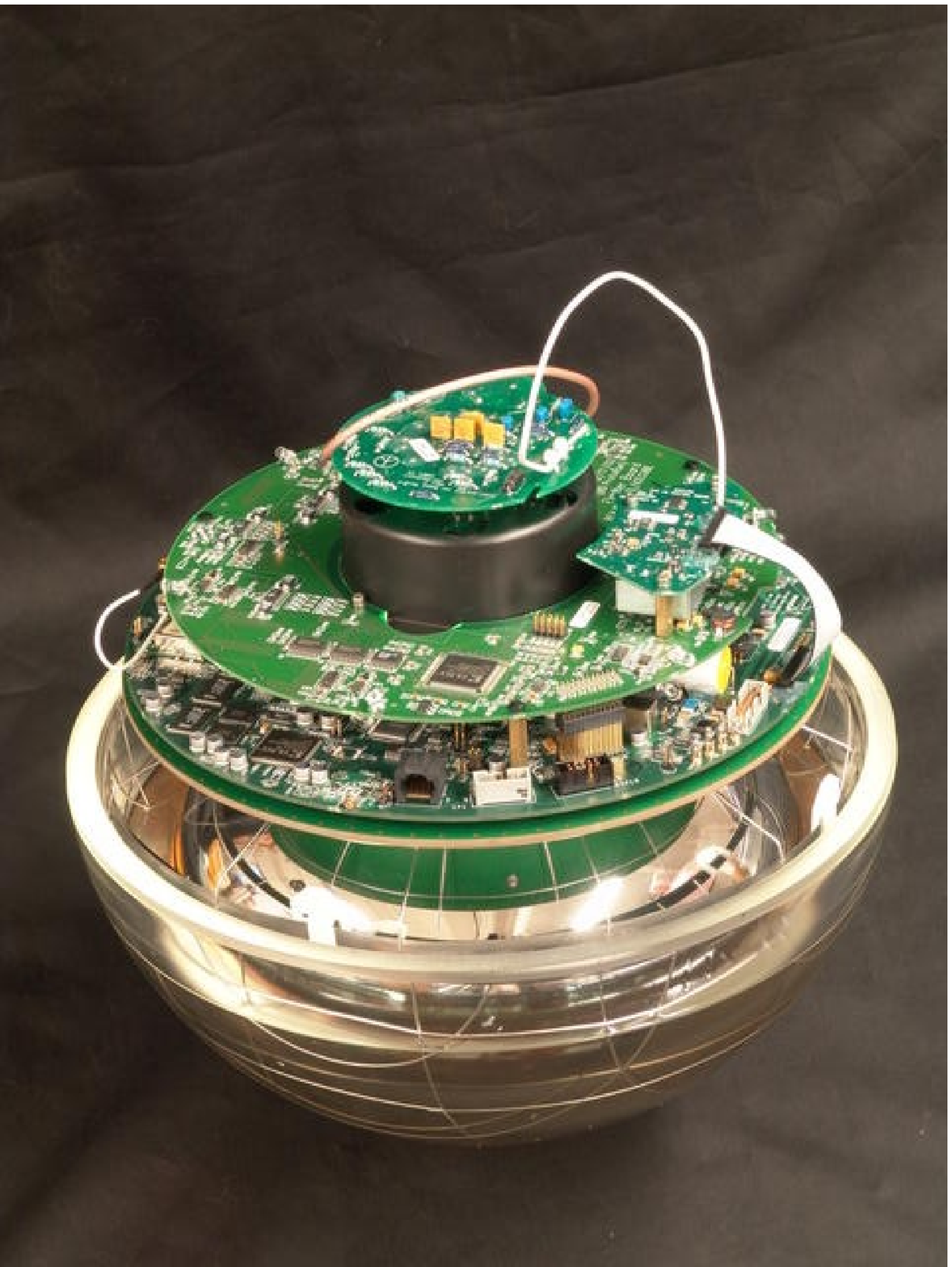}
\end{minipage}
\vspace{0.25cm}
\caption{The Digital Optical Module (DOM).}
\label{TheDOM}
\end{figure}

\begin{figure}
\centering
\includegraphics[width=0.8\textwidth]{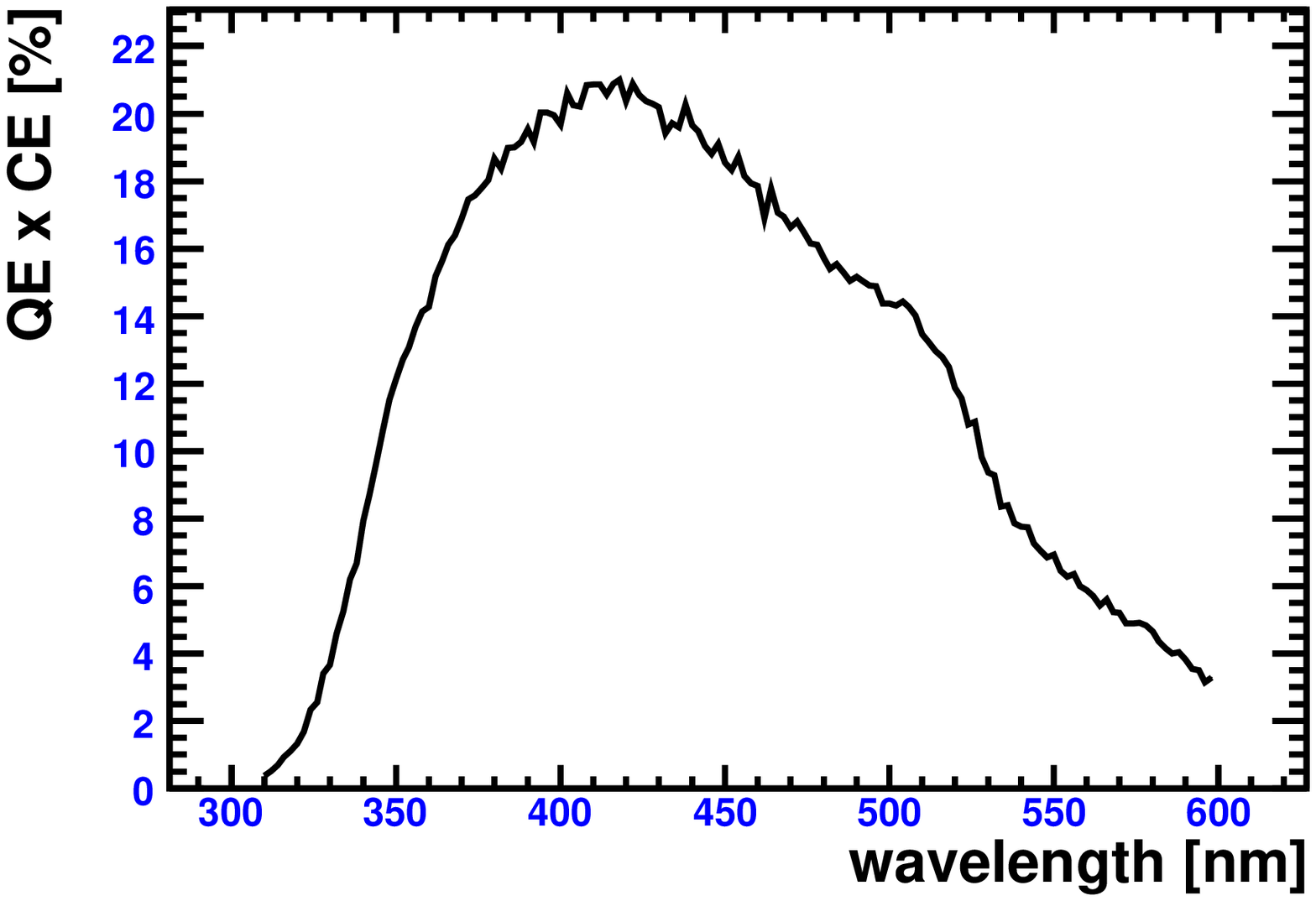}
\vspace{0.5cm} %To get a little bit of space between the figures
\includegraphics[width=0.8\textwidth]{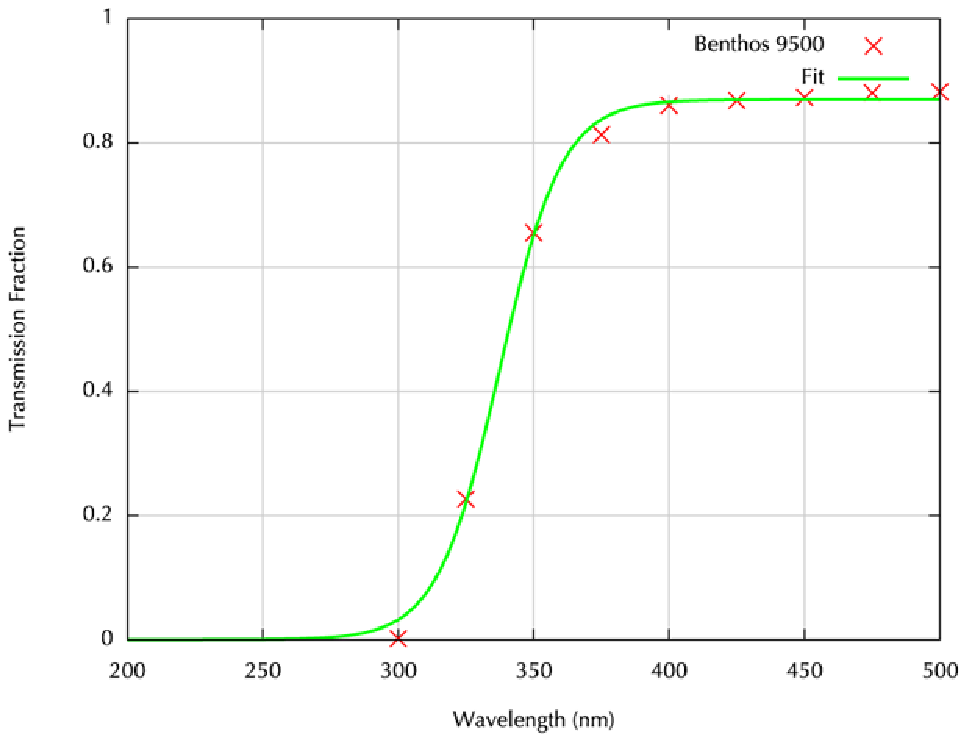}
\vspace{0.25cm}
\caption{Wavelength acceptance of the PMT (top) and glass sphere (bottom).}
\label{WavelengthAcceptance}
\end{figure}

Reference \cite{DOMPaper} gives a detailed description of the hardware and electronics found in the IceCube Digital Optical Module (DOM).  A DOM schematic and photograph are shown in figure~\ref{TheDOM}.  The DOM's light-sensing element is a 10~in. photomultiplier tube (Hamamatsu R7081-02) that achieves a gain of $10^7$ over its 10 dynode stages running at a supply voltage around 1500 V.  Its bialkali photocathode is sensitive between 300 nm and 600 nm with maximum sensitivity around 420 nm.  Figure~\ref{WavelengthAcceptance} shows the measured PMT wavelength acceptance (the product of the quantum efficiency and the collection efficiency) as a function of wavelength.  Dark noise rates are typically around 540 Hz.  

The PMT and its accompanying electronics are housed in a 13~in. diameter, 0.5~in. thick glass pressure sphere so that they can withstand the high pressures (up to 10,000 PSI) in the water column during deployment and refreezing.  The glass was formulated specifically for IceCube to maximize the transmission in the region of PMT sensitivity.  The measured glass transmission as a function of wavelength is also shown in figure~\ref{WavelengthAcceptance}.  The PMT is mechanically and optically coupled to the glass with a silicone gel (GE6156).

The current pulse from the PMT travels down several parallel paths.  One part is sent to a discriminator which decides whether or not the DOM has triggered.  The other part is sent down a 75 ns delay line while the discriminator makes it choice.  Typically the discriminator threshold is set to 0.25 photoelectrons.  If the discriminator fires, it initiates digitization of the signal by several analog-to-digital converters at the end of the delay line.

The Analog Transient Waveform Digitizer (ATWD) \cite{ATWDPaper} is a low-power, fast-sampling ADC.  It samples at 300 MSPS over a time window of 400 ns.  Three different amplifiers are placed before the ATWD inputs (x16, x2, and x0.25) to provide large dynamic range.  If any sample in the x16 channel saturates, the x2 channel is digitized.  If any sample in that channel saturates, then the x0.25 channel is digitized.  These channels are referred to as ATWD0, ATWD1, and ATWD2.  Offline analysis combines these channels into one calibrated waveform.  Each DOM has two separate ATWD's, and it ping-pongs back and forth between them to minimize dead time.

The second ADC, the flash ADC (FADC) provides sparser sampling over a longer time window.  It is meant for particularly long, bright signals.  It samples at 40 MSPS over 6.4 $\mu s$.

Each DOM is equipped with 12 on-board ``flasher'' LED's that operate at 405~nm.  The flashers are spaced around the azimuth of the DOM, 6 pointing in the horizontal direction and 6 tilted upwards at an angle $\sim45^\circ$.  Both the total brightness and the length of the flash are configurable.  The flashers are used for calibration and verification tasks (measuring ice properties, verifying the timing and geometry of the array) as well as more physics-centric tasks.  Since the flashers are essentially point sources of light, they are useful for verifying the position and energy reconstruction algorithms that are used for neutrino-induced cascades.

\subsection{Local Coincidence}

Each DOM also has circuitry for implementing a local coincidence requirement between neighboring DOM's.  When a DOM's discriminator fires, it sends signals directly to its two nearest neighbors to see if they have also fired.  These DOM's can relay the signal on to their neighbors to probe the behavior of more distant DOM's on the string.  Typically, local coincidence conditions are imposed that require at least one of a DOM's two nearest neighbors or two next-to-nearest neighbors to have fired in order for data from that DOM to be sent to the surface.  The local coincidence time window is typically $\pm 1 \mu s$ around the launch of a DOM.

Local coincidence significantly reduces the number of accidental noise hits that enter the datastream.  If a DOM fires at $r=540$ Hz from dark noise, we can calculate the probability that one of its four nearest neighbors will also have a noise hit in the local coincidence window $t=2 \mu s$.  The probability that there are no hits in any of the four DOM's is 

$$
(e^{-rt})^4 = e^{-4rt} \approx 1-4rt
$$ 

\noindent The probability of at least one hit in those four DOM's is thus

$$
1-(1-4rt)=4rt
$$

\noindent So the rate at which a DOM and at least one of its four nearest neighbors trigger from noise is $4r^2 t$.  In the case of the numbers given above this is $\approx 0.6$ Hz, three orders of magnitude below the dark noise rate.

\subsection{Triggering}

IceCube's triggering system is very flexible and can accommodate multiple triggers at the same time.  It can be configured to use a simple multiplicity trigger (SMT), which fires when a given number of DOM's are hit within a given time window.  It can be configured to use a so-called string trigger, which fires when a given number of DOM's on a single string are hit within a given time window.  The string trigger is useful for low-energy atmospheric or WIMP-annihilation neutrinos.  The triggering system also has the ability to search for more complicated event topologies in real-time.  For the neutrino-induced cascade analysis in this dissertation, the simple majority trigger was used that required 8 hit DOM's in a time window of 5 $\mu$s.   

\section{Drilling and Deployment}

Each string is deployed in a vertically-drilled hole in the glacial ice.  Drilling proceeds in two stages.  First, a firn drill melts through the loosely-packed snow and air that make up the $\sim$100 m thick firn.  It operates by running heated fluid through a closed loop of coils which are in contact with the snow.  Hot-water drilling, which is used for the deep ice drilling, is not appropriate for the firn as water leaks into the loosely packed snow rather than remaining in the hole. 

After firn drilling, the remainder of the hole is drilled by IceCube's dedicated 5~MW hot-water drill.  An array of heaters is used to heat water, which is then pumped to the drill-head.  The drill-head is attached to a heavy weight stack and shoots a high-pressure jet of hot water into the bottom of the hole.  A pump returns water back to the heaters for re-heating.  

The entire drilling process takes 2 to 3 days.  When it is complete, a column of water is left in the hole, which is approximately half a meter in diameter.  Next, a team of deployers makes the mechanical and electrical connections for each of the 60 DOM's to the string, and the string is then lowered into the hole.  Deployment takes $\sim$ 12 hours and is completed well before the hole begins to refreeze.  The final depth of the string is measured by several pressure sensors which measure pressure in the water column.  

\begin{figure}
\begin{center}
\includegraphics[width=1.0\linewidth]{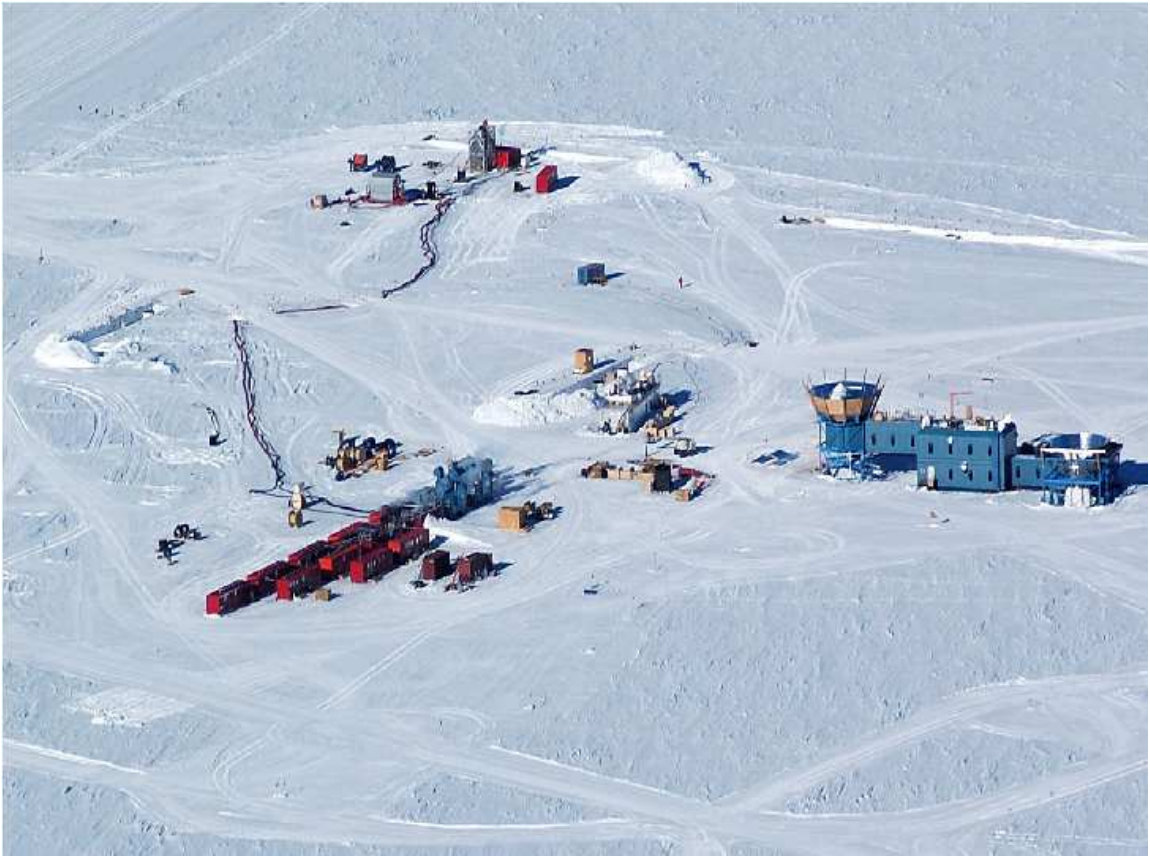}
\caption{IceCube's 5~MW hot-water drill at work.  In the lower left, a series of red trailers houses heaters and pumps.  Hot water flows from these trailers through the black hose which snakes up to the tower structure in the upper center.  This tower is situated over the hole which was being drilled at the time.}
\label{EHWD}
\end{center}
\end{figure}

Figure~\ref{EHWD} shows the 5~MW hot-water drill in action.  In the lower left of the image, a series of red trailers houses the heaters and pumps.  Hot water flows from these trailers through the black hose which snakes up to the tower structure in the upper center of the image.  This tower is situated over the hole which was being drilled at the time.

\section{Status}

During the austral summer of 2004-2005, the first IceCube string was successfully deployed and commissioned at the South Pole.  Since then, the pace of deployment has been rapidly accelerating with each season.  In the 2005-2006 season, 8 strings were installed and in 2006-2007, 13 more were deployed for a total of 22.  This 22-string array, known as IC-22, is the array used for the analysis in this dissertation.  The 2007-2008 season saw the installation of 18 more strings, bringing the array up to 40 strings.  IC-40 just completed its one year data-taking phase.  This past season, 2008-2009, 19 strings were deployed.  Currently, IceCube is taking data in its IC-59 configuration.  The full array is due for completion in 2010-2011.

\begin{figure}
\begin{center}
\includegraphics[width=0.8\textwidth]{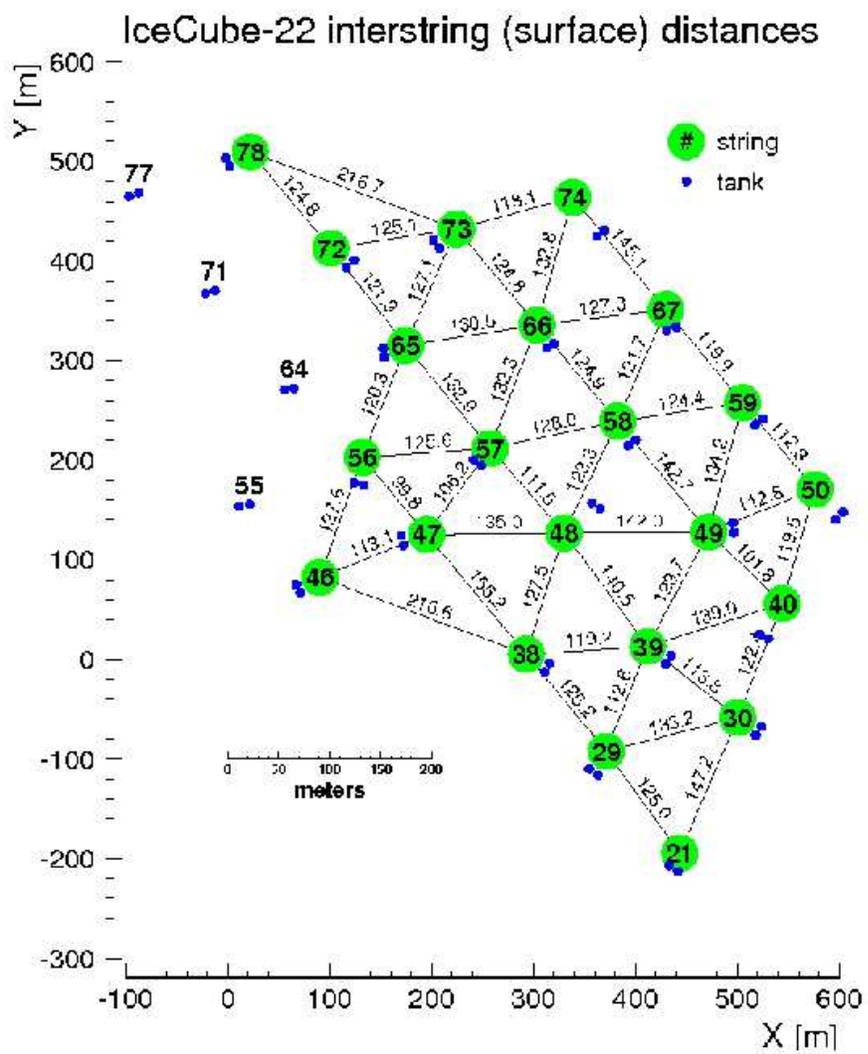}
\caption{Surface map of the 22-string IceCube array (IC-22).}
\label{IC22Distances}
\end{center}
\end{figure}

Figure~\ref{IC22Distances} shows a surface map of the string layout of IC-22.  Several salient features of this string layout are important for a search for neutrino-induced cascades.  As will become clear later, the main background for a neutrino-induced cascade search consists of muons which have a large radiative energy loss inside the detector that mimics a cascade.  In order to veto these events, it is desirable to have a contained region of the detector which is surrounded by many strings to capture early light from the muon.  For IC-22, there are only 6 strings (57, 48, 39, 66, 58, and 49) which have a single layer of strings surrounding them.  This means that the veto region is small, and vetoing muon events with large radiative losses will be a challenge.  Second, the geometry is quite irregular, with several strings (78, 72, and 46) jutting out from the main body of array.  These strings are likely to be less useful for a cascade search, and so the effective size of the detector will be reduced.

%% file: chapters/chapter4.tex
\chapter{Reconstruction of Neutrino-Induced Cascades}\label{chapter:chapter4}

As we have seen, a particle shower or ``cascade'' from a charged-current or neutral-current neutrino interaction is a relatively local deposition of energy.  In a sparsely instrumented detector like IceCube, these cascades look effectively like point sources of Cherenkov light.  For each candidate cascade event in our detector, we would like to reconstruct its physical parameters.  This chapter discusses the methods for doing so.  Event reconstruction will allow us to separate neutrino-induced events from air shower muon background events and to make sense of the resulting neutrino sample.

\section{Reconstruction Basics}

A neutrino-induced cascade is characterized by a set of unknown parameters  $\{{\mathbf a}\}$:

$$
\{{\mathbf a}\} \equiv \{\vec{r}_v, t_v, E, \hat{n}\}
$$ 

\noindent where $\vec{r}_v$ is the cascade vertex position, $t_v$ is the vertex time, $E$ is the visible energy of the cascade, and $\hat{n}$ is its direction.   To reconstruct these parameters from our observational data, we make use of maximum likelihood estimation \cite{AMANDAReconstruction}.  For a given set of observational data $\{{\mathbf x}\}$, we ask which cascade parameters $\{{\mathbf a}\}$ maximize a given likelihood function $\mathscr{L} ( {\mathbf x} | {\mathbf a} ) $.  If the $x_i$ components of the data $\{{\mathbf x}\}$ are independent, we can write

$$
\mathscr{L} ( {\mathbf x} | {\mathbf a} ) = \prod_{i} p(x_i | {\mathbf a} )
$$

\noindent where  $p(x_i | {\mathbf a} )$ gives the probability of our experiment measuring the quantity $x_i$ given the cascade parameters $\{{\mathbf a}\}$.  For example, $p(x_i | {\mathbf a} )$ could tell us the probability of observing a single photoelectron at a given time in a single DOM from a cascade with parameters $\{{\mathbf a}\}$.  Or, it could tell us the probability of observing a total of N photoelectrons in a single DOM from a cascade with parameters $\{{\mathbf a}\}$.  Maximizing the likelihood function $\mathscr{L} ( {\mathbf x} | {\mathbf a} )$ corresponds to finding the set of cascade parameters that makes the observed data most likely to have occurred.  

Various likelihood functions can be constructed for neutrino-induced cascades.  The most important uses photoelectron timing information to reconstruct the position and time of the cascade vertex.  This is discussed in section~\ref{SSVertexReconstructions}.  In addition, two different likelihood functions are available to reconstruct the visible energy of the cascade.  The traditional method does not take into account the depth-dependent optical properties of the ice because of the conceptual and technical difficulties of doing so.  It is described in section~\ref{SSPHitPNoHitEnergy}.  For this analysis, a better performing depth-dependent energy reconstruction was developed which is described in section~\ref{SDepthDependentEnergy}.  

Directional reconstruction of cascades has proven a difficult challenge.  While the light emission from a cascade starts out with a directional anisotropy due to the Cherenkov light emission, scattering in the ice tends to isotropize it very quickly.  For this reason, directional reconstruction was not performed at any stage of this analysis.  Very recent work on improved reconstruction methods suggests that it may be possible to achieve a directional resolution of $30^{\circ}$--$35^{\circ}$ for 10 TeV--10 PeV cascades in the near future \cite{EikeICRC}.

In practice, it is numerically easier to minimize a function rather than to maximize it.  Consequently, we work with the negative log of the likelihood function instead of the the likelihood function itself.  In addition, because the likelihood landscapes are often quite complex, it is desirable to have a set of seed parameters $\{\hat{{\mathbf a}}\}$ that we believe lies in the vicinity of the true global minimum of the negative log likelihood function.  This seed is usually provided by a quick-running ``first-guess'' algorithm.  The first-guess algorithm for the cascade vertex and time which is used to seed the likelihood reconstructions is described in the next section.

\section{The {\tt cfirst} First-Guess Algorithm}
The {\tt cfirst} program provides a first guess for the cascade vertex by making an analogy between photoelectrons observed in DOM's (also known as ``hits'') and masses distributed in space \cite{AMANDAReconstruction}.  We take as our first guess the center-of-gravity (COG) of the hits defined by:

$$
\overrightarrow{COG} \equiv \sum_{i=1}^{N_{\mbox{\fontsize{8}{14}\selectfont ch}}}(a_i)^{w} \cdot \vec{r_i}
$$

\noindent where $a$ is the number of photoelectrons recorded by a DOM at position $\vec{r_i}$
, $w$ is an arbitrary amplitude weighting factor,  and $N_{\mbox{\fontsize{8}{14}\selectfont ch}}$ is the number of ``hit'' DOM's (that is, the number of DOM's which received photoelectrons in the event).  

In general, this first guess works well.  In the x and y directions the COG is symmetric about the true vertex position with a resolution of around 25~m.  In the z direction, the 17~m resolution reflects the denser DOM spacing along the string.  However, in the z direction the COG tends to lie 10~m above the true vertex.  This is because a DOM's PMT points downwards, and so the DOM's above a cascade see more light than the DOM's below .  The z value of the first guess also tends to be poor near layers of dirty ice in the detector, since in these regions light may be preferentially absorbed in one direction.  Finally, the COG must by definition lie within the instrumented volume, so it is not a good guess for events which are truly outside of the detector.

After the COG position is found, it is used to calculate a first guess for the vertex time.  
The time calculation relies on the fact that hits very close to the cascade vertex have a good chance of having arrived unscattered or close to unscattered.  We choose our vertex time to maximize the number of these direct hits.  

For a hit at time $t_i$ at a distance $d_i$ from the COG, we start by assuming that it came directly from the COG and therefore take as our trial vertex time $\hat{t}_v = t_i - \frac{d_i}{c_{ice}}$.  Next, we define a time residual variable which tells us how delayed from direct travel another hit  $t_j$ is, given our trial vertex time:

$$
t_r \equiv t_j - t_{\mbox{\fontsize{8}{14}\selectfont direct}} = t_j - (\hat{t}_v + \frac{d_j}{c_{\mbox{\fontsize{8}{14}\selectfont ice}}}) = t_j - (t_i - \frac{d_i}{c_{\mbox{\fontsize{8}{14}\selectfont ice}}} + \frac{d_j}{c_{\mbox{\fontsize{8}{14}\selectfont ice}}}) =  (t_j- \frac{d_j}{c_{\mbox{\fontsize{8}{14}\selectfont ice}}})-(t_i- \frac{d_i}{c_{\mbox{\fontsize{8}{14}\selectfont ice}}})
$$ 

\noindent Positive values of $t_r$ indicate that the light was scattered on its way from the COG to the hit DOM and so arrived later.  We calculate the residuals for all of the other hits in the event given our trial vertex time and count how many of them lie in a [0 ns, 200 ns] window.  These are the hits that are close to direct for our trial vertex time.  

We repeat this procedure taking the trial vertex time from each hit in the event.  Our final vertex time is chosen to be the earliest trial time that results in at least $N_{\mbox{\fontsize{8}{14}\selectfont trigg}}=3$ hits in our [0 ns, 200 ns] direct time window from this procedure.  In case the $N_{\mbox{\fontsize{8}{14}\selectfont trigg}}=3$ condition is not met, the vertex time is estimated as the time of the first hit in the event.  This first guess method for the vertex time has a resolution of around 90 ns, though by ignoring scattering it tends to push the vertex time $\approx 35$ ns too late.

\section{Log-Likelihood Reconstructions}
This first-guess for the cascade vertex and time can then be used to seed a full likelihood reconstruction of the position, time, and energy.

\subsection{Vertex Reconstructions}
\label{SSVertexReconstructions}
The vertex likelihood functions use photoelectron arrival time information to reconstruct the cascade position and time.  The individual pdf's $p(x_i | {\mathbf a})$ are parameterized in terms of the time residual defined earlier:

$$
t_r \equiv t_j - t_{\mbox{\fontsize{8}{14}\selectfont direct}}
$$

\noindent The pdf $p(t_r | {\mathbf a})$ gives the probability of measuring a single photoelectron with time residual $t_r$ in a given DOM  from a cascade with parameters $\{{\mathbf a}\}$.

If South Pole glacial ice were not a strongly scattering medium, this time residual pdf would be a simple delta function at $t_r=0$.  However, scattering and absorption in the ice significantly complicate the situation.  For a vertex at a given position in the detector, light will propagate through a layered system of scatterers and absorbers.  Very close to the vertex, we still expect the residual function to peak around zero, as there is still a high probability of unscattered light.  But now the distribution should have a long tail towards large residuals as well.  Far from the vertex, we expect very little unscattered light.

Traditionally, AMANDA and IceCube researchers have used an analytic function called the ``Pandel function'' as a convenient parameterization of the time residual pdf as a function of the distance $d$ from the cascade vertex \cite{PandelThesis}.  The Pandel function is given by:

$$
p(t_r | d) \equiv \frac{1}{N(d)} \frac{\tau^{(-d/\lambda)} {t_{r}}^{(d/\lambda-1)}}{\Gamma(d/\lambda)} \times 
 e^{-\left(t_r(\frac{1}{\tau} + \frac{c_{\mbox{\fontsize{8}{14}\selectfont ice}}}{\lambda_a}) + \frac{d}{\lambda_a}\right)}
$$

\noindent where the normalization $N(d)$ is

$$
N(d) \equiv e^{-d/\lambda_a} \left( 1+\frac{\tau c_{\mbox{\fontsize{8}{14}\selectfont ice}}}{\lambda_a} \right)^{-d/\lambda}
$$

\noindent Here, $\lambda_a$ is the absorption length, $\lambda$ and $\tau$ are free parameters, and $\Gamma(x)$ is the standard gamma function.  The free parameters $\lambda$ and $\tau$ are determined by fitting this functional form to time residual distributions from a full Monte Carlo simulation of scattering and absorption in layered glacial ice.  The values used for the reconstructions in this analysis are $\tau = 450$~ns, $\lambda = 47.0$~m, and $\lambda_a = 98.0$~m.  They have been chosen to fit the widest range of time residual distributions and are not related to actual physical parameters.

\begin{figure}
\begin{minipage}[b]{0.48\linewidth} % A minipage that covers half the page
\centering
\includegraphics[width=7.6cm]{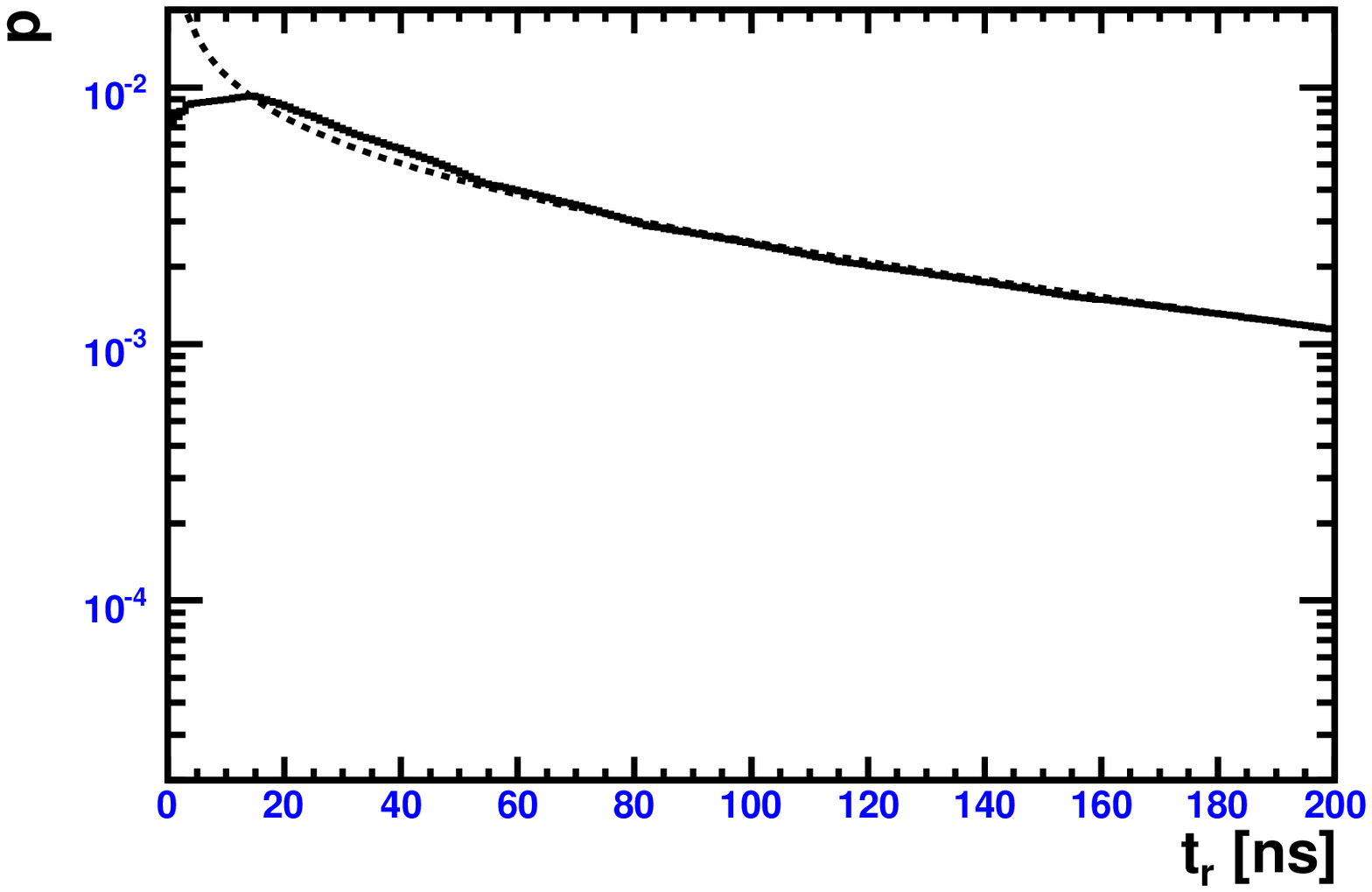}
\end{minipage}
\hspace{0.5cm} %To get a little bit of space between the figures
\begin{minipage}[b]{0.48\linewidth}
\centering
\includegraphics[width=7.6cm]{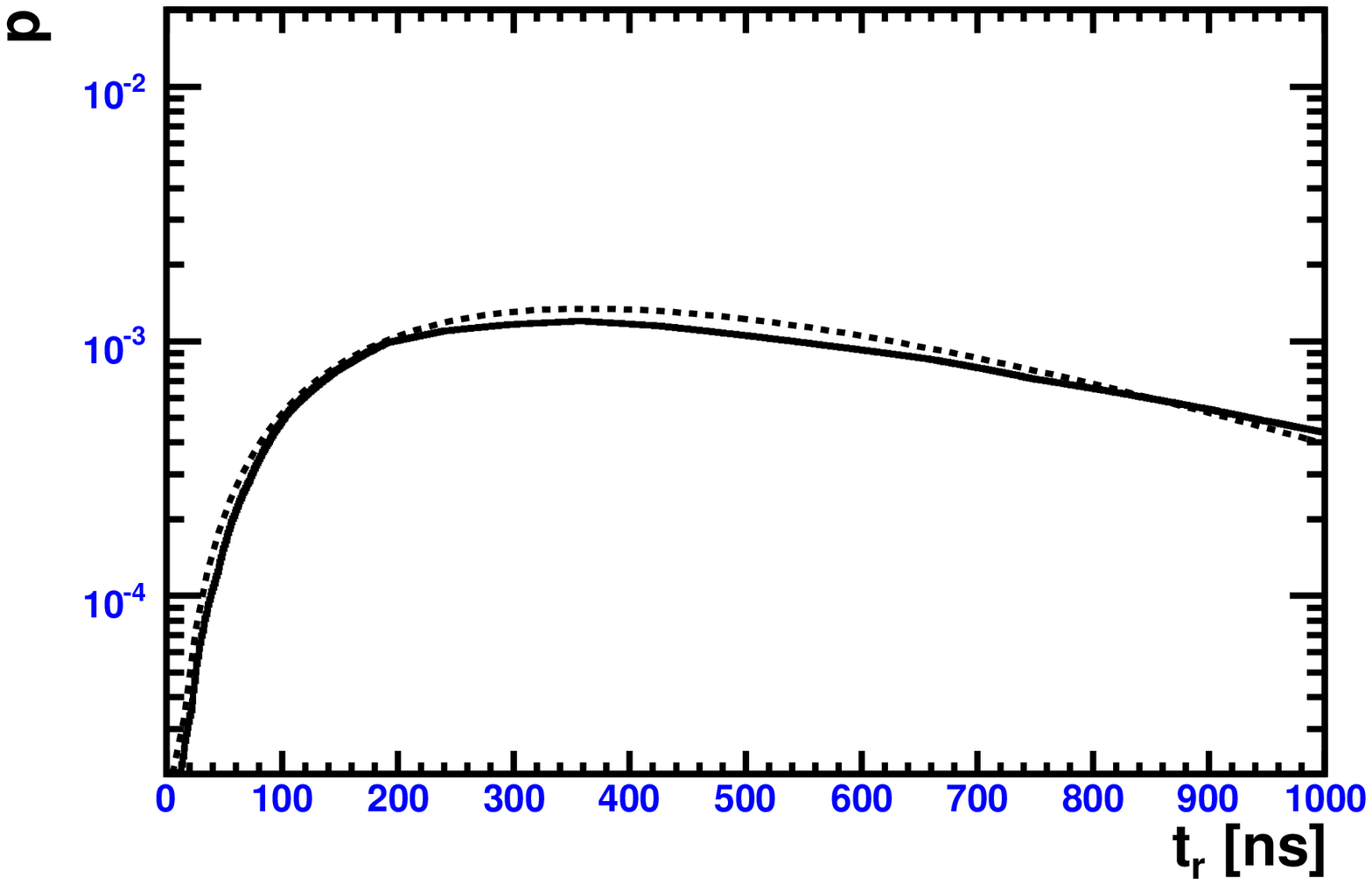}
\end{minipage}
\vspace{0.25cm}
\caption{Pandel time residual pdf (dashed) compared to {\tt photonics} \cite{PhotonicsPaper} Monte Carlo simulation (solid) at distances of 25 m (left) and 125 m (right) from a cascade vertex located at a depth of 1948 m.}
\label{PhotonicsPandel}
\end{figure}

Time residual distributions at two different distances are shown in figure~\ref{PhotonicsPandel}.  For small distances, the factor ${t_{r}}^{(d/\lambda-1)}$ causes the probability to rise sharply to a pole at zero time residual.  This reflects the high probability for unscattered photons at short distances from the vertex.  At a critical distance $d_{\mbox{\fontsize{8}{14}\selectfont crit}} \equiv \lambda$, the exponent of this factor changes sign and the distribution no longer peaks at zero time residual.  This reflects the fact that there is very little chance of getting direct, unscattered light far from the vertex.

Two additional features must be added to this pdf to make it a full description of the detector.  First, the DOM PMT and electronics don't have perfect timing resolution.  To account for this, we smear the Pandel residual distribution above with a gaussian representing our finite timing resolution.  The gaussian was chosen to have mean $\mu=0$ ns and $\sigma=15$ ns.  Second, we account for noise hits by adding a small, flat noise probability term of $10^{-10}$ to the pdf.

The final pdf incorporating all of these elements is known as the UPandel function $p_U(t_r | d)$ and has the following form:

$$
p_U(t_r | d) \equiv 
\begin{cases} 
N(0, \sigma) & \text{for $t_r < 0$} 
\\ 
P_3(t_r | d) & \text{for $0 < t_r < \sqrt{2\pi}\sigma$} 
\\
p(t_r | d) & \text{for $t_r > \sqrt{2\pi}\sigma$} 
\end{cases}
$$

\noindent where $N(0, \sigma)$ is the gaussian representing our timing resolution, $p(t_r | d)$ is the normal Pandel function defined above, and $P_3(t_r | d)$ is a third-order polynomial interpolation function.   Our full likelihood function is then:

$$
\mathscr{L}_{\mbox{\fontsize{8}{14}\selectfont UPandel}} ( {\mathbf x} | {\mathbf a} ) =  \prod_{\mbox{\fontsize{8}{14}\selectfont hits}} p_U(t_r | {\mathbf a} )
$$

\begin{figure}
\centering
\includegraphics[width=0.65\textwidth]{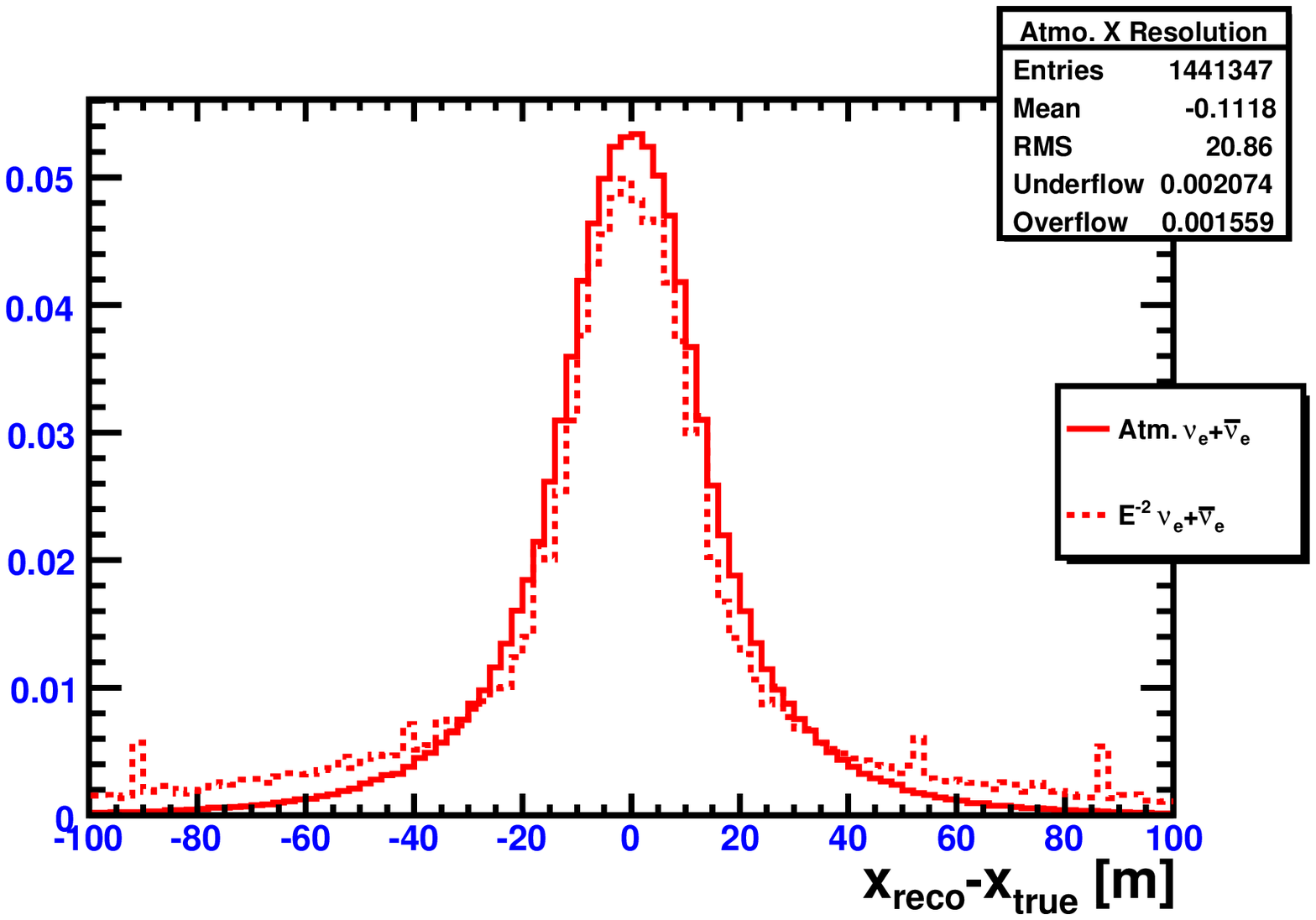}
\includegraphics[width=0.65\textwidth]{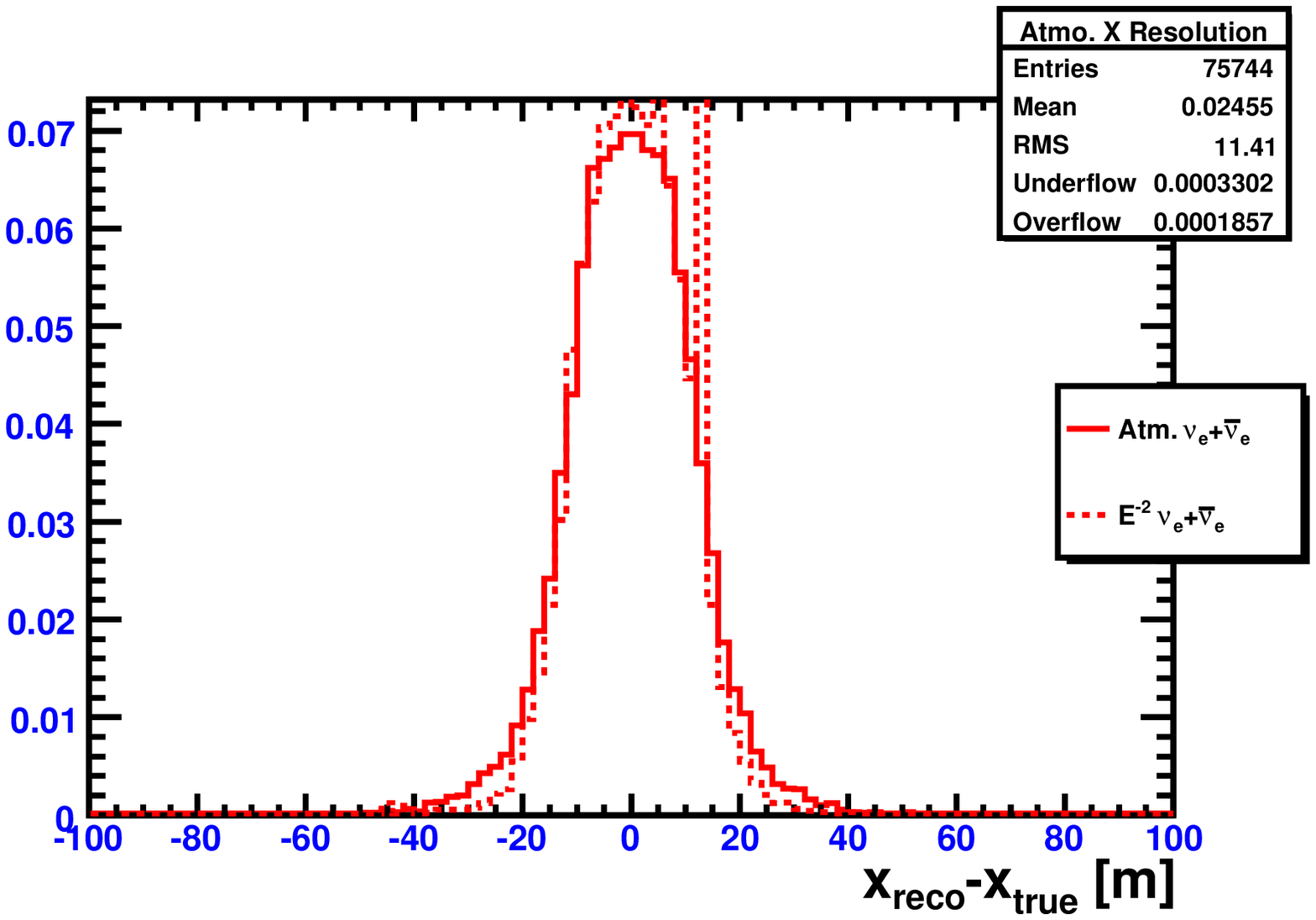}
\includegraphics[width=0.65\textwidth]{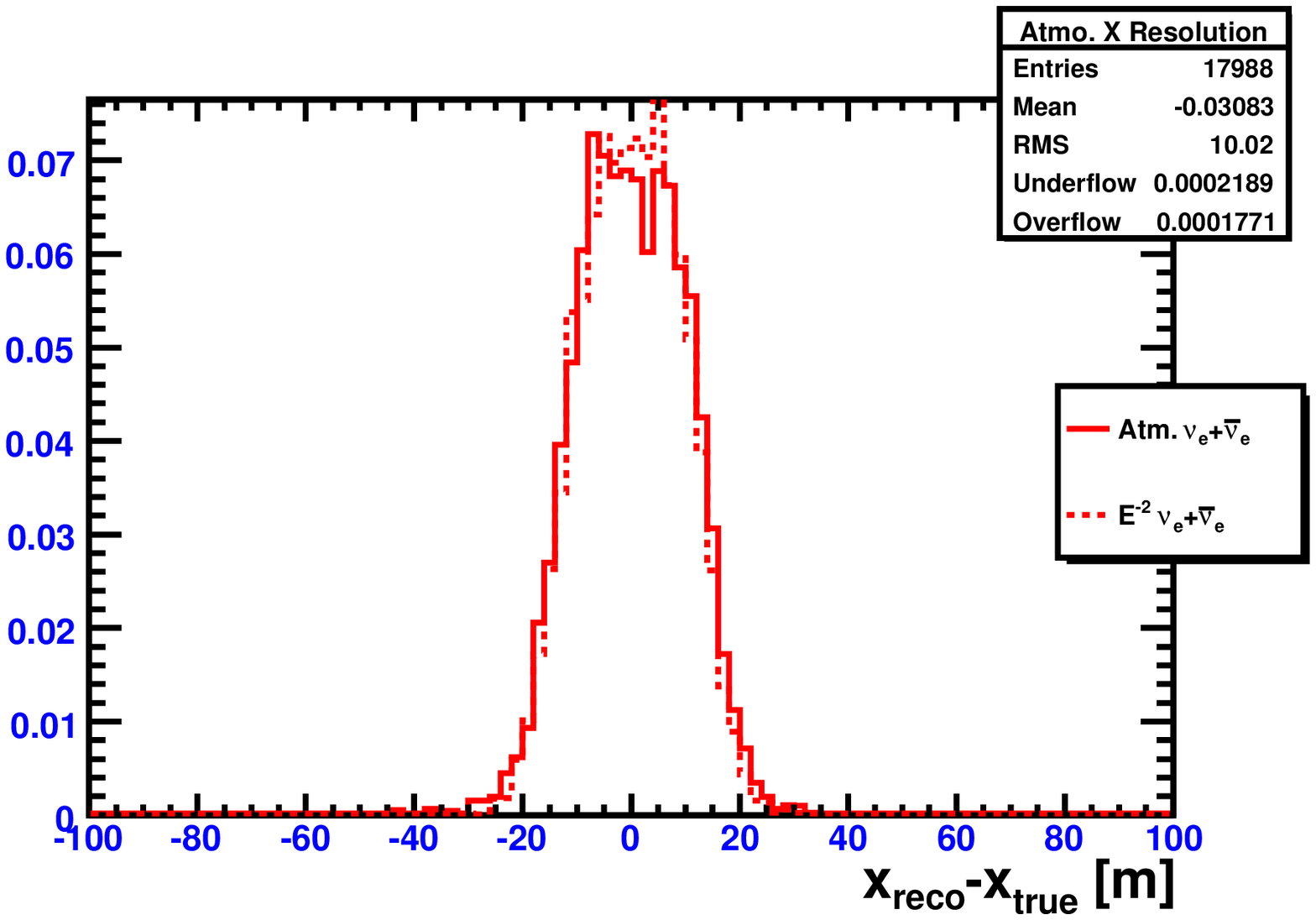}
\caption{x vertex position resolutions for $E^{-2}$ (dashed) and atmospheric (solid) spectra at level 3a (top), level 4a (middle) and final level 1 (bottom) of this analysis.}
\label{VertexResolutionsX}
\end{figure}

\begin{figure}
\centering
\includegraphics[width=0.65\textwidth]{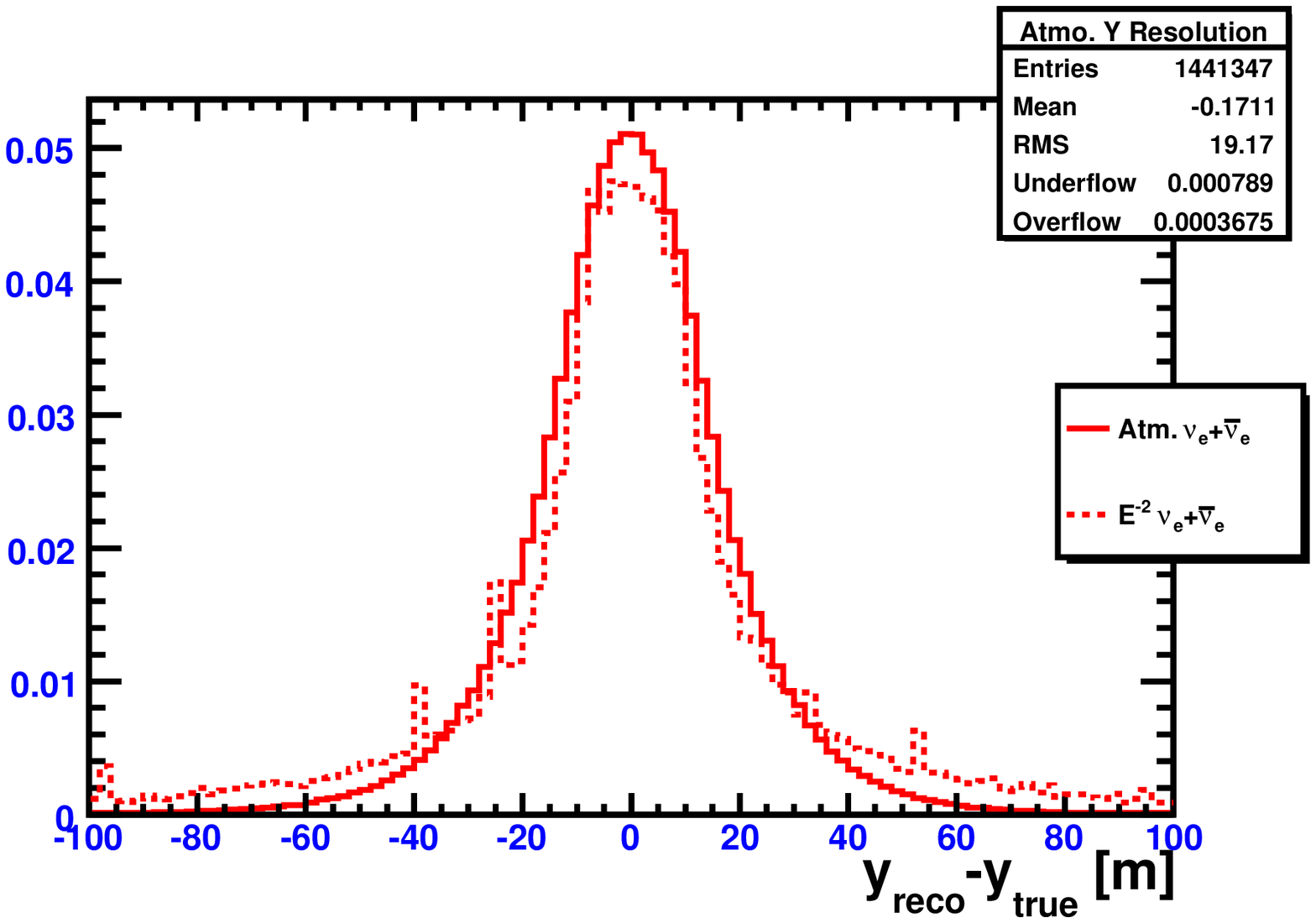}
\includegraphics[width=0.65\textwidth]{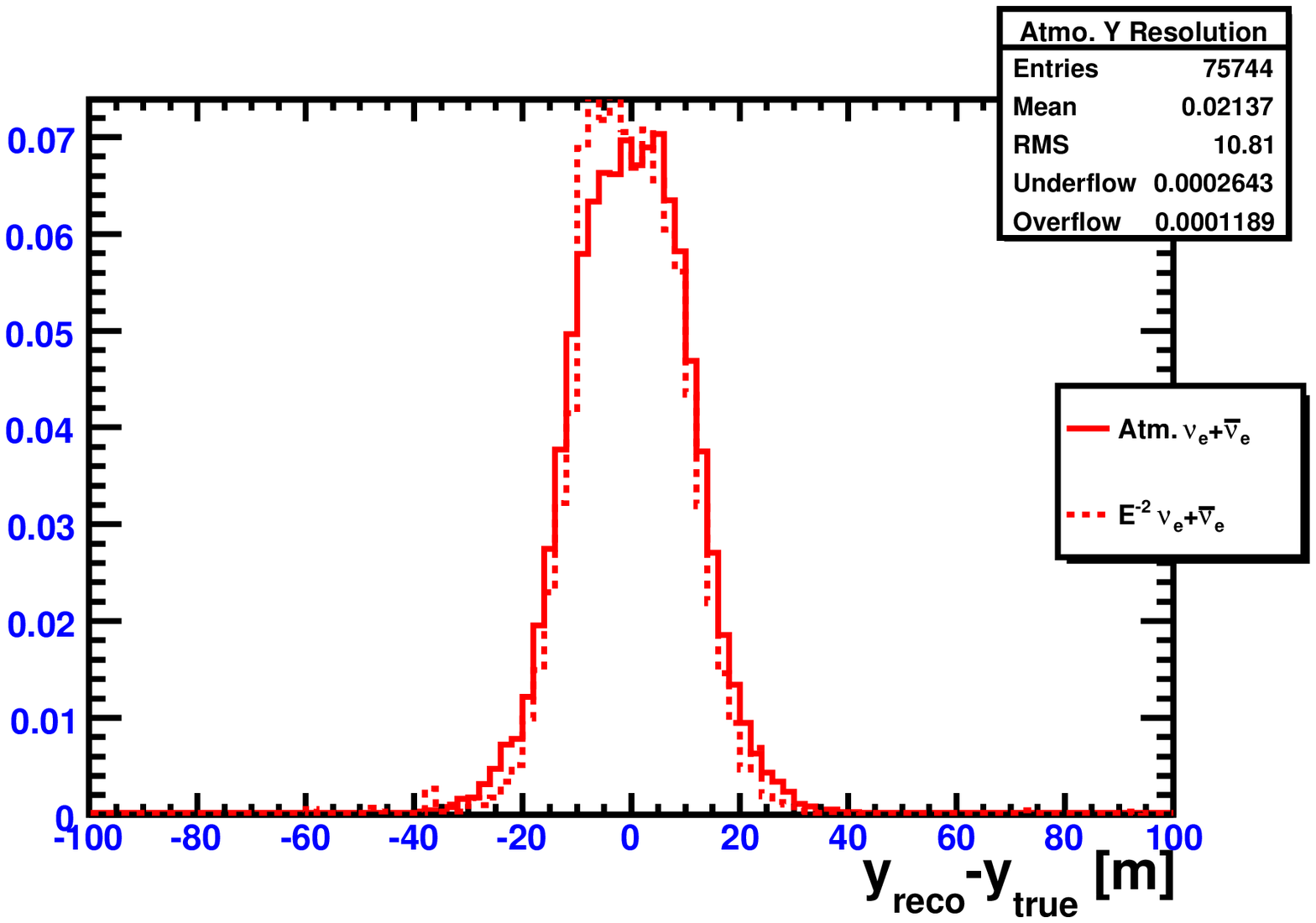}
\includegraphics[width=0.65\textwidth]{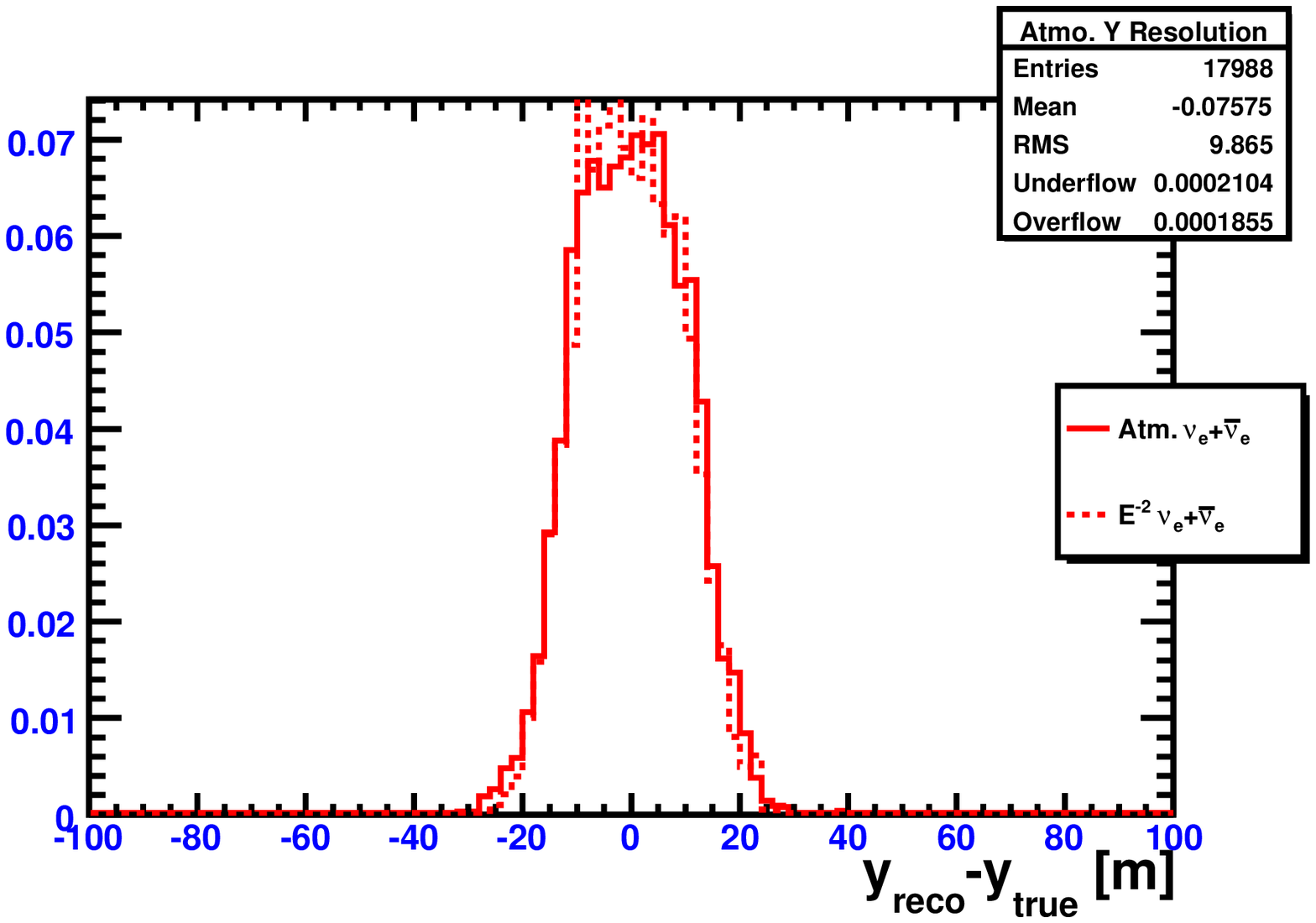}
\caption{y vertex position resolutions for $E^{-2}$ (dashed) and atmospheric (solid) spectra at level 3a (top), level 4a (middle) and final level 1 (bottom) of this analysis.}
\label{VertexResolutionsY}
\end{figure}

\begin{figure}
\centering
\includegraphics[width=0.65\textwidth]{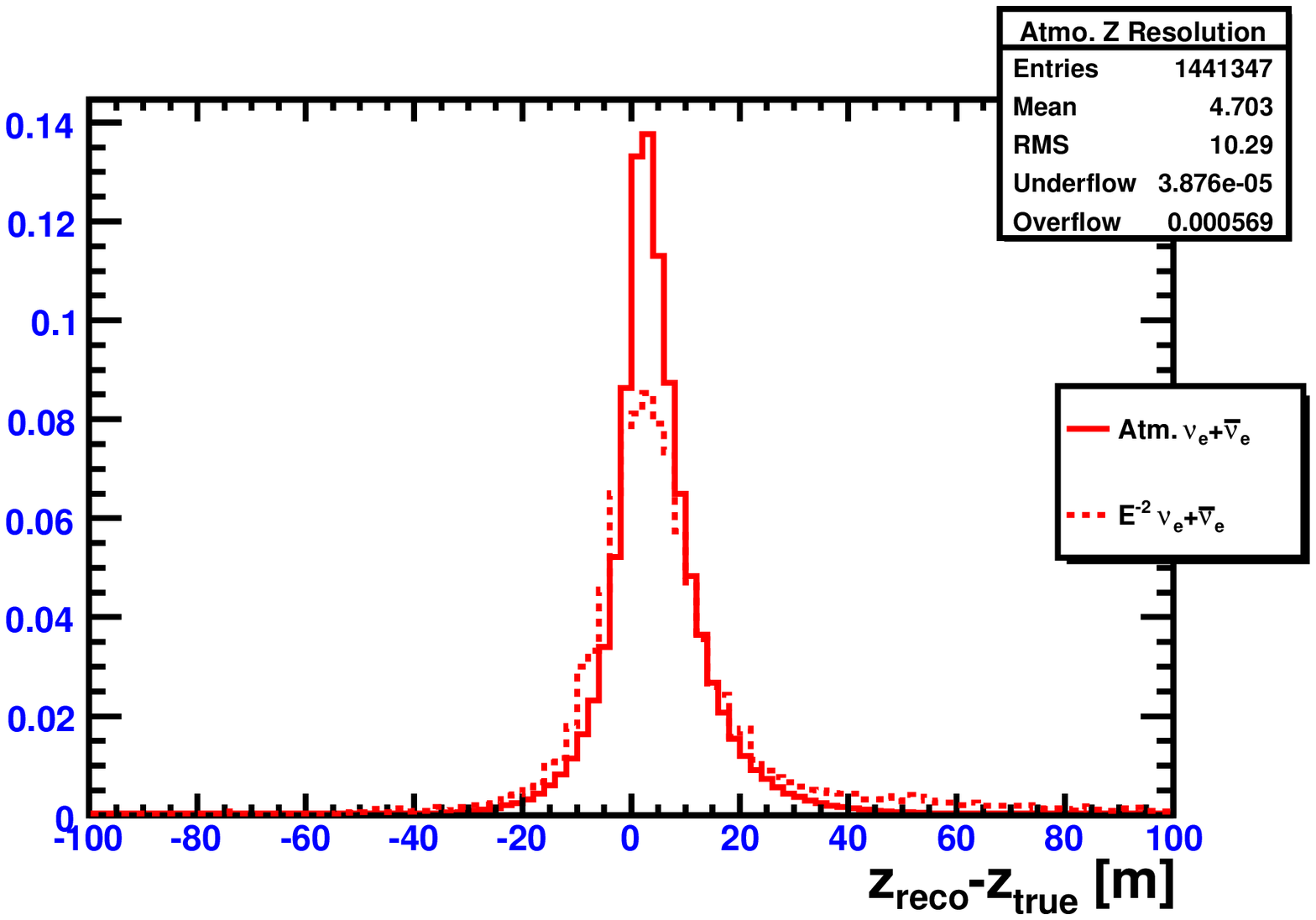}
\includegraphics[width=0.65\textwidth]{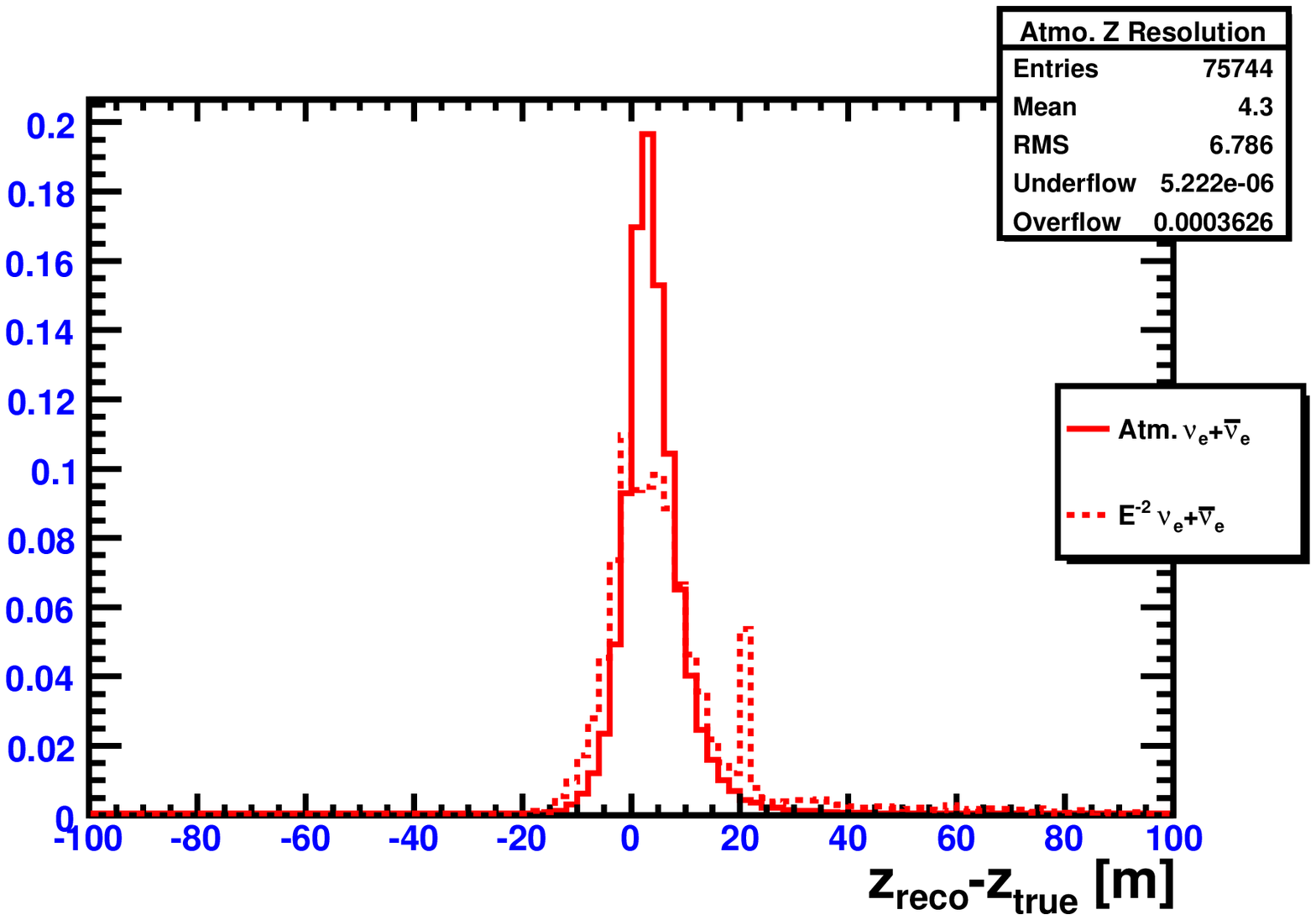}
\includegraphics[width=0.65\textwidth]{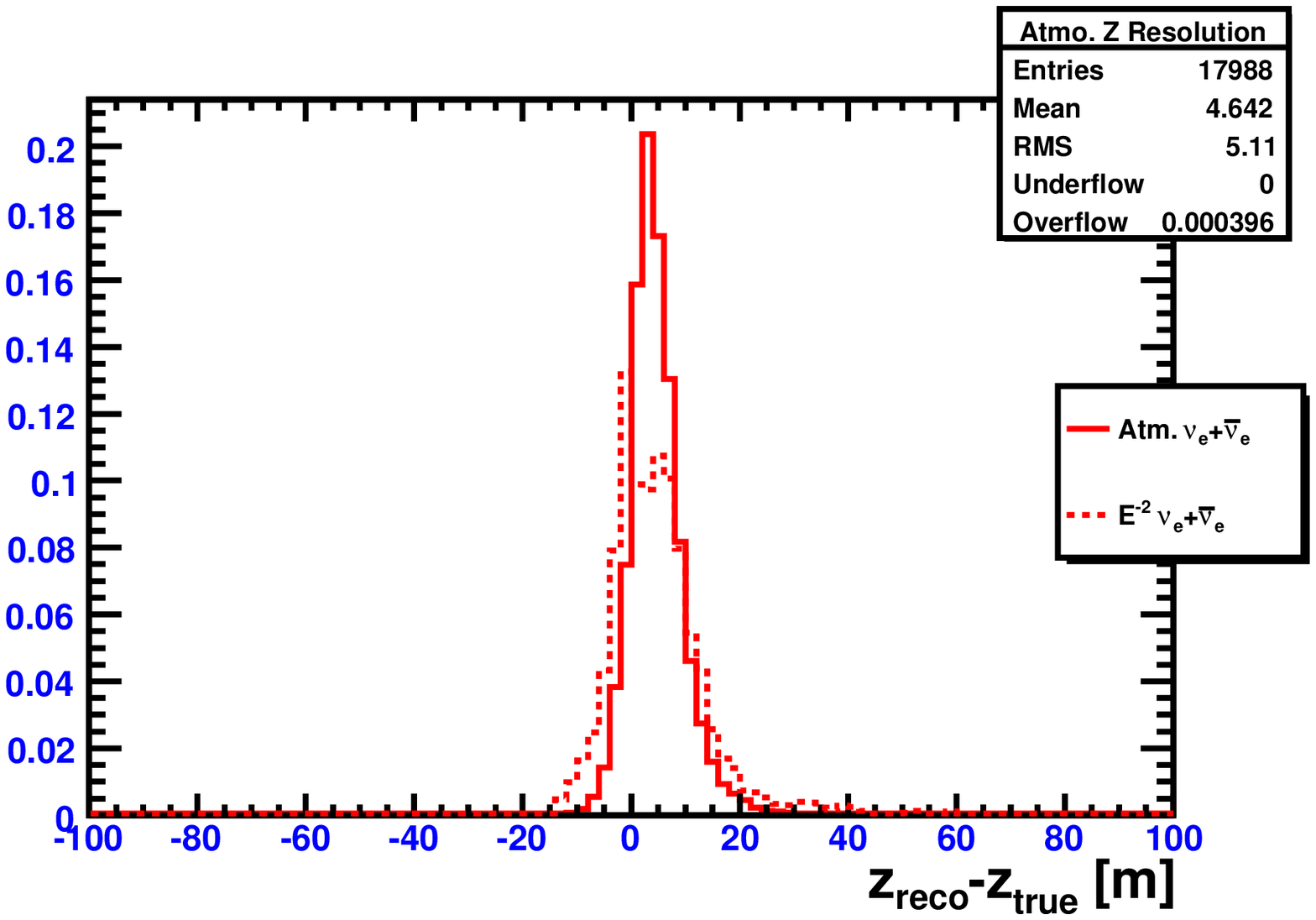}
\caption{z vertex position resolutions for $E^{-2}$ (dashed) and atmospheric (solid) spectra at level 3a (top), level 4a (middle) and final level 1 (bottom) of this analysis.}
\label{VertexResolutionsZ}
\end{figure}

\noindent Two versions of the UPandel vertex reconstruction were run as part of this analysis.  The first used only the earliest hit recorded in each DOM.  A more time-intensive reconstruction used all of the hits recorded in each DOM and was the best-performing fit.  Position resolutions from the all-hit reconstruction are shown in figures~\ref{VertexResolutionsX}--\ref{VertexResolutionsZ} for various levels of this analysis.  See chapter 6 for a definition of the analysis levels.  Table~\ref{PositionResolutionTable} lists the vertex resolutions as a function of analysis level.

\begin{table}
\caption{Vertex position resolutions in meters at various levels of this analysis weighted to an atmospheric energy spectrum.}
\vspace{0.25cm}
\centering
\begin{tabular}{ | c | c | c | c |}
\hline
Analysis Level & x & y & z \\
\hline
Level 3a & 20.87 & 19.17 & 10.30 \\
\hline
Level 4a & 11.48 & 10.84 & 6.75 \\
\hline
Final Level 1 & 10.02 & 9.85 & 5.09 \\
\hline
Final Level 2 & 9.71 & 9.58 & 5.47 \\
\hline
Final Level 3 & 9.37 & 9.29 & 5.82 \\
\hline
\end{tabular}
\label{PositionResolutionTable}
\end{table}

\newpage

\clearpage
\newpage

\clearpage
\newpage

\clearpage
\newpage

\subsection{PHit-PNoHit Energy Reconstruction}
\label{SSPHitPNoHitEnergy}
Because it relies on photoelectron arrival times, our UPandel vertex reconstruction only takes advantage of information from DOM's which were hit.  However, there is also information contained in the fact that a DOM was not hit.  This information is incorporated into the ``PHit-PNoHit'' energy reconstruction.

To construct the PHit-PNoHit likelihood function, we need to know the probability that a DOM is hit given a cascade vertex and energy.  To do so, we make several assumptions.  First, we assume that the cascade is an isotropic point source of light.  Next, we make the diffusive scattering approximation (see section~\ref{SDiffusiveRegime}).  Finally, we make use of the facts that the total track length and therefore the total light intensity are proportional to the cascade energy.  Given these assumptions we can write an expression for the mean number of photoelectrons at a distance d from the cascade \cite{MarekDissertation}:

$$
\mu \approx I_0 \frac{E}{d} e^{-d/\lambda_{p}}
$$

\noindent where $I_0$ is a normalization constant determined from fits to Monte Carlo simulations and the propagation length is defined by  $\lambda_p=\sqrt{\frac{\lambda_e \lambda_a}{3}}$.  Typically, $I_0 = 1.4 \mbox{{ GeV}}^{-1}$~m and $\lambda_{p} = 25$~m.   

From Poisson statistics, we know that the probability of observing zero photoelectrons is $e^{-\mu}$.  Therefore we can write:

$$
{P^{\mbox{\fontsize{8}{14}\selectfont casc}}}_{\mbox{\fontsize{8}{14}\selectfont hit}} = 1-{P^{\mbox{\fontsize{8}{14}\selectfont casc}}}_{\mbox{\fontsize{8}{14}\selectfont nohit}} = 1-e^{-\mu}
$$

\begin{figure}
\begin{center}
\includegraphics[width=0.8\textwidth]{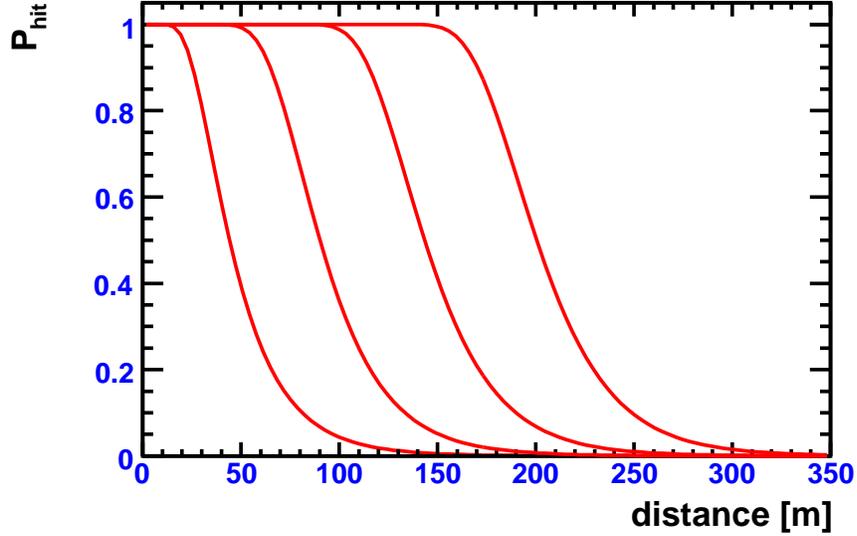}
\caption{From left to right, hit probability for a 100 GeV, 1 TeV, 10 TeV, and 100 TeV cascade as a function of distance from the cascade vertex.}
\label{PHit}
\end{center}
\end{figure}

\noindent This hit probability for several cascade energies is plotted in figure~\ref{PHit}.

To incorporate noise, we add a constant noise hit probability term:

$$
P_{\mbox{\fontsize{8}{14}\selectfont hit}} = 1-P_{\mbox{\fontsize{8}{14}\selectfont nohit}} = {P^{\mbox{\fontsize{8}{14}\selectfont casc}}}_{\mbox{\fontsize{8}{14}\selectfont hit}} + P_{\mbox{\fontsize{8}{14}\selectfont noise}} - {P^{\mbox{\fontsize{8}{14}\selectfont casc}}}_{\mbox{\fontsize{8}{14}\selectfont hit}} P_{\mbox{\fontsize{8}{14}\selectfont noise}} 
$$

\noindent where the subtracted term avoids double counting the source of a hit.  Using this expression for $P_{\mbox{\fontsize{8}{14}\selectfont hit}}$, we have our final PHit-PNoHit energy likelihood function:

$$
\mathscr{L}_{\mbox{\fontsize{8}{14}\selectfont PHit-PNoHit}} ( E ) = \prod_{\mbox{\fontsize{8}{14}\selectfont all hit DOM's}} P_{\mbox{\fontsize{8}{14}\selectfont hit}}( E,d ) \times  \prod_{\mbox{\fontsize{8}{14}\selectfont all un-hit DOM's}} P_{\mbox{\fontsize{8}{14}\selectfont nohit}}( E,d ) 
$$

It is traditional in neutrino astronomy to express energy resolution in terms of $\mbox{log}(E_{\mbox{\fontsize{8}{14}\selectfont reco}})-\mbox{log}(E_{\mbox{\fontsize{8}{14}\selectfont visible}})$.  In AMANDA, the PHit-PNoHit reconstruction resulted in an energy resolution between 0.1 and 0.2 in $\mbox{log(E)}$.  

This energy reconstruction method has several drawbacks, however.  First, it assumes that we are in the diffusive regime, which may not hold for cascades relatively close to a DOM or a string.  Second, by assuming an average propagation length, it ignores the dramatic depth variation of the optical properties of the ice.  A cascade at one depth can result in significantly more or less detected light depending on the local ice properties.  For example, a parallel IC-22 cascade analysis which made use of the PHit-PNoHit energy reconstruction found that cascades at the top of the detector were shifted in energy by -0.5 in $\mbox{log(E)}$ while cascades at the bottom of the detector were shifted in energy by +0.5 in $\mbox{log(E)}$.

To overcome these issues, we developed a depth-dependent energy reconstruction discussed in the next section.   

\section{Depth-Dependent Energy Reconstruction}
\label{SDepthDependentEnergy}
In conjunction with a colleague, I developed and implemented a fast, analytic cascade energy reconstruction that takes the depth variation of the optical properties of glacial ice into account.  This reconstruction program, {\tt AtmCscdEnergyReco}, takes a reconstructed cascade vertex and outputs a reconstructed energy given that vertex hypothesis.

To account for the depth variation of the optical properties, {\tt AtmCscdEnergyReco} makes use of photon tables generated by a dedicated Monte Carlo package called {\tt photonics} \cite{PhotonicsPaper}.  {\tt photonics} tracks photons from the cascade through the dust layers of the glacial ice, simulating the full wavelength dependent scattering and absorption along the way.  The result is a set of tables which give photon amplitudes and time delay distributions at a given location in the ice from a given cascade vertex position.  These tables have built into them assumptions about the refrozen ice which surrounds the IceCube DOM's as well as the glass transmission and PMT quantum efficiency.  To get the full amplitude, the table amplitude is scaled linearly by the cascade energy. 

These tables, known as ``photorec'' tables, tell us the expected number of photoelectrons in a given DOM for a cascade located at a particular depth in the detector.  They are used in the likelihood functions described below.

\subsection{{\tt AtmCscdEnergyReco}: Noiseless}

If we ignore noise hits, we can build an energy likelihood function as a product of Poissonian terms for the total charge seen by a DOM:

$$
 \mathscr{L}(E) = \prod_{\mbox{\fontsize{8}{14}\selectfont Hit+UnHit DOM's}}{\frac{{\mu(E)}^{n_{\mbox{\fontsize{8}{14}\selectfont pe}}} e^{-\mu(E)}}{n_{\mbox{\fontsize{8}{14}\selectfont pe}}!}} 
$$

\noindent where $\mu(E)$ is the expected number of photoelectrons in a given DOM from a cascade of energy E and $n_{\mbox{\fontsize{8}{14}\selectfont pe}}$ is the observed number of photoelectrons.  Taking the negative log of this expression we have:

$$
-\mbox{log}\mathscr{L}(E) = -\sum
\left( n_{\mbox{\fontsize{8}{14}\selectfont pe}} \mbox{log}(\mu(E)) - \mu(E)-\mbox{log}(n_{\mbox{\fontsize{8}{14}\selectfont pe}}!) \right)
$$

\noindent Differentiating with respect to $E$ and setting the result equal to 0 we have:

\begin{align*}
&\frac{-d\mbox{log}\mathscr{L}(E)}{dE} =  -\sum
\left( \frac{n_{\mbox{\fontsize{8}{14}\selectfont pe}}}{\mu(E)} \frac{d\mu(E)}{dE} - \frac{d\mu(E)}{dE} \right) = 0 \\
&\sum \frac{d\mu(E)}{dE} = \sum  \frac{n_{\mbox{\fontsize{8}{14}\selectfont pe}}}{\mu(E)} \frac{d\mu(E)}{dE} \\
\end{align*}

\noindent Remembering that the amplitude scales linearly with the cascade energy E, we can write $\mu(E)=\mu_0(\vec{r}_v, \vec{r}_{DOM})E$ where $\mu_0(\vec{r}_v, \vec{r}_{DOM})$ is the amplitude from the photorec table which depends on the cascade vertex position $\vec{r}_v$ and DOM position $\vec{r}_{DOM}$ but is independent of energy.  E is the cascade energy in GeV.  Using this expression we can write the above equation as

$$
\sum \mu_0(\vec{r}_v, \vec{r}_{DOM}) = \sum  \frac{n_{\mbox{\fontsize{8}{14}\selectfont pe}}}{ \mu_0(\vec{r}_v, \vec{r}_{DOM}) E} \times \mu_0(\vec{r}_v, \vec{r}_{DOM}) 
$$

\noindent which implies that 

$$
E[\mbox{GeV}]=\frac{\sum{n_{\mbox{\fontsize{8}{14}\selectfont pe}}}}{\sum{ \mu_0(\vec{r}_v, \vec{r}_{DOM}) }}
$$

\noindent So to reconstruct the cascade energy, we simply sum the observed charges and divide by the sum of the photorec table amplitudes for the given cascade vertex hypothesis.

Finally, we can calculate a reduced negative log-likelihood

$$
\mbox{rlog}\mathscr{L}  = \frac{-\mbox{log}\mathscr{L} _{\mbox{\fontsize{8}{14}\selectfont min}}}{N_{\mbox{\fontsize{8}{14}\selectfont ch}}}
$$

\noindent where $\mathscr{L} _{\mbox{\fontsize{8}{14}\selectfont min}}$ is the likelihood function evaluated at the reconstructed energy.  For a ``perfect'' cascade where each DOM sees exactly as much charge as the photorec tables say that it should, the reduced negative log-likelihood tends to one.  This parameter should be useful for separating signal from background as muons with extra hits should have worse (that is, larger) reduced negative log-likelihood parameters. 

\subsection{{\tt AtmCscdEnergyReco}: With Noise}

With the addition of a noise term, the expected charge becomes $\mu(E)+rT$ where $r$ is the noise rate in Hertz and $T$ is the readout window length in seconds.  Following the same logic as above, we arrive at the following equation:

$$
-\sum_i{\frac{n_{\mbox{\fontsize{8}{14}\selectfont pe}}^i}{E+\frac{rT}{ \mu_0(\vec{r}_v, \vec{r}_{DOM}) }}}+\sum_i{ \mu_0(\vec{r}_v, \vec{r}_{DOM}) }=0
$$

\noindent The presence of the noise term means that the solution for the energy is no longer analytic.  However, it can be found by an iterative root finding algorithm.  A unique root and hence a unique solution for the energy always exists. The root-finding algorithm has been implemented with the GSL library \cite{GSL}.

\subsection{Performance}
\label{SPerformance}

\begin{figure}
\centering
\includegraphics[width=0.65\textwidth]{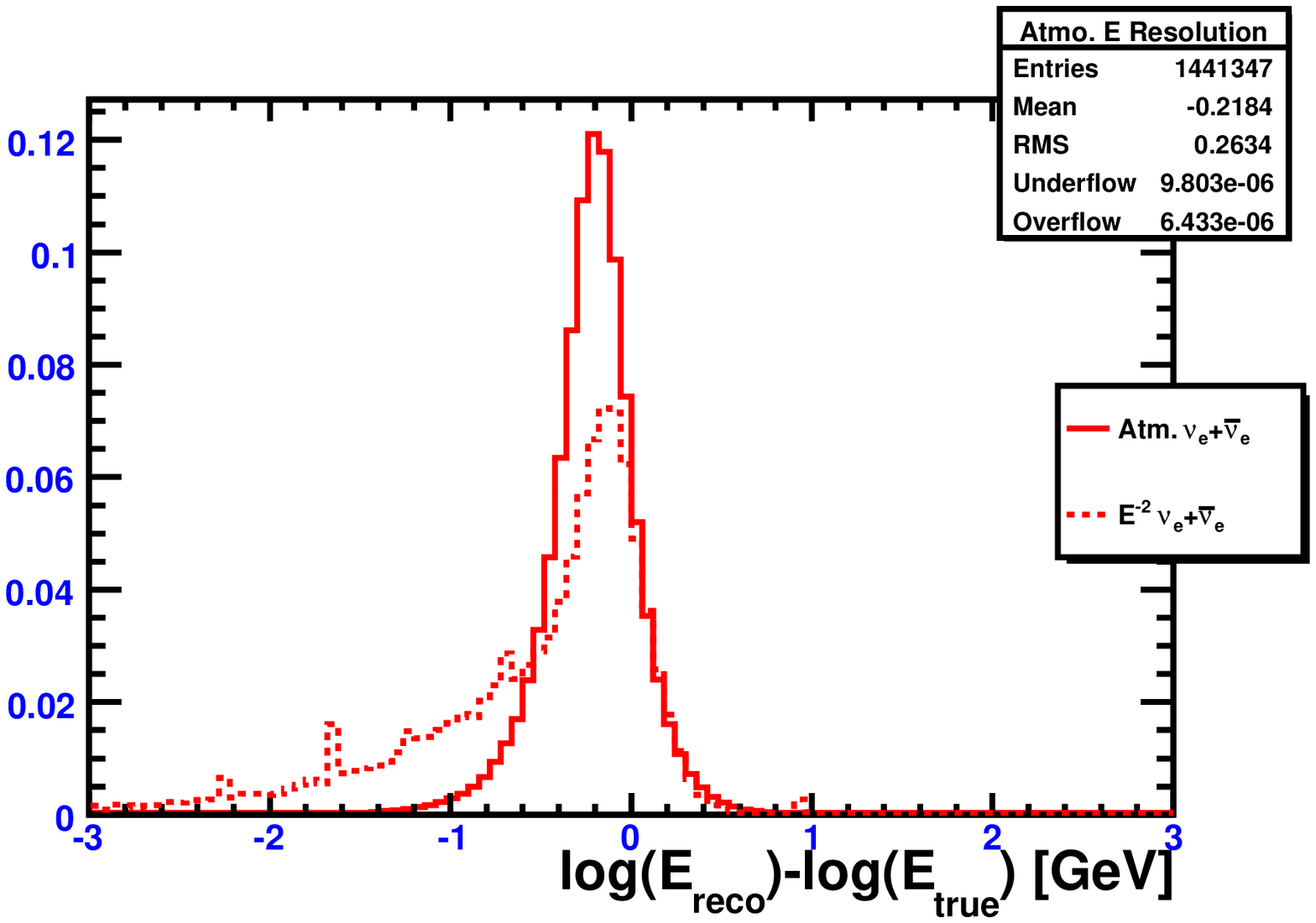}
\includegraphics[width=0.65\textwidth]{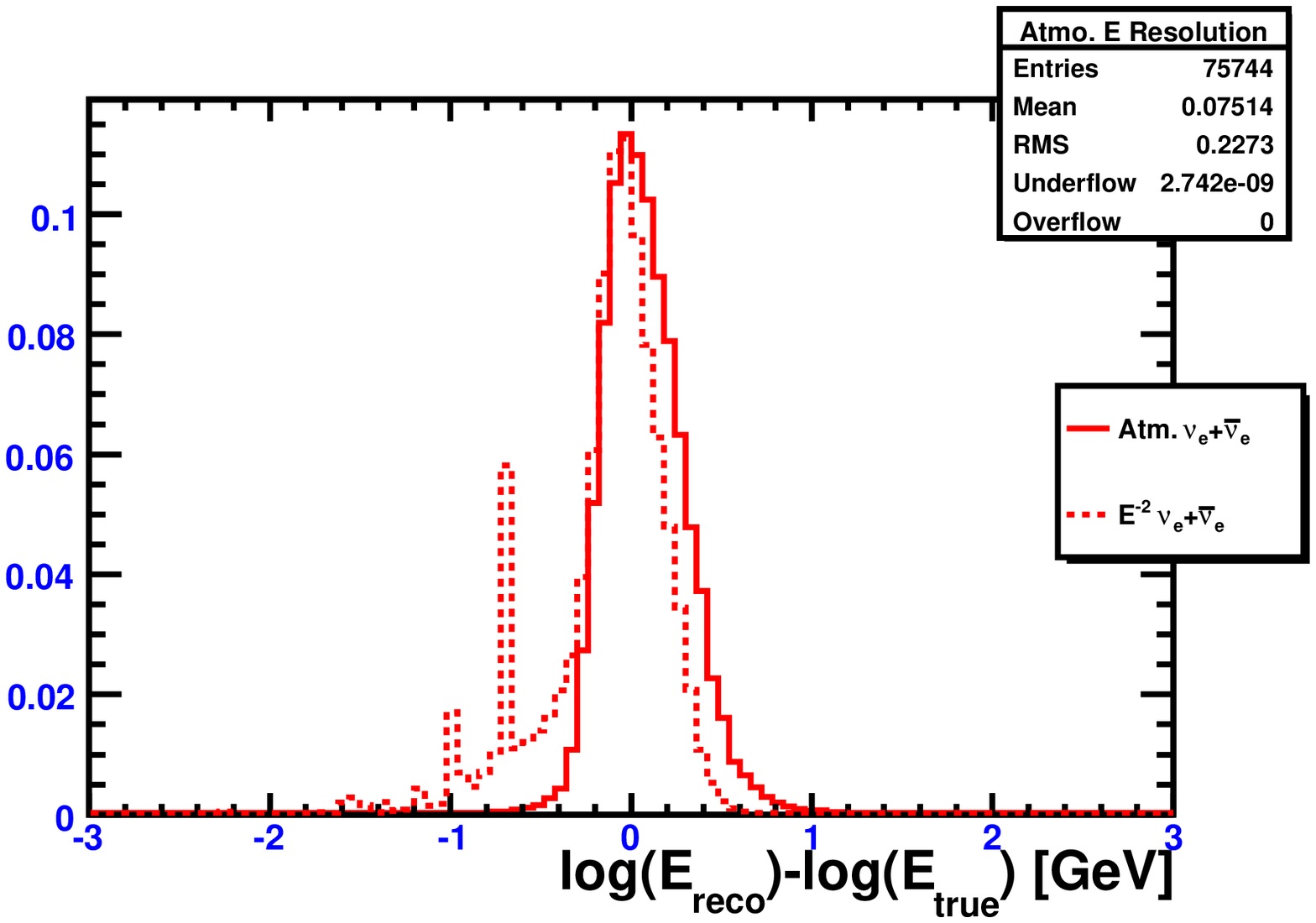}
\includegraphics[width=0.65\textwidth]{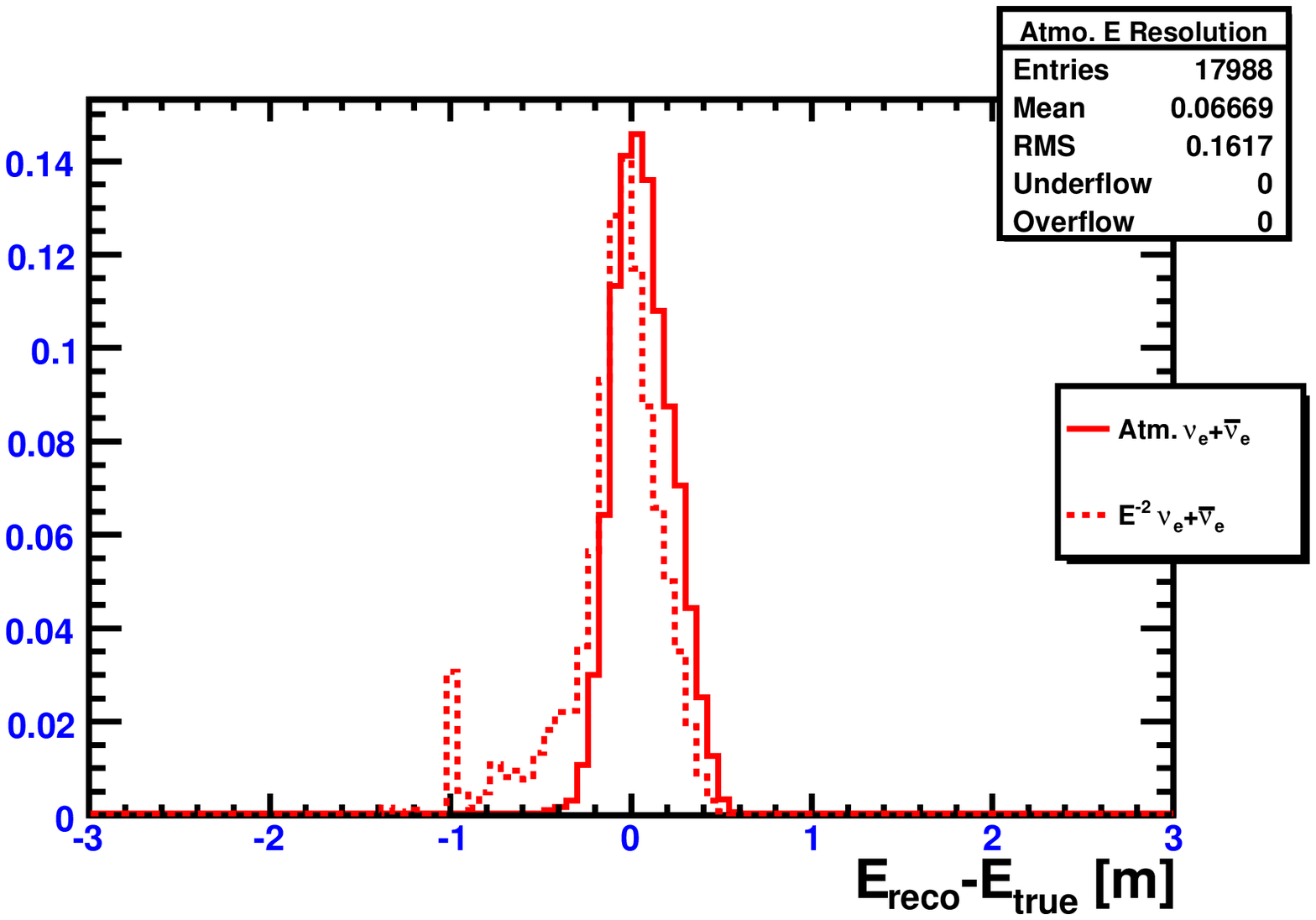}
\caption{From top to bottom, energy resolutions for $E^{-2}$ (dashed) and atmospheric (solid) spectra at level 3a, level 4a, and final level 1 of this analysis.}
\label{EnergyResolutions}
\end{figure}

\begin{figure}
\begin{center}
\includegraphics[width=0.8\textwidth]{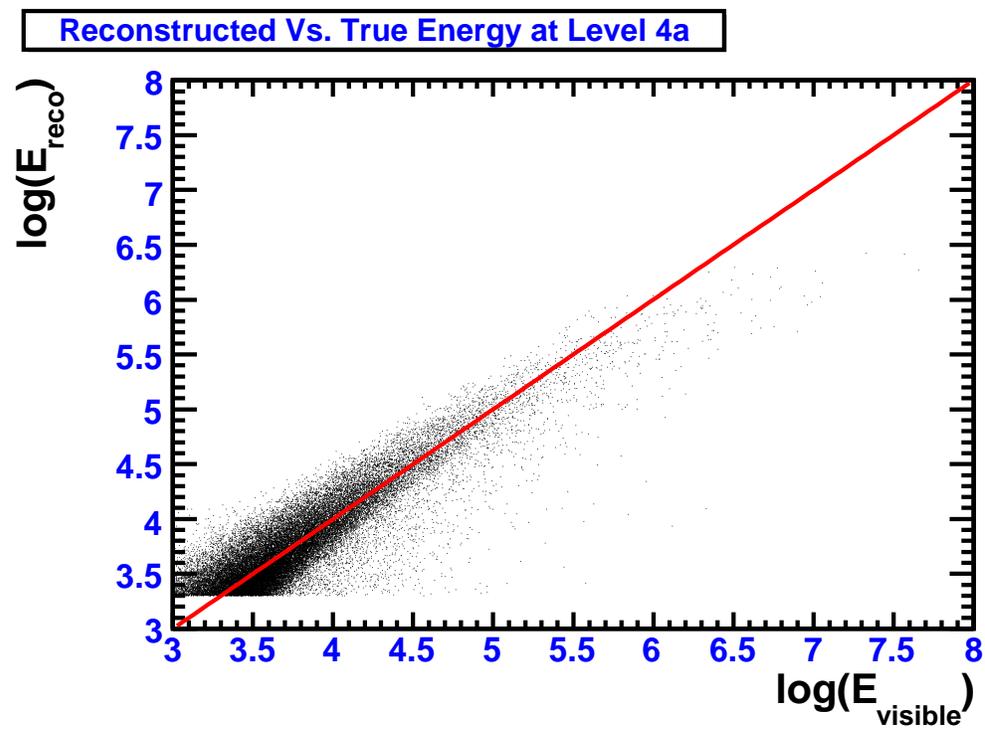}
\vspace{1.75cm}
\caption{Reconstructed energy vs. visible energy in the ice at level 4a of this analysis.}
\label{RecoVsVisibleEnergy}
\end{center}
\end{figure}

{\tt AtmCscdEnergyReco} is currently the best-performing cascade energy reconstruction in use by the IceCube collaboration.  Energy resolution plots for various levels of this analysis are shown in figure~\ref{EnergyResolutions}.  Figure~\ref{RecoVsVisibleEnergy} shows a scatter plot of reconstructed energy as a function of visible energy in the ice.  Table~\ref{EnergyResolutionTable} lists the energy resolutions as a function of analysis level.

\begin{figure}
\centering
\includegraphics[width=0.8\textwidth]{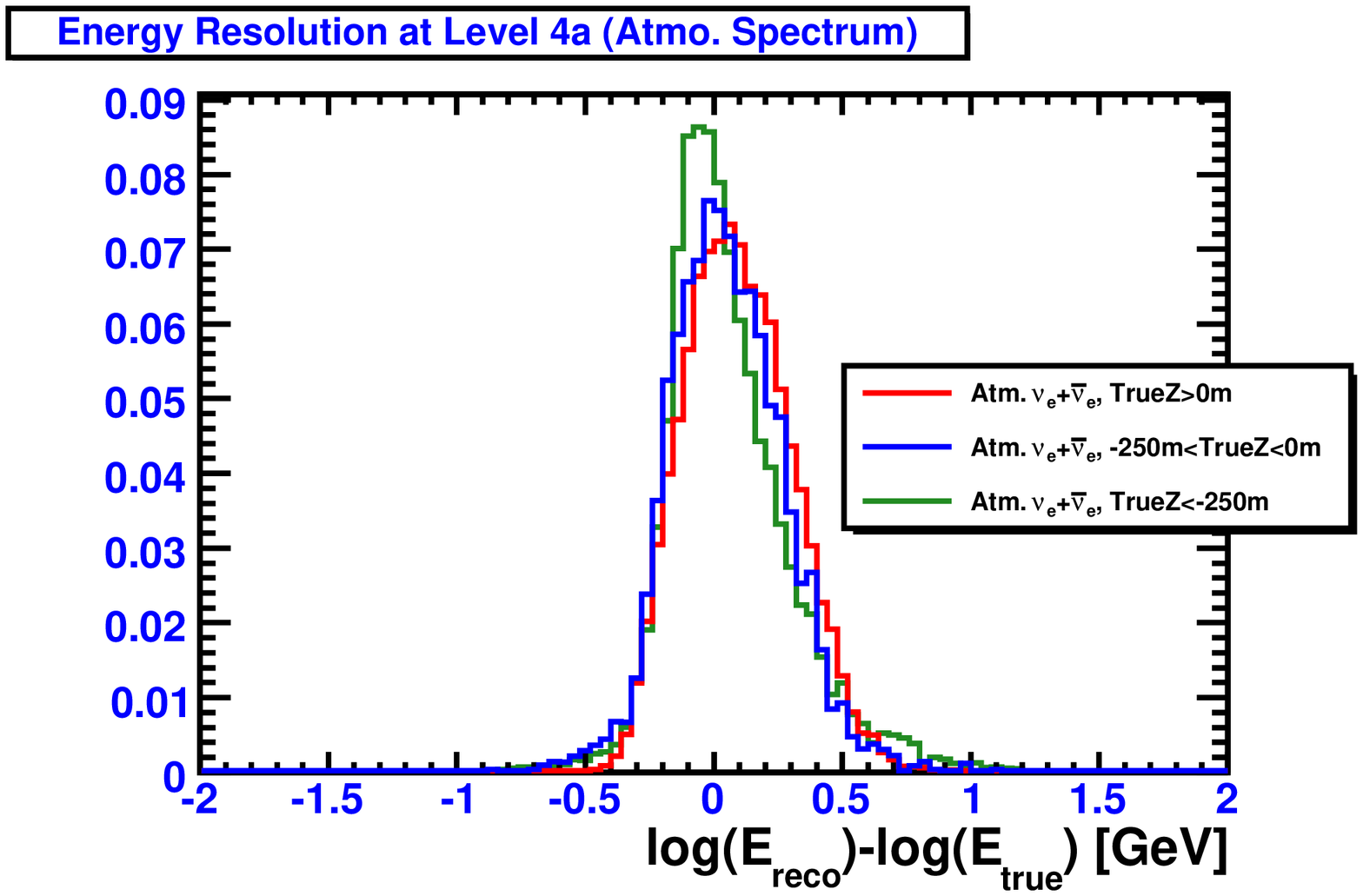}
\includegraphics[width=0.8\textwidth]{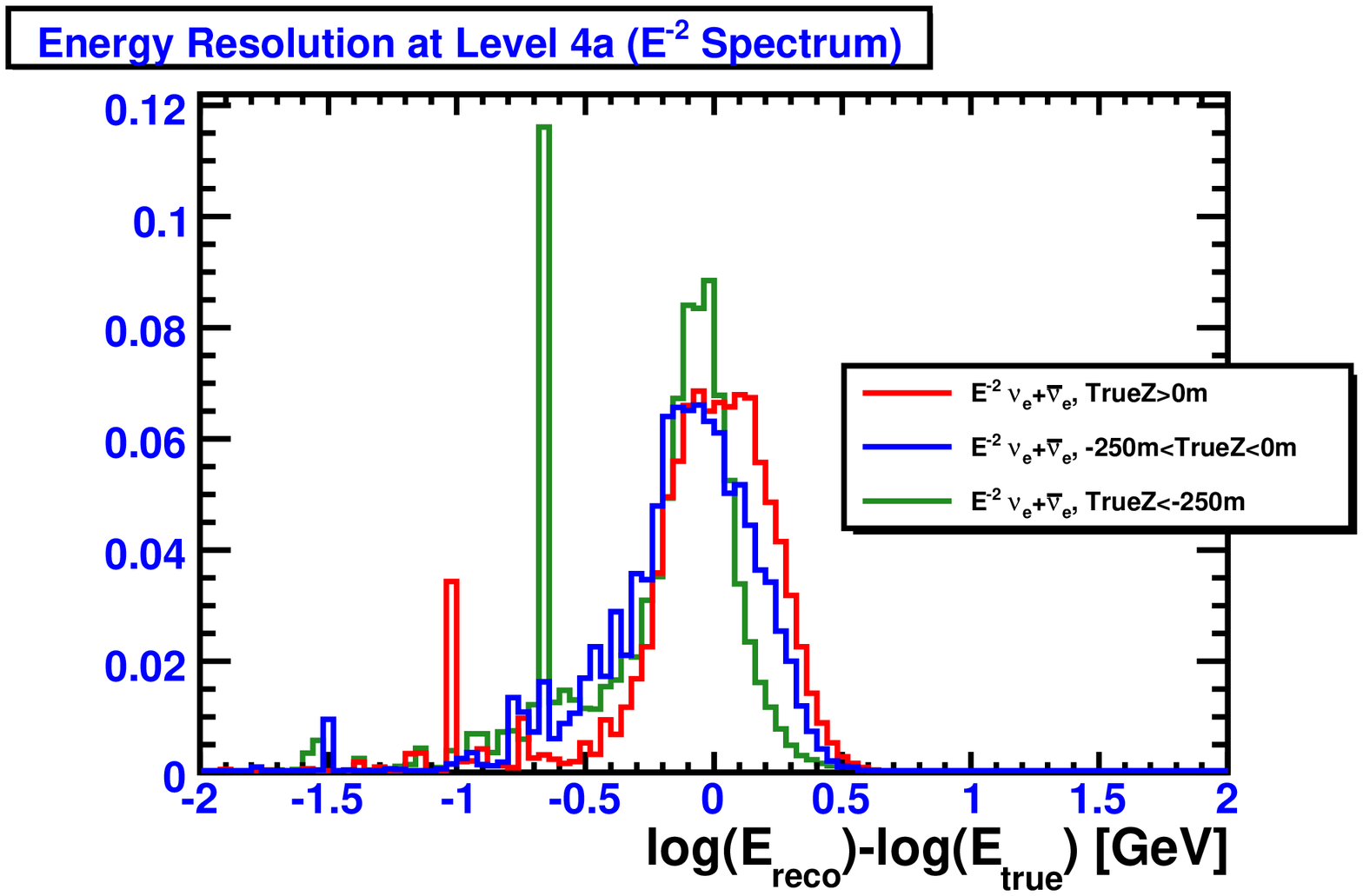}
\caption{Energy resolution at the top (red), middle (blue), and bottom (green) of the detector at level 4a of this analysis for an atmospheric spectrum (top) and an $E^{-2}$ spectrum (bottom).}
\label{EnergyDepthSplit}
\end{figure}

Because it makes use of the photorec tables which have knowledge of the depth dependence of the optical properties of the glacial ice, {\tt AtmCscdEnergyReco} performs well as a function of depth in the detector.  Figure~\ref{EnergyDepthSplit} compares the energy resolution in the top, middle, and bottom of the detector.

\begin{figure}
\centering
\includegraphics[width=0.8\textwidth]{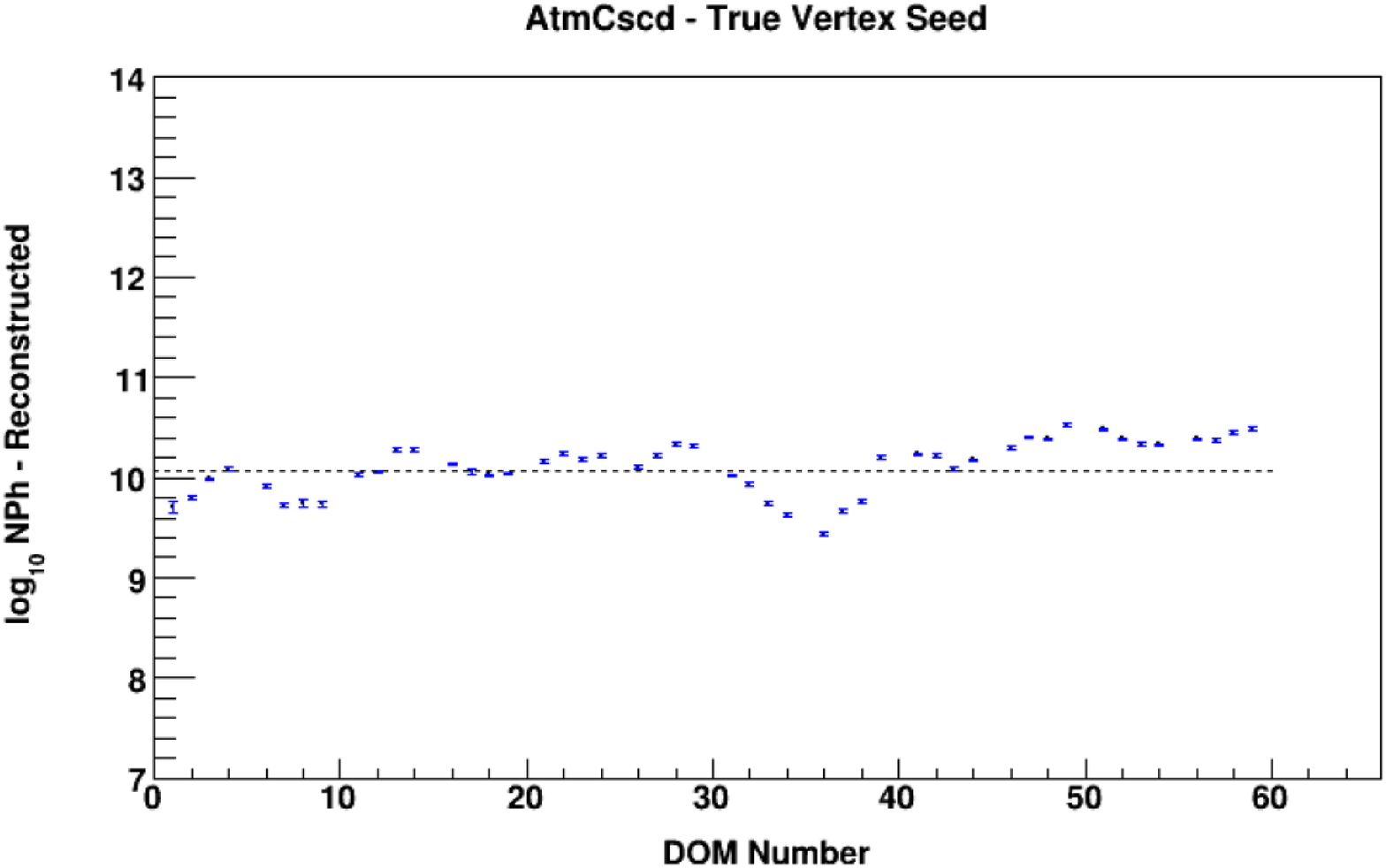}
\vspace{0.5cm}
\includegraphics[width=0.8\textwidth]{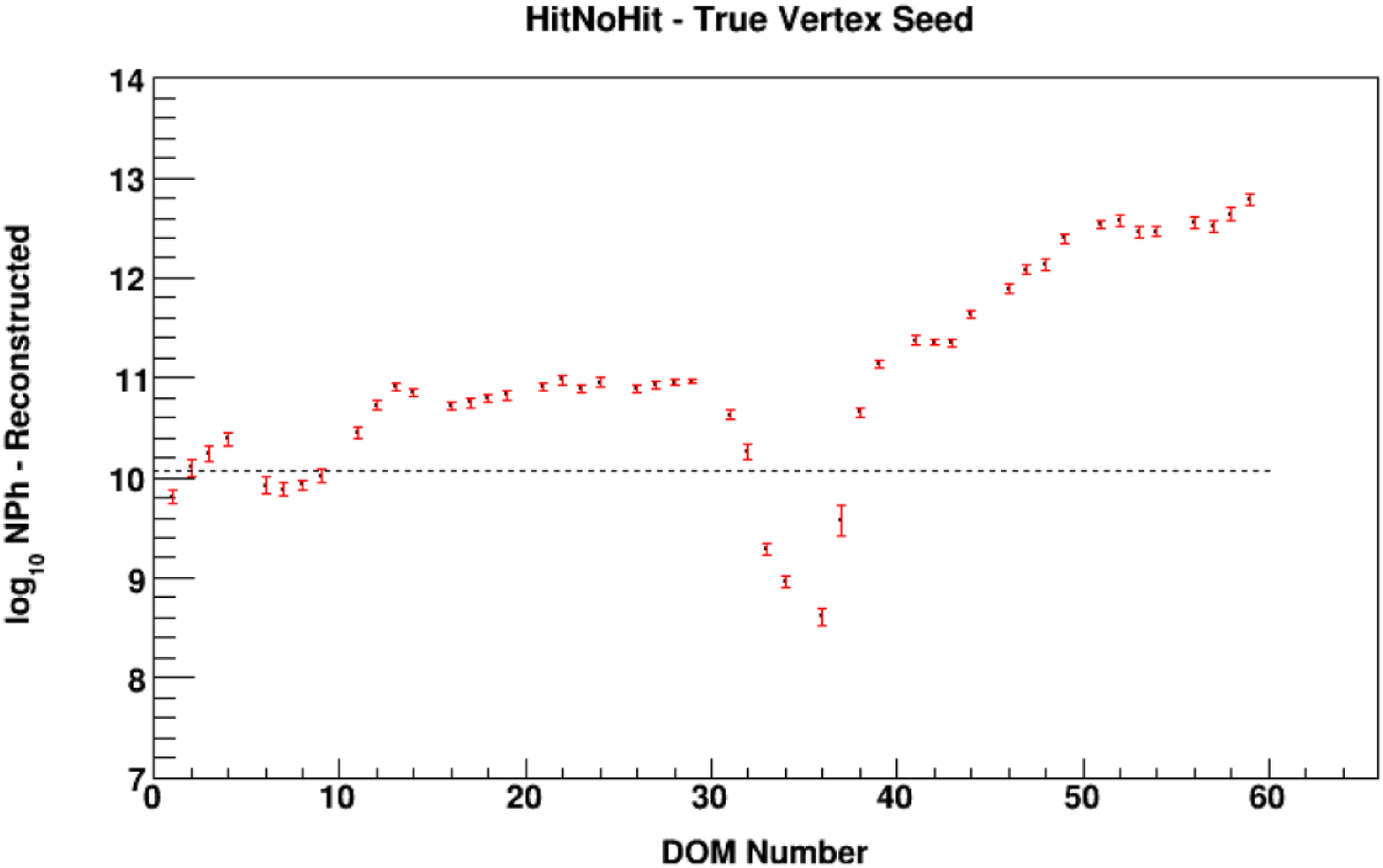}
\vspace{0.25cm}
\caption{Reconstructed number of photons per flash for flasher LED's located at various depths in the ice from \cite{JoesThesis} for {\tt AtmCscdEnergyReco} (top) and PHit-PNoHit (bottom).  The broken line shows the true number of photons per flash.  {\tt AtmCscdEnergyReco} does a much better job of smoothing out the effects of local ice properties on the reconstructed energy.  The residual structure may indicate remaining deficiencies in the modeling of the ice properties.}
\label{McCartinHalfBright}
\end{figure}

Finally, studies were carried out to verify the performance of {\tt AtmCscdEnergyReco} using the in-situ flasher LED's located in each DOM.  Figure~\ref{McCartinHalfBright} shows the reconstructed number of photons per flash for LED's located at various depths in the ice for this reconstruction and the PHit-PNoHit reconstruction.  Our {\tt AtmCscdEnergyReco} algorithm does a much better job at reducing the influence of local optical properties on the reconstructed energy. 

\newpage

\begin{table}
\caption{Energy resolutions in log(E) at various levels of this analysis weighted to an atmospheric energy spectrum.}
\vspace{0.25cm}
\centering
\begin{tabular}{ | c | c | c | c |}
\hline
Analysis Level & RMS of $\mbox{log}(E_{\mbox{\fontsize{8}{14}\selectfont reco}})-\mbox{log}(E_{\mbox{\fontsize{8}{14}\selectfont visible}})$\\
\hline
Level 3a & 0.26 \\
\hline
Level 4a & 0.23 \\
\hline
Final Level 1 & 0.16 \\
\hline
Final Level 2 & 0.17 \\
\hline
Final Level 3 & 0.17 \\
\hline
\end{tabular}
\label{EnergyResolutionTable}
\end{table}

\clearpage

\newpage

\clearpage
\newpage

\clearpage
\newpage

\clearpage
\newpage

%% file: chapters/chapter5.tex
\chapter{Simulation}\label{chapter:chapter5}

\begin{figure}
\begin{center}
\includegraphics[width=0.75\linewidth]{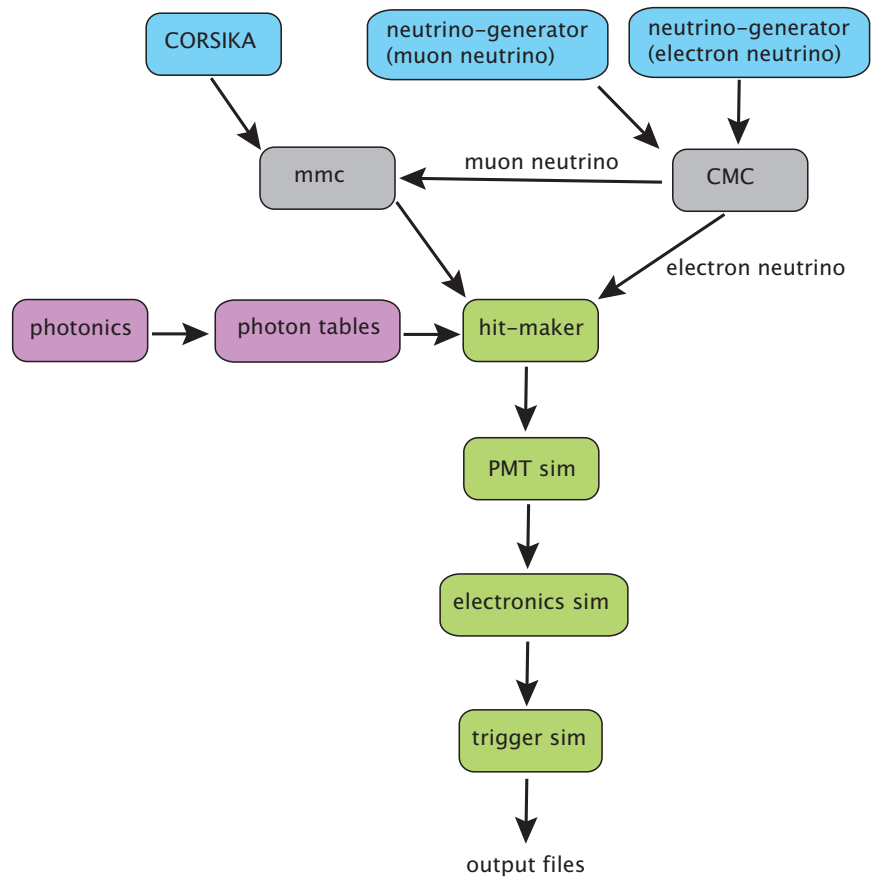}
\caption{Schematic of the various programs that go into simulating signal and background events.  Event generators are shaded blue and the elements of the detector simulation are shaded green.}
\label{SimulationSchematic}
\end{center}
\end{figure}

Analysis in IceCube is very heavily dependent on Monte Carlo simulations to model both the muon background from cosmic ray air showers and the neutrino signal.  This chapter discusses the simulation of both classes of events.  First, air shower and neutrino generation will be described.  Then, the detector simulation chain, common to both types of simulation, will be explained.  Figure~\ref{SimulationSchematic} is a schematic of the various programs that go into simulating signal and background events.

\section{Background Simulation}

\subsection{{\tt CORSIKA}}

The main background for all IceCube analyses consists of downgoing muons from cosmic ray air showers.  To simulate this background, we utilize the Monte Carlo simulation program {\tt CORSIKA} (COsmic Ray SImulations for KAscade, version 6720) \cite{CORSIKA}.

{\tt CORSIKA} propagates cosmic ray primaries through the atmosphere to their point of interaction with an air nucleus.  Various high energy hadronic interaction models can then be used to simulate the interaction, including VENUS \cite{VENUS}, QGSJET \cite{QGSJET}, DPMJET \cite{DPMJET}, or SIBYLL \cite{SIBYLL}.  For this analysis, SIBYLL was used.  The secondary particles produced in the interaction are propagated and allowed to further interact or decay.

The background simulations utilize a primary cosmic ray energy spectrum known as the Hoerandel polygonato model (Greek for ``many knees'') \cite{Hoerandel}.  In the polygonato model, the energy spectrum of each primary element (p, He, C, N, O, etc.) consists of two power laws patched together at a cutoff energy called the knee.  The cutoff energy is an increasing function of nuclear charge Z.  Individual primaries up to iron were simulated from $600$ GeV to $10^{11}$ GeV and then added together according to the Hoerandel model.

The simulation of each air shower results in a list of muons at the surface of the earth.  These muons are then handed off to the detector simulation which is described in section~\ref{SDetectorSimulation}.

\subsection{Energy Weighting}
The cosmic ray primary spectrum is a power law $E^{-\gamma}$ with index $\gamma \approx 2.7$ steepening to $\gamma \approx 3$ around an energy of 4 PeV.  Experience from previous analyses as well as preliminary studies for this analysis have suggested that primaries from the higher energy end of the spectrum are more likely to result in the energetic muons with large radiative losses that survive to high cut levels of a cascade analysis.  Consequently, it is more efficient to simulate a harder energy spectrum rather than the natural cosmic ray spectrum in order to oversample these higher energy air showers.  

The bulk of the Monte Carlo used for the analysis in this dissertation was generated with a primary energy spectrum that was one power harder than the natural spectrum.  Event weights were stored and later applied to reweight back to the natural spectrum.

\subsection{Coincident Events}
In a detector as large as IceCube, there is a non-negligible probability that muons from two or more independent air showers may enter the detector volume together.  These so-called coincident events are especially pernicious for muon neutrino analyses.  Muons from one shower may cause hits at the bottom of the detector while muons from the other shower cause slightly later hits in the upper part, thereby mimicking an upward-going neutrino-induced muon.  

To find ways to eliminate this class of background, doubly coincident and triply coincident air shower events were simulated.  First, individual {\tt CORSIKA} air showers were simulated, and events which resulted in at least one hit DOM in the detector were written to a file.  Then, these events were paired together with a relative time difference less than $\delta t = 21 \mu s$.  A calculation was then carried out to assign the proper livetime to this collection of coincident events.  That is, if we have N showers and we construct N/2 pairs, we need to calculate the true amount of time we would have to wait to see N/2 coincident events \cite{KlasInternalReport}.

\subsection{Livetime}
{\tt CORSIKA} air shower simulation is an extremely computationally intensive task.  Even with the resources of the entire IceCube collaboration running for $\sim 6$ months, we were only able to generate $\sim 20$ days of effective detector livetime.  Various optimization schemes were tried but ultimately proved unsuccessful on the timescale of this atmospheric neutrino-induced cascade analysis.  Producing enough background Monte Carlo will continue to be a significant challenge for future cascade analyses and is currently a very active research area within the IceCube collaboration.

Appendix A has details on the calculation of effective livetime for a set of weighted {\tt CORSIKA} events.

\section{Signal Neutrino Simulation}
\subsection{{\tt neutrino-generator}}
\begin{figure}
\begin{minipage}[b]{0.48\linewidth} % A minipage that covers half the page
\centering
\includegraphics[width=7.6cm]{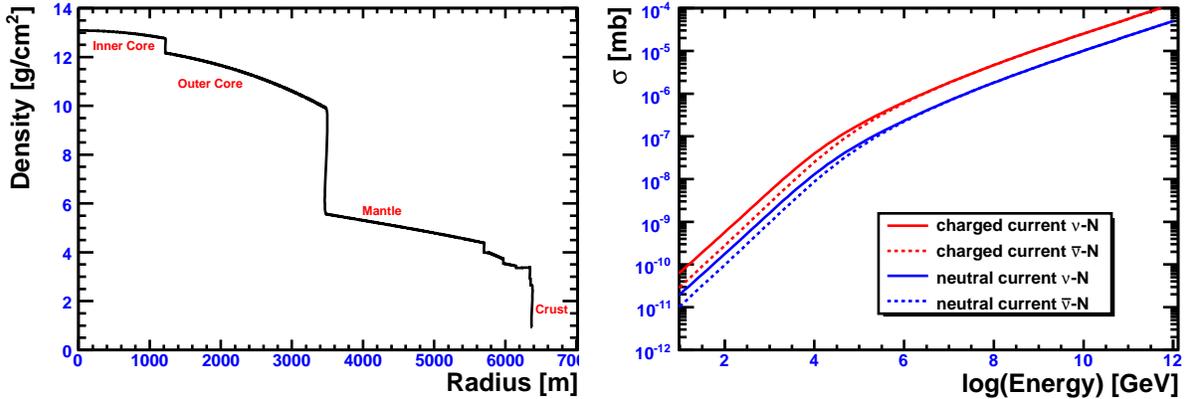}
\end{minipage}
\hspace{0.5cm} %To get a little bit of space between the figures
\begin{minipage}[b]{0.48\linewidth}
\centering
\includegraphics[width=7.6cm]{figures/thesis/eps/CTEQ5}
\end{minipage}
\vspace{0.25cm}
\caption{Earth density model and neutrino cross sections used by {\tt neutrino-generator}.  The cross section also includes the Glashow resonance, which is not shown in the figure.}
\label{NugenFigures}
\end{figure}

Neutrino simulation is done within the IceCube simulation framework {\tt IceSim} and is handled by the {\tt neutrino-generator} program, which in turn is based on the {\tt ANIS} (All Neutrino Interaction Simulation) code \cite{ANIS}.  {\tt neutrino-generator} was used to generate neutrino events distributed uniformly over both hemispheres of the earth.  The neutrinos were then propagated through the earth taking into account earth absorption (where the neutrino suffers a charged-current interaction somewhere deep inside the earth) and neutral-current regeneration (where the neutrino undergoes a neutral-current interaction resulting in a lower energy neutrino).  For electron antineutrinos, the resonant scattering off electrons (the Glashow resonance) is also included.  The Preliminary Reference Earth Model \cite{PREM} is used to model the density profile of the earth and is shown in figure~\ref{NugenFigures}.  The CTEQ5 (Coordinated Theoretical-Experimental Project on QCD) parton distribution functions \cite{CTEQ1, CTEQ2}  were used for the neutrino cross sections which are also shown in figure~\ref{NugenFigures}.

When the neutrino gets within a certain distance of the detector, an interaction of a given type is forced to occur, and the probability for that interaction is stored for event reweighting.  In order to get absolute event rates for the atmospheric neutrino flux, the probability weight is combined with appropriate flux weights to take into account the atmospheric flux variation as a function of energy and zenith angle.

The output particles from the forced neutrino interaction are handed off to the {\tt CMC} program, described in the next section.  The output of {\tt CMC}, in turn, is handed off to the detector simulation which is described in section~\ref{SDetectorSimulation}.

For this analysis, 4,930 electron neutrino files and 2,494 muon neutrino files were simulated.  Each file contained 200,000 generated events from 40 GeV to $10^{10}$ GeV with an $E^{-2}$ energy spectrum for the primary neutrino.  The interaction vertices were distributed around the detector in a cylinder of radius 1200~m and length 1700~m.

\subsection{{\tt CMC}}
\label{SCMC}
We often refer to cascades as point sources of light because of the sparse instrumentation of IceCube.  In reality, cascades do have a longitudinal profile and at energies above a TeV can extend over several meters.  To simulate the longitudinal development of electromagnetic and hadronic cascades, the program {\tt CMC} (Cascade Monte Carlo) was run on the {\tt neutrino-generator} output.

Above an energy threshold of 1 TeV, {\tt CMC} splits cascades into several smaller, lower energy sub-cascades to simulate the longitudinal development of the shower.  The sub-cascades are spaced three radiation lengths apart, where $X_0 \approx 41$ cm for ice.  Each sub-cascade points in the same direction as the original cascade and has a time delay appropriate for light travel in vacuum.  The energies of the sub-cascades are taken from the parameterization of the shower profile described in section~\ref{SEMCascades}.  For hadronic cascades, {\tt CMC} also applies an energy-dependent correction factor to account for the fact that hadronic cascades produce less Cherenkov light than their electromagnetic counterparts (see section \ref{SHadronicCascades}).

\section{Detector Simulation}
\label{SDetectorSimulation}
The result of {\tt CORSIKA} air shower simulation is a list of muons at the surface of the earth.  For {\tt neutrino-generator} muon neutrino events that have undergone a charged-current interaction, we also have a muon.  In the first step of the detector simulation, these muons are propagated through the ice by the program {\tt mmc} \cite{DimaMMC}.
\subsection{Muon Propagation with {\tt mmc}}
\begin{figure}
\begin{center}
\includegraphics[width=0.6\linewidth]{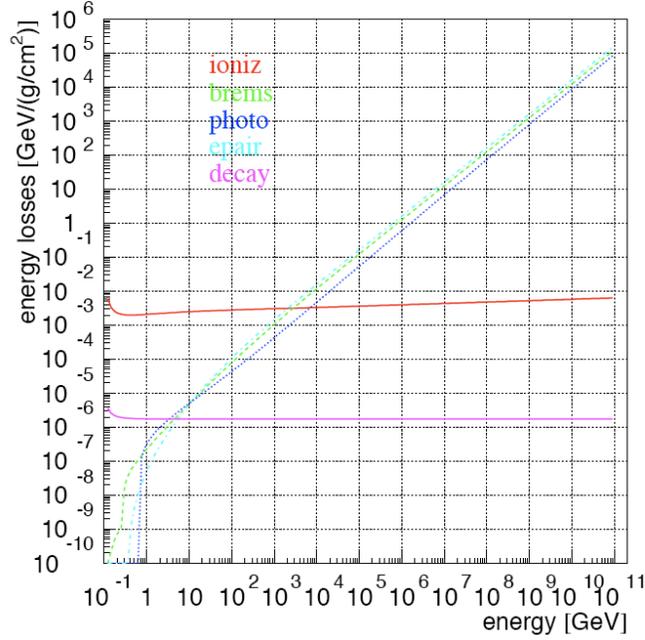}
\caption{Muon energy loss processes simulated by {\tt mmc}.  From \cite{DimaMMC}.}
\label{MMCLosses}
\end{center}
\end{figure}

\begin{figure}
\begin{center}
\includegraphics[width=0.6\linewidth]{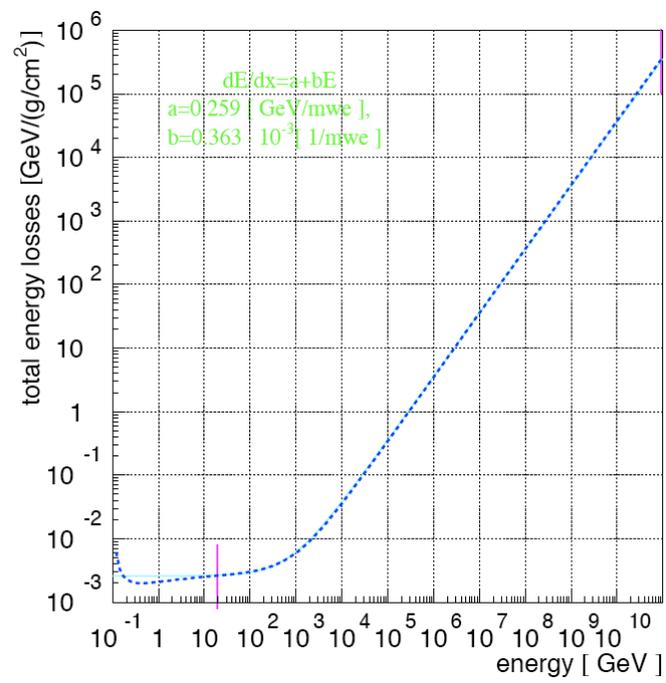}
\caption{Total muon energy loss as a function of energy simulated by {\tt mmc}.  From \cite{DimaMMC}.}
\label{MMCLosses2}
\end{center}
\end{figure}

{\tt mmc} simulates the energy losses experienced by muons as they traverse the glacial ice at the South Pole.  It takes into account ionization, bremsstrahlung, photo-nuclear interactions, and pair production.  These energy losses in ice are summarized in figures~\ref{MMCLosses} and \ref{MMCLosses2}.  The output of {\tt mmc} is a list of muon track segments, along with their energies, and the secondary particles from the stochastic energy losses.

\subsection{Photon Propagation with {\tt photonics}}
Next, we simulate the Cherenkov light emission and propagation from the muons and secondaries from {\tt mmc} and from the electromagnetic and hadronic cascades from {\tt CMC}.  Direct tracking of the Cherenkov photons through the layered glacial ice from source to DOM is too computationally intensive for IceCube's Monte Carlo production.  To speed up the process, a dedicated Monte Carlo package called {\tt photonics} \cite{PhotonicsPaper} is run beforehand to build lookup tables which are then used during the detector simulation.  {\tt photonics} tracks photons from a track or a cascade through the dust layers of the glacial ice, simulating the full wavelength-dependent scattering and absorption along the way.  The result is a set of tables which give photon amplitudes and time delay distributions at a given location in the ice from a given source type and location.  These tables have built into them assumptions about the refrozen ice which surrounds the IceCube DOM's as well as the glass transmission and PMT quantum efficiency.

\subsection{PMT and DOM Simulation}
The simulation chain ends with the hardware simulation.  For each DOM in the array, the photon tables are queried by a program called {\tt hit-maker} to find the mean number of photons that particular DOM would receive from each source in the event (muon, radiative loss, electromagnetic cascade, etc.).  A Poisson distribution with this mean is sampled to determine how many photons the DOM actually receives.  Next, the time delay distributions from the photon tables are sampled to determine the arrival time of each photon.  The result is a list of photons with their arrival times for each DOM in the array.  

Finally, this list of photons is propagated through a PMT simulation which produces the output PMT current.  This is then run through a simulation of the digitization electronics, local coincidence signaling, and event triggering.

The output of an {\tt IceSim} background {\tt CORSIKA} or signal neutrino simulation is a file identical to the experimental data produced by the detector at the South Pole.  This file can then be run through the same processing and analysis scripts that are run over the experimental data.

%% file: chapters/chapter6.tex
\chapter{Data Analysis}\label{chapter:chapter6}

\section{Strategy}
In its 22-string configuration (IC-22), IceCube expected $\sim 10^5$ atmospheric neutrino-induced cascade events in its one year dataset.  These signal events are buried in $\sim 10^{10}$ downgoing muon events from cosmic ray air showers.  The challenge is to develop a series of cuts to eliminate the background muon events while retaining sensitivity to the neutrino-induced cascade signal.  While most of these muons should be rather easy to eliminate, we will eventually reach the truly difficult background for neutrino-induced cascade searches---muons that suffer a large radiative energy loss inside of the detector.  These large radiative losses are indistinguishable from cascades unless there is additional early light from the muon track.  In this chapter, we will discuss the cuts and the analysis strategies developed to recognize and eliminate these muons.

Data analysis in IceCube is conducted in a blind fashion.  That is, cuts are developed using Monte Carlo simulations of signal and background, and a representative 10\% data sample is used for validation of the background Monte Carlo.  Once a set of cuts is developed that seems to reach the target sensitivity, the full data sample is ``unblinded'' and examined.  In this way, we hope to avoid hidden biases on the part of the analyzer.  Below we present the search strategy for atmospheric neutrino-induced cascades as it was developed on Monte Carlo and the 10\% data sample.  The next chapter contains the results of the full, unblinded one year dataset.

\begin{figure}
\begin{center}
\includegraphics[width=0.75\linewidth]{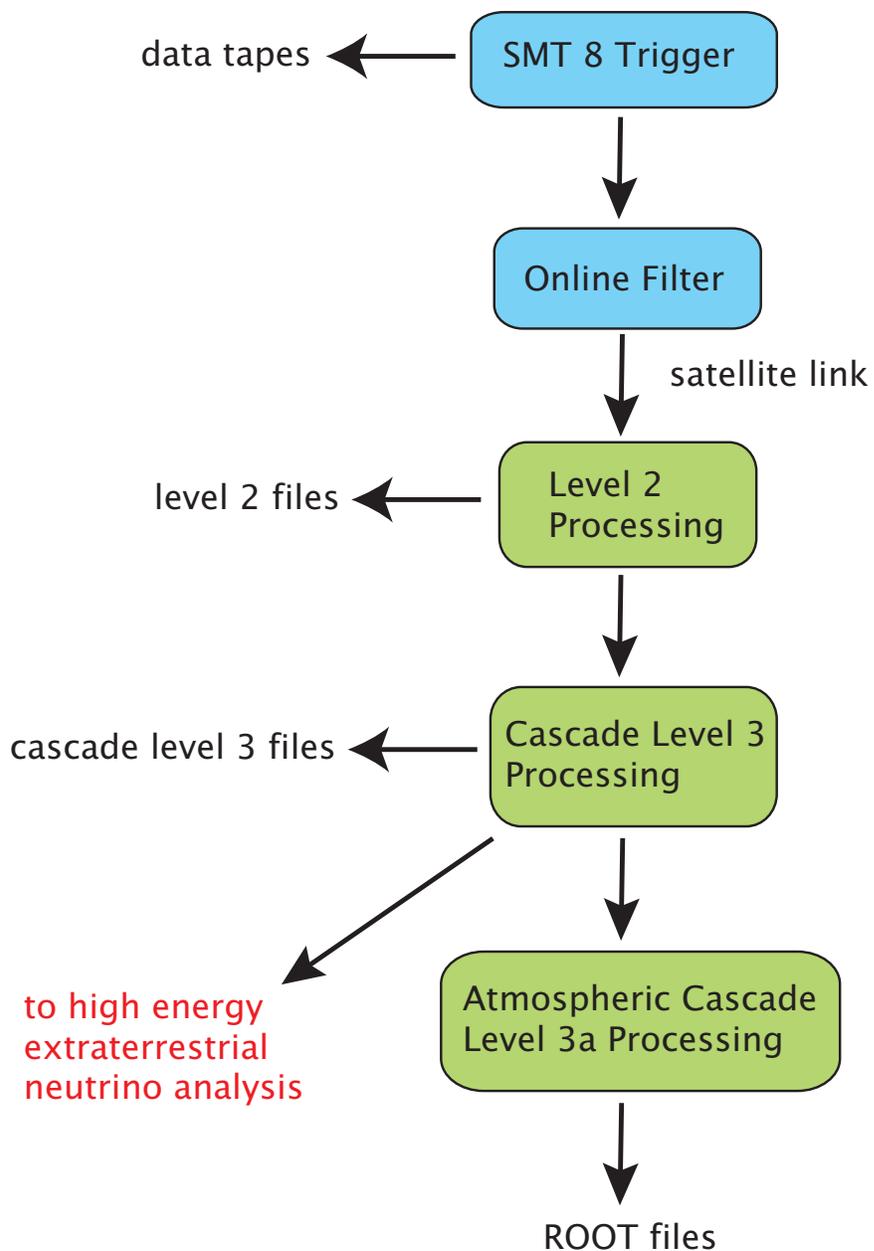}
\caption{Schematic of the stages of the atmospheric neutrino-induced cascade analysis.  Stages that take place at the South Pole are shaded blue and those that take place in the Northern Hemisphere are shaded green.  Also indicated is the point where this analysis splits off from a parallel search for a diffuse flux of high energy extraterrestrial neutrino-induced cascades.}
\label{AnalysisSchematic}
\end{center}
\end{figure}

Figure~\ref{AnalysisSchematic} shows a schematic of the different data processing stages, and table~\ref{CutTable} summarizes the various cuts.  They will be useful guides through the rest of the chapter.

\newpage

\clearpage

\newpage

\begin{landscape}

\begin{table}
\caption{Summary table of the various cuts made in this analysis.}
\vspace{0.25cm}
\centering
\begin{tabular}{ | c | l | c | c | }
\hline
&  & \multicolumn{2}{  c | }{{\bf Rate After Cut (Hz)}} \\
\hline
{\bf Cut Stage} & {\bf Cut} & {\bf Background} & {\bf Signal ($\nue+\numu$)} \\
\hline
Trigger & SMT 8 in 5~$\mu\mbox{s}$ & $558.4$ & $(1.6+87.)\times 10^{-4}$ \\
\hline
Online Filter & EvalRatio$>$0.109+LFVel$<$0.25 & $14.2$ & $(6.0+38.)\times 10^{-5}$ \\
\hline
Level3 & TrackFitZenith$>$1.27+LlhRatio$<$-16.2 & $3.1$ & $(4.9+5.6)\times 10^{-5}$ \\
\hline
Level4a & E$>$2 TeV+ParallelogramDist$<$1.4+L3aMLP$>$0.4 & $1.8\times 10^{-3}$ & $(1.1+1.4)\times 10^{-6}$ \\
\hline
Final Level 1 & E$>$5 TeV+MLPProd$>$0.73+FullAllHitSPEZ$<$440+FullAllHitSPEReducedLlh$<$10 & --- & $(1.2+5.2)\times 10^{-7}$ \\
\hline
Final Level 2 & E$>$10 TeV+MLPProd$>$0.72+FullAllHitSPEZ$<$440+FullAllHitSPEReducedLlh$<$10 & --- & $(0.5+2.4)\times 10^{-7}$ \\
\hline
Final Level 3 & E$>$15 TeV+MLPProd$>$0.69+FullAllHitSPEZ$<$440+FullAllHitSPEReducedLlh$<$10 & --- & $(0.3+1.4)\times 10^{-7}$ \\
\hline
\end{tabular}
\label{CutTable}
\end{table}

\end{landscape}

\clearpage

\newpage

\section{Online Filter at the South Pole}
The IC-22 simple multiplicity trigger (SMT) required 8 hit DOM's in a time window of 5 $\mu$s.  Throughout the year, the SMT rate varied between 500 and 600 Hz as atmospheric conditions changed with the seasons and short term weather events \cite{SerapICRC}.  All triggered data was written to tape and retrieved from the South Pole at the end of the Austral winter.

In order to proceed with analysis in a more reasonable time frame, promising candidate events were selected by several online filters for immediate satellite transmission to the Northern hemisphere.  These online filters were required to run quickly so that they could keep up with the realtime data rate.  In addition, data volume had to be reduced significantly in order to meet satellite bandwidth quotas.

To select a stream of interesting events for cascade analysis, I developed and implemented a general purpose online cascade filter suitable for both low energy and high energy cascade studies.   It was based on two quickly-calculated analytic variables: the tensor-of-inertia eigenvalue ratio and the linefit velocity.  

\subsection{Tensor-of-Inertia}

The tensor-of-inertia calculation makes an analogy between photoelectrons observed in DOM's and masses distributed in space \cite{AMANDAReconstruction}.  First, we calculate the center-of-gravity (COG) of the hits:

$$
\overrightarrow{COG} \equiv \sum_{i=1}^{N_{\mbox{\fontsize{8}{14}\selectfont ch}}}(a_i)^{w} \cdot \vec{r_i}
$$

\noindent where $a$ is the number of photoelectrons recorded by a DOM at position $\vec{r_i}$
, $w$ is an arbitrary amplitude weighting factor,  and $N_{\mbox{\fontsize{8}{14}\selectfont ch}}$ is the number of hit DOM's.

The COG is used as the position origin for the tensor-of-inertia, which is defined by

$$
I^{kl} \equiv \sum_{i=1}^{N_{\mbox{\fontsize{8}{14}\selectfont ch}}}(a_i)^{w} \cdot [\delta^{kl} \cdot |\vec{r_i}|^2 - r_i^k \cdot r_i^l]
$$ 

\noindent Continuing the analogy with a rigid body, we calculate the principal moments or eigenvalues of the tensor-of-inertia.  

\begin{figure}
\begin{minipage}[b]{0.48\linewidth} % A minipage that covers half the page
\centering
\includegraphics[width=7.6cm]{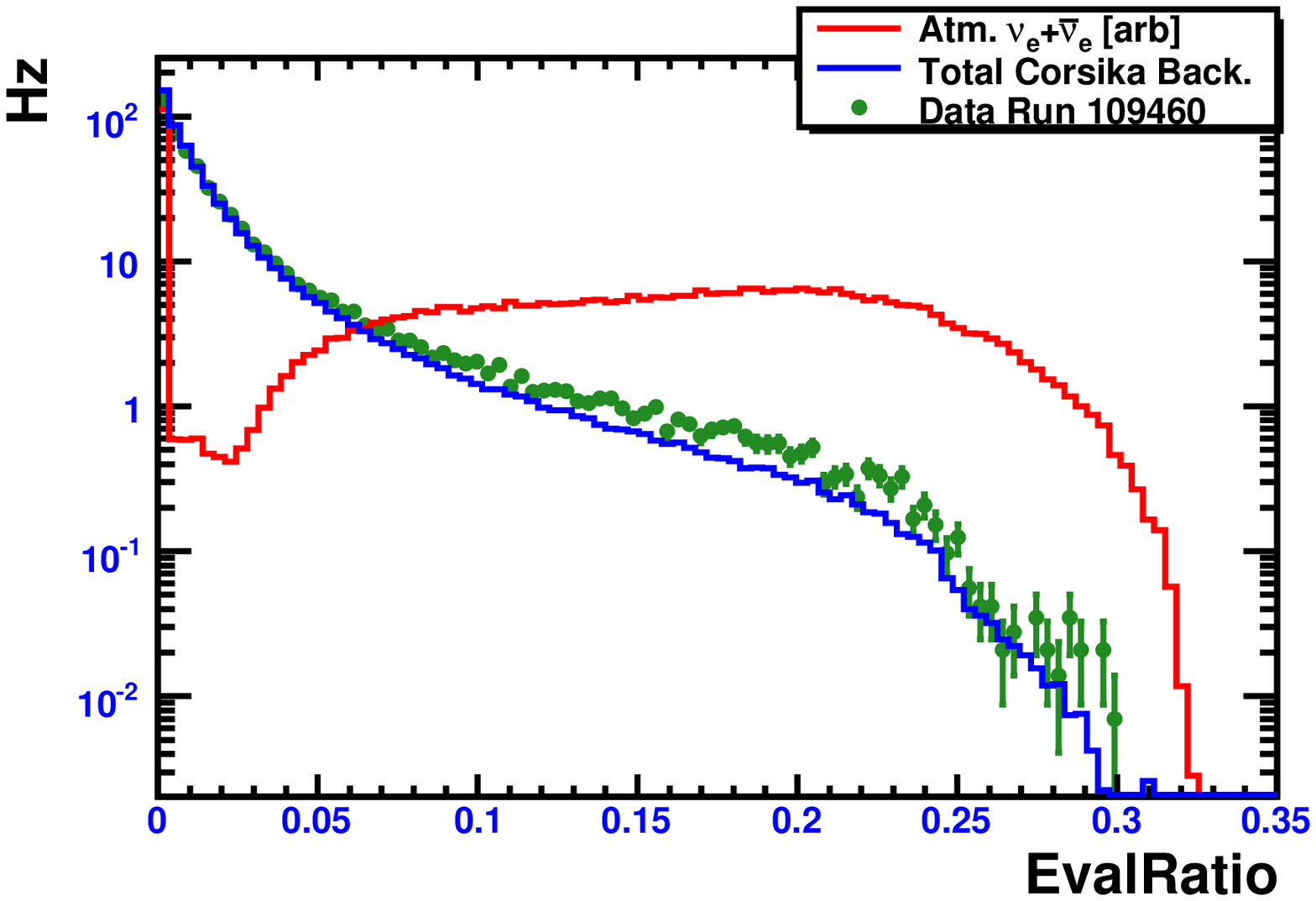}
\end{minipage}
\hspace{0.5cm} %To get a little bit of space between the figures
\begin{minipage}[b]{0.48\linewidth}
\centering
\includegraphics[width=7.6cm]{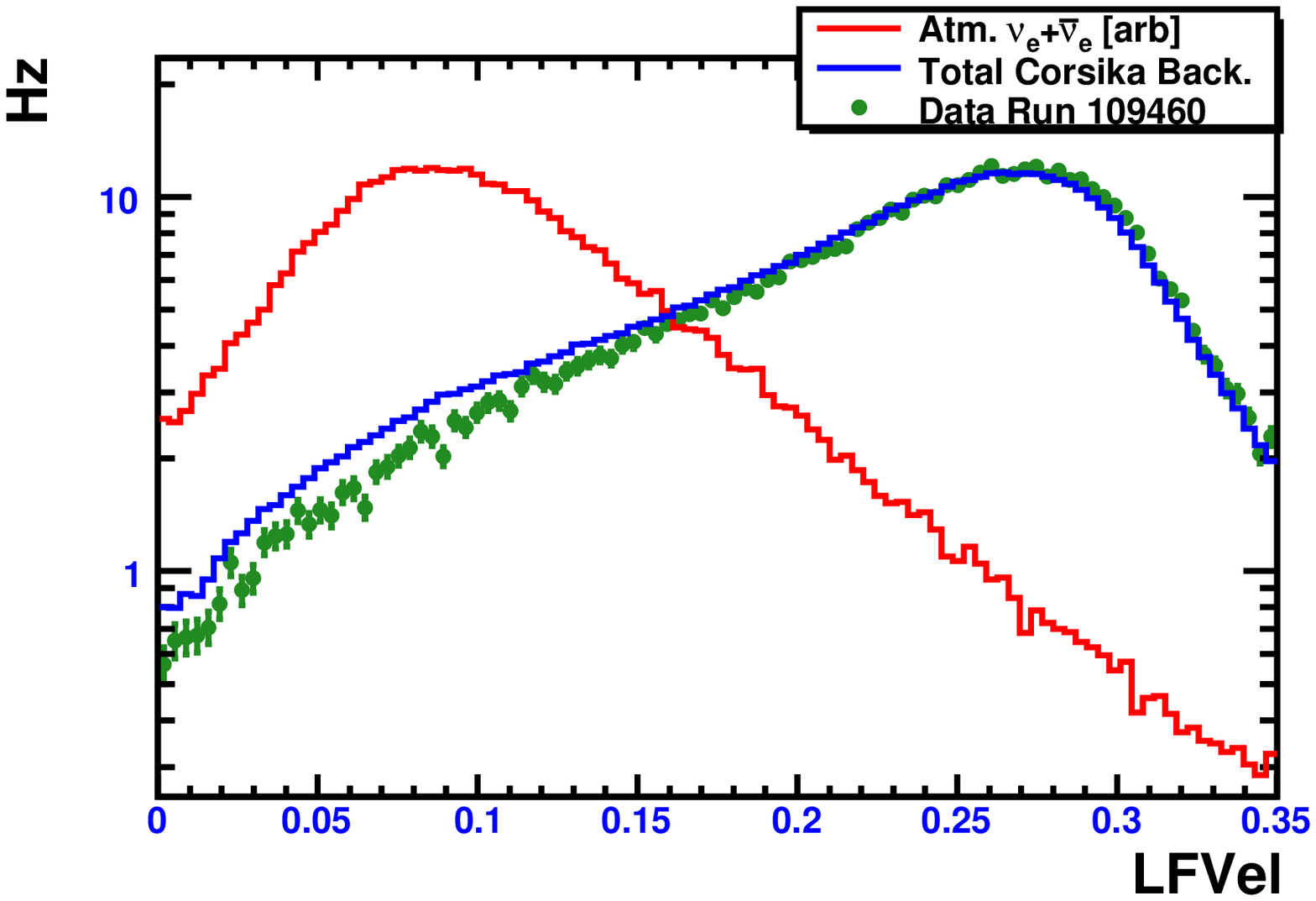}
\end{minipage}
\vspace{0.25cm}
\caption{Online filter variables, absolutely normalized.  EvalRatio measures how spherical an event is, and LFVel measures how fast a track moves through the detector.  Cuts were placed to keep events with EvalRatio$>$0.109 and LFVel$<$0.25.}
\label{OnlineFilter}
\end{figure}

Geometrically, we know that cascades look spherical in a sparsely instrumented detector like IceCube (see figure~\ref{Signatures}). So, if we calculate the three eigenvalues of the tensor-of-inertia they should have roughly equal values.  That is, the eigenvalue ratio, defined as the minimum eigenvalue divided by the sum of the three eigenvalues, should be large and close to 1/3. For a long muon track, on the other hand, we expect the rigid body to have one long principal axis and two very short principal axes.  The minimum eigenvalue, and hence the eigenvalue ratio, should be very small. The distribution of this variable for 8 hours of trigger level data is shown in figure~\ref{OnlineFilter}. A cut was placed that passes events with EvalRatio$>$0.109.

\subsection{Linefit}

The linefit algorithm is a simple muon track-fitting algorithm.  It ignores the geometry of Cherenkov emission and assumes a planar wavefront traveling at a velocity $\vec{v}$ through the detector.  Essentially, a straight line is drawn through the hit DOM's:

$$
\vec{r_i} \approx \vec{r} + \vec{v} \cdot t_i
$$

\noindent where $\vec{r_i}$ is the location of the DOM which is hit at time $t_i$.  We want to find the position $\vec{r}$ and velocity $\vec{v}$ of the track that minimize the chi-squared:

$$
\chi^2 \equiv  \sum_{i=1}^{N_{\mbox{\fontsize{8}{14}\selectfont ch}}} |\vec{r_i}-\vec{r}-\vec{v} \cdot t_i|^2
$$

\noindent The analytical solution to this equation is \cite{AMANDAReconstruction}

\begin{align*}
&\vec{r} = \langle \vec{r_i} \rangle - \vec{v} \cdot  \langle t_i \rangle \\
&\vec{v} = \frac{\langle \vec{r_i} \cdot t_i \rangle -  \langle \vec{r_i} \rangle \cdot  \langle t_i \rangle}
	{\langle t_i^2 \rangle - {\langle t_i \rangle}^2} \\
\end{align*}

\noindent where angle brackets denote an average over all hits (photoelectrons), e.g.

$$
\langle t_i \rangle \equiv \frac{1}{N_{\mbox{\fontsize{8}{14}\selectfont hit}}} \sum_{i=1}^{N_{\mbox{\fontsize{8}{14}\selectfont hit}}} t_i
$$

Since a muon track travels at relativistic velocities through the ice, the speed of the linefit should be close to the speed of light.  However, for a cascade, where the light slowly diffuses out from a single point, the linefit velocity is expected to be small. The distribution of this variable is shown in figure~\ref{OnlineFilter}. A cut was placed that passes events with LFVel$<$0.25.

Together, these two cuts rejected 97\% of the background while passing 37\% of the atmospheric $\nue$ events and 4\% of the atmospheric $\numu$ events. The $\numu$ efficiency appears low because, at this level, we're including all $\numu$ events, even long throughgoing tracks where the hadronic cascade at the interaction vertex is not inside the detector.  The data rate from the online cascade filter varied between 17 and 22 Hz throughout the year.

As a technical note, these reconstructions were run at the South Pole on hits that were required to have local coincidence span one.  That is, a DOM was used in the calculations only if one its two nearest neighbors on a string was also hit.  In contrast, the SMT imposed local coincidence span two, reading out a DOM only if one of its two nearest or two next-to-nearest neighbors  on a string was also hit.  Tightening the local coincidence for the online cascade filter was found to improve the signal-to-noise ratio.

All subsequent stages of this analysis used only events that passed the general purpose online cascade filter.  In the standard IceCube data files, the general purpose cascade filter is denoted by the key ``{\tt CascadeFilter}'' in the FilterMask.  

\section{Level 2 Processing in the Northern Hemisphere}
After satellite transmission to the Northern hemisphere, more time-intensive likelihood reconstructions were run on each event.  This stage of processing is known as the level 2 standard processing and was done on a collaboration-wide basis over all data streams.  

For the cascade stream, I selected several cascade-specific reconstructions that would allow us to further pare down the dataset.  First, the cfirst center-of-gravity first guess algorithm was run and used to seed a likelihood reconstruction.  This likelihood reconstruction assumed a point-like cascade hypothesis and reconstructed the vertex and time using only the leading edge timing information from the waveforms and a UPandel probability density function.  It is denoted by the name ``{\tt CascadeLlhVertexFit}'' in the standard IceCube data files.  See chapter 4 for details on these first guess and likelihood reconstructions. 

Both of these reconstructions were inadvertently run on hits that were required to have local coincidence span one.  This oversight was corrected at a later processing stage.

Finally, every event was subjected to a single iteration likelihood reconstruction assuming a muon track hypothesis.  This fit is denoted by the name ``{\tt TrackLlhFit}''  in the standard IceCube data files.  

No cuts were applied during level 2 processing.  The quantities calculated here were used during the cascade-specific level 3 processing, where cuts were placed and additional reconstructions were run.

\section{Cascade Level 3 Processing}
The cascade level 3 processing started from the collaboration-wide level 2 files. We cut on two variables that satisfied both the needs of this atmospheric neutrino analysis and a parallel analysis searching for a high energy extraterrestrial neutrino signal.  Additional reconstructions were run on the surviving events. The resulting files are referred to as the cascade level 3 files.

\begin{figure}
\begin{minipage}[b]{0.48\linewidth} % A minipage that covers half the page
\centering
\includegraphics[width=7.6cm]{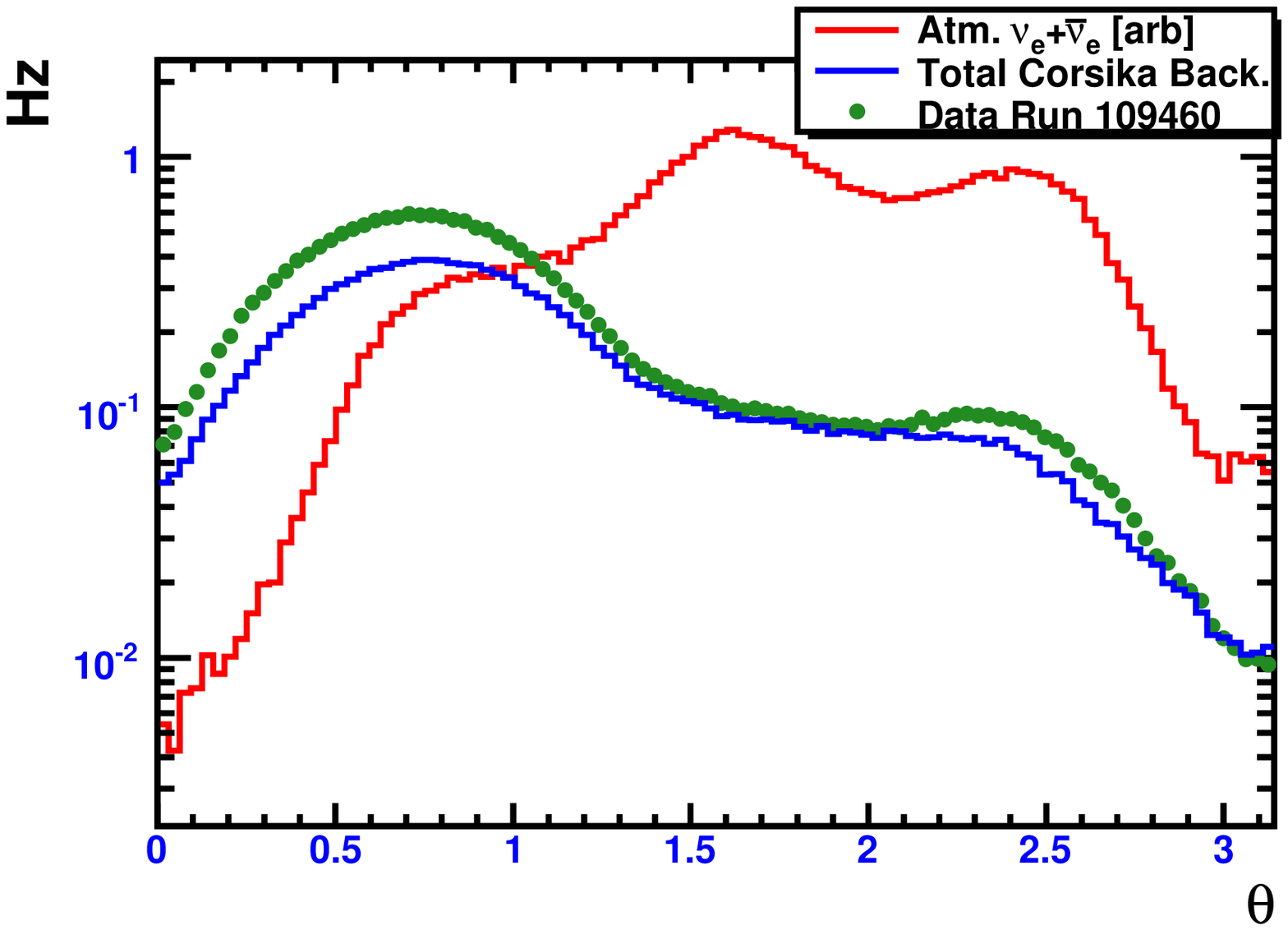}
\end{minipage}
\hspace{0.5cm} %To get a little bit of space between the figures
\begin{minipage}[b]{0.48\linewidth}
\centering
\includegraphics[width=7.6cm]{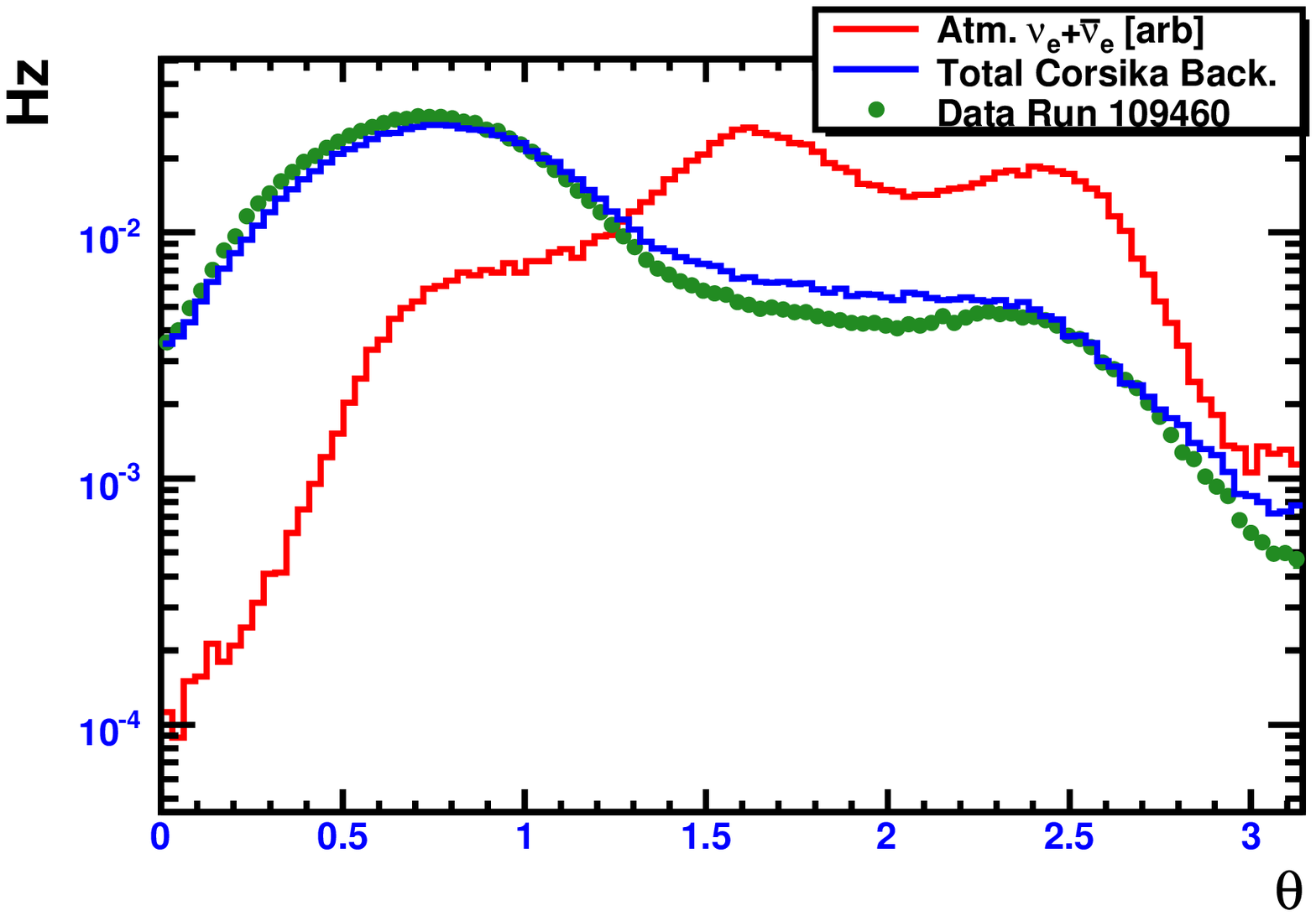}
\end{minipage}
\vspace{0.25cm}
\caption{Reconstructed track zenith angle at filter level, absolutely normalized (left) and normalized to the same area (right).  A zenith angle of zero is perfectly downgoing.  A cut was placed that keeps events with TrackFitZenith$>$1.27.}
\label{Zenith}
\end{figure}

First, a cut was placed on the zenith angle of the muon track reconstruction. This cut throws out events which reconstruct as obvious downgoing muons from cosmic ray air showers.  The zenith angle distributions are shown in figure~\ref{Zenith} for Monte Carlo and 8 hours of data, both absolutely normalized (left) and normalized to the same area (right). A cut was placed that passes events with TrackFitZenith$>$1.27.

\begin{figure}
\begin{minipage}[b]{0.48\linewidth} % A minipage that covers half the page
\centering
\includegraphics[width=7.6cm]{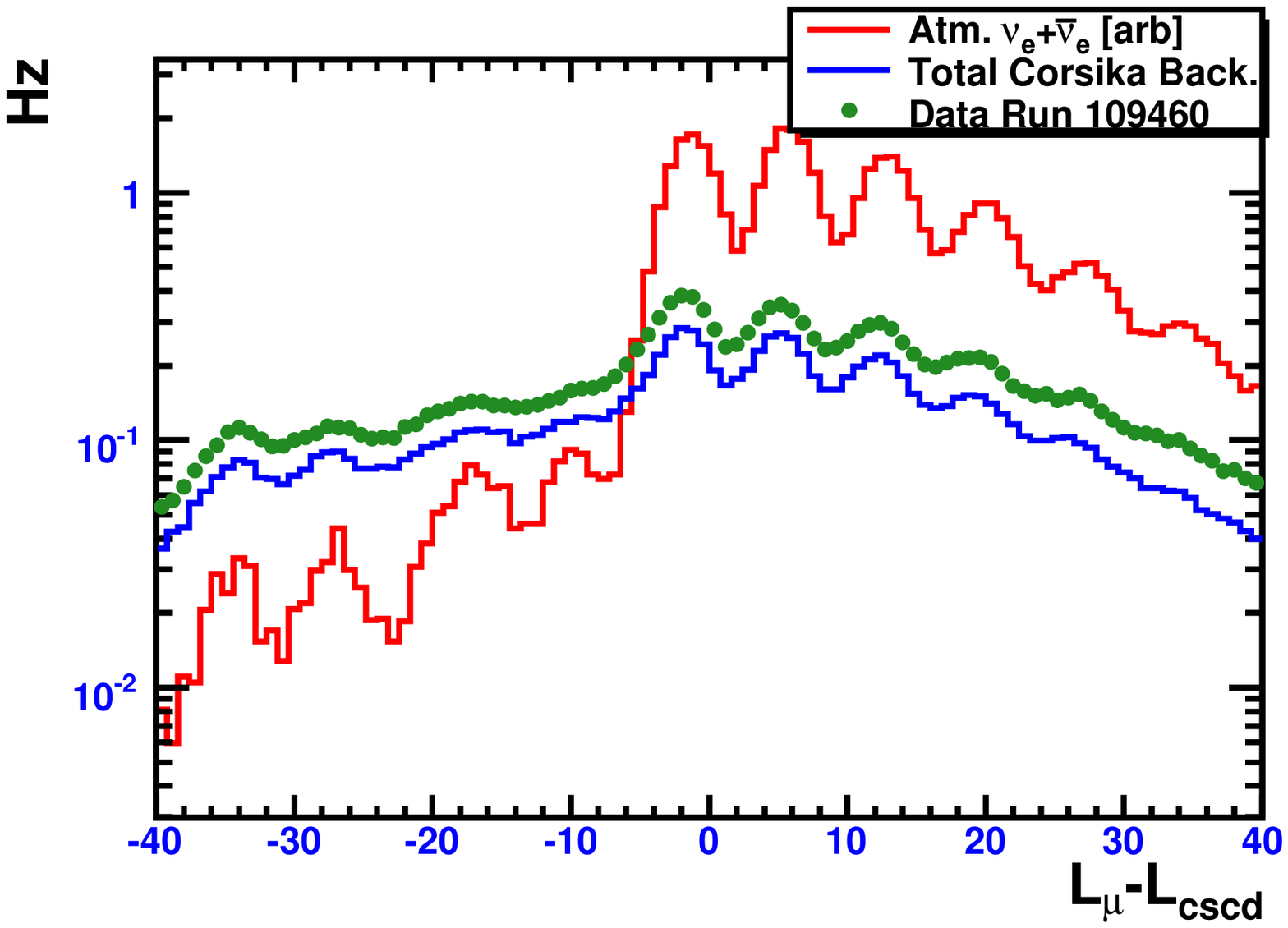}
\end{minipage}
\hspace{0.5cm} %To get a little bit of space between the figures
\begin{minipage}[b]{0.48\linewidth}
\centering
\includegraphics[width=7.6cm]{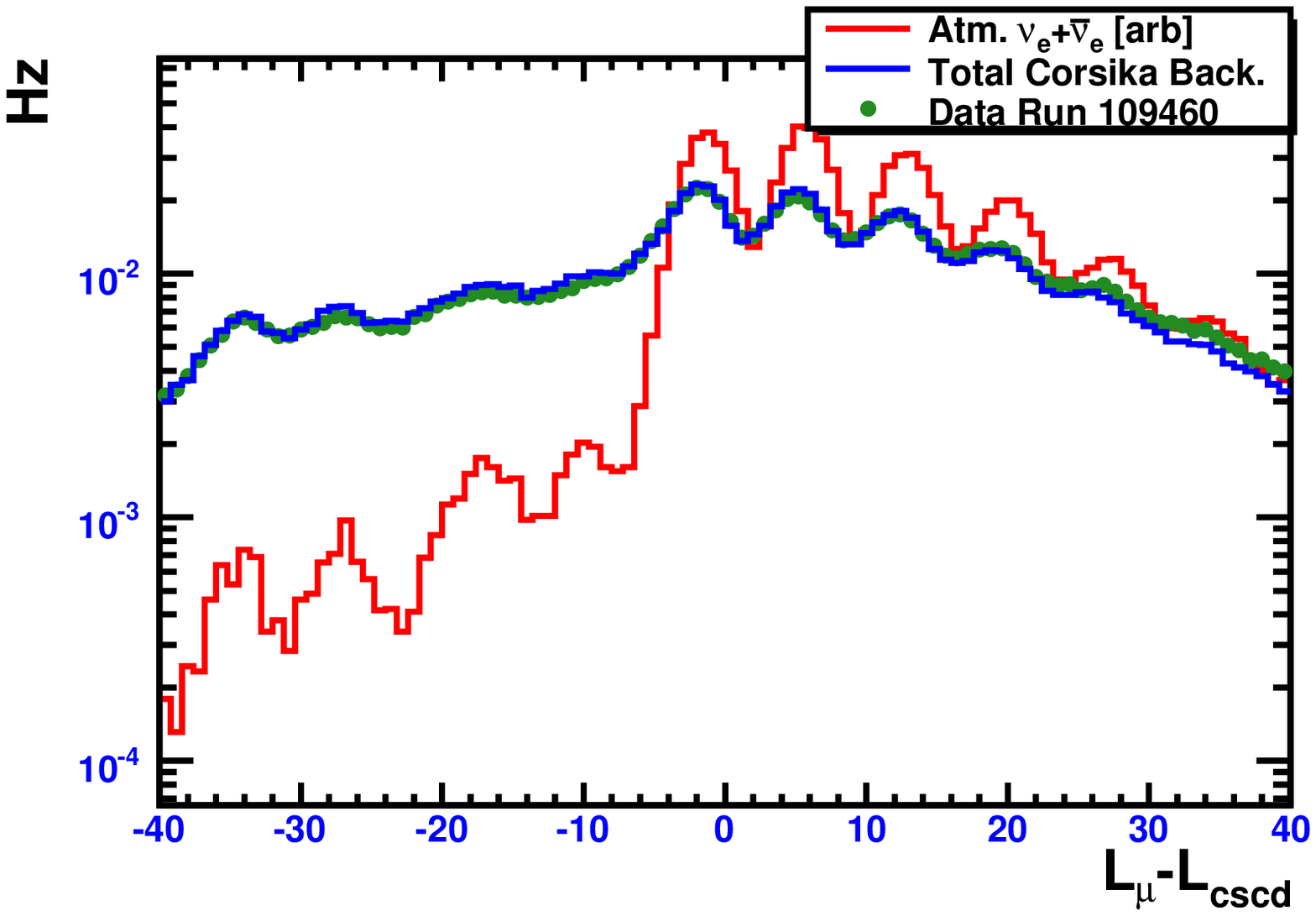}
\end{minipage}
\vspace{0.25cm}
\caption{Likelihood ratio at filter level, absolutely normalized (left) and normalized to the same area (right).  A cut was placed that keeps events with LlhRatio$>$-16.2.}
\label{LikelihoodRatio}
\end{figure}

Second, a likelihood ratio was formed between the track fit and the cascade vertex fit run at level 2. The likelihood ratio was defined as the negative log of the track likelihood minus the negative log of the cascade likelihood.  A muon track should have a small negative log-likelihood for the track fit and a high negative log-likelihood for the cascade fit.  Therefore the likelihood ratio should be small.  The opposite holds for cascades.  The distributions are shown in figure~\ref{LikelihoodRatio}, absolutely normalized (left) and normalized to the same area (right).  The oscillations are an understood artifact.  A soft cut was placed that passes events with LlhRatio$>$-16.2.

Together, these two cuts rejected another 78\% of the background while passing 82\% of the atmospheric $\nue$ events and 1\% of the atmospheric $\numu$ events. Again, the $\numu$ efficiency appears low here because, at this level, the $\numu$ events are still dominated by long throughgoing tracks where the hadronic cascade at the interaction vertex is not inside the detector.

After these cuts were placed, additional reconstructions were run.  The first was the best-performing likelihood reconstruction of the cascade vertex.  This uses all of the photoelectrons from the full set of DOM's at local coincidence span two.  It's denoted by the name ``{\tt FullAllHitSPECascadeLlh}''.  

Next, a split cascade reconstruction was run.  The photoelectrons were split into two sets based on arrival time---an early set of hits and a late set of hits.  The cascade vertex for each set was individually reconstructed.  For a cascade which originates at a single point in the detector, the two sets of hits should form an inner sphere and an outer spherical shell around the vertex.  Therefore both vertex reconstructions should reconstruct to the same point in space.  

Finally, a 32-fold iterative likelihood reconstruction with a track hypothesis was run.  This is the best track reconstruction and is denoted by ``{\tt TrackLlhFit32}''.

\section{Atmospheric Cascade Level 3a Processing}
At this point, this atmospheric neutrino analysis and the parallel high energy extraterrestrial analysis split and developed separate cuts. For my atmospheric analysis, I ran one more additional set of reconstructions on all events to produce a set of level 3a files and level 3a ROOT files.

Several important reconstructions were run at this level. The first was the parallelogram distance calculation, which is a measure of how contained a reconstructed cascade vertex is within the detector.  This is described in section~\ref{SParallelogramDistance}.  Next, I ran the fill-ratio calculation, which is described in section~\ref{SFillRatio}.  Finally, I ran an analytic cascade energy reconstruction that takes into account the depth variation of the optical properties of the ice (``{\tt AtmCscdEnergyReco}'').  This energy reconstruction was developed specifically for the atmospheric neutrino analysis and was described in detail in chapter 4.

In addition, several basic quality parameters were calculated for each event.  Of particular importance is the number of direct hits.  A direct hit is defined to be one that arrives in a time window of -15~ns to 250~ns around the direct travel time from the cascade vertex to a DOM.  High quality cascades which have well-reconstructed vertices generally have a large number of direct hits.  On the other hand, muons which have a large stochastic energy loss are likely to leave early hits.  These early hits pull the vertex time reconstruction toward them.  Consequently, most of the hits show up as late rather than direct.

\subsection{Parallelogram Distance}
\label{SParallelogramDistance}
In AMANDA, the radial distance of a reconstructed cascade vertex from the center of the detector was traditionally used as a measure of containment.  However,  the lack of cylindrical symmetry in IC-22 reduces the utility of such a measure. To better capture the string layout of IC-22, I defined a new quantity which I denote the ``parallelogram distance''.

\begin{figure}
\begin{center}
\includegraphics[width=0.8\textwidth]{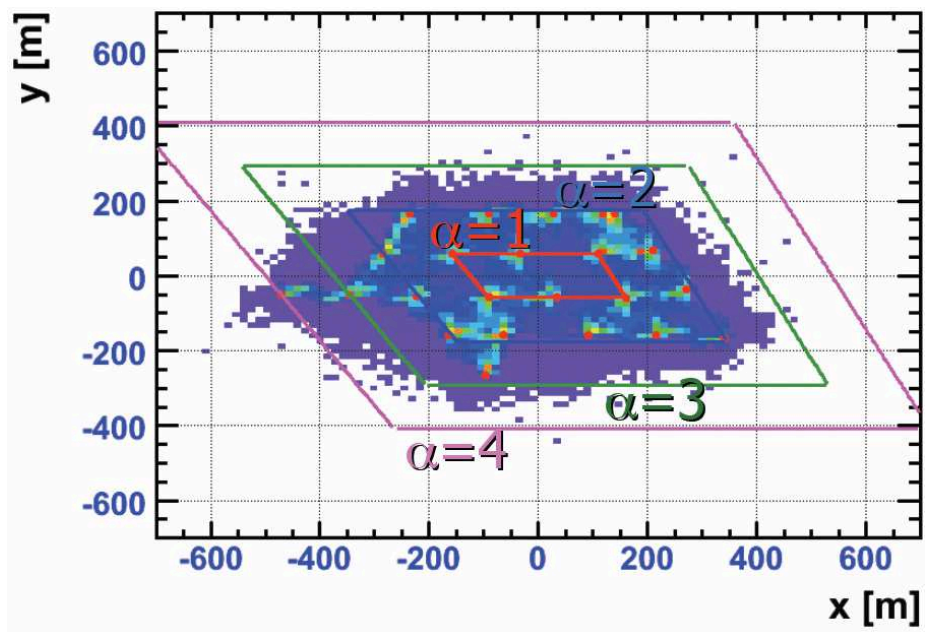}
\caption{Geometry of the parallelogram distance containment construction.}
\label{ParallelogramDistanceConstruction}
\end{center}
\end{figure}

Figure~\ref{ParallelogramDistanceConstruction} illustrates the construction. It shows a top view of IC-22. The solid red circles are the locations of the 22 strings, and the density plot shows reconstructed cascade vertex positions for data.

The construction begins by defining a parallelogram that hits the innermost layer of 6 strings in IC-22. This parallelogram is shown in red in the figure. By definition vertices that fall on this parallelogram have parallelogram distance equal to one.

Next, for a given vertex position, this parallelogram is stretched until it hits the vertex. The stretching is parameterized by a single number $\alpha$.  Let $Y_u$ ($Y_l$) denote the y-coordinate of the midpoint of the upper (lower) line of this innermost parallelogram and let $X_l$ ($X_r$) denote the x-coordinate of the midpoint of the left (right) line of this innermost parallelogram.  The transformation takes

\begin{align*}
&Y_u \rightarrow Y_u+2(\alpha-1)Y_u \\
&Y_l \rightarrow Y_l+2(\alpha-1)Y_l \\
&X_l \rightarrow \alpha X_l \\
&X_r \rightarrow \alpha X_r \\
\end{align*}

\noindent The transformation is defined in such a way that integer values of the parallelogram distance correspond to string layers of IC-22 (in general the factor need not be an integer though). The blue parallelogram in the figure corresponds to parallelogram distance equal to two and hits the main outer string layer. Parallelogram distances three and four are in green and magenta in the figure and hit outlying strings.

In general, we expect that our analysis will need to focus on well-contained cascade events. These would be events with parallelogram distance less than 1.5 or so, as will be determined at later stages of the analysis.

\begin{figure}
\begin{minipage}[b]{0.48\linewidth} % A minipage that covers half the page
\centering
\includegraphics[width=7.6cm]{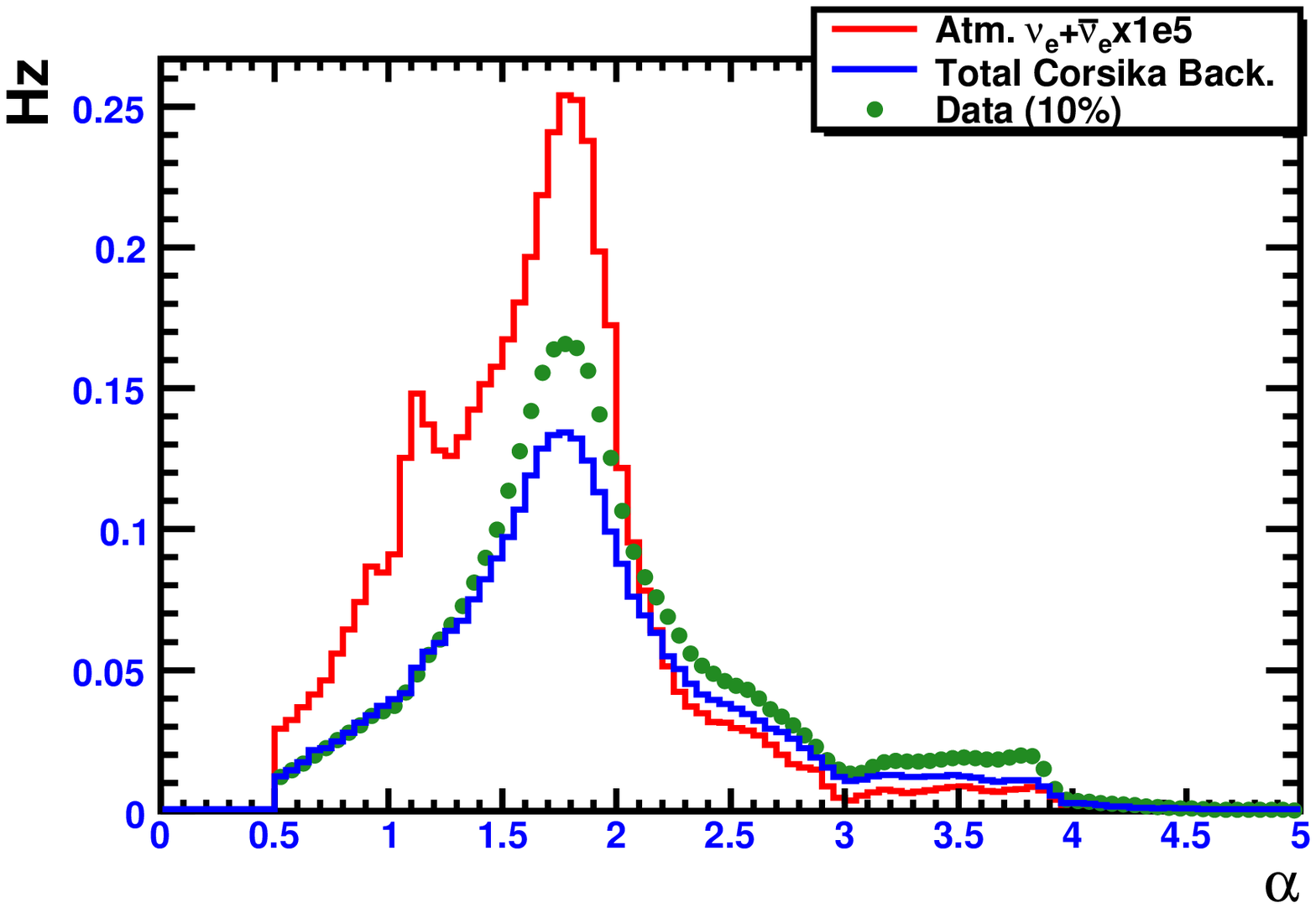}
\end{minipage}
\hspace{0.5cm} %To get a little bit of space between the figures
\begin{minipage}[b]{0.48\linewidth}
\centering
\includegraphics[width=7.6cm]{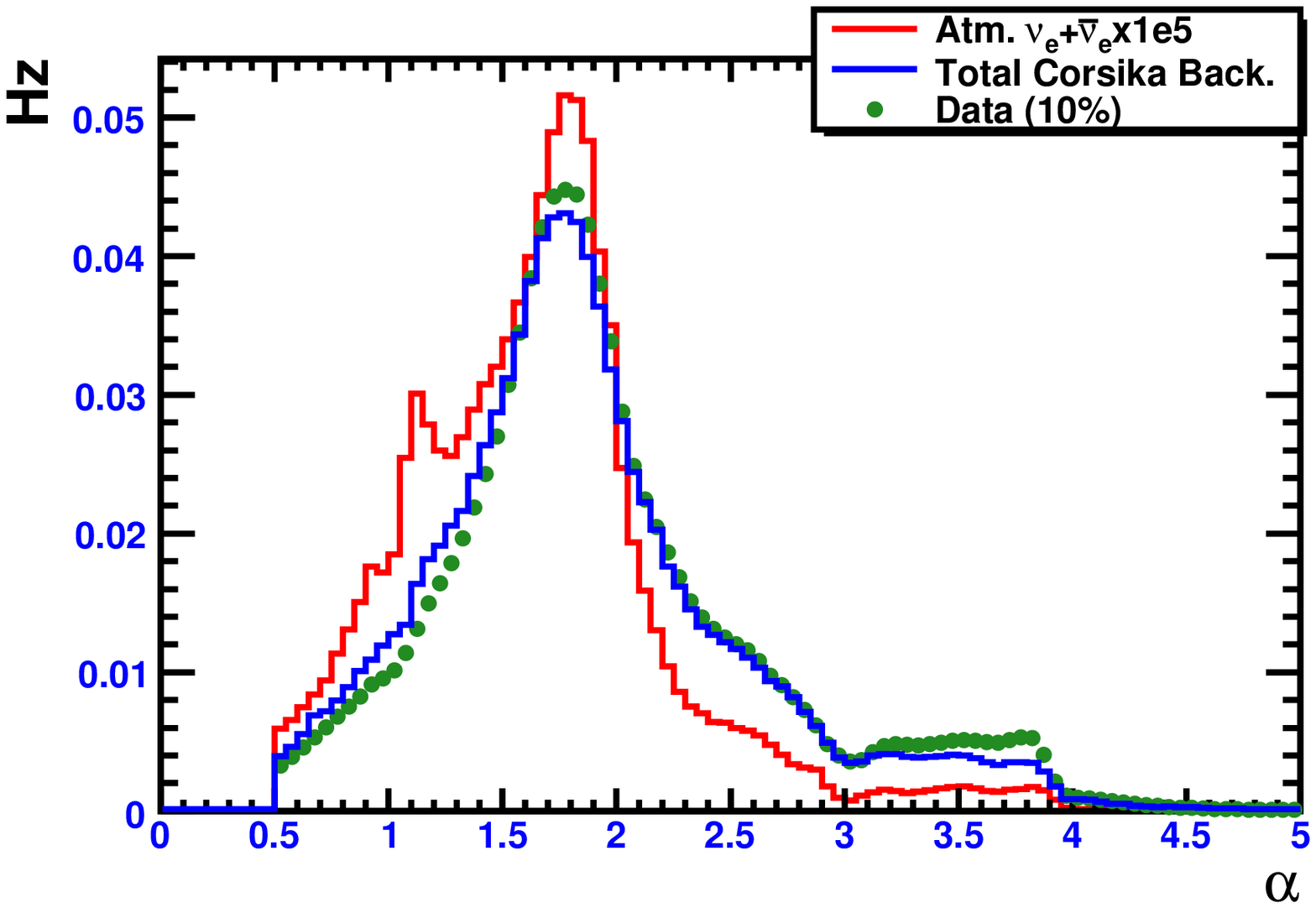}
\end{minipage}
\vspace{0.25cm}
\caption{Parallelogram distance at level 3a, absolutely normalized (left) and normalized to the same area (right).}
\label{ParallelogramDistance}
\end{figure}

The distribution of this parameter at level 3a is shown in figure~\ref{ParallelogramDistance}, absolutely normalized (left) and normalized to the same area (right).

\subsection{Fill-Ratio}
\label{SFillRatio}
The fill-ratio calculation is intended to exploit the spherical versus track topology of cascades and muons. It begins with a reconstructed cascade vertex and calculates the distance from this vertex to all of the hit DOM's in the event. It builds the distribution of these distances and calculates the mean and RMS distance from the vertex. 

Next, a sphere is drawn around the reconstructed vertex.  The sphere has a radius which is a multiple of the mean or RMS hit distance.  The ratio of the number of hit DOM's within the sphere to the total number of DOM's within the sphere is calculated. The two parameters are called the fill-ratio from mean and the fill-ratio from RMS.

The sensitivity of this cut parameter comes from two things. First, cascades are rather spherical while muons are long thin tracks. A cascade is therefore expected to fill a larger proportion of its sphere than a muon is. However, our main background consists of large stochastic energy losses from muons, which are also spherical. As the energy of the stochastic increases though, the probability of there being an extra, early hit from the track increases too. This extra hit pushes the sphere to have a larger radius, so the proportion of hit DOM's within the sphere should go down.

\begin{figure}
\begin{minipage}[b]{0.48\linewidth} % A minipage that covers half the page
\centering
\includegraphics[width=7.6cm]{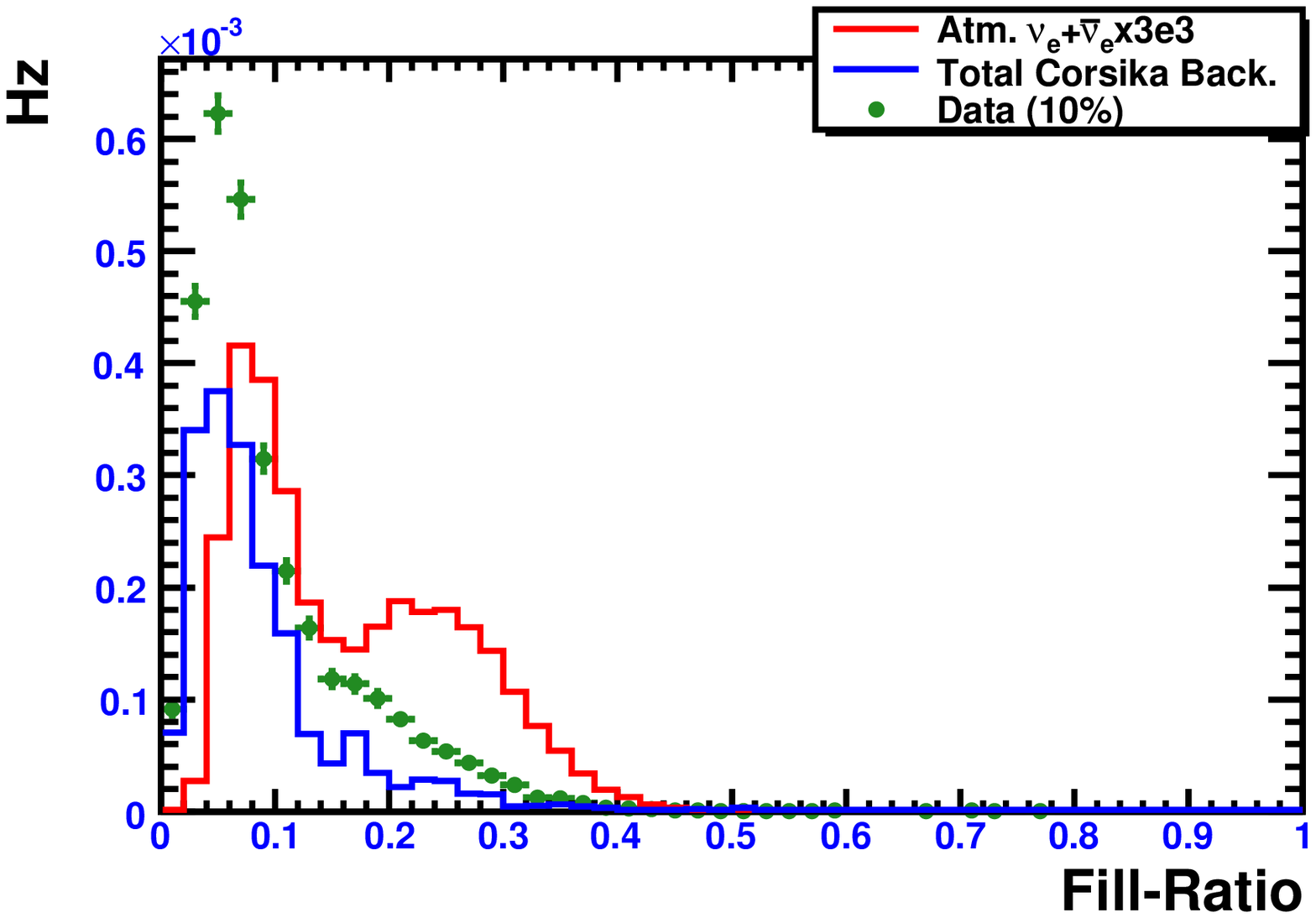}
\end{minipage}
\hspace{0.5cm} %To get a little bit of space between the figures
\begin{minipage}[b]{0.48\linewidth}
\centering
\includegraphics[width=7.6cm]{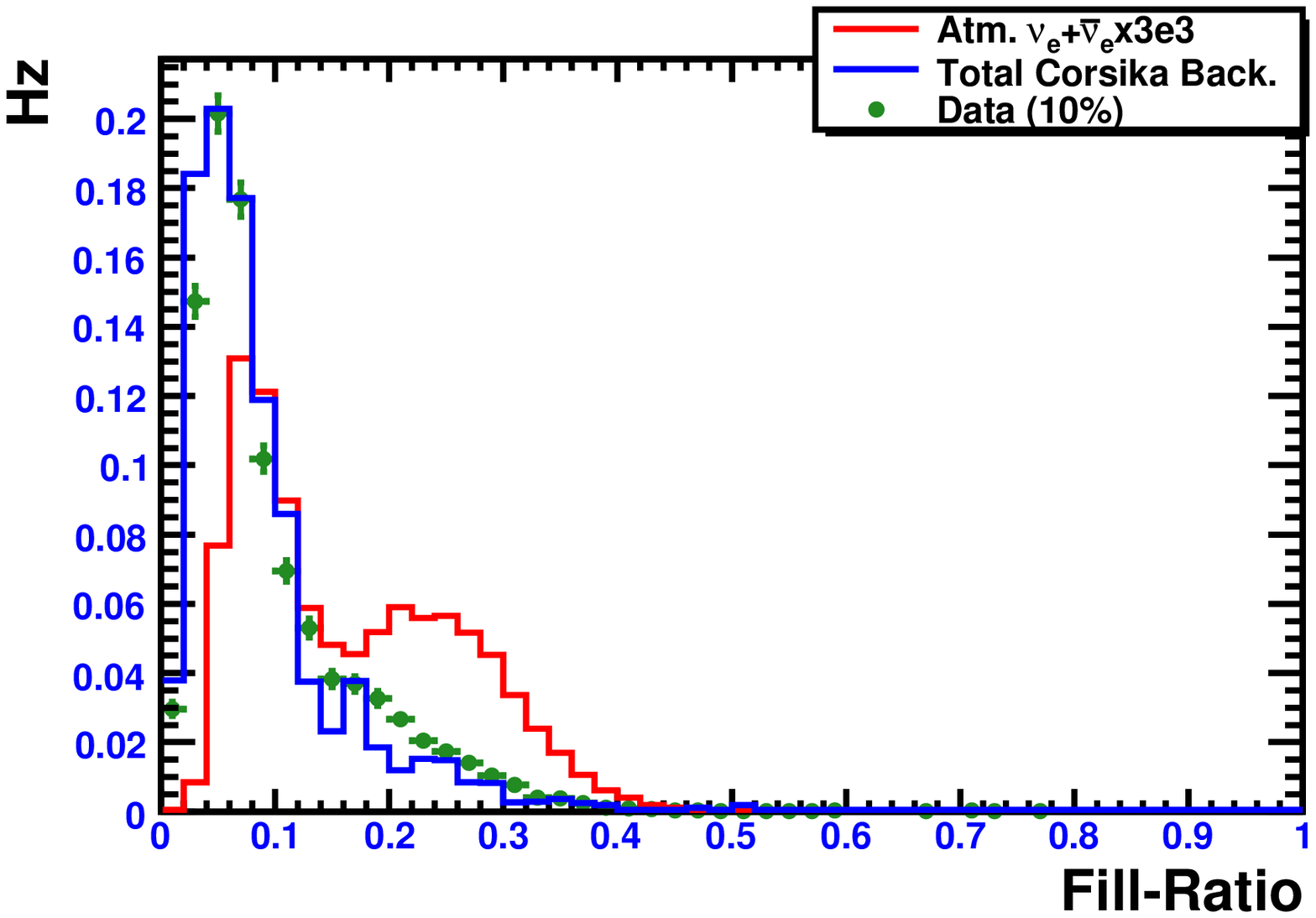}
\end{minipage}
\vspace{0.25cm}
\caption{The fill-ratio calculated from the mean hit distance at level 4a (defined below), absolutely normalized (left) and normalized to the same area (right).}
\label{FillRatioFromMean}
\end{figure}

\begin{figure}
\begin{minipage}[b]{0.48\linewidth} % A minipage that covers half the page
\centering
\includegraphics[width=7.6cm]{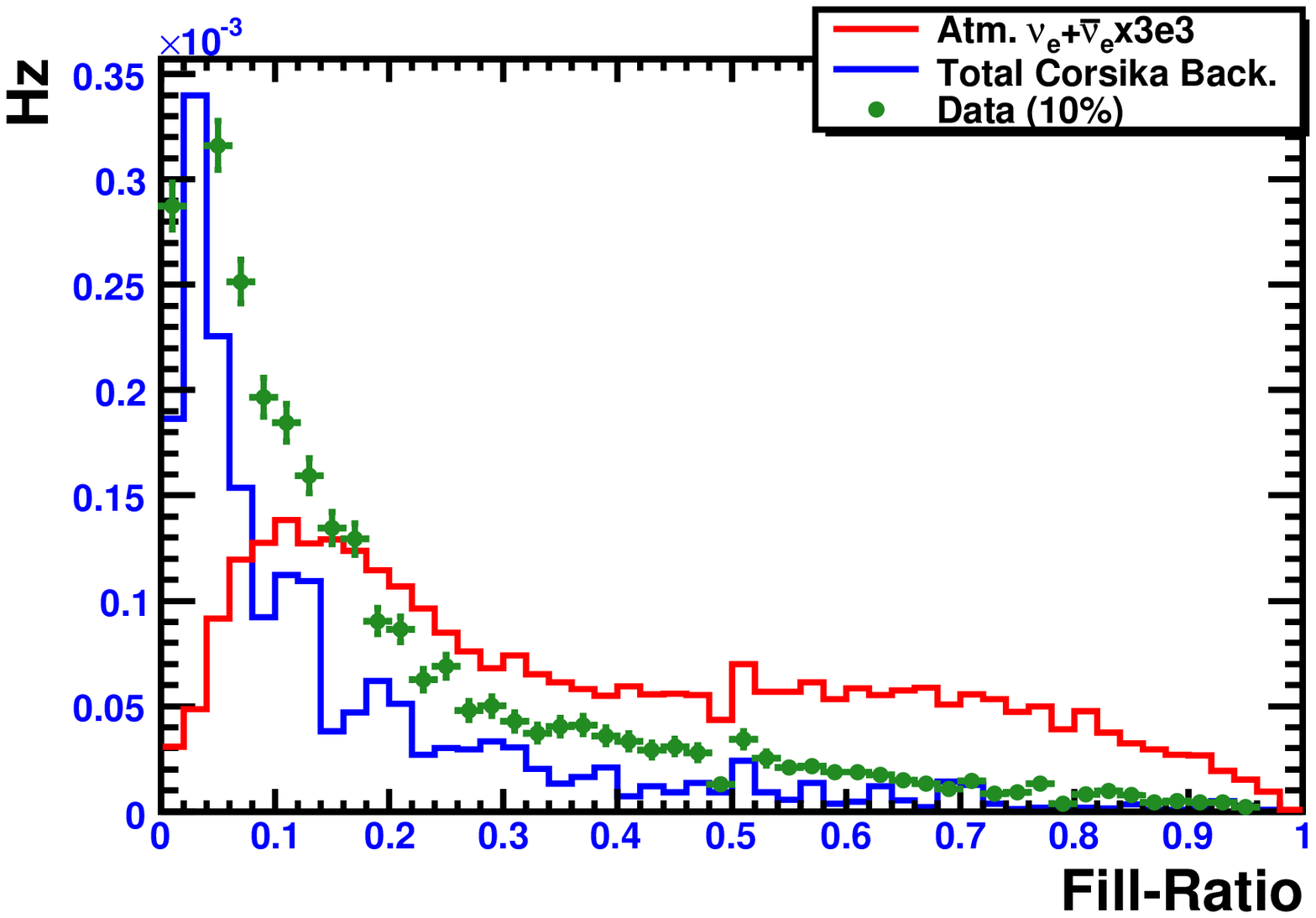}
\end{minipage}
\hspace{0.5cm} %To get a little bit of space between the figures
\begin{minipage}[b]{0.48\linewidth}
\centering
\includegraphics[width=7.6cm]{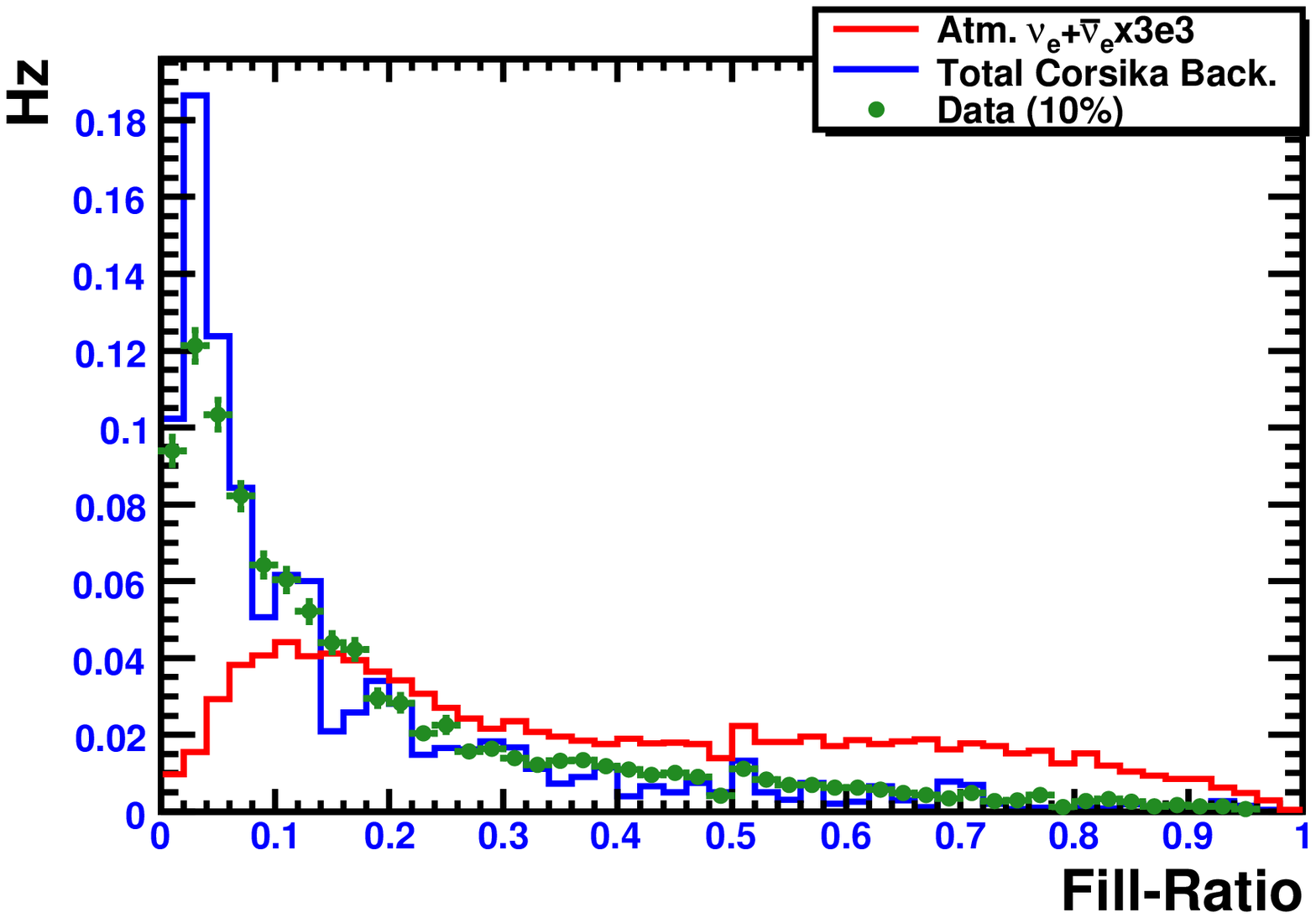}
\end{minipage}
\vspace{0.25cm}
\caption{The fill-ratio calculated from the RMS hit distance at level 4a (defined below), absolutely normalized (left) and normalized to the same area (right).}
\label{FillRatioFromRMS}
\end{figure}

The distribution of the fill-ratio from mean at level 4a is shown in figure~\ref{FillRatioFromMean}, absolutely normalized (left) and normalized to the same area (right).  The distribution of the fill-ratio from RMS at level 4a is shown in figure~\ref{FillRatioFromRMS}, absolutely normalized (left) and normalized to the same area (right).

\clearpage

\section{Neural Networks}
After the level 3a processing, the calculation of all of our discriminating variables is complete.  We want to find the optimal combination of these variables, perhaps in several stages, in order to reject the most muon background while retaining the largest amount of our atmospheric neutrino-induced cascade signal.  To accomplish this task, we use artificial neural networks (ANN) as implemented in ROOT's TMVA package \cite{TMVAPackage}.

\subsection{Neural Networks as Multivariate Classifiers}
Let's assume that our events are characterized by N discriminating variables and that we want to use these variables to characterize them as signal or background.  We can think of each event as a point in an N-dimensional space, where the coordinate along each axis is the value of that particular discriminating variable.  If our variables are selected well, signal events should cluster in a certain area of this N-dimensional space and background events in another.  The classification problem then becomes a problem of drawing a surface in this N-dimensional space that optimally separates the signal region from the background region with the least amount of leakage from one side to the other.

There are many different multivariate classifiers in the literature, and many or all of them have been applied to high energy physics analyses.  Examples include likelihood methods, nearest-neighbor methods, support vector machines, and neural networks.  ROOT's TMVA package is a flexible framework for training, testing, and evaluating a large number of these multivariate classifiers.  In testing, its neural networks outperformed all other methods for this analysis.

Essentially, a neural network is just a nonlinear function.  Given the N input discriminating variables, it outputs a real number.  If we denote the values of the discriminating variables by $\{x_1, x_2, ... x_N\}$ then:

$$
f_{\mbox{\fontsize{8}{14}\selectfont ANN}}(x_1, x_2, ... x_N) = y_{\mbox{\fontsize{8}{14}\selectfont ANN}}  \in \mathbb{R}
$$

\begin{figure}
\begin{center}
\includegraphics[width=0.8\textwidth]{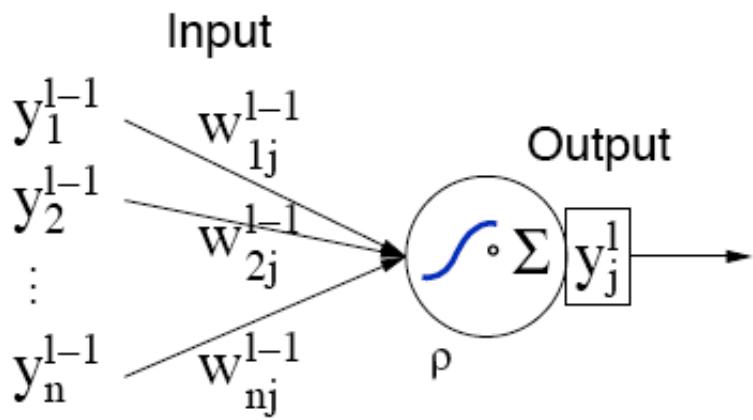}
\caption{Schematic of a single neuron from a neural network.  From \cite{TMVAManual}.}
\label{SingleNeuron}
\end{center}
\end{figure}

This nonlinear function is constructed by considering a collection, or network, of artificial neurons.  Each neuron integrates the information from a set of inputs and produces an output.  Figure~\ref{SingleNeuron} shows a schematic illustration of neuron $j$ of a neural network.  The inputs $\{y^{l-1}_1, y^{l-1}_2, ... y^{l-1}_n\}$ are connected to the neuron with weights $\{w^{l-1}_{1j}, w^{l-1}_{2j}, ... w^{l-1}_{nj}\}$.  The neuron uses its ``synapse function'' to integrate these weighted inputs.  The synapse function can take several forms.  In this analysis, a simple sum was used:

$$
x_j \equiv w^{l-1}_{0j}+\sum_{i=1}^{n} y^{l-1}_i w^{l-1}_{ij}
$$

Next, the neuron takes the integrated input from this synapse function and uses its ``activation function'' to produce an output value.  Again, the activation function can take several forms.  In this analysis, a hyperbolic tangent function was used:

$$
x_j \rightarrow y^l_j = \frac{e^{x_j}-e^{-x_j}}{e^{x_j}+e^{-x_j}}
$$

\begin{figure}
\begin{center}
\includegraphics[width=0.8\textwidth]{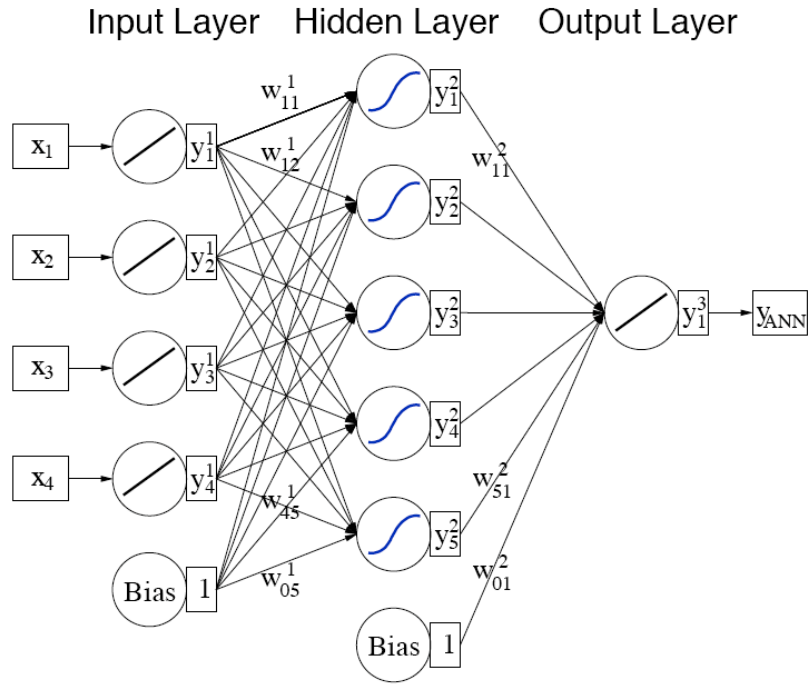}
\caption{General layout of a multilayer perceptron (MLP).  From \cite{TMVAManual}.}
\label{NeuralNetArchitecture}
\end{center}
\end{figure}

A neural network, then, is a collection of these single neurons.  Each neuron's output can be connected to the input of other neurons in the network.  In general, $Q$ neurons can have $Q^2$ connections between them.  However, it is simplest to organize the neurons by layers and to only allow connections from one layer to the next.  This type of network is known as a ``multilayer perceptron'' and is the type used for this analysis.  It is indicated schematically in figure~\ref{NeuralNetArchitecture}.

In the schematic, we have $N=4$ discriminating variables $\{x_1, x_2, x_3, x_4\}$.  They enter the network from the left and are connected through a set of weights to the neurons in the hidden layer.  These neurons, in turn, are connected by another set of weights to the output neuron which produces the final output variable $y_{\mbox{\fontsize{8}{14}\selectfont ANN}}$.

All neural networks used in this analysis had two hidden layers, the first with N+1 neurons and the second with N neurons where, again, N is the number of discriminating variables.

The sets of weights connecting the layers of the neural network are chosen during the training process.  A series of signal and background events (usually from Monte Carlo) are fed to the network.  The weights are adjusted in an attempt to push the output variable $y_{\mbox{\fontsize{8}{14}\selectfont ANN}}$ as close as possible to 1 for signal events and  to -1 for background events. 

\subsection{Level 3a Neural Network}
A first neural network was trained on five variables.  These variables are listed below by decreasing separation power: 

\begin{itemize}
\item{{\bf 32FoldZenith}: The track zenith angle of the 32-fold iterative muon reconstruction.  While downgoing muons were cut with a zenith cut during level 3 processing, this better, more time-intensive likelihood reconstruction recovers some of the previously mis-reconstructed downgoing muons.}
\item{{\bf FullAllHitSPEReducedLlh}: The reduced likelihood parameter from the best performing cascade vertex reconstruction.  A lower value of this variable indicates a better fit to the cascade hypothesis.}
\item{{\bf N1HitFrac}:  The number of DOM's which receive a single photoelectron divided by the total number of photoelectrons seen by all DOM's.  Muons are expected to have more single hits and so a higher value of this parameter.}
\item{{\bf NDirE}:  The number of direct hits from the reconstructed cascade vertex.  This should be large for a cascade.  For a muon, early hits from the muon track skew the vertex time.  A muon, then, should have a small number of direct hits.}
\item{{\bf SplitDistZ}: The difference in z vertex positions for the two split cascade reconstructions.  It should be close to zero for cascade events, since the early hits and the late hits should both reconstruct to the same position.}
\end{itemize}

\begin{figure}
\begin{minipage}[b]{0.48\linewidth} % A minipage that covers half the page
\centering
\includegraphics[width=7.6cm]{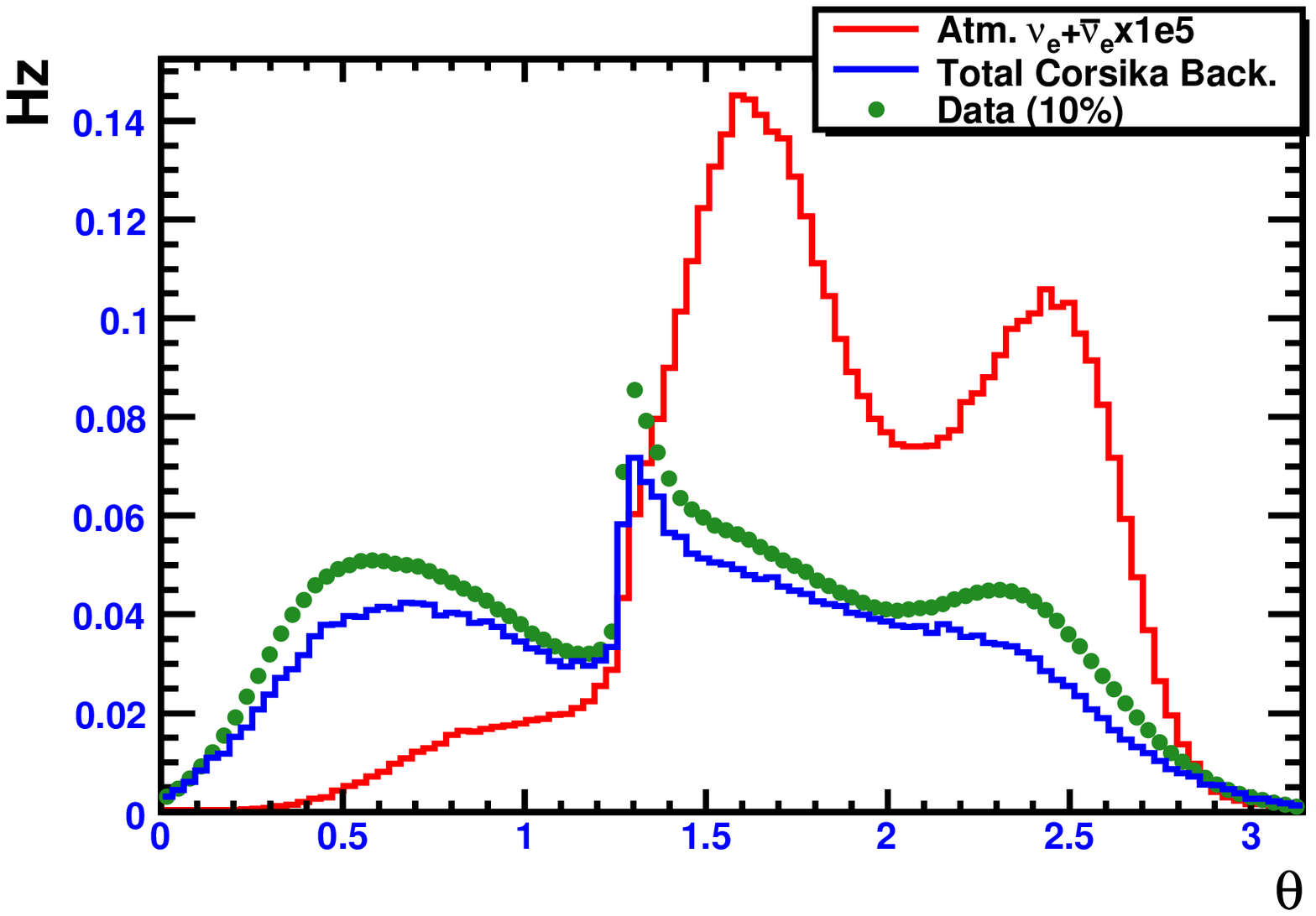}
\label{32-Fold Zenith}
\end{minipage}
\hspace{0.5cm} %To get a little bit of space between the figures
\begin{minipage}[b]{0.48\linewidth}
\centering
\includegraphics[width=7.6cm]{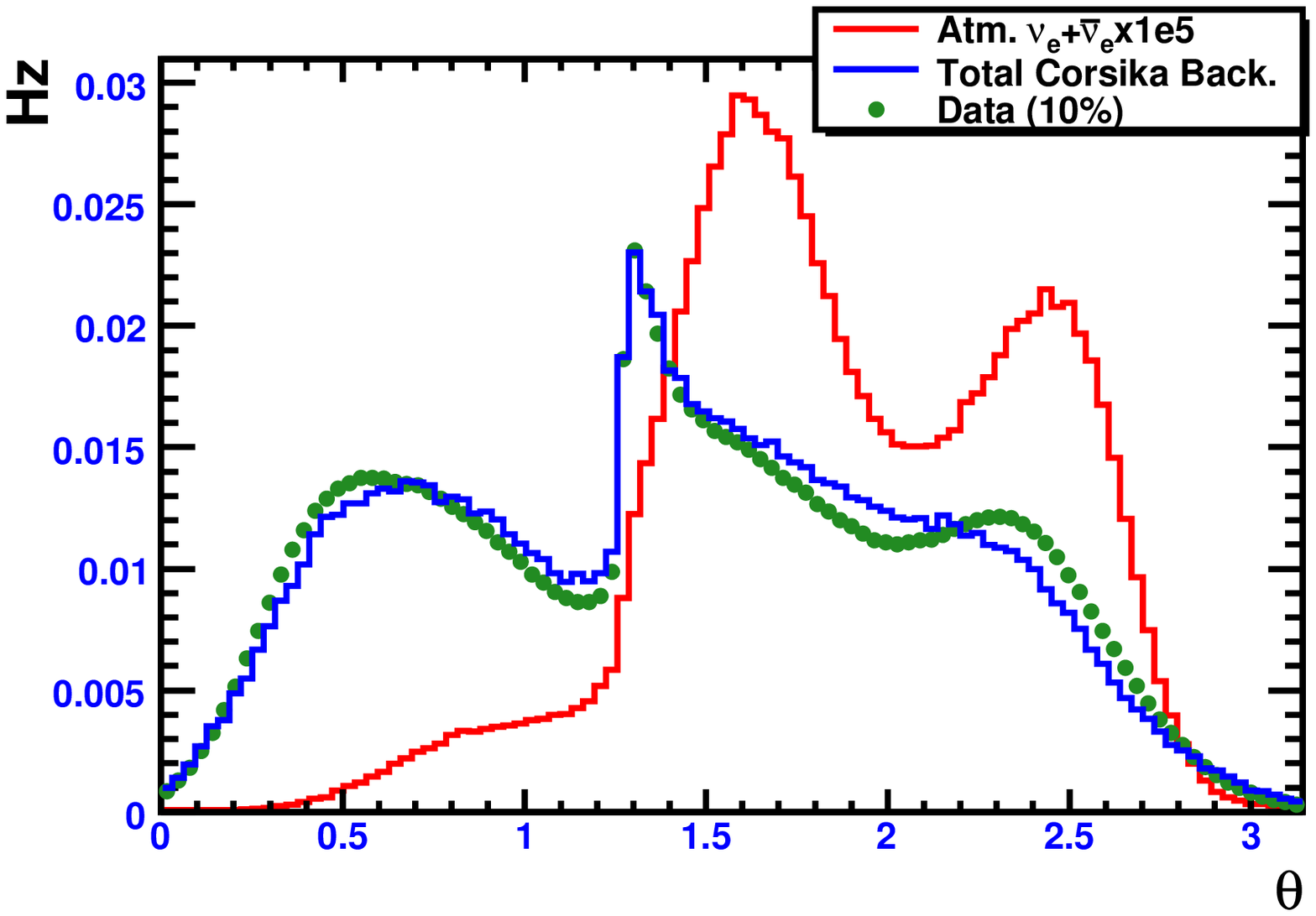}
\label{32-Fold Zenith (norm)}
\end{minipage}
\vspace{0.25cm}
\caption{Reconstructed track zenith angle from the 32-fold iterative reconstruction at level 3a, absolutely normalized (left) and normalized to the same area (right).}
\end{figure}

\begin{figure}
\begin{minipage}[b]{0.48\linewidth} % A minipage that covers half the page
\centering
\includegraphics[width=7.6cm]{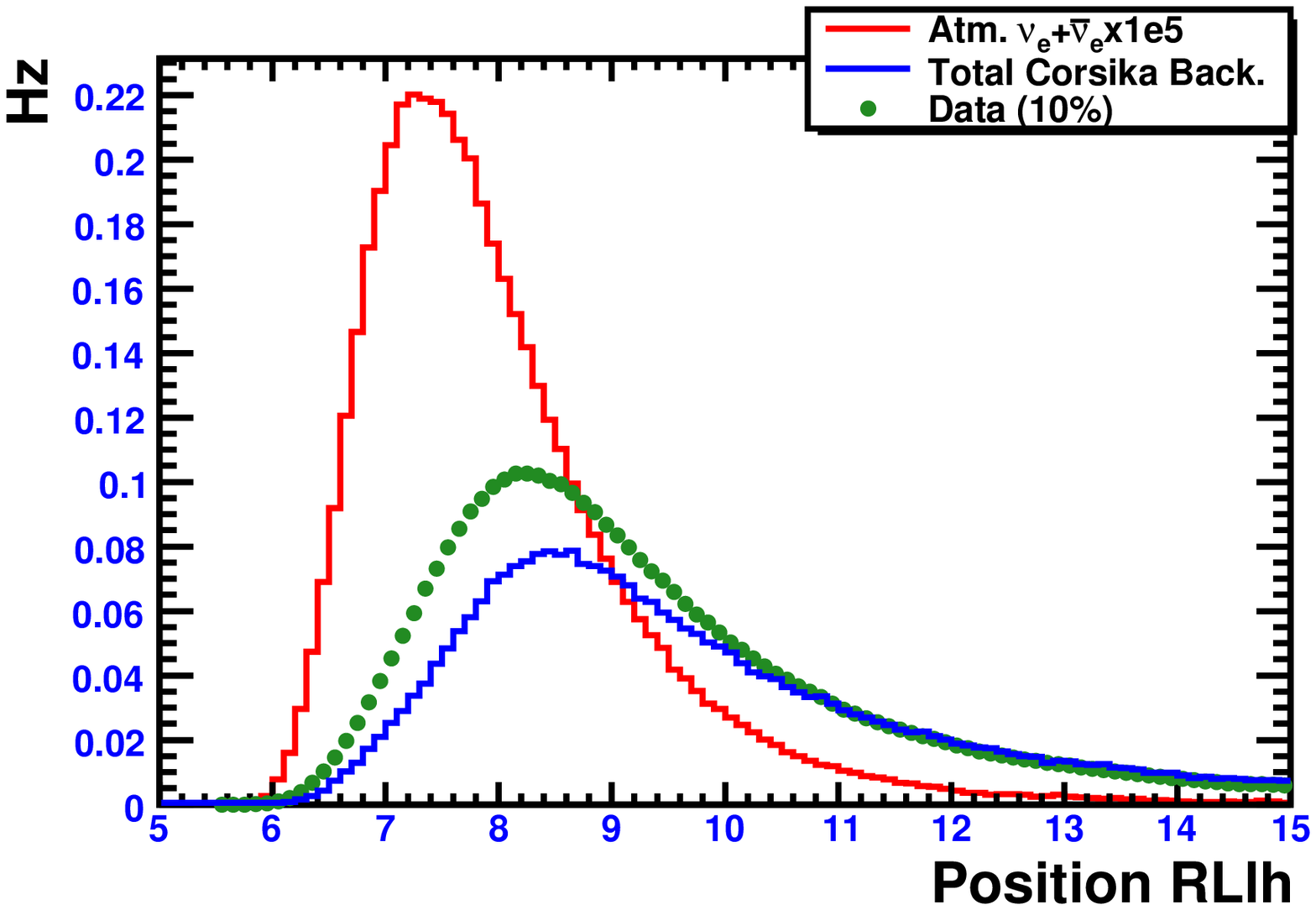}
\label{FullAllHitSPEReducedLlh}
\end{minipage}
\hspace{0.5cm} %To get a little bit of space between the figures
\begin{minipage}[b]{0.48\linewidth}
\centering
\includegraphics[width=7.6cm]{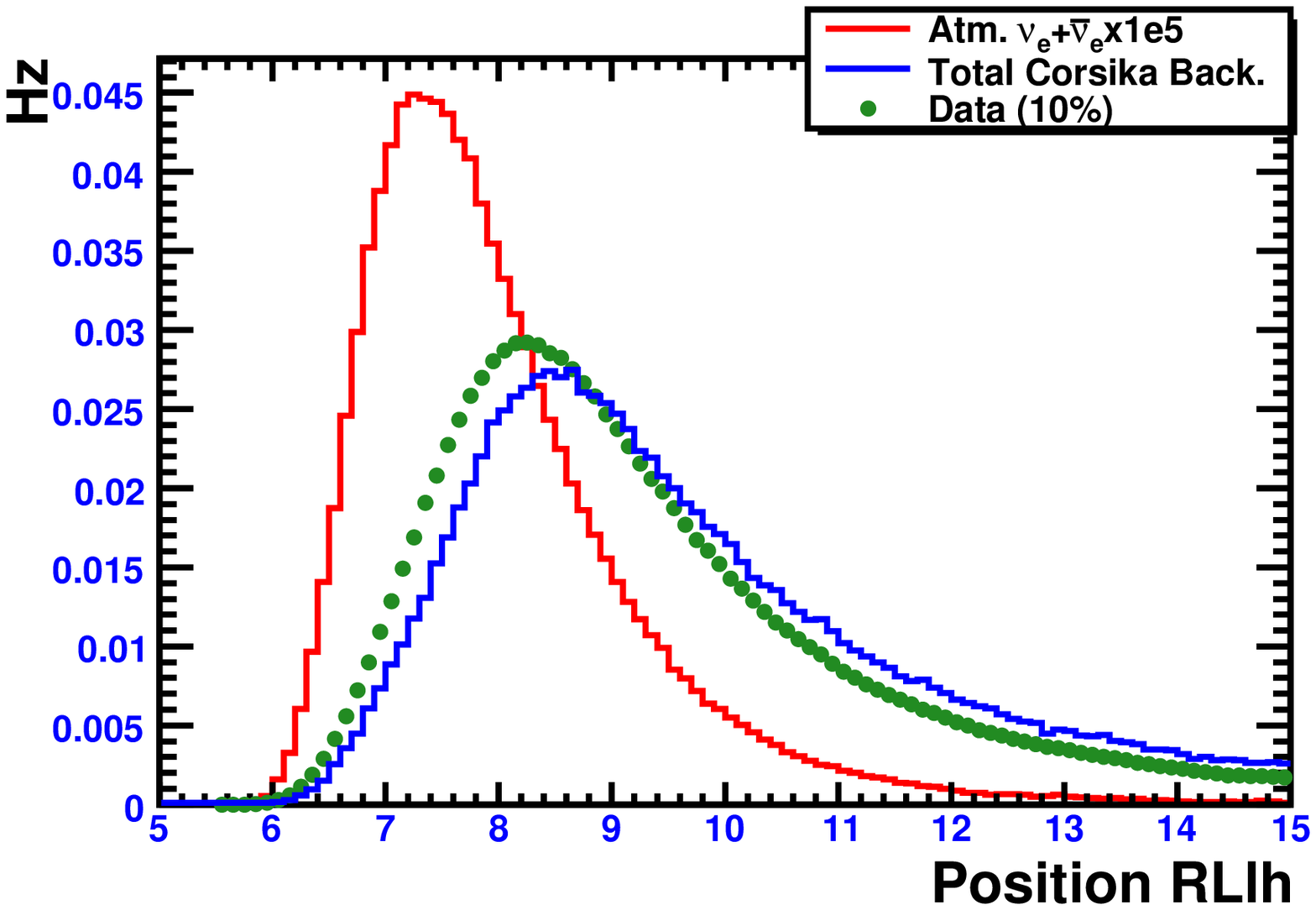}
\label{FullAllHitSPEReducedLlh (norm)}
\end{minipage}
\vspace{0.25cm}
\caption{Reduced likelihood from the FullAllHitSPEReducedLlh vertex reconstruction at level 3a, absolutely normalized (left) and normalized to the same area (right).}
\end{figure}

\begin{figure}
\begin{minipage}[b]{0.48\linewidth} % A minipage that covers half the page
\centering
\includegraphics[width=7.6cm]{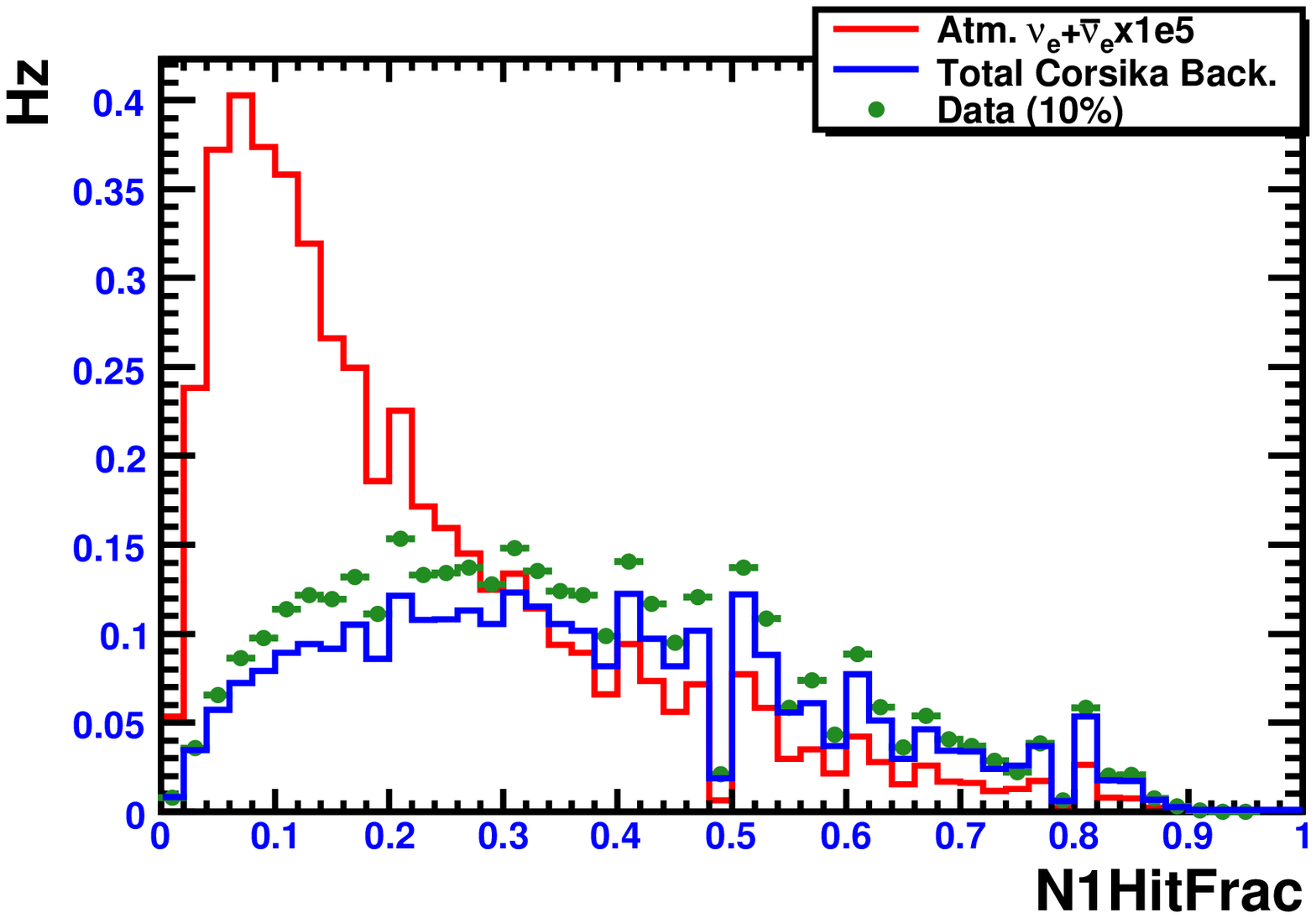}
\label{N1HitFrac}
\end{minipage}
\hspace{0.5cm} %To get a little bit of space between the figures
\begin{minipage}[b]{0.48\linewidth}
\centering
\includegraphics[width=7.6cm]{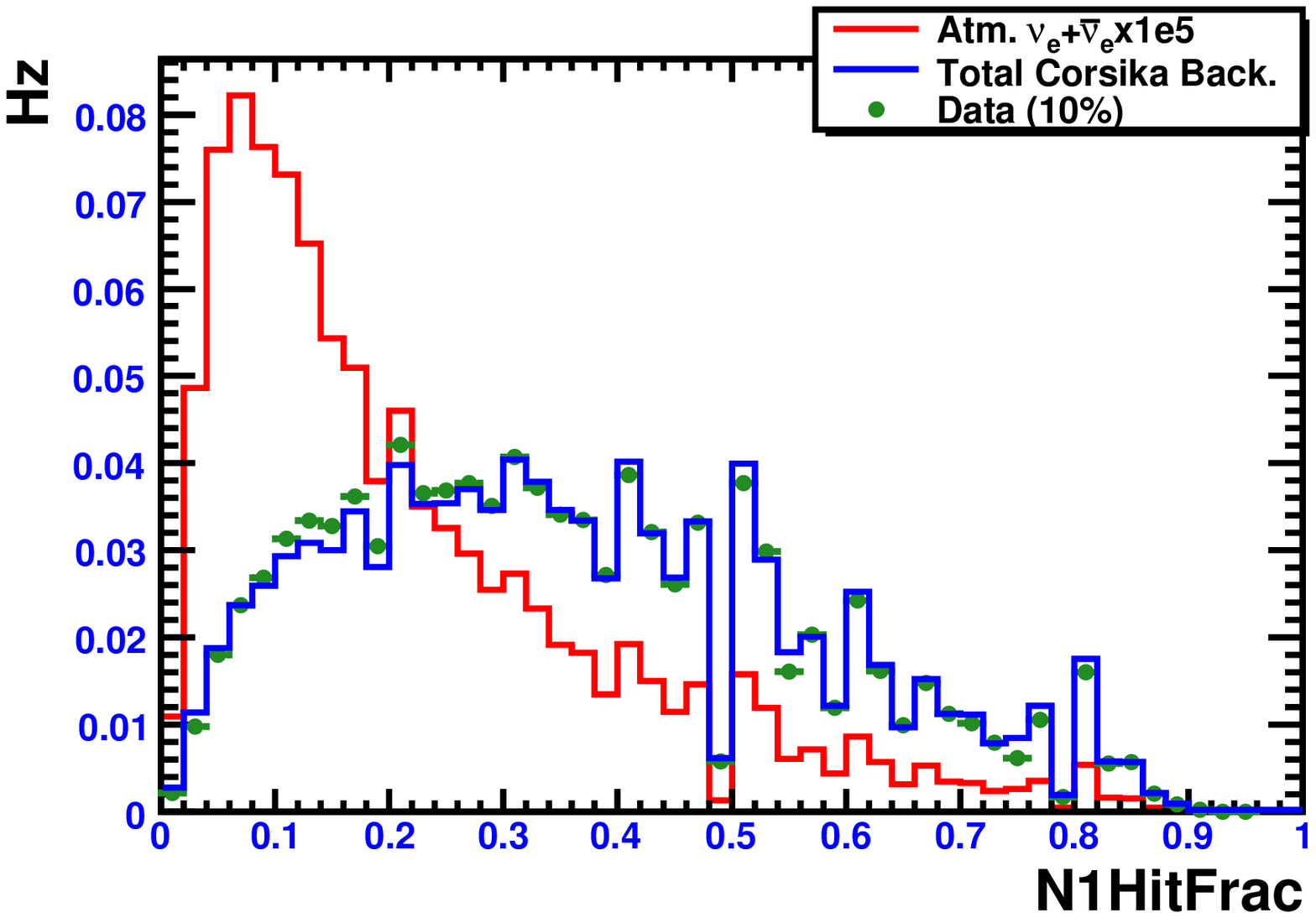}
\label{N1HitFrac (norm)}
\end{minipage}
\vspace{0.25cm}
\caption{N1HitFrac at level 3a, absolutely normalized (left) and normalized to the same area (right).}
\end{figure}

\begin{figure}
\begin{minipage}[b]{0.48\linewidth} % A minipage that covers half the page
\centering
\includegraphics[width=7.6cm]{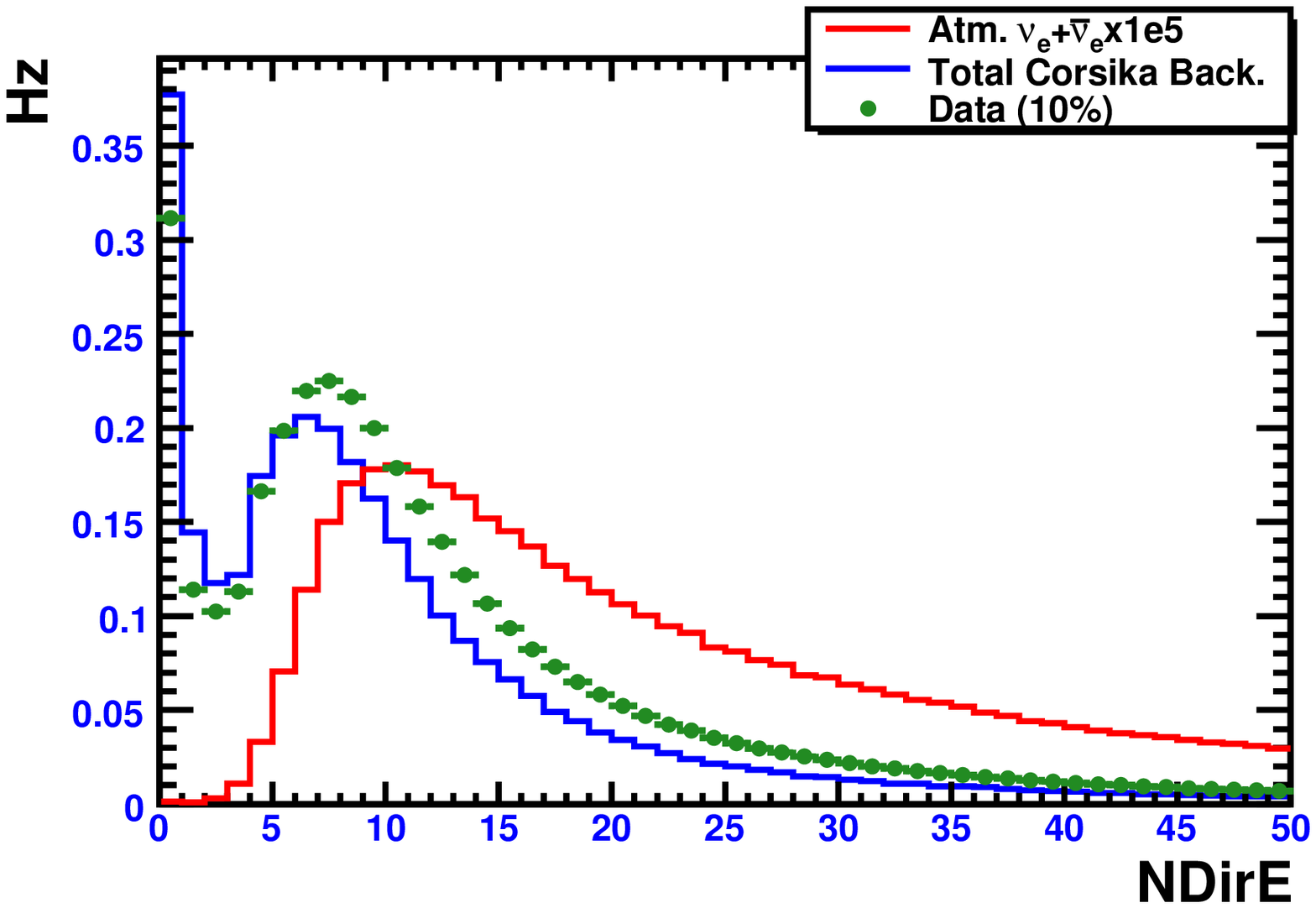}
\label{NDirE}
\end{minipage}
\hspace{0.5cm} %To get a little bit of space between the figures
\begin{minipage}[b]{0.48\linewidth}
\centering
\includegraphics[width=7.6cm]{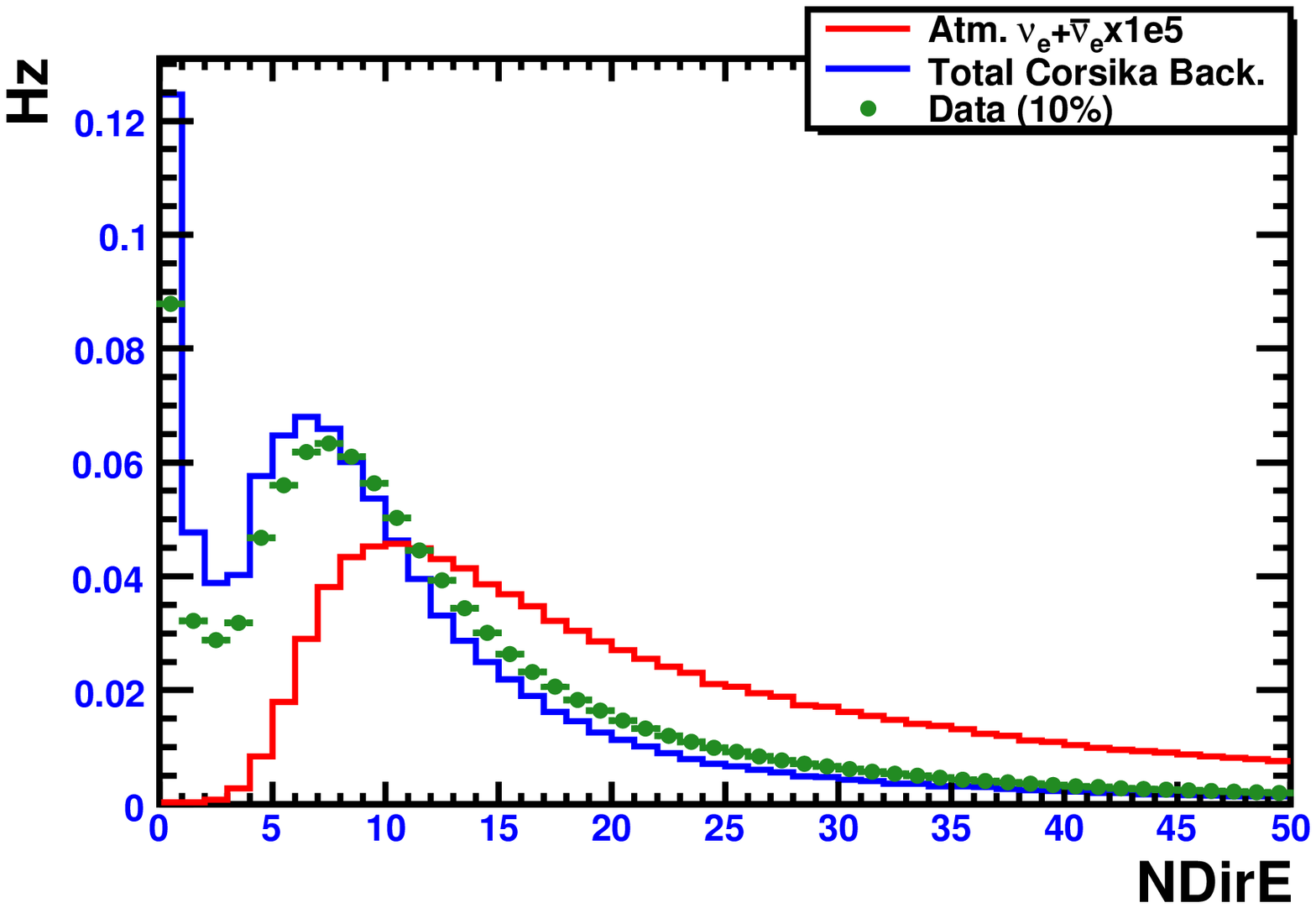}
\label{NDirE (norm)}
\end{minipage}
\vspace{0.25cm}
\caption{Number of direct hits NDirE at level 3a, absolutely normalized (left) and normalized to the same area (right).}
\end{figure}

\begin{figure}
\begin{minipage}[b]{0.48\linewidth} % A minipage that covers half the page
\centering
\includegraphics[width=7.6cm]{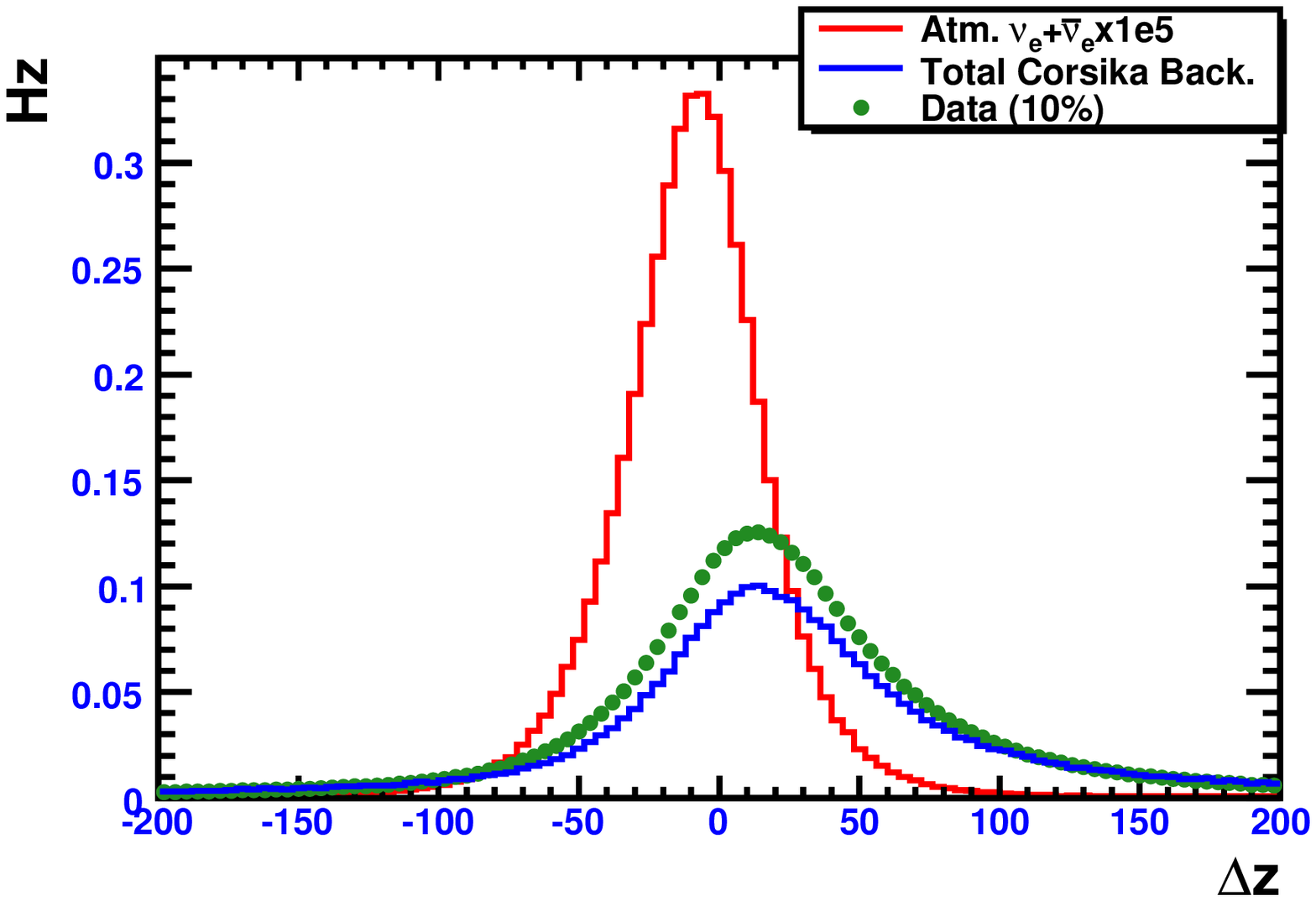}
\label{SplitDistZ}
\end{minipage}
\hspace{0.5cm} %To get a little bit of space between the figures
\begin{minipage}[b]{0.48\linewidth}
\centering
\includegraphics[width=7.6cm]{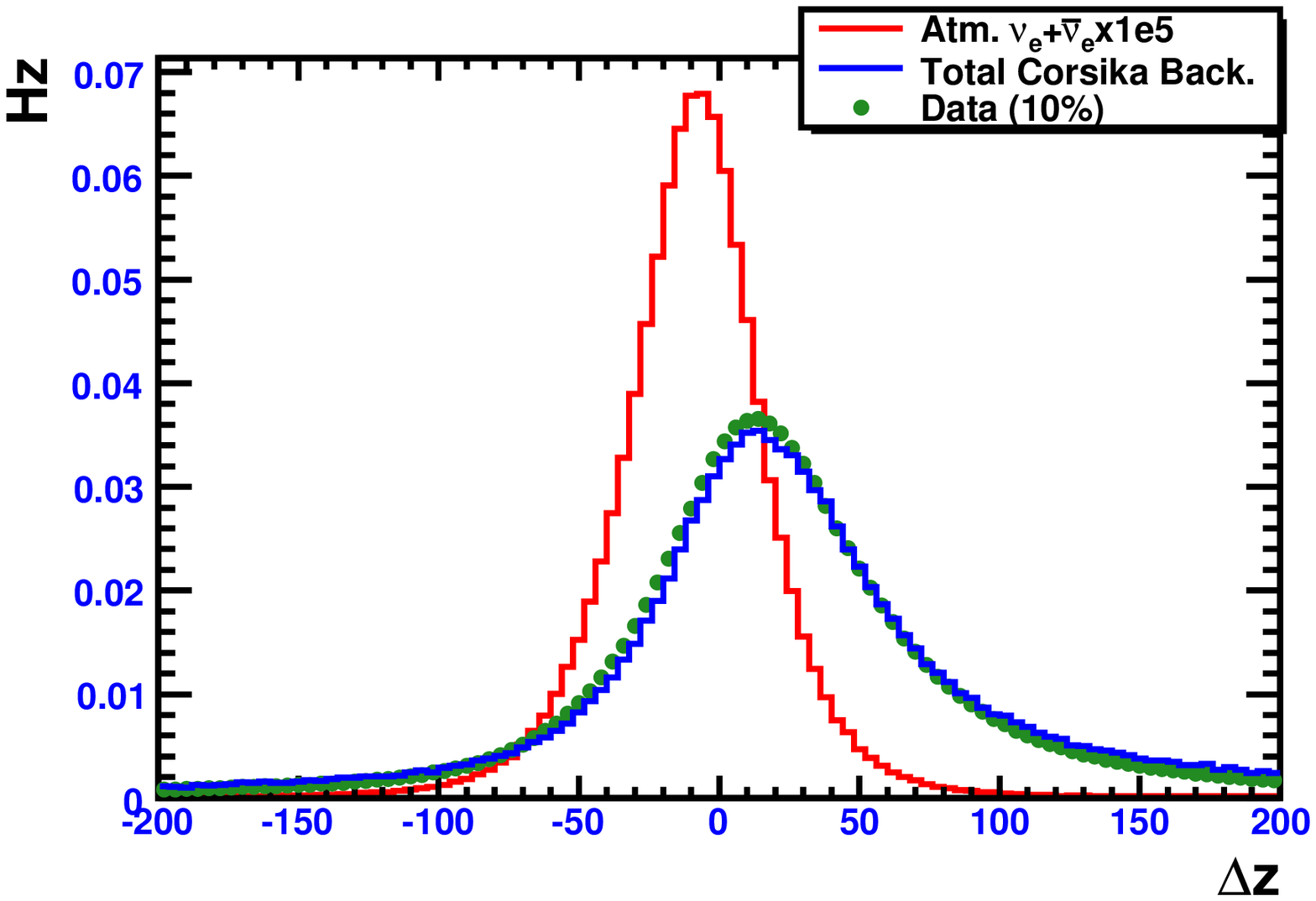}
\label{SplitDistZ (norm)}
\end{minipage}
\vspace{0.25cm}
\caption{Split reconstruction z vertex difference SplitDistZ at level 3a, absolutely normalized (left) and normalized to the same area (right).}
\end{figure}

\begin{figure}
\begin{minipage}[b]{0.48\linewidth} % A minipage that covers half the page
\centering
\includegraphics[width=7.6cm]{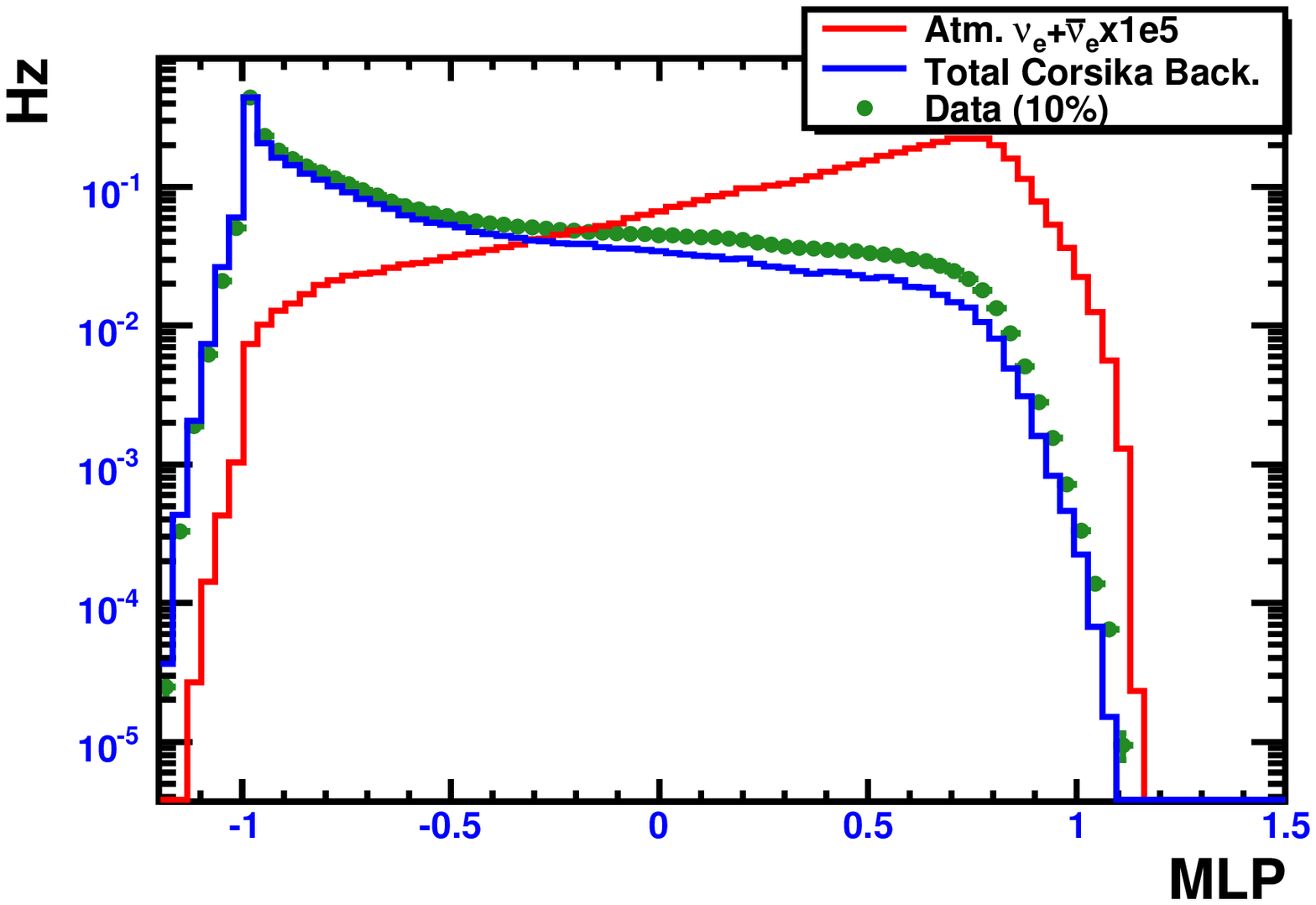}
\label{L3aMLP}
\end{minipage}
\hspace{0.5cm} %To get a little bit of space between the figures
\begin{minipage}[b]{0.48\linewidth}
\centering
\includegraphics[width=7.6cm]{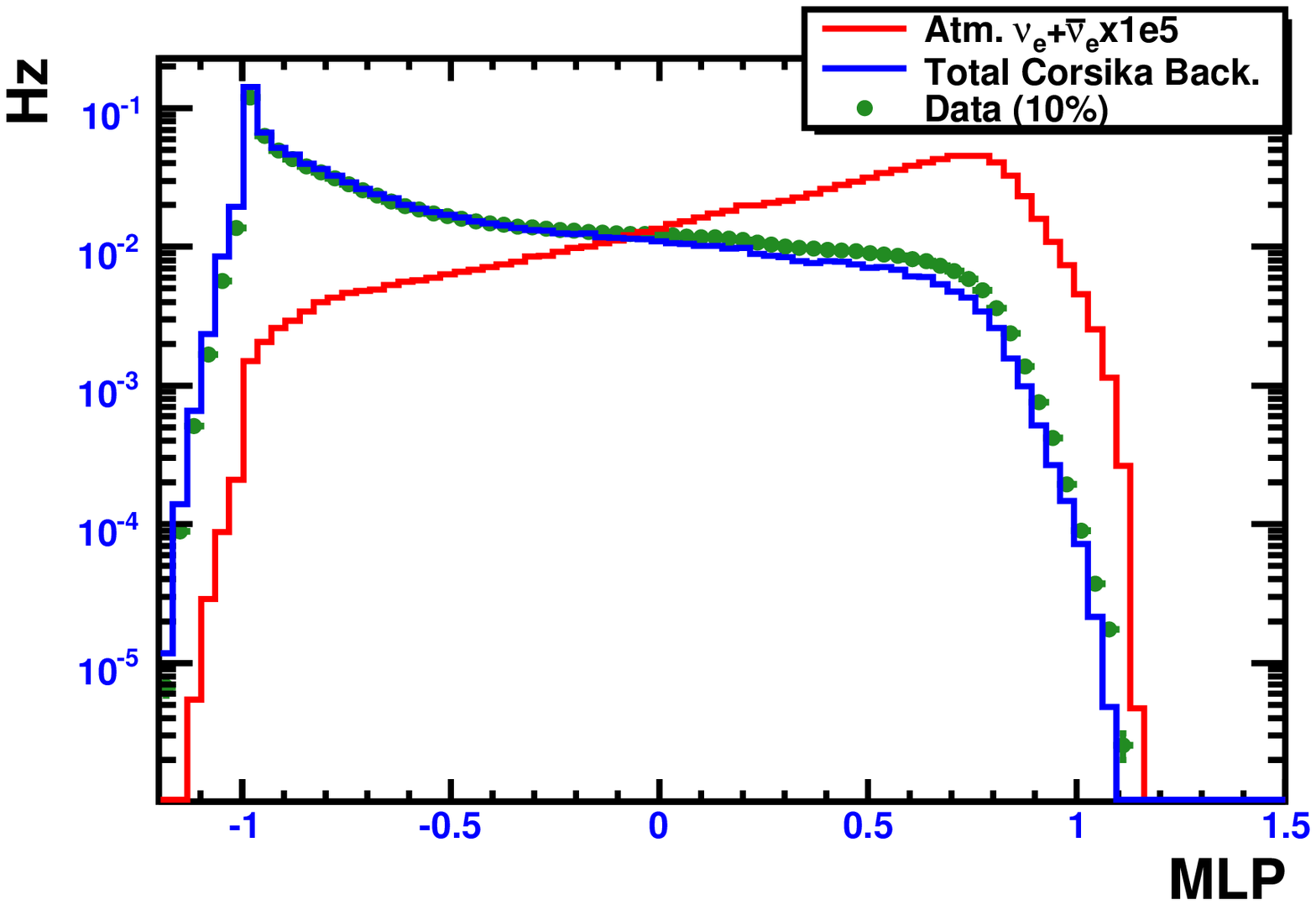}
\label{L3aMLP (norm)}
\end{minipage}
\vspace{0.25cm}
\caption{Level 3a neural net output variable L3aMLP at level 3a, absolutely normalized (left) and normalized to the same area (right).}
\end{figure}

\noindent Figures~\ref{32-Fold Zenith (norm)}--\ref{L3aMLP (norm)} show these five input variables and the output neural net classifier variable called L3aMLP.  Plots in the right column are normalized to the same area.

\newpage

\clearpage

\clearpage

\clearpage

\subsection{Level 4a Neural Networks}
A cut was placed on the level 3 neural network requiring L3aMLP$>$0.4.  In addition, soft cuts were placed on the reconstructed energy and the containment of the reconstructed cascade vertex requiring Energy$>$2 TeV and ParallelogramDist$<$1.4.  We expect that we will eventually have to tighten both the energy cut and the containment cut at later stages, but by cutting some of these events out now, further neural network training should be more efficient.  These three cuts bring the analysis from level 3a to level 4a.  

At level4a, two individual neural networks were trained separately. Because we have very limited training statistics, this should result in better performance. 

The first was trained on the four variables that are the most strongly correlated with energy and with each other.  These variables are listed below by decreasing separation power: 

\begin{itemize}
\item{{\bf NDirE}: The number of direct hits from the reconstructed cascade vertex still retains discriminating power.  Again, it should be larger for cascades than for muons.}
\item{{\bf NHit}: The total number of observed photoelectrons in all DOM's.  This is correlated with energy.}
\item{{\bf LlhRatio}: The likelihood from the best cascade vertex reconstruction minus the 32-fold iterative muon track reconstruction.  This tells us which hypothesis is a better fit to the observed data.}
\item{{\bf Energy}: the reconstructed energy from the {\tt AtmCscdEnergyReco} fit.  We expect that it should be easier to reject muons with higher energy radiative losses because the muons are more likely to leave early light.}
\end{itemize}

\begin{figure}
\begin{minipage}[b]{0.48\linewidth} % A minipage that covers half the page
\centering
\includegraphics[width=7.6cm]{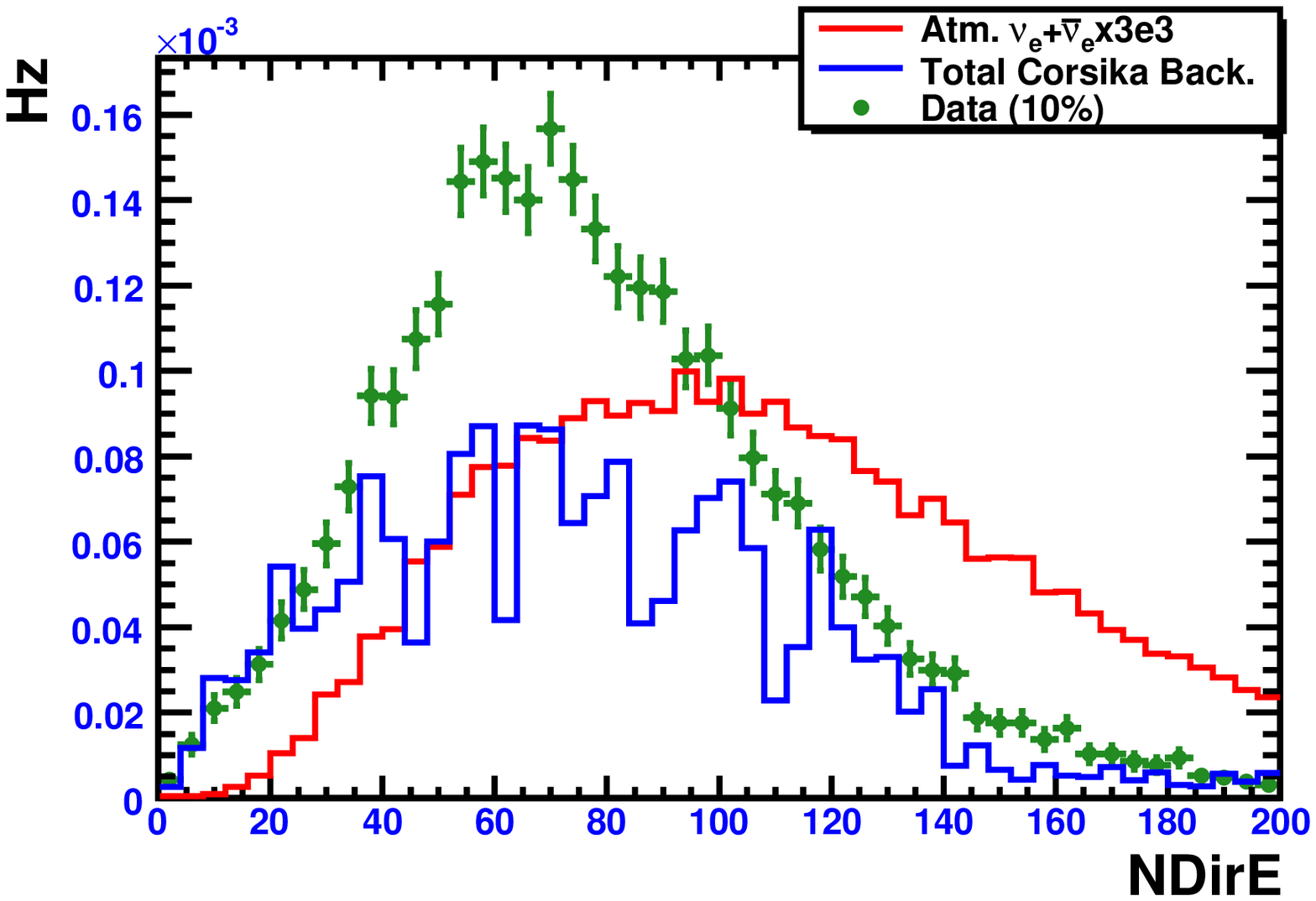}
\label{NDirE at Level4a}
\end{minipage}
\hspace{0.5cm} %To get a little bit of space between the figures
\begin{minipage}[b]{0.48\linewidth}
\centering
\includegraphics[width=7.6cm]{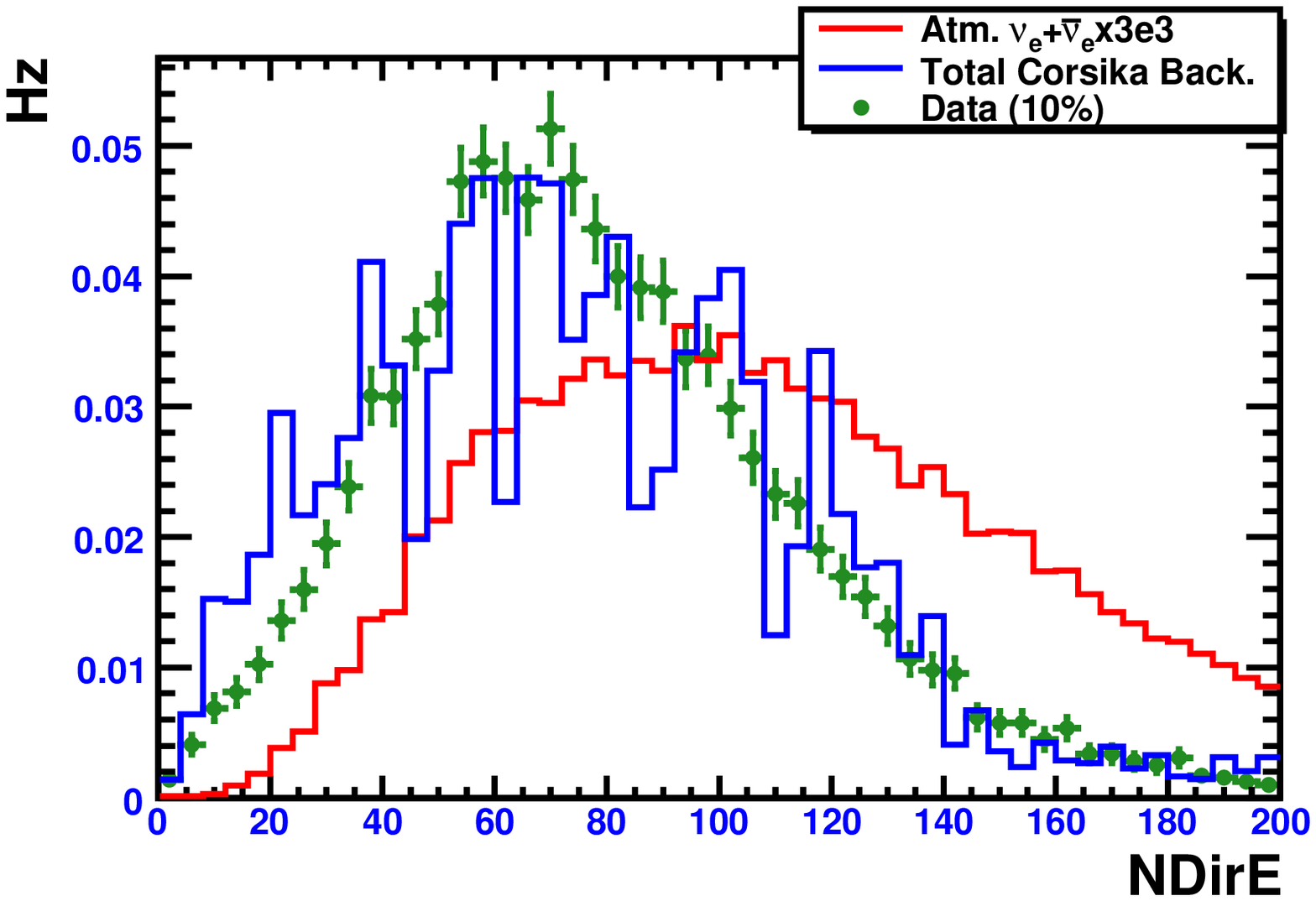}
\label{NDirE at Level4a (norm)}
\end{minipage}
\vspace{0.25cm}
\caption{Number of direct hits NDirE at level 4a, absolutely normalized (left) and normalized to the same area (right).}
\end{figure}

\begin{figure}
\begin{minipage}[b]{0.48\linewidth} % A minipage that covers half the page
\centering
\includegraphics[width=7.6cm]{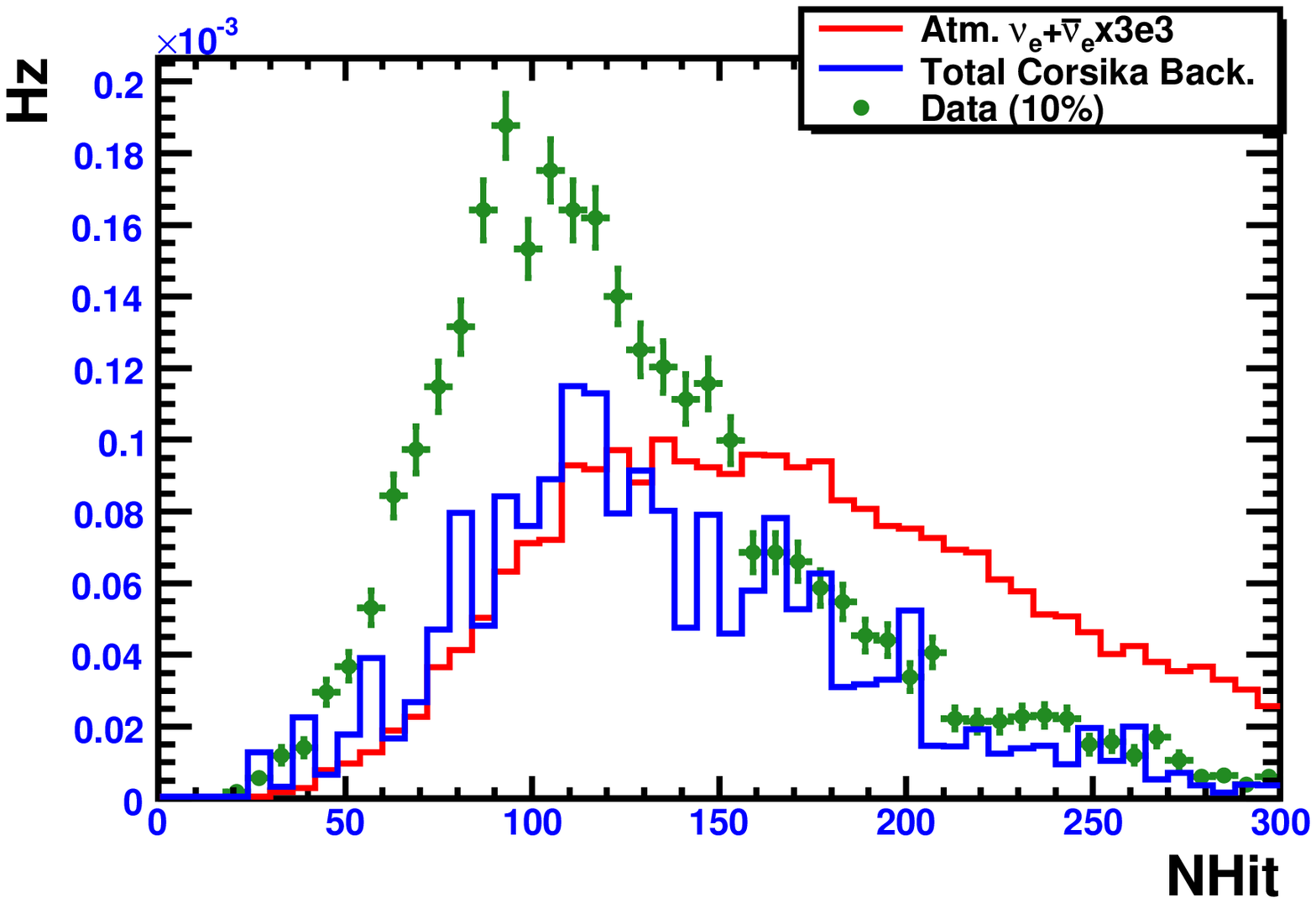}
\label{NHit at Level4a}
\end{minipage}
\hspace{0.5cm} %To get a little bit of space between the figures
\begin{minipage}[b]{0.48\linewidth}
\centering
\includegraphics[width=7.6cm]{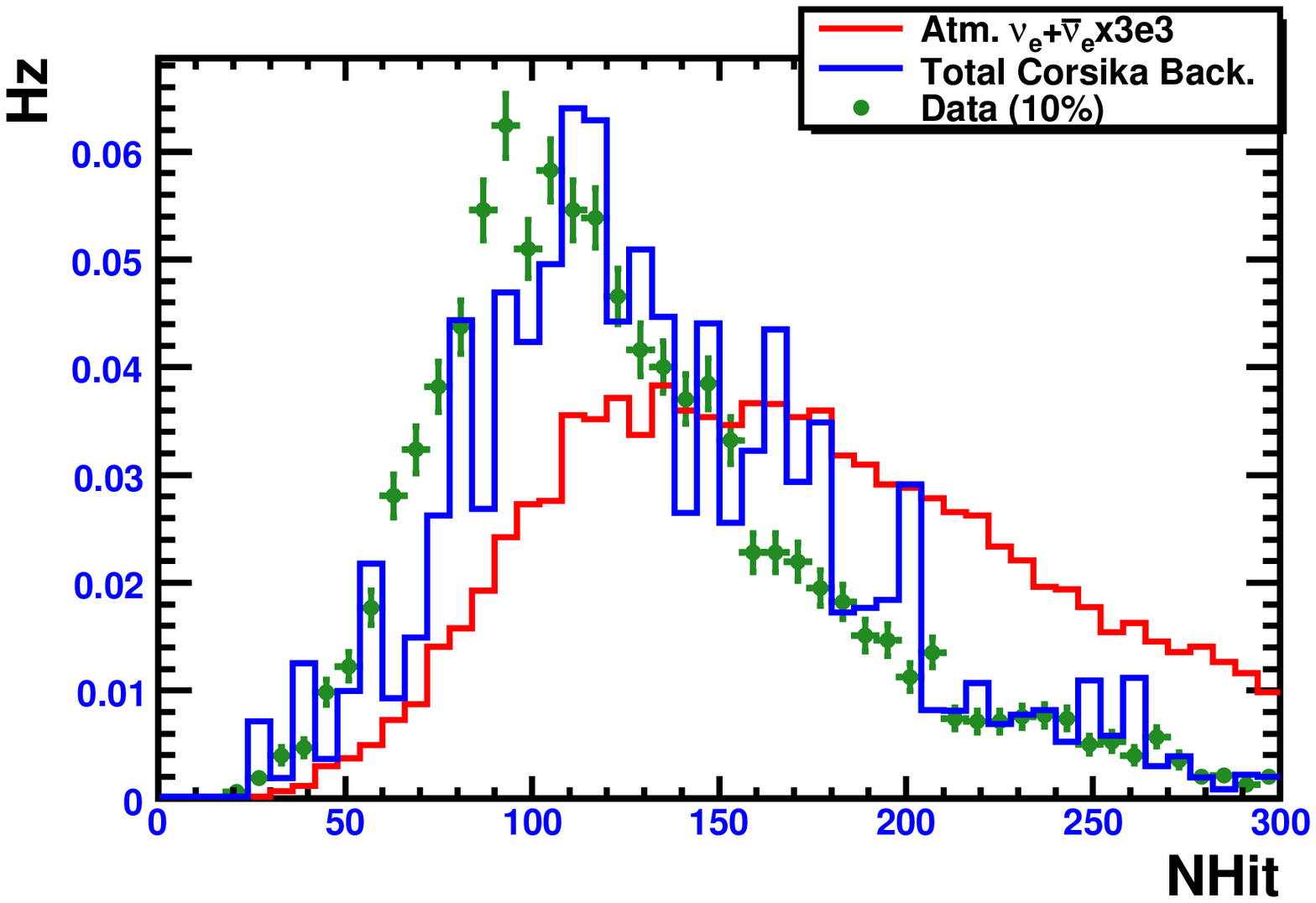}
\label{NHit at Level4a (norm)}
\end{minipage}
\vspace{0.25cm}
\caption{Total number of photoelectrons NHit at level 4a, absolutely normalized (left) and normalized to the same area (right).}
\end{figure}

\begin{figure}
\begin{minipage}[b]{0.48\linewidth} % A minipage that covers half the page
\centering
\includegraphics[width=7.6cm]{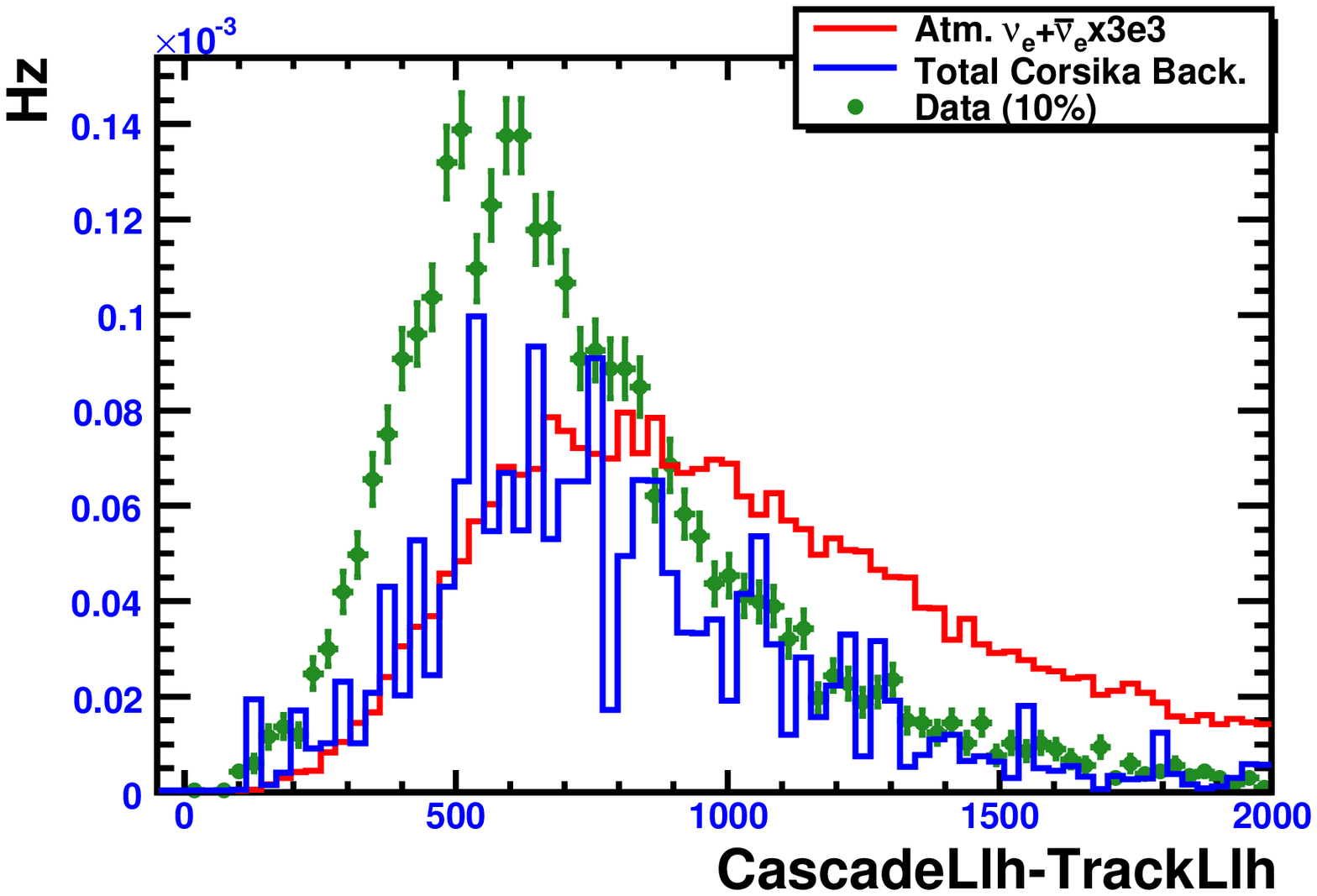}
\label{LlhRatio at Level4a}
\end{minipage}
\hspace{0.5cm} %To get a little bit of space between the figures
\begin{minipage}[b]{0.48\linewidth}
\centering
\includegraphics[width=7.6cm]{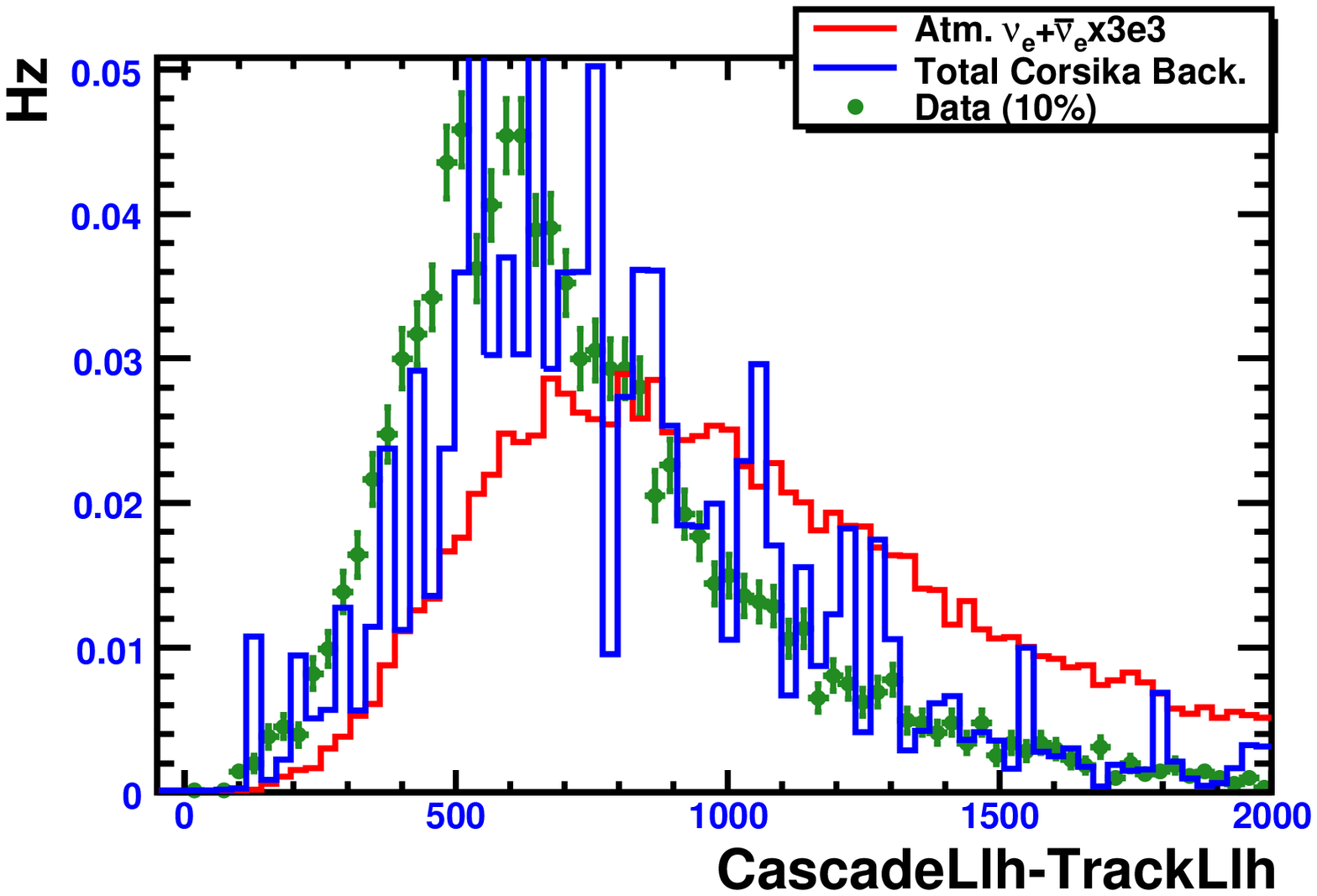}
\label{LlhRatio at Level4a (norm)}
\end{minipage}
\vspace{0.25cm}
\caption{Likelihood ratio at level 4a, absolutely normalized (left) and normalized to the same area (right).}
\end{figure}

\begin{figure}
\begin{minipage}[b]{0.48\linewidth} % A minipage that covers half the page
\centering
\includegraphics[width=7.6cm]{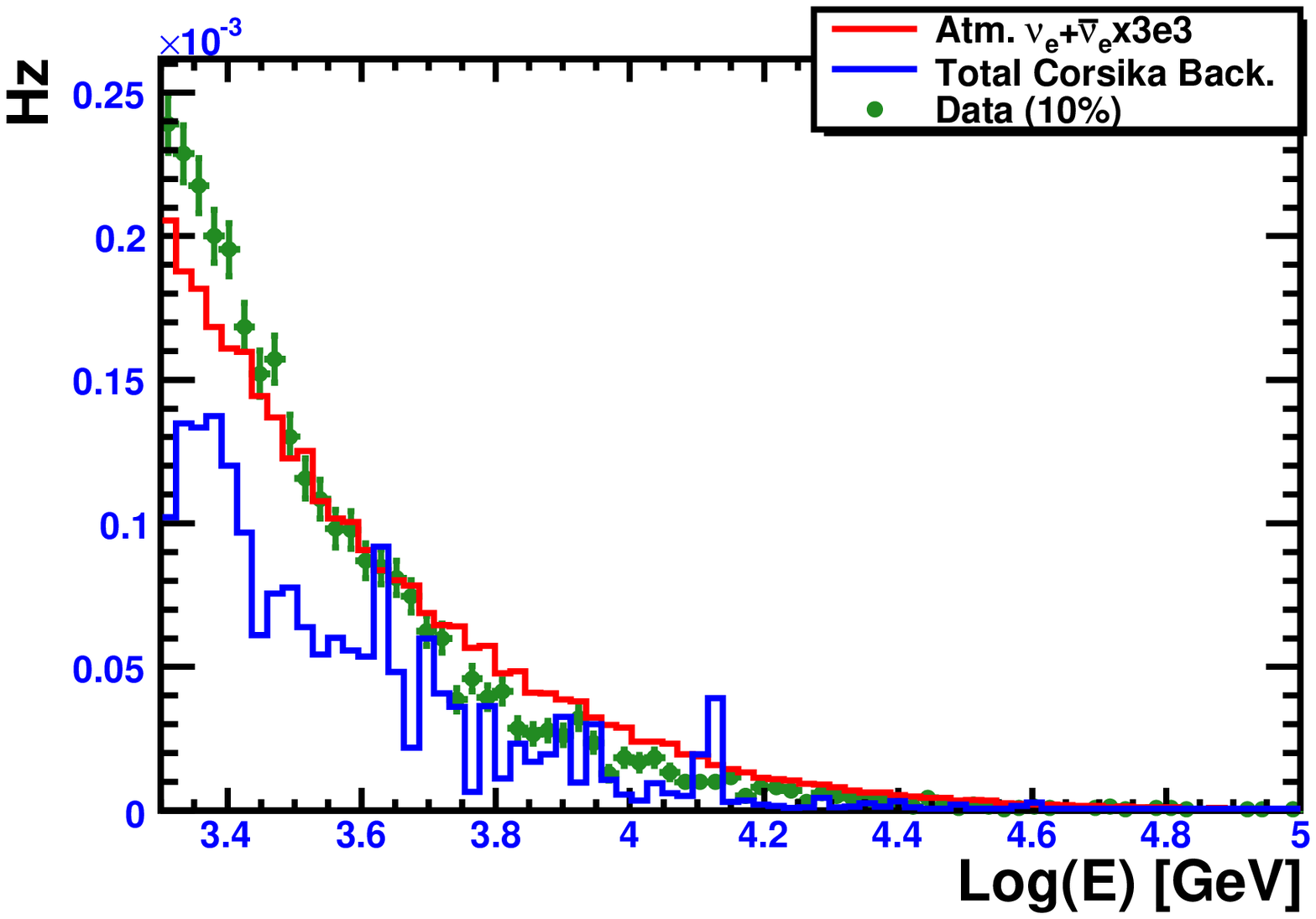}
\label{Energy at Level 4a}
\end{minipage}
\hspace{0.5cm} %To get a little bit of space between the figures
\begin{minipage}[b]{0.48\linewidth}
\centering
\includegraphics[width=7.6cm]{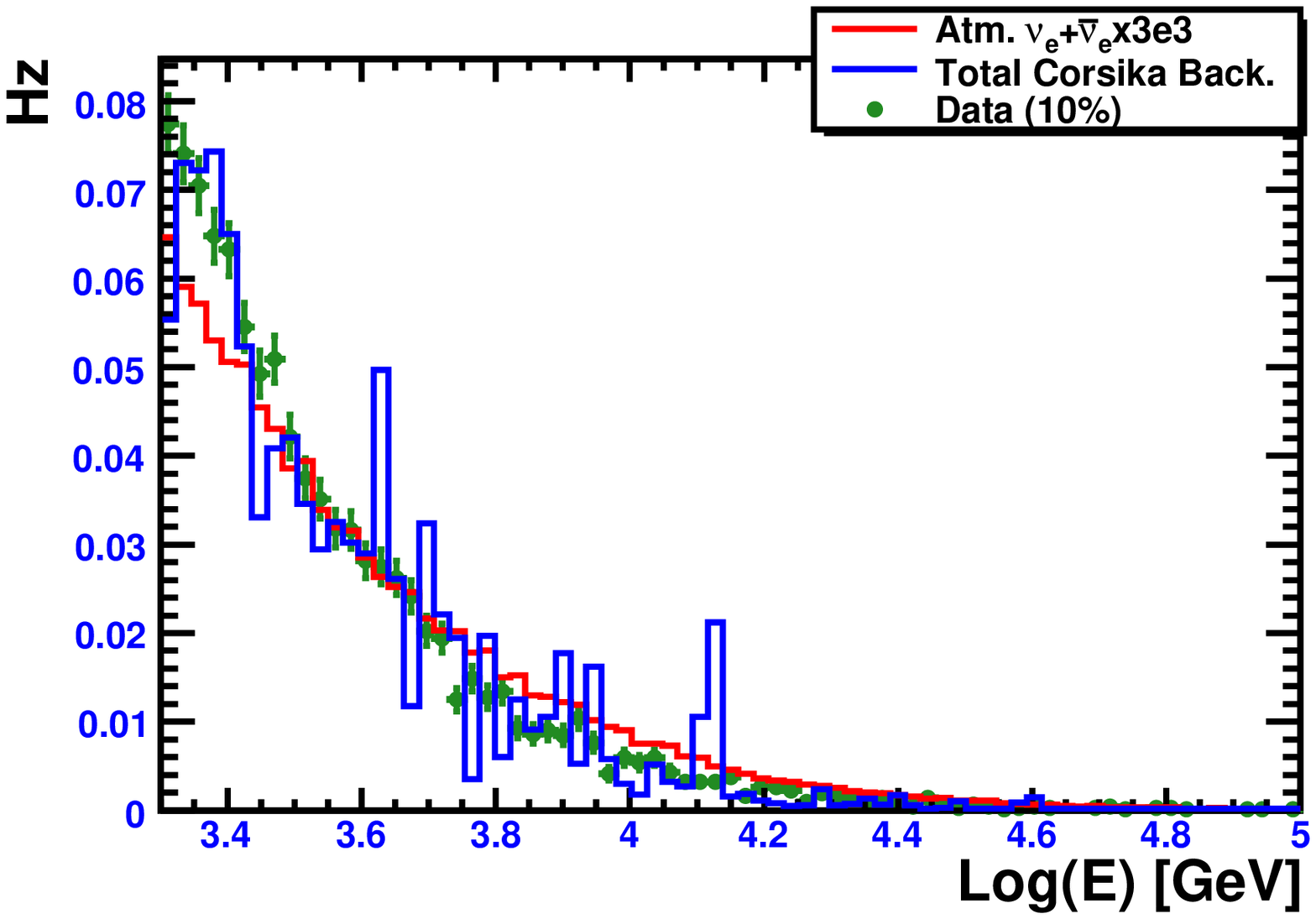}
\label{Energy at Level 4a (norm)}
\end{minipage}
\vspace{0.25cm}
\caption{Reconstructed energy at level 4a, absolutely normalized (left) and normalized to the same area (right).}
\end{figure}

\begin{figure}
\begin{minipage}[b]{0.48\linewidth} % A minipage that covers half the page
\centering
\includegraphics[width=7.6cm]{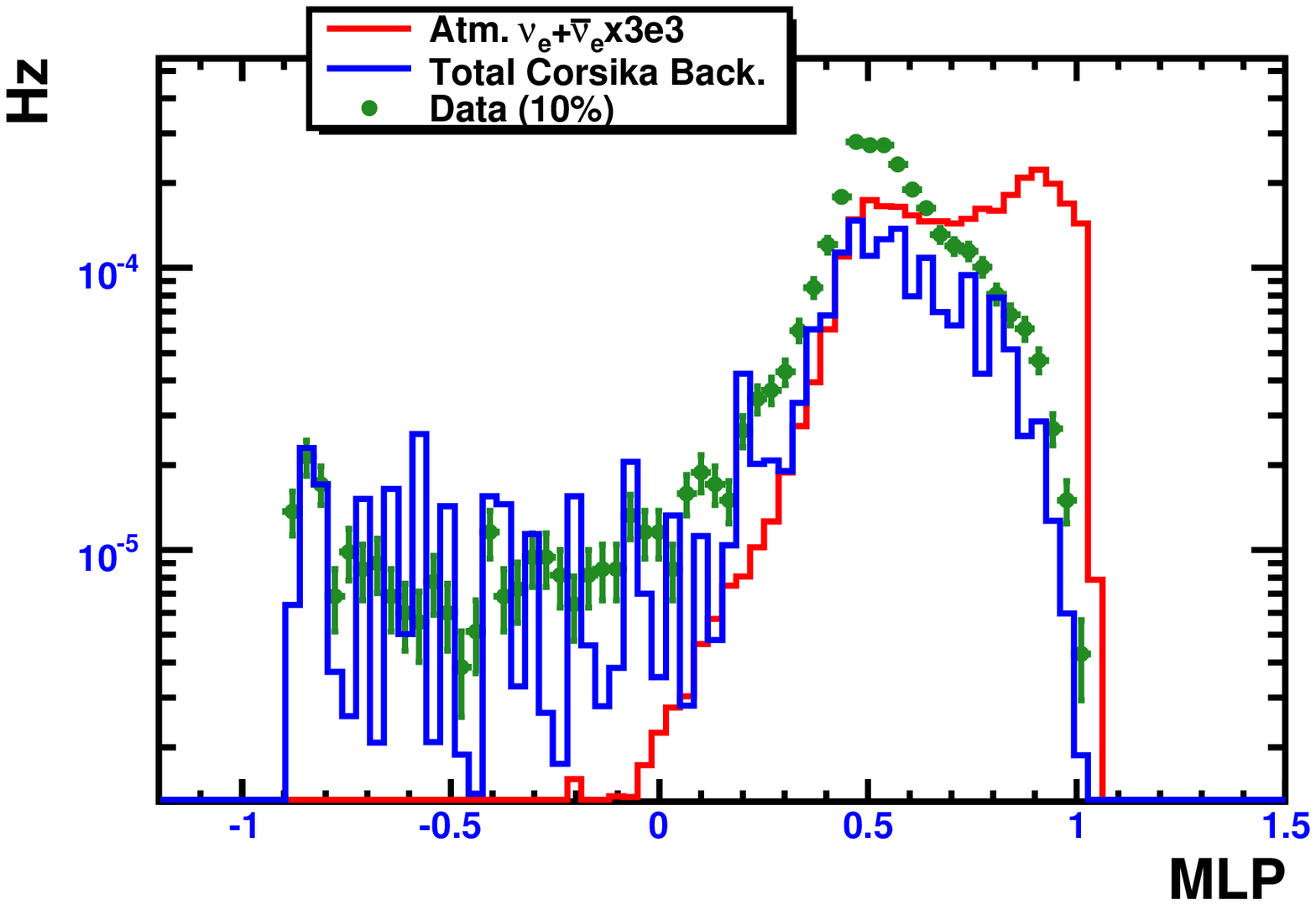}
\end{minipage}
\hspace{0.5cm} %To get a little bit of space between the figures
\begin{minipage}[b]{0.48\linewidth}
\centering
\includegraphics[width=7.6cm]{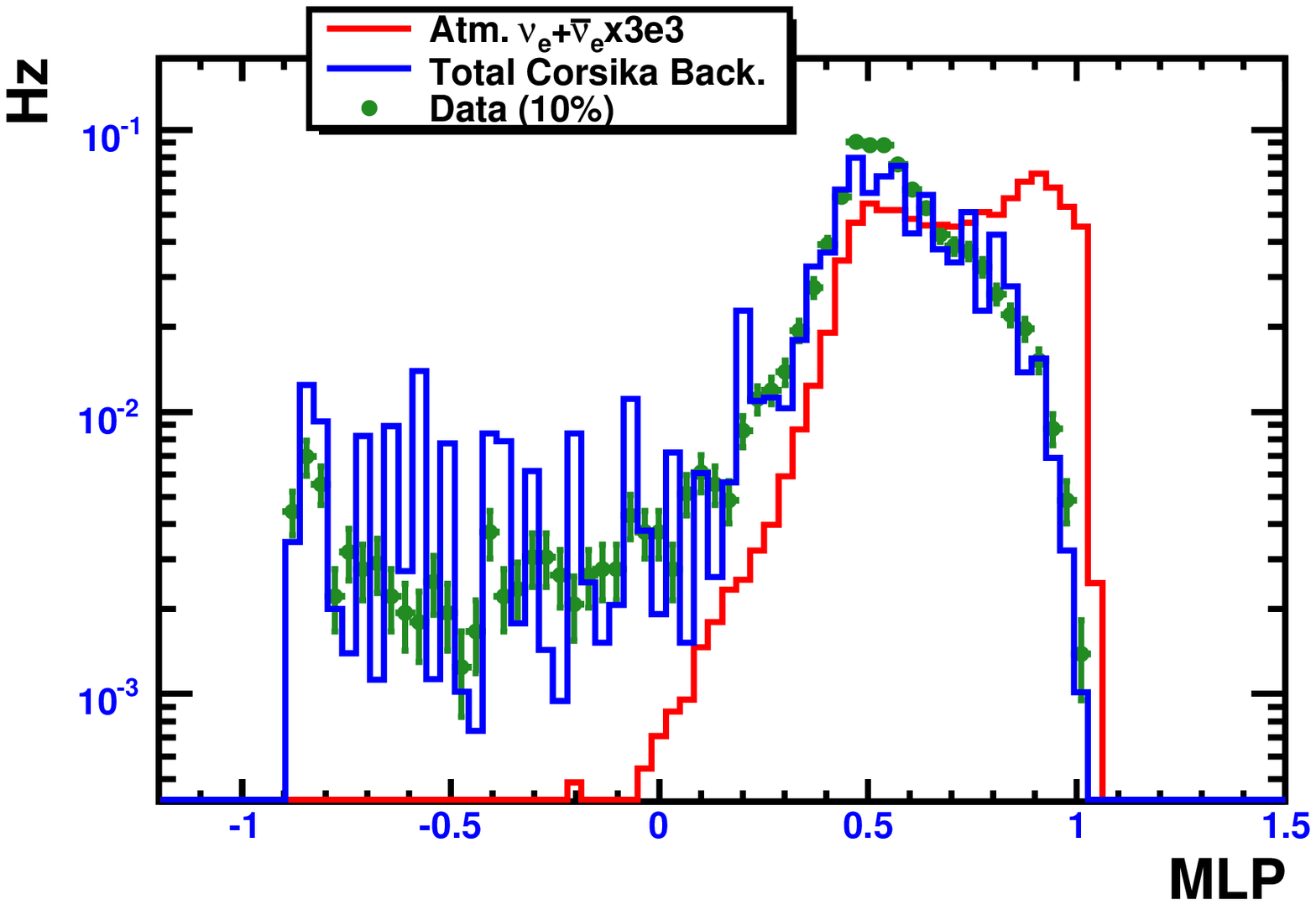}
\end{minipage}
\vspace{0.25cm}
\caption{First level 4a neural net output variable L4aMLP1 at level 4a, absolutely normalized (left) and normalized to the same area (right).}
\label{L4aMLP1Level4a}
\end{figure}

\noindent Figures~\ref{NDirE at Level4a (norm)}--\ref{L4aMLP1Level4a} show these four input variables and the output neural net classifier variable called L4aMLP1.  Plots in the right column are normalized to the same area. 

\newpage

\clearpage

\clearpage

\clearpage

The second neural net was trained on seven variables.  These variables are listed below by decreasing separation power: 
 
\begin{itemize}
\item{{\bf FirstHistDist}: The distance from the cascade vertex to the first hit in the event.  For cascades, the first hit should be close to the reconstructed vertex.  For muons which leave early hits and then have a large stochastic energy loss, this distance should be larger.}
\item{{\bf ParallelogramDist}: A measure of the cascade vertex containment described in section~\ref{SParallelogramDistance}.}
\item{{\bf FRFillRatioFromRMS}: The fill-ratio calculated from the RMS hit distance described in section~\ref{SFillRatio}.}
\item{{\bf FullAllHitSPEReducedLlh}: The reduced likelihood parameter from the best performing cascade vertex reconstruction still retains discriminating power.  A lower value of this variable indicates a better fit to the cascade hypothesis.}
\item{{\bf SplitDistZ}: The difference in z vertex positions for the two split cascade reconstructions still retains discriminating power.  It should be close to zero for cascade events.}
\item{{\bf 32FoldZenith}:  The track zenith angle of the 32-fold iterative muon reconstruction still retains discriminating power.} 
\item{{\bf FRFillRatioFromMean}: The fill-ratio calculated from the mean hit distance described in section~\ref{SFillRatio}.}
\end{itemize}

\begin{figure}
\begin{minipage}[b]{0.48\linewidth} % A minipage that covers half the page
\centering
\includegraphics[width=7.6cm]{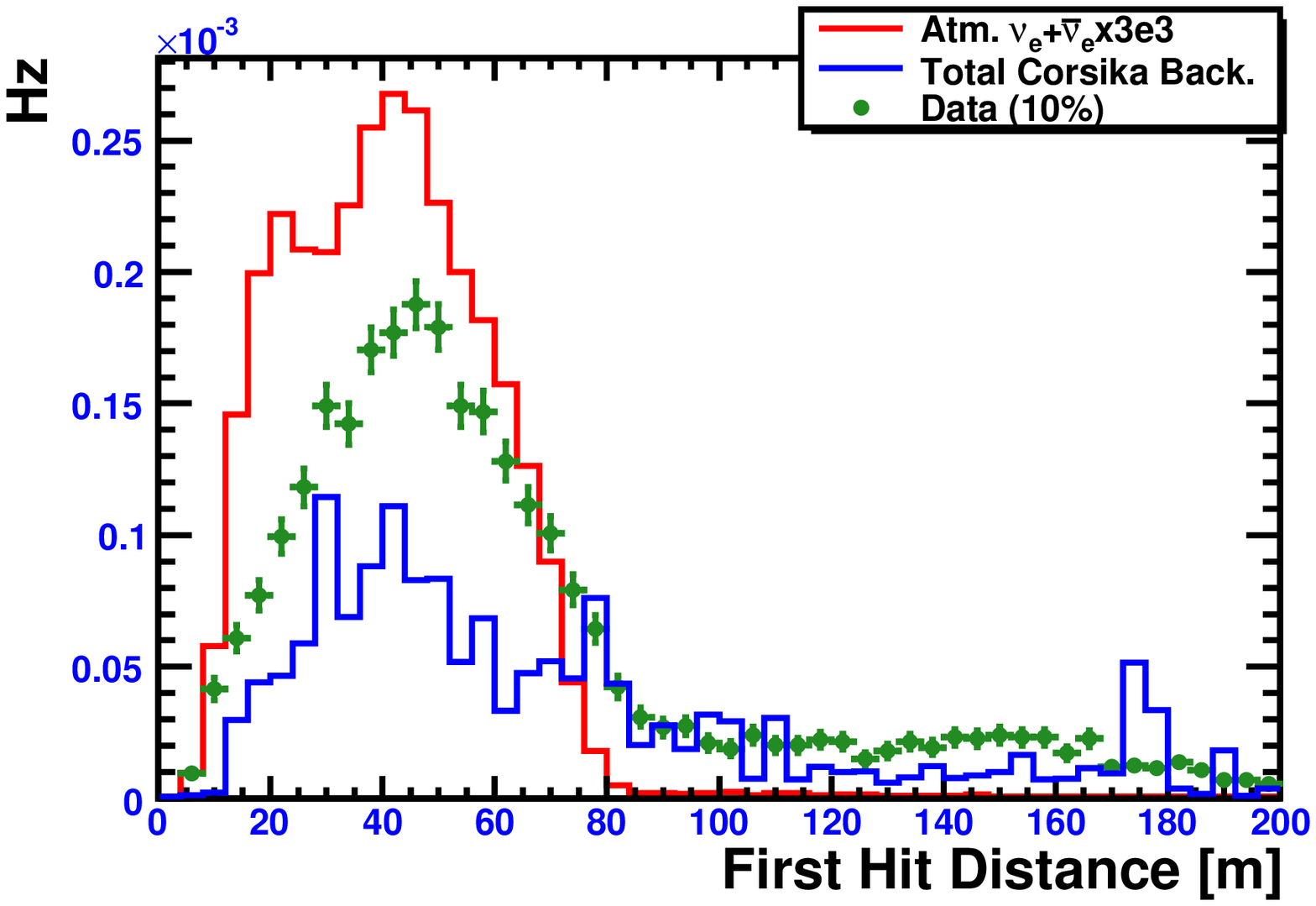}
\label{FirstHitDist}
\end{minipage}
\hspace{0.5cm} %To get a little bit of space between the figures
\begin{minipage}[b]{0.48\linewidth}
\centering
\includegraphics[width=7.6cm]{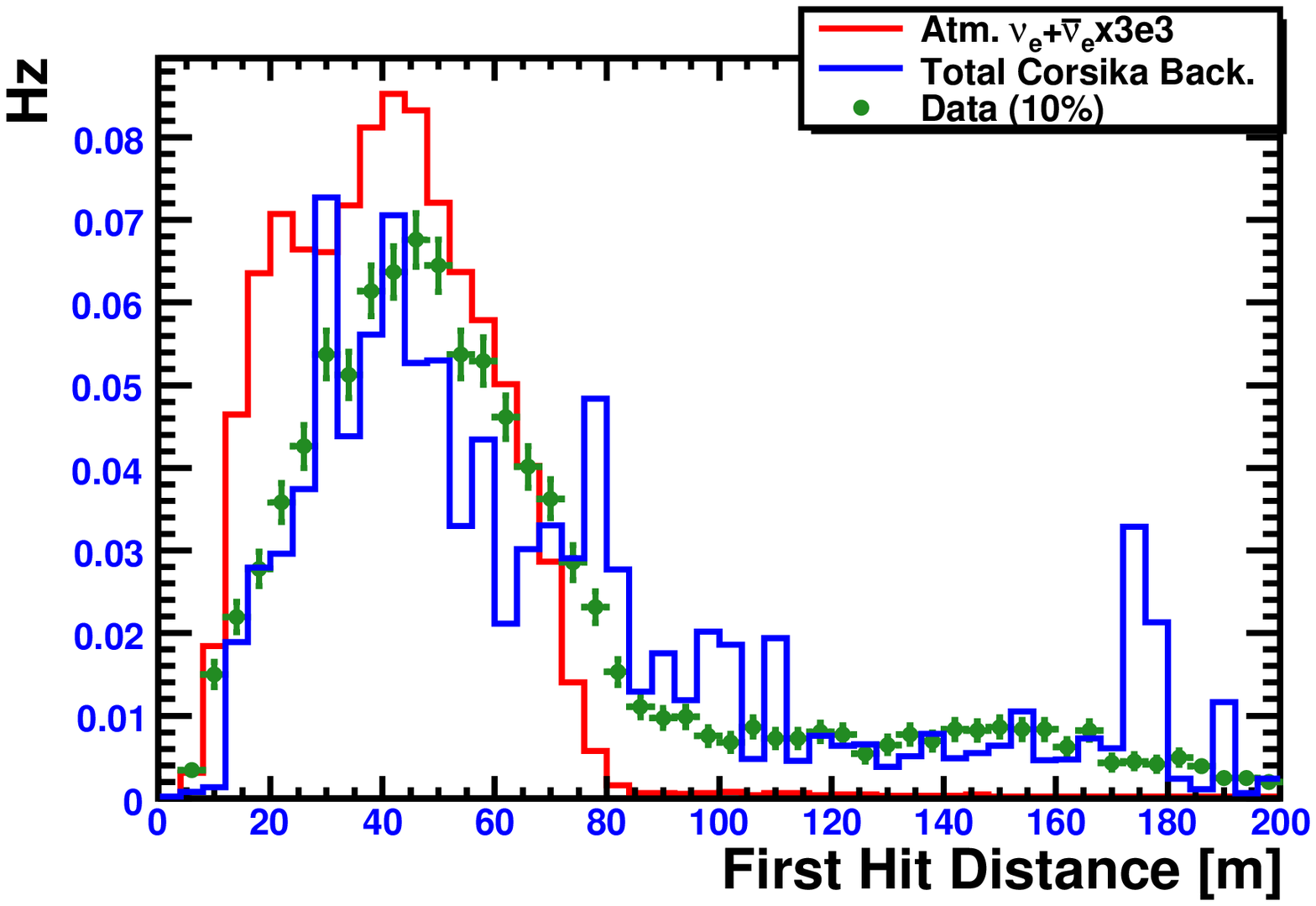}
\label{FirstHitDist (norm)}
\end{minipage}
\vspace{0.25cm}
\caption{FirstHitDist at level 4a, absolutely normalized (left) and normalized to the same area (right).}
\end{figure}

\begin{figure}
\begin{minipage}[b]{0.48\linewidth} % A minipage that covers half the page
\centering
\includegraphics[width=7.6cm]{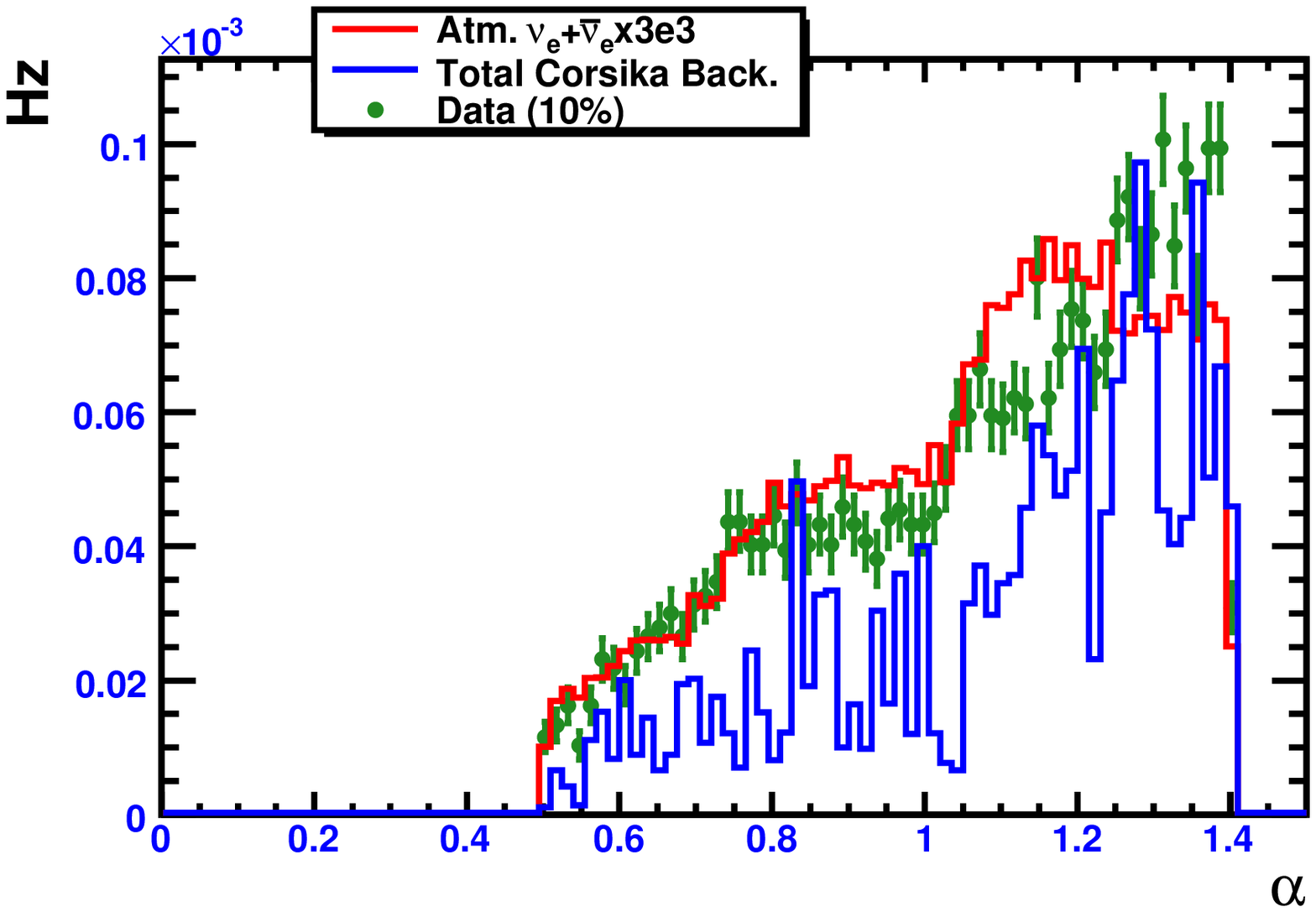}
\label{ParallelogramDist at Level 4a}
\end{minipage}
\hspace{0.5cm} %To get a little bit of space between the figures
\begin{minipage}[b]{0.48\linewidth}
\centering
\includegraphics[width=7.6cm]{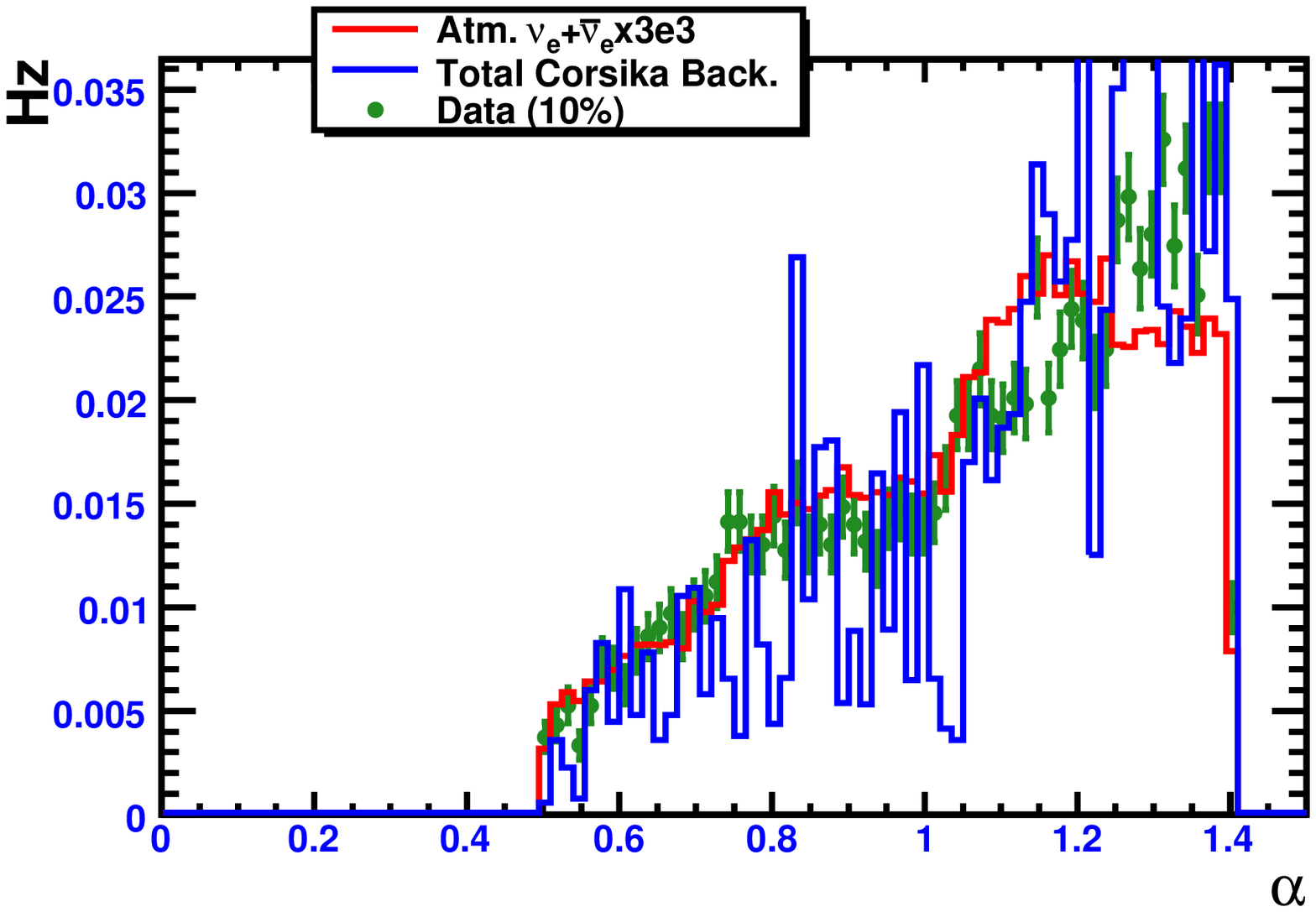}
\label{ParallelogramDist at Level 4a (norm)}
\end{minipage}
\vspace{0.25cm}
\caption{ParallelogramDist at level 4a, absolutely normalized (left) and normalized to the same area (right).}
\end{figure}

\begin{figure}
\begin{minipage}[b]{0.48\linewidth} % A minipage that covers half the page
\centering
\includegraphics[width=7.6cm]{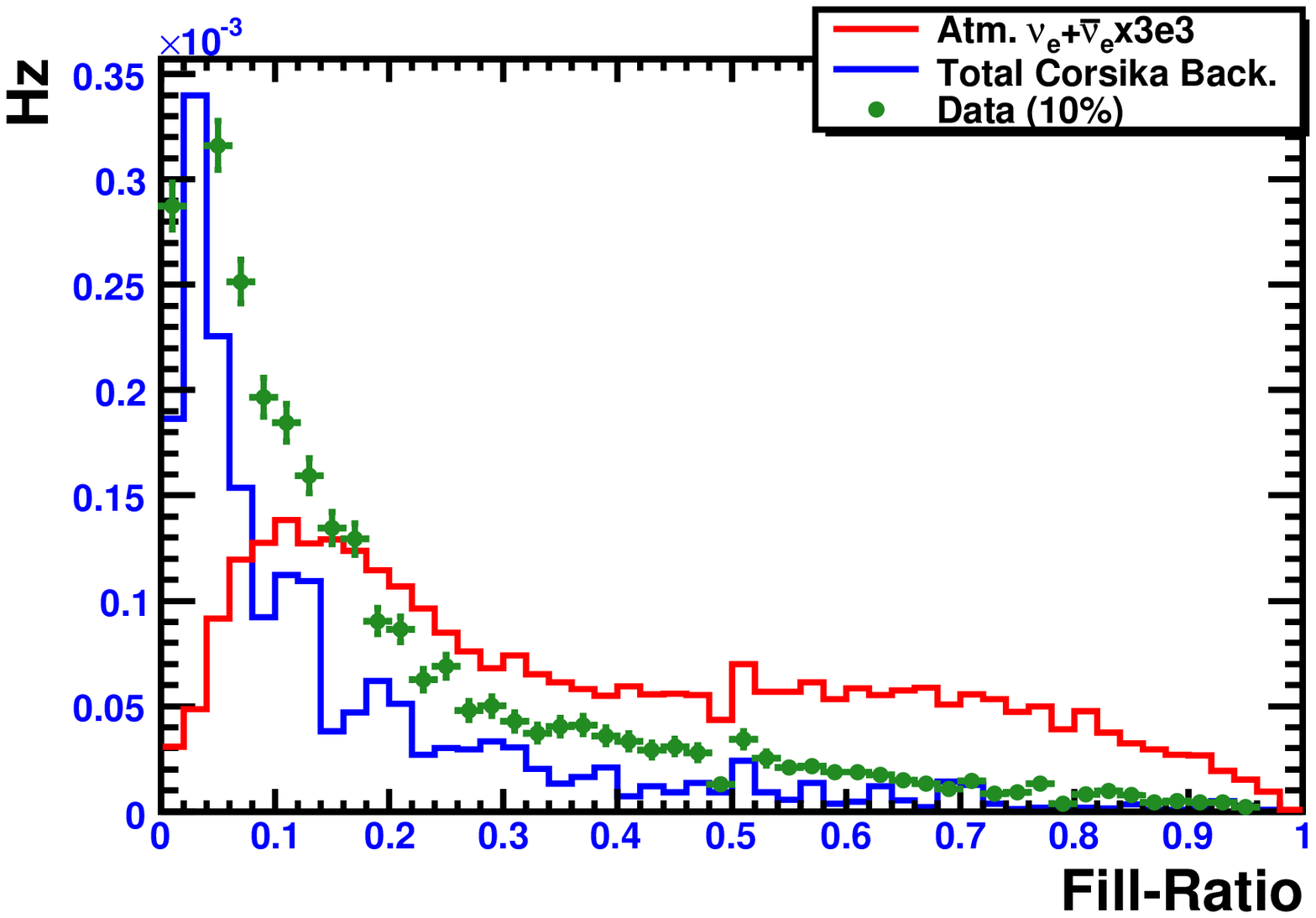}
\label{Fill-Ratio from RMS at Level 4a}
\end{minipage}
\hspace{0.5cm} %To get a little bit of space between the figures
\begin{minipage}[b]{0.48\linewidth}
\centering
\includegraphics[width=7.6cm]{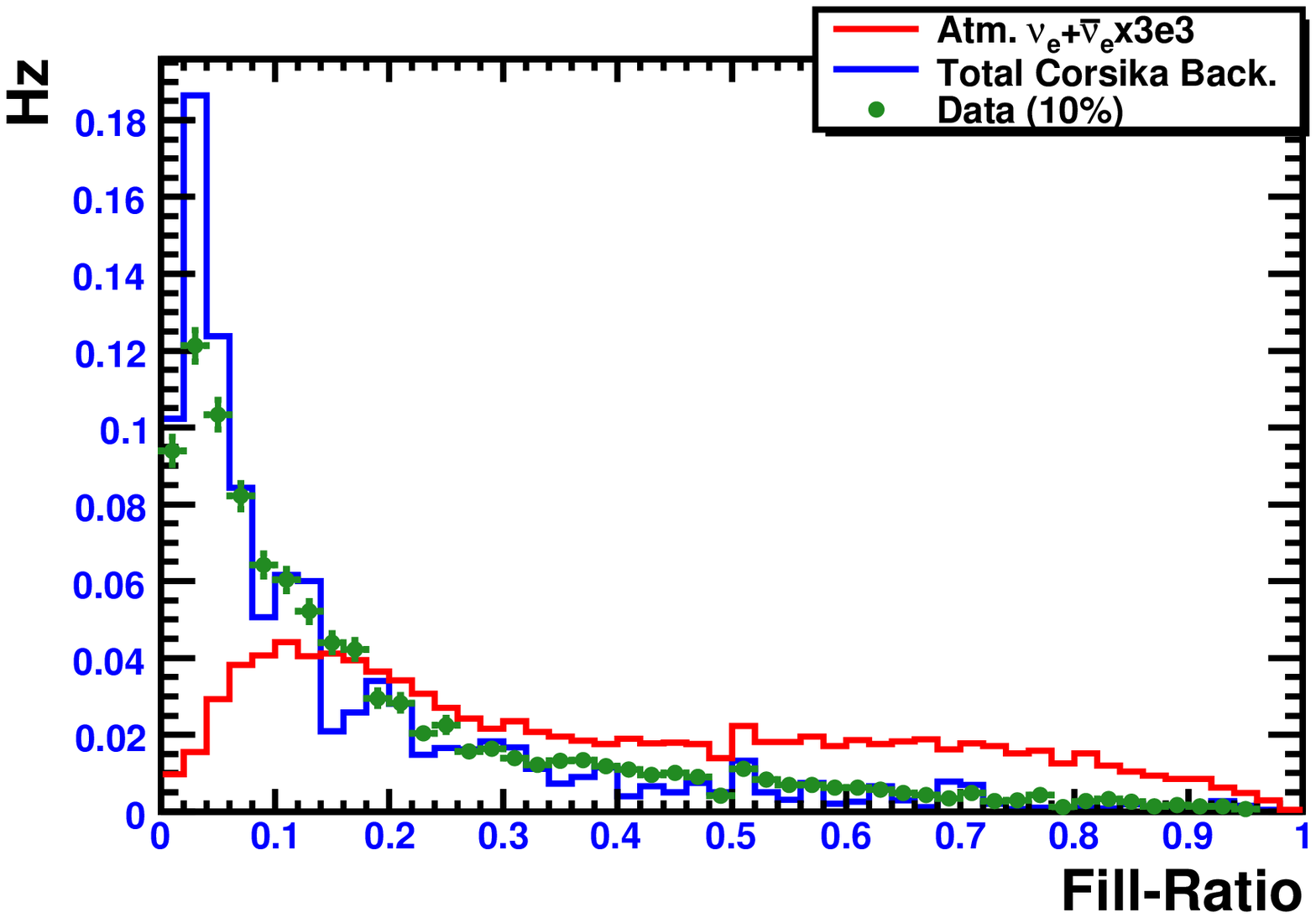}
\label{Fill-Ratio from RMS at Level4a (norm)}
\end{minipage}
\vspace{0.25cm}
\caption{FRFillRatioFromRMS at level 4a, absolutely normalized (left) and normalized to the same area (right).}
\end{figure}

\begin{figure}
\begin{minipage}[b]{0.48\linewidth} % A minipage that covers half the page
\centering
\includegraphics[width=7.6cm]{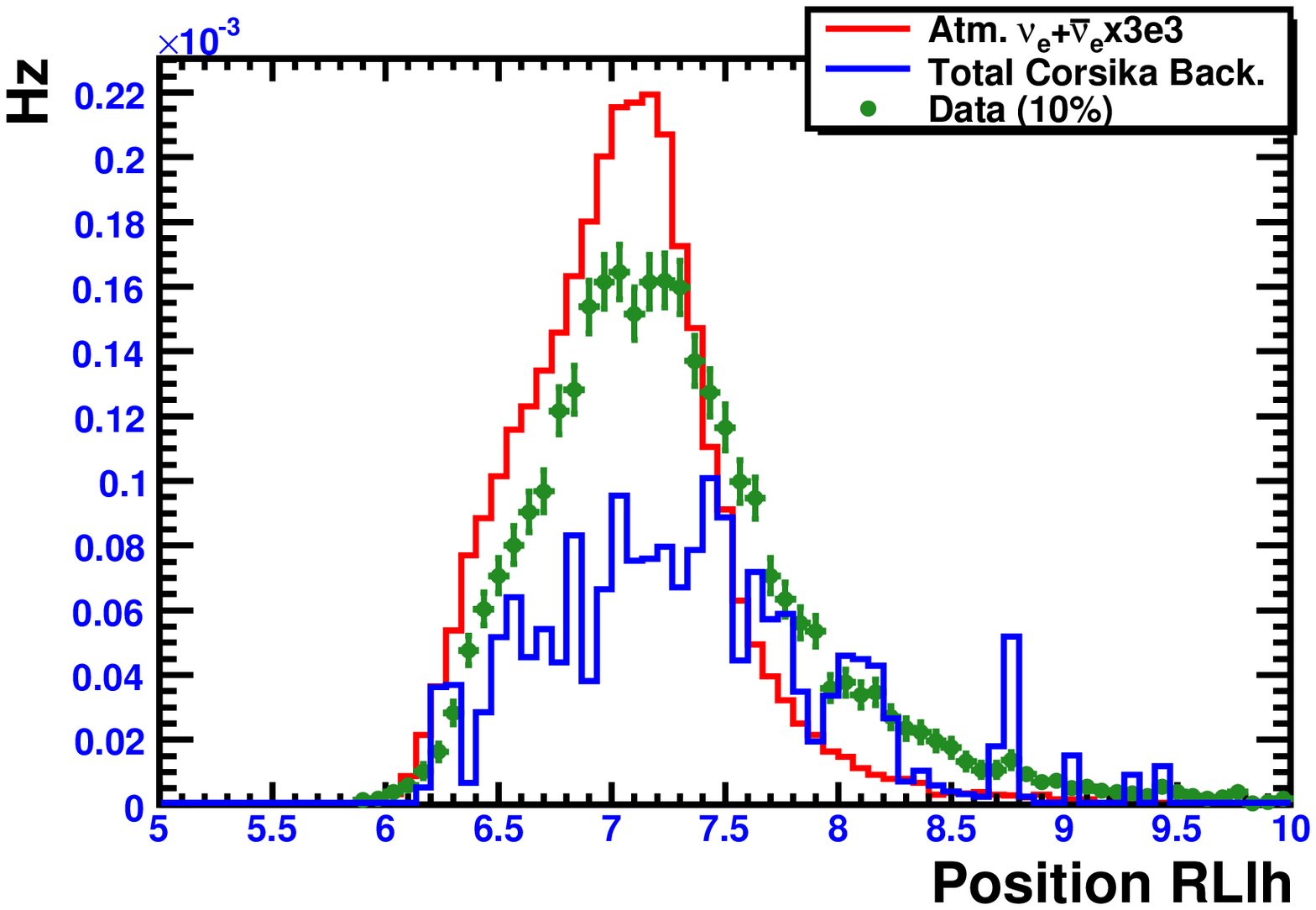}
\label{FullAllHitSPEReducedLlh at Level 4a}
\end{minipage}
\hspace{0.5cm} %To get a little bit of space between the figures
\begin{minipage}[b]{0.48\linewidth}
\centering
\includegraphics[width=7.6cm]{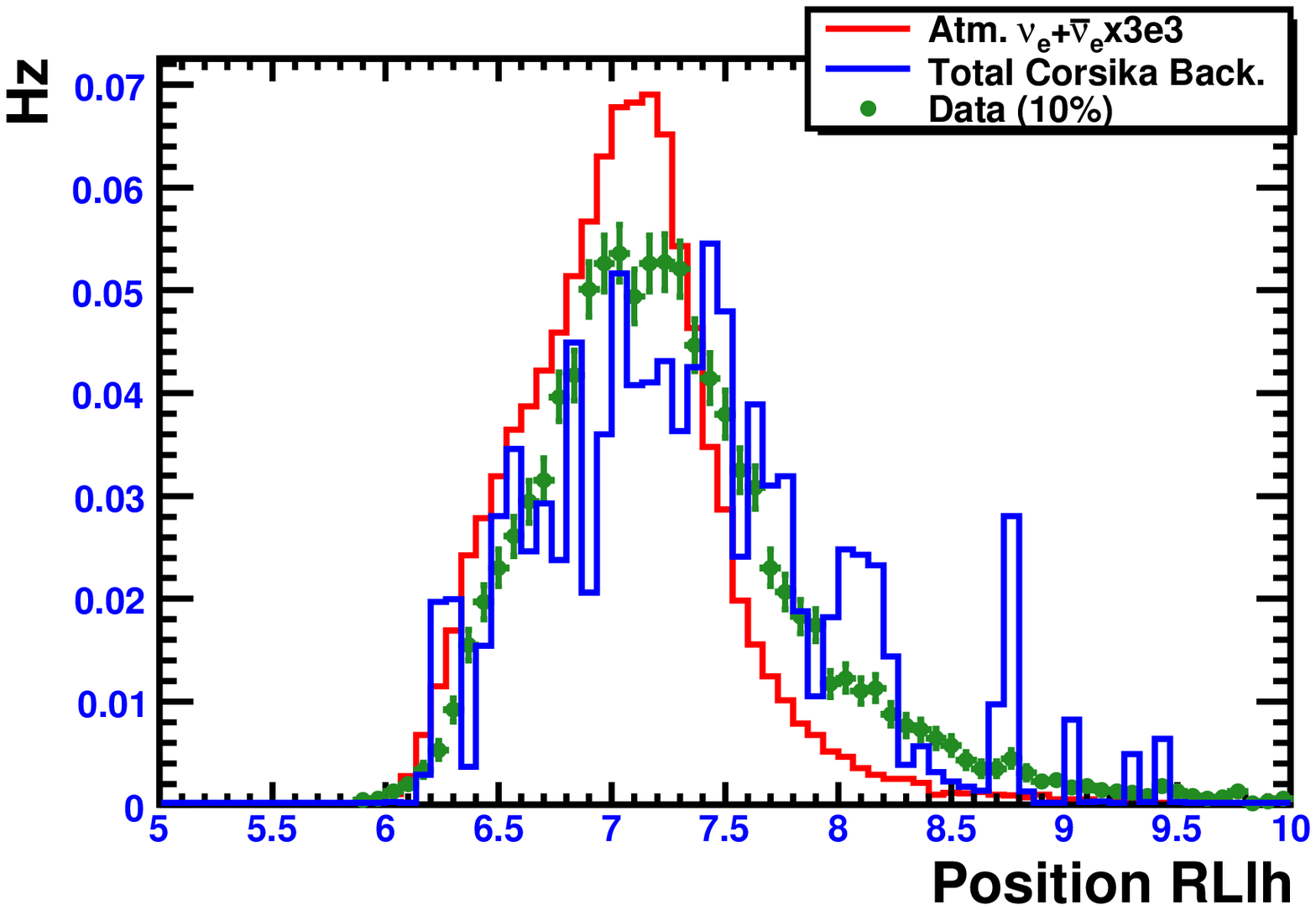}
\label{FullAllHitSPEReducedLlh at Level 4a (norm)}
\end{minipage}
\vspace{0.25cm}
\caption{FullAllHitSPEReducedLlh at level 4a, absolutely normalized (left) and normalized to the same area (right).}
\end{figure}

\begin{figure}
\begin{minipage}[b]{0.48\linewidth} % A minipage that covers half the page
\centering
\includegraphics[width=7.6cm]{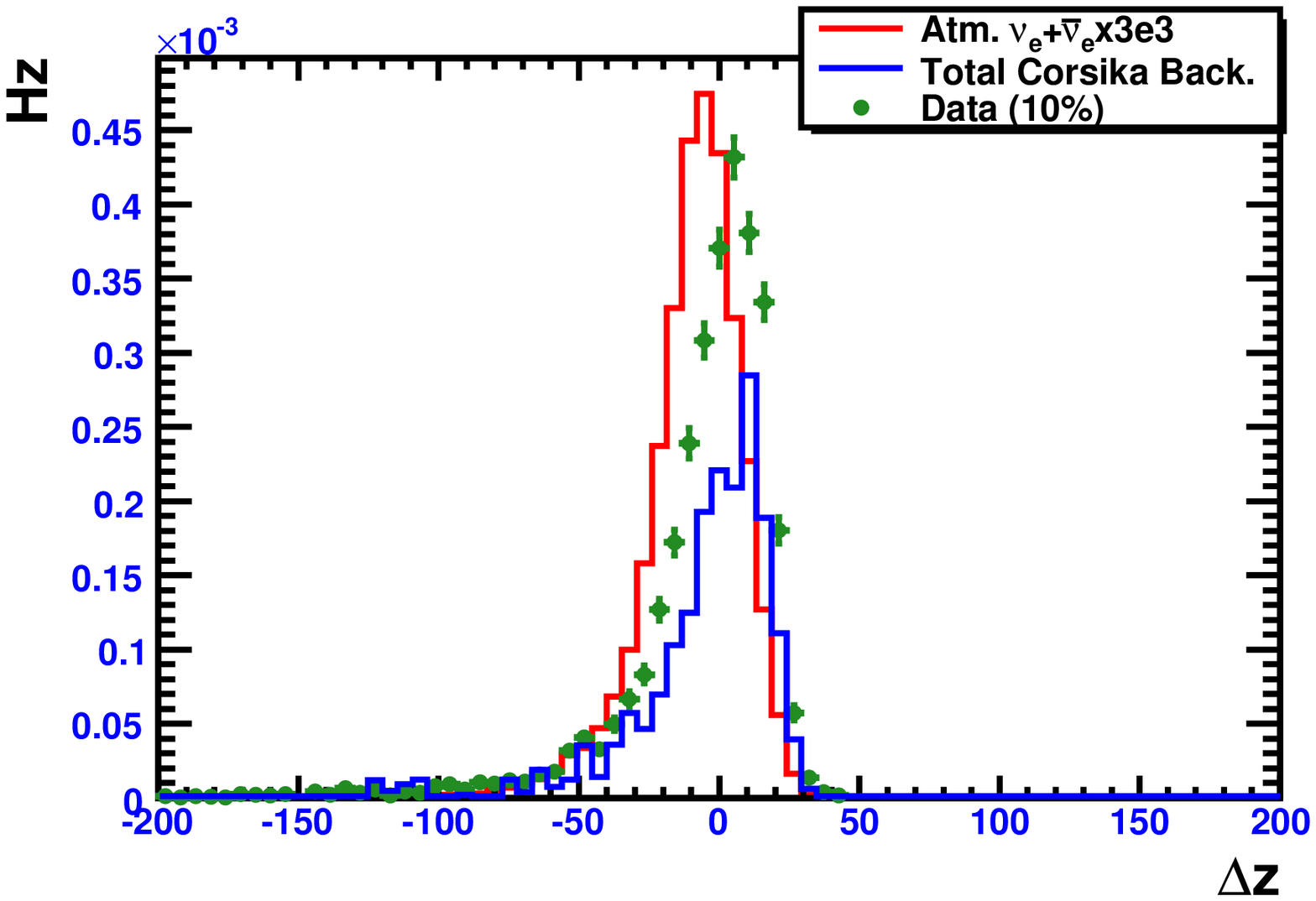}
\label{SplitDistZ at Level4a}
\end{minipage}
\hspace{0.5cm} %To get a little bit of space between the figures
\begin{minipage}[b]{0.48\linewidth}
\centering
\includegraphics[width=7.6cm]{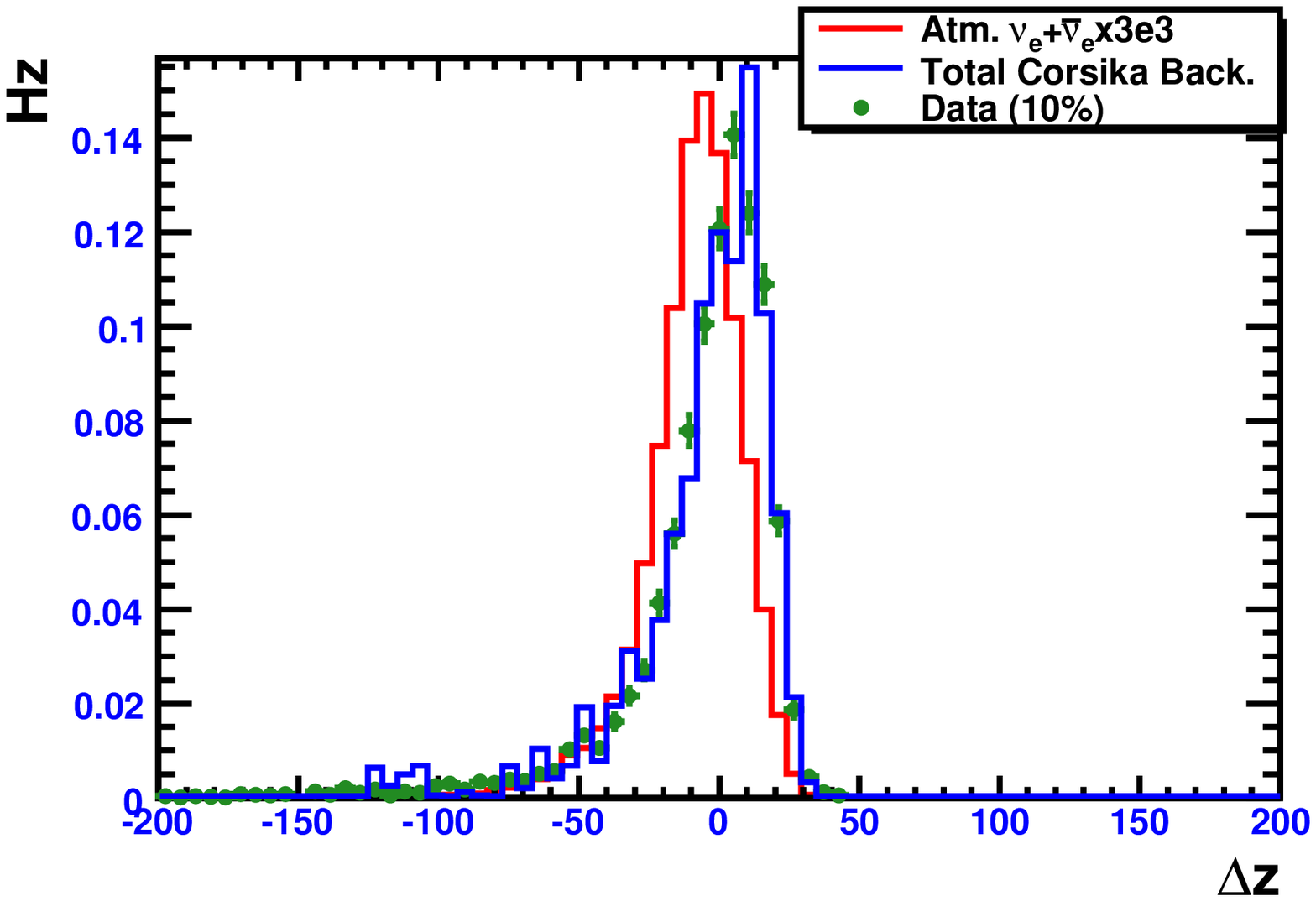}
\label{SplitDistZ at Level4a (norm)}
\end{minipage}
\vspace{0.25cm}
\caption{SplitDistZ at level 4a, absolutely normalized (left) and normalized to the same area (right).}
\end{figure}

\begin{figure}
\begin{minipage}[b]{0.48\linewidth} % A minipage that covers half the page
\centering
\includegraphics[width=7.6cm]{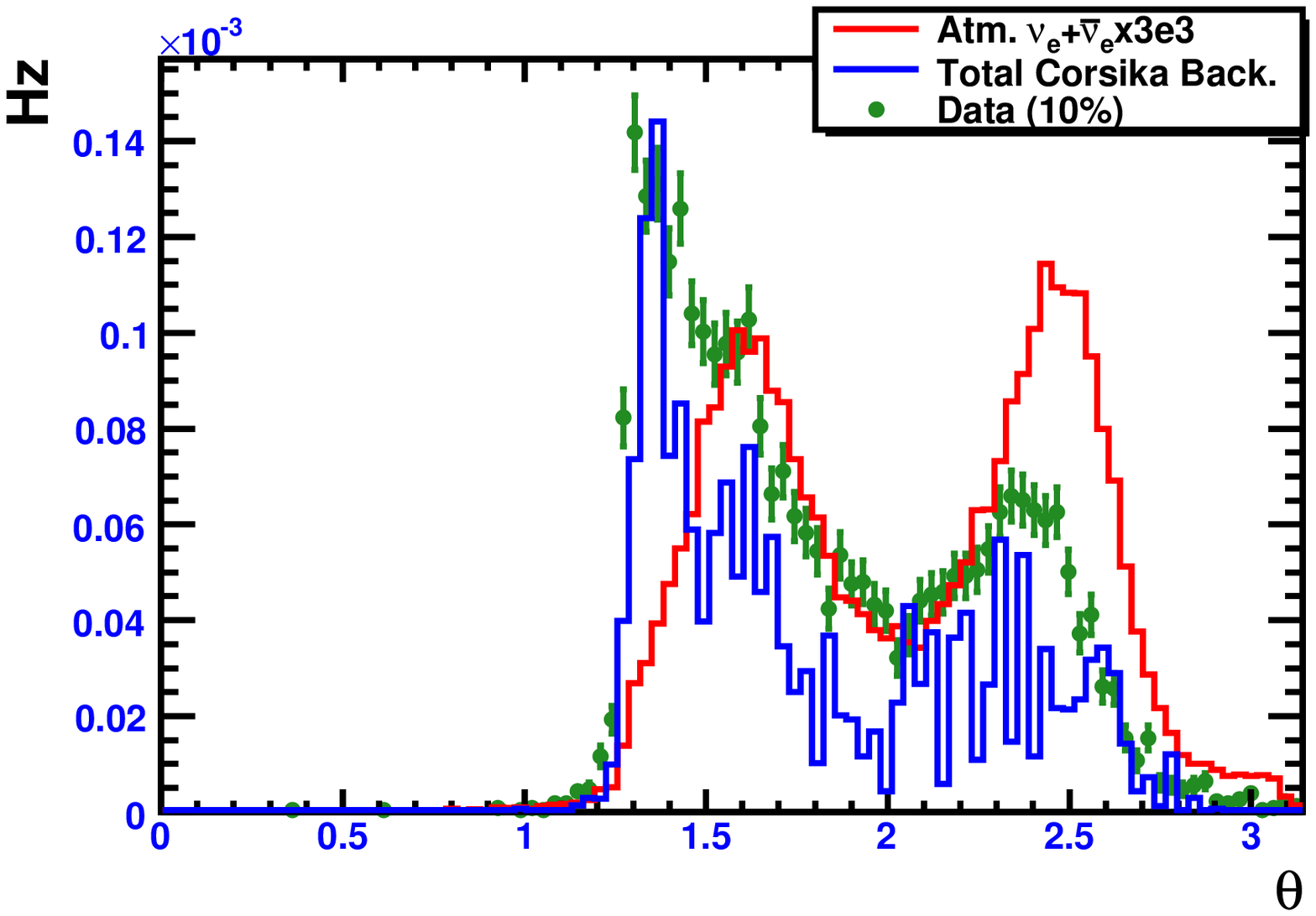}
\label{32FoldZenith at Level4a}
\end{minipage}
\hspace{0.5cm} %To get a little bit of space between the figures
\begin{minipage}[b]{0.48\linewidth}
\centering
\includegraphics[width=7.6cm]{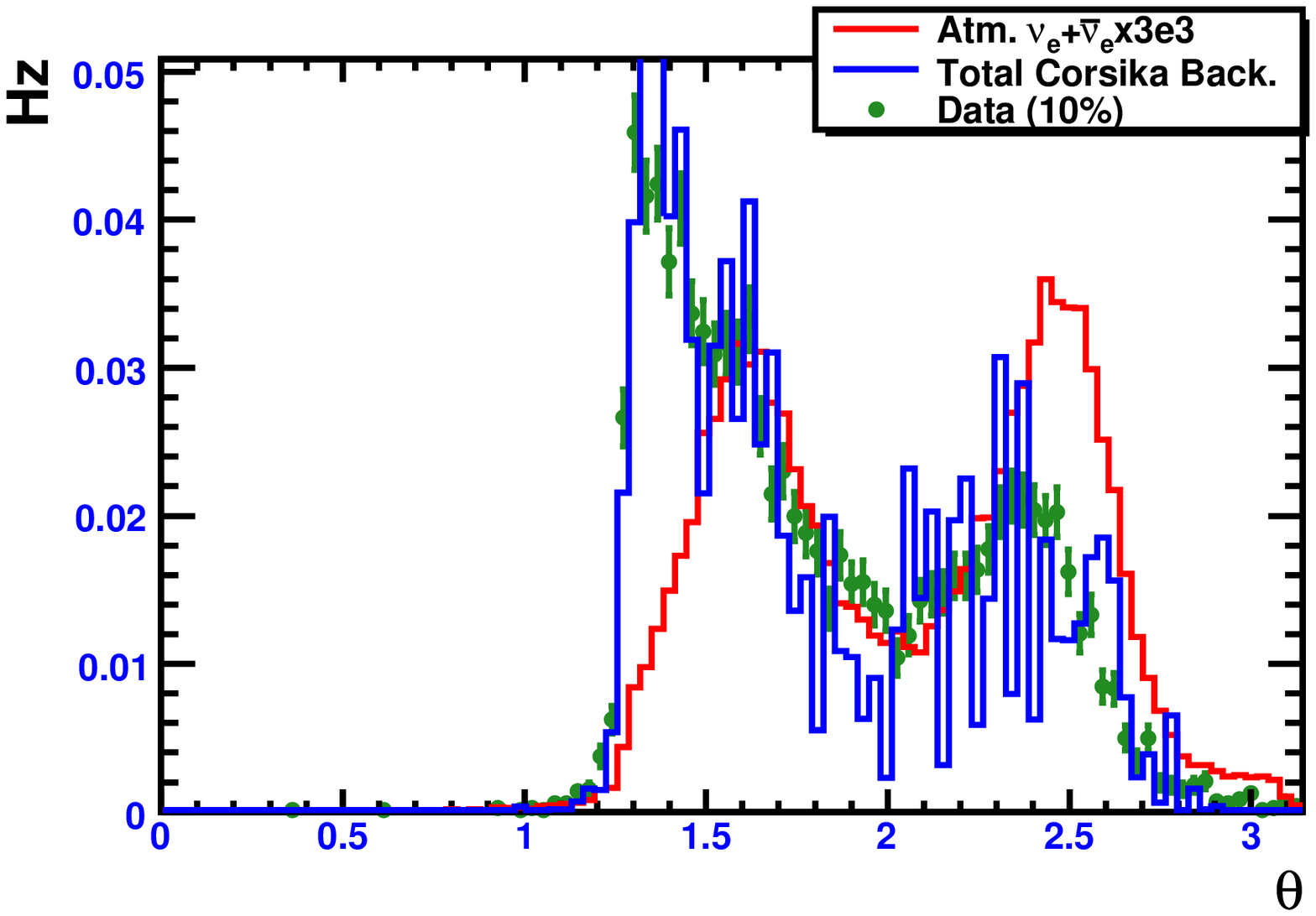}
\label{32FoldZenith at Level4a (norm)}
\end{minipage}
\vspace{0.25cm}
\caption{32FoldZenith at level 4a, absolutely normalized (left) and normalized to the same area (right).}
\end{figure}

\begin{figure}
\begin{minipage}[b]{0.48\linewidth} % A minipage that covers half the page
\centering
\includegraphics[width=7.6cm]{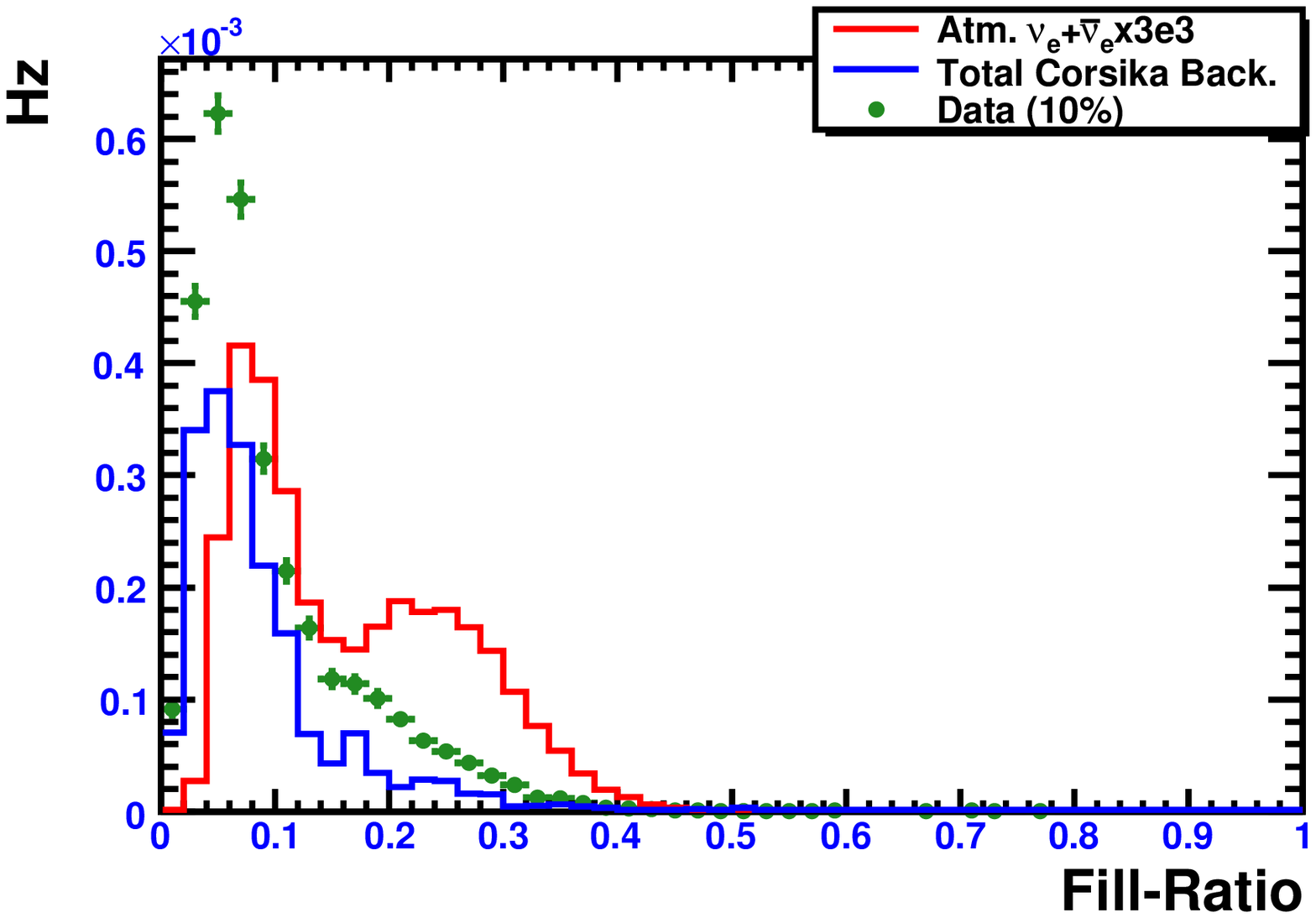}
\label{Fill-Ratio from Mean at Level 4a}
\end{minipage}
\hspace{0.5cm} %To get a little bit of space between the figures
\begin{minipage}[b]{0.48\linewidth}
\centering
\includegraphics[width=7.6cm]{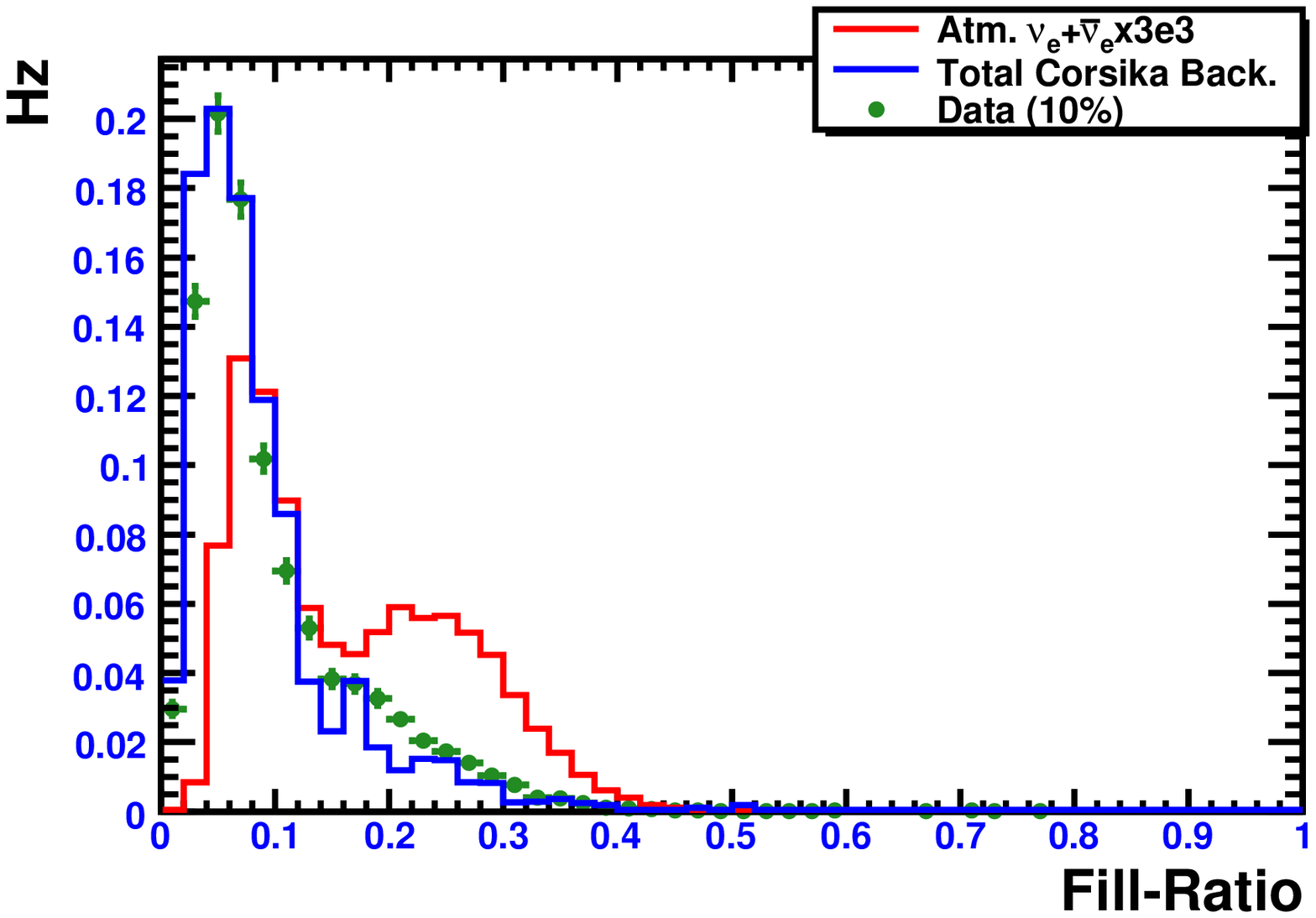}
\label{Fill-Ratio from Mean at Level 4a (norm)}
\end{minipage}
\vspace{0.25cm}
\caption{FRFillRatioFromMean at level 4a, absolutely normalized (left) and normalized to the same area (right).}
\end{figure}

\begin{figure}
\begin{minipage}[b]{0.48\linewidth} % A minipage that covers half the page
\centering
\includegraphics[width=7.6cm]{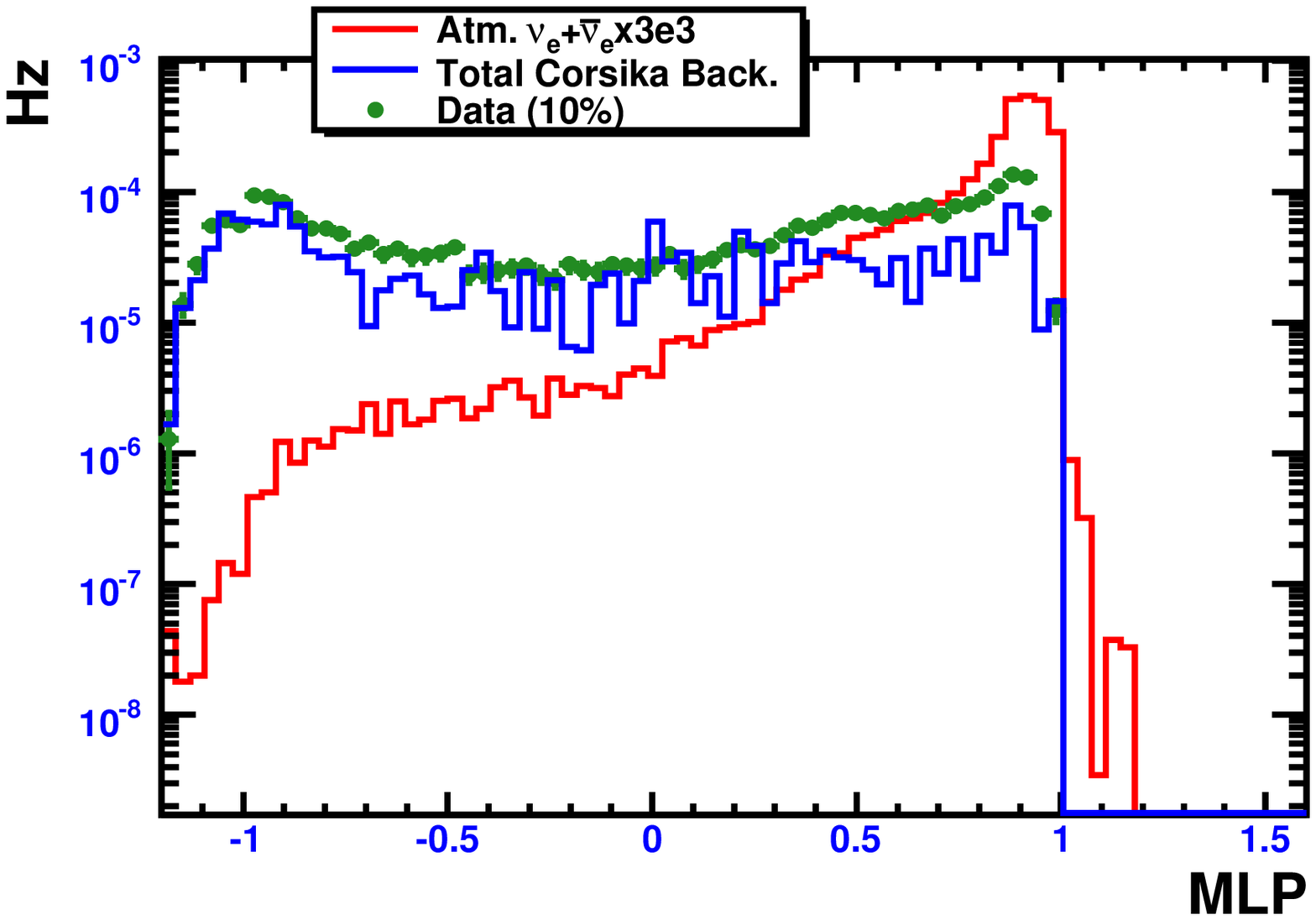}
\end{minipage}
\hspace{0.5cm} %To get a little bit of space between the figures
\begin{minipage}[b]{0.48\linewidth}
\centering
\includegraphics[width=7.6cm]{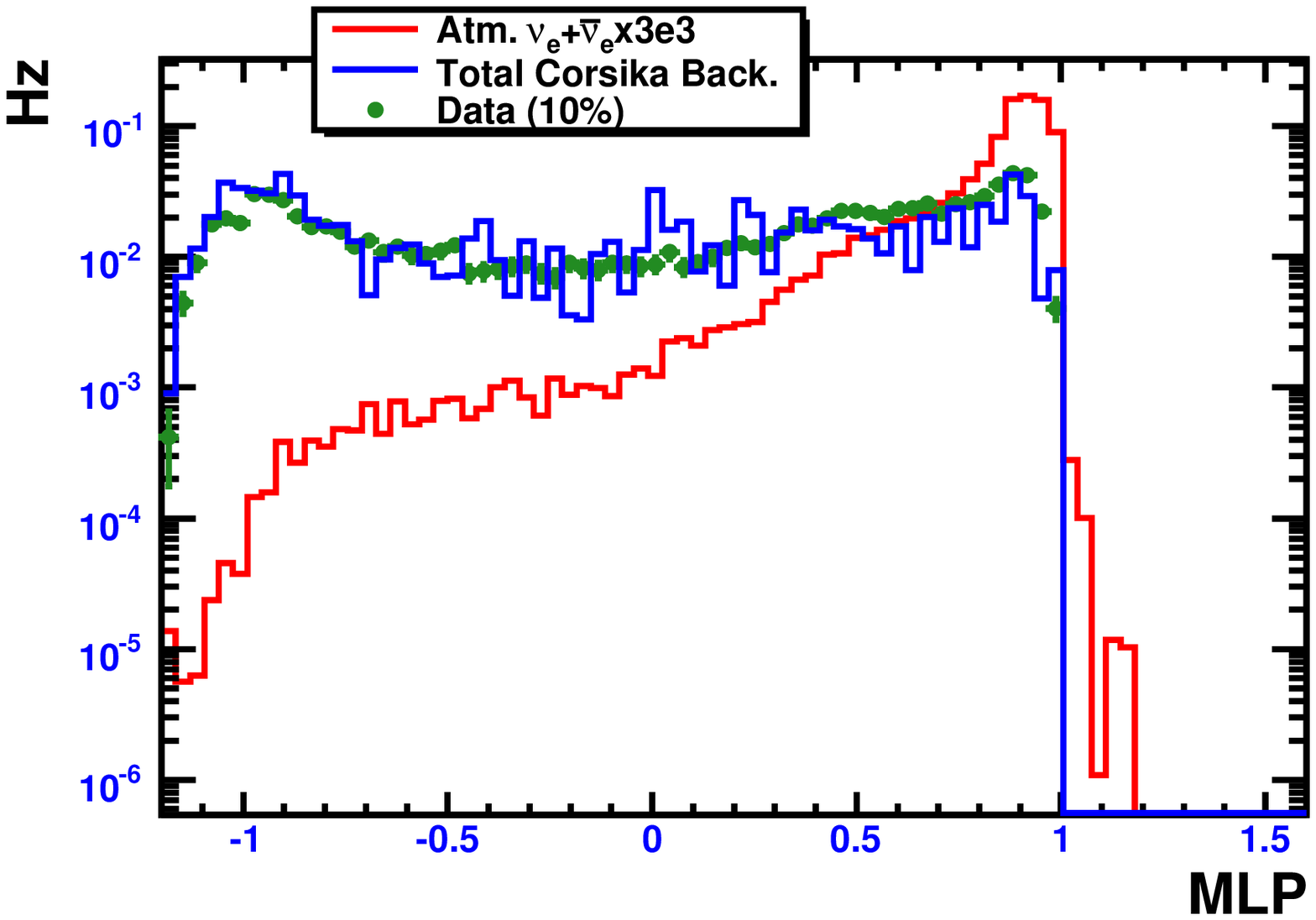}
\end{minipage}
\vspace{0.25cm}
\caption{Second level 4a neural net output variable L4aMLP1 at level 4a, absolutely normalized (left) and normalized to the same area (right).}
\label{L4aMLP2Level4a}
\end{figure}

\noindent Figures~\ref{FirstHitDist (norm)}--\ref{L4aMLP2Level4a} show these four input variables and the output neural net classifier variable called L4aMLP2.  Plots in the right column are normalized to the same area. 

\newpage

\clearpage

\clearpage

\clearpage

\clearpage

\begin{figure}
\begin{minipage}[b]{0.48\linewidth} % A minipage that covers half the page
\centering
\includegraphics[width=7.6cm]{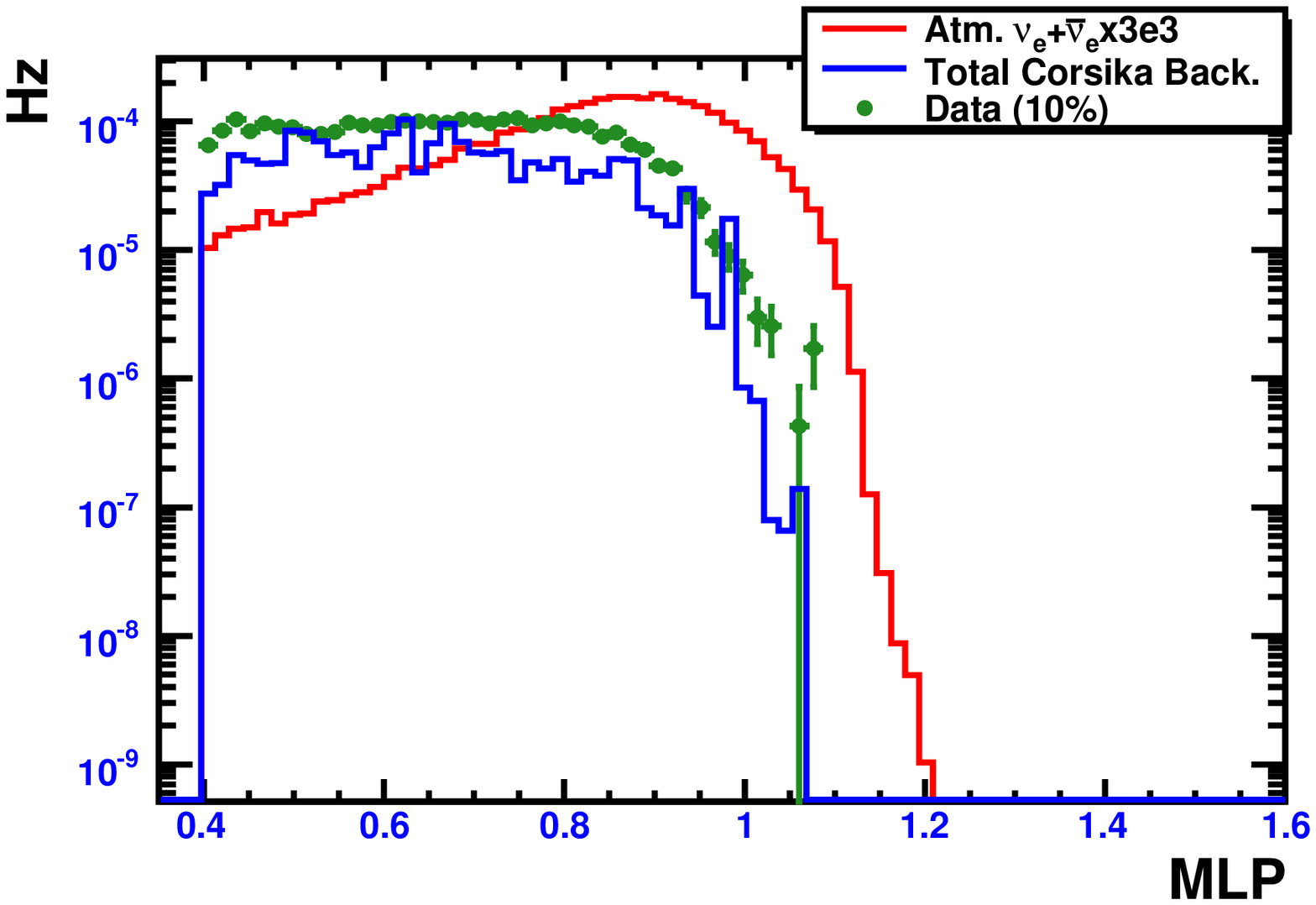}
\end{minipage}
\hspace{0.5cm} %To get a little bit of space between the figures
\begin{minipage}[b]{0.48\linewidth}
\centering
\includegraphics[width=7.6cm]{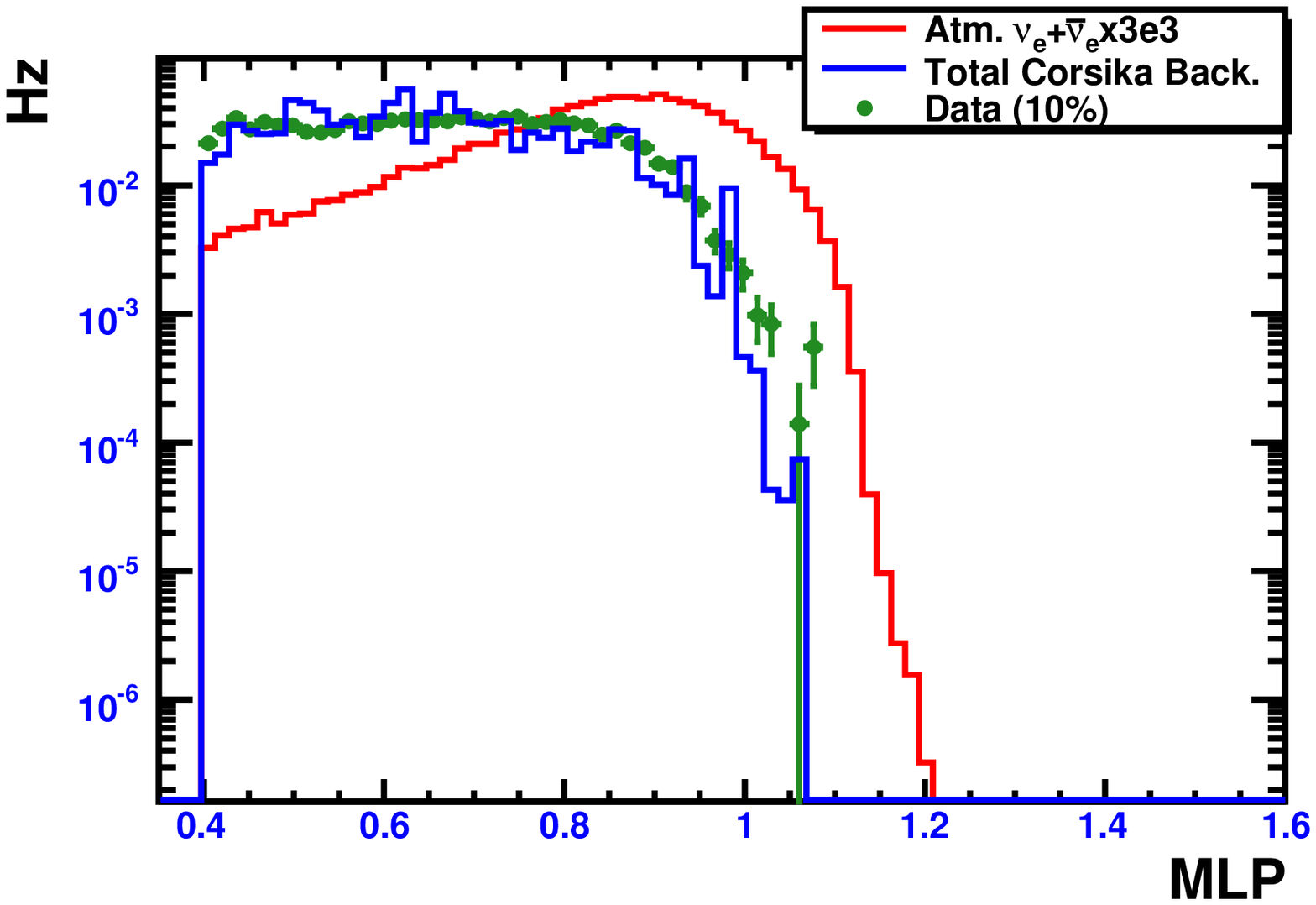}
\end{minipage}
\vspace{0.25cm}
\caption{L3aMLP at level 4a, absolutely normalized (left) and normalized to the same area (right).}
\label{L3aMLPLevel4a}
\end{figure}

In addition, the neural network output from level 3a, on which a soft cut was placed, still has discriminating power. It's shown in figure~\ref{L3aMLPLevel4a}, absolutely normalized (left) and normalized to the same area (right).

\clearpage

\section{Final Cuts}
\begin{figure}
\begin{minipage}[b]{0.48\linewidth} % A minipage that covers half the page
\centering
\includegraphics[width=7.6cm]{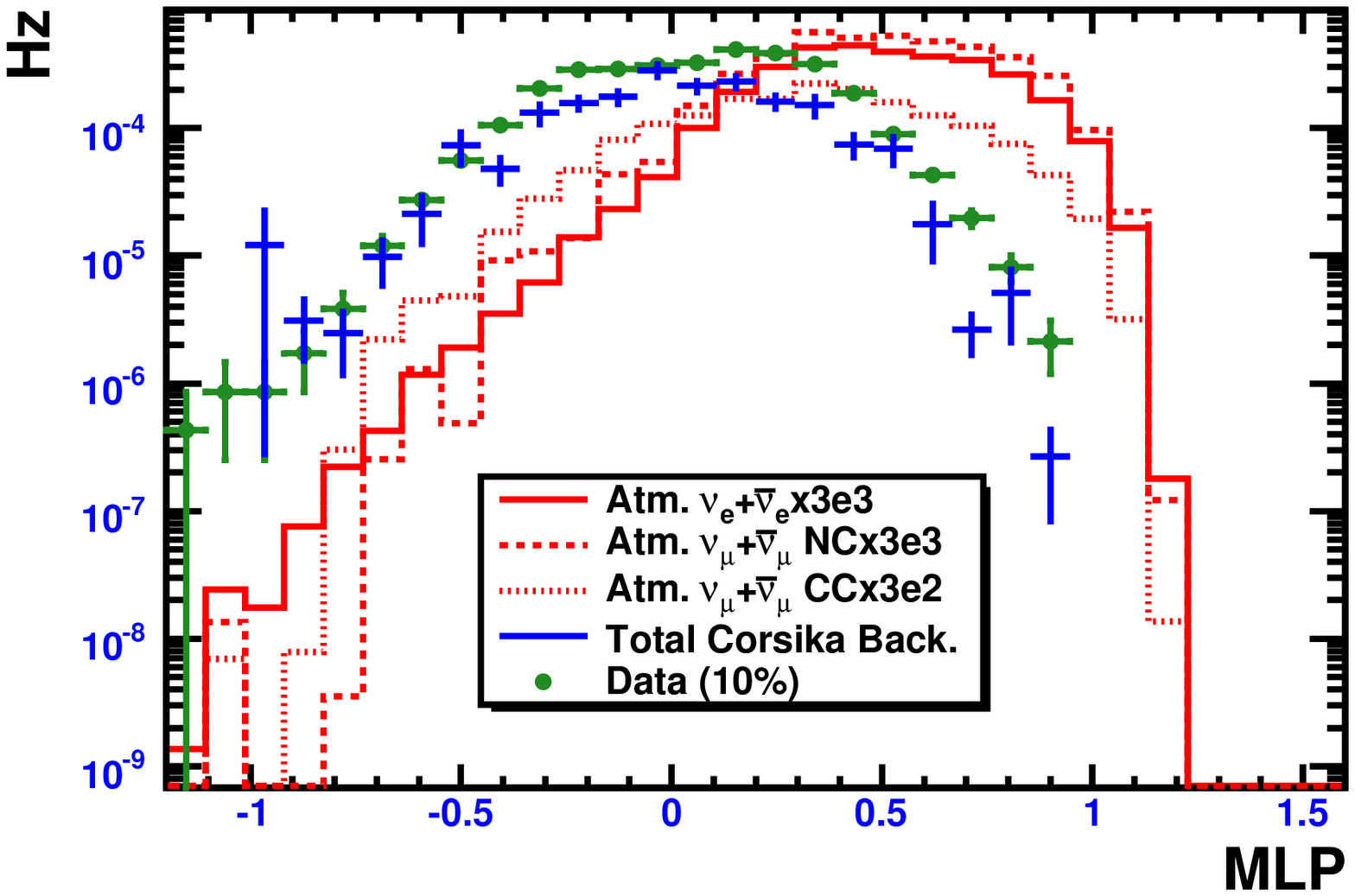}
\end{minipage}
\hspace{0.5cm} %To get a little bit of space between the figures
\begin{minipage}[b]{0.48\linewidth}
\centering
\includegraphics[width=7.6cm]{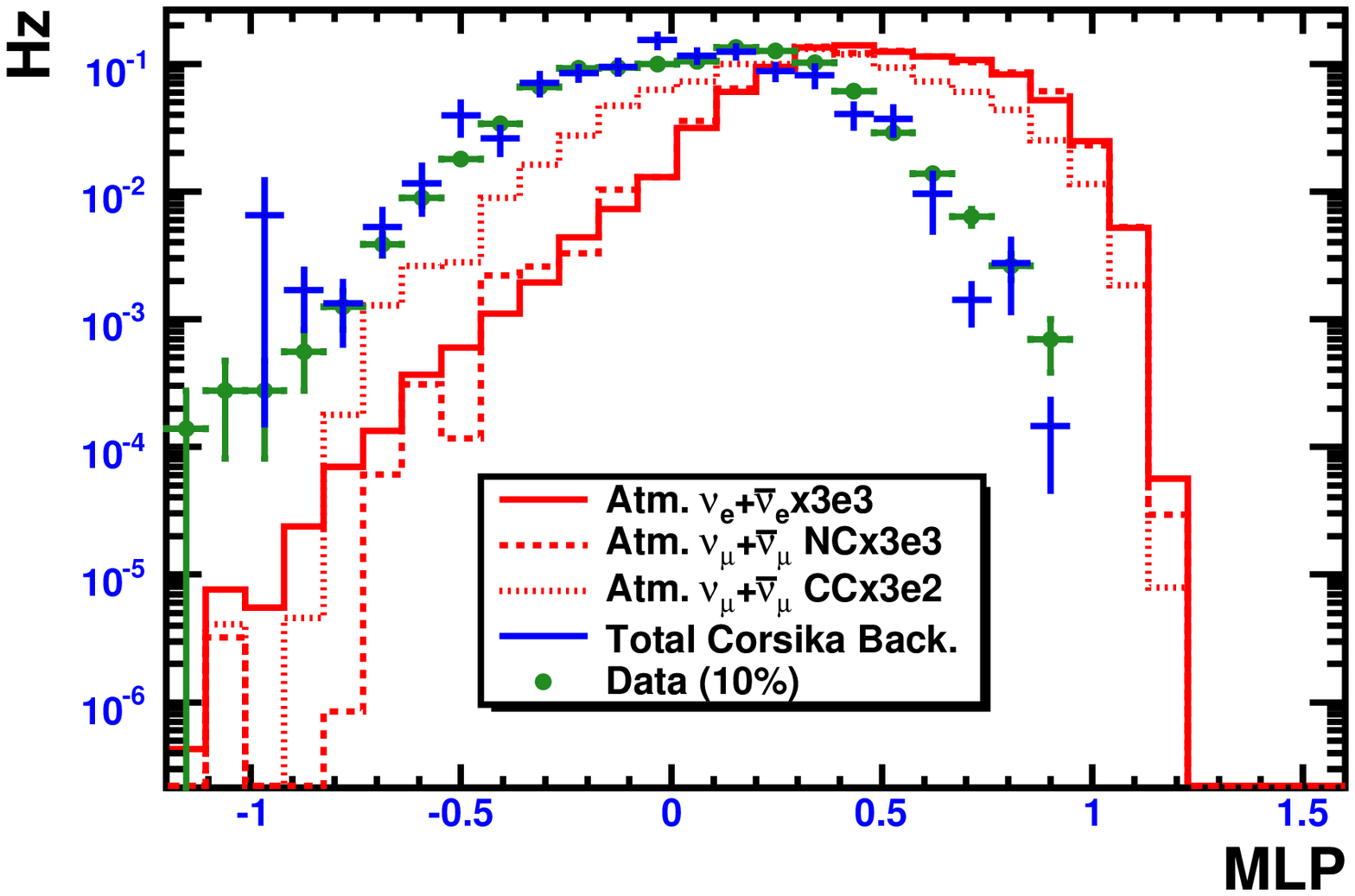}
\end{minipage}
\vspace{0.25cm}
\caption{The product of L3aMLP, L4aMLP1, and L4aMLP2 at level 4a, absolutely normalized (left) and normalized to the same area (right).  This will be our final multivariate classifier.}
\label{MLPProdLevel4a}
\end{figure}

We have three good discriminating neural network variables at level 4a: L3aMLP, L4aMLP1, and L4aMLP2.  These three variables are shown above in figures~\ref{L3aMLPLevel4a}, \ref{L4aMLP1Level4a}, and \ref{L4aMLP2Level4a}.  For the final cut variable, it was decided to take a simple product of these three variables. The resulting distributions are shown in figure~\ref{MLPProdLevel4a}, absolutely normalized (left) and normalized to the same area (right).  When normalized to the same area, the agreement between data and Monte Carlo is excellent.  

In figure~\ref{MLPProdLevel4a} the contribution from atmospheric $\numu$ is plotted separately.  As expected, cascades from neutral-current $\numu$ events look identical to cascades from $\nue$ events.  However, charged-current $\numu$ events are more complicated.  These are so-called starting events, where the hadronic cascade at the $\numu$ interaction vertex is inside the detector.  Below cut values of 0 or so they look more background-like than the $\nue$ or the neutral-current $\numu$ events.  A visual scan of these events indicates that they are ones where the outgoing muon contributes light to the event. As the cut variable is tightened, the hadronic cascade at the interaction vertex dominates the light output and they become true cascade events.

Two additional cuts were placed before optimization of this final variable. First, a soft depth cut was placed on the reconstructed cascade vertex to eliminate events at the very top of the detector.  This cut required that FullAllHitSPEZ$<$440~m.  Next, a cut was placed on the reduced likelihood parameter from the best cascade vertex reconstruction.  This was found to eliminate background from two or more muons from coincident cosmic ray air showers.  It required that FullAllHitSPEReducedLlh$<$10.

\subsection{Optimization}

We want to determine the optimal cut in energy and final multivariate output variable. For each potential energy cut, we plot a cumulative distribution that shows the absolute number of surviving signal and background events expected for 275.46 days of livetime (the full year dataset) as a function of the final multivariate cut variable.  For each plot, CORSIKA background is normalized up to data at a cut value of 0.4.  The goal is to look for a transition from a background-dominated region to a signal-dominated region before we run out of signal.  However, as we will see, the background Monte Carlo statistics become the limiting factor.

\begin{figure}
\begin{minipage}[b]{0.48\linewidth} % A minipage that covers half the page
\centering
\includegraphics[width=7.6cm]{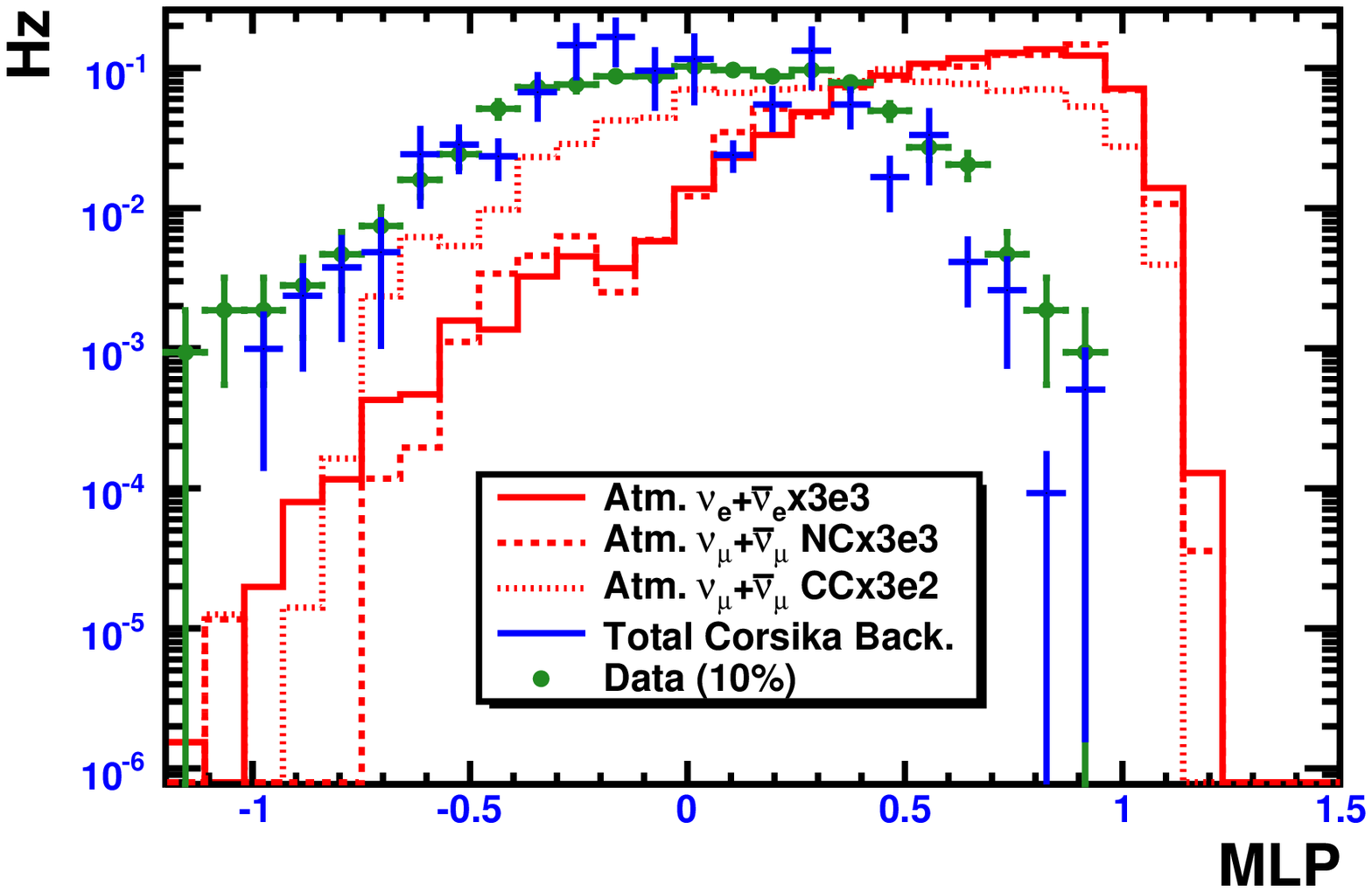}
\end{minipage}
\hspace{0.5cm} %To get a little bit of space between the figures
\begin{minipage}[b]{0.48\linewidth}
\centering
\includegraphics[width=7.6cm]{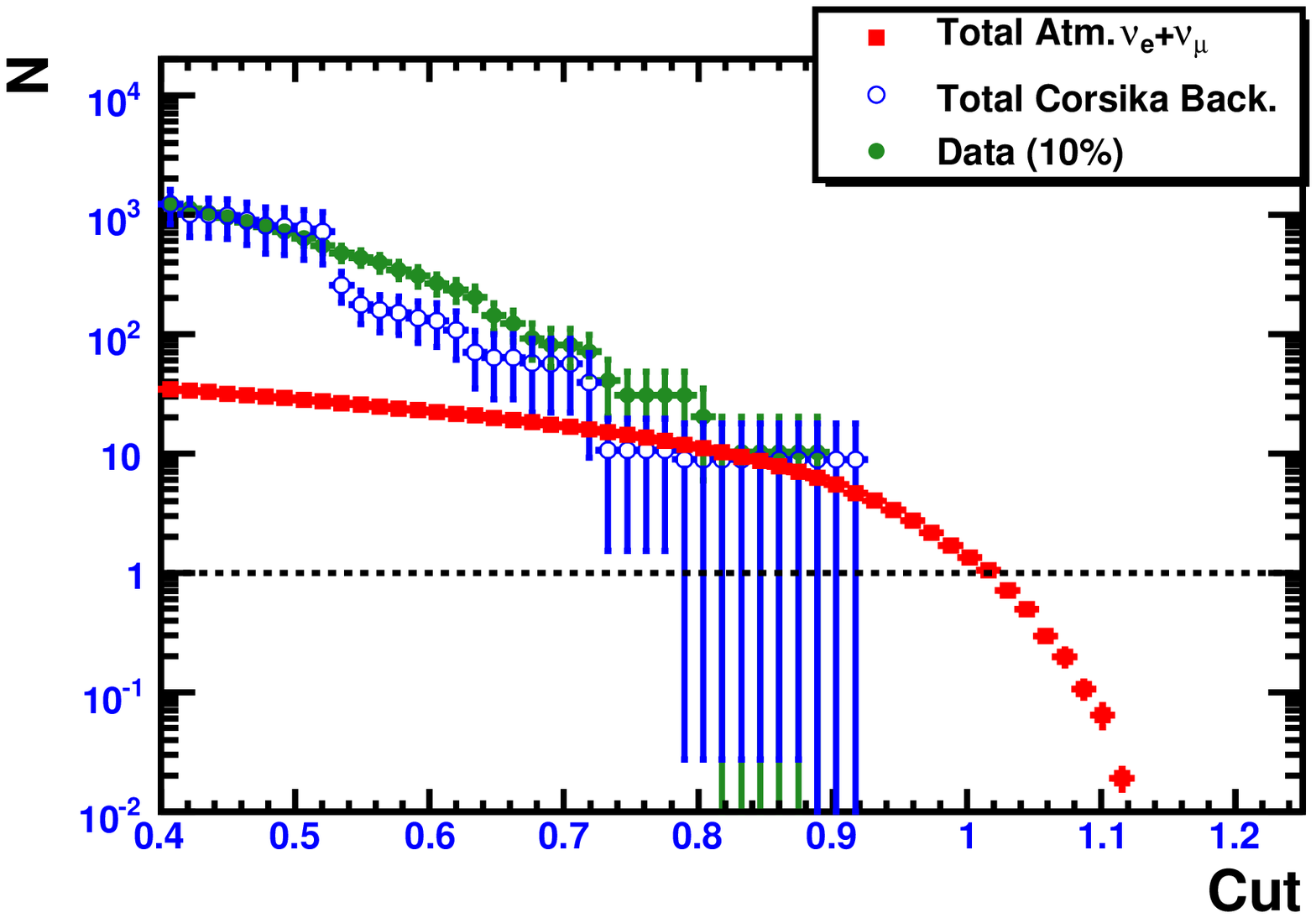}
\end{minipage}
\vspace{0.25cm}
\caption{Final variable distribution (left) and cumulative distribution (right) for an energy cut of 5 TeV.}
\label{MLPProdAbove5TeV}
\end{figure}

\begin{figure}
\begin{minipage}[b]{0.48\linewidth} % A minipage that covers half the page
\centering
\includegraphics[width=7.6cm]{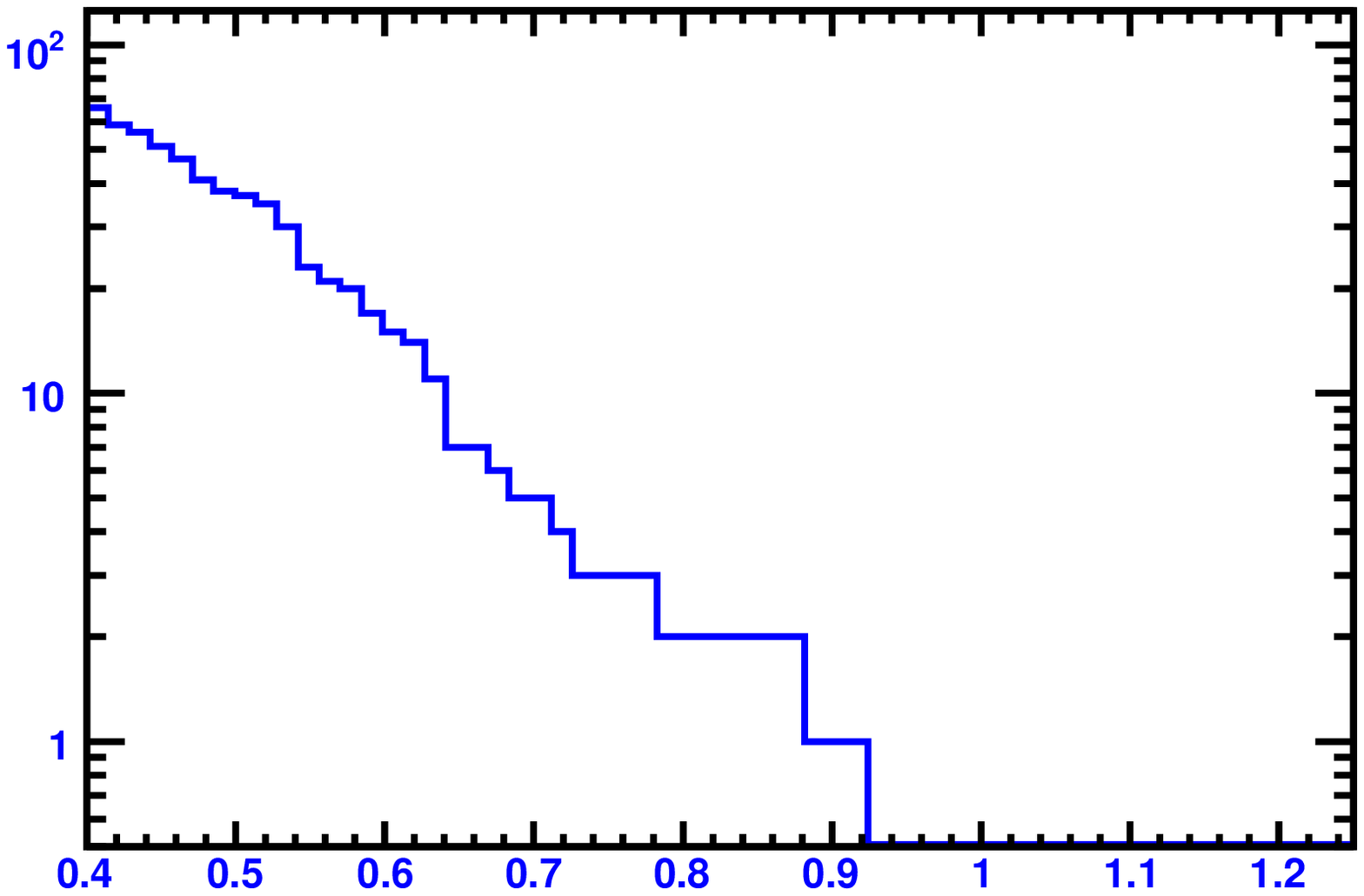}
\end{minipage}
\hspace{0.5cm} %To get a little bit of space between the figures
\begin{minipage}[b]{0.48\linewidth}
\centering
\includegraphics[width=7.6cm]{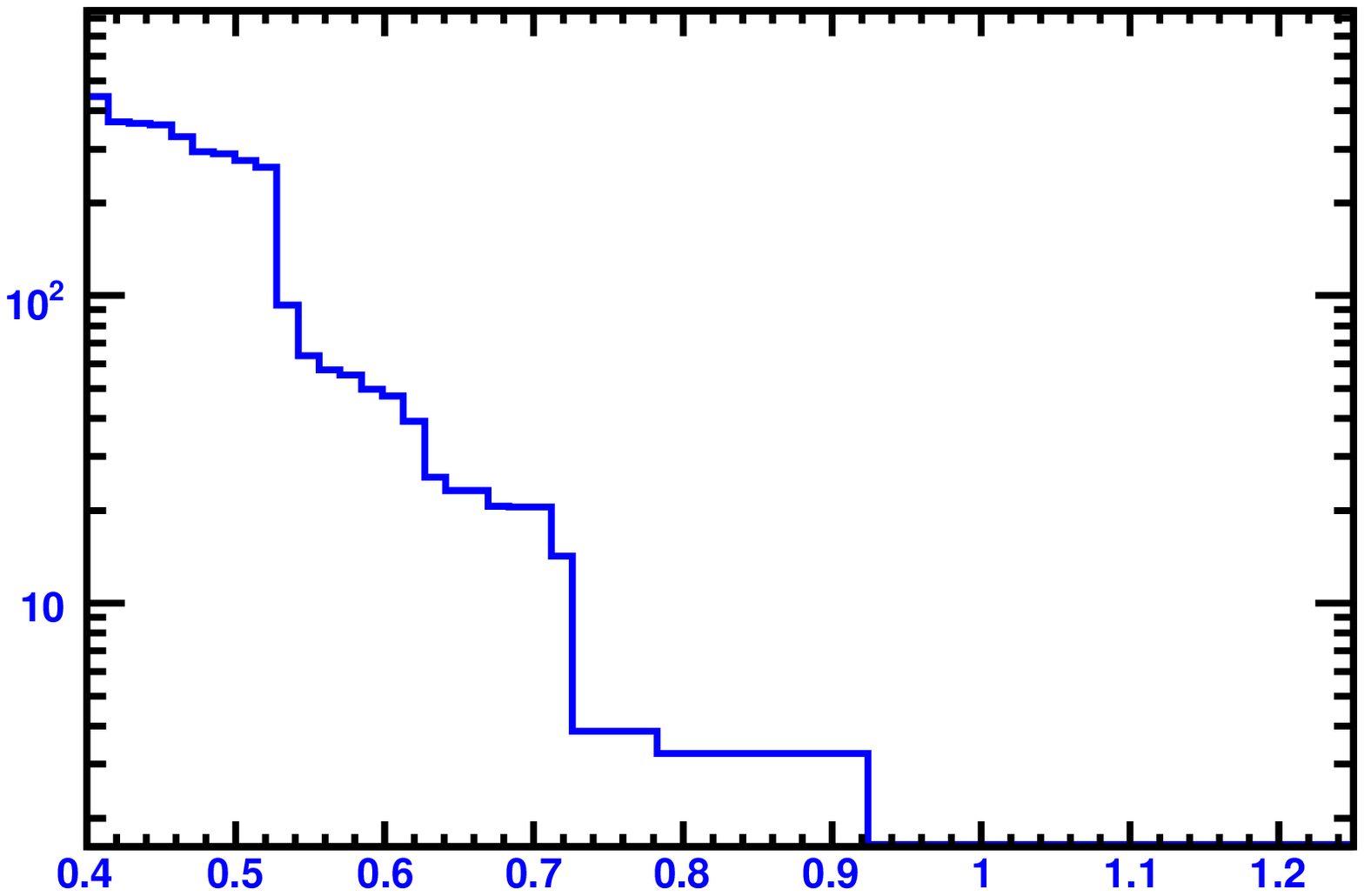}
\end{minipage}
\vspace{0.25cm}
\caption{CORSIKA background Monte Carlo raw number of events (left) and sum of event weights (right) for an energy cut of 5 TeV.}
\label{DiagnosticsMLPProdAbove5TeV}
\end{figure}

For an energy cut of 5 TeV, the final level distributions are shown in figure~\ref{MLPProdAbove5TeV}.  Two diagnostic plots are  also shown in figure~\ref{DiagnosticsMLPProdAbove5TeV}.  The first is a cumulative plot of the raw number of weighted CORSIKA background Monte Carlo events.  This tells us the number of events before we have applied their weights.  The second is a zoomed-in histogram of the weighted CORSIKA trace after the weights have been applied.

\clearpage

\noindent The cumulative distribution in figure~\ref{MLPProdAbove5TeV} indicates a promising trend---the data decreases towards the signal prediction as the final cut variable is tightened.  However, the diagnostic plots tell us that there is a problem with the background Monte Carlo prediction.  In the region of highest significance, where the data is approaching the signal, we have only 2 or 3 surviving weighted background Monte Carlo events.  This causes large fluctuations in the background prediction, which can be seen in the large steps in the second diagnostic plot and which come from the loss of individual high weight events.  These large fluctuations mean that we won't be able to accurately predict the background from our Monte Carlo, and a rigorous statistical detection will not be possible.

\begin{figure}
\begin{minipage}[b]{0.48\linewidth} % A minipage that covers half the page
\centering
\includegraphics[width=7.6cm]{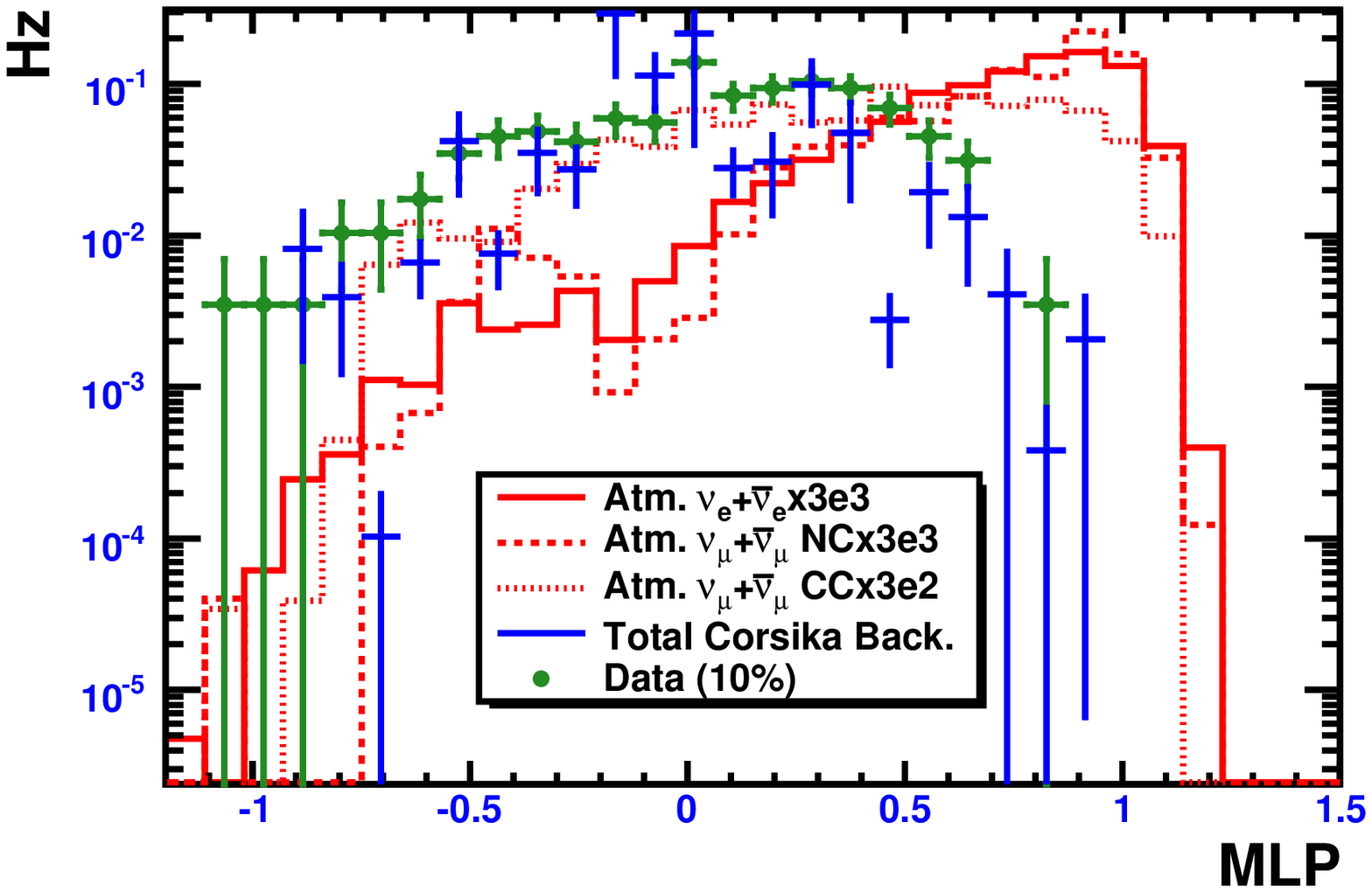}
\end{minipage}
\hspace{0.5cm} %To get a little bit of space between the figures
\begin{minipage}[b]{0.48\linewidth}
\centering
\includegraphics[width=7.6cm]{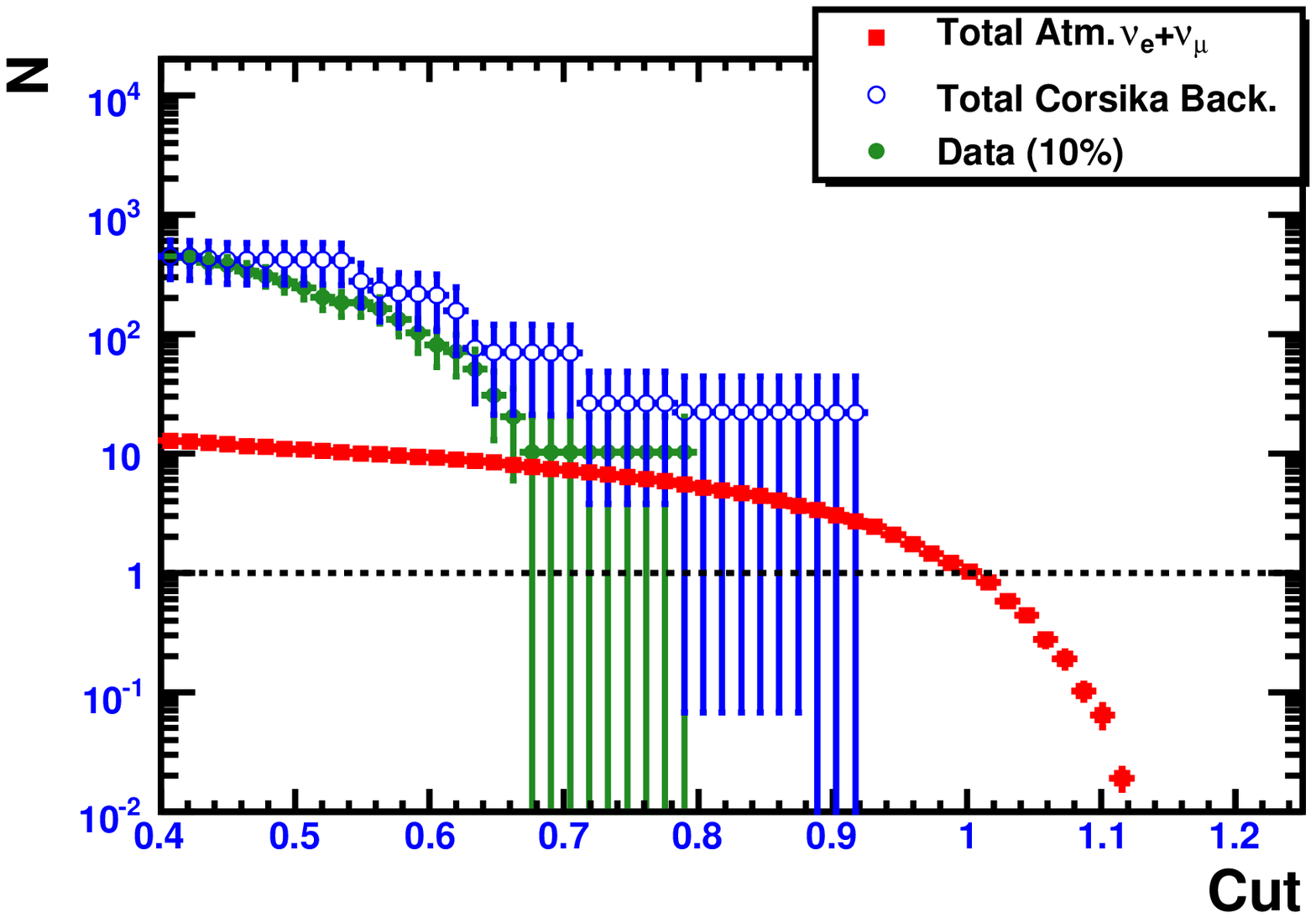}
\end{minipage}
\vspace{0.25cm}
\caption{Final variable distribution (left) and cumulative distribution (right) for an energy cut of 10 TeV.}
\label{MLPProdAbove10TeV}
\end{figure}

\begin{figure}
\begin{minipage}[b]{0.48\linewidth} % A minipage that covers half the page
\centering
\includegraphics[width=7.6cm]{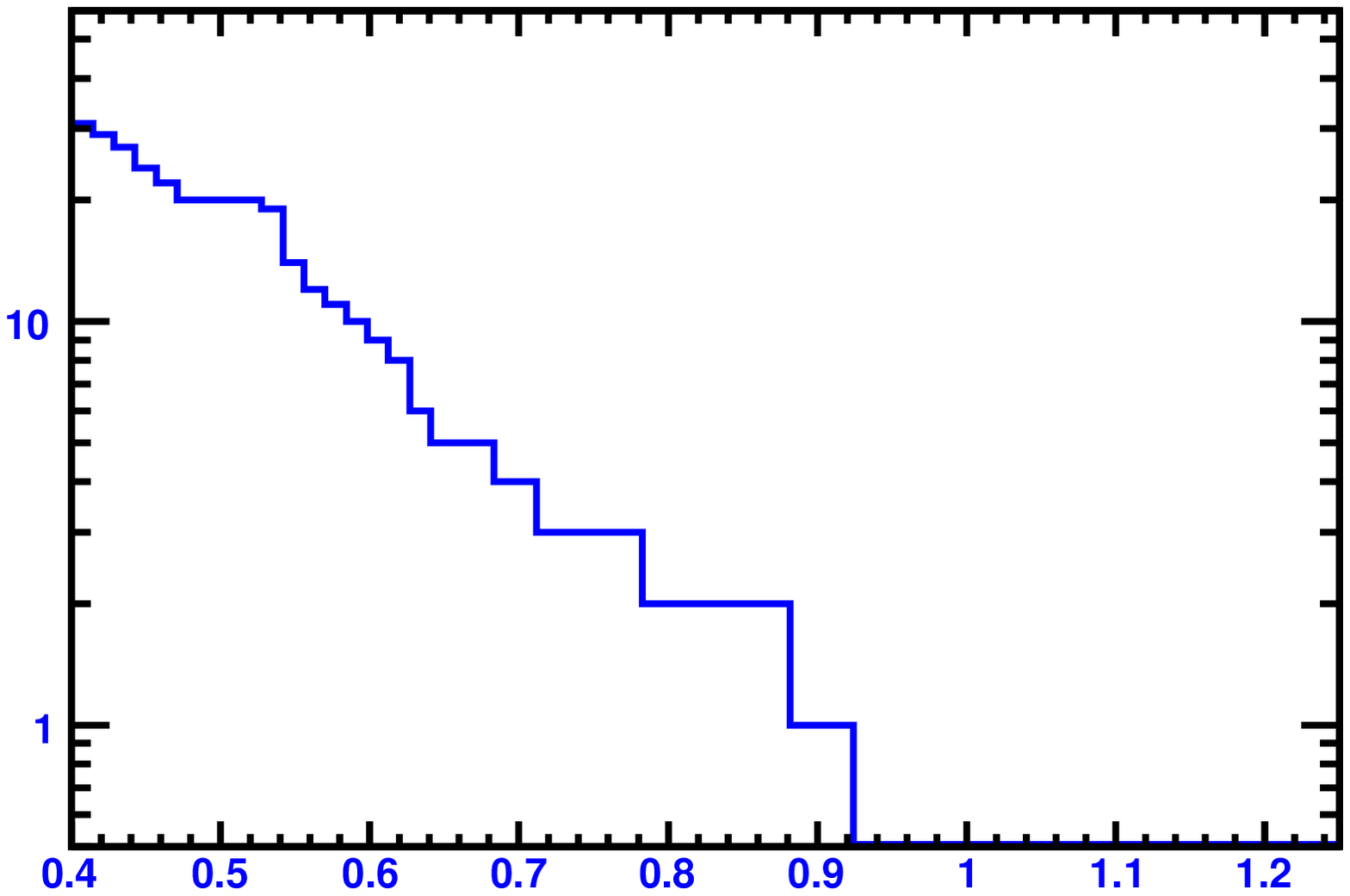}
\end{minipage}
\hspace{0.5cm} %To get a little bit of space between the figures
\begin{minipage}[b]{0.48\linewidth}
\centering
\includegraphics[width=7.6cm]{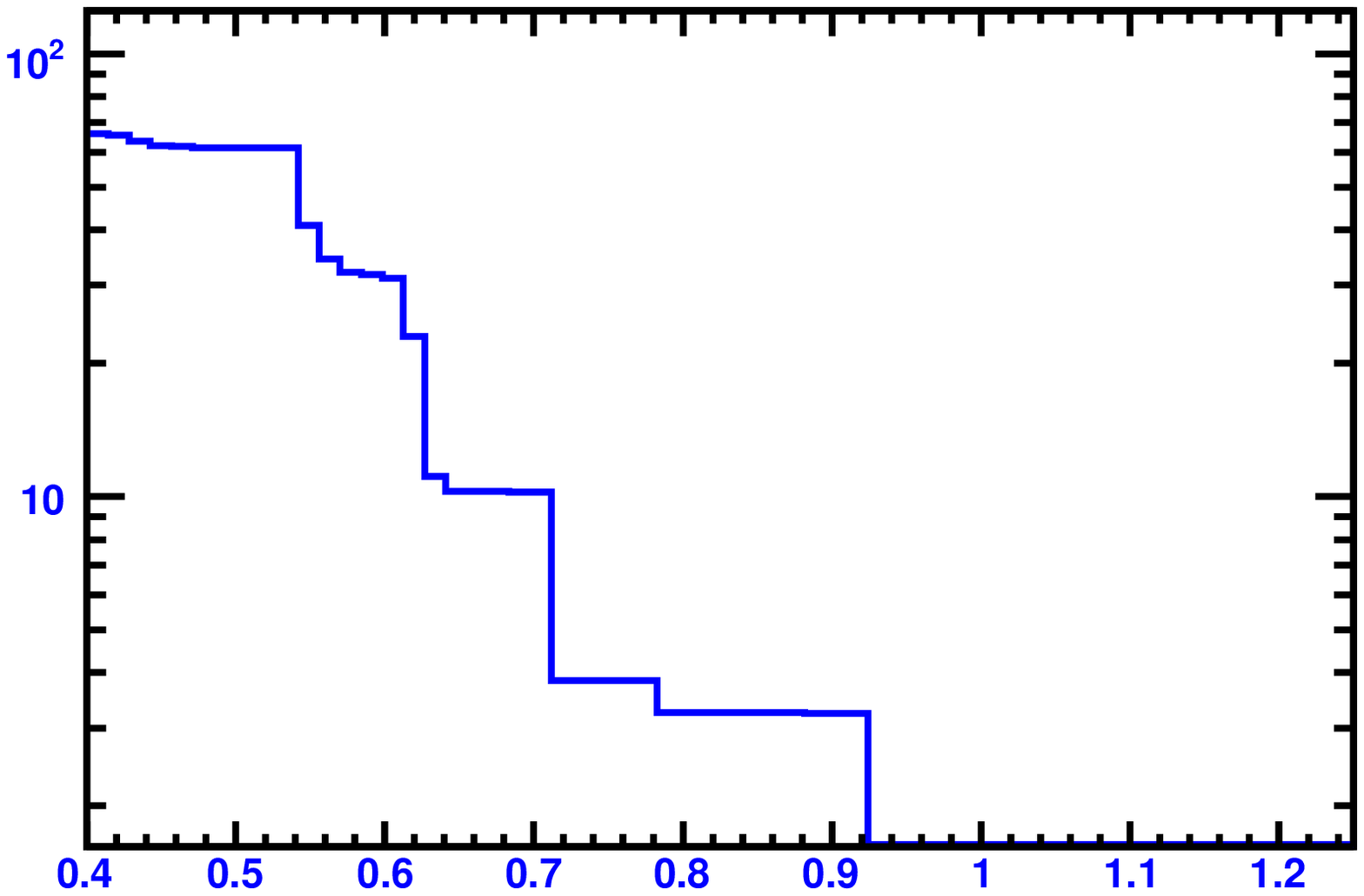}
\end{minipage}
\vspace{0.25cm}
\caption{Final variable distribution (left) and cumulative distribution (right) for an energy cut of 15 TeV.}
\label{DiagnosticsMLPProdAbove10TeV}
\end{figure}

\begin{figure}
\begin{minipage}[b]{0.48\linewidth} % A minipage that covers half the page
\centering
\includegraphics[width=7.6cm]{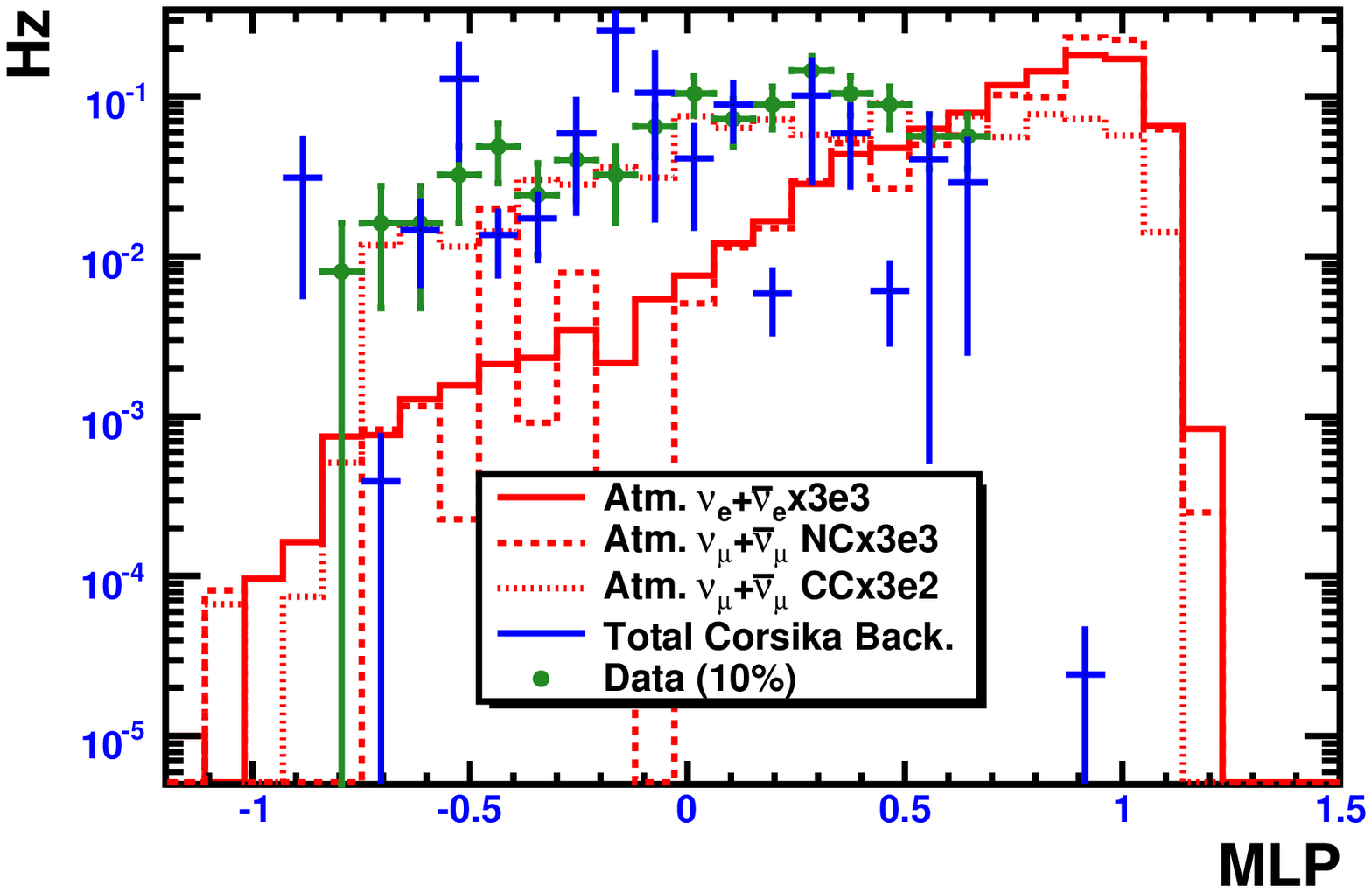}
\end{minipage}
\hspace{0.5cm} %To get a little bit of space between the figures
\begin{minipage}[b]{0.48\linewidth}
\centering
\includegraphics[width=7.6cm]{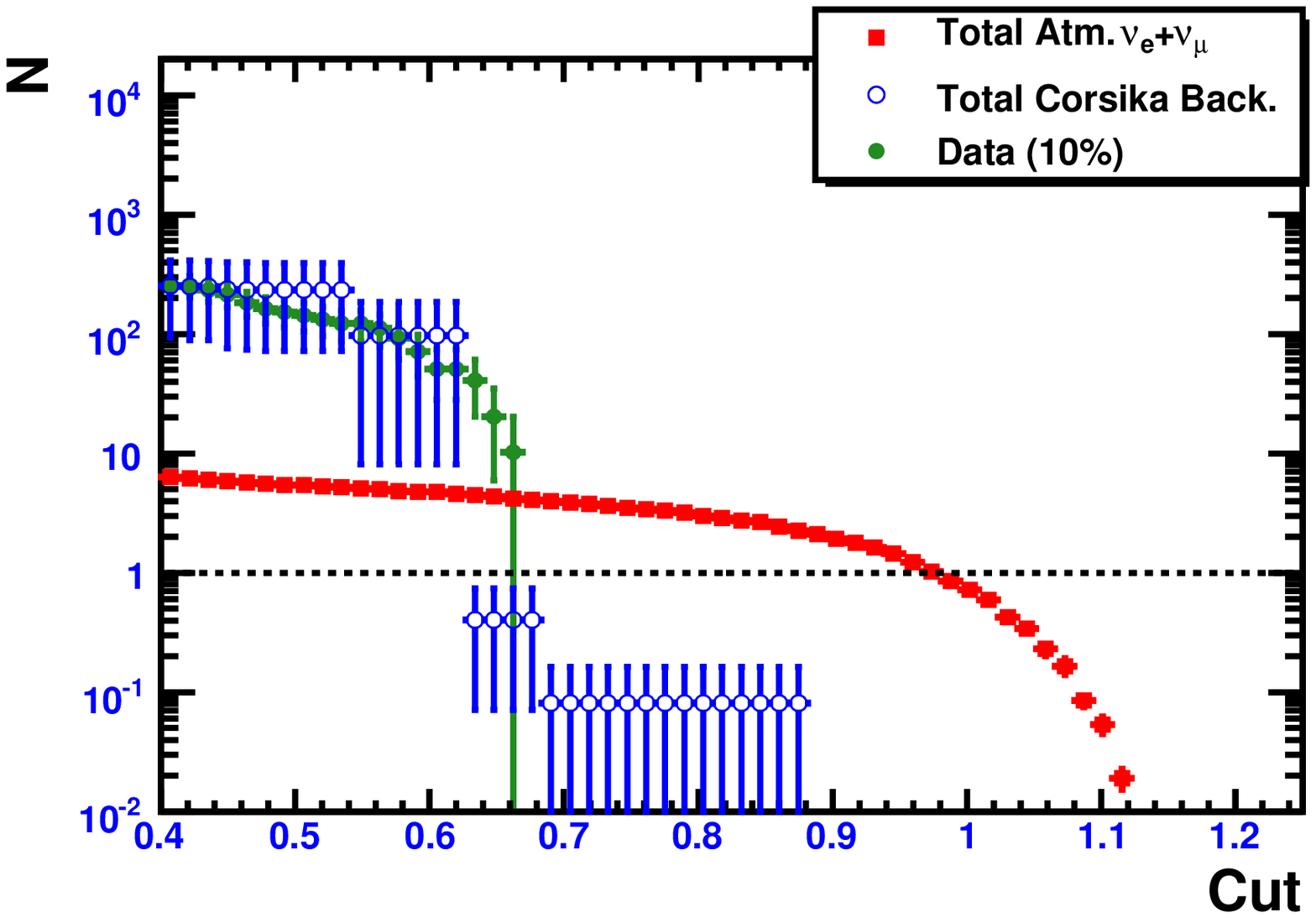}
\end{minipage}
\vspace{0.25cm}
\caption{CORSIKA background Monte Carlo raw number of events (left) and sum of event weights (right) for an energy cut of 5 TeV.}
\label{MLPProdAbove15TeV}
\end{figure}

\begin{figure}
\begin{minipage}[b]{0.48\linewidth} % A minipage that covers half the page
\centering
\includegraphics[width=7.6cm]{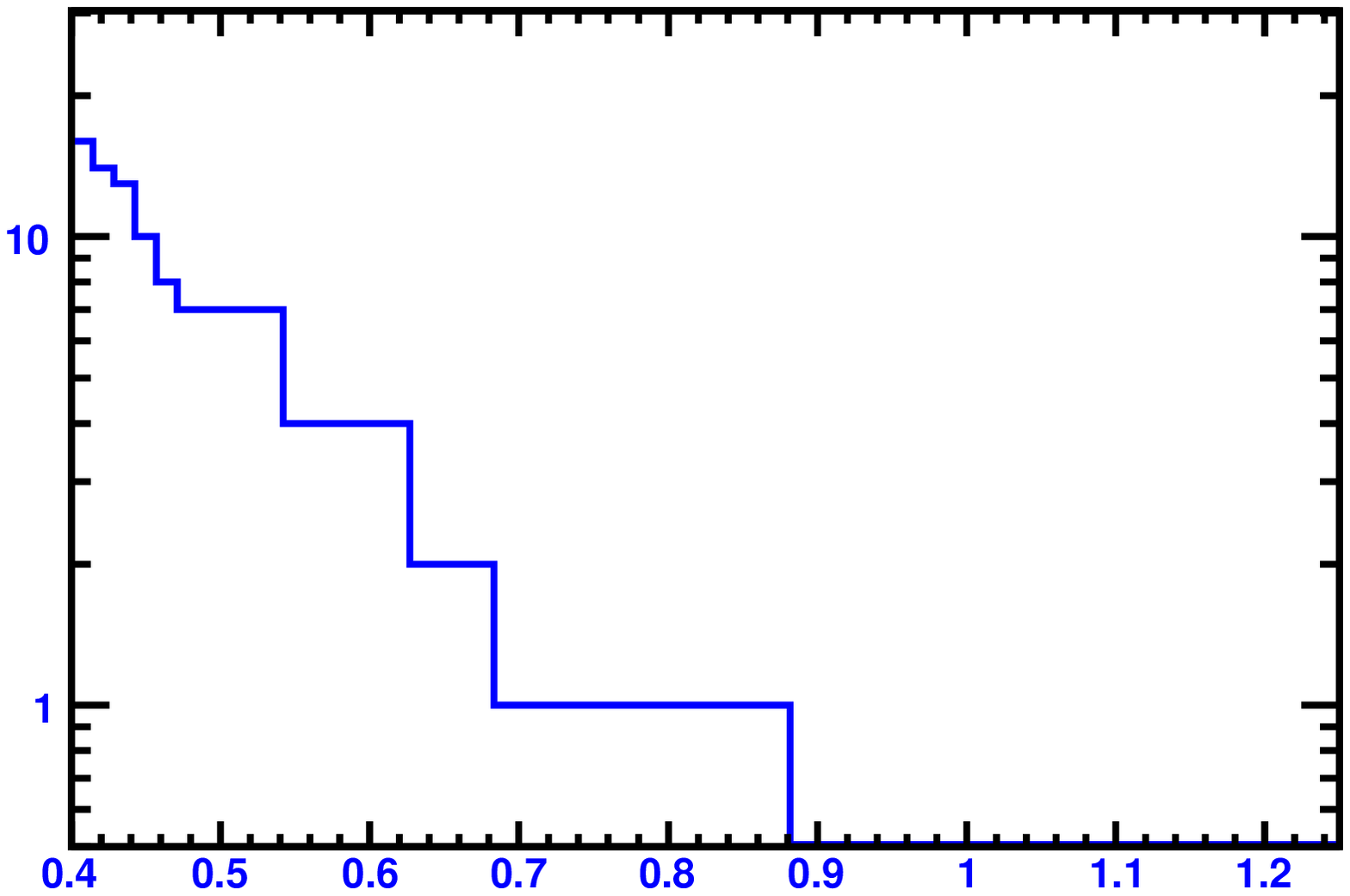}
\end{minipage}
\hspace{0.5cm} %To get a little bit of space between the figures
\begin{minipage}[b]{0.48\linewidth}
\centering
\includegraphics[width=7.6cm]{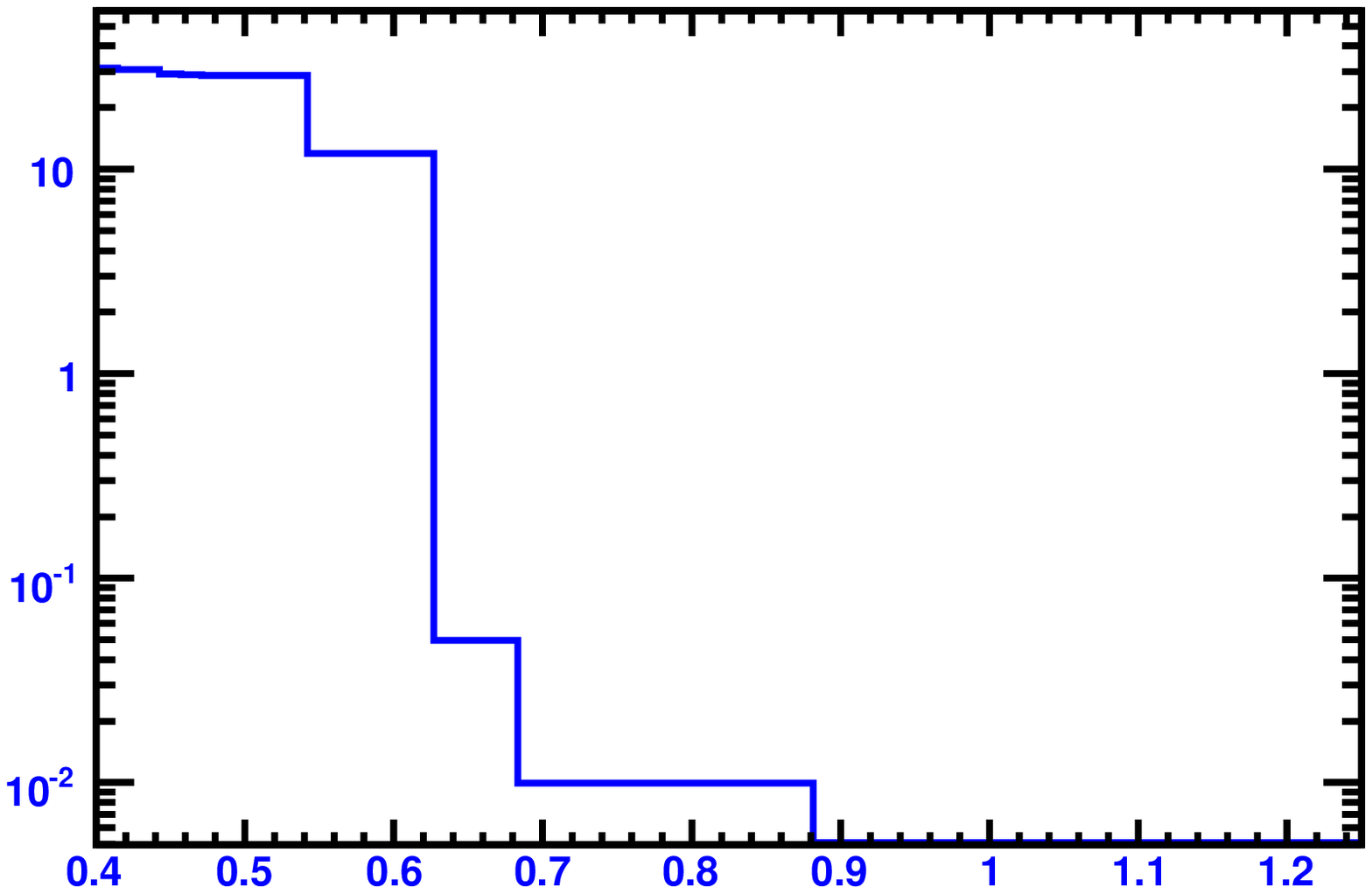}
\end{minipage}
\vspace{0.25cm}
\caption{CORSIKA background Monte Carlo raw number of events (left) and sum of event weights (right) for an energy cut of 5 TeV.}
\label{DiagnosticsMLPProdAbove15TeV}
\end{figure}

The same conclusions hold for energy cuts of 10 TeV and 15 TeV---the data approaches the signal prediction as the final cut variable is tightened, but the diagnostic plots show that we don't have enough background Monte Carlo to accurately predict the background.  The plots are displayed in figures~\ref{MLPProdAbove10TeV}--\ref{DiagnosticsMLPProdAbove15TeV}.

\clearpage

\newpage

\clearpage

\newpage

\subsection{Final Analysis Plan}

With such limited statistics, the CORSIKA background Monte Carlo can really only be treated as a guide.  The 10\% validation data sample is the better background predictor. The signal prediction, coming from a known calibration source, is relatively robust. Looking at the data sample and the signal simulation, the trend is clear: the data should reach a signal dominated region as the final multivariate cut variable is tightened.  Accordingly, we proposed to unblind three sets of cuts:

\begin{itemize}
\item{{\bf Final Level 1: E$>$5TeV + MLPProd$>$0.73}}
\item{{\bf Final Level 2: E$>$10TeV + MLPProd$>$0.72}}
\item{{\bf Final Level 3: E$>$15TeV + MLPProd$>$0.69}}
\end{itemize}

\noindent After unblinding the full year's dataset with these cuts, we planned to examine the full cumulative distributions to look for an obvious transition from background-dominated behavior to signal-dominated behavior beyond the final cut points. 

Permission was granted from the full IceCube collaboration to proceed with this analysis plan.  The next chapter presents the results from the full 2007-2008 IC-22 dataset.

%% file: chapters/chapter7.tex
\chapter{Results}\label{chapter:chapter7}

After a lengthy discussion and review by the IceCube collaboration, approval was granted to proceed with the analysis proposal outlined in the last chapter.  Because background Monte Carlo statistics were severely limited, and because the enormous computational resources to produce more on a reasonable timescale did not exist, we decided to compare the observed data to the atmospheric neutrino-induced cascade signal prediction to look for a transition from a background-dominated region to a region where the data qualitatively agreed with the signal prediction.

\section{Bartol Atmospheric Flux Comparison}  

\begin{figure}
\begin{minipage}[b]{0.48\linewidth} % A minipage that covers half the page
\centering
\includegraphics[width=7.6cm]{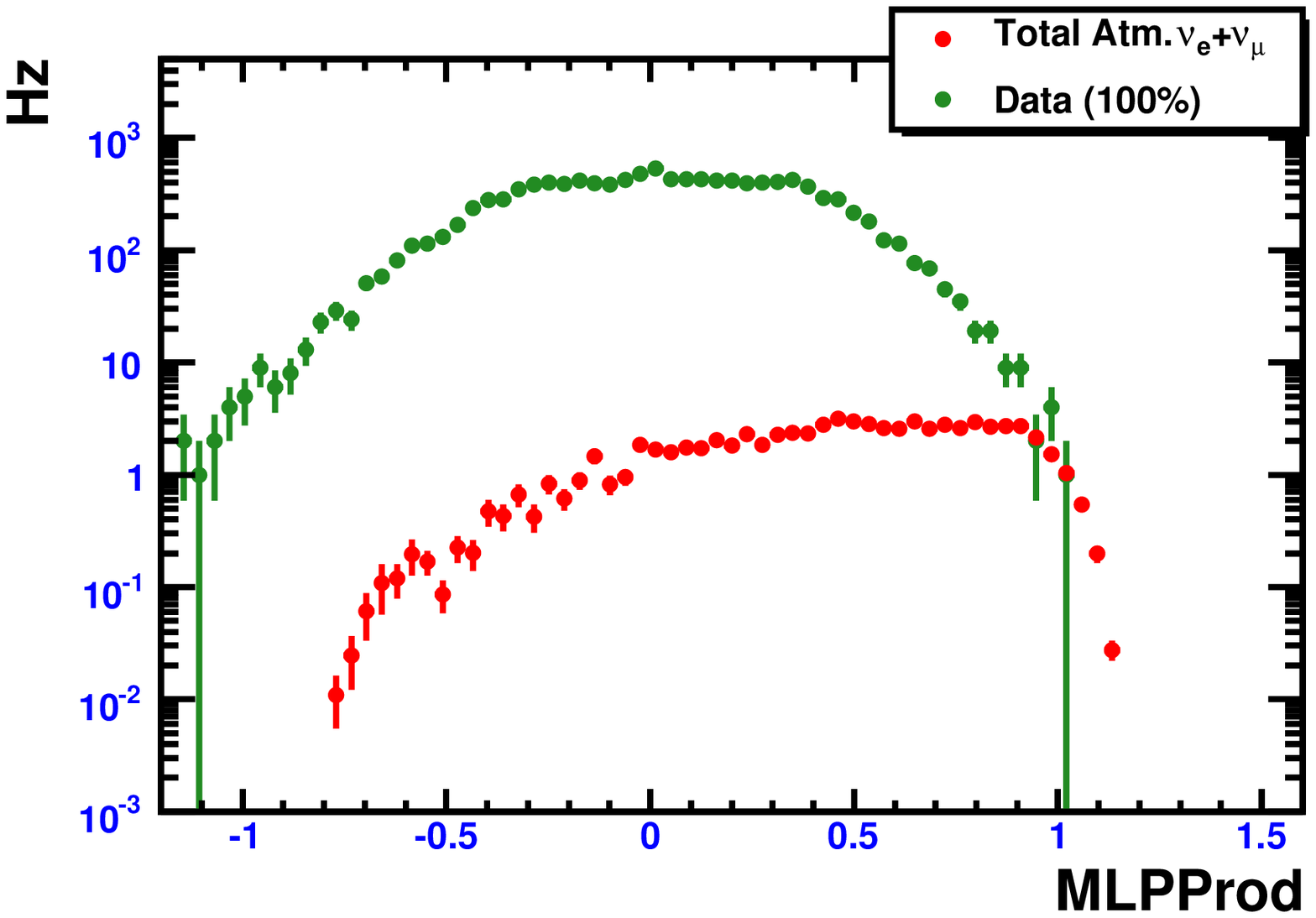}
\end{minipage}
\hspace{0.5cm} %To get a little bit of space between the figures
\begin{minipage}[b]{0.48\linewidth}
\centering
\includegraphics[width=7.6cm]{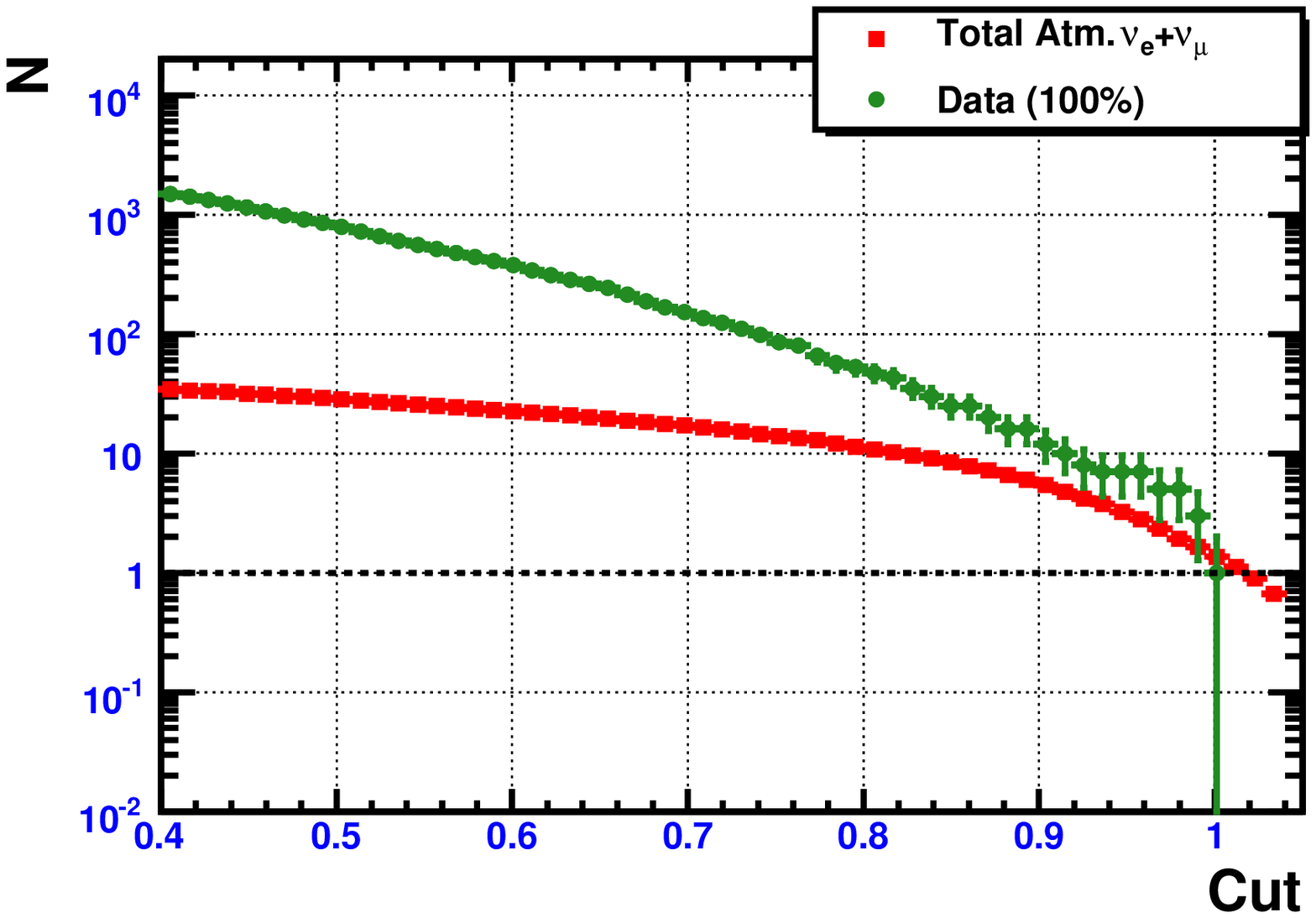}
\end{minipage}

\vspace{1.0cm}

\begin{minipage}[b]{0.48\linewidth} % A minipage that covers half the page
\centering
\includegraphics[width=7.6cm]{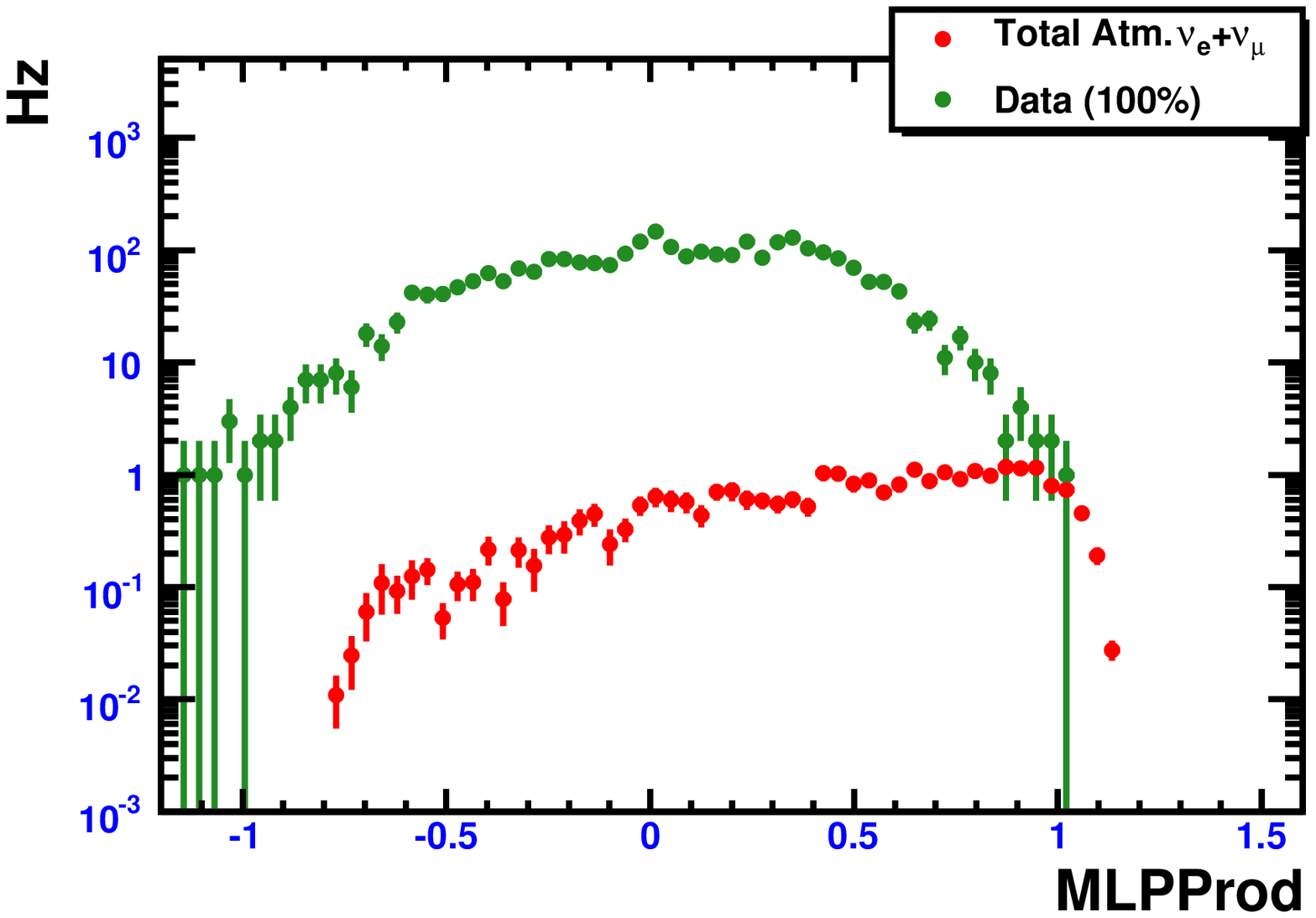}
\end{minipage}
\hspace{0.5cm} %To get a little bit of space between the figures
\begin{minipage}[b]{0.48\linewidth}
\centering
\includegraphics[width=7.6cm]{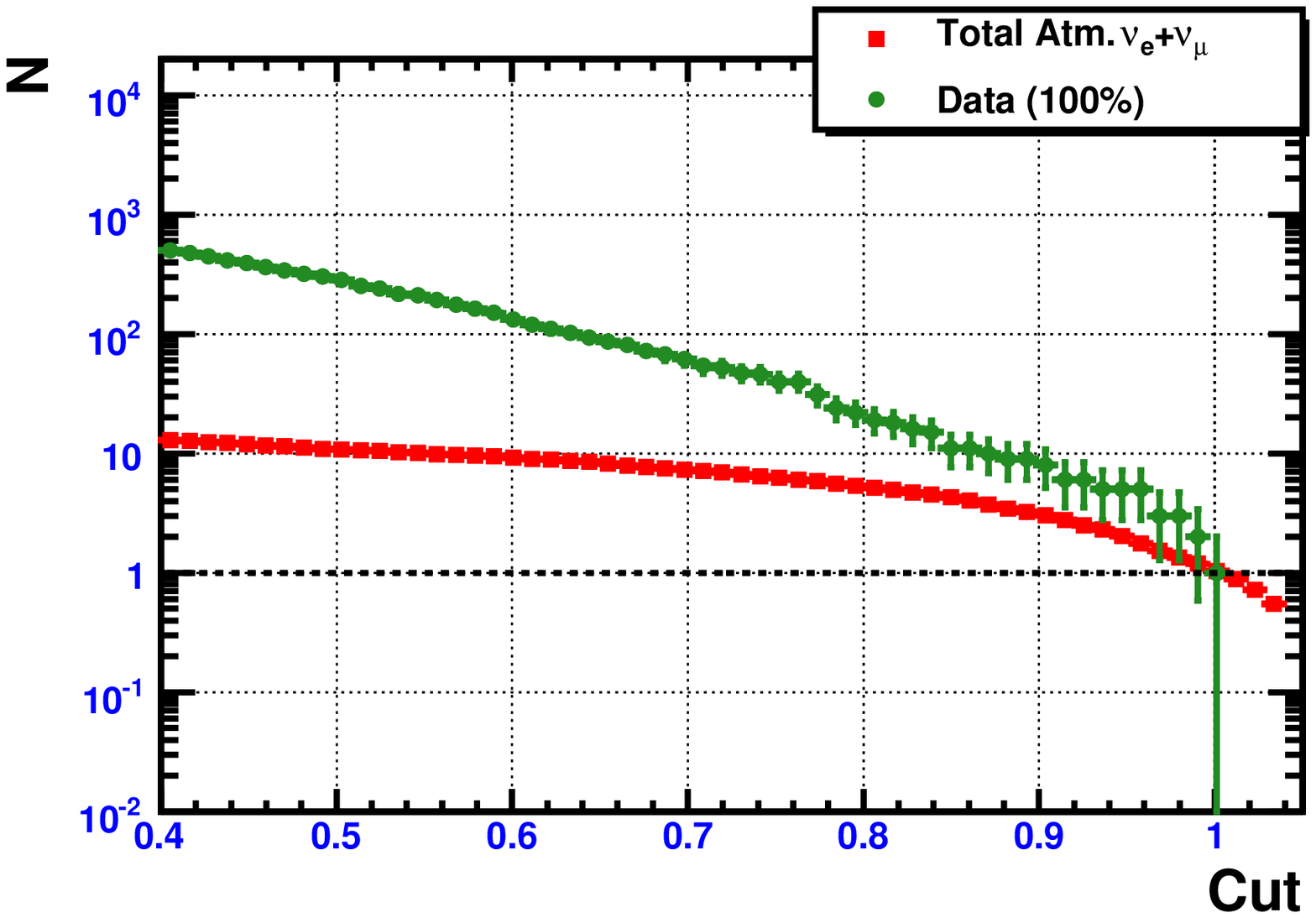}
\end{minipage}

\vspace{1.0cm}

\begin{minipage}[b]{0.48\linewidth} % A minipage that covers half the page
\centering
\includegraphics[width=7.6cm]{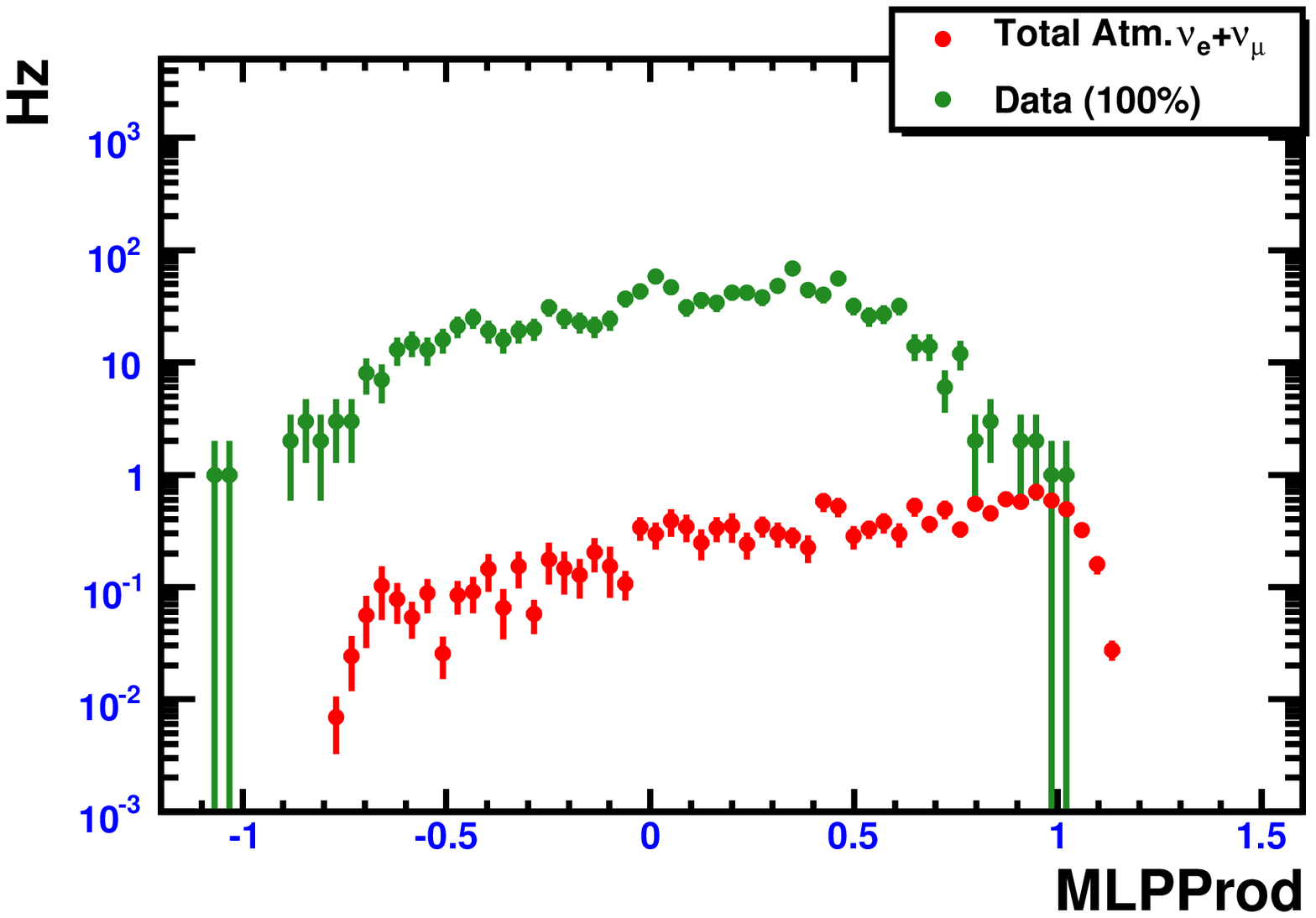}
\end{minipage}
\hspace{0.5cm} %To get a little bit of space between the figures
\begin{minipage}[b]{0.48\linewidth}
\centering
\includegraphics[width=7.6cm]{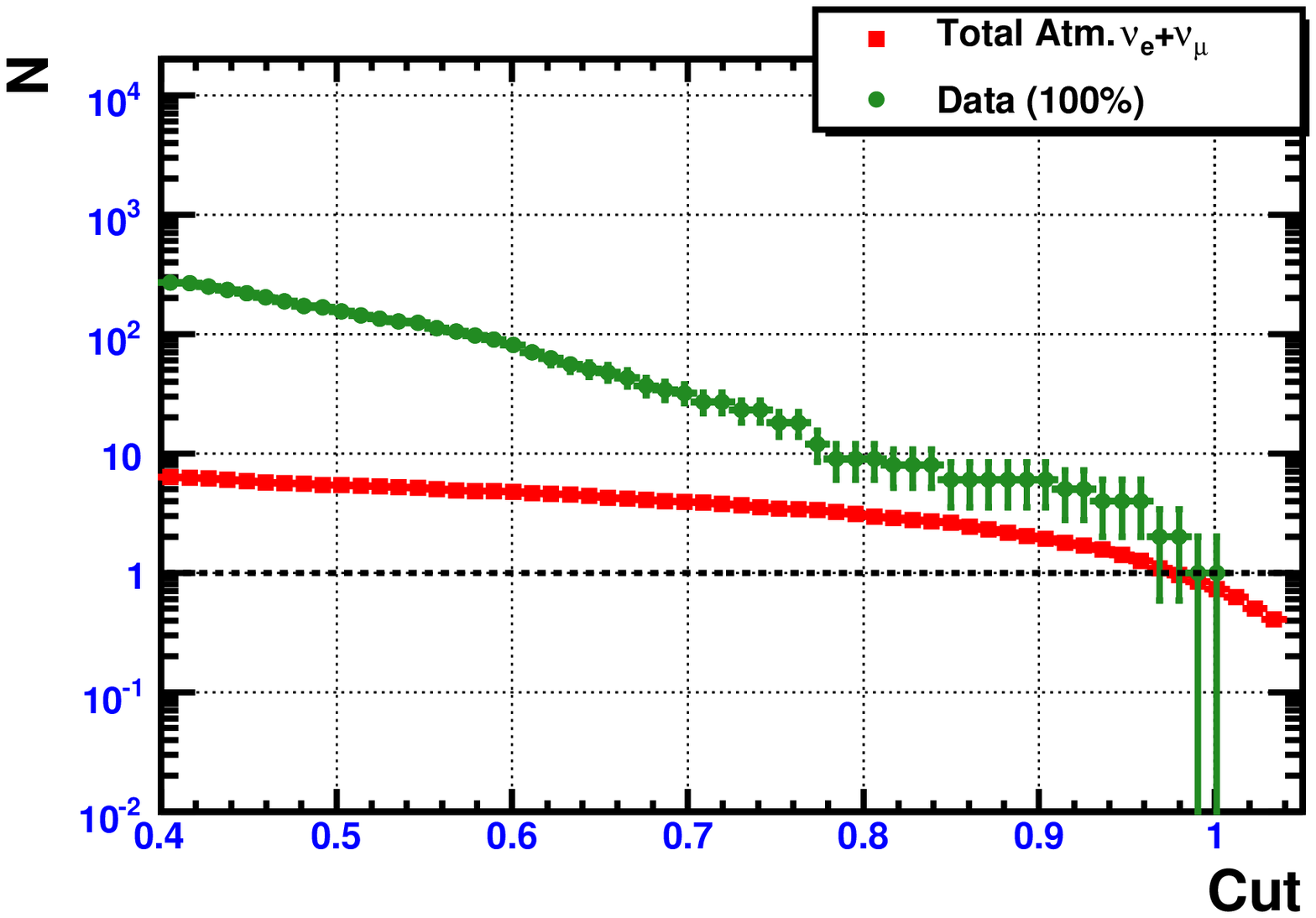}
\end{minipage}
\vspace{0.25cm}
\caption{Final multivariate classifier distribution (left) and cumulative distribution showing the number of events surviving beyond a given classifier cut (right) for the full IC-22 dataset with a 5 TeV (top), 10 TeV (middle), and 15 TeV (bottom) energy cut.}
\label{UNBLINDEDPlots}
\end{figure}

\begin{figure}
\centering
\includegraphics[width=0.55\linewidth]{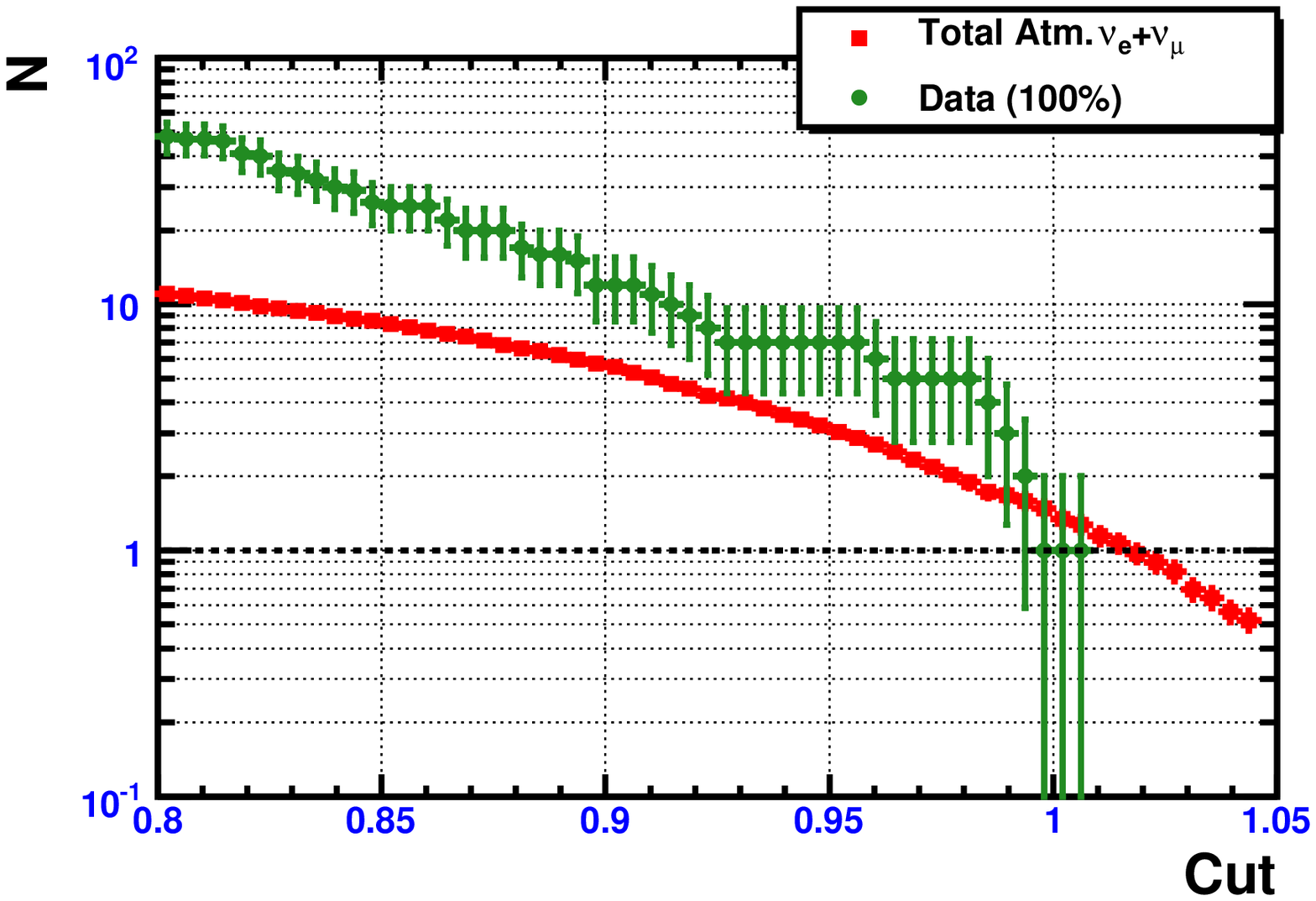}

\vspace{1.2cm}

\centering
\includegraphics[width=0.55\linewidth]{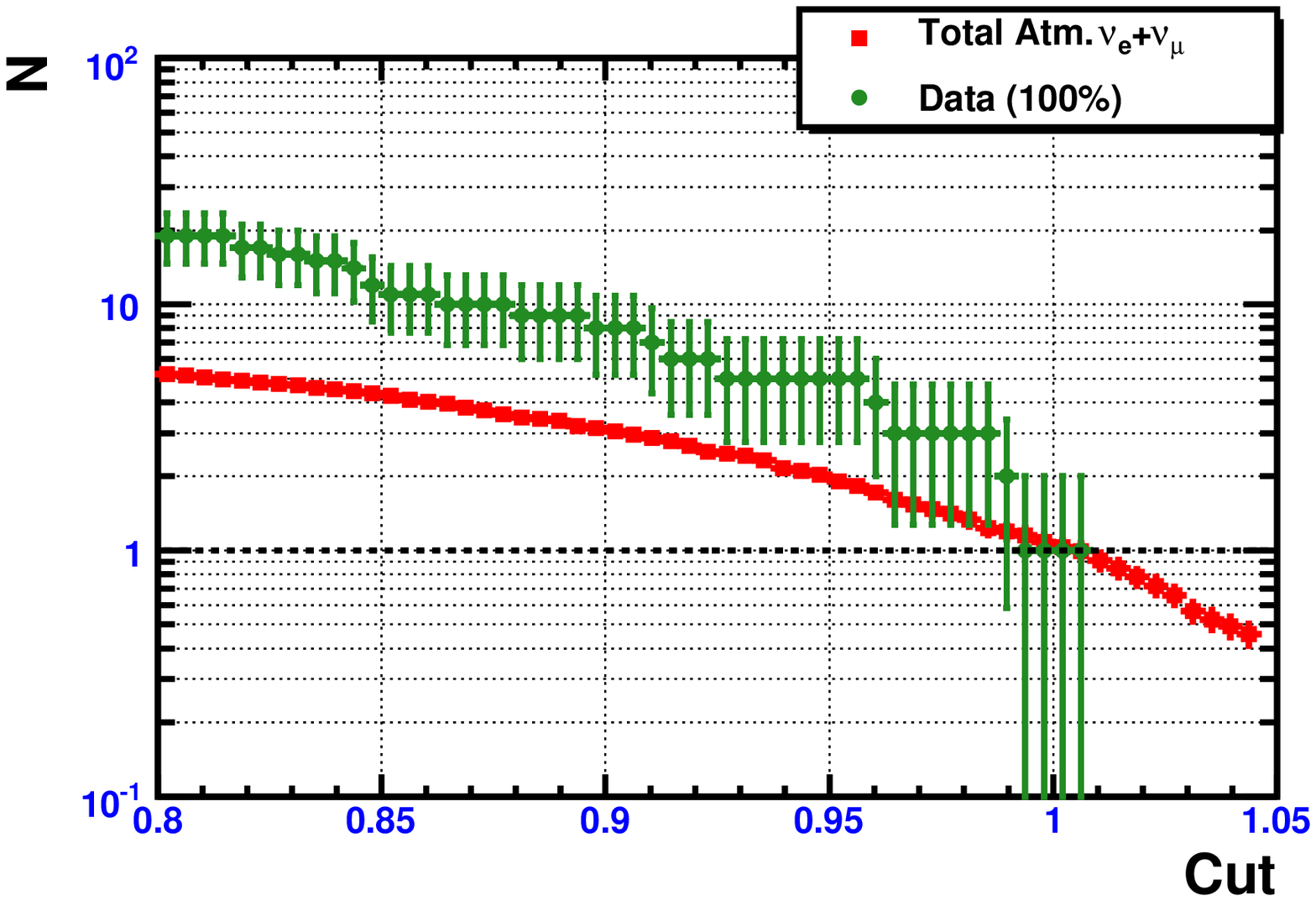}

\vspace{1.2cm}

\centering
\includegraphics[width=0.55\linewidth]{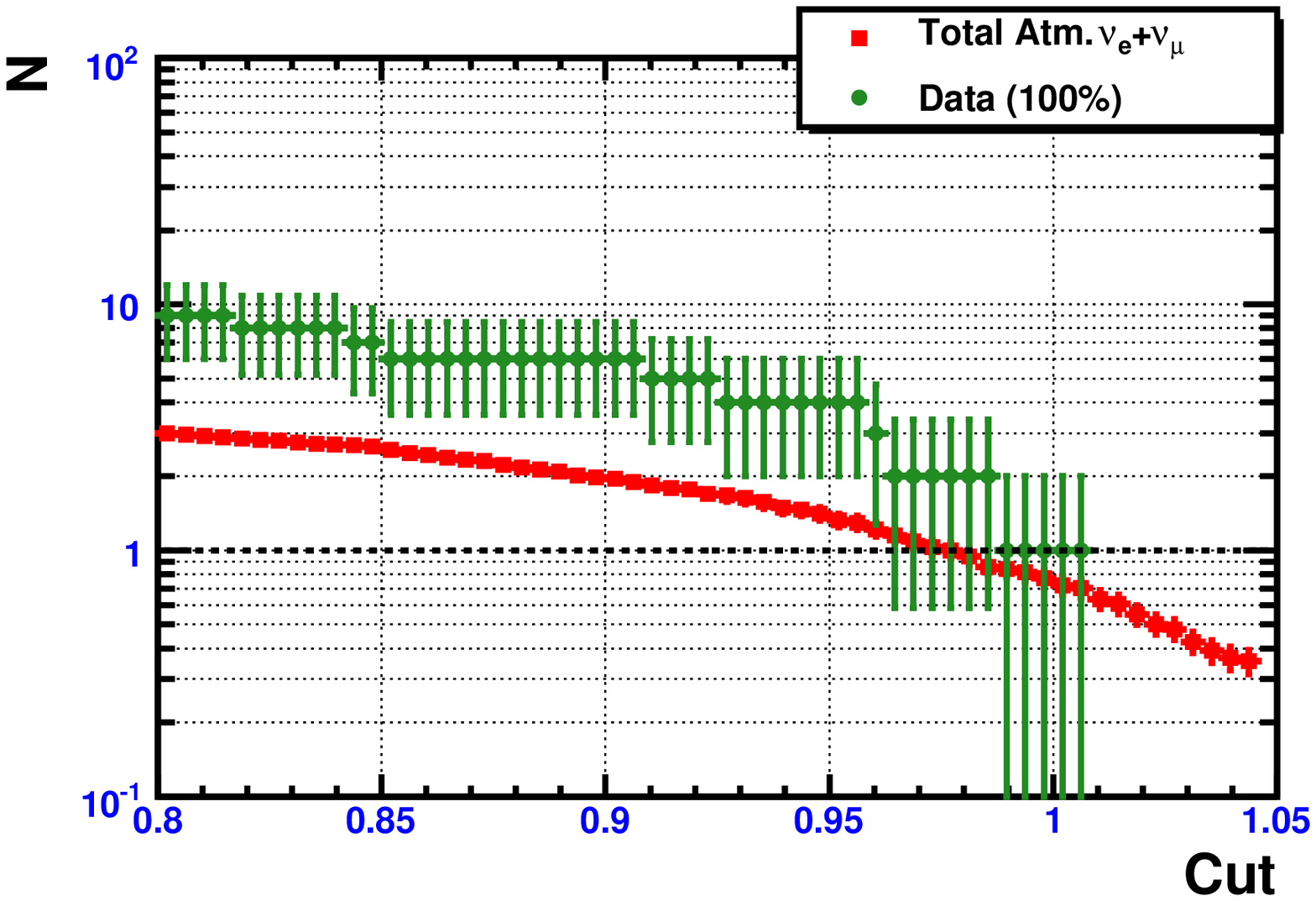}
\vspace{0.8cm}
\caption{Zoomed-in cumulative distribution showing the number of events surviving beyond a given classifier cut for the full IC-22 dataset with a 5 TeV (top), 10 TeV (middle), and 15 TeV (bottom) energy cut.}
\label{UNBLINDEDZoomedPlots}
\end{figure}

Figure~\ref{UNBLINDEDPlots} shows the unblinded distributions and cumulative distributions of the final multivariate cut parameter for reconstructed energy cuts of 5 TeV, 10 TeV, and 15 TeV.  These plots make it clear that the full one year dataset does not transition to the signal prediction around the final cut values anticipated in the last chapter.  However, zooming into the distributions we can see that the data does indeed start to approach the cascade signal prediction around a final multivariate classifier value of 0.9.  Figure~\ref{UNBLINDEDZoomedPlots} shows these zoomed-in cumulative distributions.  Steps in the distributions indicate the loss of a single event as the cut is tightened.

\clearpage 

\newpage

\clearpage

\newpage

\clearpage

\newpage

\section{AMANDA Flux Comparison}

In comparing the observed data to the atmospheric neutrino-induced cascade signal prediction, we need to take into account the uncertainties in the atmospheric neutrino flux.  Up to this point, all plots in this dissertation have used the canonical Bartol atmospheric neutrino flux \cite{BarrBartol1, BarrBartol2} for signal normalization.  In \cite{BarrUncertainties}, the uncertainties in this flux were evaluated up to energies of 1 TeV.  At 1 TeV, the uncertainty, which comes mainly from knowledge of the primaries and of the hadronic interactions, is $40\%$ at $1\sigma$.  Recently, IceCube's predecessor array AMANDA actually measured the atmospheric neutrino flux at the energies relevant for this analysis and found it to be higher than the Bartol flux \cite{JohnKelleyPRD}.  The AMANDA measurements were best fit by a flux given by

$$
\frac{d\Phi_{\mbox{\fontsize{8}{14}\selectfont best-fit}}}{dE} = (1.1 \pm 0.1) \left( \frac{E}{640\mbox{ GeV}} \right)^{0.056} \cdot \frac{d\Phi_{\mbox{\fontsize{8}{14}\selectfont Bartol}}}{dE}
$$

\begin{figure}
\begin{center}
\includegraphics[width=0.65\linewidth]{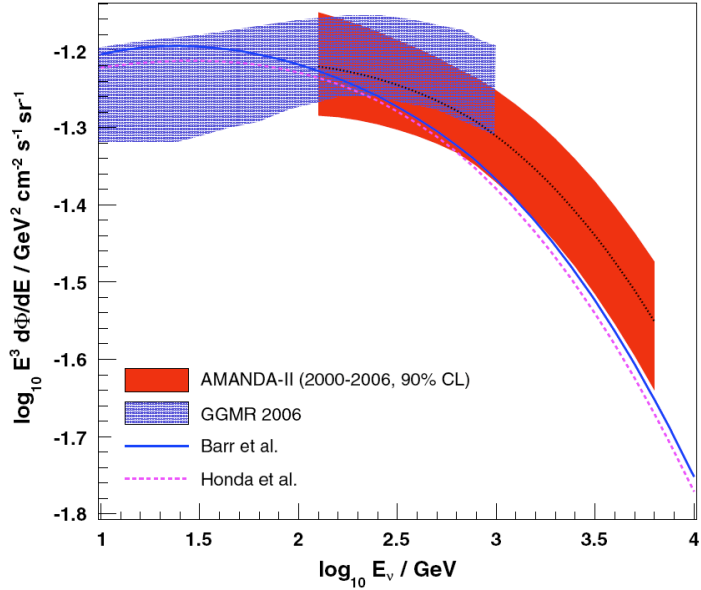}
\caption{The red band shows the best-fit to the AMANDA atmospheric neutrino flux measurement.  In blue, labeled ``Barr et al.", is the canonical Bartol atmospheric neutrino flux.  The blue band is a flux measurement by Super-Kamiokande.  From \cite{JohnKelleyPRD}.}
\label{AMANDAFlux}
\end{center}
\end{figure}

\noindent This flux and its 90\% error region are shown in figure~\ref{AMANDAFlux}.

\begin{figure}
\begin{minipage}[b]{0.48\linewidth} % A minipage that covers half the page
\centering
\includegraphics[width=7.6cm]{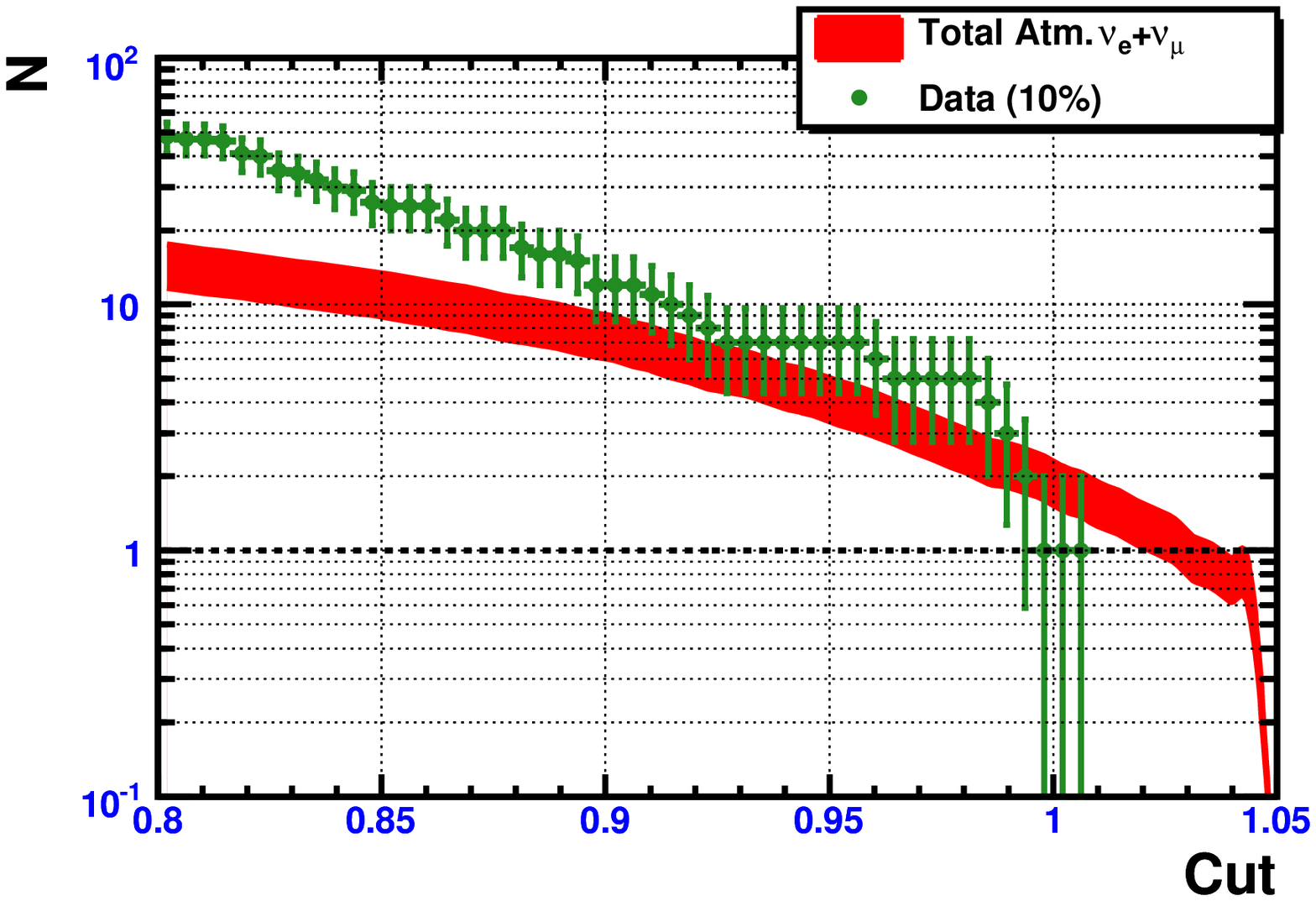}
\end{minipage}
\hspace{0.65cm} %To get a little bit of space between the figures
\begin{minipage}[b]{0.46\linewidth}
\centering
\includegraphics[width=7.6cm]{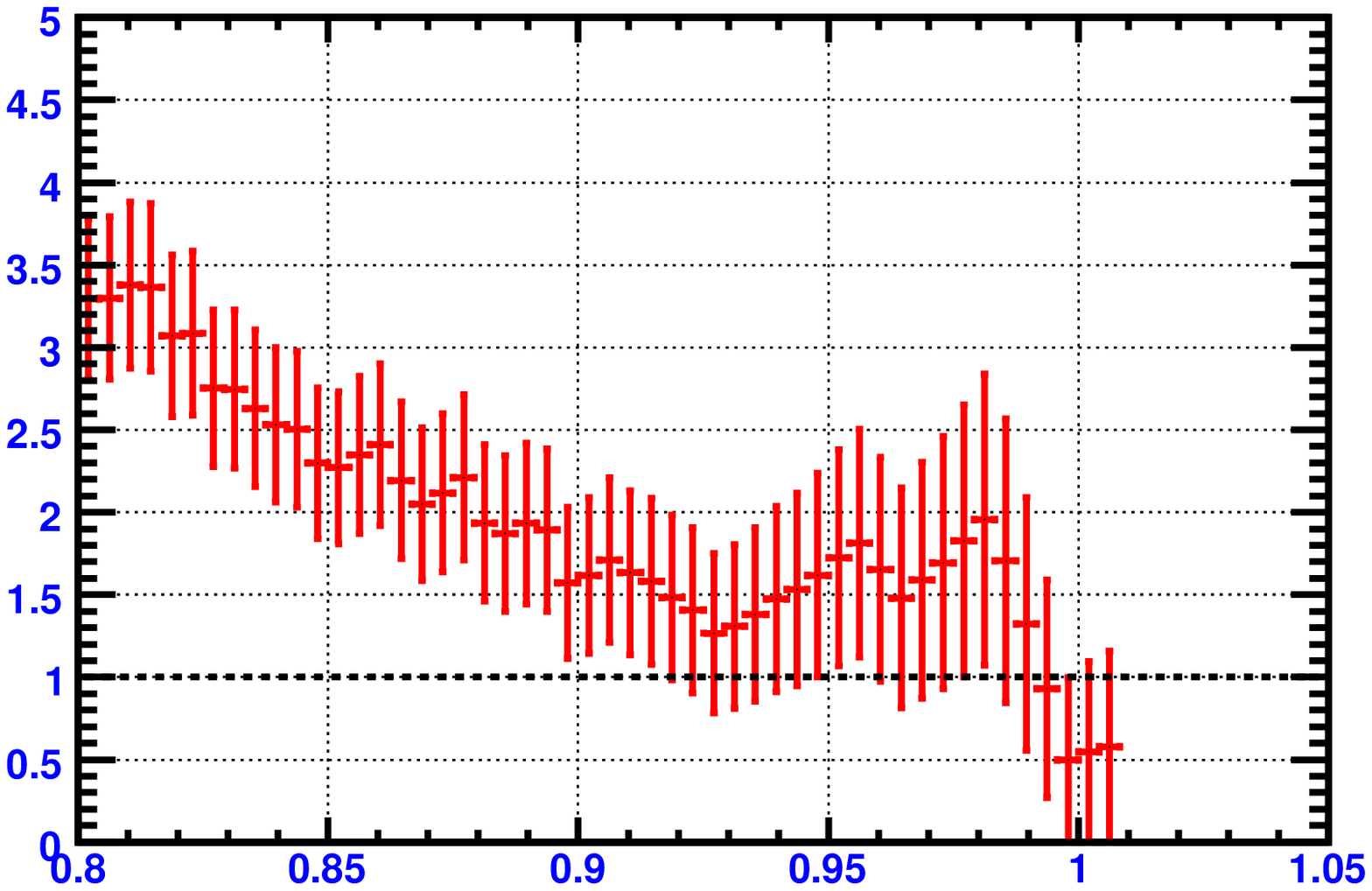}
\end{minipage}

\vspace{1.0cm}

\begin{minipage}[b]{0.48\linewidth} % A minipage that covers half the page
\centering
\includegraphics[width=7.6cm]{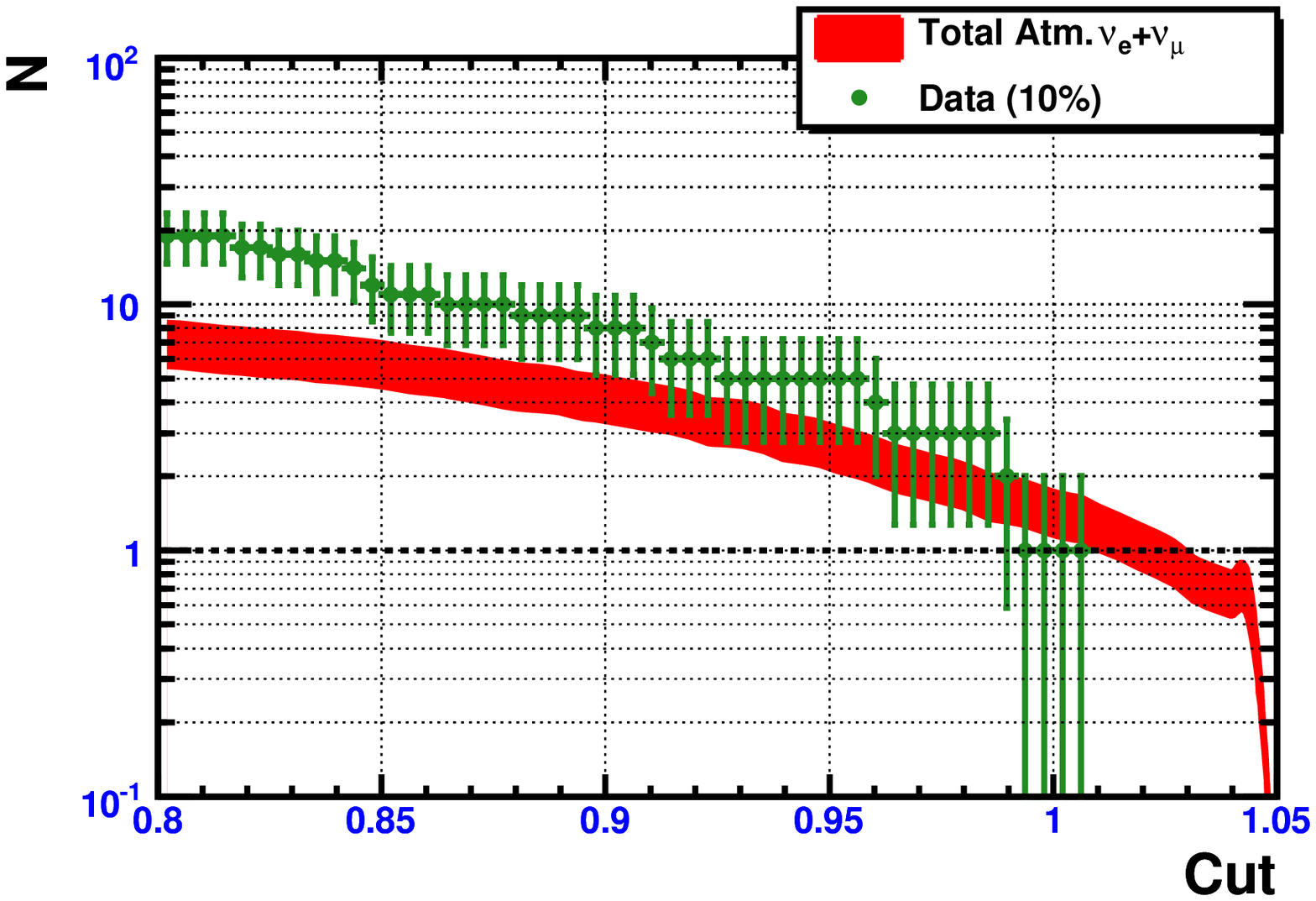}
\end{minipage}
\hspace{0.65cm} %To get a little bit of space between the figures
\begin{minipage}[b]{0.46\linewidth}
\centering
\includegraphics[width=7.6cm]{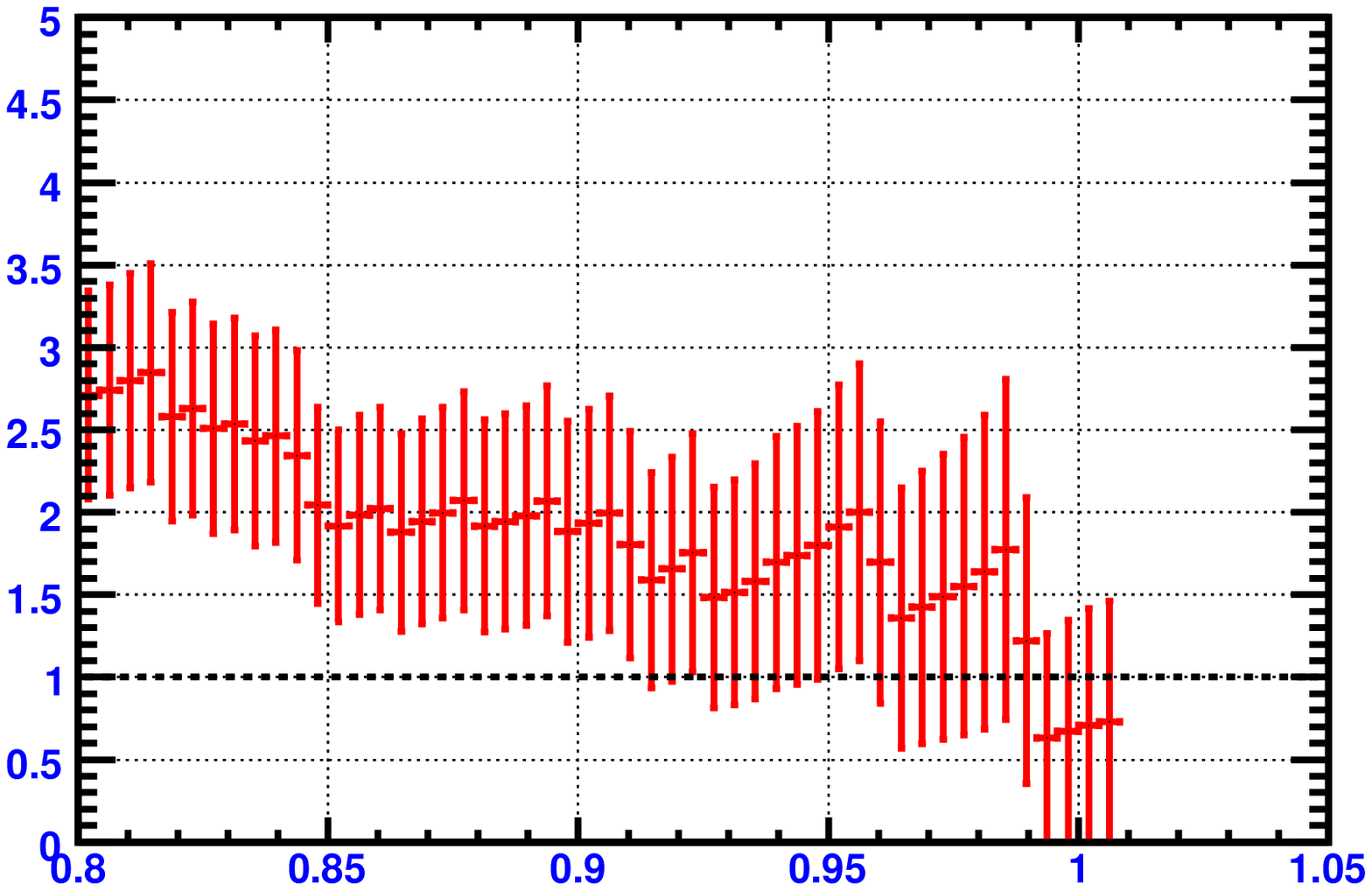}
\end{minipage}

\vspace{1.0cm}

\begin{minipage}[b]{0.48\linewidth} % A minipage that covers half the page
\centering
\includegraphics[width=7.6cm]{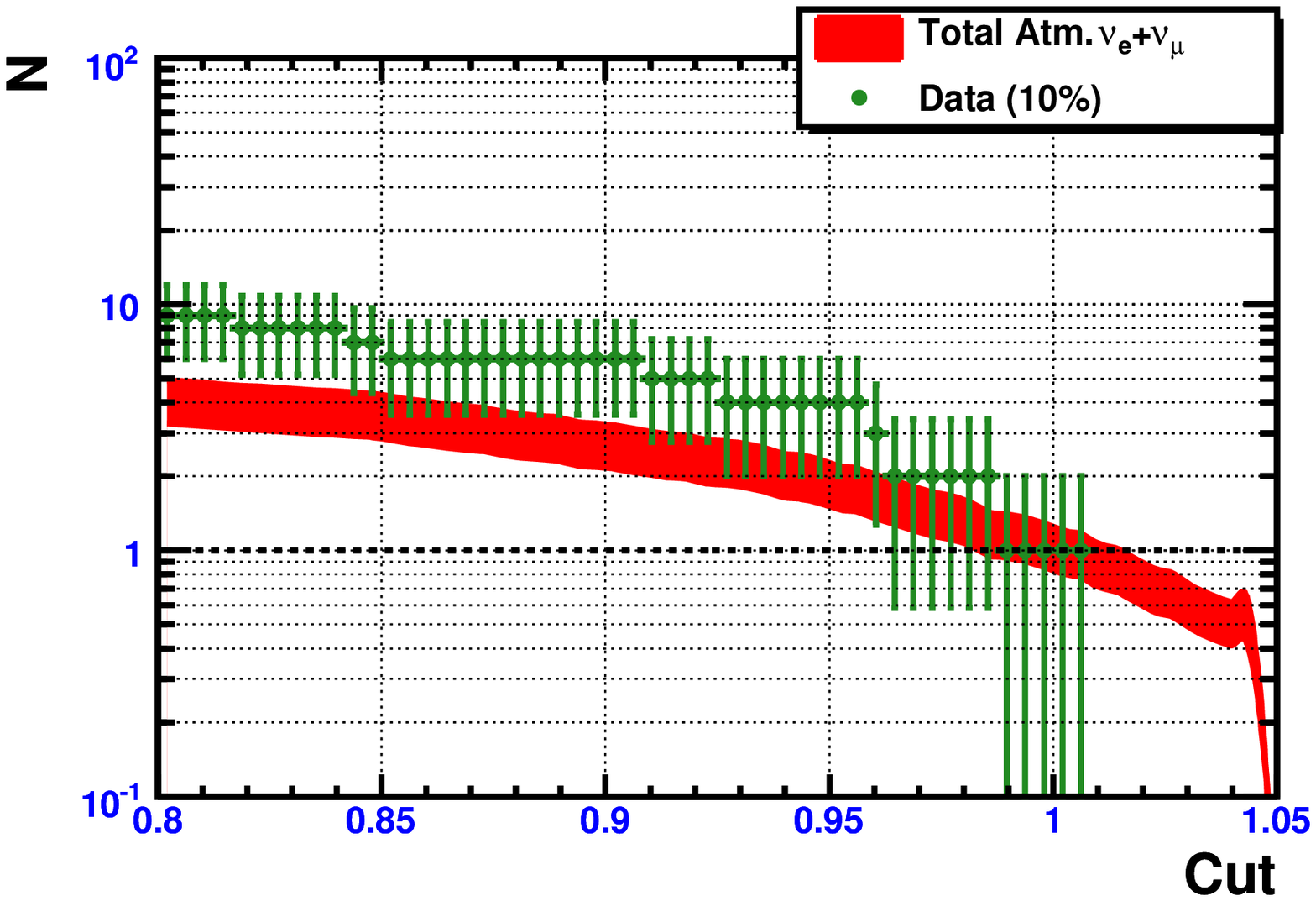}
\end{minipage}
\hspace{0.65cm} %To get a little bit of space between the figures
\begin{minipage}[b]{0.46\linewidth}
\centering
\includegraphics[width=7.6cm]{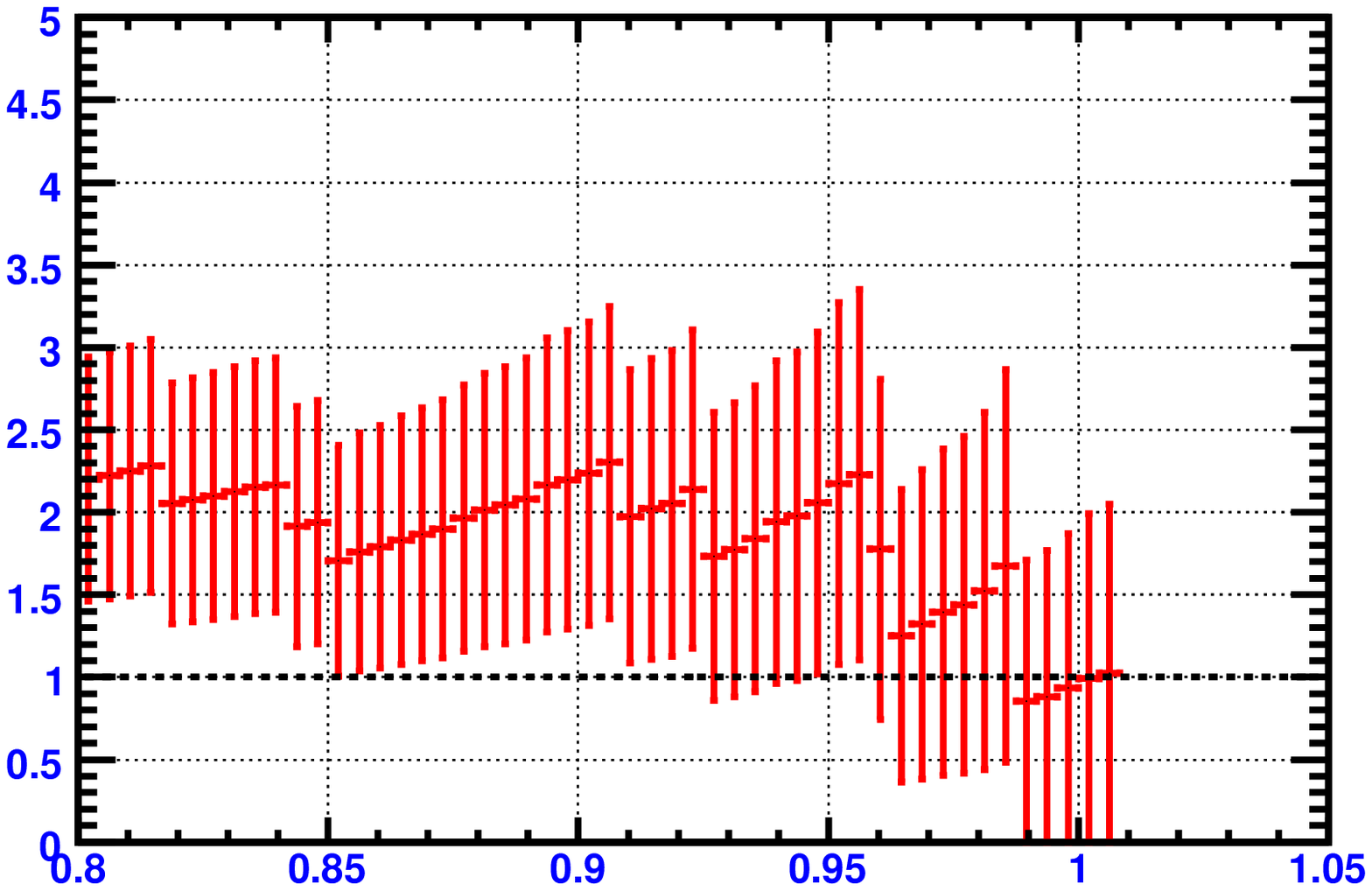}
\end{minipage}
\vspace{0.25cm}
\caption{Zoomed-in cumulative distribution (left) showing the number of events surviving beyond a given classifier cut for the full IC-22 dataset and the signal expectation from the measured AMANDA atmospheric neutrino flux with a 5 TeV (top), 10 TeV (middle), and 15 TeV (bottom) energy cut with   Also displayed is the ratio of data to the neutrino signal prediction (right).}
\label{UNBLINDEDAMAPlots}
\end{figure}

Figure~\ref{UNBLINDEDAMAPlots} shows the zoomed-in cumulative distributions for the full one-year dataset compared to the signal prediction using the best-fit AMANDA atmospheric neutrino flux measurement.  The $90\%$ confidence region is approximated by a $\pm20\%$ error on the best-fit flux.  The ratio of the data to the signal prediction is also displayed.  From these plots we can conclude that the data is much more consistent with the measured AMANDA atmospheric neutrino flux than with the canonical Bartol flux.  The data-to-signal ratios at high cut value are generally 2 or better, indicating a signal to background ratio $\sim1$.

\clearpage

\newpage

\clearpage

\newpage

\section{Summary Tables}

Tables~\ref{SummaryTable5TeV}--\ref{SummaryTable15TeV} summarize the observed and predicted event numbers by class for the Bartol and AMANDA fluxes.  The Bartol flux predictions have at least a $\pm40\%$ error at $1\sigma$ and the AMANDA flux predictions have a $\pm20\%$ error at $90\%$ confidence which are not displayed in the tables.

\vspace{2.0cm}

\begin{table}[h]
\caption{Summary table of observed and expected event numbers above 5 TeV.}
\vspace{0.25cm}
\centering
\begin{tabular}{ | c | c | c | c | c | c | c | c | c | c |  }
\hline
&  & \multicolumn{4}{  c | }{Bartol Flux} & \multicolumn{4}{  c | }{AMANDA Flux}  \\
\hline
Cut & {\bf \color{red} Observed} & $\numu \mbox{ CC}$ & $\numu \mbox{ NC}$ & $\nue\mbox{ }$ & {\bf \color{red} Total $\nu\mbox{ }$} & $\numu \mbox{ CC}$ & $\numu \mbox{ NC}$ & $\nue\mbox{ }$ & {\bf \color{red} Total $\nu\mbox{ }$} \\
\hline
0.90 & {\bf \color{red} 12} & 2.87 & 1.61 & 1.15 & {\bf \color{red} 5.63} & 3.82 & 2.15 & 1.50 & {\bf \color{red} 7.48} \\
\hline
0.91 & {\bf \color{red} 11} & 2.61 & 1.45 & 1.05 & {\bf \color{red} 5.11} & 3.48 & 1.94 & 1.37 & {\bf \color{red} 6.80} \\
\hline
0.92 & {\bf \color{red} 8} & 2.26 & 1.27 & 0.95 & {\bf \color{red} 4.49} & 3.02 & 1.71 & 1.25 & {\bf \color{red} 5.98} \\
\hline
0.93 & {\bf \color{red} 7} & 2.06 & 1.14 & 0.85 & {\bf \color{red} 4.05} & 2.75 & 1.53 & 1.12 & {\bf \color{red} 5.41} \\
\hline
0.94 & {\bf \color{red} 7} & 1.82 & 0.96 & 0.77 & {\bf \color{red} 3.55} & 2.44 & 1.29 & 1.01 & {\bf \color{red} 4.74} \\
\hline
0.95 & {\bf \color{red} 7} & 1.60 & 0.85 & 0.68 & {\bf \color{red} 3.13} & 2.15 & 1.16 & 0.89 & {\bf \color{red} 4.20} \\
\hline
0.96 & {\bf \color{red} 6} & 1.38 & 0.74 & 0.59 & {\bf \color{red} 2.71} & 1.86 & 1.00 & 0.79 & {\bf \color{red} 3.65} \\
\hline
0.97 & {\bf \color{red} 5} & 1.13 & 0.66 & 0.52 & {\bf \color{red} 2.31} & 1.53 & 0.90 & 0.68 & {\bf \color{red} 3.11} \\
\hline
0.98 & {\bf \color{red} 5} & 0.92 & 0.56 & 0.44 & {\bf \color{red} 1.92} & 1.25 & 0.77 & 0.59 & {\bf \color{red} 2.60} \\
\hline
0.99 & {\bf \color{red} 3} & 0.77 & 0.52 & 0.38 & {\bf \color{red} 1.67} & 1.05 & 0.71 & 0.50 & {\bf \color{red} 2.26} \\
\hline
1.0 & {\bf \color{red} 1} & 0.67 & 0.43 & 0.32 & {\bf \color{red} 1.42} & 0.90 & 0.59 & 0.43 & {\bf \color{red} 1.93} \\
\hline
\end{tabular}
\label{SummaryTable5TeV}
\end{table}

\clearpage

\newpage

\begin{table}
\caption{Summary table of observed and expected event numbers above 10 TeV.}
\vspace{0.25cm}
\centering
\begin{tabular}{ | c | c | c | c | c | c | c | c | c | c |  }
\hline
&  & \multicolumn{4}{  c | }{Bartol Flux} & \multicolumn{4}{  c | }{AMANDA Flux}  \\
\hline
Cut & {\bf \color{red} Observed} & $\numu \mbox{ CC}$ & $\numu \mbox{ NC}$ & $\nue\mbox{ }$ & {\bf \color{red} Total $\nu\mbox{ }$} & $\numu \mbox{ CC}$ & $\numu \mbox{ NC}$ & $\nue\mbox{ }$ & {\bf \color{red} Total $\nu\mbox{ }$} \\
\hline
0.90 & {\bf \color{red} 8} & 1.52 & 0.92 & 0.63 & {\bf \color{red} 3.07} & 2.07 & 1.25 & 0.85 & {\bf \color{red} 4.16} \\
\hline
0.91 & {\bf \color{red} 7} & 1.40 & 0.90 & 0.60 & {\bf \color{red} 2.90} & 1.91 & 1.22 & 0.80 & {\bf \color{red} 3.93} \\
\hline
0.92 & {\bf \color{red} 6} & 1.28 & 0.82 & 0.56 & {\bf \color{red} 2.65} & 1.74 & 1.11 & 0.74 & {\bf \color{red} 3.60} \\
\hline
0.93 & {\bf \color{red} 5} & 1.21 & 0.73 & 0.51 & {\bf \color{red} 2.45} & 1.65 & 1.00 & 0.69 & {\bf \color{red} 3.34} \\
\hline
0.94 & {\bf \color{red} 5} & 1.07 & 0.61 & 0.47 & {\bf \color{red} 2.16} & 1.47 & 0.84 & 0.63 & {\bf \color{red} 2.95} \\
\hline
0.95 & {\bf \color{red} 5} & 0.99 & 0.58 & 0.43 & {\bf \color{red} 2.00} & 1.36 & 0.79 & 0.58 & {\bf \color{red} 2.73} \\
\hline
0.96 & {\bf \color{red} 4} & 0.84 & 0.50 & 0.39 & {\bf \color{red} 1.73} & 1.15 & 0.69 & 0.52 & {\bf \color{red} 2.37} \\
\hline
0.97 & {\bf \color{red} 3} & 0.70 & 0.47 & 0.35 & {\bf \color{red} 1.52} & 0.96 & 0.65 & 0.47 & {\bf \color{red} 2.09} \\
\hline
0.98 & {\bf \color{red} 3} & 0.61 & 0.42 & 0.31 & {\bf \color{red} 1.34} & 0.84 & 0.58 & 0.42 & {\bf \color{red} 1.84} \\
\hline
0.99 & {\bf \color{red} 2} & 0.52 & 0.40 & 0.27 & {\bf \color{red} 1.19} & 0.72 & 0.55 & 0.37 & {\bf \color{red} 1.63} \\
\hline
1.0 & {\bf \color{red} 1} & 0.45 & 0.36 & 0.24 & {\bf \color{red} 1.05} & 0.63 & 0.50 & 0.32 & {\bf \color{red} 1.45} \\
\hline
\end{tabular}
\label{SummaryTable10TeV}
\end{table}

\clearpage

\newpage

\begin{table}
\caption{Summary table of observed and expected event numbers above 15 TeV.}
\vspace{0.25cm}
\centering
\begin{tabular}{ | c | c | c | c | c | c | c | c | c | c |  }
\hline
&  & \multicolumn{4}{  c | }{Bartol Flux} & \multicolumn{4}{  c | }{AMANDA Flux}  \\
\hline
Cut & {\bf \color{red} Observed} & $\numu \mbox{ CC}$ & $\numu \mbox{ NC}$ & $\nue\mbox{ }$ & {\bf \color{red} Total $\nu\mbox{ }$} & $\numu \mbox{ CC}$ & $\numu \mbox{ NC}$ & $\nue\mbox{ }$ & {\bf \color{red} Total $\nu\mbox{ }$} \\
\hline 
0.90 & {\bf \color{red} 6} & 0.96 & 0.62 & 0.39 & {\bf \color{red} 1.97} & 1.33 & 0.85 & 0.53 & {\bf \color{red} 2.71} \\
\hline
0.91 & {\bf \color{red} 5} & 0.88 & 0.60 & 0.37 & {\bf \color{red} 1.85} & 1.22 & 0.83 & 0.50 & {\bf \color{red} 2.55} \\
\hline
0.92 & {\bf \color{red} 5} & 0.84 & 0.57 & 0.34 & {\bf \color{red} 1.76} & 1.16 & 0.79 & 0.47 & {\bf \color{red} 2.42} \\
\hline
0.93 & {\bf \color{red} 4} & 0.79 & 0.55 & 0.32 & {\bf \color{red} 1.66} & 1.09 & 0.75 & 0.44 & {\bf \color{red} 2.28} \\
\hline
0.94 & {\bf \color{red} 4} & 0.74 & 0.45 & 0.30 & {\bf \color{red} 1.49} & 1.02 & 0.63 & 0.41 & {\bf \color{red} 2.06} \\
\hline
0.95 & {\bf \color{red} 4} & 0.68 & 0.43 & 0.28 & {\bf \color{red} 1.39} & 0.95 & 0.60 & 0.38 & {\bf \color{red} 1.92} \\
\hline
0.96 & {\bf \color{red} 3} & 0.59 & 0.38 & 0.26 & {\bf \color{red} 1.22} & 0.82 & 0.52 & 0.35 & {\bf \color{red} 1.70} \\
\hline
0.97 & {\bf \color{red} 2} & 0.50 & 0.35 & 0.23 & {\bf \color{red} 1.08} & 0.70 & 0.49 & 0.32 & {\bf \color{red} 1.50} \\
\hline
0.98 & {\bf \color{red} 2} & 0.43 & 0.31 & 0.21 & {\bf \color{red} 0.95} & 0.60 & 0.43 & 0.29 & {\bf \color{red} 1.32} \\
\hline
0.99 & {\bf \color{red} 1} & 0.36 & 0.30 & 0.18 & {\bf \color{red} 0.84} & 0.50 & 0.41 & 0.25 & {\bf \color{red} 1.17} \\
\hline
1.0 & {\bf \color{red} 1} & 0.32 & 0.26 & 0.16 & {\bf \color{red} 0.74} & 0.45 & 0.37 & 0.22 & {\bf \color{red} 1.03} \\
\hline
\end{tabular}
\label{SummaryTable15TeV}
\end{table}

\clearpage

\newpage

We can get a rough, intuitive feel for the relative ratios of the different neutrino-induced cascade signal classes displayed in the previous tables with some simple arguments.  To have the same visible energy as a given electron neutrino charged-current event, a muon neutrino neutral-current event must come from a higher energy neutrino because of the hadronic visible energy correction (see section~\ref{SHadronicCascades}).  This correction is around 0.8 at 1 TeV, so the muon neutrino must be a factor 1.25 higher in energy.  The flux falls as $E^{-3.7}$, but the muon neutrino flux is a factor $\sim20$ higher in normalization at these energies.  In addition,  the neutral-current cross section is 1/3 of the charged-current cross section.  Putting all these factors together, we expect roughly 2 times more muon neutrino neutral-current events than electron neutrino charged-current events.  The ratio from the full Monte Carlo is $\sim$1.4, which confirms this.

In the same fiducial volume, muon neutrino charged-current events should occur 3 times more frequently than muon neutrino neutral-current events because of the ratio of cross sections.  However, this is an upper bound, since many of these charged-current events will have bright outgoing muons and so will not survive our cascade cuts.  The ratio from the full Monte Carlo is around 1.8, which again makes sense given our rough estimate.

\section{Event Images and Waveforms}

\begin{figure}
\begin{minipage}[b]{0.48\linewidth} % A minipage that covers half the page
\centering
\includegraphics[width=7.6cm]{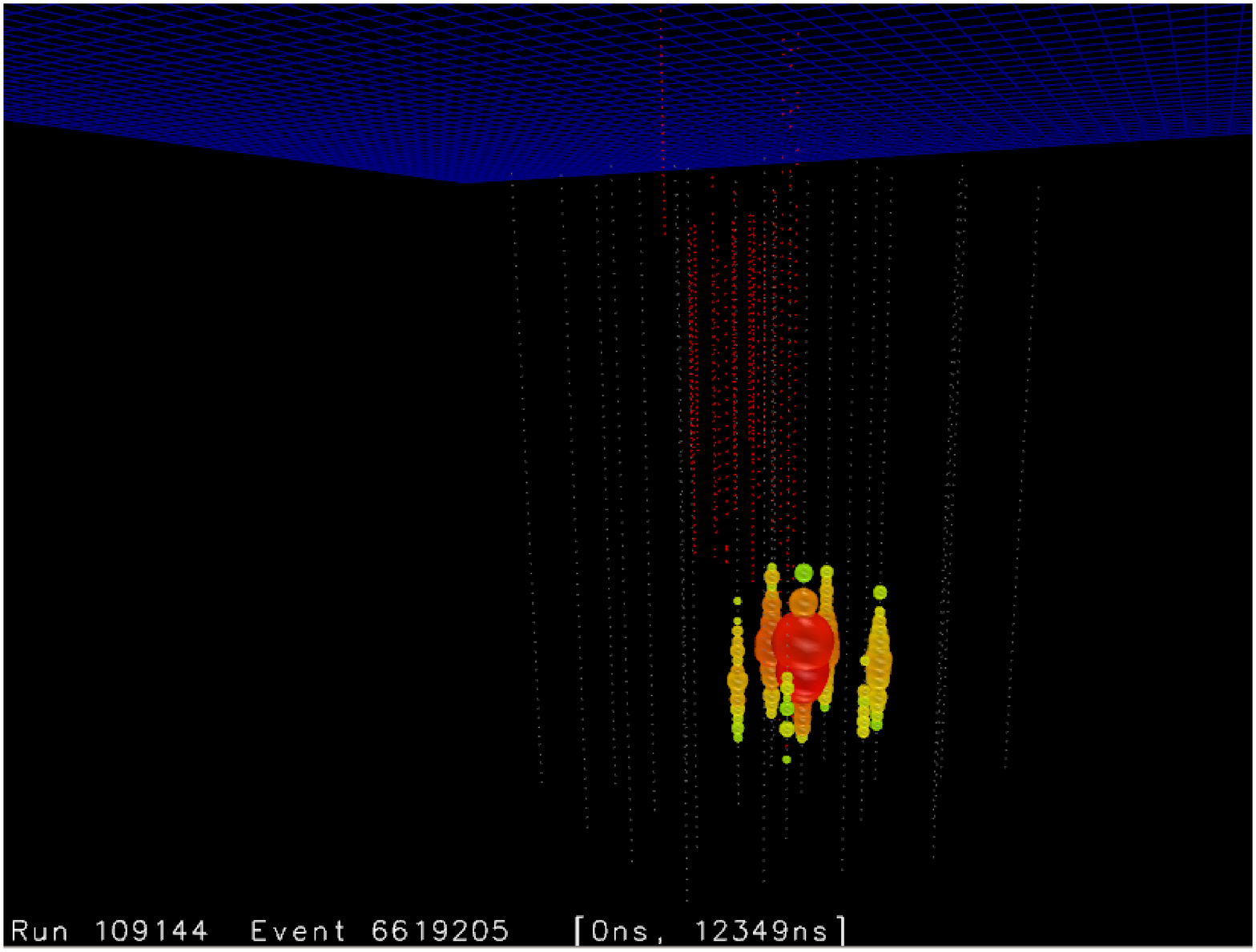}
\end{minipage}
\hspace{0.5cm} %To get a little bit of space between the figures
\begin{minipage}[b]{0.48\linewidth}
\centering
\includegraphics[width=7.6cm]{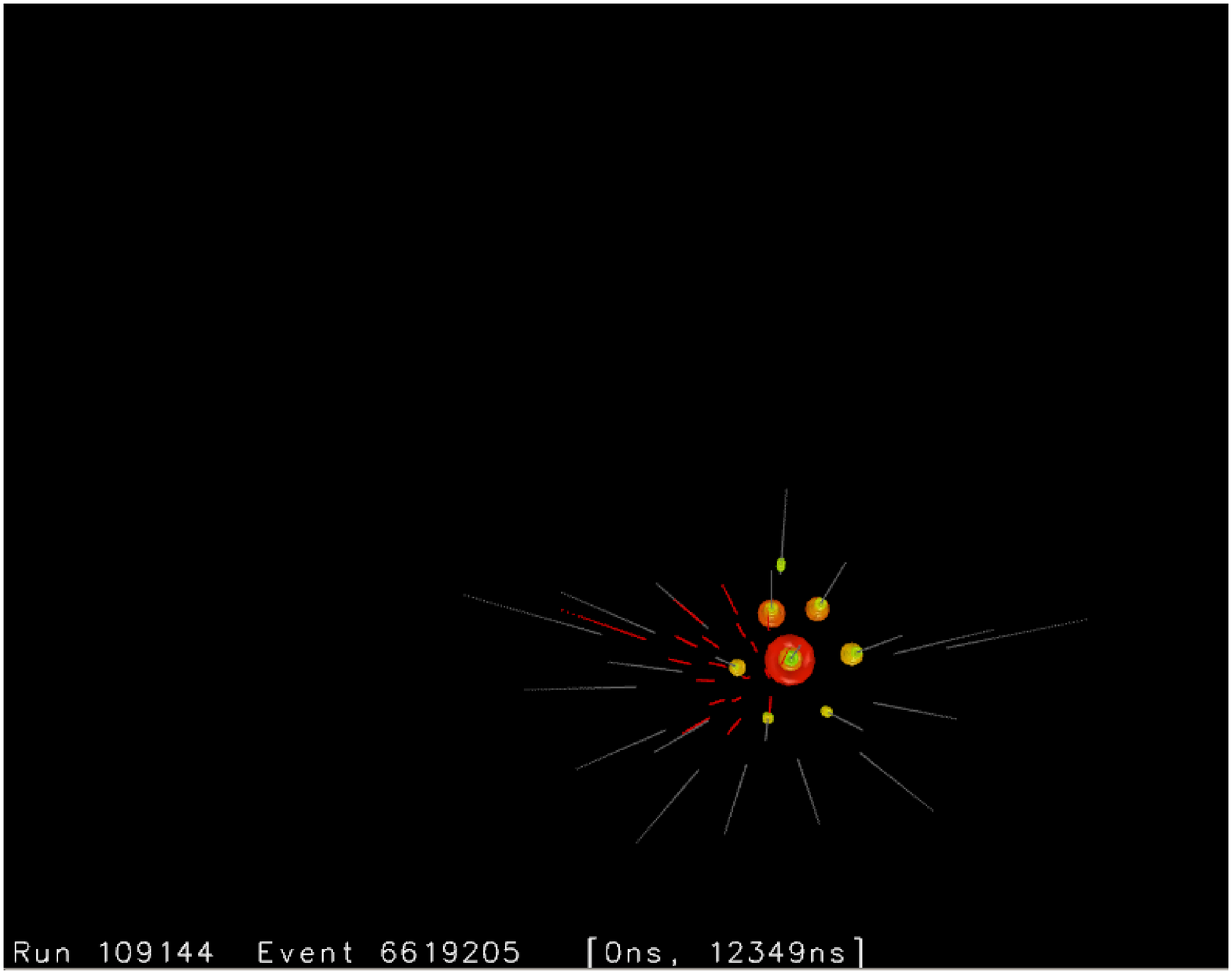}
\end{minipage}
\vspace{0.25cm}
\caption{Run 109144, Event 6619205, E=26.9 TeV}
\label{Run109144Image}
\end{figure}

\begin{figure}
\begin{minipage}[b]{0.48\linewidth} % A minipage that covers half the page
\centering
\includegraphics[width=7.6cm]{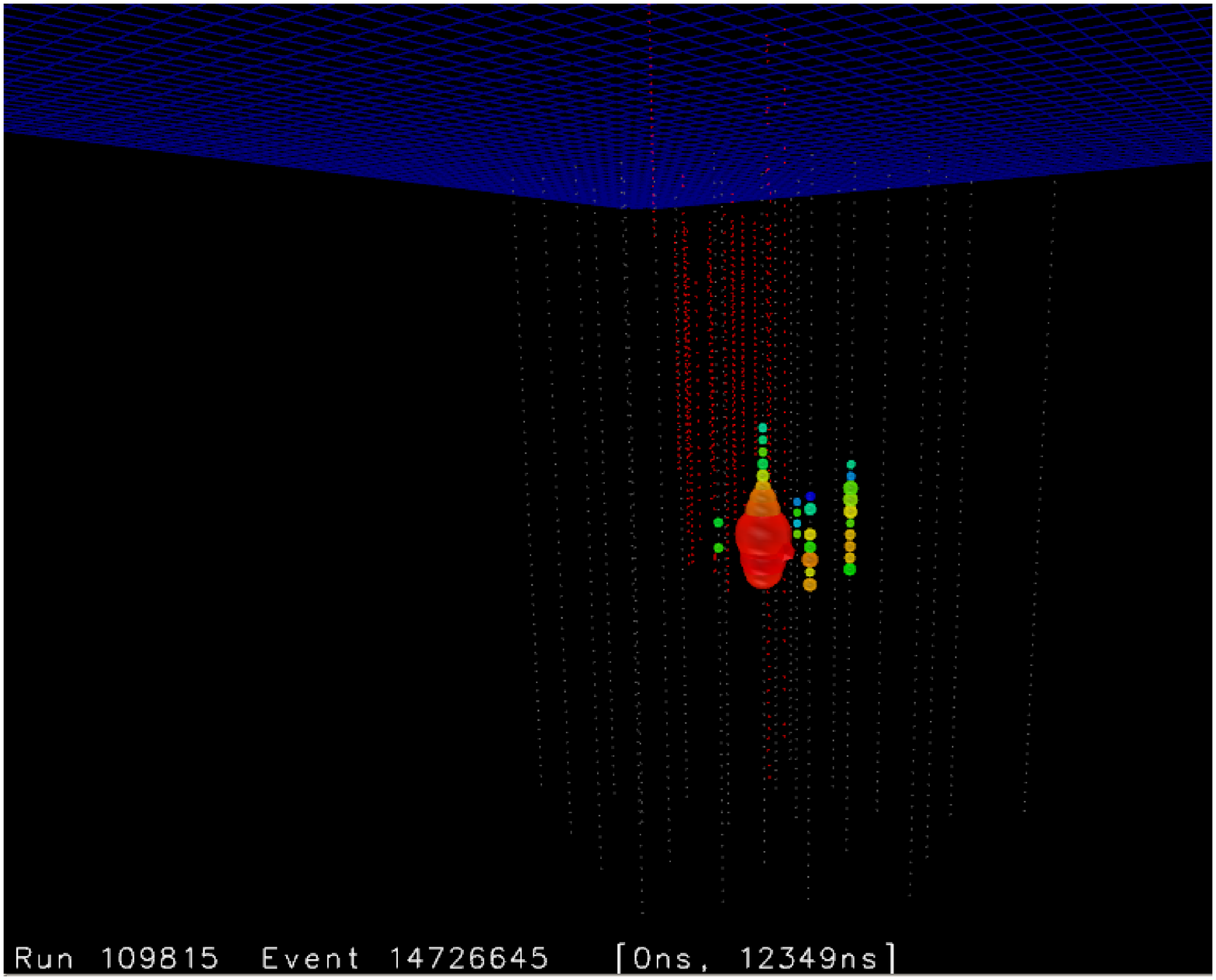}
\end{minipage}
\hspace{0.5cm} %To get a little bit of space between the figures
\begin{minipage}[b]{0.48\linewidth}
\centering
\includegraphics[width=7.6cm]{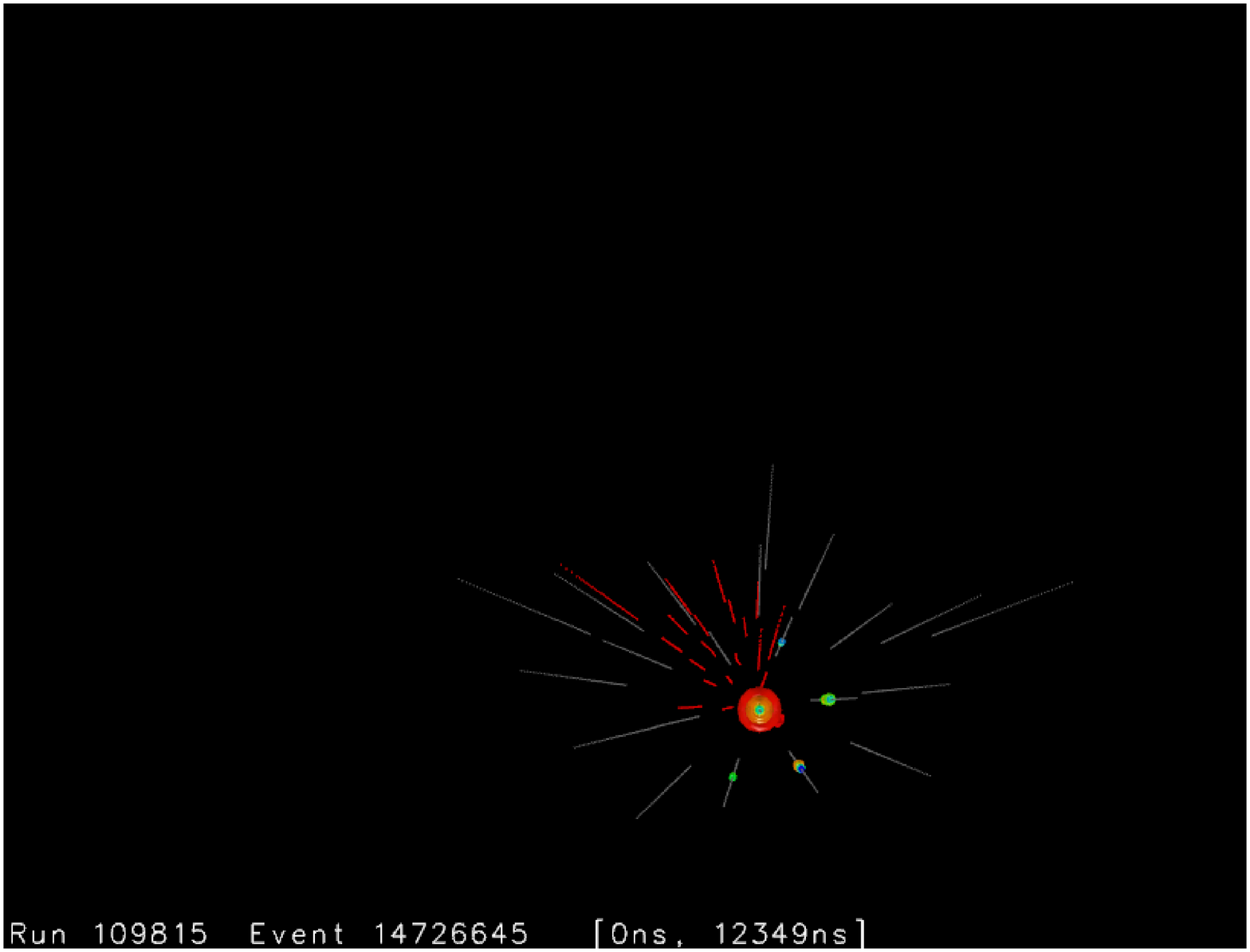}
\end{minipage}
\vspace{0.25cm}
\caption{Run 109815, Event 14726645, E=11.2 TeV}
\label{Run109815Image}
\end{figure}

\begin{figure}
\begin{minipage}[b]{0.48\linewidth} % A minipage that covers half the page
\centering
\includegraphics[width=7.6cm]{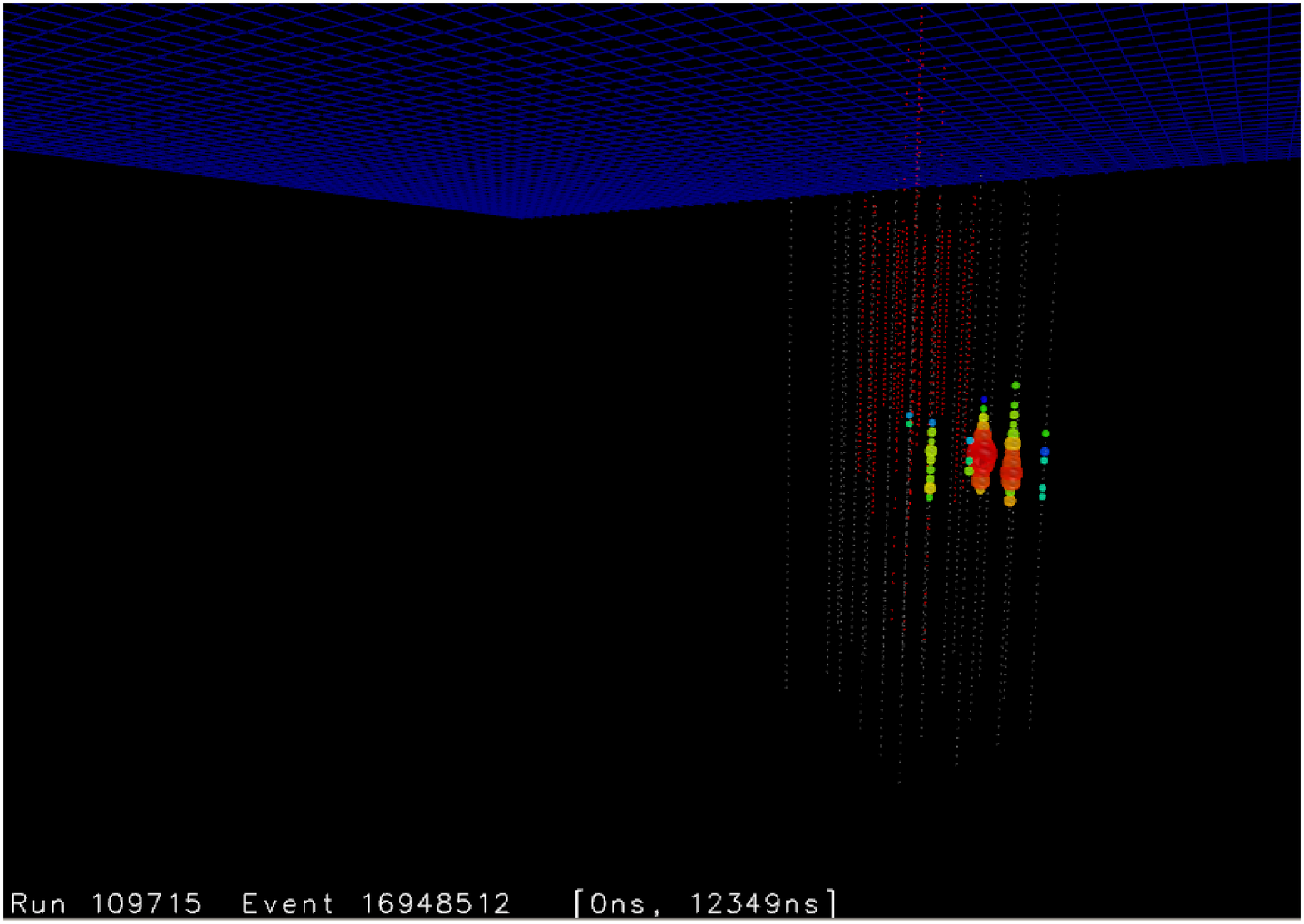}
\end{minipage}
\hspace{0.5cm} %To get a little bit of space between the figures
\begin{minipage}[b]{0.48\linewidth}
\centering
\includegraphics[width=7.6cm]{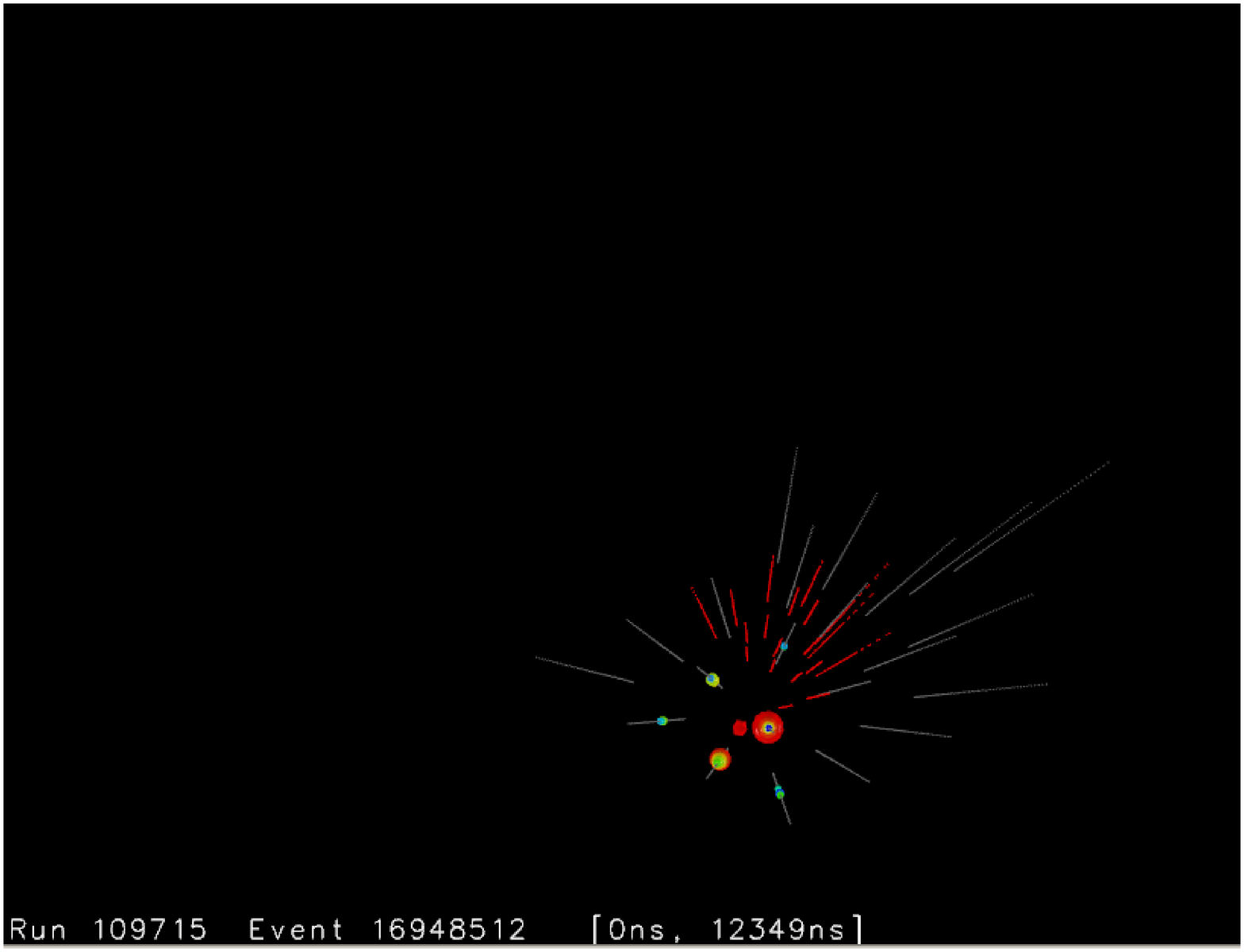}
\end{minipage}
\vspace{0.25cm}
\caption{Run 109715, Event 16948512, E=5.4 TeV}
\label{Run109715Image}
\end{figure}

\begin{figure}
\begin{minipage}[b]{0.48\linewidth} % A minipage that covers half the page
\centering
\includegraphics[width=7.6cm]{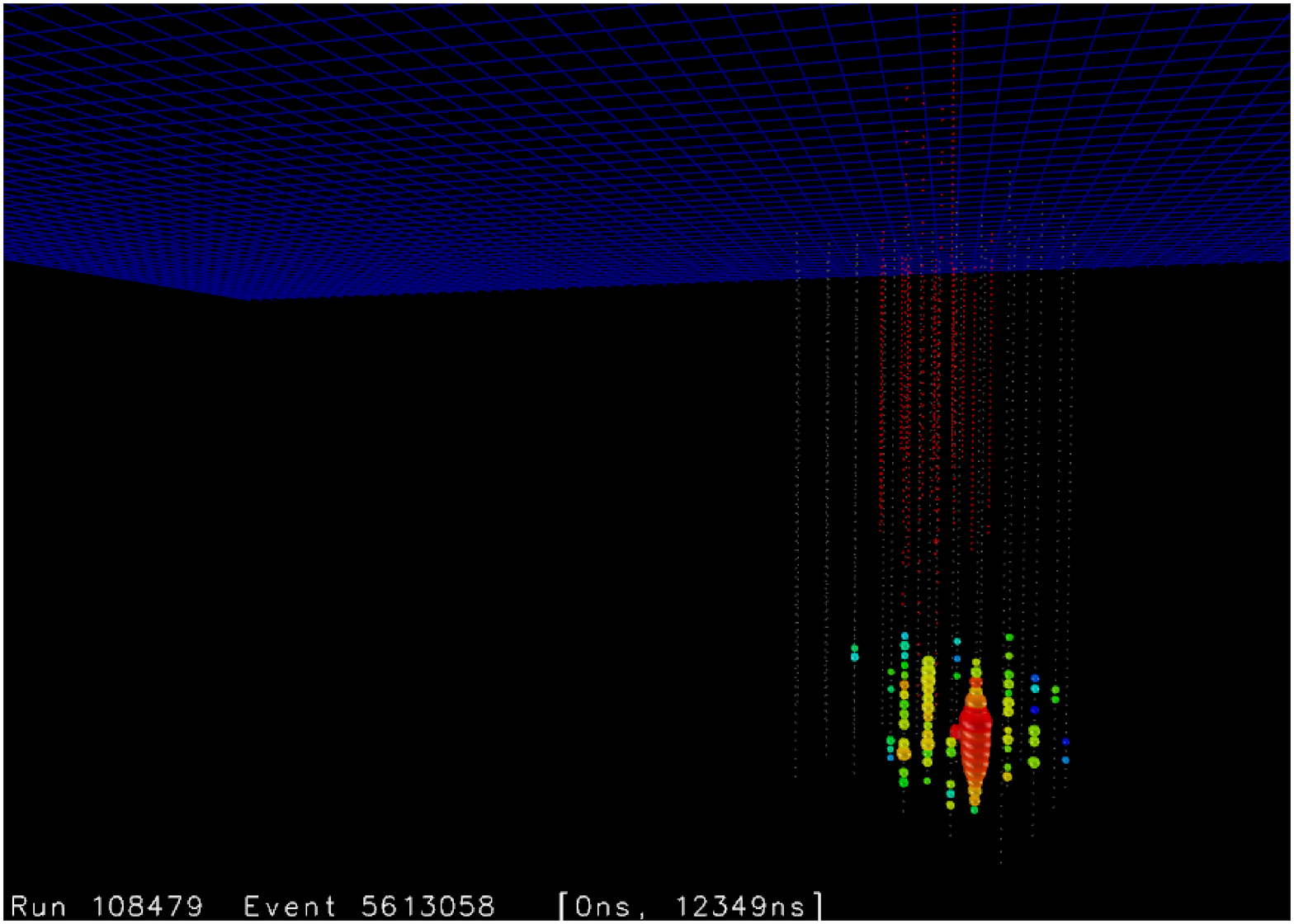}
\end{minipage}
\hspace{0.5cm} %To get a little bit of space between the figures
\begin{minipage}[b]{0.48\linewidth}
\centering
\includegraphics[width=7.6cm]{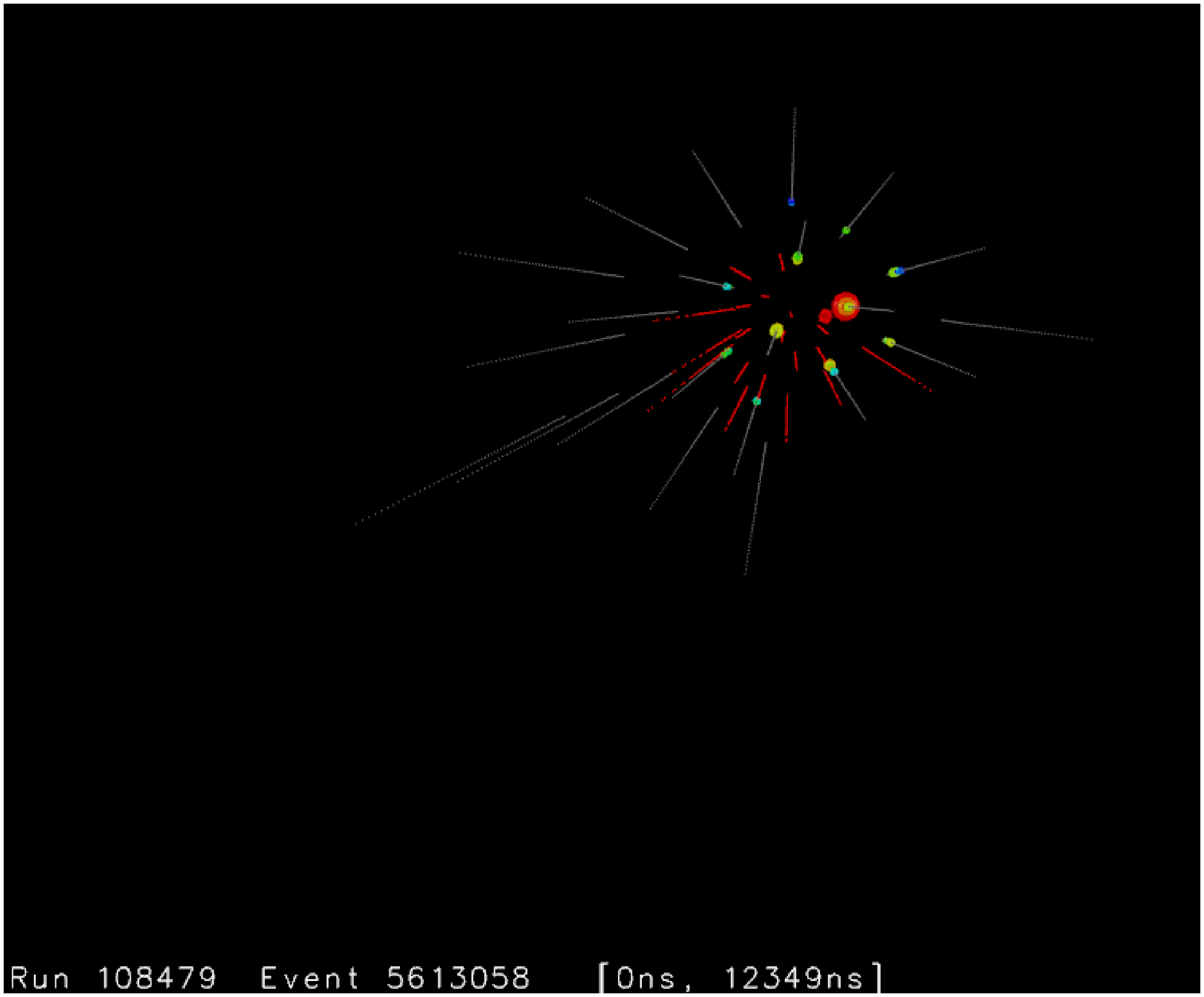}
\end{minipage}
\vspace{0.25cm}
\caption{Run 108479, Event 5613058, E=5.3 TeV}
\label{Run108479Image}
\end{figure}

\begin{figure}
\begin{minipage}[b]{0.48\linewidth} % A minipage that covers half the page
\centering
\includegraphics[width=7.6cm]{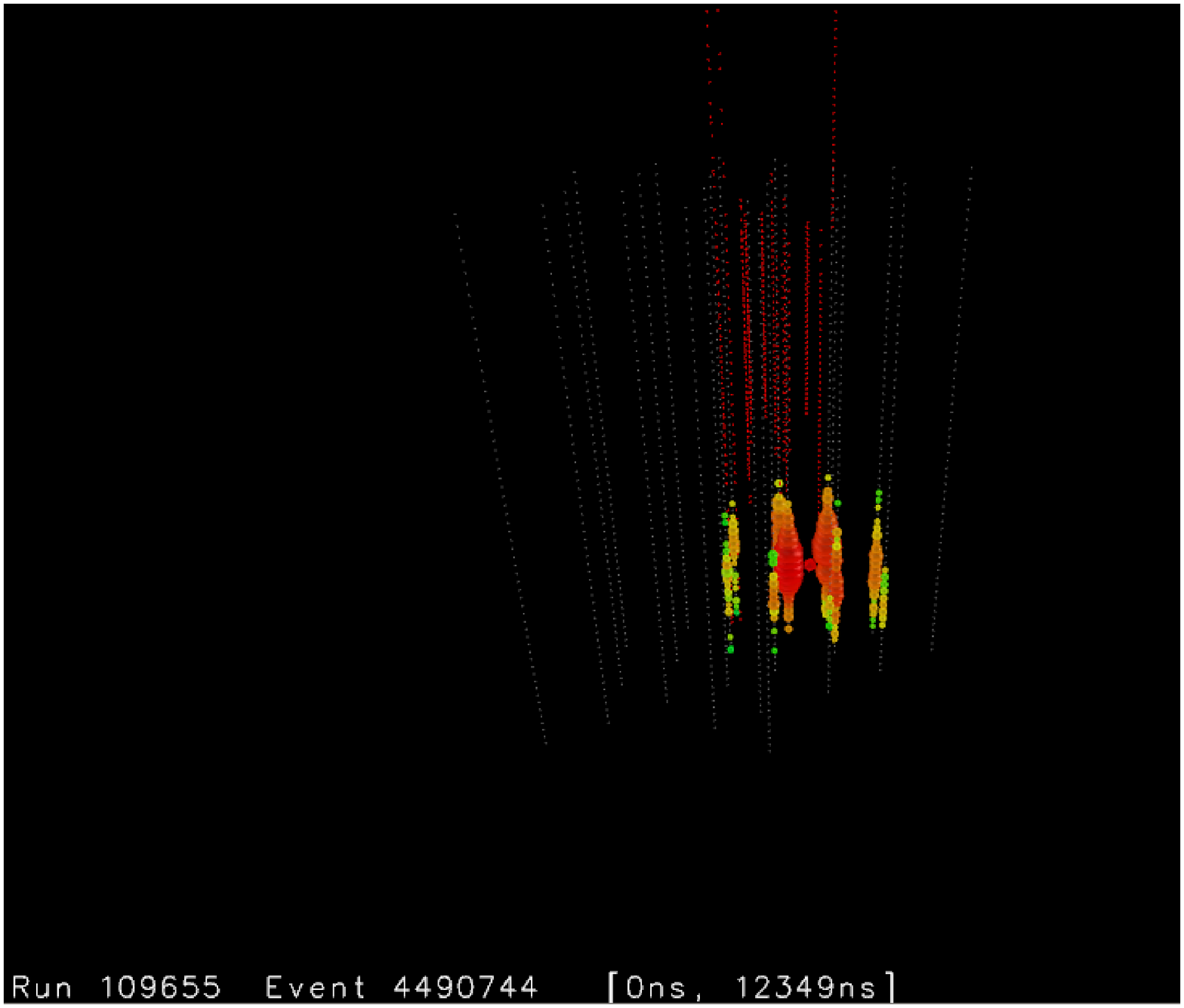}
\end{minipage}
\hspace{0.5cm} %To get a little bit of space between the figures
\begin{minipage}[b]{0.48\linewidth}
\centering
\includegraphics[width=7.6cm]{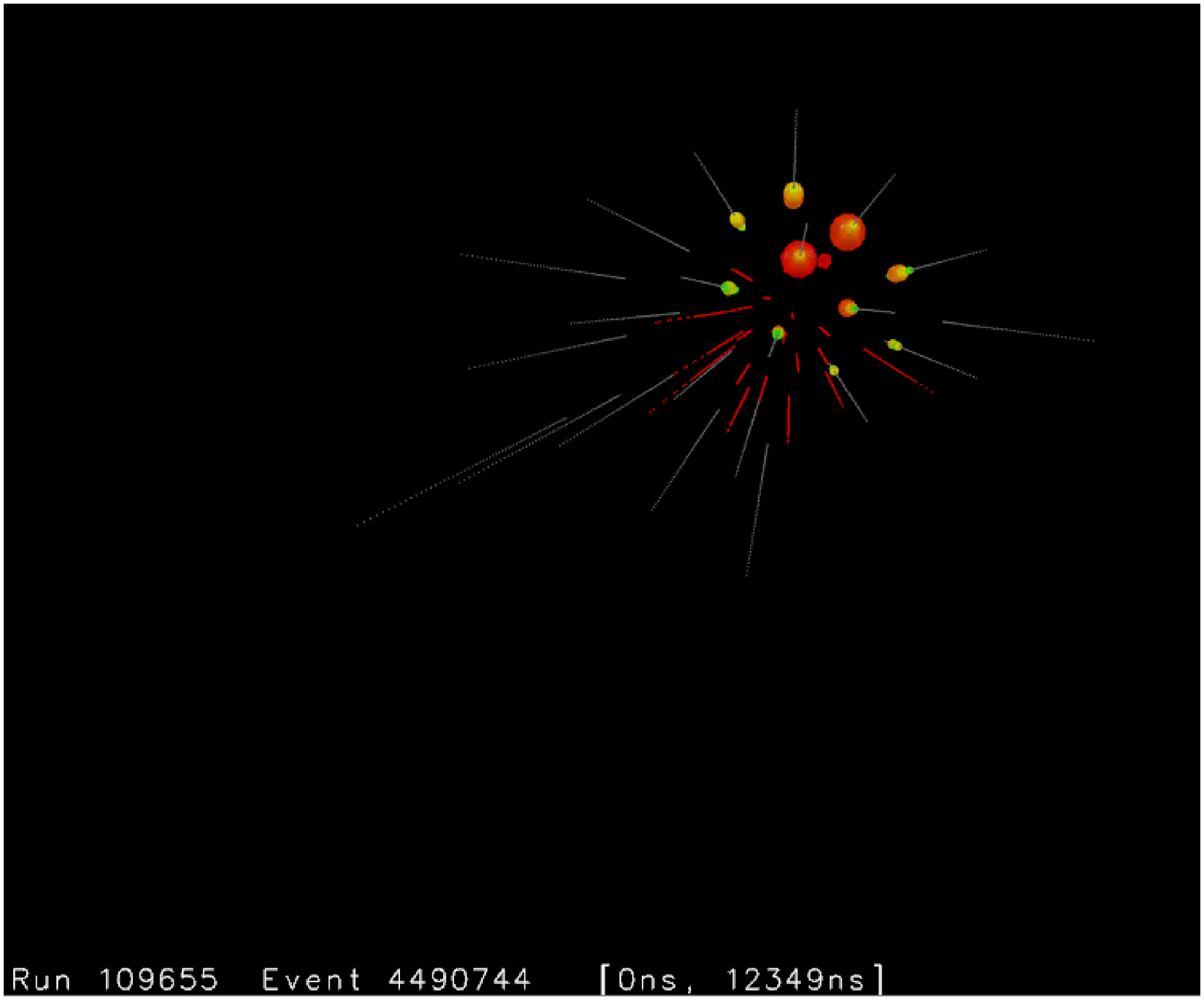}
\end{minipage}
\vspace{0.25cm}
\caption{Run 109655, Event 4490744, E=22.2 TeV}
\label{Run109655Image}
\end{figure}

\begin{figure}
\begin{minipage}[b]{0.48\linewidth} % A minipage that covers half the page
\centering
\includegraphics[width=7.6cm]{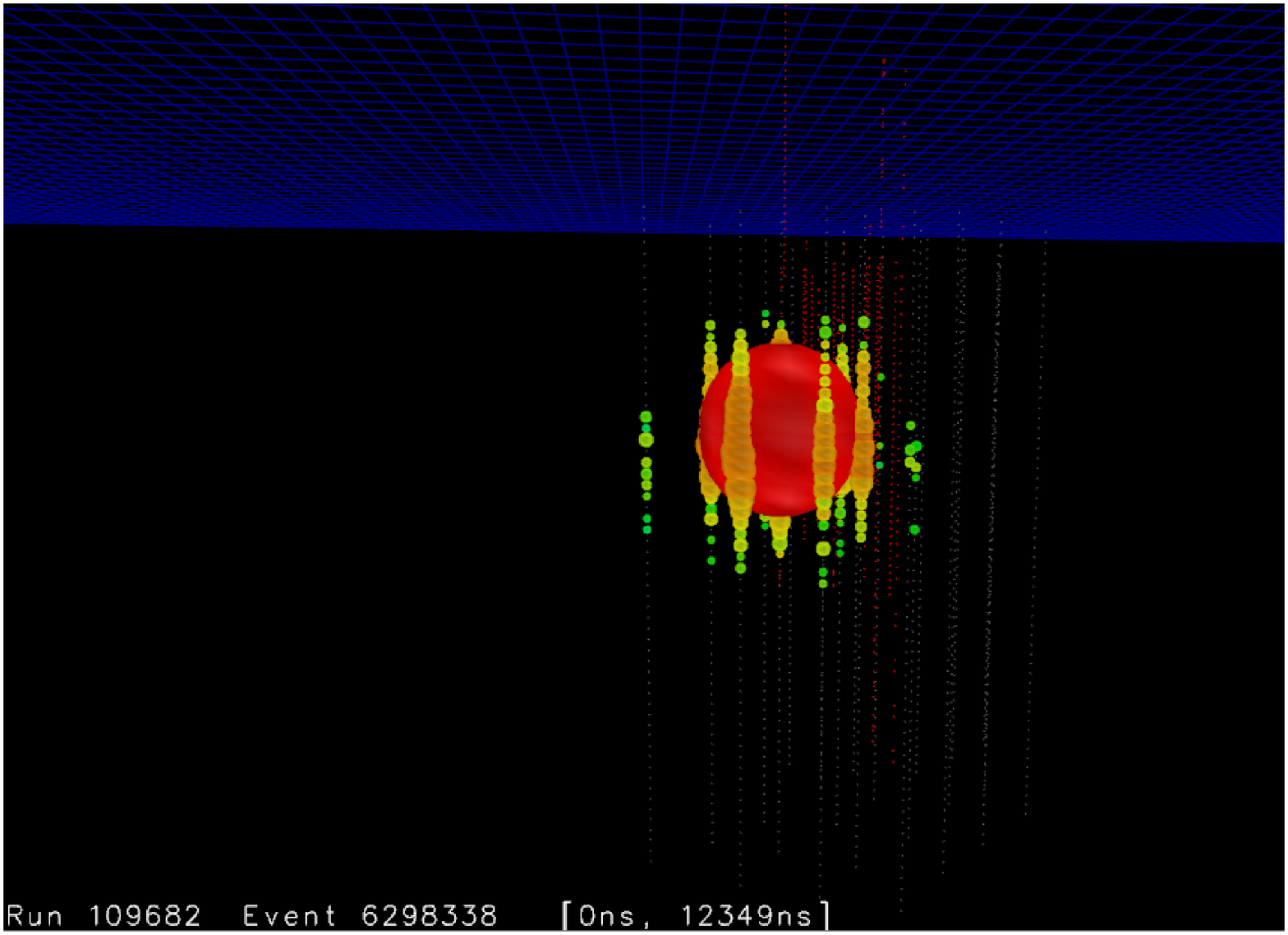}
\end{minipage}
\hspace{0.5cm} %To get a little bit of space between the figures
\begin{minipage}[b]{0.48\linewidth}
\centering
\includegraphics[width=7.6cm]{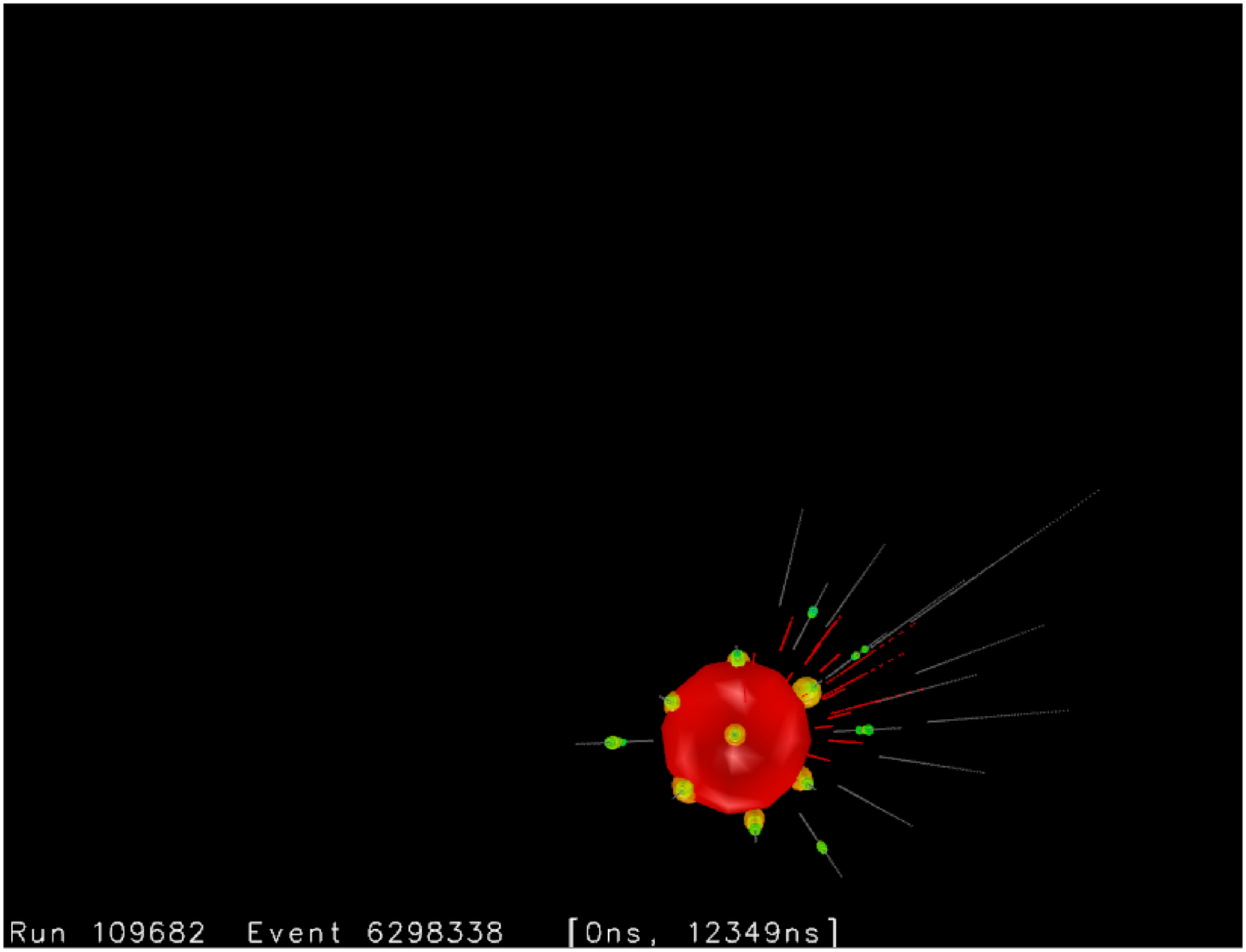}
\end{minipage}
\vspace{0.25cm}
\caption{Run 109682, Event 6298338, E=133.7 TeV}
\label{Run109682Image}
\end{figure}

\begin{figure}
\begin{minipage}[b]{0.48\linewidth} % A minipage that covers half the page
\centering
\includegraphics[width=7.6cm]{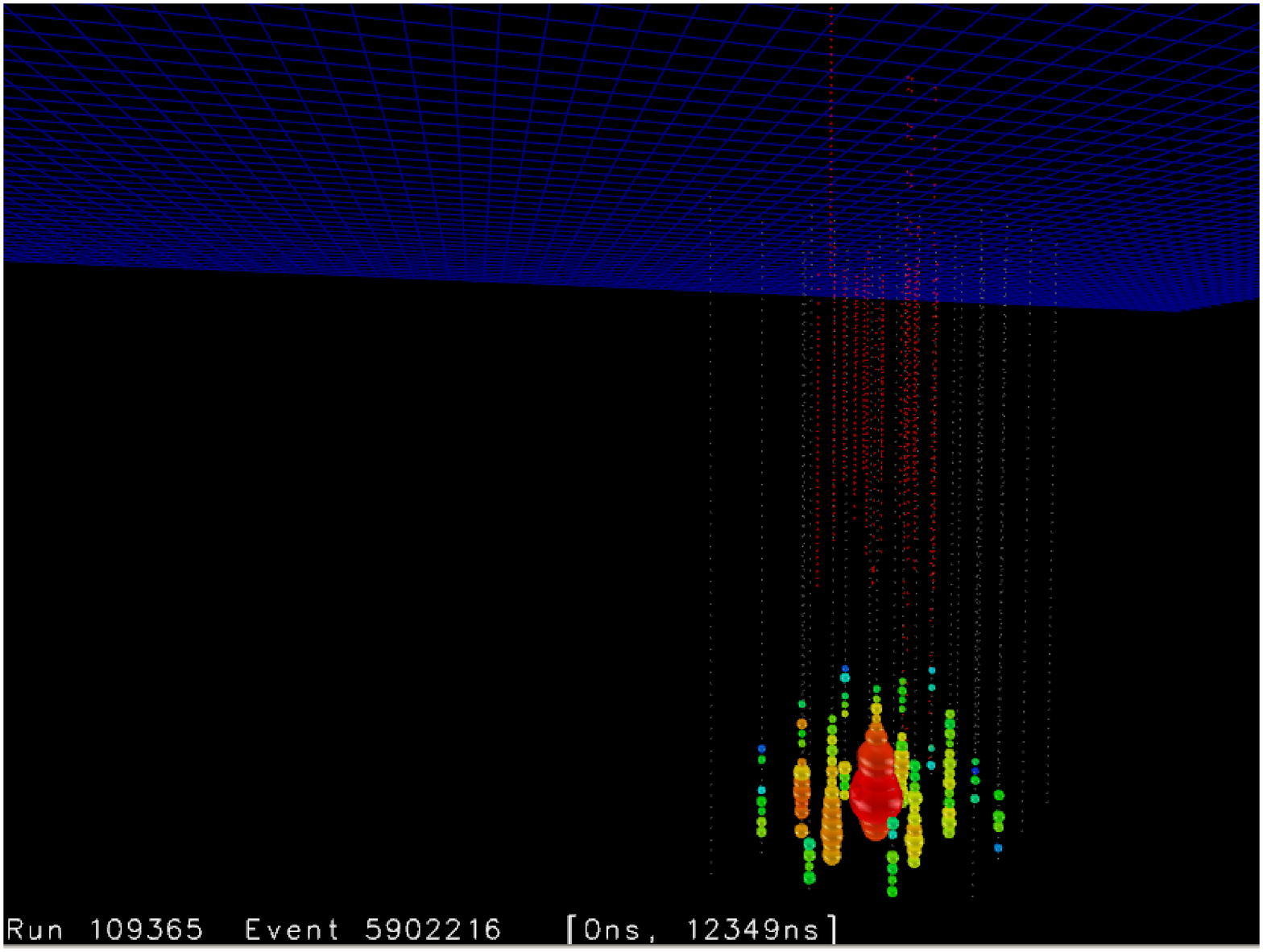}
\end{minipage}
\hspace{0.5cm} %To get a little bit of space between the figures
\begin{minipage}[b]{0.48\linewidth}
\centering
\includegraphics[width=7.6cm]{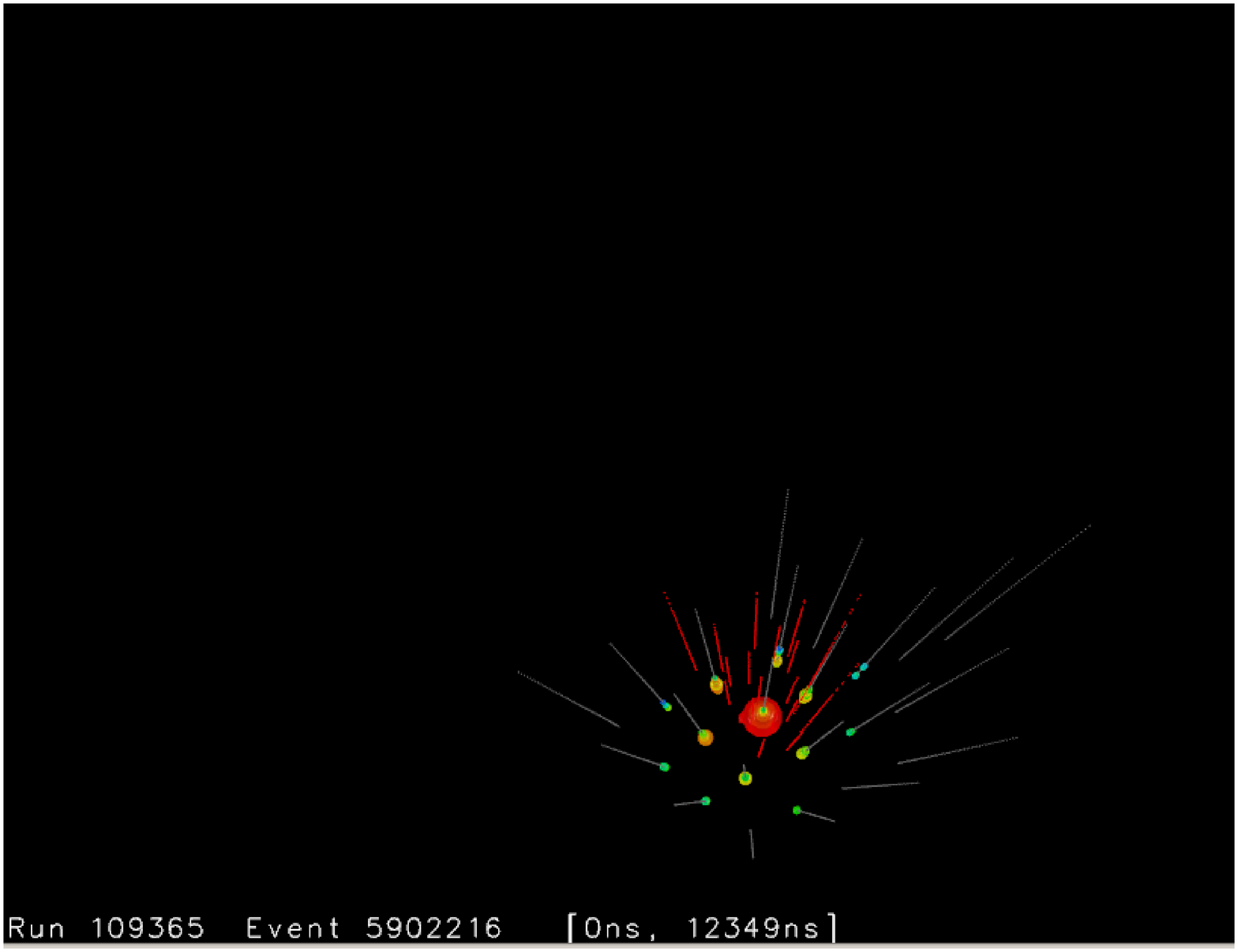}
\end{minipage}
\vspace{0.25cm}
\caption{Run 109365, Event 5902216, E=19.5 TeV}
\label{Run109365Image}
\end{figure}

\begin{figure}
\begin{minipage}[b]{0.48\linewidth} % A minipage that covers half the page
\centering
\includegraphics[width=7.6cm]{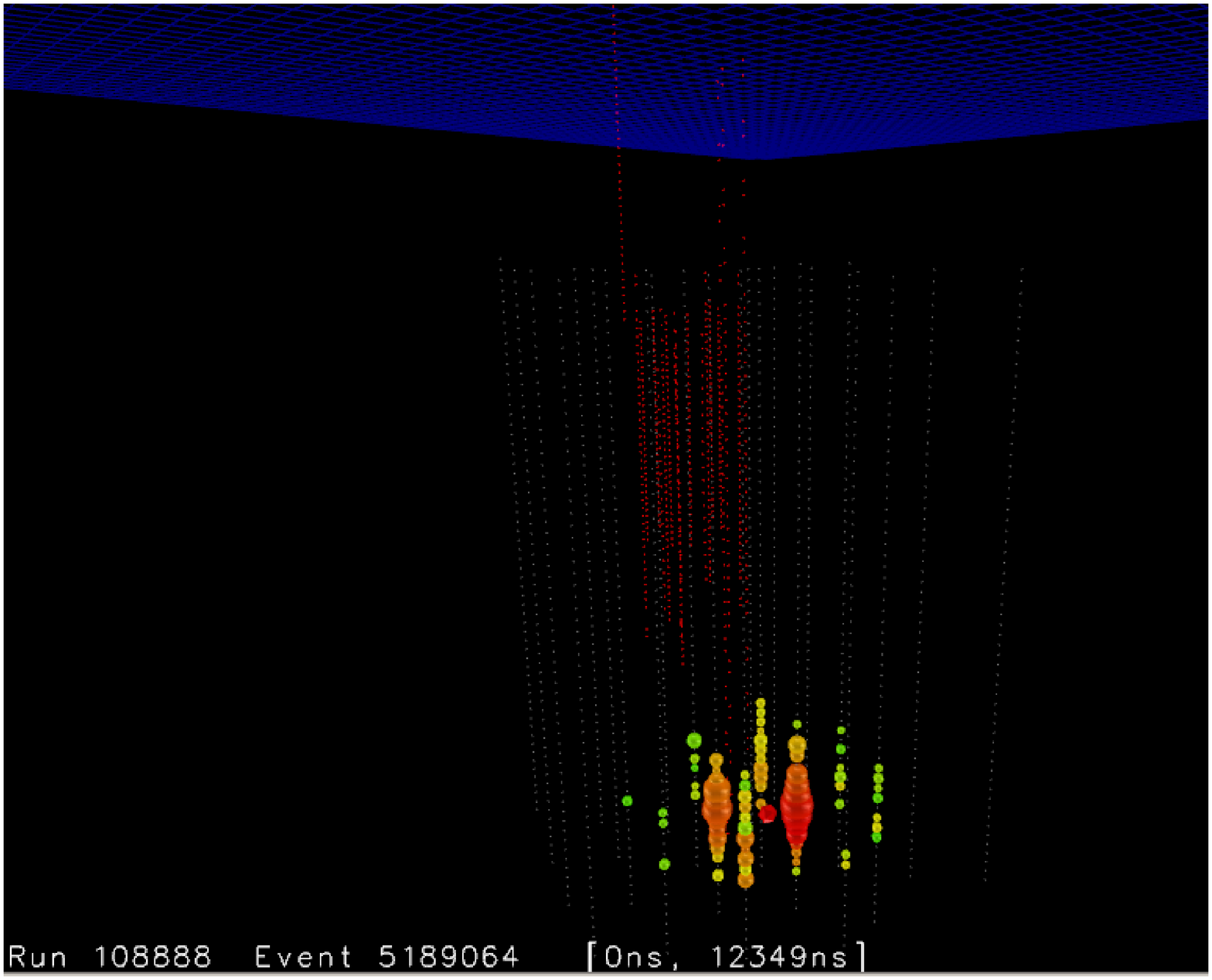}
\end{minipage}
\hspace{0.5cm} %To get a little bit of space between the figures
\begin{minipage}[b]{0.48\linewidth}
\centering
\includegraphics[width=7.6cm]{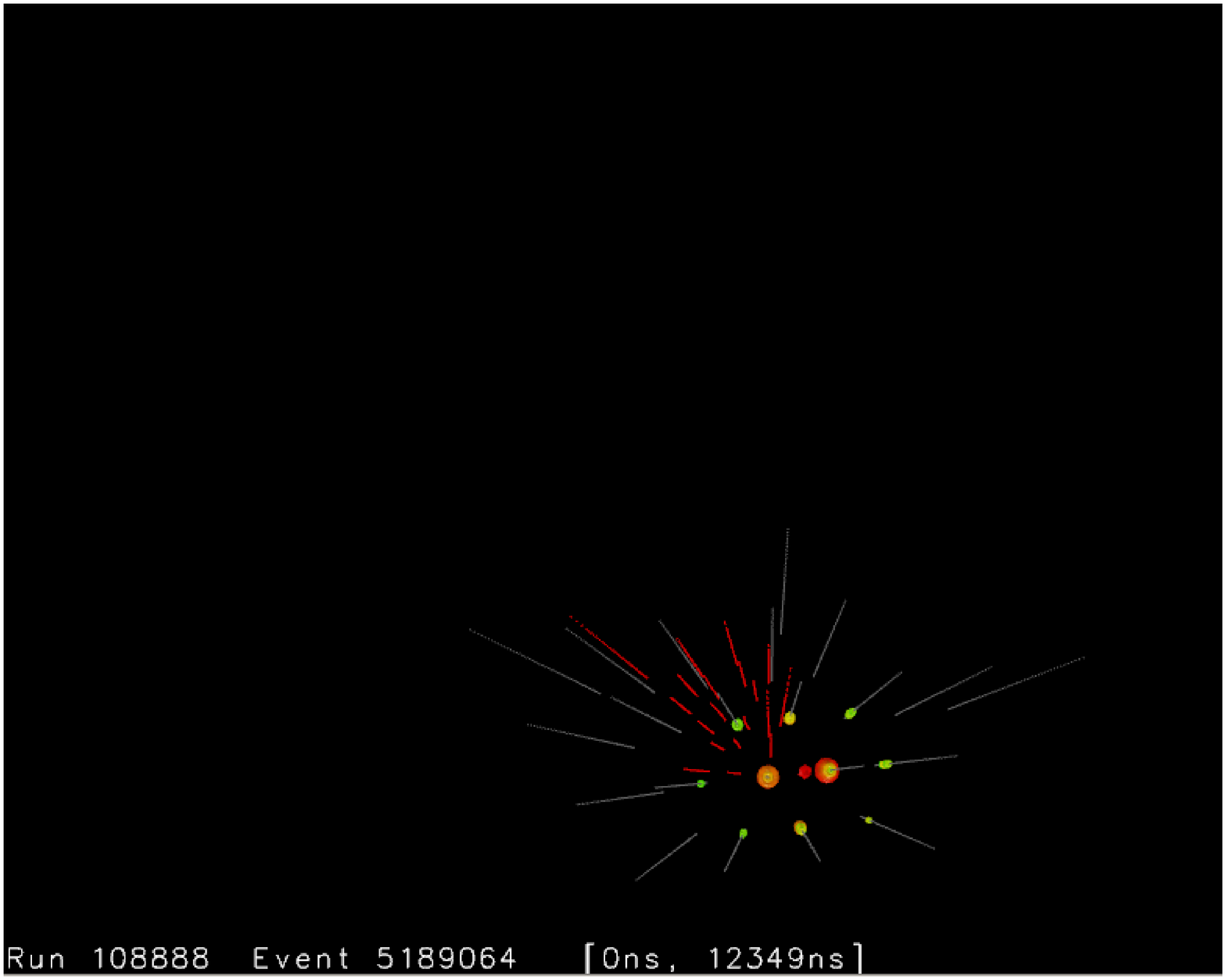}
\end{minipage}
\vspace{0.25cm}
\caption{Run 108888, Event 5189064, E=6.5 TeV}
\label{Run108888Image}
\end{figure}

\begin{figure}
\begin{minipage}[b]{0.48\linewidth} % A minipage that covers half the page
\centering
\includegraphics[width=7.6cm]{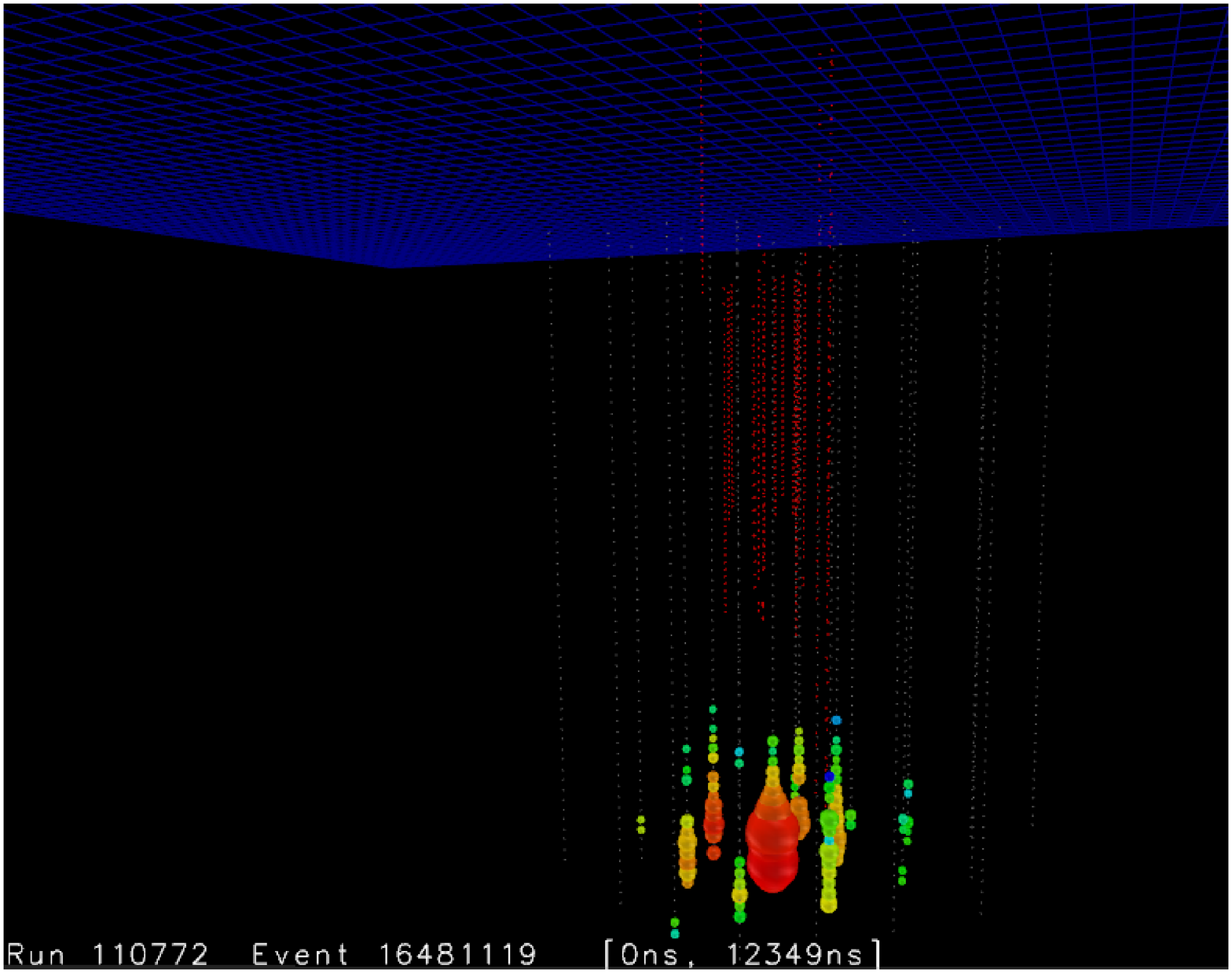}
\end{minipage}
\hspace{0.5cm} %To get a little bit of space between the figures
\begin{minipage}[b]{0.48\linewidth}
\centering
\includegraphics[width=7.6cm]{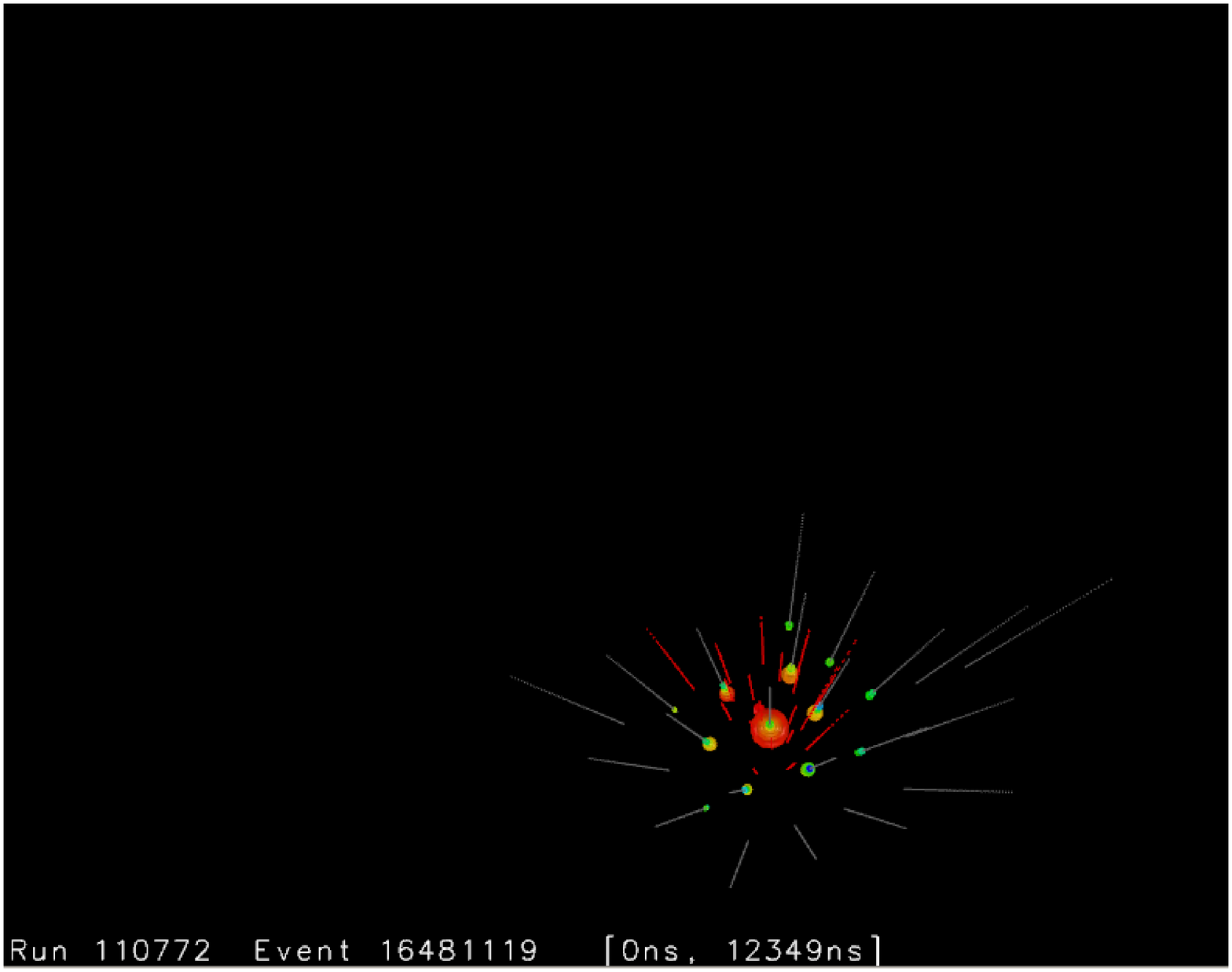}
\end{minipage}
\vspace{0.25cm}
\caption{Run 110772, Event 16481119, E=31.3 TeV}
\label{Run110772Image}
\end{figure}

\begin{figure}
\begin{minipage}[b]{0.48\linewidth} % A minipage that covers half the page
\centering
\includegraphics[width=7.6cm]{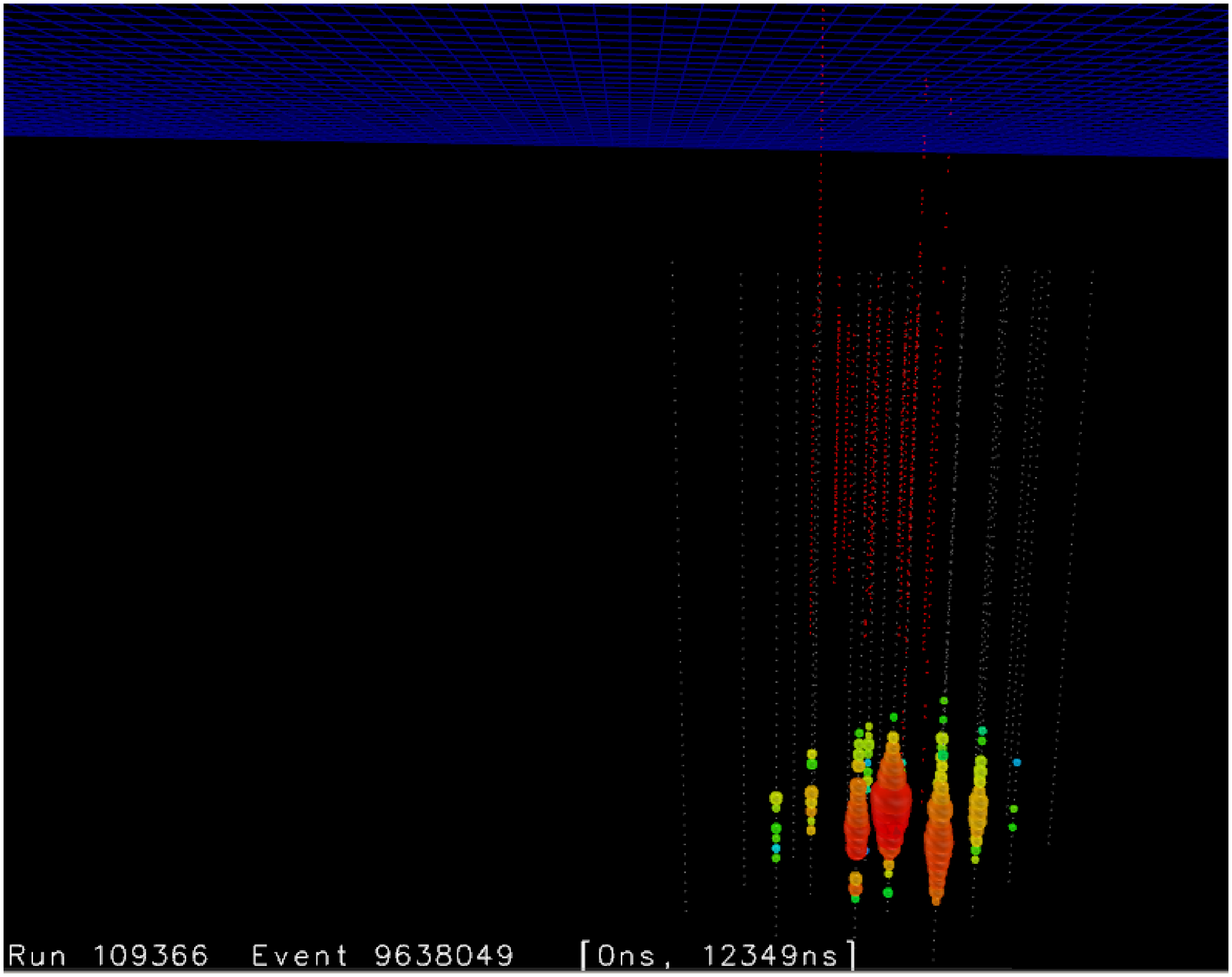}
\end{minipage}
\hspace{0.5cm} %To get a little bit of space between the figures
\begin{minipage}[b]{0.48\linewidth}
\centering
\includegraphics[width=7.6cm]{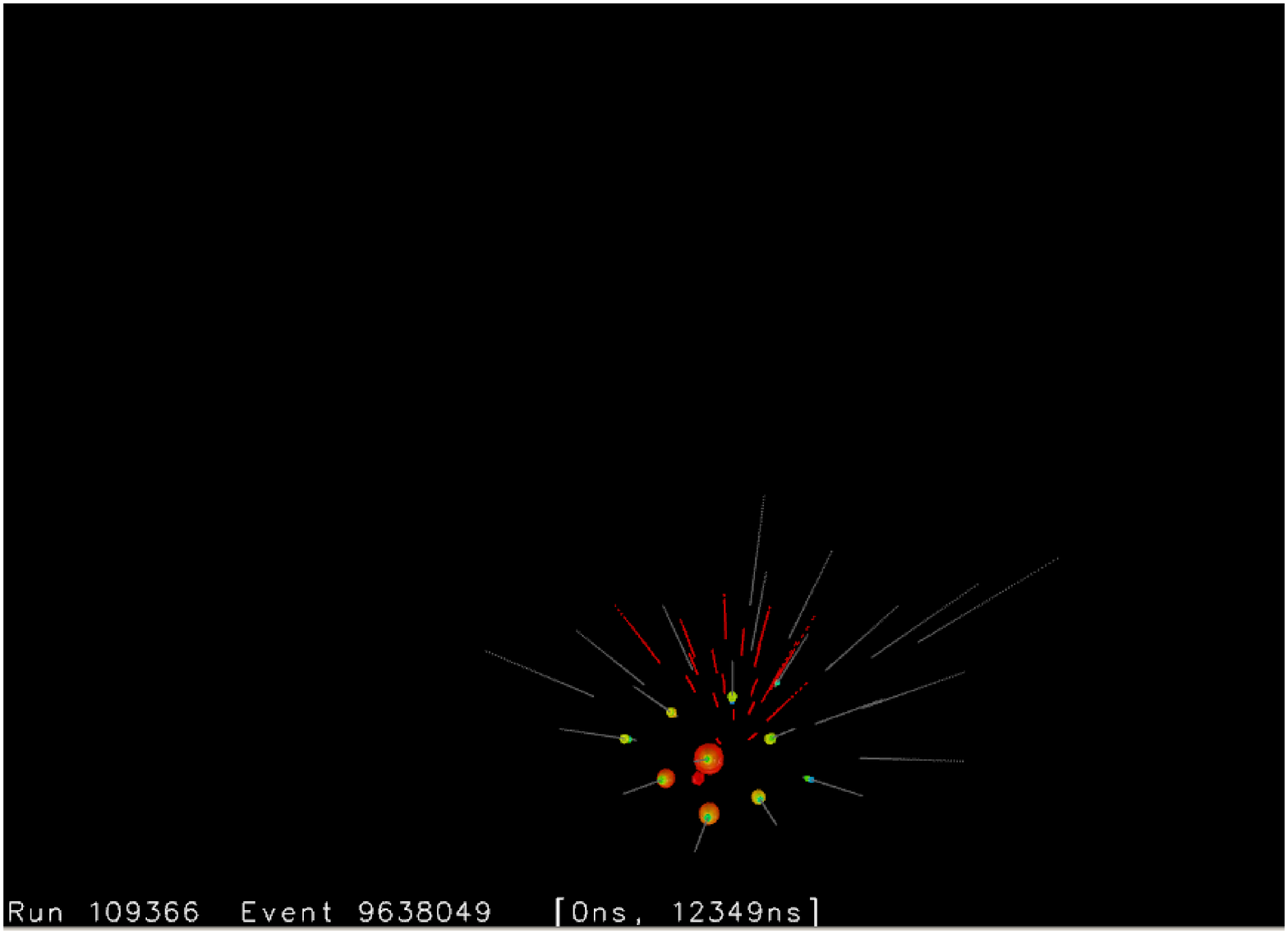}
\end{minipage}
\vspace{0.25cm}
\caption{Run 109366, Event 9638049, E=10.5 TeV}
\label{Run109366Image}
\end{figure}

\begin{figure}
\begin{minipage}[b]{0.48\linewidth} % A minipage that covers half the page
\centering
\includegraphics[width=7.6cm]{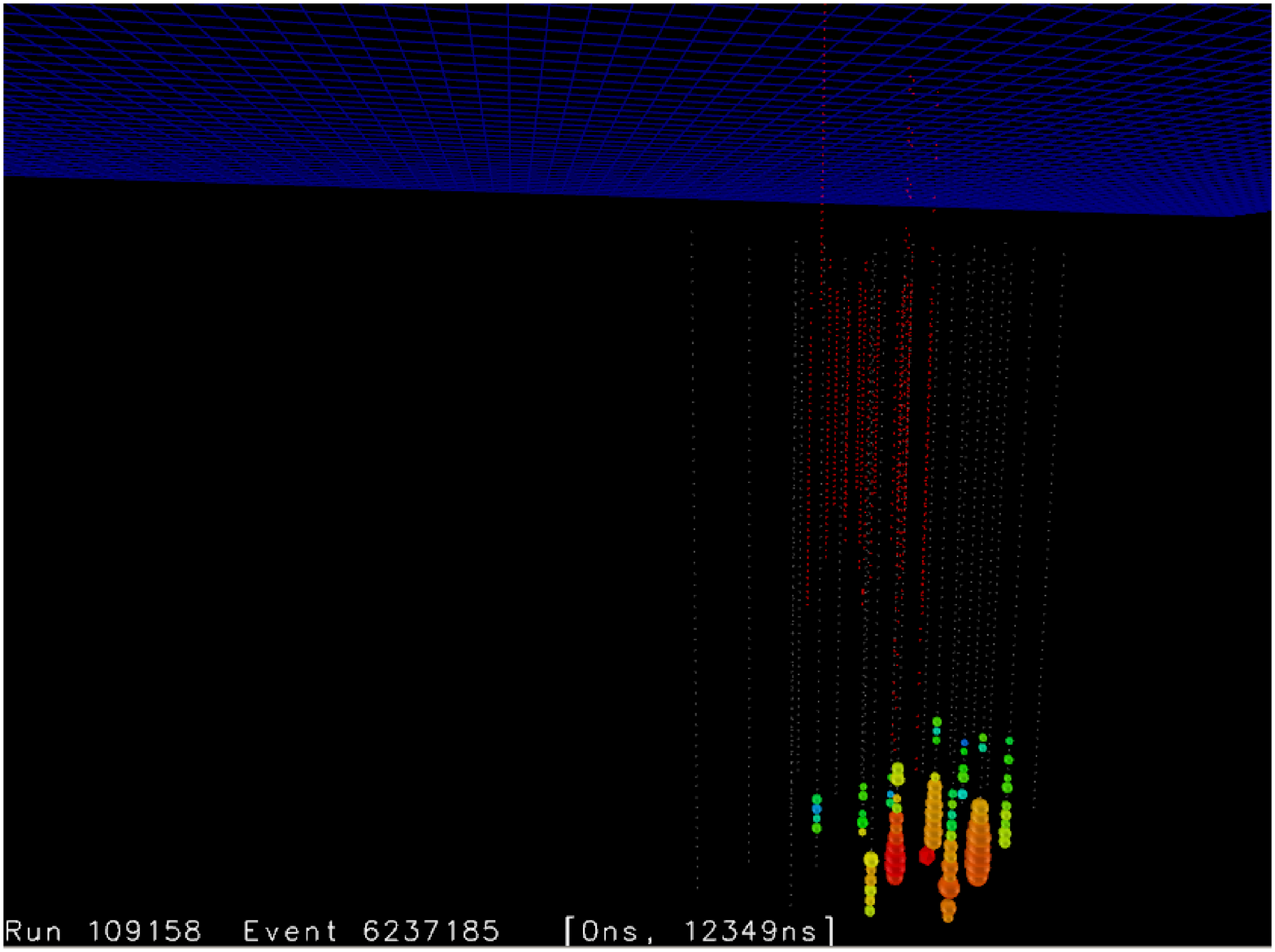}
\end{minipage}
\hspace{0.5cm} %To get a little bit of space between the figures
\begin{minipage}[b]{0.48\linewidth}
\centering
\includegraphics[width=7.6cm]{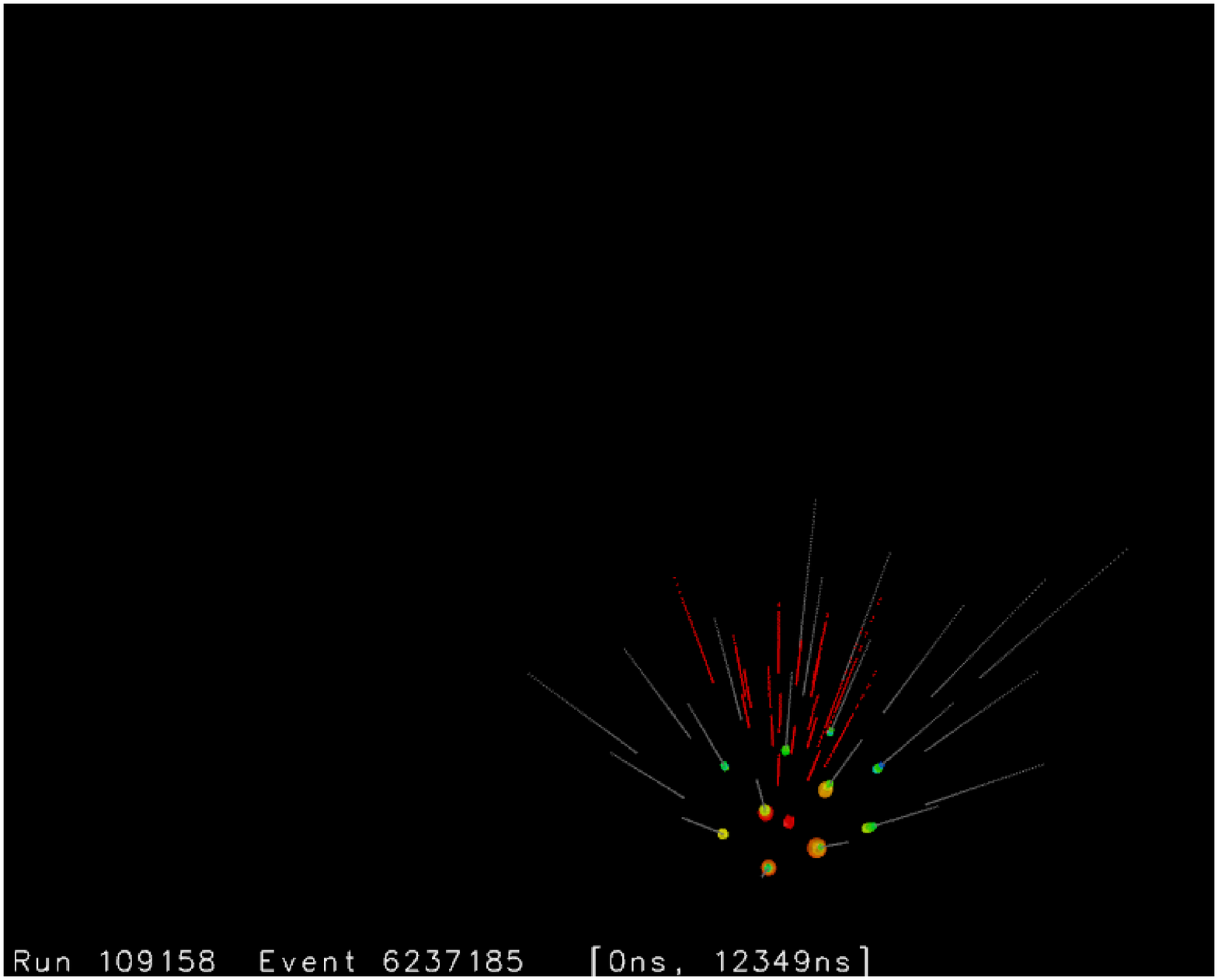}
\end{minipage}
\vspace{0.25cm}
\caption{Run 109158, Event 6237185, E=6.7 TeV}
\label{Run109158Image}
\end{figure}

\begin{figure}
\begin{minipage}[b]{0.48\linewidth} % A minipage that covers half the page
\centering
\includegraphics[width=7.6cm]{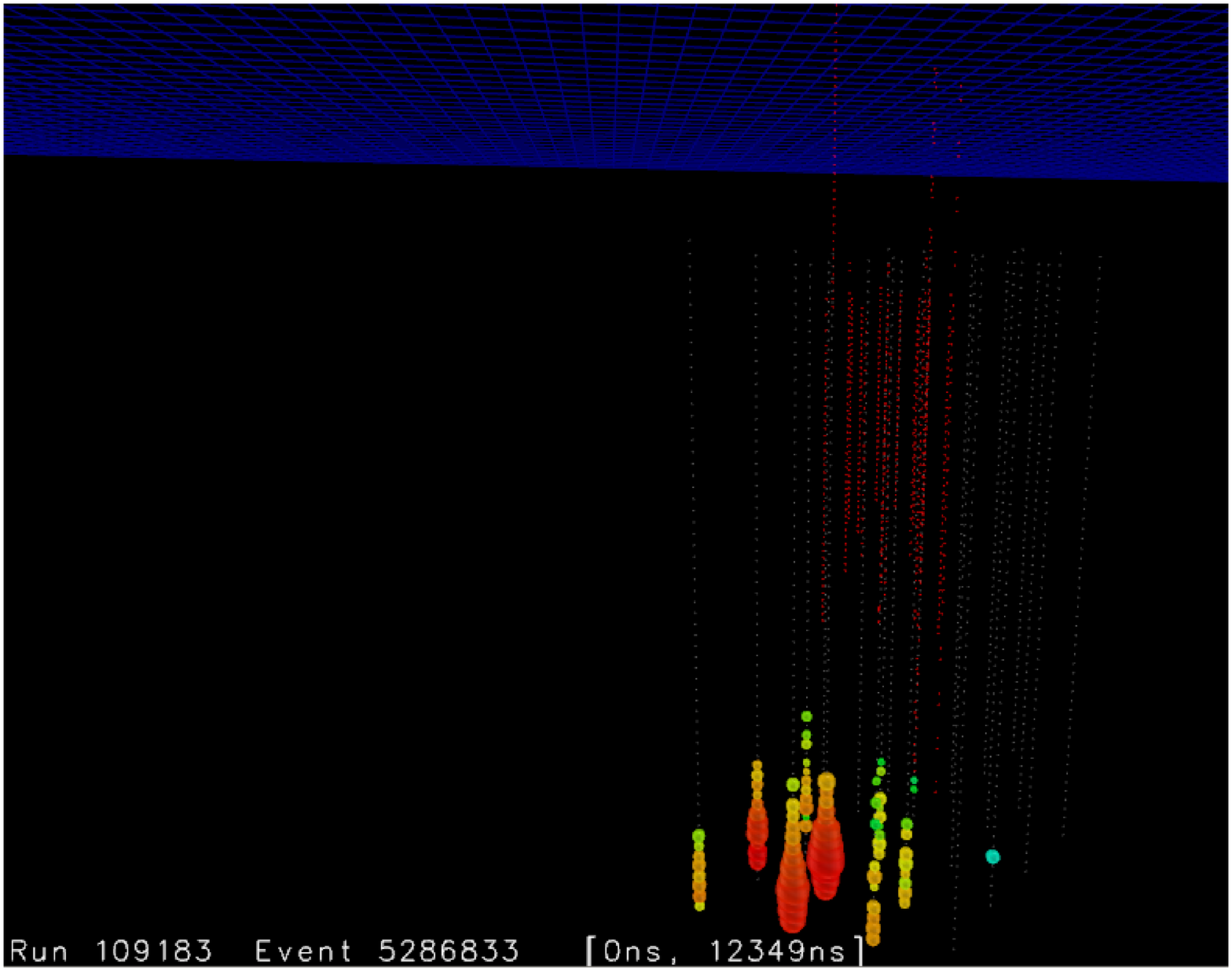}
\end{minipage}
\hspace{0.5cm} %To get a little bit of space between the figures
\begin{minipage}[b]{0.48\linewidth}
\centering
\includegraphics[width=7.6cm]{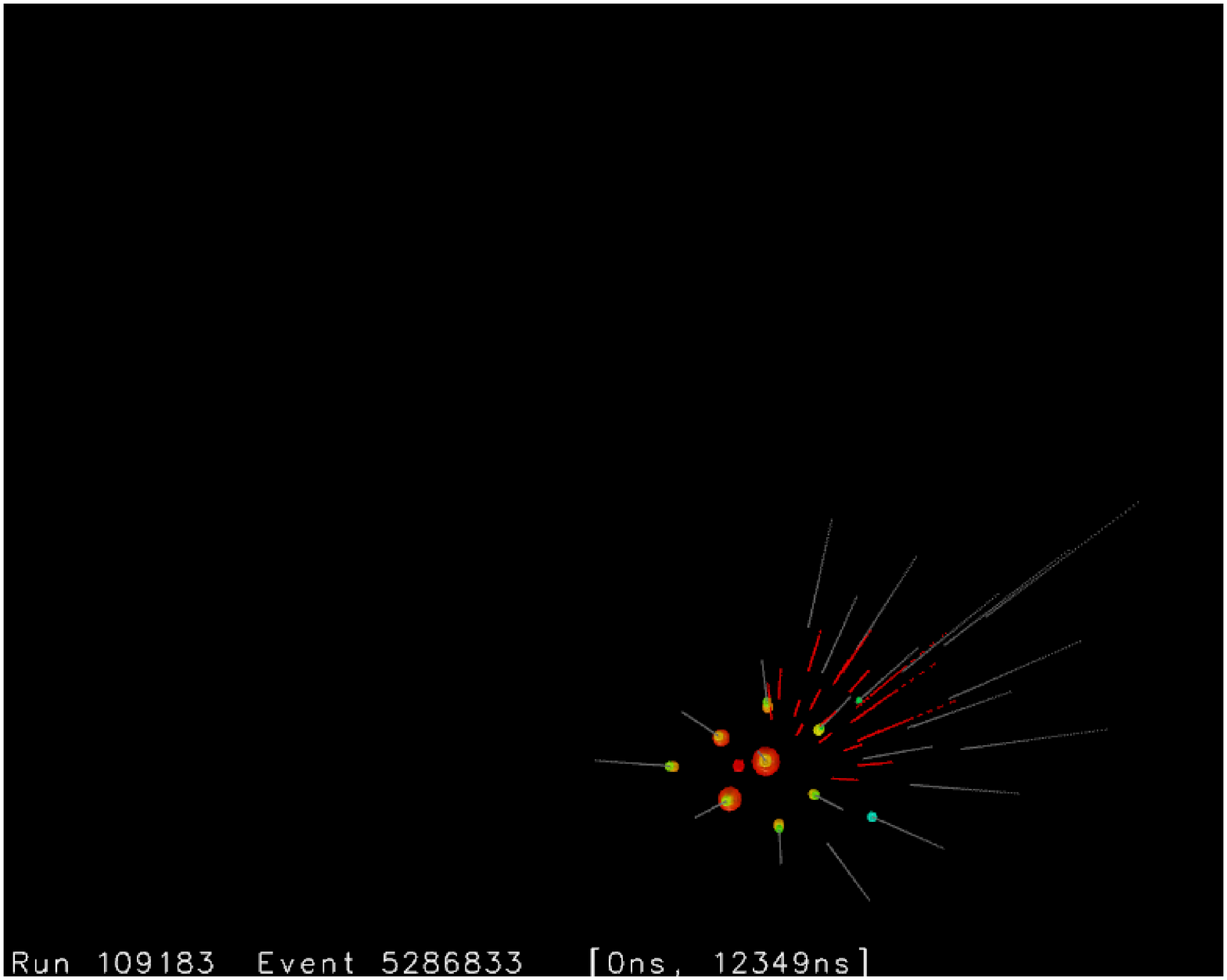}
\end{minipage}
\vspace{0.25cm}
\caption{Run 109183, Event 5286833, E=15.6 TeV}
\label{Run109183Image}
\end{figure}

From the cumulative distributions and tables, we see that the number of observed events is slightly higher than, but relatively close to, the neutrino signal prediction.  Next, we would like to examine the 12 surviving events above the final cut value of 0.90 to see if they show any characteristics of muon background events.  Figures~\ref{Run109144Image}--\ref{Run109183Image} on the next pages show IceCube event-viewer images of these 12 surviving candidate events along with their reconstructed energies, arranged in order of increasing ``quality'' as defined by the final multivariate cut variable.  Balloon sizes are proportional to the amount of observed light, and the balloon color indicates hit timing---from the earliest hits in red to the latest hits in blue.  

A detailed examination of these events did not uncover any muon-like characteristics.  None of the events had early or out-of-time hits that were consistent with a muon track plus a large radiative energy loss.  This was a necessary but not sufficient check to prove these events are indeed neutrino-induced cascades.

\clearpage

\newpage

\clearpage

\newpage

\clearpage

\newpage

\clearpage

\newpage

\clearpage

\newpage

\clearpage

\newpage

\clearpage

\newpage

\begin{figure}
\begin{minipage}[b]{0.48\linewidth} % A minipage that covers half the page
\centering
\includegraphics[width=7.6cm]{figures/eventviewer_images/Run109682_SIDE}
\end{minipage}
\hspace{0.5cm} %To get a little bit of space between the figures
\begin{minipage}[b]{0.48\linewidth}
\centering
\includegraphics[width=7.6cm]{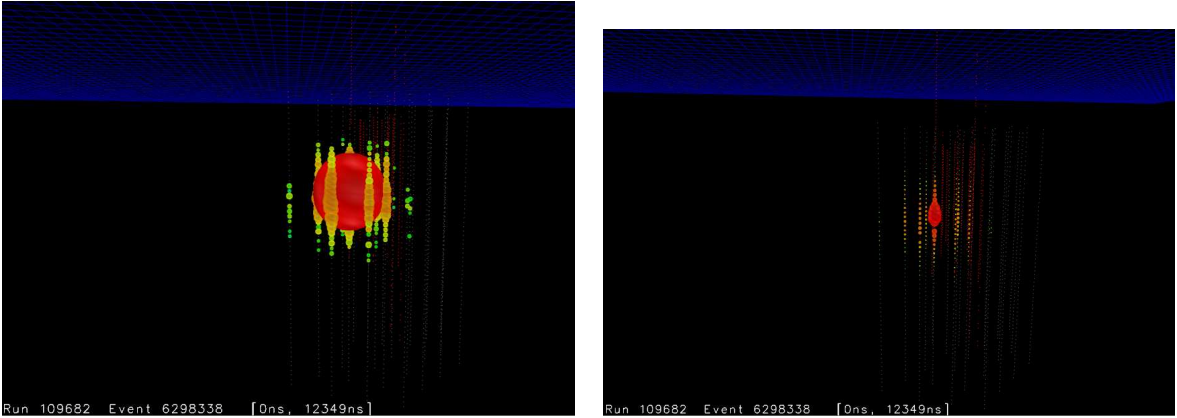}
\end{minipage}
\vspace{0.25cm}
\caption{Run 109682, Event 6298338, E=133.7 TeV with normal scaling (left) and shrunken scaling to show details of the innermost DOM's (right).  The balloon sizes are proportional to the amount of detected light.}
\label{Run109682ImageShrunk}
\end{figure}

Of these 12 events, one is a very intriguing outlier.  This event is Run 109682 Event 6298338, which reconstructs at an energy E=133.7 TeV.  The event is reproduced in figure~\ref{Run109682ImageShrunk} along with an image where the balloon sizes, which are proportional to the amount of detected light, have been scaled down.  The down-scaled image illustrates that several DOM's very close to the vertex received a tremendous amount of light in the event (these DOM's could not be resolved in the first image).

Many different cross-checks were performed on this event to make sure that it is indeed a genuine physics event and not a detector artifact.  First, monitoring and detector records showed that the detector was in a stable operating mode for this run---nothing out of the ordinary was reported and all diagnostic distributions look normal.  No in-situ light runs were scheduled anywhere near this run, either before or after.  

\begin{figure}
\begin{minipage}[b]{0.48\linewidth} % A minipage that covers half the page
\centering
\includegraphics[width=7.6cm]{figures/eventviewer_images/Run109682_SIDE}
\end{minipage}
\hspace{0.5cm} %To get a little bit of space between the figures
\begin{minipage}[b]{0.48\linewidth}
\centering
\includegraphics[width=7.6cm]{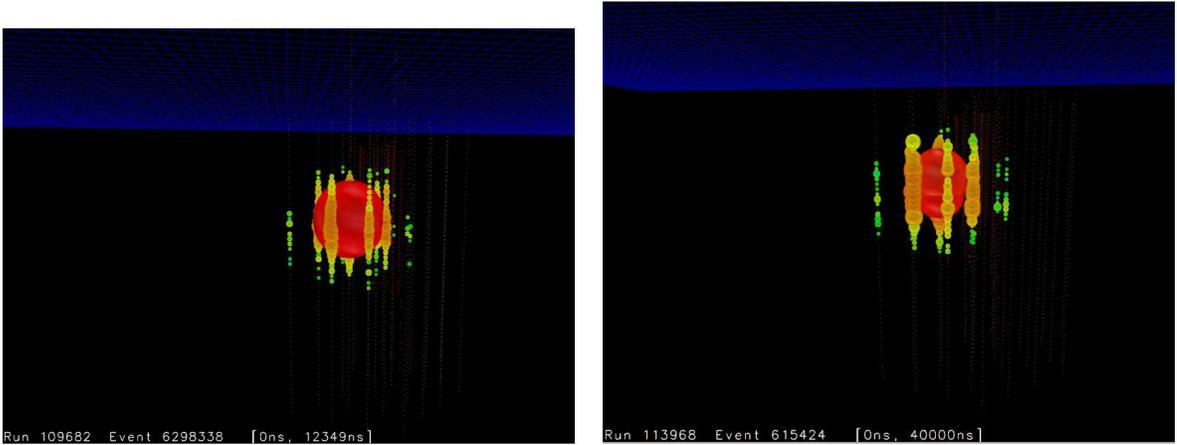}
\end{minipage}
\vspace{0.25cm}
\caption{Run 109682, Event 6298338, E=133.7 TeV compared to an in-situ flasher event.}
\label{Run109682ComparedToFlasherImage}
\end{figure}

To exclude the possibility of an accidental discharge of an in-situ light source, a dedicated ``flasher'' run was taken where we flashed all 12 in-situ LED's in DOM 39-19, which is the closest DOM to the reconstructed cascade vertex for this event.  Figure~\ref{Run109682ComparedToFlasherImage} compares the 133 TeV event to one of these flasher events.

Several features distinguish the flashers from the 133 TeV event. First, there is a pronounced up-down asymmetry in the flasher events. The DOM's above the flasher see much more charge than do the DOM's below. For the 133 TeV event, however, DOM's 39-19 and 39-20 see almost the same amount of charge.

\begin{figure}
\begin{minipage}[b]{0.32\textwidth} % A minipage that covers half the page
\centering
\includegraphics[width=5.55cm]{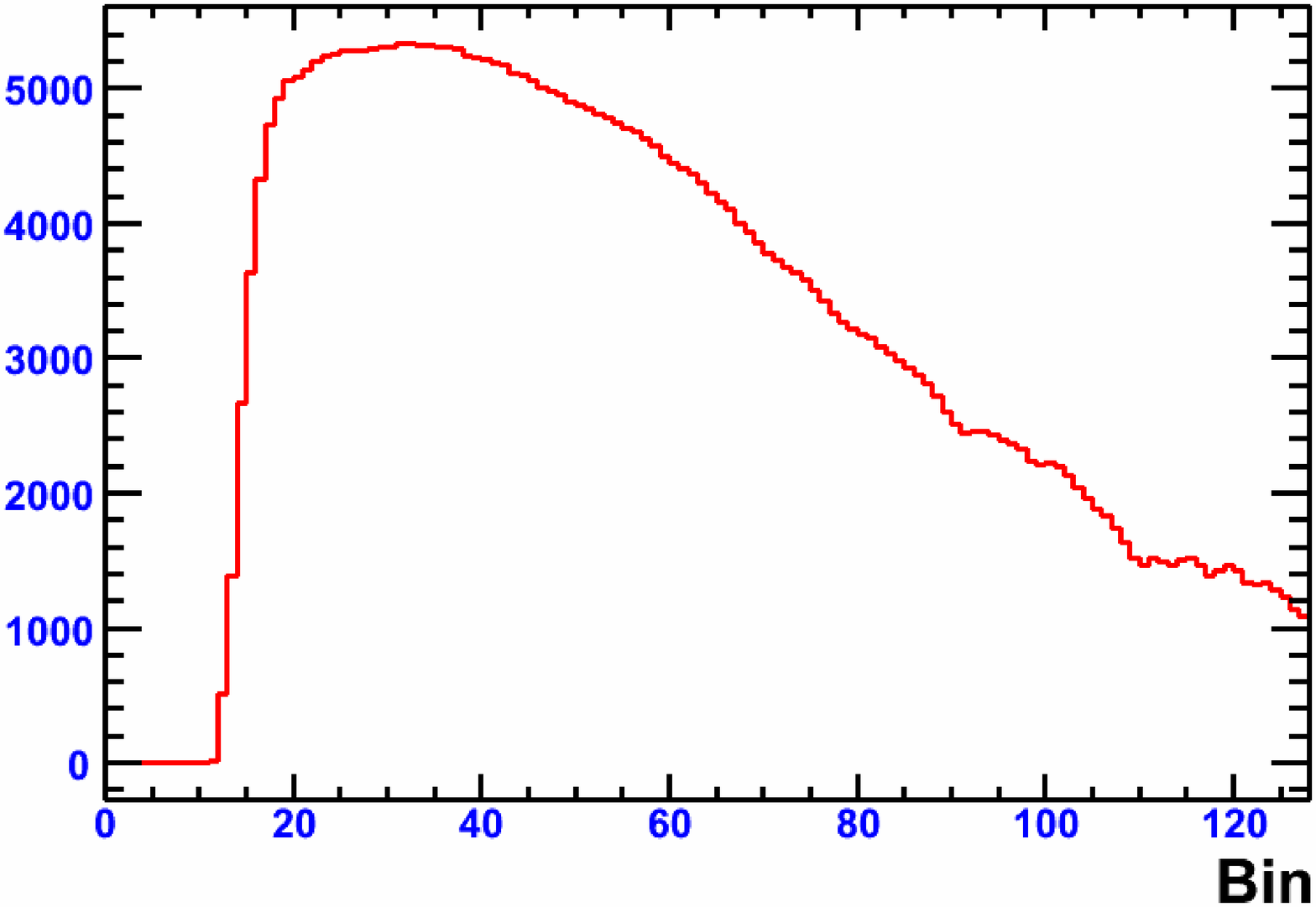}
\end{minipage}
\begin{minipage}[b]{0.32\textwidth}
\centering
\includegraphics[width=5.55cm]{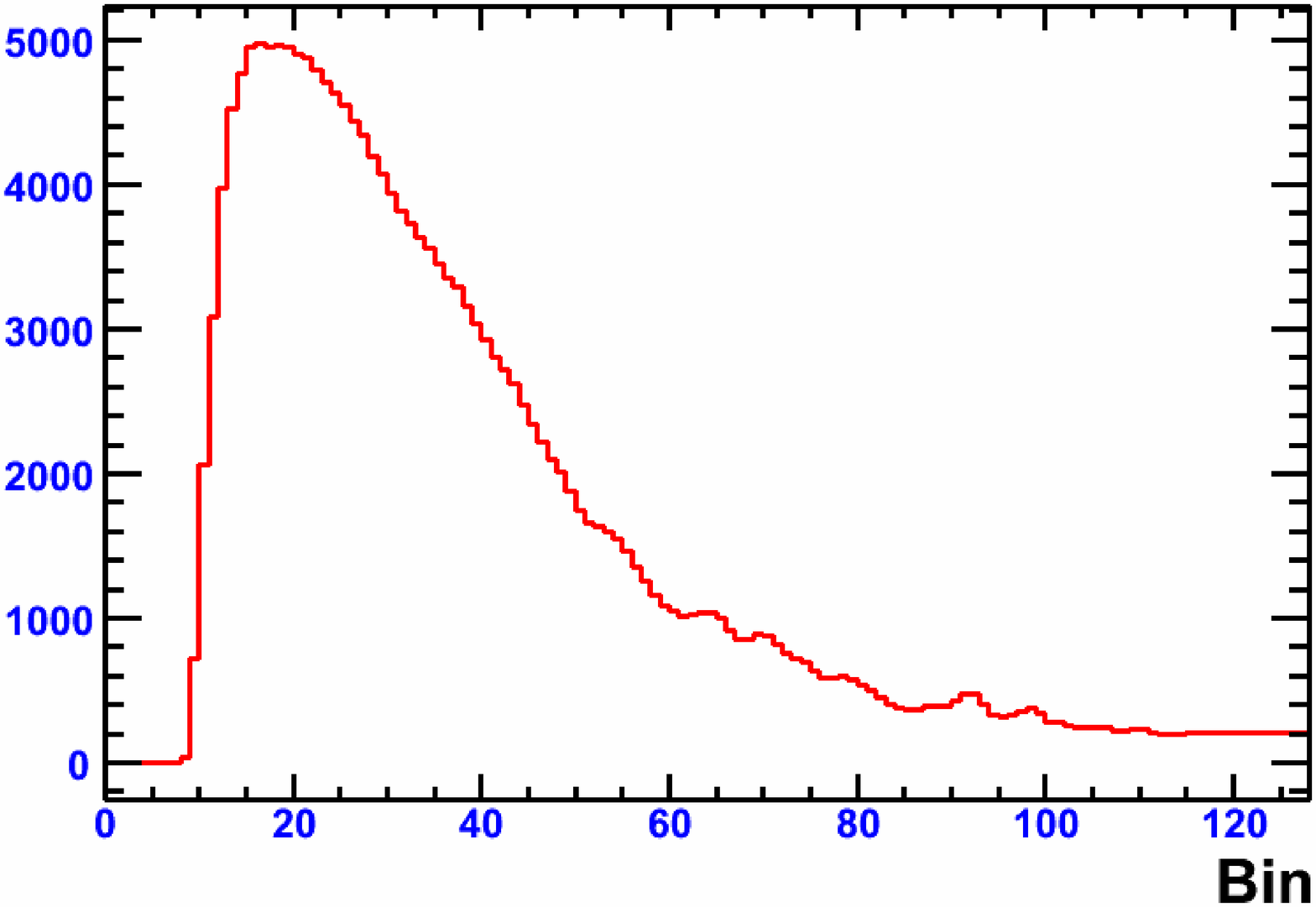}
\end{minipage}
\begin{minipage}[b]{0.32\textwidth}
\centering
\includegraphics[width=5.55cm]{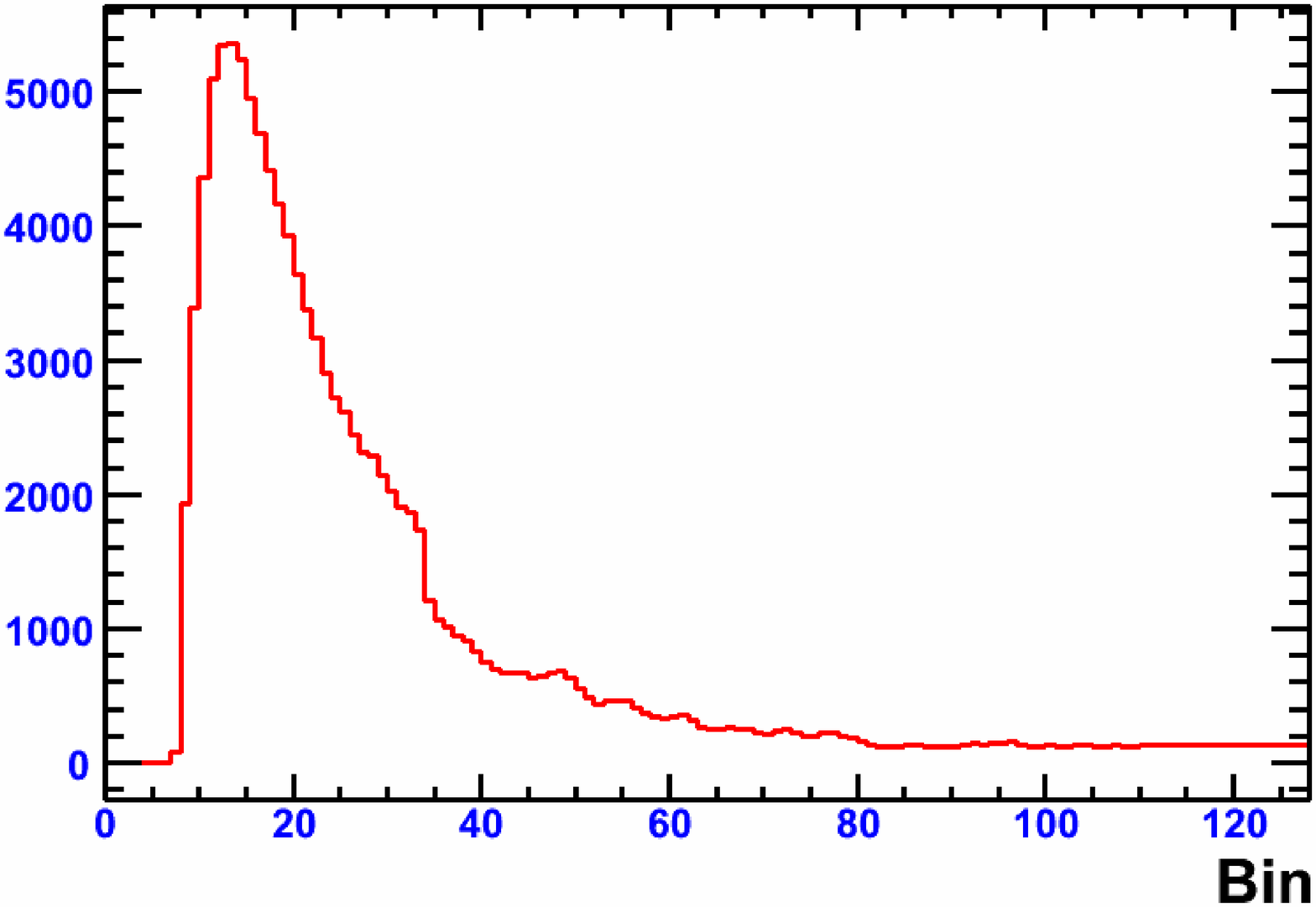}
\end{minipage}

\vspace{0.65cm}

\begin{minipage}[b]{0.32\textwidth} % A minipage that covers half the page
\centering
\includegraphics[width=5.55cm]{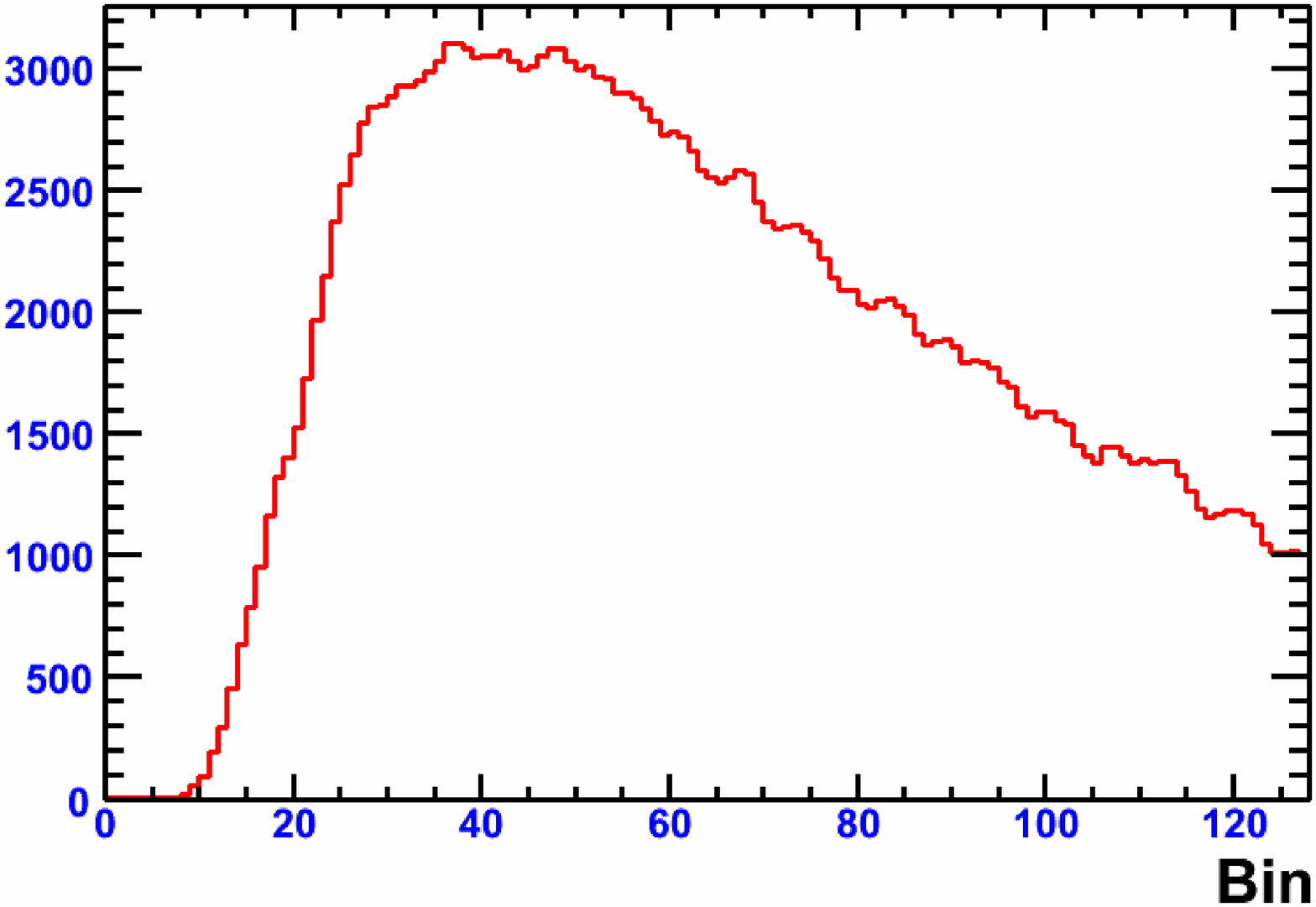}
\end{minipage}
\begin{minipage}[b]{0.32\textwidth}
\centering
\includegraphics[width=5.55cm]{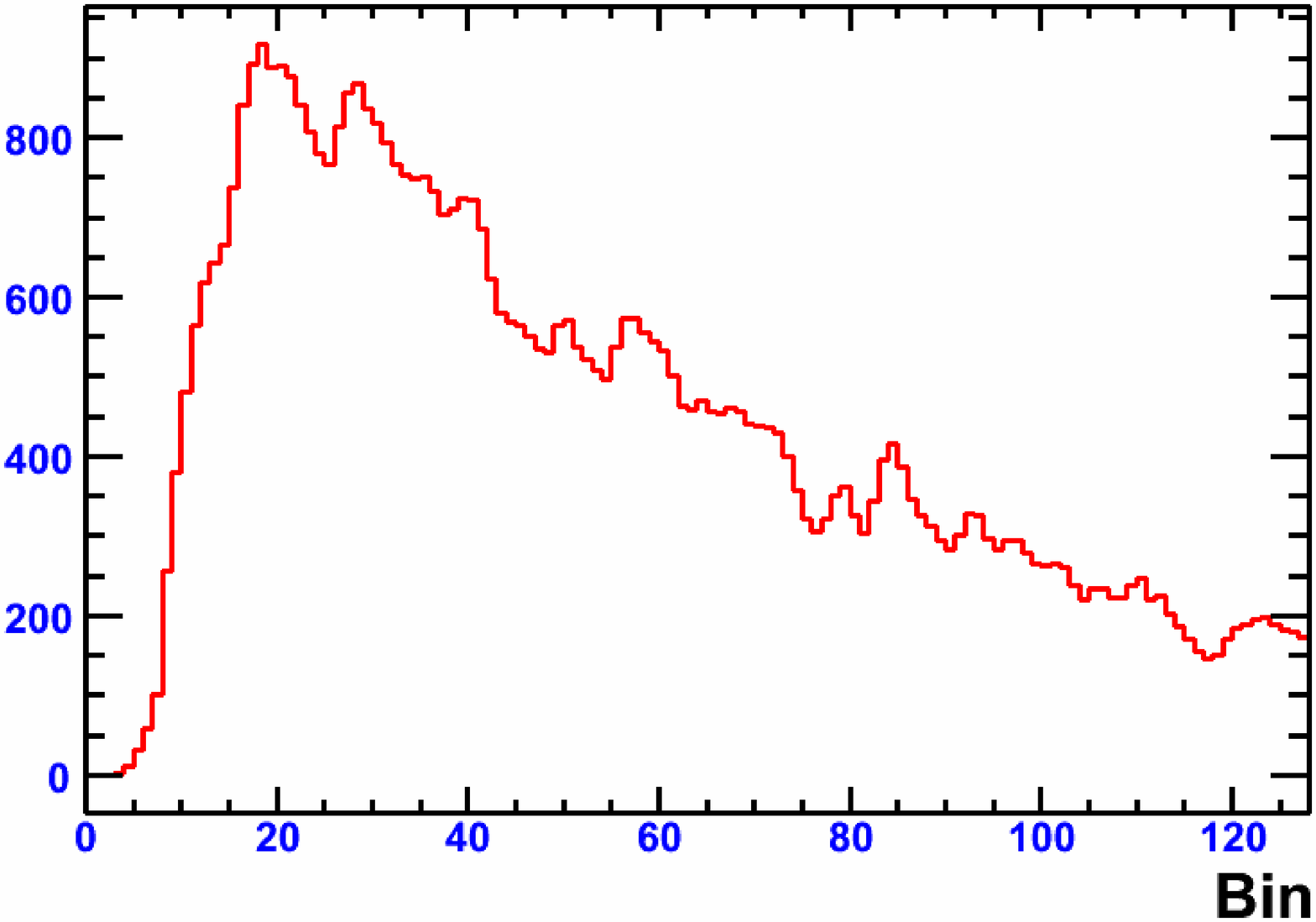}
\end{minipage}
\begin{minipage}[b]{0.32\textwidth}
\centering
\includegraphics[width=5.55cm]{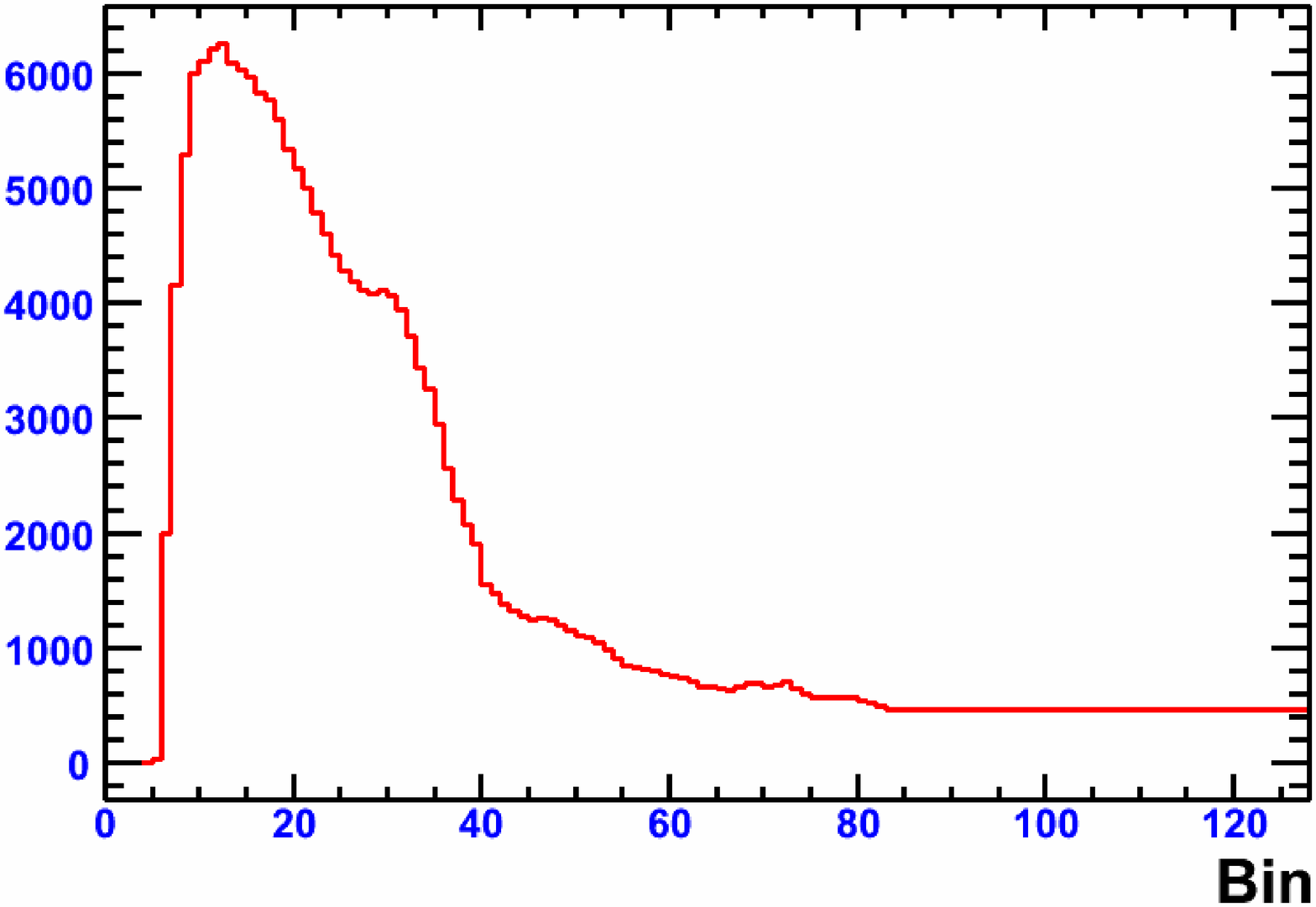}
\end{minipage}

\vspace{0.65cm}

\begin{minipage}[b]{0.32\textwidth} % A minipage that covers half the page
\centering
\includegraphics[width=5.55cm]{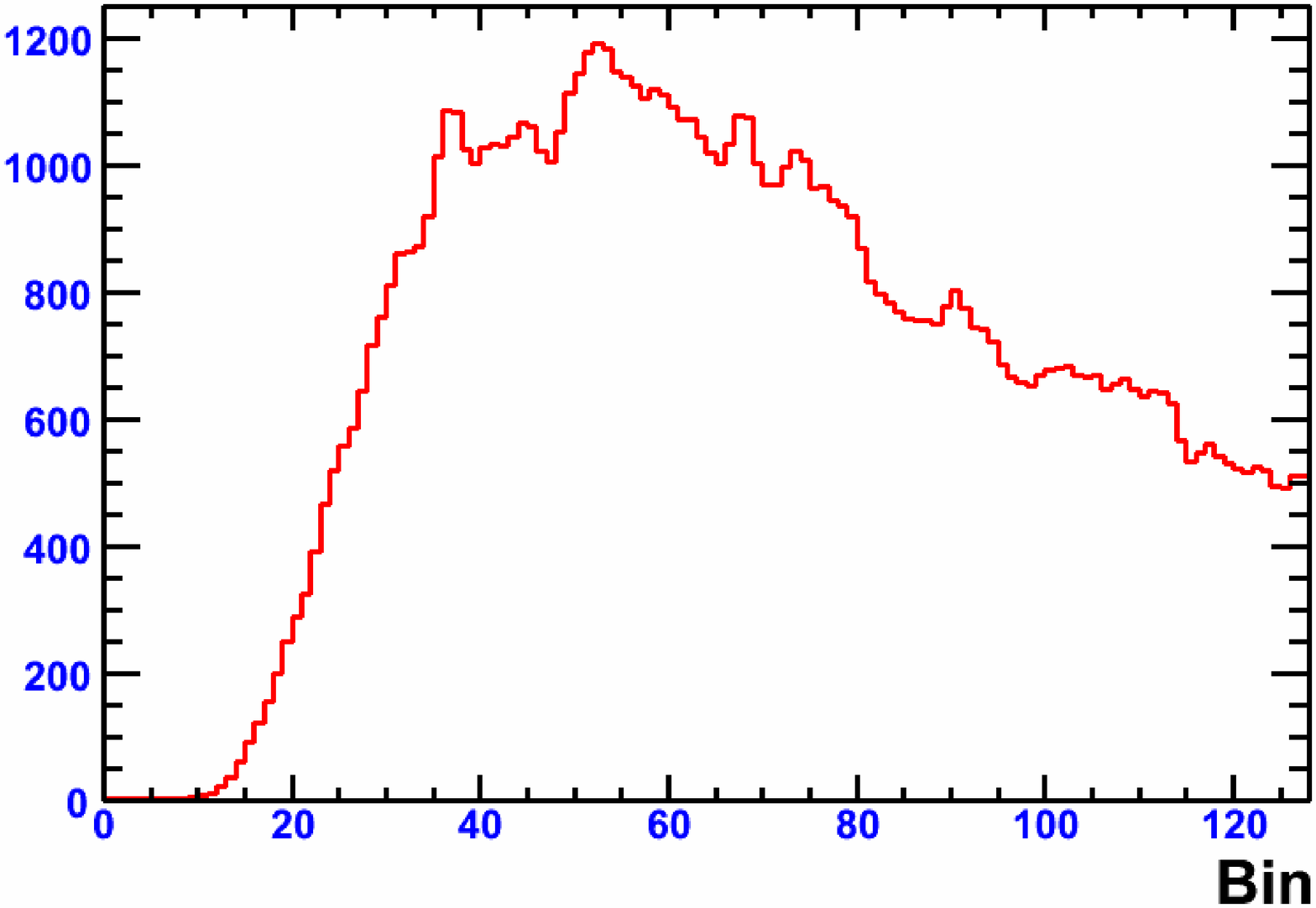}
\end{minipage}
\begin{minipage}[b]{0.32\textwidth}
\centering
\includegraphics[width=5.55cm]{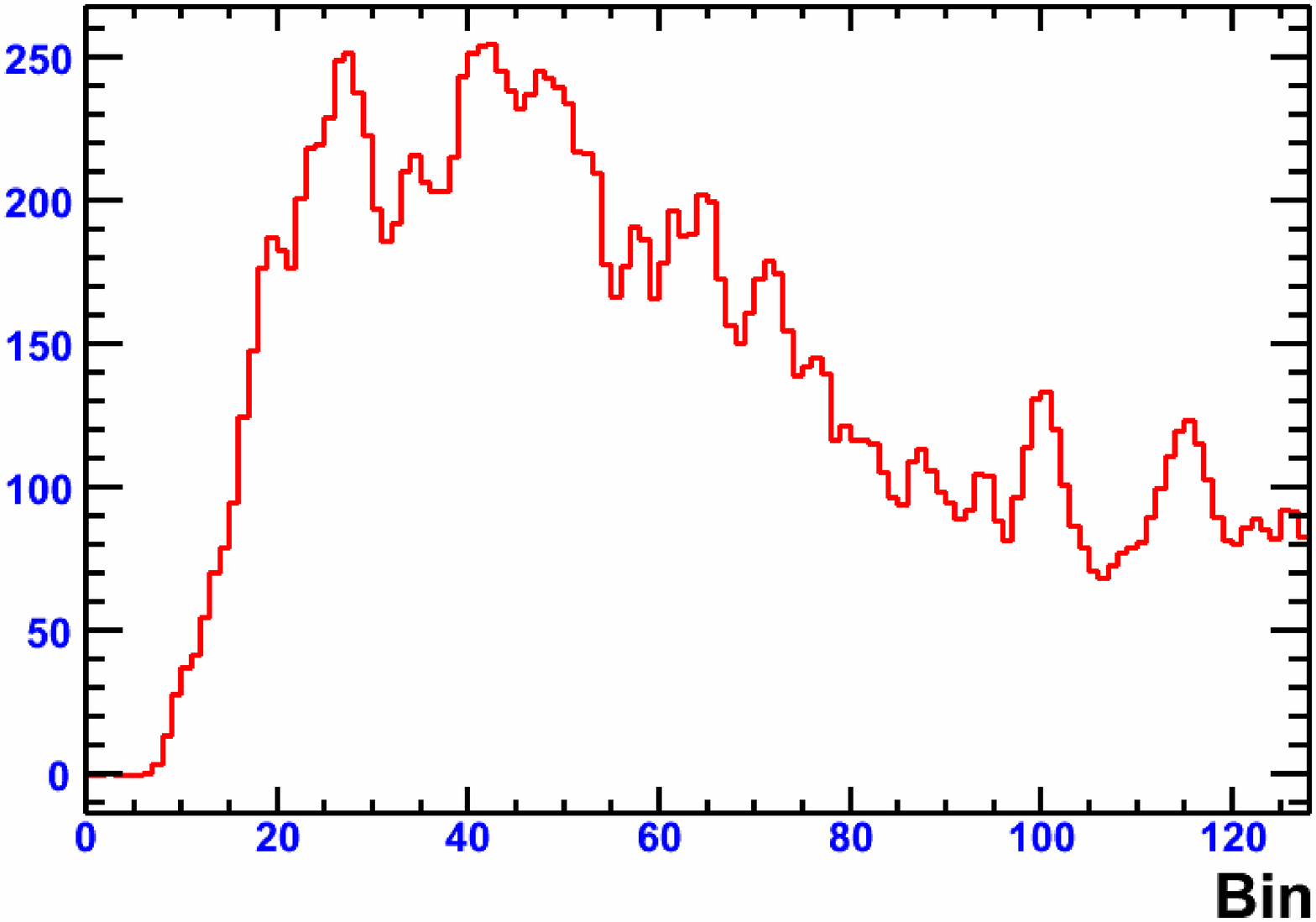}
\end{minipage}
\begin{minipage}[b]{0.32\textwidth}
\centering
\includegraphics[width=5.55cm]{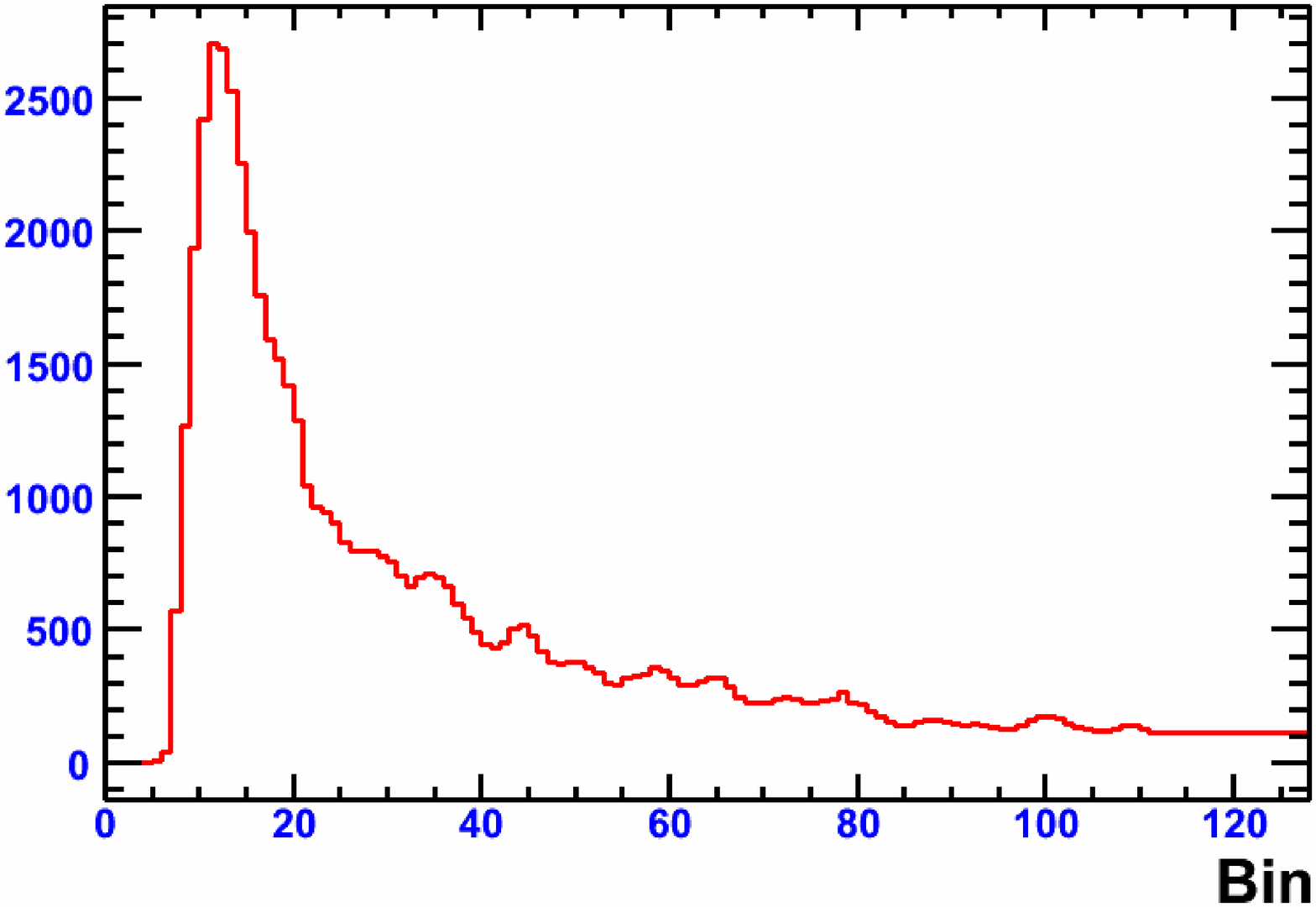}
\end{minipage}
\vspace{0.1cm}
\caption{ATWD waveforms for flasher events compared to the 133 TeV data event.  A flasher event at the widest setting (left), a flasher event at the narrowest setting (center), and the 133 TeV data event (right) are compared for DOM's 39-18 (top), 39-20 (middle), and 39-21 (bottom).  The waveform bin width is 3.3 ns in all plots.}
\label{DOMFlasherWaveformsCompared}
\end{figure}

Next, the pulses in the 133 TeV event are much narrower in time than even the narrowest flasher pulse.  These waveforms are reproduced in figure~\ref{DOMFlasherWaveformsCompared} for a flasher event with the broadest and the narrowest current pulse to the LED's.  

Finally, the timing is not consistent with an in-situ light source. For the flashers, there is a pattern to the hits: first the DOM above the flasher is hit, then the one below the flasher. Next comes the DOM two places above the flasher and then the DOM two places below the flasher. If two DOM's are neighbors, the time difference between them is never less than the direct travel time for 17 m. This, of course, is forbidden by causality.

However, for the 133 TeV event, DOM 39-20 is hit first, and its leading edge time is at 10,000 ns plus $\sim5$ ATWD samples. DOM 39-19's leading edge time is at 10,014.2 ns plus $\sim5$ ATWD samples. The time difference of $\sim14$ ns is far below the direct travel time of 75 ns for 17 m. This means that the source of the light could not have been at the position of a DOM. It had to have either been between DOM's 39-19 and 39-20 on the string, or else somewhere off of the string entirely.  So, we can conclude that the 133 TeV event could not have been a flasher, even an accidentally discharged flasher. 

\newpage

\clearpage

\newpage

\clearpage

These various lines of evidence indicate that this 133 TeV event is truly an in-ice particle physics event and not a detector artifact or an accidental discharge of light from a DOM or an in-situ light source.  

All experience from this analysis as well as previous cascade analyses suggests that it is highly unlikely for a muon with energy greater than 133 TeV to have a large radiative energy loss inside the detector without first leaving early hits on an outside string.  However, until further background Monte Carlo generation is completed, this statement cannot be made more quantitative. 

An obvious question that arises is the following: How likely is such an event from the known atmospheric neutrino flux? The expected number of events above 133 TeV for the livetime of this dataset and for a final cut greater than 0.9 is $0.02 \numu+0.006 \nue=0.03$. The probability of a Poissonian with a mean of 0.03 fluctuating up to 1 or more counts is $2.9\%$. So this event is not incompatible with the conventional atmospheric neutrino flux.

\subsection{Double Pulse Waveforms: A Tau Neutrino?}

\begin{figure}
\begin{minipage}[b]{0.49\textwidth} % A minipage that covers half the page
\centering
\includegraphics[width=7.6cm]{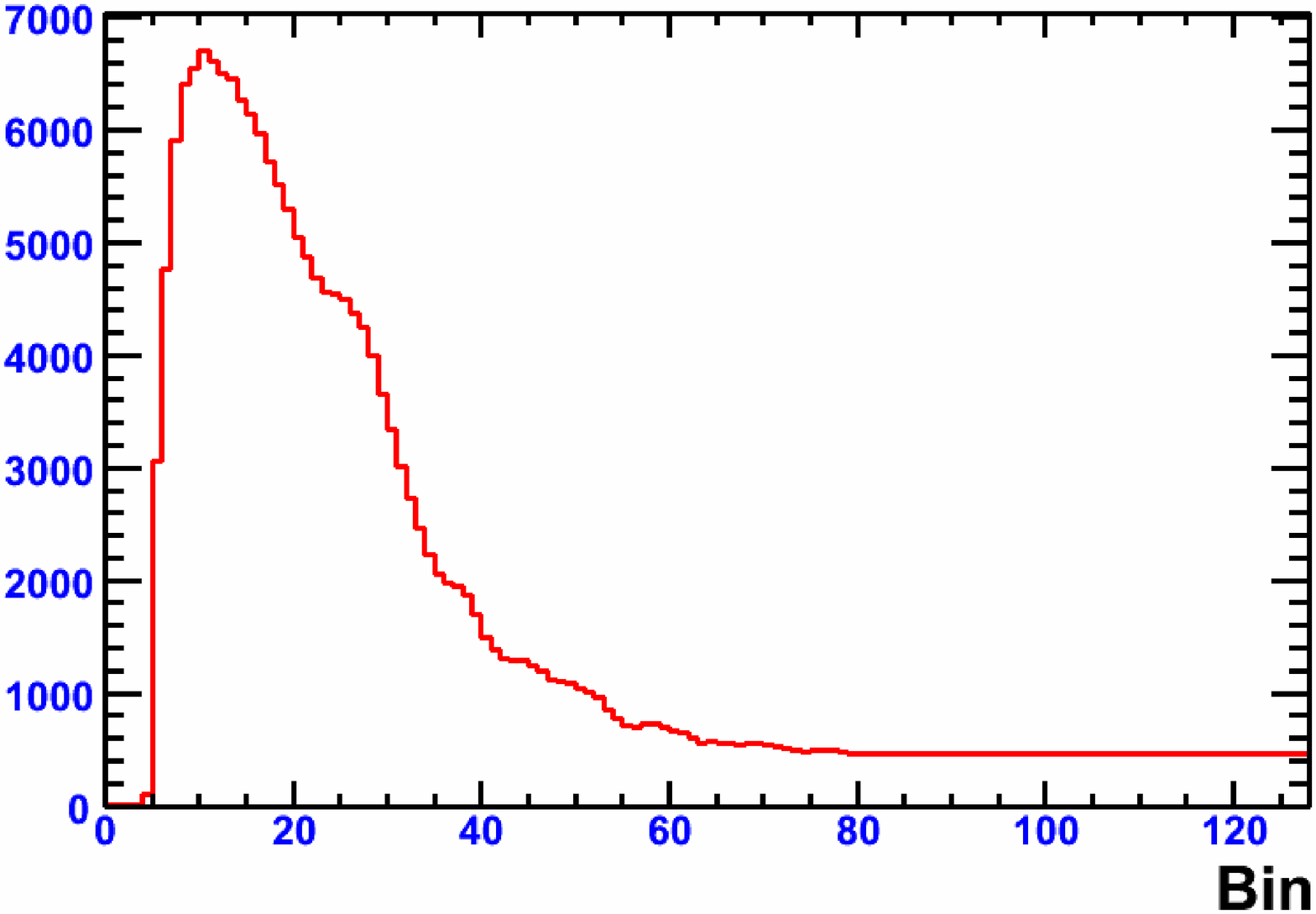}
\end{minipage}
\hspace{0.5cm} %To get a little bit of space between the figures
\begin{minipage}[b]{0.49\textwidth}
\centering
\includegraphics[width=7.6cm]{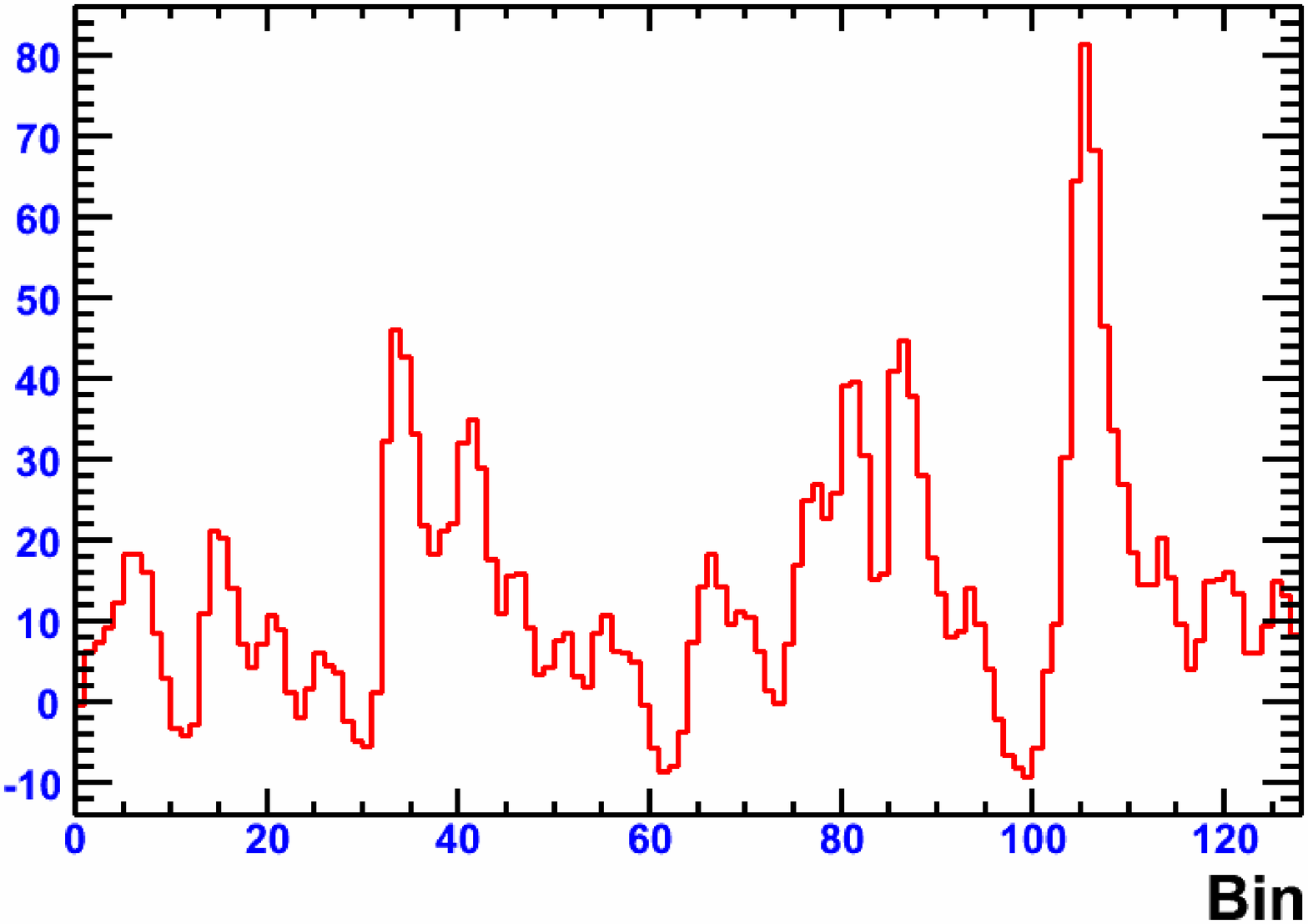}
\end{minipage}

\vspace{0.65cm}

\begin{minipage}[b]{0.49\textwidth} % A minipage that covers half the page
\centering
\includegraphics[width=7.6cm]{figures/eventviewer_images/waveforms/new/DOM_39_20_ATWD_LAUNCH1}
\end{minipage}
\hspace{0.5cm} %To get a little bit of space between the figures
\begin{minipage}[b]{0.49\textwidth}
\centering
\includegraphics[width=7.6cm]{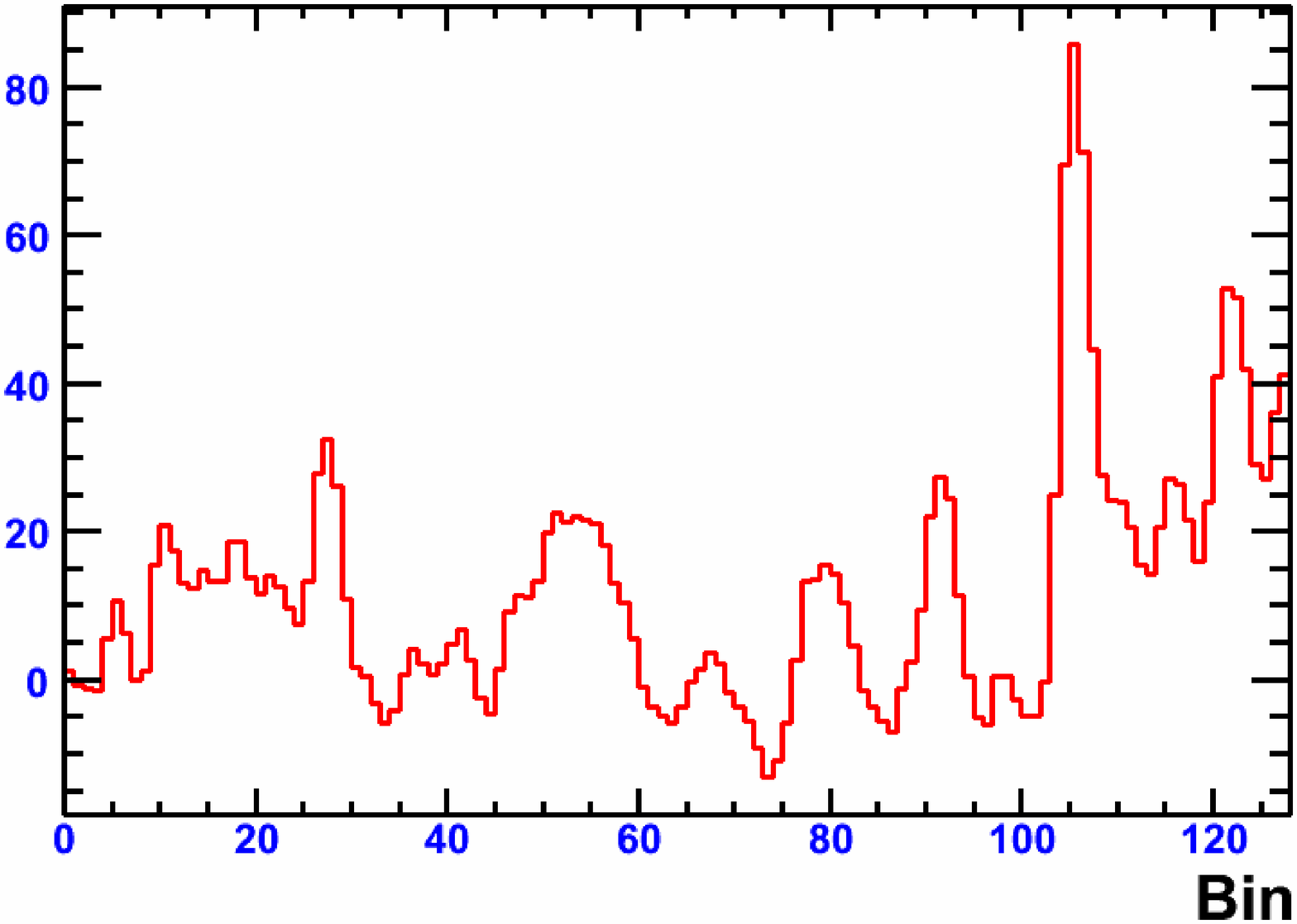}
\end{minipage}
\vspace{0.25cm}
\caption{ATWD waveforms for the 133 TeV event for the first, main DOM launch (left) and the second, later DOM launch (right) for DOM's 39-19 (top) and 39-20 (bottom).  The waveform bin width is 3.3 ns.}
\label{DOM39-19And39-20Waveforms}
\end{figure}

Finally, there is one additional, tantalizing feature of this outlier event.  The waveforms from the two DOM's which receive the most charge in the event are displayed in figure~\ref{DOM39-19And39-20Waveforms}.  Both waveforms have a shoulder, a hint of a second pulse after the main pulse.  This ``double pulse'' signature is one of the main signatures of a tau neutrino double bang.  

\clearpage

\newpage

\clearpage

\newpage

This shoulder is not present in any other DOM's in the event.  It's also not present in the flasher waveforms, which suggests that it's not an artifact of ice layering.  The two pulses are separated by approximately 50 ns.  To check the feasibility of a tau neutrino double bang, we can estimate the energetics of the tau track.  In the best case scenario (the one with the lowest energy tau), the tau track points straight towards the DOM and the two bangs are separated by 50 ns.  To live this long before decaying, the tau would have needed a gamma factor of $\gamma = \delta t/\tau = 50/(2.9\times 10^{-4}) \approx 1.7\times 10^{5}$.  This leads to a tau of energy $E = \gamma m_{\tau} = (1.7\times 10^{5}) \cdot 1.776 \mbox{ GeV} \approx 300 \mbox{ TeV}$.  For an $E^{-2}$ spectrum above 100 TeV with our final cuts, our energy reconstruction has a resolution of 0.3 in log(E) with an offset of the reconstructed energy of -0.2.  The offset would transform a 133 TeV event into a 210 TeV event, and the resolution means the energy would lie between 149 and 297 TeV at one standard deviation.  The energetics, then, are at least on the right scale.  

As exciting as this event is, a full assessment of the backgrounds must be conducted.  It is currently unknown whether two radiative losses along a cosmic ray muon track can mimic a tau, or whether the longitudinal development of a high energy cascade can do so.  Another possible background is a starting muon neutrino event where the outgoing muon immediately suffers a large radiative loss.  Within IceCube, the likelihood reconstruction tools for tau neutrino analysis are just now being completed and tested.  A dedicated tau analysis on the final event sample from this analysis, taking into consideration the possible backgrounds, will be conducted in the coming months by a group at Pennsylvania State University.

\section{Comparison to Other Fluxes}

\begin{figure}
\centering
\includegraphics[width=0.55\linewidth]{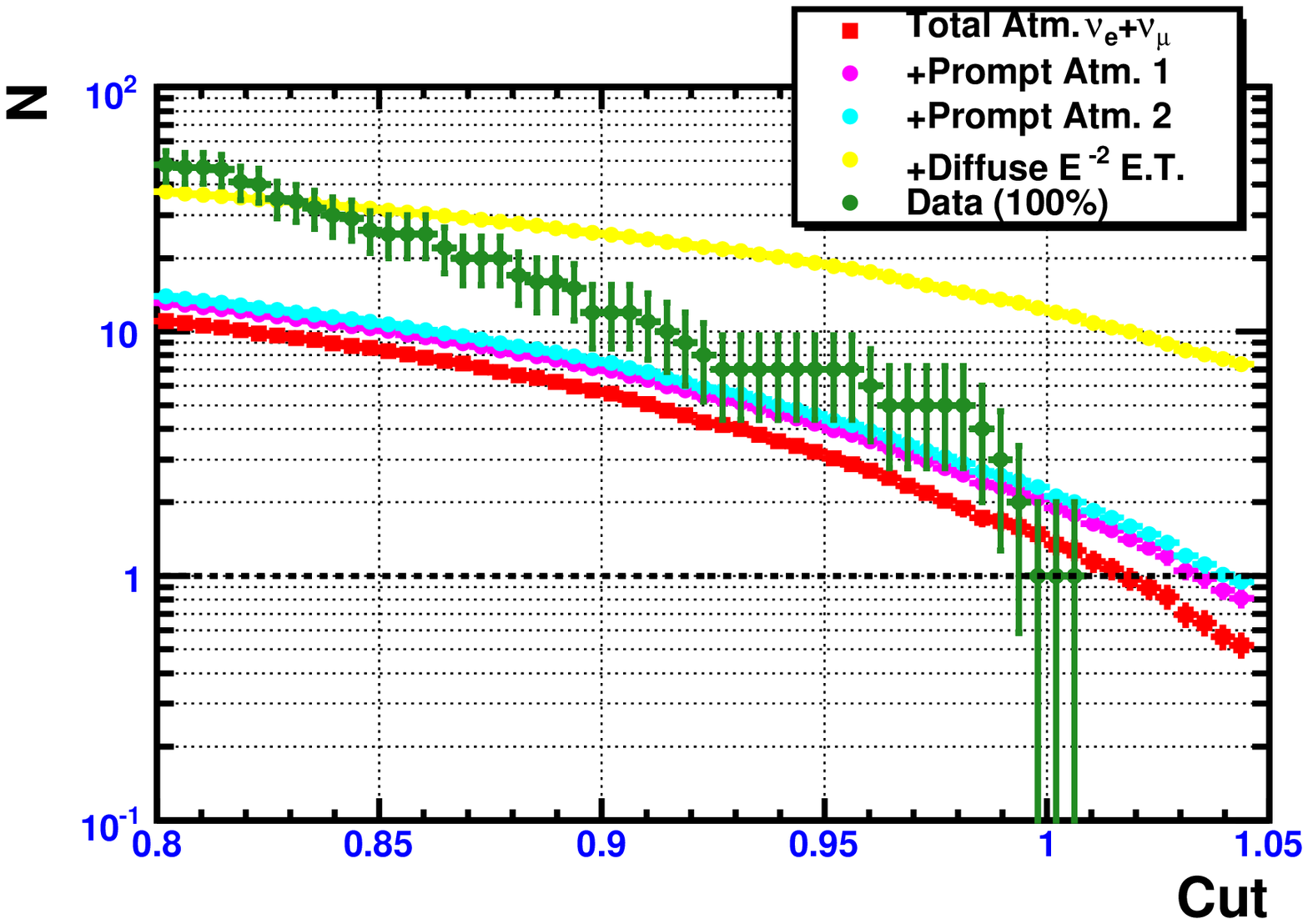}

\vspace{1.2cm}

\centering
\includegraphics[width=0.55\linewidth]{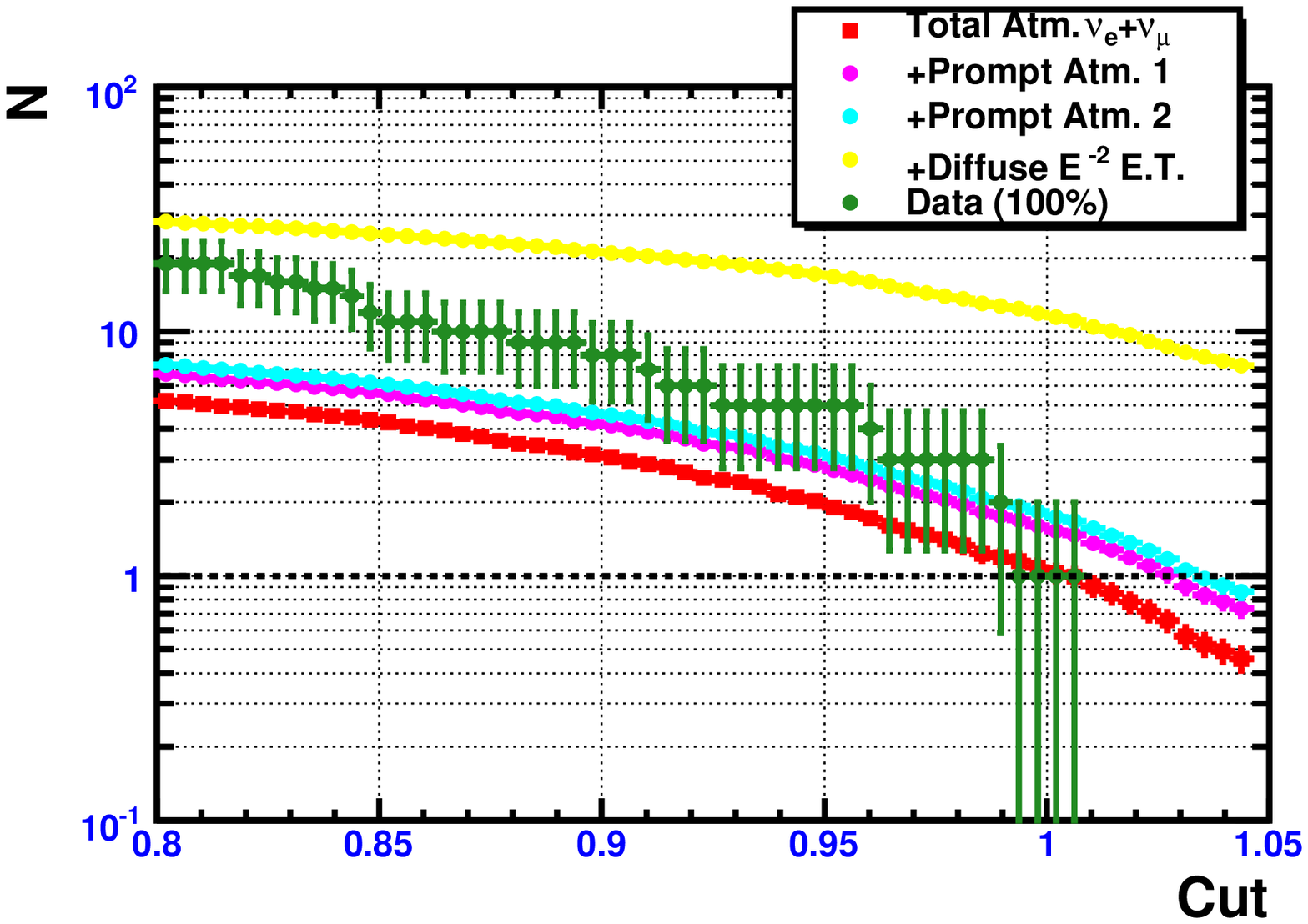}

\vspace{1.2cm}

\centering
\includegraphics[width=0.55\linewidth]{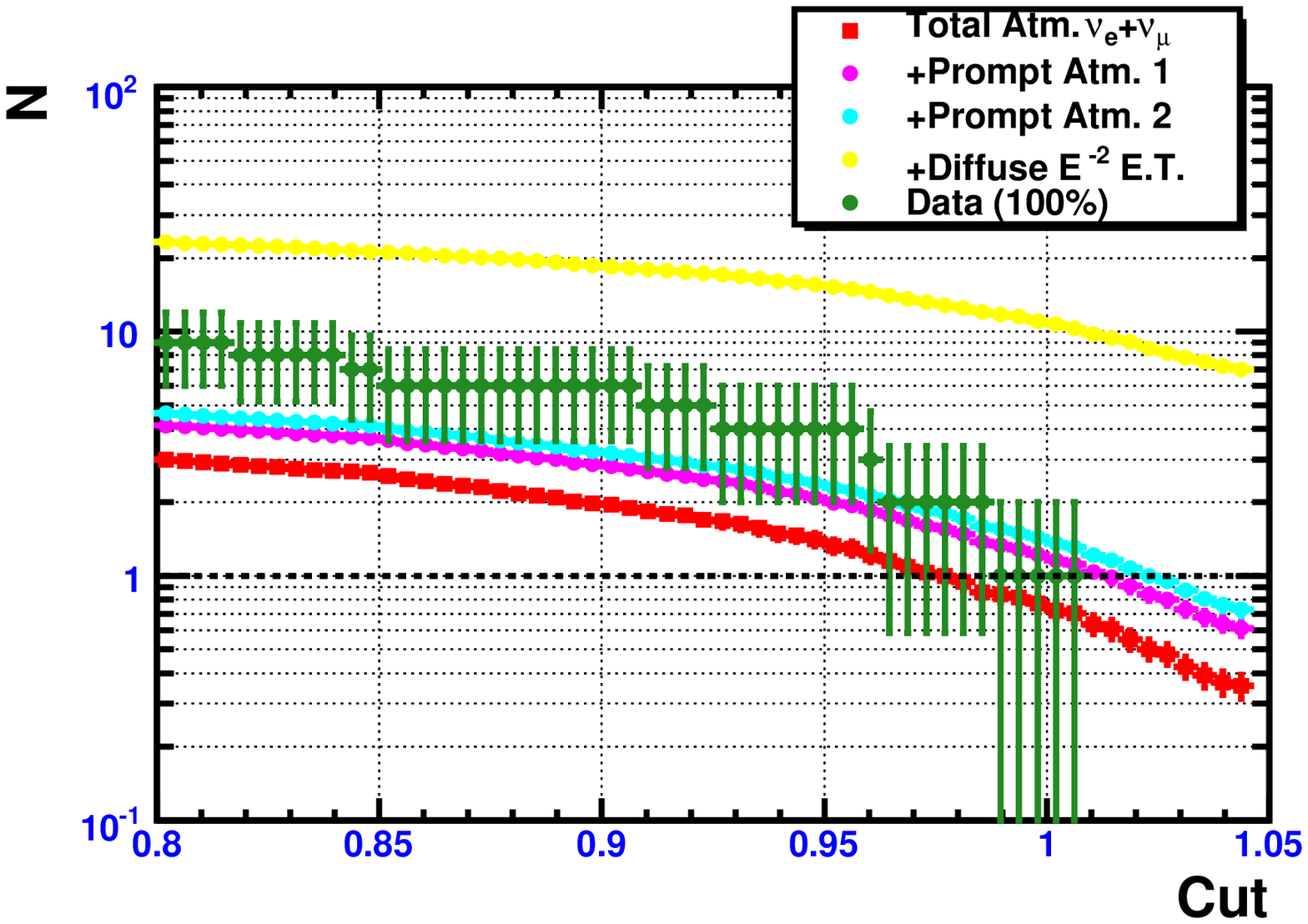}
\vspace{0.8cm}
\caption{Cumulative distributions showing the number of events surviving beyond a given classifier cut for the full IC-22 dataset for 5 TeV (top), 10 TeV (middle), and 15 TeV (bottom) energy cuts.  The signal traces show conventional atmospheric neutrinos (red), conventional plus prompt atmospheric neutrinos according to the models of \cite{NaumovPrompt} (cyan) and \cite{SarcevicPrompt} (magenta), and for conventional atmospheric neutrinos plus an extraterrestrial $E^{-2}$ component at the current best limit \cite{OxanaICRC} (yellow).}
\label{AllFluxesPlot}
\end{figure}

Figure~\ref{AllFluxesPlot} compares the observed data to several neutrino flux predictions.  Red traces show the conventional Bartol atmospheric flux.  Cyan and magenta traces show the sum of the Bartol conventional flux plus the contribution from two prompt neutrino models \cite{SarcevicPrompt, NaumovPrompt}.  Prompt neutrinos do not have a resolvable effect within the statistics of this dataset, but should become resolvable as IceCube increases its instrumented volume.

Finally, the yellow trace shows the prediction for an extraterrestrial $E^{-2}$ neutrino flux at the level of the current best neutrino-induced cascade limit from 5 years of AMANDA data \cite{OxanaICRC}.  The trace is the sum of all muon, electron, and tau neutrino contributions to the cascade signal.  It is clear that our data excludes an extraterrestrial flux at this level and that this analysis is at least as sensitive as the most-sensitive extraterrestrial analysis to date.  However, limit setting is beyond the scope of this dissertation and will be handled by a separate, dedicated extraterrestrial neutrino-induced cascade analysis.  

\clearpage

\newpage

\clearpage

\newpage

\section{Systematic Uncertainties}

One remaining task is to quantify the systematic uncertainties on our atmospheric neutrino-induced cascade signal prediction.  The dominant source of uncertainty is the absolute flux normalization discussed above ($\pm 40\%$ on the Bartol model and $\pm 20\%$ on the measured AMANDA flux).  Additional sources of uncertainty include:

\begin{itemize}
\item{{\bf Ice Properties}: There are uncertainties in our knowledge of the scattering and absorption coefficients of the glacial ice as a function of depth.  Past analyses have found that this contributes an overall uncertainty of $\sim 15\%$.   To quantify the effect of these uncertainties on this analysis, dedicated photon tables and cascade signal simulations will be generated and propagated through the analysis chain.}  

\item{{\bf DOM Efficiencies}: Our understanding of the glass, gel, and PMT efficiencies is based on lab measurements of a handful of DOM's.  The simulation assumes an average value, and variations around this average are expected to be around $\pm 10\%$.  Again, to quantify the effect of this uncertainty, cascade signal simulations will be generated with varied DOM efficiencies and propagated through the analysis chain.}  

\item{{\bf Cross Sections}: The uncertainty on the neutrino-nucleon cross sections has been estimated to be $\sim 5\%$ \cite{Pumplin}.}

\item{{\bf Absolute Energy Scale}: The performance of the energy reconstruction was validated using studies of in-situ flasher light sources (see section~\ref{SPerformance}).  However, additional work must be conducted to show that the energy reconstruction performs the same way on data and simulation.}
\end{itemize}

%% file: chapters/conclusion.tex
\chapter{Conclusions and Implications}\label{chapter:conclusion}

With this analysis, we now have the first evidence that IceCube has the ability to see atmospheric neutrino-induced cascades.  Above a reconstructed energy of 5~TeV, 12 events were observed in the full dataset.  The signal expectation from the canonical Bartol atmospheric neutrino flux model is $5.63\pm2.25$ events, while the expectation from the atmospheric neutrino flux as measured by IceCube's predecessor array AMANDA is $7.48\pm1.50$ events.  Quoted errors include the uncertainty on the flux only.  These results are consistent with the atmospheric flux prediction plus a small amount of residual background contamination.  Additional background Monte Carlo is needed in order to assess the purity of the final event sample.  However, such a time intensive undertaking (six months or more on a computer cluster consisting of hundreds of cores) could not be completed on the timescale of this dissertation.

The techniques developed in this analysis will prove useful for searches for astrophysical neutrinos that interact through the cascade channel.  Already, the online filter and early cuts developed for this analysis have been used for a search for cascades from a high energy diffuse flux of astrophysical neutrinos \cite{JoannaICRC}.  Additional cuts developed for this analysis are currently being used to search the full sky for cascades from high energy neutrinos from gamma ray bursts \cite{GRBICRC}.  Work is underway to loosen the final cuts from this analysis to provide an event sample to search for correlations with flares from active galactic nuclei over the full sky.  Finally, the possible tau neutrino candidate event will be the subject of a dedicated tau analysis in the coming months.  

\section{Implications for Future Detectors}
These results are also encouraging for future cascade analyses with IceCube.  At the time of this writing, the 40-string configuration of IceCube (IC-40) has already completed its physics run.  IC-40 has twice the instrumented volume of IC-22, and the dataset has 25\% more detector uptime.  That should give a factor of 2.25 improvement even without further refinement of the analysis techniques.  The IC-40 analysis could expect to have around 50 neutrino candidate events in the cascade channel.  

The work in this dissertation has also convinced the IceCube collaboration of the urgency of producing more background Monte Carlo.  That work is already underway for the IC-40 analysis.   

At the time of this writing, the 59-string configuration of IceCube (IC-59) is currently taking physics data.  The data acquisition system for IC-59 was upgraded and should result in a significant improvement for cascade analysis.  The IC-22 and IC-40 physics runs contained a hard local coincidence condition in the triggering.  That is, DOM's sent data to the surface only if one of their two nearest or two next-to-nearest neighbors on the string also received light in the event.  For IC-59, the physics run is being taken with a soft local coincidence condition in the triggering.  That is, if a DOM receives light but none of its neighbors do, it sends a reduced subset of the event information to the surface instead of the full waveforms.  This subset of information includes a time stamp and the total received charge.  

Monte Carlo studies conducted for this analysis suggest that soft local coincidence effectively lowers the threshold for muons to leave early hits on outer strings of the detector.  This should significantly improve our ability to veto the muons with large stochastic energy losses that are the main background for cascade searches.

In its final 80-string configuration, IceCube will be considerably better at detecting cascades.  In addition to its quadrupled volume, superior uptime, and soft local coincidence data acquisition, its shape will lead to non-linear increases in background rejection ability.  The amount of contained volume in IC-22, IC-40, and even IC-59 is not very large.  When IC-80 is complete, there will be a large inner region surrounded by at least two layers of strings.  This should dramatically improve our ability to veto muons with large radiative losses.  Using the reconstruction and analysis techniques developed in this dissertation, IC-80 could expect to see many hundreds of atmospheric neutrino-induced cascades per year.  This should allow IceCube to probe models of charm production in the atmosphere.  And in the event that IceCube detects astrophysical neutrinos, measurements in the cascade channel should allow us to probe the flavor ratio, perhaps telling us something about new physics or the conditions in the violent astrophysical sources that produce neutrinos.

%% file: appendices/appendix1.tex
\chapter{Oscillations and the Astrophysical Flavor Ratio}\label{appendix:appendix1}

With some simple arguments, we can understand how neutrino oscillations take an astrophysical flux ratio of

$$
(\numu:\nue:\nutau)_{\mbox{\fontsize{8}{14}\selectfont source}} \approx 2:1:0
$$.  

\noindent and transform it to

$$
(\numu:\nue:\nutau)_{\mbox{\fontsize{8}{14}\selectfont earth}} \approx 1:1:1
$$

First, we're talking about distances on the scales of kiloparsecs and higher ($1\mbox{ kpc}\approx 3\times 10^{16}\mbox{ km}$) and neutrinos on the order of hundreds of GeV or higher.  The  measured values of $\Delta m^2$ are on the order of $10^{-3}$ and $10^{-5}$.  So the phases of the $\sin$ terms in the oscillation probability are

\begin{align*}
&\frac{1.27 \cdot 10^{-3} \cdot 3 \times 10^16}{10^2} \sim 10^{11} \\
&\frac{1.27 \cdot 10^{-5} \cdot 3 \times 10^16}{10^2} \sim 10^{9}  \\
\end{align*}

\noindent  These phases are extremely large, so we can safely replace the $\sin^2$ term with its average value, $1/2$.  

Second, we know that $\theta_{13}$ is very small, so we can safely take it to be zero.  All imaginary CP-violating phases in the mixing matrix are proportional to $\sin \theta_{13}$, so that means that we're taking the mixing matrix to be fully real.  Second, we know that $\sin^2 2\theta_{23}=\sin^2 2\theta_{atm} \approx 1$, which means that $\stwothreesq=\ctwothreesq = 1/2$.  Also, $\tan^2 \theta_{12} \approx 0.47 \approx 1/2$, so 

\begin{align*}
1+\tan^2 \theta_{12} = \sec^2 \theta_{12} \Rightarrow & \conetwosq = 2/3 \\
                                                                                                & \sonetwosq = 1/3 \\
\end{align*}

Now we need to calculate the terms $\Re(U^*_{\alpha i} U_{\beta i} U_{\alpha j} U^*_{\beta j})= U_{\alpha i} U_{\beta i} U_{\alpha j} U_{\beta j}$ (since $U$ is real).  For $\alpha=\mu, \beta=\mu$ we have

\begin{align*}
U_{\mu i} U_{\mu i} U_{\mu j} U_{\mu j} &  \\
& = U_{\mu 2} U_{\mu 2} U_{\mu 1} U_{\mu 1}  + U_{\mu 3} U_{\mu 3} U_{\mu 1} U_{\mu 1}  + U_{\mu 3} U_{\mu 3} U_{\mu 2} U_{\mu 2}  \\
& = (\conetwo \ctwothree \sonetwo \ctwothree)^2 + (\stwothree \sonetwo \ctwothree)^2 + (\stwothree \conetwo \ctwothree)^2  \\
& = (\conetwo \ctwothree \sonetwo \ctwothree)^2+(\ctwothree \stwothree)^2  \\
& = \ctwothreesq(\ctwothreesq \sonetwosq \conetwosq+\stwothree)^2 \\
& = \frac{1}{2}(\frac{1}{2}\frac{1}{3}\frac{2}{3}+\frac{1}{2}) = \frac{11}{36}
\end{align*}

\noindent Since $\sonethree=0$, for $\alpha=\mu, \beta=e$ which is the same as $\alpha=e, \beta=\mu$ we have

\begin{align*}
U_{\mu i} U_{e i} U_{\mu j} U_{e j} &  \\
& = U_{\mu 2} U_{e 2} U_{\mu 1} U_{e 1} \\
& = -\conetwo \ctwothree \sonetwo \sonetwo \ctwothree \conetwo \\
& = -\conetwosq \ctwothreesq \sonetwosq \\
& = -\frac{2}{3}\frac{1}{2}\frac{1}{3} \\
& = -\frac{1}{9}
\end{align*}

\noindent For the same reason, for $\alpha=e, \beta=e$ we have

\begin{align*}
U_{e i} U_{e i} U_{e j} U_{e j} &  \\
& = U_{e 2} U_{e 2} U_{e 1} U_{e 1} \\
& = \sonetwo \sonetwo \conetwo \conetwo \\
& = \sonetwosq \conetwosq \\
& = \frac{2}{3}\frac{1}{3} \\
& = -\frac{2}{9}
\end{align*}

\noindent We can now write down the $\numu$ and $\nue$ oscillation probabilities:

\begin{align*}
&P(\numu \rightarrow \numu) = 1-4 \frac{1}{2} \frac{11}{36} =  \frac{7}{18}  \\
&P(\numu \rightarrow \nue) = P(\nue \rightarrow \numu) = -4 \frac{1}{2} -\frac{1}{9} =  \frac{4}{18}  \\
&P(\numu \rightarrow \nutau) = 1- P(\numu \rightarrow \numu) - P(\numu \rightarrow \nue) = 1- \frac{7}{18}  - \frac{4}{18} =  \frac{7}{18}  \\ 
&P(\nue \rightarrow \nue) = 1-4 \frac{1}{2} \frac{2}{9} =  \frac{10}{18}  \\
&P(\nue \rightarrow \nutau) = 1- P(\nue \rightarrow \nue) - P(\numu \rightarrow \numu) = 1- \frac{10}{18}  - \frac{4}{18} =  \frac{4}{18}  \\ 
\end{align*}

\noindent So, if we start with two muon neutrinos and one electron neutrino, we'll have

\begin{align*}
&N_{\numu} = 2P(\numu \rightarrow \numu) + P(\nue \rightarrow \numu) = 2 \frac{7}{18} + \frac{4}{18} = 1 \\
&N_{\nue} = 2P(\numu \rightarrow \nue) + P(\nue \rightarrow \nue) = 2 \frac{4}{18} \frac{10}{18} = 1 \\
&N_{\nutau} = 2P(\numu \rightarrow \nutau) + P(\nue \rightarrow \nutau) = 2 \frac{7}{18} \frac{4}{18} = 1 \\
\end{align*}

\noindent Therefore the final flux ratio after oscillations is $1:1:1$.

%% file: appendices/appendix2.tex
\chapter{Effective Livetime For Weighted {\tt CORSIKA} Monte Carlo}\label{appendix:appendix2}

There's been some confusion and discussion within the collaboration lately about how exactly to define an effective livetime for weighted corsika.  This note is my attempt to summarize my thinking on the subject.

\section{Fundamental Definition}
For a given number of weighted corsika files, it's handy to define an effective livetime as a figure of merit.  That way, one can say, for example, that a certain amount of weighted Monte Carlo corresponds to several hundred days of effective livetime at a given energy.

The basic definition is this: The effective livetime is the amount of unweighted simulation you'd need to get the same relative error bars as the weighted simulation gives you.

\section{Simple Example}

For me, the easiest way to understand this is by considering a simple example.  Suppose that I simulate $T_{\mbox{\fontsize{8}{14}\selectfont uw}}=900s$ of detector livetime with unweighted corsika.  This results in $N_{\mbox{\fontsize{8}{14}\selectfont uw}}=9$ events passing a certain set of analysis cuts.  This can be in a given bin when binned with respect to some quantity (e.g. energy) or summed over all bins to get a total number of events.  The relative error bar is given by

$$
\frac{\sqrt{N_{\mbox{\fontsize{8}{14}\selectfont uw}}}}{N_{\mbox{\fontsize{8}{14}\selectfont uw}}}=\frac{1}{\sqrt{N_{\mbox{\fontsize{8}{14}\selectfont uw}}}}=\frac{1}{3}
$$

Now, let's assume that we introduce some weighting scheme.  This results in $N_{\mbox{\fontsize{8}{14}\selectfont w}}=9000$ events passing our cuts, each with a weight $w_{i}=10^{-3}$.  Our prediction for the number of events stays the same, since we have

$$
\sum_{i}w_{i}=9000\times10^{-3}=9
$$

\noindent The error bar is given by

$$
\sqrt{\sum_{i}w_{i}^{2}}=\sqrt{9000\times10^{-6}}\approx0.1
$$

\noindent The relative error bar is given by

$$
\frac{\sqrt{\sum_{i}w_{i}^{2}}}{\sum_{i}w_{i}}\approx0.01
$$

\noindent As we can see, the weighting has reduced our relative error bar, which, after all, was the entire point of introducing the weighting in the first place.

To determine the effective livetime of this weighted Monte Carlo, we ask ourselves the following question:  How many unweighted events $N_{\mbox{\fontsize{8}{14}\selectfont eff}}$ would we need to give us the same relative error bar as we have with our weighted Monte Carlo?  Setting the relative error bars equal to each other, we have

$$
\frac{1}{\sqrt{N_{\mbox{\fontsize{8}{14}\selectfont eff}}}}=\frac{\sqrt{\sum_{i}w_{i}^{2}}}{\sum_{i}w_{i}} \Rightarrow
N_{\mbox{\fontsize{8}{14}\selectfont eff}}=\left(\frac{\sum_{i}w_{i}}{\sqrt{\sum_{i}w_{i}^{2}}}\right)^{2}
$$

For the example we're considering here, $N_{\mbox{\fontsize{8}{14}\selectfont eff}}\approx10,000$.  Now it's clear how to get the effective livetime: it's the time it would take to get $N_{\mbox{\fontsize{8}{14}\selectfont eff}}$ events out of an unweighted simulation which generates $N_{\mbox{\fontsize{8}{14}\selectfont uw}}$ events in $T_{\mbox{\fontsize{8}{14}\selectfont uw}}$ seconds.  So we finally have the result we were after:

%\begin{equation}
\boxedeqn{
T_{\mbox{\fontsize{8}{14}\selectfont eff}}=N_{\mbox{\fontsize{8}{14}\selectfont eff}}\left(\frac{T_{\mbox{\fontsize{8}{14}\selectfont uw}}}{N_{\mbox{\fontsize{8}{14}\selectfont uw}}}\right)=\left(\frac{\sum_{i}w_{i}}{\sqrt{\sum_{i}w_{i}^{2}}}\right)^{2}\left(\frac{T_{\mbox{\fontsize{8}{14}\selectfont uw}}}{N_{\mbox{\fontsize{8}{14}\selectfont uw}}}\right)
}
%\end{equation}

\noindent For our example, $T_{\mbox{\fontsize{8}{14}\selectfont eff}}\approx10^{6}s$.

This result scales with the number of weighted corsika files we have, as one would expect.  If the sum of weights in one file is $\sum_{i}w_{i}$ then for $n$ files it's $\approx n\times\sum_{i}w_{i}$.  Likewise, if the sum of the squares of the weights for one file is $\sum_{i}w_{i}^{2}$, then for $n$ files it's $\approx n\times\sum_{i}w_{i}^{2}$.  So the livetime for $n$ files is

$$
T_{\mbox{\fontsize{8}{14}\selectfont eff}}=\left(\frac{n\times\sum_{i}w_{i}}{\sqrt{n\times\sum_{i}w_{i}^{2}}}\right)^{2}\frac{T_{\mbox{\fontsize{8}{14}\selectfont uw}}}{N_{\mbox{\fontsize{8}{14}\selectfont uw}}}
=n\times\left(\frac{\sum_{i}w_{i}}{\sqrt{\sum_{i}w_{i}^{2}}}\right)^{2}\frac{T_{\mbox{\fontsize{8}{14}\selectfont uw}}}{N_{\mbox{\fontsize{8}{14}\selectfont uw}}}
$$

\noindent In terms of {\tt IceTray}-speak, the weights $w_{i}$ are given by

$$
w_{i}=\frac{\mbox{weight} \times\mbox{DiplopiaWeight}}{\mbox{TimeScale}}
$$

\noindent or simply by

$$
w_{i}=\mbox{weight} \times \mbox{DiplopiaWeight}
$$

\noindent since the constant $\mbox{TimeScale}$ cancels in the numerator and denominator of equation (B.1).

\section{Addendum}
There's actually a cancellation in equation (B.1) that results in a simpler expression for the effective livetime.  Since the {\tt IceTray} weights 

$$
w_{i}=\frac{\mbox{weight} \times \mbox{DiplopiaWeight}}{\mbox{TimeScale}}
$$

\noindent are defined by

$$
\sum_{i}w_{i} = \sum_{i}\frac{\mbox{weight} \times \mbox{DiplopiaWeight}}{\mbox{TimeScale}}= \mbox{rate in Hz}=\frac{N_{\mbox{\fontsize{8}{14}\selectfont uw}}}{T_{\mbox{\fontsize{8}{14}\selectfont uw}}}
$$

\noindent we can write equation (1) as

$$
T_{\mbox{\fontsize{8}{14}\selectfont eff}}=\left(\frac{\sum_{i}w_{i}}{\sqrt{\sum_{i}w_{i}^{2}}}\right)^{2}\left(\frac{T_{\mbox{\fontsize{8}{14}\selectfont uw}}}{N_{\mbox{\fontsize{8}{14}\selectfont uw}}}\right)
=\left(\frac{\sum_{i}w_{i} \times \sum_{i}w_{i}}{\sum_{i}w_{i}^{2}}\right)\left(\frac{T_{\mbox{\fontsize{8}{14}\selectfont uw}}}{N_{\mbox{\fontsize{8}{14}\selectfont uw}}}\right)
=\left(\frac{\sum_{i}w_{i}}{\sum_{i}w_{i}^{2}}\right)\left(\frac{\cancel{N_{\mbox{\fontsize{8}{14}\selectfont uw}}}}{\cancel{T_{\mbox{\fontsize{8}{14}\selectfont uw}}}}\right)\left(\frac{\cancel{T_{\mbox{\fontsize{8}{14}\selectfont uw}}}}{\cancel{N_{\mbox{\fontsize{8}{14}\selectfont uw}}}}\right)
$$

\noindent from the above definition of the {\tt IceTray} weights.  So we finally have an equivalent, simpler formulation of the effective livetime that doesn't rely on any unweighted Monte Carlo:

\boxedeqn{
T_{\mbox{\fontsize{8}{14}\selectfont eff}}=\left(\frac{\sum_{i}w_{i}}{\sum_{i}w_{i}^{2}}\right)
}

\begin{figure}
\begin{center}
\includegraphics[width=0.8\textwidth]{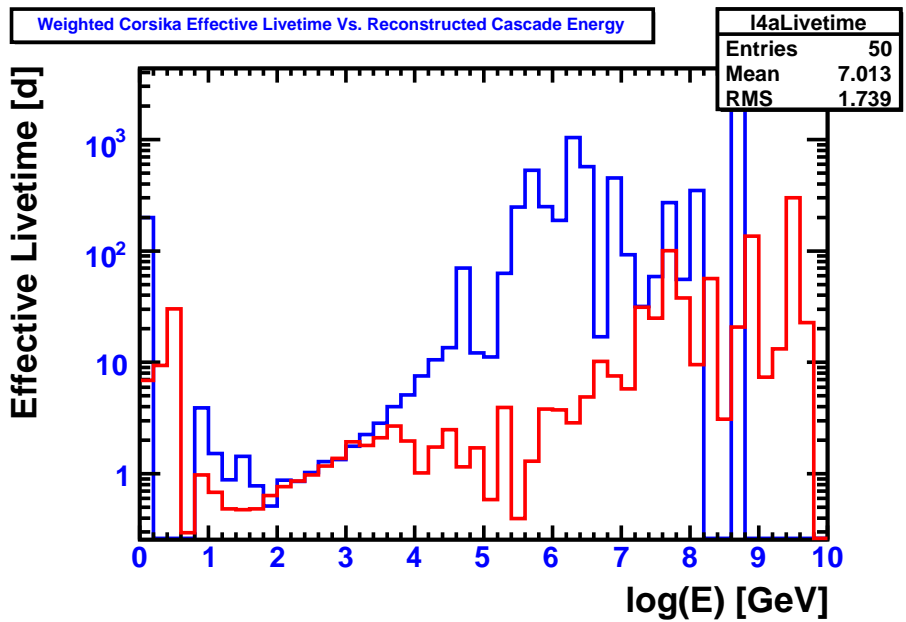}
\caption{This figure shows the effective livetime for 53,512 files from the weighted corsika dataset 1541 as a function of reconstructed cascade energy.  The red trace is for the common cascade level3.  The blue trace is for level4 of the atmospheric analysis.}
\end{center}
\end{figure}

\begin{figure}
\begin{center}
\includegraphics[width=0.8\textwidth]{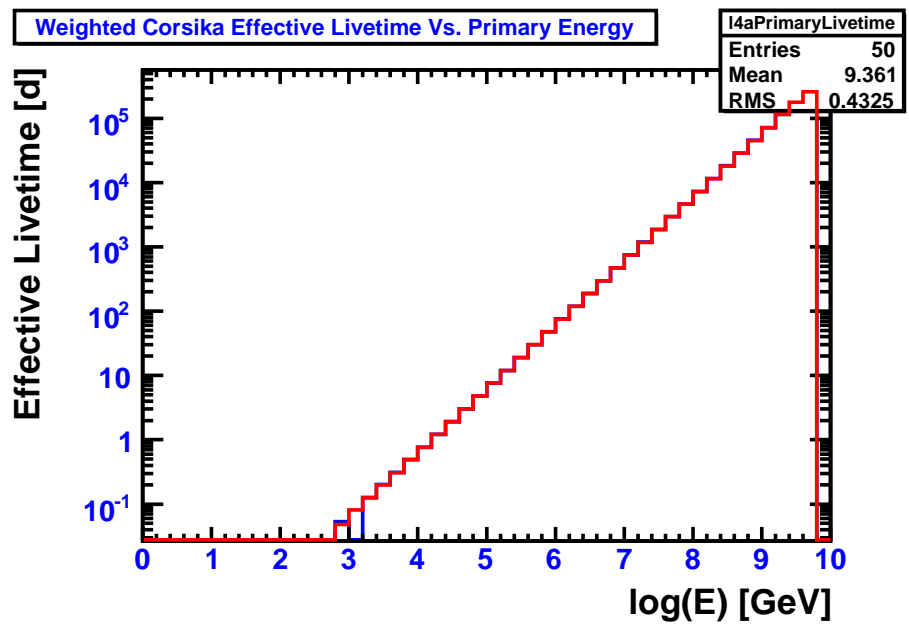}
\caption{Same as above, but plotted as a function of primary energy.}
\end{center}
\end{figure}

%% file: appendices/appendix3.tex
\chapter{Monte Carlo Event Weighting}\label{appendix:appendix3}

In order to normalize neutrino Monte Carlo distributions to the actual number of predicted events from a given neutrino flux for a given observation period, it is necessary to apply a weight to each Monte Carlo event.  This appendix discusses that event weighting procedure.

In general, the expression for number of expected events from a time-independent flux $\Phi(\Omega, E)$ is given by

$$
N = T \int d\Omega~dE~\Phi(\Omega, E) \times A_{\mbox{\fontsize{8}{14}\selectfont eff}}(\Omega, E)
$$

\noindent where $T$ is the observation time, $\Omega$ is solid angle, and $A_{\mbox{\fontsize{8}{14}\selectfont eff}}$ is the so-called effective area of the detector.  Letting $x = \log E$ we have

$$
E = 10^x \Rightarrow \ln E = x~\ln 10 \Rightarrow x = \frac{\ln E}{\ln 10} \Rightarrow dx = \frac{1}{E~\ln 10}dE \Rightarrow dE = E~\ln 10~d(\log E)
$$

\noindent  Assuming the flux is independent of the azimuth $\phi$ and using the result above to transform to $\log E$ we have

$$
N = 2\pi T\int d(\cos\theta)~d(\log E)~E~\ln 10~\Phi(\cos\theta, \log E) \times A_{\mbox{\fontsize{8}{14}\selectfont eff}}(\cos\theta, \log E)
$$

The effective area $A_{\mbox{\fontsize{8}{14}\selectfont eff}}$ represents how well our detector captures neutrinos as a function of energy and angle and is determined from Monte Carlo simulations.  We ``throw'' neutrinos of an energy $E$ from an angle $\theta$ at our detector, forcing them to interact, and see how many of them trigger the detector.  The effective area is then defined as

$$
A_{\mbox{\fontsize{8}{14}\selectfont eff}}(\cos\theta, \log E) = \frac{dN_{\mbox{\fontsize{8}{14}\selectfont trigg}}}{dN_{\mbox{\fontsize{8}{14}\selectfont thrown}}} (\cos\theta, \log E) \times W
$$

\noindent where $W$ is an interaction probability weight that encapsulates how likely our forced neutrino interaction is to occur.  We can re-write this expression as

$$
A_{\mbox{\fontsize{8}{14}\selectfont eff}}(\cos\theta, \log E) = \frac{dN_{\mbox{\fontsize{8}{14}\selectfont trigg}}/d(\cos\theta)d(\log E)}{dN_{\mbox{\fontsize{8}{14}\selectfont thrown}}/d(\cos\theta)d(\log E)} (\cos\theta, \log E) \times W
$$

\noindent Substituting this expression for $A_{\mbox{\fontsize{8}{14}\selectfont eff}}$ into the integral for $N$ above, the factor $d(\cos\theta)d(\log E)$ in the numerator will cancel the variables of integration.  We're left with an integral over $dN_{\mbox{\fontsize{8}{14}\selectfont trigg}}$:

$$
N = 2\pi T\int dN_{\mbox{\fontsize{8}{14}\selectfont trigg}}~\left[\frac{E~\ln 10 \times W \times \Phi(\cos\theta, \log E)}{dN_{\mbox{\fontsize{8}{14}\selectfont thrown}}/d(\cos\theta)d(\log E)}\right]
$$

\noindent It's clear now that the expression in square brackets is our event weight, which we apply to each triggered event before we sum over all triggered events to get the actual number of expected events.  

The factor in the denominator of the weight is determined by the parameters of the thrown Monte Carlo.  Typically, this thrown Monte Carlo has either an $E^{-1}$ or $E^{-2}$ energy spectrum.  For a thrown Monte Carlo spectrum of $E^{-1}$ we have

$$
\frac{dN_{\mbox{\fontsize{8}{14}\selectfont thrown}}}{d(\cos\theta)d(\log E)} = A
$$

\noindent $A$ here is the normalization constant.  Integrating over $\cos\theta$ from $-1$ to $1$ and over $\log E$ from $\log E_{\mbox{\fontsize{8}{14}\selectfont min}}$ to $\log E_{\mbox{\fontsize{8}{14}\selectfont max}}$ and requiring the total number of thrown events to be $N_{\mbox{\fontsize{8}{14}\selectfont thrown}}$ we get an expression for A:

$$
\frac{dN_{\mbox{\fontsize{8}{14}\selectfont thrown}}}{d(\cos\theta)d(\log E)} = A = \frac{N_{\mbox{\fontsize{8}{14}\selectfont thrown}}}{2~\log\frac{E_{\mbox{\fontsize{8}{14}\selectfont max}}}{E_{\mbox{\fontsize{8}{14}\selectfont min}}}}
$$

One more subtlety remains.  Monte Carlo generation typically throws an equal number of $\nu$ and $\bar{\nu}$ together.  If our flux is expressed in terms of the sum of $\nu+\bar{\nu}$, this expression for $A$ is fine.  However, if our flux is expressed in terms of $\nu$ and $\bar{\nu}$ separately, as is the case for atmospheric fluxes, then we need to add a factor of 2 to account for the fact that we've actually thrown $N_{\mbox{\fontsize{8}{14}\selectfont thrown}}/2$ $\nu$ and $N_{\mbox{\fontsize{8}{14}\selectfont thrown}}/2$ $\bar{\nu}$.  In this case $A$ is given by

$$
\frac{dN_{\mbox{\fontsize{8}{14}\selectfont thrown}}}{d(\cos\theta)d(\log E)} = A = \frac{N_{\mbox{\fontsize{8}{14}\selectfont thrown}}}{2\times 2~\log\frac{E_{\mbox{\fontsize{8}{14}\selectfont max}}}{E_{\mbox{\fontsize{8}{14}\selectfont min}}}}
$$

For a thrown Monte Carlo spectrum of $E^{-2}$ we have

$$
\frac{dN_{\mbox{\fontsize{8}{14}\selectfont thrown}}}{d(\cos\theta)d(\log E)} = \frac{B~\ln 10}{E}
$$

\noindent Again, integrating over $\cos\theta$ and $\log E$ we get an expression for B:

$$
B = \frac{N_{\mbox{\fontsize{8}{14}\selectfont thrown}}}{2}\frac{E_{\mbox{\fontsize{8}{14}\selectfont min}}~E_{\mbox{\fontsize{8}{14}\selectfont max}}}{E_{\mbox{\fontsize{8}{14}\selectfont max}}-E_{\mbox{\fontsize{8}{14}\selectfont min}}}
$$

\noindent and again, if the flux is expressed separately in terms of $\nu$ and $\bar{\nu}$ we get an extra factor of two:

$$
B = \frac{N_{\mbox{\fontsize{8}{14}\selectfont thrown}}}{2\times 2}\frac{E_{\mbox{\fontsize{8}{14}\selectfont min}}~E_{\mbox{\fontsize{8}{14}\selectfont max}}}{E_{\mbox{\fontsize{8}{14}\selectfont max}}-E_{\mbox{\fontsize{8}{14}\selectfont min}}}
$$

\noindent Our expression for $\frac{dN_{\mbox{\fontsize{8}{14}\selectfont thrown}}}{d(\cos\theta)d(\log E)}$ is thus

$$
\frac{dN_{\mbox{\fontsize{8}{14}\selectfont thrown}}}{d(\cos\theta)d(\log E)} = \frac{B~\ln 10}{E} = \frac{N_{\mbox{\fontsize{8}{14}\selectfont thrown}}~\ln 10}{2E}\frac{E_{\mbox{\fontsize{8}{14}\selectfont min}}~E_{\mbox{\fontsize{8}{14}\selectfont max}}}{E_{\mbox{\fontsize{8}{14}\selectfont max}}-E_{\mbox{\fontsize{8}{14}\selectfont min}}}
$$

or

$$
\frac{dN_{\mbox{\fontsize{8}{14}\selectfont thrown}}}{d(\cos\theta)d(\log E)} = \frac{B~\ln 10}{E} = \frac{N_{\mbox{\fontsize{8}{14}\selectfont thrown}}~\ln 10}{2\times 2E}\frac{E_{\mbox{\fontsize{8}{14}\selectfont min}}~E_{\mbox{\fontsize{8}{14}\selectfont max}}}{E_{\mbox{\fontsize{8}{14}\selectfont max}}-E_{\mbox{\fontsize{8}{14}\selectfont min}}}
$$

\noindent We now have all the ingredients that go into the event weight to properly normalize event number expectations for different types of Monte Carlo.

%% file: appendices/appendix4.tex
\chapter{List of Publications}\label{appendix:appendix4}

\noindent M. D'Agostino.  Search for atmospheric neutrino-induced cascades with IceCube.  Prepared for 31st International Cosmic Ray Conference (ICRC 2009), Merida, L\'{o}d\'{z}, Poland, 7-15 Jul 2009.
\\
\\
\noindent E. Middell and M. D'Agostino.  Improved reconstruction of cascade-like events in IceCube.  Prepared for 31st International Cosmic Ray Conference (ICRC 2009), Merida, L\'{o}d\'{z}, Poland, 7-15 Jul 2009.
\\
\\
\noindent Michelangelo D'Agostino.  ``A Physicist At Work'' in \emph{The Physical Universe}, Arthur Beiser and Konrad Krauskopf.  New York: McGraw Hill Higher Education Press, 13th ed. (2009)
\\
\\
\noindent Ignacio Taboada and Michelangelo V. D'Agostino.  Correlating prompt GRB photons with neutrinos.  \emph{astro-ph/0711.2277}, 2007.
\\
\\
\noindent J. Kiryluk, M. V. D'Agostino, S. R. Klein, C. Song, and D. R. Williams.  IceCube performance with artificial light sources: The road to cascade analyses.  Prepared for 30th International Cosmic Ray Conference (ICRC 2007), Merida, Yucatan, Mexico, 3-11 Jul 2007.

\section{IceCube Publications}

\noindent R. Abbasi et al.  Limits on a muon flux from neutralino annihilations in the Sun with the IceCube 22-string detector.  \emph{Phys. Rev. Lett.}, 102:201302, 2009.
\\
\\
\noindent R. Abbasi et al.  Search for point sources of high energy neutrinos with final data from AMANDA-II.  \emph{Phys. Rev.}, D79:062001, 2009.
\\
\\
\noindent R. Abbasi et al.  Determination of the atmospheric neutrino flux and searches for new physics with AMANDA-II.  \emph{Phys. Rev.}, D79:102005, 2009.
\\
\\
\noindent R. Abbasi et al.  The IceCube data acquisition system: Signal capture, digitization, and timestamping.  \emph{Nucl. Instrum. Meth.}, A601:294-316, 2009.
\\
\\
\noindent M. Ackermann et al.  Search for ultra high-energy neutrinos with AMANDA-II.  \emph{Astrophys. J.}, 675:1014, 2008.
\\
\\
\noindent A. Achterberg et al.  Five years of searches for point sources of astrophysical neutrinos with the AMANDA-II neutrino telescope.  \emph{Phys. Rev.}, D75:102001, 2007.
\\
\\
\noindent A. Achterberg et al.  Detection of atmospheric muon neutrinos with the IceCube 9-string detector.  \emph{Phys. Rev.}, D76:027101, 2007.
\\
\\
\noindent A. Achterberg et al.  Multi-year search for a diffuse flux of muon neutrinos with AMANDA-II.  \emph{Phys. Rev.}, D76:042008, 2007.
\\
\\
\noindent A. Achterberg et al.  Search for neutrino-induced cascades from gamma-ray bursts with AMANDA.  \emph{Astrophys. J.}, 664:397, 2007.
\\
\\
\noindent Mathieu Ribordy et al.  From AMANDA to Icecube.  Invited talk at 5th International Conference on Non-accelerator New Physics (NANP 05), Dubna, Russia, 20-25 Jun 2005.  \emph{Phys. Atom. Nucl.}, 69:1899-1907, 2006.
\\
\\
\noindent A. Achterberg et al.  Contributions to 2nd TeV Particle Astrophysics Conference (TeV PA II). Madison, Wisconsin 28-31 Aug 2006.  \emph{J. Phys. Conf. Ser.}  \emph{astro-ph/0611597}.
\\
\\
\noindent A. Achterberg et al.  Limits on the muon flux from neutralino annihilations at the center of the Earth with AMANDA.  \emph{Astropart. Phys.}, 26:129-139, 2006. 
\\
\\
\noindent A. Achterberg et al.  First year performance of the IceCube neutrino telescope.  \emph{Astropart. Phys.}, 26:155-173, 2006.
\\
\\
\noindent A. Achterberg et al.  On the selection of AGN neutrino source candidates for a source stacking analysis with neutrino telescopes.  \emph{Astropart. Phys.}, 26:282-300, 2006.
\\
\\
\noindent A. Achterberg et al.  Limits on the high-energy gamma and neutrino fluxes from the SGR 1806-20 giant flare of December 27th, 2004 with the AMANDA-II detector.  \emph{Phys. Rev. Lett.}, 97:221101, 2006.
\\
\\
\noindent A. Achterberg et al.  Multi-messenger studies with AMANDA/IceCube: observations and strategies.  To appear in the proceedings of 7th Workshop on Towards a Network of Atmospheric Cherenkov Detectors 2005, Palaiseau, France, 27-29 Apr 2005.  \emph{astro-ph/0509396}.
\\
\\
\noindent A. Achterberg et al.  The IceCube collaboration: Contributions to the 29th International Cosmic Ray Conference (ICRC 2005), Pune, India, Aug. 2005.  \emph{astro-ph/0509330}.

\section{Articles For the Popular Press}

\noindent ``Techs and the city'', \emph{The Economist}.  July 27, 2009. 
\\
\\
\noindent ``The electric slide'', \emph{Physical Review Focus}.  June 8, 2009.
\\
\\
\noindent ``The price of prejudice'', \emph{The Economist}.   January 17, 2009.
\\
\\
\noindent ``Blowin' in the wind'', \emph{The Economist}.  November 27, 2008.
\\
\\
\noindent ``Schr\"{o}dinger's drum'', \emph{Physical Review Focus}.  November 11, 2008.
\\
\\
\noindent ``Spot prices'', \emph{The Economist}.  September 17, 2008.
\\
\\
\noindent ``Antimatter bounces off matter'', \emph{Physical Review Focus}.  August 11, 2008.
\\
\\
\noindent ``Thermal explosions on film'', \emph{Physical Review Focus}.  June 9, 2008.
\\
\\
\noindent ``Doctor on call'', \emph{The Economist}.  May 15, 2008.
\\
\\
\noindent ``Wanted: Einstein Jr.'', \emph{The Economist}.  March 8, 2008.
\\
\\
\noindent ``Of ice and men'', \emph{The Economist}.  February 18, 2008.
\\
\\
\noindent ``Snow place like home'', \emph{The Economist}.  January 19, 2008.
\\
\\
\noindent ``A lithium imbalance'', \emph{The Economist}.  June 9, 2007
\\
\\
\noindent ``To coldly go'', \emph{The Economist}.  March 31, 2007.
\\
\\
\noindent ``Concrete possibilities'', \emph{The Economist},  Technology Quarterly.  September 23, 2006.
\\
\\
\noindent ``Accidence and substance'', \emph{The Economist}.  April 6, 2006.
\\
\\
\noindent ``Faster, higher, smarter'', \emph{The Economist}.  February 11, 2006.
\\
\\
\noindent ``ESP for ESP's'', \emph{The Economist}.  December 17, 2005.
\\
\\
\noindent ``Much ado about nothing'', \emph{The Economist}.  October 22, 2005.
\\
\\
\noindent ``Websites of mass description'', \emph{The Economist}, Technology Quarterly.  September 17th, 2005.
\\
\\
\noindent ``Dermatology for 'droids'', \emph{The Economist}.  August 20th, 2005.
\\
\\
\noindent ``The strange way pasta paysÓ, \emph{Marketplace}. National Public Radio. Broadcast 16 August 2005.
\\
\\
\noindent ``Pasta alla fisica'', \emph{The Economist}.  August 13th, 2005.
\\
\\
\noindent ``To be, or not to be'', \emph{The Economist}.  August 6th, 2005.
\\
\\
\noindent ``Rules of engagement'', \emph{The Economist}.  July 23, 2005.
\\
\\
\noindent ``Obituary: Jack Kilby'', \emph{The Economist}.  July 9, 2005.
\\
\\
\noindent ``Dispatches from the void'', \emph{The Economist}.  July 9, 2005.
\\
\\
\noindent ``Star-spangled slammer'', \emph{The Economist}.  July 2, 2005.
\\
\\
\noindent ``Ripples in the sands of time'', \emph{The Economist}.  June 25, 2005.
\\
\\
\noindent ``I'm on the train'', \emph{The Economist}.  June 18, 2005.
\\
\\
\noindent ``Sail of the century'', \emph{The Economist}.  June 18, 2005.
\\
\\
\noindent ``The truth will out'', \emph{The Economist}.  June 11, 2005.
\\
\\
\noindent ``Paying through the nose'', \emph{The Economist}.  June 4, 2005.
\\
\\